\documentclass[11pt,a4paper]{article}

\usepackage{graphicx}
\usepackage{float}
\usepackage{afterpage}
\usepackage{epsfig,cite}
\usepackage{amssymb}
\usepackage{amsmath}
\usepackage{dsfont}
\usepackage{multirow}
\usepackage{url,hyperref}

\usepackage{url}
\usepackage{hyperref}

\newcommand{\be}{\begin{equation}}
\newcommand{\ee}{\end{equation}}
\newcommand{\bea}{\begin{eqnarray}}
\newcommand{\eea}{\end{eqnarray}}
\newcommand{\bi}{\begin{itemize}}
\newcommand{\ei}{\end{itemize}}
\newcommand{\ben}{\begin{enumerate}}
\newcommand{\een}{\end{enumerate}}

\newcommand{\lc}{\left[}
\newcommand{\rc}{\right]}
\newcommand{\lp}{\left(}
\newcommand{\rp}{\right)}

\def\frac#1#2{{{#1}\over {#2}}}
\def\gsim{\mathrel{\rlap{\lower4pt\hbox{\hskip1pt$\sim$}}
    \raise1pt\hbox{$>$}}}         
\def\lsim{\mathrel{\rlap{\lower4pt\hbox{\hskip1pt$\sim$}}
    \raise1pt\hbox{$<$}}}         

\newcommand{\rep}{\mathrm{rep}}

\newcommand{\Mll}{M_{ll}}

\newcommand{\draft}[1]{}

\def\beq{\begin{equation}}  
\def\eeq{\end{equation}}  

\def \n0{N_j^{(0)}}

\def\lapprox{\lower .7ex\hbox{$\;\stackrel{\textstyle <}{\sim}\;$}}
\def\gapprox{\lower .7ex\hbox{$\;\stackrel{\textstyle >}{\sim}\;$}}

\numberwithin{equation}{section}
\numberwithin{figure}{section}
\numberwithin{table}{section}

\begin{document}
\begin{figure}[h]
  \includegraphics[width=0.32\textwidth]{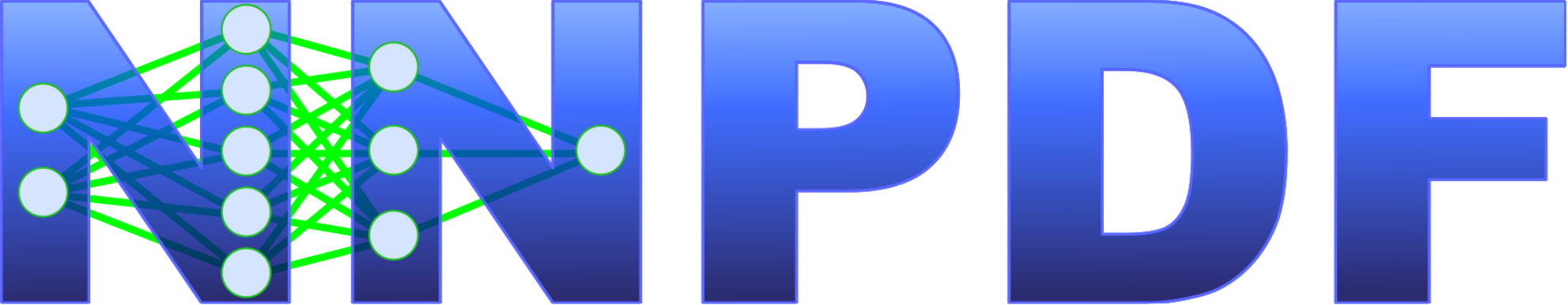}
\end{figure}
\vspace{-2.5cm}
\begin{flushright}
CAVENDISH-HEP-17-06\\
CERN-TH-2017-077\\
Edinburgh 2017/08\\
Nikhef/2017-006\\
OUTP-17-04P\\
TIF-UNIMI-2017-3
\end{flushright}
\vspace{2cm}

\begin{center}
{\Large \bf Parton distributions from high-precision collider data}
\vspace{3.0cm}

{\small
  {\bf  The NNPDF Collaboration:} \\
  Richard~D.~Ball,$^{1}$ Valerio~Bertone,$^{2}$ 
Stefano~Carrazza,$^{3}$ Luigi~Del~Debbio,$^{1}$ 
Stefano~Forte,$^4$ Patrick~Groth-Merrild,$^{1}$  Alberto~Guffanti,$^{5}$
Nathan~P.~Hartland,$^{2}$ Zahari~Kassabov,$^{4,5}$ Jos\'e~I.~Latorre,$^{6,7}$ Emanuele~R.~Nocera,$^{8}$
Juan~Rojo,$^{2}$  Luca~Rottoli,$^{8}$ Emma~Slade,$^{8}$ and Maria~Ubiali$^{9}$}

\vspace{1.0cm}
{\it \small ~$^1$ The Higgs Centre for Theoretical Physics, University of Edinburgh,\\
  JCMB, KB, Mayfield Rd, Edinburgh EH9 3JZ, Scotland\\
~$^2$ Department of Physics and Astronomy, VU University, NL-1081 HV Amsterdam,\\
and Nikhef Theory Group, Science Park 105, 1098 XG Amsterdam, The Netherlands\\
  ~$^3$ Theoretical Physics Department, CERN, CH-1211 Geneva, Switzerland\\
~$^4$ Tif Lab, Dipartimento di Fisica, Universit\`a di Milano and\\
INFN, Sezione di Milano, Via Celoria 16, I-20133 Milano, Italy\\
~$^5$ Dipartimento di Fisica, Universit\`a
di Torino and\\ INFN, Sezione di Torino,
Via P. Giuria 1, I-10125, Turin, Italy\\
~$^6$
Departament de F\'isica Qu\`antica i Astrof\'isica, Universitat de Barcelona,\\ Diagonal 645, 08028 Barcelona, Spain\\
~$^7$ Center for Quantum Technologies, National University of Singapore\\
  ~$^8$ Rudolf Peierls Centre for Theoretical Physics, 1 Keble
Road,\\ University of Oxford, OX1 3NP Oxford, United Kingdom\\
~$^9$ Cavendish Laboratory, HEP group,
University of Cambridge,\\ J.J. Thomson Avenue,
Cambridge CB3 0HE, United Kingdom
}


\clearpage

{\bf \large Abstract}

\end{center}
We present a new set of parton distributions, NNPDF3.1, which updates
NNPDF3.0, the first global set of PDFs determined using a methodology validated 
by a closure test.
The update is motivated by recent progress in methodology
and available data, and involves both. On the methodological side,
we now parametrize and determine 
the charm PDF alongside the light quarks and gluon ones,
thereby increasing from seven to eight the number of independent PDFs.
On the data side, we now include the D0 electron and muon $W$ asymmetries 
from the final Tevatron dataset, the complete
LHCb measurements of $W$ and $Z$ production in the forward
region at 7 and 8 TeV, and new ATLAS and CMS measurements of inclusive
jet and electroweak boson production. We also include
for the first time top-quark pair differential distributions
and the transverse momentum  of the $Z$ bosons from
ATLAS and CMS.
We investigate the impact of parametrizing charm and provide evidence that
the accuracy and stability of the PDFs are thereby improved. We study
the impact of the new data  by producing a variety
of determinations based on reduced datasets.
We find that both improvements have a
significant impact on the PDFs, with some substantial reductions in
uncertainties, but with the new PDFs
generally in agreement with the previous set at the one sigma level.
The most significant changes are seen in the light-quark flavor separation, 
and in increased precision in the determination of the gluon.
We explore the implications of NNPDF3.1 for LHC phenomenology at Run
II, compare with  recent
LHC measurements at 13 TeV, provide updated predictions for Higgs production
cross-sections and discuss the strangeness and charm content of the
proton in light of our improved dataset and methodology. The NNPDF3.1 PDFs
are delivered for the first time both as Hessian sets, and as
optimized Monte
Carlo sets with a compressed number  of replicas.

\clearpage

\tableofcontents

\clearpage

\section{Introduction}
\label{sec:introduction}

A precise understanding of parton
distributions~\cite{Forte:2010dt,Forte:2013wc,Ball:2015oha} (PDFs) has 
played a major role in the discovery of the Higgs boson and will be a 
key ingredient in searches for new physics at the LHC~\cite{Rojo:2015acz}. In recent years a new
generation of PDF sets~\cite{Ball:2014uwa,Dulat:2015mca,Harland-Lang:2014zoa,Alekhin:2017kpj,Abramowicz:2015mha,Jimenez-Delgado:2014twa,Accardi:2016qay} have 
been developed for use at the LHC Run II.
Some of these have been used in the construction of the PDF4LHC15 
combined sets,  
recommended for new physics searches and for the assessment of PDF uncertainties on
precision observables~\cite{Butterworth:2015oua}.  These PDF4LHC15 sets are
obtained by means of statistical combination of the three global
sets~\cite{Ball:2014uwa,Dulat:2015mca,Harland-Lang:2014zoa}: this is
justified by the improved level of agreement in the global
determinations, with differences between 
them largely consistent with statistical fluctuation. 

Despite these developments, there remains a need for improvements in the precision and
reliability of PDF determinations. Precision measurements at the LHC, 
such as in the search for new physics through Higgs
coupling measurements, will eventually require a systematic knowledge of PDFs at the percent 
level in order to fully exploit the LHC's potential. 
The NNPDF3.0 PDF set~\cite{Ball:2014uwa}, which is one of the sets entering the PDF4LHC15
combination, is unique in being a PDF set based on a methodology 
systematically validated by means of closure tests, which ensure the
statistical consistency of the procedure used to extract the PDFs from data.
The goal of this paper is to present NNPDF3.1, an update of the
NNPDF3.0 set, and  
a first step towards PDFs with percent-level uncertainties. Two directions of
progress are required in order to reach
this goal, the motivation for an update being accordingly twofold.

On the one hand, bringing the precision of PDFs down to the percent
level needs  a larger and more precise dataset, with correspondingly 
precise theoretical predictions. In the time since the release of NNPDF3.0, a significant 
number of new experimental measurements have become available.
From the Tevatron, we now have the final measurements of the 
$W$ boson asymmetries with the electron and muon final states 
based upon the complete
Run II dataset~\cite{Abazov:2013rja,D0:2014kma}. At the LHC, 
the ATLAS, CMS and LHCb experiments have released a wide variety
of measurements on inclusive jet production, gauge boson production and 
top production. Finally, the combined legacy measurements of DIS 
structure functions from HERA have also become available~\cite{Abramowicz:2015mha}.
In parallel with the experimental developments, an impressive number of new high-precision 
QCD calculations of hadron collider processes with direct sensitivity to PDFs have
recently been completed, enabling their use in the determination of PDFs 
at NNLO. These include differential distributions in top quark pair
production~\cite{Czakon:2015owf,Czakon:2016dgf}, the transverse 
momentum of the $Z$ and $W$ bosons~\cite{Boughezal:2015dva,Ridder:2015dxa},
and inclusive jet production~\cite{Currie:2013dwa,Currie:2016bfm}, for all
of which precision ATLAS and CMS datasets are available.

All of these new datasets and calculations have been incorporated into
NNPDF3.1. The inclusion of the new data presents new challenges. Given
the large datasets on which some of these measurements are based,
uncorrelated experimental uncertainties are often at the permille
level. Achieving a good fit then requires an unprecedented control of
both correlated systematics and of the numerical accuracy of
theoretical predictions. 

On the other hand, with uncertainties at the percent level, accuracy
issues related to theoretical uncertainties hitherto not included
in PDF determinations become relevant. Whereas the comprehensive inclusion of theoretical
uncertainties in PDF determination will require further study, we
have recently argued that a significant 
source of theoretical bias arises from the conventional assumption 
that charm is generated entirely perturbatively from gluons and light quarks. 
A methodology which allows for the inclusion
of a parametrized 
heavy quark PDFs within the FONLL matched general-mass variable
flavor number scheme has been developed~\cite{Ball:2015tna,Ball:2015dpa}, and 
implemented in an NNPDF PDF determination~\cite{Ball:2016neh}.
It was found that when the charm PDF is parametrized and determined
from the data  alongside the other PDFs, 
much of the uncertainty related to the value of the charm mass becomes part of
the standard PDF uncertainty, while any bias related to the assumption 
that the charm PDF is purely perturbative is eliminated~\cite{Ball:2016neh}.
In NNPDF3.1 charm is therefore parametrized as an independent PDF, 
in an equivalent manner to light quarks and the gluon. 
We will show that this leads to improvements in fit quality without an increase
in uncertainty, and that it stabilises the dependence of PDFs on the charm mass, 
all but removing it in the light quark PDFs. 

The NNPDF3.1 PDf sets are released at LO, NLO, and NNLO accuracy. For the first
time, all NLO and NNLO 
PDFs are delivered both as Hessian sets and as Monte Carlo
replicas, exploiting recent powerful methods for the construction of
optimal Hessian representations of 
PDFs~\cite{Carrazza:2015aoa}. Furthermore, and also 
for the first time, the default
PDF sets are provided as compressed Monte Carlo
sets~\cite{Carrazza:2015hva}. Therefore despite being presented
as sets of only 100 Monte Carlo replicas, they exhibit many of
the statistical properties of a much larger set, 
reducing observable computation time without loss of
information. A further improvement in computational efficiency can be
obtained by means of the {\tt SM-PDF} tool~\cite{Carrazza:2016htc}, which
allows for the selection of optimal subsets of Hessian eigenvectors
for the computation of uncertainties on specific processes or classes
of processes, and which is available as a web
interface~\cite{Carrazza:2016wte} now  also including the NNPDF3.1 sets.
A variety
of PDF sets based on subsets of data are also provided (as standard
100 replica Monte Carlo sets), which may be
useful for specific applications such as new physics searches, or measurements of
standard model parameters.

The outline of this paper is as follows.
First, in Sect.~\ref{sec:expdata} we discuss the
experimental aspects and the relevant theoretical
issues of the new datasets. 
We then turn in Sect.~\ref{sec:results} to a detailed
description of the baseline NNPDF3.1 PDF sets, with a specific discussion of 
the impact of methodological improvements, specifically the fact that
the charm PDF is now independently parametrized and determined like
all other PDFs. In
Sect.~\ref{sec:impactnewdata} we discuss the impact of the
new data by comparing PDF sets based upon various data subsets, and also
discuss PDF sets based on more conservative data subsets.
In Sect.~\ref{sec:pheno} we summarise the status of uncertainties on PDFs
and luminosities, and
specifically discuss the strange and charm content of the proton in
light of our results, and  present
first phenomenological studies at the LHC.
Finally, a summary of the PDFs being delivered in various formats  is
provided in 
Sect.~\ref{sec:conclusion}, together with links to repositories whence
more detailed sets  of plots may be downloaded.


%
\section{Experimental and theoretical input}
\label{sec:expdata}

The NNPDF3.1 PDF sets include a wealth of new experimental data.  We
have augmented our dataset with improved determinations of observables
already included in NNPDF 3.0 (such as $W$ and $Z$ rapidity
distributions) as well as two new process: top quark differential
distributions, and the $Z$ transverse momentum distribution, which is
include for the first time in a global PDF determination.

In this Section we discuss the NNPDF3.1 dataset in detail. After a
general overview, each observable 
will be examined: we describe the individual measurements, and 
address specific theoretical and phenomenological issues related to
their inclusion, particularly in relation to the use of recent NNLO results.

In NNPDF3.1 only LHC data from Run I, taken at centre-of-mass
energies of 2.76~TeV, 7~TeV and 8~TeV (with one single exception), are included. The more recent 13~TeV
dataset is reserved for phenomenological comparison purposes in Sect.~\ref{sec:pheno}.
Available and upcoming LHC Run II data at 13~TeV will be part of future NNPDF releases.

\subsection{Experimental data: general overview}

The NNPDF3.0 global analysis involved data from deep-inelastic scattering (DIS) experiments, fixed-target Drell-Yan data, and
collider measurements from the Tevatron and LHC.
The fixed-target and collider DIS datasets included measurements from NMC~\cite{Arneodo:1996kd,Arneodo:1996qe}, 
BCDMS~\cite{bcdms1,bcdms2} and SLAC~\cite{Whitlow:1991uw}; the combined HERA-I 
inclusive structure function dataset~\cite{Aaron:2009aa} and HERA-II inclusive measurements
from H1 and ZEUS~\cite{Aaron:2012qi,Collaboration:2010ry,ZEUS:2012bx,Collaboration:2010xc};
the HERA combined measurements of the charm production cross-section $\sigma_{c}^{\rm NC}$~\cite{Abramowicz:1900rp};
CHORUS inclusive neutrino DIS~\cite{Onengut:2005kv}, and NuTeV dimuon production 
data~\cite{Goncharov:2001qe,MasonPhD}.
From the Tevatron, CDF~\cite{Aaltonen:2010zza} and D0~\cite{Abazov:2007jy} $Z$ rapidity distributions; and
CDF~\cite{Aaltonen:2008eq} Run-II one-jet inclusive cross-sections were used. Constraints from fixed-target Drell-Yan came from the 
E605~\cite{Moreno:1990sf} and E866~\cite{Webb:2003ps,Webb:2003bj,Towell:2001nh} experiments. 
LHC measurements included electroweak boson production data from ATLAS~\cite{Aad:2011dm,Aad:2013iua,Aad:2011fp}, 
CMS~\cite{Chatrchyan:2012xt,Chatrchyan:2013mza,Chatrchyan:2013tia} and LHCb~\cite{Aaij:2012vn,Aaij:2012mda};
one-jet inclusive cross-sections from ATLAS~\cite{Aad:2011fc,Aad:2013lpa} and CMS~\cite{Chatrchyan:2012bja};
the differential distributions for $W$ production in association with charm quarks from CMS~\cite{Chatrchyan:2013uja};
and total cross-section measurements for top quark pair production data from ATLAS and CMS at 7 and 8~TeV~\cite{ATLAS:2012aa,ATLAS:2011xha,TheATLAScollaboration:2013dja, Chatrchyan:2013faa,Chatrchyan:2012bra,Chatrchyan:2012ria}.

For NNPDF3.1 we have made a number of improvements to the NNPDF3.0
dataset. Firstly we  have included the final
datasets for several experiments which have now concluded, replacing superseded data in the NNPDF3.0 analysis.
The HERA-I data and the H1 and ZEUS HERA-II inclusive structure functions have been replaced by the final 
HERA combination~\cite{Abramowicz:2015mha}.
The HERA dataset has also been enlarged by the inclusion of H1 and ZEUS measurements of the 
bottom structure function $F_2^b(x,Q^2)$~\cite{Aaron:2009af,Abramowicz:2014zub}, which may prove useful in 
specific applications such as in the determination of the bottom quark mass $m_b$.
In order to perform dedicated studies of the charm content of the
proton, we have constructed a PDF set also 
including the EMC measurements of charm structure functions at
large-$x$~\cite{Aubert:1982tt}, which 
will be discussed in Sect.~\ref{sec:phenocharm}. However, these measurements are not included in the standard dataset.
The legacy $W$ lepton asymmetries from D0 using the complete Tevatron luminosity, both in the electron~\cite{D0:2014kma}
and in the muon~\cite{Abazov:2013rja} channels have been added.
These precise weak gauge boson production measurements provide important information on the quark flavor separation at large-$x$, 
as demonstrated in~\cite{Camarda:2015zba}.

Aside from the updated legacy datsets, in NNPDF3.1 a large number of recent measurements from ATLAS, CMS and LHCb are included.
For ATLAS, we now include the $Z$ boson $(p_T^Z,y_Z)$ and $(p_T^Z,M_{ll})$ double differential distributions measured at 8~TeV~\cite{Aad:2015auj}; the inclusive $W^+$, $W^-$ and $Z$ rapidity distributions at 
7~TeV from the 2011 dataset~\cite{Aaboud:2016btc}, the top-quark pair production normalized $y_t$ distribution at 8~TeV~\cite{Aad:2015mbv}; total cross-sections for top quark pair production at 7, 8 and 13~TeV~\cite{Aad:2014kva,Aaboud:2016pbd}; inclusive jet cross-sections at 7~TeV from the 2011 dataset~\cite{Aad:2014vwa}; and 
finally low mass Drell-Yan $M_{ll}$ distributions at 7~TeV from the 2010 run~\cite{Aad:2014qja}. The transverse momentum 
spectrum at 7~TeV (2011 dataset)~\cite{Aad:2014xaa} will be studied in Sec.~\ref{sec:impactzpt} but it is not included in the default 
set. The total top cross-section is the only data point at 13~TeV which
is included.
For CMS, NNPDF3.1 includes the $W^+$ and $W^-$ rapidity distributions at 8~TeV~\cite{Khachatryan:2016pev},
together with their cross-correlations; the inclusive jet production
cross-sections at 2.76~TeV~\cite{Khachatryan:2015luy};
top-quark pair production normalized $y_{t\bar{t}}$ distributions at 8~TeV~\cite{Khachatryan:2015oqa}, total inclusive $t\bar{t}$ cross-sections at 7, 8
and 13~TeV~\cite{Khachatryan:2016mqs}; the distribution of the
$Z$ boson double differentially in $(p_T,y_Z)$ at 8~TeV~\cite{Khachatryan:2015oaa}. The double-differential
distributions $(y_{ll},M_{ll})$ in Drell-Yan production at 8~TeV~\cite{CMS:2014jea} will be studied in
Sect.~\ref{sec:cms8tevimpact} below, but it is not included in the default
PDF determination.
For LHCb, NNPDF3.1 includes the
complete 7 and 8~TeV measurements of inclusive $W$ and $Z$
production in the muon
channel~\cite{Aaij:2015gna,Aaij:2015zlq}, which supersedes all previous
measurements in the same final state.

An overview of the data included in NNPDF3.1 is presented in
Tables~\ref{tab:completedataset},~\ref{tab:completedataset2},
and~\ref{tab:completedataset3}, for the DIS structure function data,
the fixed target and Tevatron Drell-Yan experiments, and the LHC
datasets, respectively.
For each dataset we indicate the corresponding published reference, the number
of data points in the NLO/NNLO PDF determinations 
before and after (in parenthesis) kinematic cuts, the kinematic
range covered in the relevant variables after cuts, and the code used
to compute the NLO and NNLO results.
Datasets included 
for the first time in NNPDF3.1 are flagged with an asterisk. The
datasets not used for the default determination are in brackets. 
The total number of data points for the default
PDF determination 
is $4175/4295/4285$ at LO/NLO/NNLO.


\begin{table}
\footnotesize
\begin{centering}
      \renewcommand{\arraystretch}{1.4}
  \begin{tabular}{|c|c|c|c|c|c|c|}
\hline
{Experiment} & Obs. & {Ref.}  & $ N_{\rm \bf dat}$ & $x$ range  & $Q$ range (GeV)  & Theory \tabularnewline
\hline
\hline
\multirow{2}{*}{NMC}
&   $F_2^d/F_2^p$ & \cite{Arneodo:1996kd}&   260 (121/121) & $ 0.012\le x \le 0.68$
& $ 2.1 \le Q \le 10 $  &\multirow{2}{*}{\tt APFEL}\tabularnewline
&   $\sigma^{\rm NC,p}$ & \cite{Arneodo:1996qe} & 292 (204/204) &
$0.012 \le x \le 0.50$ & $ 1.8 \le Q \le 7.9 $  &  \tabularnewline
\hline
\multirow{2}{*}{SLAC} &   $F_2^p$ & \cite{Whitlow:1991uw} &  211 (33/33)
& $ 0.14\le x \le 0.55$  &
$ 1.9 \le Q \le 4.4 $  & \multirow{2}{*}{\tt APFEL}\tabularnewline
&   $F_2^d$ & \cite{Whitlow:1991uw}& 211 (34/34) &
$ 0.14 \le x \le 0.55 $ & $ 1.9\le Q \le 4.4 $  & \tabularnewline
\hline
\multirow{2}{*}{BCDMS} &   $F_2^p$ & \cite{bcdms1}&  351 (333/333) & $0.07 \le x \le 0.75$ & $
2.7\le Q \le 15.1 $  &
\multirow{2}{*}{\tt APFEL}\tabularnewline
&   $F_2^d$ & \cite{bcdms2}&  254 (248/248) & $0.07 \le x \le 0.75$ & $ 3.0\le  Q \le
15.1$  & \tabularnewline
\hline
\multirow{2}{*}{CHORUS} &  $\sigma^{\rm CC,\nu}$  & \cite{Onengut:2005kv} & 607 (416/416) &
$0.045 \le x \le 0.65 $  &
$ 1.9\le Q \le 9.8 $  &
\multirow{2}{*}{\tt APFEL} \tabularnewline
&   $\sigma^{\rm CC,\bar{\nu}}$ & \cite{Onengut:2005kv} & 607 (416/416)  & $0.045 \le x \le 0.65$  &
$ 1.9 \le Q \le 9.8 $  & \tabularnewline
\hline
\multirow{2}{*}{NuTeV} &    $\sigma_\nu^{cc}$ & \cite{Goncharov:2001qe,MasonPhD} & 45 (39/39)
&$ 0.02 \le x \le 0.33 $ &
$ 2.0 \le Q \le 10.8  $ 
& \multirow{2}{*}{\tt APFEL} \tabularnewline
&   $\sigma_{\bar{\nu}}^{cc}$ & \cite{Goncharov:2001qe,MasonPhD} & 45 (37/37) &
$ 0.02 \le x \le 0.21$&
$ 1.9 \le Q \le 8.3 $ 
&\tabularnewline
\hline
\multirow{3}{*}{HERA}  &  $\sigma_{\rm NC,CC}^{p}$ {\bf (*)} & \cite{Abramowicz:2015mha}
&  1306 (1145/1145)  & $4\cdot 10^{-5}\le x \le 0.65$ & $ 1.87 \le Q \le 223 $   & \multirow{3}{*}{\tt APFEL}
 \tabularnewline
  & $\sigma_{\rm NC}^{c}$    & \cite{Abramowicz:1900rp} & 52 (47/37) & $7 \cdot 10^{-5} \le x \le 0.05$  & $2.2 \le Q \le 45$  &
 \tabularnewline
 &  $F_2^b$   {\bf (*)} & \cite{Aaron:2009af,Abramowicz:2014zub} & 29 (29/29) & $ 2\cdot 10^{-4}\le x \le
 0.5 $ & $ 2.2 \le Q \le 45$  &
 \tabularnewline
 \hline
 EMC
 &   [ $F_2^c$ ] {\bf (*)}  & \cite{Aubert:1982tt} & 21 (16/16) &$0.014\le x \le 0.44$  & $2.1 \le Q \le 8.8 $ & \tt APFEL
 \tabularnewline
  \hline
\end{tabular}
\par\end{centering}
\caption{\small Deep-inelastic scattering
data included in  NNPDF3.1. The EMC $F_2^c$ data are in brackets because
they are only included in a dedicated set but not in the default dataset.
New datasets, not included in NNPDF3.0, are
denoted {\bf (*)}. 
The kinematic range covered in each variable is given after cuts are applied. 
The total number of DIS data points after cuts is 
$3102/3092$ for the NLO/NNLO PDF determinations (not including the EMC $F_2^c$ data).
\label{tab:completedataset}
}
\end{table}

\begin{table}
\footnotesize
\begin{centering}
    \renewcommand{\arraystretch}{1.4}
\begin{tabular}{|c|c|c|c|c|c|c|}
\hline
{Exp.} & Obs. & {Ref.}  & $ N_{\rm \bf dat}$ & Kin$_1$  & Kin$_2$ (GeV) & Theory \tabularnewline
\hline
\hline
\multirow{2}{*}{E866} &   $\sigma_{\rm DY}^d/\sigma_{\rm DY}^p$ & \cite{Towell:2001nh} & 15 (15/15)
& 0.07 $\le y_{ll}\le 1.53$ &
$ 4.6 \le M_{ll}\le 12.9 $  & {\tt APFEL+Vrap}\tabularnewline
&  $\sigma_{\rm DY}^p$ & \cite{Webb:2003ps,Webb:2003bj} & 184 (89/89) & $0 \le y_{ll}\le 1.36 $ &
$  4.5 \le M_{ll}\le 8.5 $  &{\tt APFEL+Vrap} \tabularnewline
 \hline
 E605 & $\sigma_{\rm DY}^p$ &  \cite{Moreno:1990sf} & 119 (85/85) & $ -0.2 \le y_{ll}\le 0.4$ &
 $ 7.1 \le M_{ll}\le 10.9 $  & {\tt APFEL+Vrap} \tabularnewline
 \hline
\multirow{2}{*}{CDF} &   $d\sigma_Z/dy_Z$ & \cite{Aaltonen:2010zza}&  29 (29/29)  & $ 0 \le y_{ll}\le 2.9 $  & $66 \le M_{ll}\le 116$   &{\tt Sherpa+Vrap} \tabularnewline
&  $k_t$ incl jets & \cite{Abulencia:2007ez} & 76 (76/76) & $ 0\le y_{\rm jet}\le 1.9$  &
 $ 58 \le p_T^{\rm jet}\le 613$  &{\tt NLOjet++} \tabularnewline
\hline
\multirow{3}{*}{D0}   &   $d\sigma_Z/dy_Z$ &  \cite{Abazov:2007jy}  & 28 (28/28) & $0 \le y_{ll}\le 2.8 $ &
 $66 \le M_{ll}\le 116$ & {\tt Sherpa+Vrap}\tabularnewline
&   $W$ electron asy {\bf (*)}  & \cite{D0:2014kma}  & 13  (13/8) &  $0 \le y_{e}\le 2.9 $& $Q=M_W$  &{\tt MCFM}+{\tt FEWZ}\tabularnewline
 &   $W$ muon asy {\bf (*)}  & \cite{Abazov:2013rja}  &  10 (10/9)   & $0 \le y_{\mu}\le 1.9 $ & $Q=M_W$ & {\tt MCFM}+{\tt FEWZ}\tabularnewline
  \hline
\end{tabular}
\par\end{centering}
\caption{\small Same as Table~\ref{tab:completedataset}
  for the Tevatron fixed-target Drell-Yan  and $W$, $Z$ and jet collider
  data. The total number of Tevatron data points after cuts is $345/339$ 
for NLO/NNLO fits.
\label{tab:completedataset2}
}
\end{table}

\begin{table}[ht]
\scriptsize
\begin{centering}
   \renewcommand{\arraystretch}{1.4}
\begin{tabular}{|c|c|c|c|c|c|c|}
  \hline
  {Exp.} & Obs. & {Ref.}  & $ N_{\rm \bf dat}$  & Kin$_1$  &  Kin$_2$ (GeV) & Theory \tabularnewline
\hline
\hline
\multirow{12}{*}{ATLAS} &   $W,Z$ 2010 & \cite{Aad:2011dm}  &  30 (30/30) & $ 0 \le |\eta_l| \le 3.2$ & $Q=M_W, M_Z$ &
{\tt MCFM}+{\tt FEWZ}\tabularnewline
&   $W,Z$ 2011 {\bf (*)}  & \cite{Aaboud:2016btc}  & 34 (34/34)   & $0 \le |\eta_l| \le 2.3$   & $Q=M_W, M_Z$ &
{\tt MCFM}+{\tt FEWZ}\tabularnewline
&    high-mass DY 2011 & \cite{Aad:2013iua}  & 11 (5/5) & $0 \le |\eta_l| \le 2.1$ & $ 116 \le M_{ll} \le 1500$  &
{\tt MCFM}+{\tt FEWZ} \tabularnewline
&    low-mass DY 2011 {\bf (*)}   & \cite{Aad:2014qja}  & 6 (4/6)  & $0 \le  |\eta_l| \le 2.1$  & $  14 \le M_{ll} \le 56  $  &
{\tt MCFM}+{\tt FEWZ}
\tabularnewline
&    [$Z$ $p_T$ 7 TeV $\lp p_T^Z,y_Z\rp$]  {\bf (*)}  & \cite{Aad:2014xaa}  & 64 (39/39)  &
$0 \le |y_{Z}| \le 2.5$  &  $30 \le p_T^Z \le 300$   & {\tt MCFM}+NNLO
\tabularnewline
&    $Z$ $p_T$ 8 TeV $\lp p_T^Z,M_{ll}\rp$  {\bf (*)}  & \cite{Aad:2015auj}  & 64 (44/44)  &  $  12 \le M_{ll} \le
150 $  GeV & $30 \le p_T^Z \le 900$  & {\tt MCFM}+NNLO
      \tabularnewline
      &    $Z$ $p_T$ 8 TeV $\lp p_T^Z,y_Z\rp$  {\bf (*)}  & \cite{Aad:2015auj}  & 120 (48/48)  & $   0.0\le |y_{Z}| \le 2.4  $  
        & $   30\le p_{T}^Z \le 150  $  &
          {\tt MCFM}+NNLO
            \tabularnewline
&    7 TeV jets 2010 & \cite{Aad:2011fc}  & 90 (90/90) &
$0 \le |y^{\rm jet}| \le 4.4 $ & $ 25 \le p_T^{\rm jet} \le 1350 $&
{\tt NLOjet++}
\tabularnewline
&      2.76 TeV jets  &  \cite{Aad:2013lpa}  &
59 (59/59) & $0 \le |y^{\rm jet}| \le 4.4$ & $20 \le p_T^{\rm jet} \le 200$ & {\tt NLOjet++}
\tabularnewline
&      7 TeV jets 2011 {\bf (*)}  &  \cite{Aad:2014vwa}  & 140 (31/31)
  & $ 0 \le |y^{\rm jet}| \le 0.5  $ & $108 \le p_T^{\rm jet} \le 1760 $ & {\tt NLOjet++}
\tabularnewline
  &    $\sigma_{\rm tot}(t\bar{t})$  & \cite{Aad:2014kva,Aaboud:2016pbd}& 
3 (3/3) & - & $Q=m_t$ & {\tt top++} \tabularnewline
 &    $(1/\sigma_{t\bar{t}})d\sigma(t\bar{t})/y_t$  {\bf (*)}   & \cite{Aad:2015mbv}& 
10 (10/10)  & $0<|y_t|<2.5$ & $Q=m_t$ & {\tt Sherpa}+NNLO \tabularnewline
 \hline
 \hline
\multirow{12}{*}{CMS} &   $W$ electron asy & \cite{Chatrchyan:2012xt}  &  11 (11/11) &
$0 \le |\eta_{\rm e}| \le 2.4  $ & $Q=M_W$  & {\tt MCFM}+{\tt FEWZ} \tabularnewline
 &   $W$ muon asy  & \cite{Chatrchyan:2013mza}&   11 (11/11) & 
$0 \le |\eta_{\mu}| \le 2.4  $  & $Q=M_W$ & {\tt MCFM}+{\tt FEWZ} \tabularnewline
&    $W+c$ total  & \cite{Chatrchyan:2013uja}& 5 (5/0) &  $0 \le |\eta_l| \le 2.1$ &
$Q=M_W$  & {\tt MCFM}
\tabularnewline
&    $W+c$ ratio  & \cite{Chatrchyan:2013uja}&  5 (5/0) & $0 \le |\eta_l| \le 2.1$ &
$Q=M_W$
& {\tt MCFM}
\tabularnewline
&   2D DY 2011 7 TeV    & \cite{Chatrchyan:2013tia} &   124 (88/110) & $0 \le |\eta_{ll}| \le 2.2$  &
$20 \le M_{ll} \le 200$  &  {\tt MCFM}+{\tt FEWZ}
\tabularnewline
&   [2D DY 2012 8 TeV]     & \cite{CMS:2014jea} &   124 (108/108) & $0 \le |\eta_{ll}| \le 2.4$  &
$20 \le M_{ll} \le 1200$  &  {\tt MCFM}+{\tt FEWZ}
\tabularnewline
&   $W^{\pm}$ rap 8 TeV  {\bf (*)}     & \cite{Khachatryan:2016pev} &  22 (22/22)   & $0 \le |\eta_{l}| \le 2.3$  &
$Q=M_W$  &  {\tt MCFM}+{\tt FEWZ}
\tabularnewline
&   $Z$ $p_T$ 8 TeV  {\bf (*)}     & \cite{Khachatryan:2015oaa} & 50 (28/28)    & $ 0.0 \le |y_{Z}| \le 1.6 $  &
$  30 \le p_T^Z \le 170 $   &  {\tt MCFM}+NNLO
\tabularnewline
 &   7 TeV jets 2011     & \cite{Chatrchyan:2012bja}&  133 (133/133) &  
$0 \le |y^{\rm jet}| \le 2.5$ & $114 \le p_T^{\rm jet} \le 2116$   &
{\tt NLOjet++}
\tabularnewline
 &   2.76 TeV jets  {\bf (*)}     & \cite{Khachatryan:2015luy}& 81 (81/81)   &  $0 \le |y_{\rm jet}| \le 2.8$
  & $80 \le p_T^{\rm jet} \le$ 570   & {\tt NLOjet++}
 \tabularnewline
&    $\sigma_{\rm tot}(t\bar{t})$   & \cite{Khachatryan:2016mqs,CMS:2016syx}& 
 3 (3/3) & - & $Q=m_t$ & {\tt top++} \tabularnewline
 &     $(1/\sigma_{t\bar{t}})d\sigma(t\bar{t})/y_{t\bar{t}}$ {\bf (*)}   & \cite{Khachatryan:2015oqa}& 
 10 (10/10)  & $-2.1<y_{t\bar{t}}<2.1$  & $Q=m_t$ & {\tt Sherpa}+NNLO \tabularnewline
 \hline
 \hline
\multirow{4}{*}{LHCb} &    $Z$ rapidity 940 pb  & \cite{Aaij:2012vn} &   9 (9/9) &
$2.0 \le \eta_l \le 4.5$& $Q=M_Z$ &
{\tt MCFM}+{\tt FEWZ} \tabularnewline
&    $Z\to ee$ rapidity 2 fb & \cite{Aaij:2012mda}&  17 (17/17) & $2.0 \le \eta_l \le 4.5$ & $Q=M_Z$  & {\tt MCFM}+{\tt FEWZ}
\tabularnewline
&    $W,Z\to \mu$ 7 TeV {\bf (*)}  & \cite{Aaij:2015gna} & 33 (33/29)  & $2.0 \le \eta_l \le 4.5$
& $Q=M_W,M_Z$ &  {\tt MCFM}+{\tt FEWZ}\tabularnewline
&    $W,Z\to \mu$ 8 TeV {\bf (*)}  & \cite{Aaij:2015zlq}  & 34 (34/30) & $2.0 \le \eta_l \le 4.5$
& $Q=M_W,M_Z$ &  {\tt MCFM}+{\tt FEWZ}  \tabularnewline
 \hline
 \end{tabular}
\par\end{centering}
\caption{\small
Same as Table~\ref{tab:completedataset}, for
ATLAS, CMS and LHCb data from the LHC Run I at
$\sqrt{s}=2.76$~TeV,  $\sqrt{s}=7$~TeV and
$\sqrt{s}=8$~TeV. The ATLAS $7$~TeV $Z$ $p_T$ and CMS 2D DY 2012 are
in  brackets because they are only included in a dedicated study
but not in the default PDF set.
The total number of LHC data points after cuts is $848/854$ 
for NLO/NNLO fits (not including ATLAS $7$~TeV $Z$ $p_T$ and CMS 2D DY 2012).
\label{tab:completedataset3}
}
\end{table}

\begin{figure}[t]
\centering
\epsfig{width=0.80\textwidth,figure=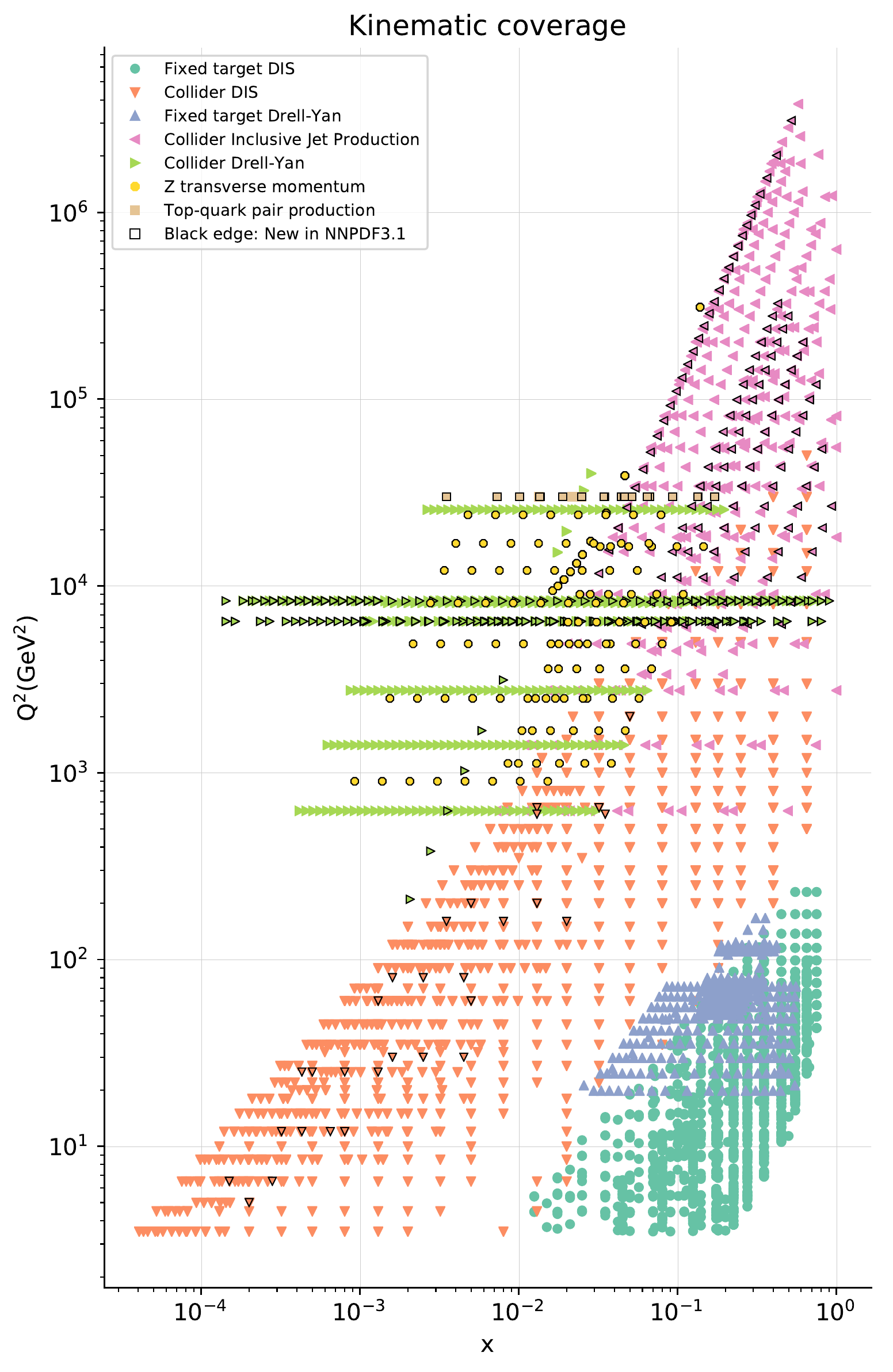}
\caption{\small 
  The kinematic coverage of the NNPDF3.1 dataset in the $\lp x,Q^2\rp$
  plane.
 \label{fig:kin31}} 
\end{figure}

In Fig.~\ref{fig:kin31} we show
the kinematic coverage of the NNPDF3.1 dataset in the $\lp x,Q^2\rp$ plane.
For hadronic data, leading-order kinematics have been assumed for
illustrative purposes, with central rapidity used when rapidity is
integrated over and the plotted value of $Q^2$ set equal to the
factorization scale. It is clear that the new data added in NNPDF3.1
are distributed in a wide range of scales and $x$, 
considerably extending the kinematic reach and coverage of the dataset.

In Table~\ref{eq:tablekincuts} we present
a summary of the kinematic cuts applied to the various processes
included in NNPDF3.1 at NLO and NNLO.
These cuts ensure that only data where theoretical calculations are reliable are
included. Specifically, we always remove from the NLO dataset points
for which the NNLO corrections exceed the statistical
uncertainty. The further cuts collected in
Table~\ref{eq:tablekincuts}, specific to individual datasets,  will be 
described
when discussing each dataset in turn. All computations are performed up
to NNLO in QCD, not including electroweak corrections. We have
checked that with the cuts described in Table~\ref{eq:tablekincuts},
electroweak corrections 
never exceed experimental uncertainties. 

The codes used to perform NLO computations will be discussed in each
subsection below. With the exception of deep-inelastic scattering,
NNLO corrections are implemented by computing at the hadron level the
bin-by-bin ratio of 
the NNLO to NLO prediction with a pre-defined PDF set, and applying
the correction to the NLO computation (see Sect.~2.3 of
Ref.~\cite{Ball:2014uwa}). For all new data included in NNPDF3.1,
the PDF set used for the computation of these
correction factors (often refereed to as $K$-factors, and in
Ref.~\cite{Ball:2014uwa} as $C$-factors) is NNPDF3.0, except for the
CMS~W~rap~8~TeV and ATLAS~W/Z~2011 entries of
Tab.~\ref{tab:completedataset3} for which published {\tt xFitter}
results have been used and
the CMS~2D~DY~2012 data for which MMHT PDFs have been
used~\cite{mmhtpriv} 
(see Sect.~\ref{data:inclusiveWZ} below); the
PDF dependence of the correction factors is much smaller than all
other relevant uncertainties as we will demonstrate explicitly in
Sect.~\ref{sec:topdata} below.

\begin{table}[t]
  \begin{center}
    \footnotesize
    \renewcommand{\arraystretch}{1.7}
    \begin{tabular}{|l|c|c|}
\hline
Dataset & NLO & NNLO \\
\hline
\hline
\multirow{2}{*}{DIS structure functions}  & $W^2\ge 12.5$ GeV$^2$ & $W^2\ge 12.5$ GeV$^2$ \\
& $Q^2\ge 3.5$ GeV$^2$ & $Q^2\ge 3.5$ GeV$^2$ \\
\hline
HERA $\sigma_c^{\rm NC}$ (in addition)  &  - &   $Q^2 \ge 8$ GeV$^2$ (fitted charm) \\
\hline
\hline
ATLAS 7 TeV inclusive jets 2011  & $|y_{\rm jet}|\le 0.4 $ &
$|y_{\rm jet}|\le 0.4 $ \\
\hline
\hline
\multirow{2}{*}{Drell-Yan E605 and E866}  & $\tau \le 0.080$ &  $\tau \le 0.080$ \\
& $|y/y_{\rm max}|\le 0.663$ &  $|y/y_{\rm max}|\le 0.663$  \\
\hline
D0 $W \to l\nu$ asymmetries  & - & $|A_{l}|\ge 0.03 $ \\
\hline
\multirow{2}{*}{CMS Drell-Yan 2D 7 TeV}
& 30 GeV $\le M_{ll}\le 200$ GeV &  $ M_{ll}\le 200$ GeV   \\
&  $|y_Z|\le 2.2$  &  $|y_Z|\le 2.2$  \\
\hline
\multirow{1}{*}{[CMS Drell-Yan 2D  8 TeV]}   &  $ M_{ll}\ge 30$ GeV &  $ M_{ll}\ge 30$ GeV   \\
\hline
LHCb 7 TeV and 8 TeV $W,Z\to \mu$  & - & $|y_{l}|\ge 2.25 $ \\
\hline
\hline
[ATLAS $Z$ $p_T$ 7 TeV]   & $30~{\rm GeV}\le p_T^Z \le 500~{\rm GeV}$ &
$30~{\rm GeV}\le p_T^Z \le 500~{\rm GeV}$ \\
\hline
ATLAS $Z$ $p_T$ 8 TeV $(p_T,M_{ll})$    & $ p_T^Z\ge30~{\rm GeV}$ &
$ p_T^Z\ge30~{\rm GeV}$ \\
\hline
ATLAS $Z$ $p_T$ 8 TeV $(p_T,y_{Z})$    & $30~{\rm GeV}\le p_T^Z \le 150~{\rm GeV}$ &
$30~{\rm GeV}\le p_T^Z \le 150~{\rm GeV}$ \\
\hline
\multirow{2}{*}{CMS $Z$ $p_T$ 8 TeV $(p_T,y_{Z})$}   & $30~{\rm GeV}\le p_T^Z \le 170~{\rm GeV}$ &
$30~{\rm GeV}\le p_T^Z \le 170~{\rm GeV}$ \\
& $|y_Z| \le 1.6$ &   $|y_Z| \le 1.6$ \\
\hline
    \end{tabular}
    \caption{\small 
Full set of kinematical cuts applied to the
      processes
      used for  NNPDF3.1 PDF determination at NLO and at NNLO.
     Only data satisfying the constraints in the table are retained.
      The experiments in brackets are not part of the global dataset and only
      used for dedicated studies.
      The cut on the HERA charm structure function data at NNLO
      is applied only when charm is fitted, and it is applied
      in addition to the other DIS kinematical cuts.
      \label{eq:tablekincuts}
}
\end{center}
\end{table}

\clearpage

\subsection{Deep-inelastic structure functions}
\label{sec:datadis}

The main difference between the NNPDF 3.0 and 3.1 DIS structure function 
datasets is the replacement of the separate HERA-I and ZEUS/H1 HERA-II inclusive 
structure function measurements by the final
legacy HERA combination~\cite{Abramowicz:2015mha}. The impact of the HERA-II data on a global fit which includes HERA-I
data is
known~\cite{Ball:2014uwa,Thorne:2015caa,Hou:2016nqm,Harland-Lang:2016yfn} to be moderate to
begin with; the further impact of replacing the separate HERA-I and
HERA-II data used in NNPDF3.0 with their combination has been studied
in~\cite{Rojo:2015nxa} and found to be completely negligible.

Additionally, the NNPDF3.1 dataset includes the H1 and ZEUS measurements of
the bottom structure function
$F_2^b(x,Q^2)$~\cite{Aaron:2009af,Abramowicz:2014zub}.
While the $F_2^b$ dataset is known to have a very limited pull,
the inclusion of this dataset is useful for applications, such as the determination of
the bottom mass~\cite{Harland-Lang:2015qea}.

While it is not included in the default NNPDF3.1 dataset, the EMC data on charm
structure functions~\cite{Aubert:1982tt} will also be used for specific
studies of the charm content of the proton in Sect.~\ref{sec:phenocharm}.
As discussed in Refs.~\cite{Ball:2016neh,Rottoli:2016lsg}, the EMC
dataset has been corrected by updating the BR$(D\to \mu)$ branching  
ratio: the value used in the original analysis~\cite{Aubert:1982tt} 
is replaced with the latest PDG value~\cite{Olive:2016xmw}. A conservative
uncertainty on this branching ratio of $\pm 15\%$ is also included.

The cuts applied to DIS data are as follows. As in NNPDF3.0, for
all structure function datasets we exclude
data with $Q^2 < 3.5$ GeV$^2$
and $W^2< 12.5$ GeV$^2$, i.e.
the region where higher twist corrections
might become relevant and the perturbative expansion may become
unreliable. 
At NNLO we also remove $F_2^c$ data with 
$Q^2 < 8$ GeV$^2$ in order to minimize the possible impact of unknown
NNLO terms related to initial-state charm (see below).

The computation of structure functions has changed in comparison to
previous NNPDF releases. Indeed, in
NNPDF3.0 the solution of the DGLAP evolution equations
and the structure functions were computed with the
internal NNPDF code {\tt FKgenerator}~\cite{Ball:2008by,Ball:2010de},
based on the Mellin space formalism.
In NNPDF3.1, as was already the case in the charm study 
of~Ref.\cite{Ball:2016neh}, PDF
evolution and DIS structure functions are computed using the {\tt
 APFEL} public code~\cite{Bertone:2013vaa}, based instead on the
$x$-space formalism.
The two codes have been extensively benchmarked against each other,
see  
App.~\ref{sec:benchmarking}.
DIS structure functions are computed at NLO in the FONLL-B general-mass 
variable flavor number scheme, and at NNLO in the FONLL-C 
scheme~\cite{Forte:2010ta}.
All computations include target mass corrections.

In NNPDF3.1 we now parametrize charm independently, and thus
the FONLL GM-VFN has been extended in order to include
initial-state heavy quarks. This is accomplished using the formalism of
Refs.~\cite{Ball:2015tna,Ball:2015dpa}. Within this formalism, a
massive correction to the charm-initiated contribution is included
alongside the contribution of fitted charm as a non-vanishing boundary
condition to PDF evolution. 
 At NNLO this correction
requires knowledge of massive 
charm-initiated contributions to the DIS coefficient functions up to
$\mathcal{O}\lp \alpha_S^2\rp$, which are currently only known
to $\mathcal{O}\lp \alpha_S\rp$~\cite{Kretzer:1998ju}.
Therefore, in the NNLO PDF determination, the NLO expression for this
correction is used: this corresponds to setting the 
 unknown $\mathcal{O}\lp
\alpha^2_S\rp$ contribution to the massive charm-initiated term
to zero. Such an approximation was used
Ref.~\cite{Ball:2016neh}, where it was shown that it is justified
by the fact that even setting to zero the full correction  (i.e. using
the LO expression for the massive correction) has an
effect which at the PDF level is much smaller
than PDF uncertainties (see in
particular Fig.~10 of Ref.\cite{Ball:2016neh}).

Finally, as in previous NNPDF studies,  
no nuclear corrections are applied to the deuteron structure function and
neutrino charged-current cross-section
data taken on heavy nuclei, in particular NuTeV and CHORUS. We will return to
this issue in Sect.~\ref{sec:nonucl}.

\subsection{Fixed-target Drell-Yan production}

In NNPDF3.1 we have included 
the same fixed-target Drell-Yan (DY) data as in NNPDF3.0,
namely the Fermilab E605 and E866 datasets; in the latter case
both the proton-proton data and the ratio of cross-sections between
deuteron and proton targets, $\sigma_{\rm DY}^d/\sigma_{\rm DY}^p$ are included.
However, the kinematic cuts applied to these two experiments differ
 from those in NNPDF3.0, based on 
the study
of~\cite{Bonvini:2015ira}, which showed that theoretical 
predictions for data points too close to the production threshold 
become unstable. Requiring reliability of the fixed-order
perturbative approximation leads to the cuts
\be
\tau \le 0.08 \, \quad {\rm and} \quad |y/y_{\rm max}|\le 0.663 \, ,
\ee
where 
$\tau=M_{ll}^2/s$ and $y_{\rm max}=-\frac{1}{2} \ln \tau$, with $M_{ll}$
the dilepton invariant mass distribution and $\sqrt{s}$ the center
of mass energy of the collision.

As in the case of DIS, NLO fixed-target Drell-Yan cross-sections were 
computed in NNPDF3.0 
using the Mellin-space {\tt FKgenerator} code, while in NNPDF3.1
they are obtained using {\tt APFEL}. The two computations are 
benchmarked in App.~\ref{sec:benchmarking}.
NNLO corrections are determined using {\tt Vrap}~\cite{Anastasiou:2003ds}.
Once more, as in previous NNPDF studies, 
no nuclear corrections are applied; again we will return to this issue
in Sect.~\ref{sec:nonucl} below.

\subsection{Single-inclusive jets}
\label{sec:datajets}

Four single-inclusive jet cross-section measurements were part of
the NNPDF3.0 dataset:
CDF Run II $k_T$~\cite{Aaltonen:2008eq}, CMS 2011~\cite{Chatrchyan:2012bja},
 ATLAS 7~TeV 2010 and ATLAS 2.76~TeV, including correlations to the 7~TeV
 data~\cite{Aad:2011fc,Aad:2013lpa}.
On top of these, in NNPDF3.1 we also include the 
ATLAS 7~TeV 2011~\cite{Aad:2014vwa} and CMS 2.76~TeV~\cite{Khachatryan:2015luy} data.
Some of these measurements are available for different values of
the jet $R$ parameter; the values used in NNPDF3.1 are listed
in Table~\ref{jetRadius}.

\begin{table}[H]
\small
\begin{centering}
    \renewcommand{\arraystretch}{1.4}
\begin{tabular}{|c|c|c|}
  \hline
  Dataset  &  Ref.  & Jet Radius \\
  \hline
  \hline
  CDF Run II $k_t$ incl jets & \cite{Abulencia:2007ez} & $R=0.7$
  \tabularnewline
   ATLAS   7 TeV jets 2010 & \cite{Aad:2011fc}  &  $R=0.4$
\tabularnewline
  ATLAS    2.76 TeV jets & \cite{Aad:2013lpa}  & $R=0.4$
\tabularnewline
ATLAS      7 TeV jets 2011   &  \cite{Aad:2014vwa}  & $R=0.6$
\tabularnewline
 CMS   7 TeV jets 2011     & \cite{Chatrchyan:2012bja}&  $R=0.7$
\tabularnewline
 CMS  2.76 TeV jets       & \cite{Khachatryan:2015luy}& $R=0.7$
 \tabularnewline
 \hline
 \end{tabular}
\par\end{centering}
\caption{\small Values of the jet $R$ parameter used for the
  jet production datasets included in NNPDF3.1.
\label{jetRadius}
}
\end{table}

No cuts are applied to any of jet datasets included in NNPDF3.1, except 
for the ATLAS 2011 7~TeV data, for which  achieving a good description 
turns out to be impossible if all five rapidity bins are
included simultaneously.
We can obtain a good agreement between data
and theory when using only the central rapidity bin,
$|\eta^{\rm jet}|<0.4$.
The origin of this state of affairs is 
not understood: we have verified
that a reasonable description can be obtained if some of the
systematic uncertainties are decorrelated, but we have no
justification for such a procedure. We have therefore chosen to only include 
in NNPDF3.1 data from the central rapidity bin,
$|\eta^{\rm jet}|<0.4$ for this set. This is also the rapidity bin with the 
largest PDF sensitivity~\cite{Rojo:2014kta}.

In NNPDF3.1, all NLO jet cross-sections are computed using
 {\tt NLOjet++}~\cite{Nagy:2001fj}
interfaced to {\tt APPLgrid}~\cite{Carli:2010rw}.
The jet $p_T$ is used as the central factorization and
renormalization scale 
in all cases, as this choice exhibits improved perturbative
convergence compared to other scale choices such as the leading
jet $p_T^{1}$~\cite{Carrazza:2014hra,Currie:2017ctp}.

\begin{figure}[t]
\centering
\epsfig{width=0.46\textwidth,figure=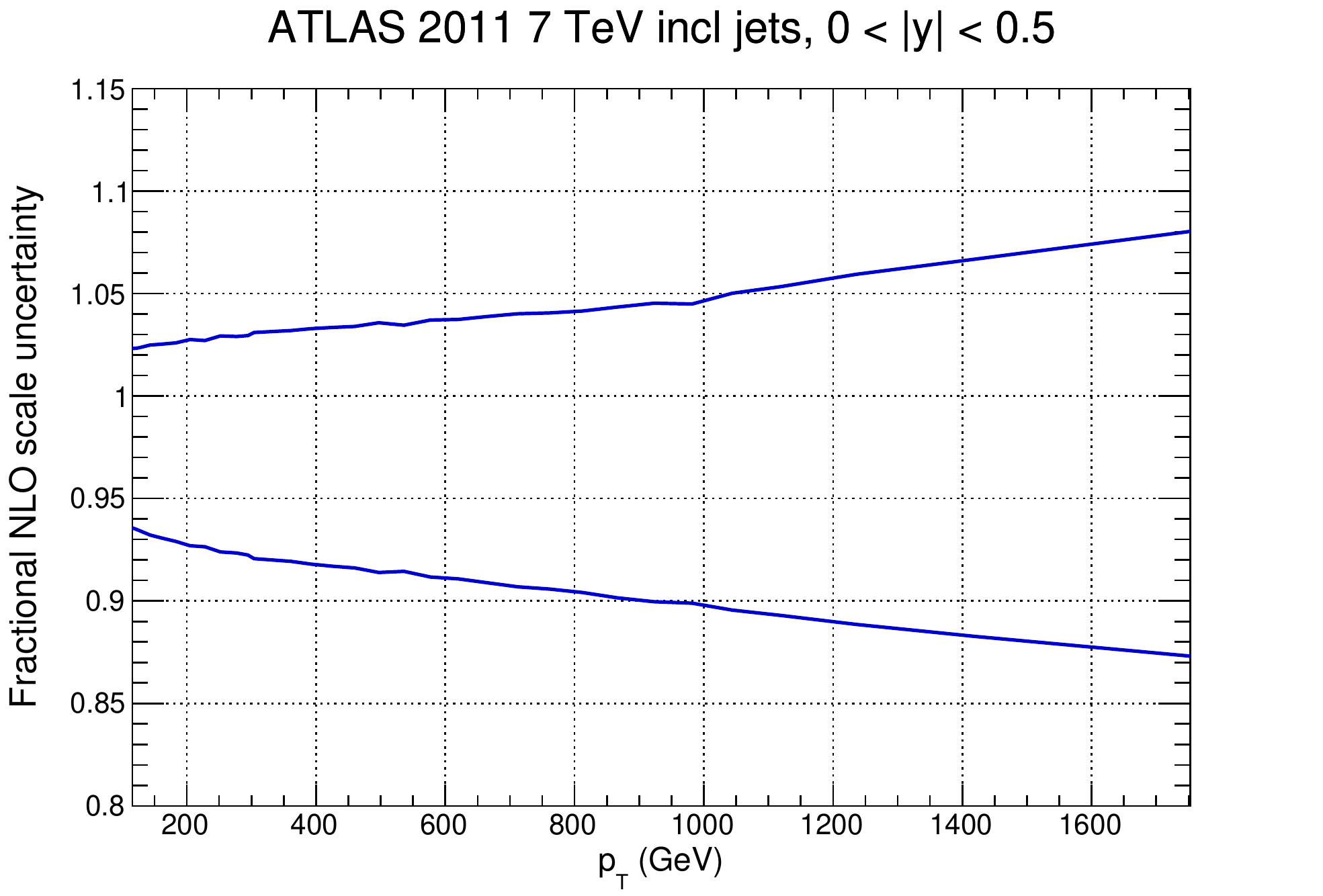}
\epsfig{width=0.46\textwidth,figure=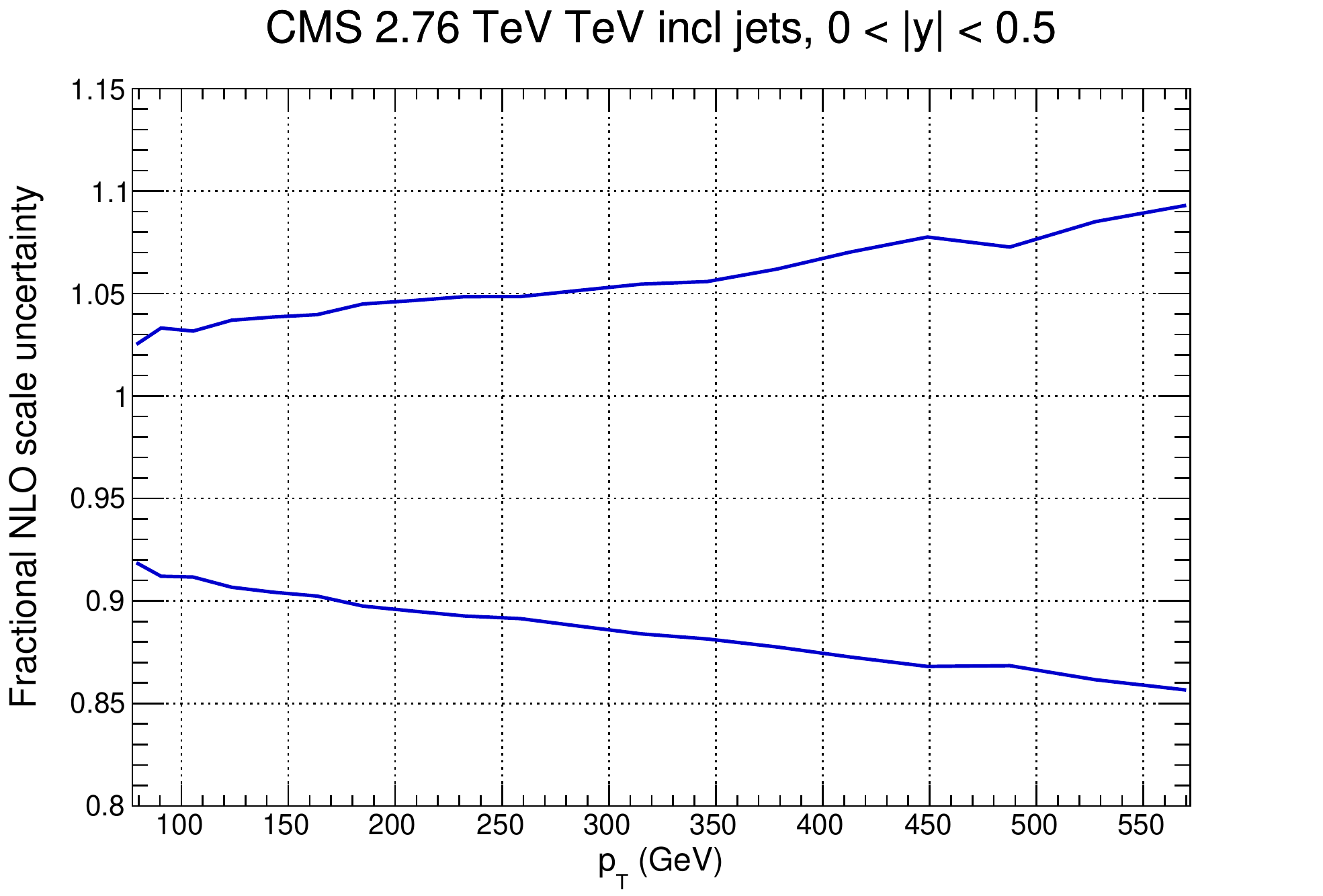}
\caption{\small 
  The fractional scale uncertainty on NLO 
  single-inclusive jet production, as a function of the
  jet $p_T$ for
  the central rapidity bins of ATLAS~7~TeV~2011
  (left) and the CMS~2.76~TeV (right).
 \label{fig:THscales}} 
\end{figure}

While the NNLO calculation of inclusive jet production has been recently
published~\cite{Currie:2016bfm,Currie:2017ctp}, results are not yet
available for all datasets included in NNPDF3.1.
Therefore, jet data are included as default in the NNPDF3.1 NNLO
determination using NNLO PDF evolution
but NLO matrix elements, while adding to the covariance matrix 
an additional fully correlated theoretical systematic 
uncertainty estimated from scale variation of the
NLO calculation.
The NLO scale variations are performed using {\tt APPLgrid} interfaced to
{\tt HOPPET}~\cite{Salam:2008qg}. We take the associated uncertainty
as the the envelope of the result of seven-point scale variation
$\mu_F \in \lc p_T/2,2p_T\rc$
and $\mu_R \in \lc p_T/2,2p_T\rc$ with 
 $1/2 \le \mu_F/\mu_R \le 2$.%
The NNLO corrections are generally well within this scale variation band when the jet $p_T$ is
chosen as a central scale~\cite{Currie:2017ctp}. 
This scale uncertainty is
shown in Fig.~\ref{fig:THscales} for ATLAS 7~TeV 2011
  and CMS 2.76~TeV as a function of the
  jet $p_T$ for the central rapidity bin. It is seen to range between a few percent
  at low $p_T$ up to around 10\% at the largest $p_T$. A similar
 behaviour is observed in other rapidity bins, with a more
 asymmetric band at forward rapidity.

\begin{figure}[t]
\begin{center}
  \includegraphics[scale=0.55]{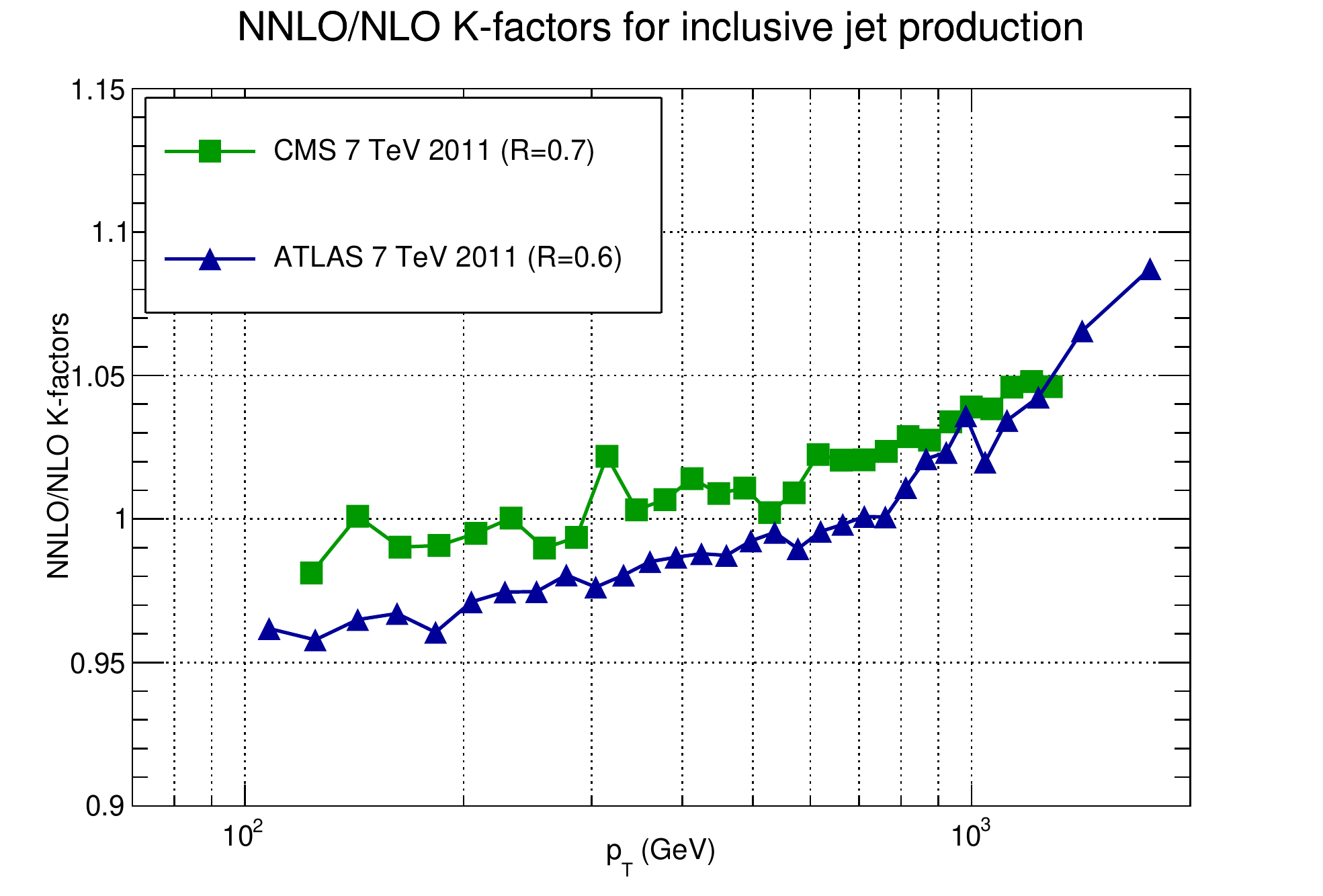}
  \caption{\small
    The NNLO/NLO cross-section ratio~\cite{curriepriv} for the central rapidity
    bin ($0\le |y_{\rm jet}|\le 0.5$) of the ATLAS
    and CMS 7~TeV 2011 jet data, with the values
    of $R$ of Tab.~\ref{jetRadius},  plotted vs. $p_T$.
    \label{fig:jetcf}}
\end{center}
\end{figure}

In order to gauge the reliability of our approximate treatment of the
jet data, we have produced a PDF determination in which all data
for which NNLO corrections are known, namely the 7~TeV ATLAS and CMS
datasets, are included
using exact NNLO theory. This will be discussed in
Sect.~\ref{sec:jetdata}. Representative NNLO corrections are shown 
in Fig.~\ref{fig:jetcf}, where we show the NNLO/NLO ratio 
for the central rapidity
    bin ($0\le |y_{\rm jet}|\le 0.5$) of the ATLAS
    and CMS 7~TeV 2011 datasets, plotted 
    as a function of $p_T$~\cite{curriepriv}: note (see 
    Table~\ref{jetRadius}) that the values of $R$ are different,
    thereby explaining the different size of the correction, which for
    CMS is $\sim -2\%$ for $p_T\sim 100$~GeV, increasing
    up to $\sim 5\%$ for $p_T\sim 2$~TeV, and for ATLAS it ranges from
    $\sim -4\%$ increasing
    up to $\sim 9\%$ as a function of $p_T$. Unlike in the case of the
    $Z$ transverse momentum  distribution, to be discussed in
    Sect.~\ref{sec:zpt},  
the lack of smoothness of
    the corrections seen in Fig.~\ref{fig:jetcf} is not problematic as
    the fluctuations are rather 
smaller than typical uncorrelated uncertainties on
    these data.

\subsection{Drell-Yan production at hadron colliders}
\label{data:inclusiveWZ}

The NNPDF3.0 determination already included a wide set of collider
Drell-Yan data, both at the $W$ and $Z$ peak and off-shell. This dataset has been
further expanded in NNPDF3.1. We discuss here invariant mass and rapidity
distributions; transverse momentum distributions will be discussed in
Sect.~\ref{sec:zpt}. 

In NNPDF3.1 we include for the first time D0 legacy $W$
asymmetry measurements based on the complete dataset in the electron~\cite{D0:2014kma} and 
muon~\cite{Abazov:2013rja} channels.
The only cut applied to this dataset is at NNLO, where we remove data with
 $\mathcal{A}_l(y_l)\le 0.03$ in both the
electron and muon channel data.  This is
due to the fact that when  the asymmetry is very close
to zero, even with high  absolute accuracy  on the 
NNLO theoretical calculation, it is difficult to achieve high
percentage accuracy, 
thereby making the NNLO correction to the asymmetry unreliable.
The NLO computation is performed using
{\tt APPLgrids} from the HERAfitter study of~\cite{Camarda:2015zba},
which we have cross-checked using {\tt Sherpa}~\cite{Gleisberg:2008ta}
interfaced to {\tt MCgrid}~\cite{DelDebbio:2013kxa}.
NNLO corrections are computed using {\tt FEWZ}~\cite{Gavin:2010az,Gavin:2012sy,Li:2012wna}.

New results are included for ATLAS, CMS and LHCb.
For ATLAS, NNPDF3.0 included 2010 $W$ and $Z$ 7~TeV rapidity distributions and their cross-correlations~\cite{Aad:2011dm}.
A recent update of the same measurement~\cite{Aaboud:2016btc}, based on the entire 7~TeV
integrated luminosity of 4.6 fb$^{-1}$ is included in NNPDF3.1, albeit
partially. 
This measurement provides differential distributions in lepton pseudo-rapidity $|\eta_l|$ in the range $0\le |\eta_l|\le 2.5$ for on-shell 
$W^+$ and $W^-$ production. For $Z/\gamma^*$  production results are
provided either with both leptons measured in the range $0\le
|\eta_l|\le 2.5$, or with one lepton with $0\le |\eta_l|\le 2.5$ and
the other with $2.5\le |\eta_l|\le 4.9$. The central rapidity data sre
given   
for three bins in the dilepton invariant mass $46<m_{ll}<66$,
$66<m_{ll}<116$ and $116<m_{ll}<150$~GeV, and the forward rapidity
data in the last two mass bins (on-peak and high-mass).
We only include the on-shell, $0\le |\eta_l|\le 2.5$ data, thereby
neglecting the two low- and high-mass $Z$ production bins in the
central rapidity region, and the on-peak and high-mass  $Z$ production bins at forward rapidity.
The 
full dataset will be included in future NNPDF releases. 
No other cuts are applied to the dataset.
Theoretical predictions are obtained using NLO {\tt APPLgrids}~\cite{Carli:2010rw} generated using {\tt MCFM}~\cite{MCFMurl}, 
while the NNLO corrections are taken from the {\tt xFitter} analysis of Ref~\cite{Aaboud:2016btc}.

Also new to NNPDF3.1 is the ATLAS low-mass Drell-Yan data from Ref.~\cite{Aad:2014qja}.
We use only the low-mass DY cross-sections in the muon channel measured from $35~\mathrm{pb}^{-1}$ 2010 dataset,
which extends down to $M_{ll}=12$ GeV.
The 2011 7~TeV data with invariant masses between 26~GeV and 66~GeV 
are not included because they are affected by large electroweak
corrections and are therefore excluded by our cuts. Furthermore, two
datapoints are removed from the NLO datasets because NNLO corrections
exceed experimental uncertainties.
Theoretical predictions are obtained at NLO using {\tt APPLgrids}~\cite{Carli:2010rw} constructed using
{\tt MCFM}, and at NNLO corrections are computed using {\tt FEWZ}.

For CMS, NNPDF3.1 includes 8~TeV $W^+$ and $W^-$ rapidity distributions, including information on
their correlation~\cite{Khachatryan:2016pev}. No cuts have been applied to this dataset. 
Theoretical predictions are obtained using the NLO {\tt APPLgrids} generated with {\tt MCFM} and the NNLO correction factors 
computed using {\tt FEWZ} in the context of the {\tt xFitter}~\cite{Alekhin:2014irh} analysis presented in Ref.~\cite{Khachatryan:2016pev}.
Double differential rapidity $y_{ll}$ and invariant mass $M_{ll}$ distributions for $Z/\gamma^*$ production from 
the 2012 8~TeV data~\cite{CMS:2014jea} have been studied by including them in a specialized PDF determination. However,
the dataset has been left out of default NNPDF3.1 dataset, for reasons to be discussed in Sect.~\ref{sec:cms8tevimpact}. 
The only cut applied to this dataset, based on a previous MMHT analysis\cite{mmhtpriv} is $M_{\ell\ell}\ge30$~GeV, because in the
lowest mass bin the leading-order prediction in this bin vanishes.
Theoretical predictions are obtained at NLO using {\tt APPLgrids} constructed using {\tt MCFM}, and at NNLO 
corrections have been computed~\cite{mmhtpriv} using {\tt FEWZ}.

\begin{figure}[t]
\begin{center}
  \includegraphics[width=0.49\linewidth]{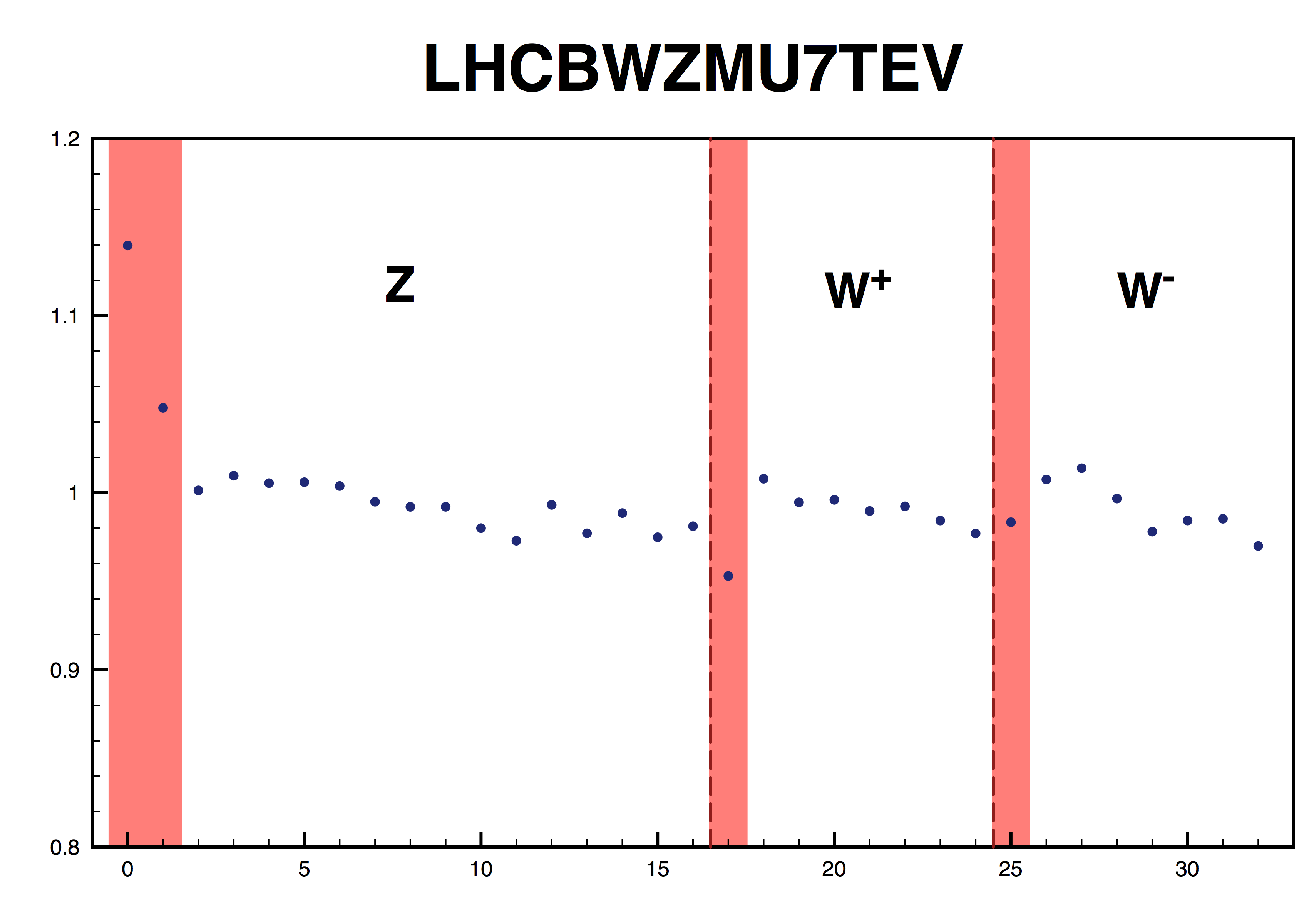}
  \includegraphics[width=0.49\linewidth]{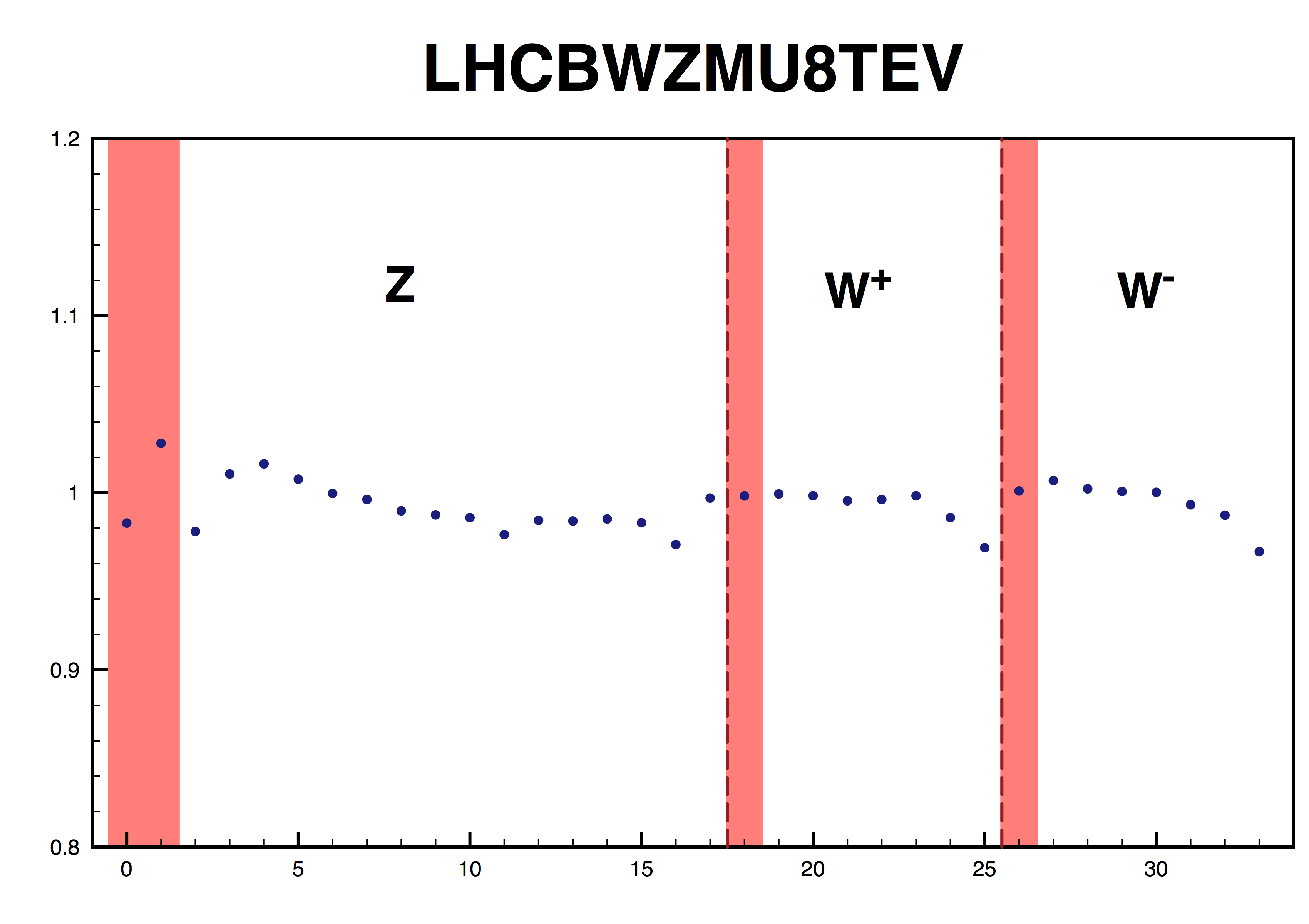}
  \caption{\small   The NNLO/NLO cross-section 
    for the LHCb $7$ (left) and $8$~TeV (right) data. The central
    rapidity region which is cut is shaded in red.
    \label{fig:lhcbcf}}
\end{center}
\end{figure}
For LHCb, previous data included in NNPDF3.0 are replaced by
the final 7~TeV and 8~TeV~$W+$, $W^-$ and $Z$ rapidity distributions in the muon
channel~\cite{Aaij:2015gna,Aaij:2015zlq}. The NNLO/NLO cross-section
ratios are shown in Fig.~\ref{fig:lhcbcf}.
The 
Data with $|y_l|\le 2.25$ from this set have been cut because the
anomalously large size of the NNLO corrections suggests that they may
be unreliable.
Theoretical predictions are obtained at NLO using
{\tt APPLgrids} constructed using
MCFM, and at NNLO 
corrections computed using {\tt FEWZ}.

\subsection{The transverse momentum of $Z$ bosons}
\label{sec:zpt}

The transverse momentum distribution of the $Z$ boson is included for the
first time in a global PDF determination thanks to the recent computation of the process
at NNLO~\cite{Ridder:2015dxa,Ridder:2016nkl,Boughezal:2015ded,Boughezal:2016isb}.
In the NNPDF3.1 determination we include recent datasets from ATLAS and
CMS  following the detailed study in Ref.~\cite{Boughezal:2017nla}.

ATLAS has published measurements of the spectrum of the $Z$ 
transverse momentum 
at 7~TeV~\cite{Aad:2014xaa} and at
$8$~TeV~\cite{Aad:2015auj}. Measurements are performed
in the $Z/\gamma^*\to e^+e^-$ and $Z/\gamma^*\to \mu^+\mu^-$ 
channels which are then combined. The 7~TeV data are based
on an integrated luminosity of 4.7~${\rm fb}^{-1}$, while the 8~TeV data are
based on an integrated luminosity of 20.3~fb$^{-1}$. We now discuss each of these two datasets in
turn.

The 7~TeV data are taken at the $Z$ peak, reaching values of the $Z$
transverse momentum of up to $p_T^Z=800$ GeV.
They are given inclusively for $Z/\gamma^*$ rapidities up to
$|y_Z|=2.4$, as well as in three separated rapidity bins
given by $0.0\le |y_Z| \le 1.0$, $1.0 \le |y_Z| \le
2.0$ and $2.0 \le |y_Z| \le 2.4$.
In order to maximize the potential constraint on PDFs, only the differential
measurement will be considered.
The measurement is presented in terms of normalized cross-sections
$(1/\sigma_Z)\,d\sigma(Z)/d p_T^Z$, where $\sigma_Z$
is the fiducial cross-section in the corresponding di-lepton rapidity
bin.
%
This dataset has been left out of default NNPDF3.1 dataset, for reasons to be discussed in Sect.~\ref{sec:impactzpt}.

The 8~TeV dataset, which reaches $p_T^Z$ values as high as $900$~GeV, 
is presented in three separate invariant mass bins:
low mass below the $Z$-peak, on-peak, and high mass above the $Z$-peak up to $\Mll=$ 150 GeV.
In addition, the measurement taken at the $Z$-peak
is provided both inclusively in the whole rapidity range
$0.0<|y_Z|<2.4$ as well as exclusively in six
separate rapidity bins $0<y_Z<0.4$,
$0.4<|y_Z|<0.8$, $0.8<|y_Z|<1.2$, $1.2<|y_Z|<1.6$, $1.6<|y_Z|<2.0$ and
$2.0<|y_Z|<2.4$.
Once again, here the more differential measurement will be used. 
In contrast to the 7~TeV data, the dataset is given both in terms of normalized and absolute distributions. 
We will use the latter, not only because of the extra information on the 
cross-section normalization, but also as problems can occur whenever
the data used to compute the normalization are provided in a range
which differs from that of the data used for PDF determination. This problem is 
discussed in detail in Ref.~\cite{Boughezal:2017nla} and described in Sect.~\ref{sec:impactzpt}.

CMS has measured the cross-sections differentially in $p_T$ and rapidity $y_Z$
at 8~TeV~\cite{Khachatryan:2015oaa}, based
on an integrated luminosity of 19.7 fb$^{-1}$ in the muon channel.
Data is provided in five rapidity bins 
 $0.0<|y_Z|<0.4$,
$0.4<|y_Z|<0.8$, $0.8<|y_Z|<1.2$, $1.2<|y_Z|<1.6$ and $1.6<|y_Z|<2.0$.
We do not consider a previous CMS measurement at 7~TeV~\cite{Chatrchyan:2011wt}, which
is based on a smaller dataset, and would constitute
double counting of the double differential distributions~\cite{Chatrchyan:2013tia} already included
in NNPDF3.0, and retained in NNPDF3.1.

Three sets of kinematic cuts are applied to the data. 
Firstly, ensuring the reliability of fixed-order perturbation theory
imposes a cut of $p_T^Z \ge$ 30 GeV
(resummation would be required for smaller $p_T$)~\cite{Boughezal:2017nla}.
Secondly, removing regions in which electroweak corrections
are large and comparable to the experimental data imposes
a cut of $p_T^Z \le 150~(170)$ GeV for the ATLAS~(CMS)
data~\cite{Boughezal:2017nla}.  
Finally, the CMS dataset in the largest rapidity bin is discarded
due to an apparent incompatibility with both the corresponding ATLAS
measurement in the same bin and the theoretical prediction. The origin of this incompatibility
remains unclear~\cite{Boughezal:2017nla}.

\begin{figure}[t]
\begin{center}
  \includegraphics[width=0.49\linewidth]{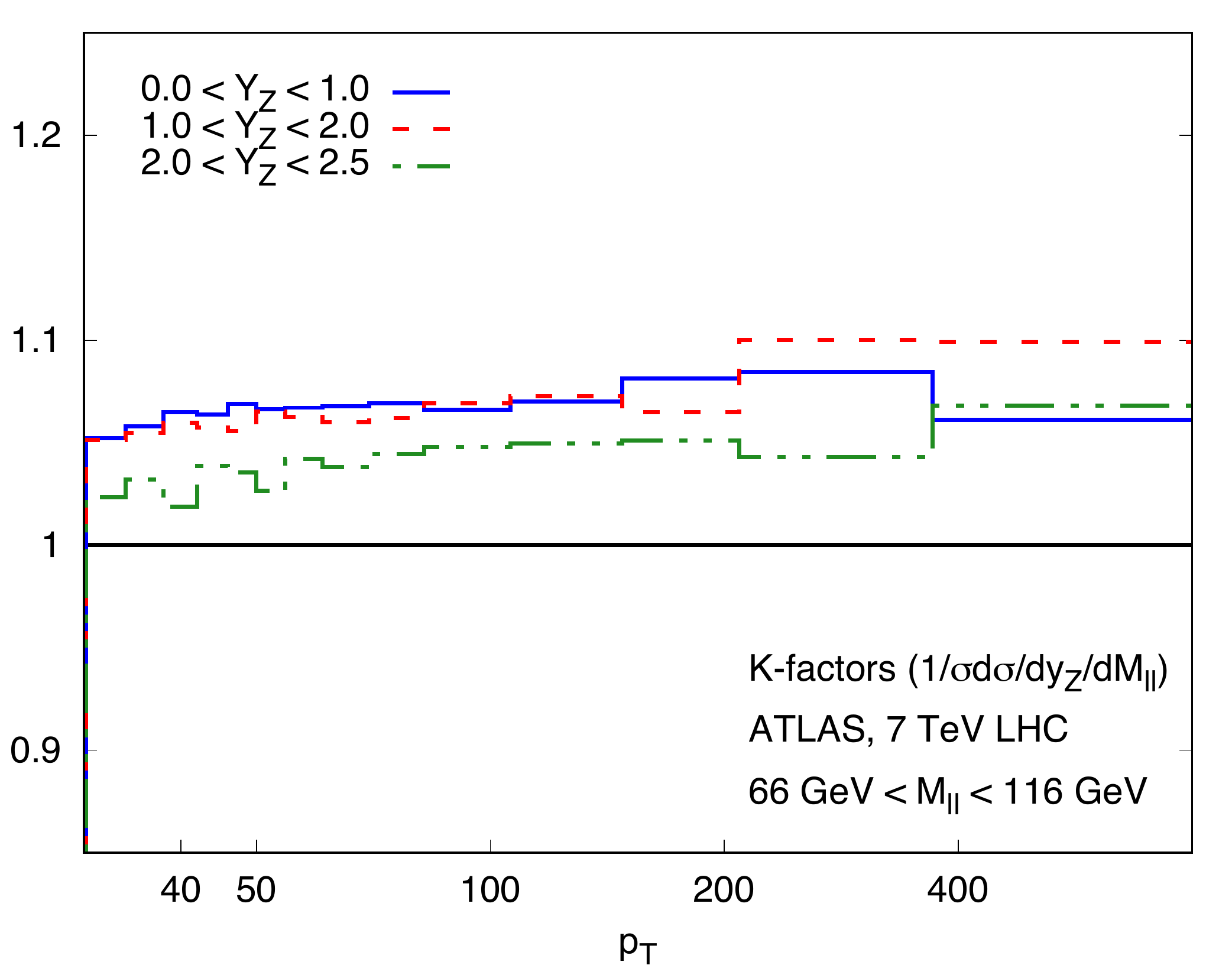}
  \includegraphics[width=0.49\linewidth]{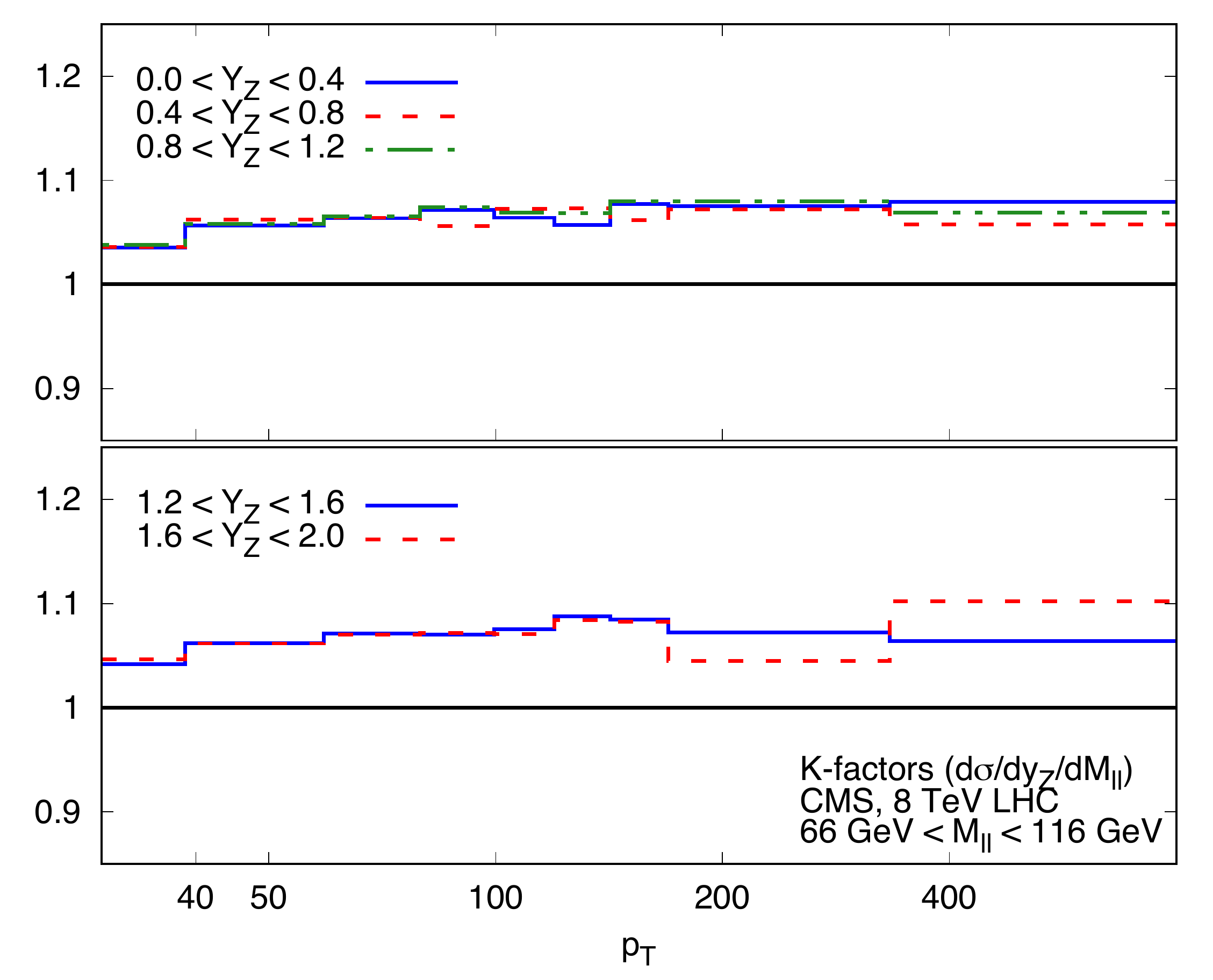}\\
  \includegraphics[width=0.49\linewidth]{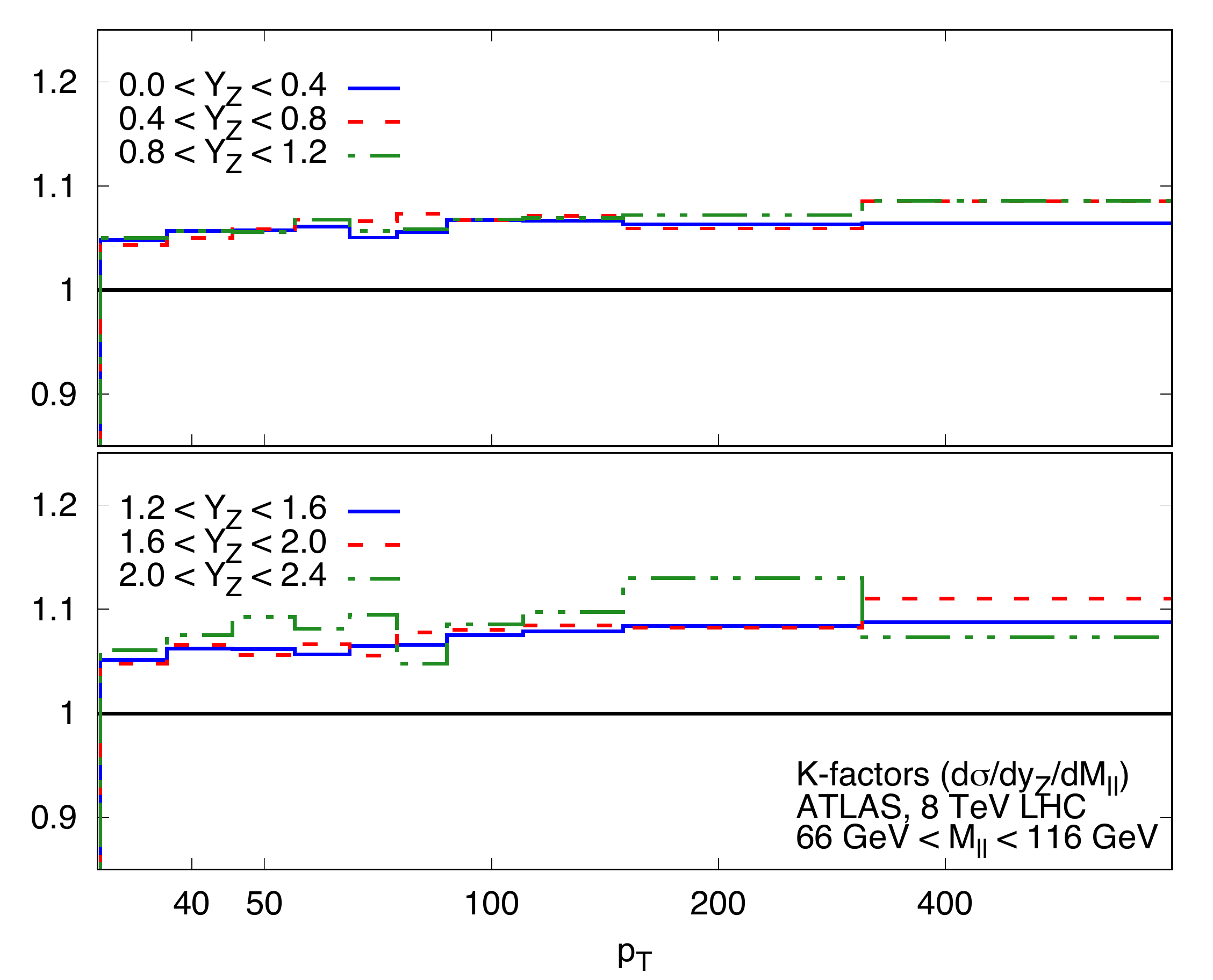}
  \includegraphics[width=0.49\linewidth]{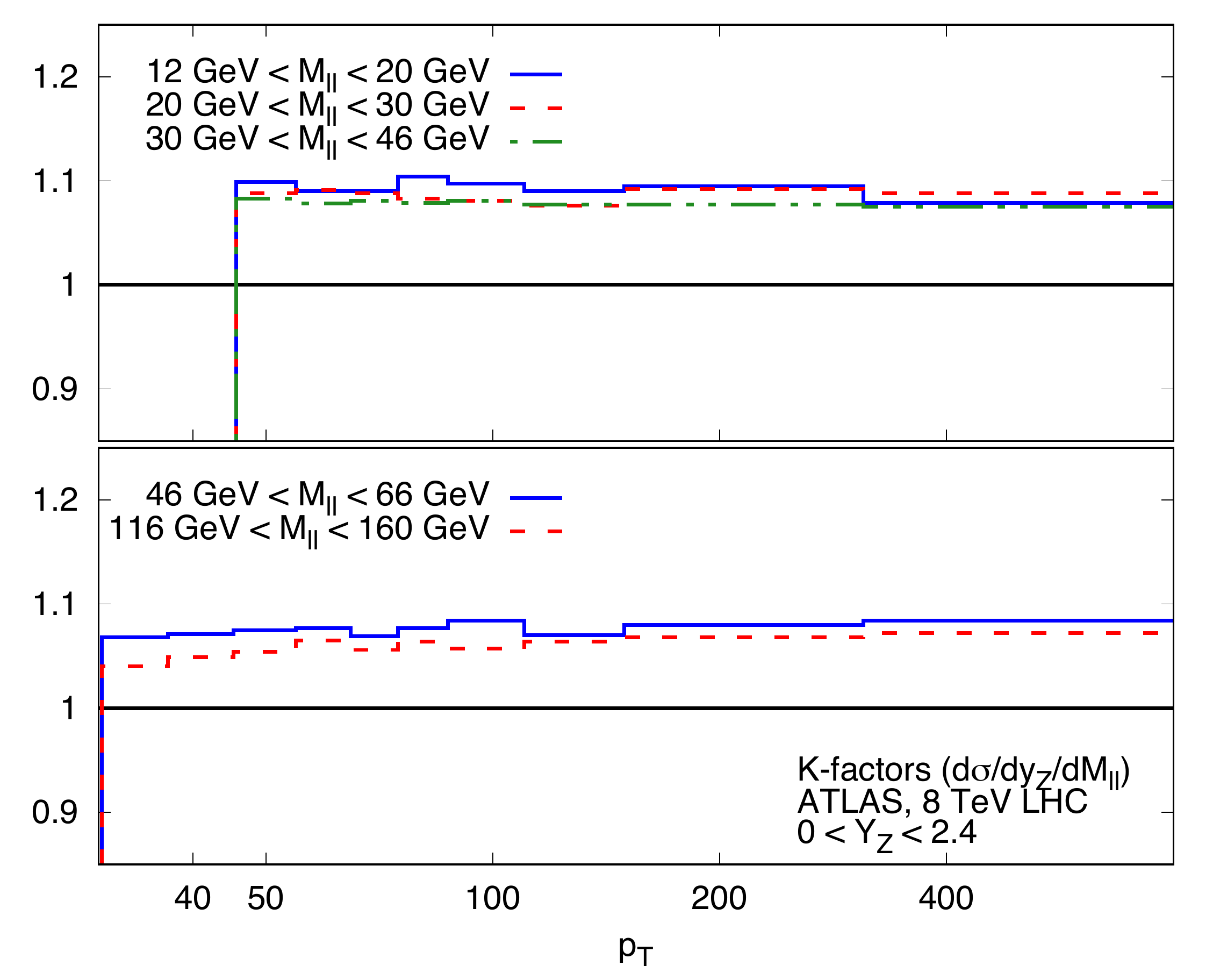}
  \caption{\small   The NNLO/NLO cross-section 
    for the $Z$ $p_T$ data
    corresponding
    to the acceptance cuts and binning
    of the ATLAS 7~TeV (top left), CMS 8~TeV (top right),
    and the ATLAS 8 TeV (bottom) rapidity (left) and invariant mass
    (right) distributions.
    \label{fig:zptcf}}
\end{center}
\end{figure}

Theoretical predictions have been obtained from
Ref.~\cite{Boughezal:2017nla}, based upon the NNLO computation of
$Z$+jet production of Refs.~\cite{Boughezal:2015ded,Boughezal:2016isb}.
Factorization and renormalization scales are chosen as
\be
\mu_R=\mu_F =
\sqrt{(p_T)^2+\Mll^2} \, ,
\ee
where $\Mll$ is the invariant mass of the
final-state lepton pair.
The calculation includes the $Z$ and $\gamma^*$ contributions, their interference and decay to
lepton pairs. The NNLO/NLO ratio is shown in
Fig.~\ref{fig:zptcf} for the observables with the ATLAS and CMS acceptance cuts, 
computed using NNPDF3.0 PDFs, with
$\alpha_s(m_Z)=0.118$; the NNLO correction varies
from around
2-3\% at low $p_T$ up to around 10\% at high $p_T$ and is therefore
required in order to describe data with sub-percent accuracy. 
%

\begin{figure}[t]
\begin{center}
  \includegraphics[scale=0.55]{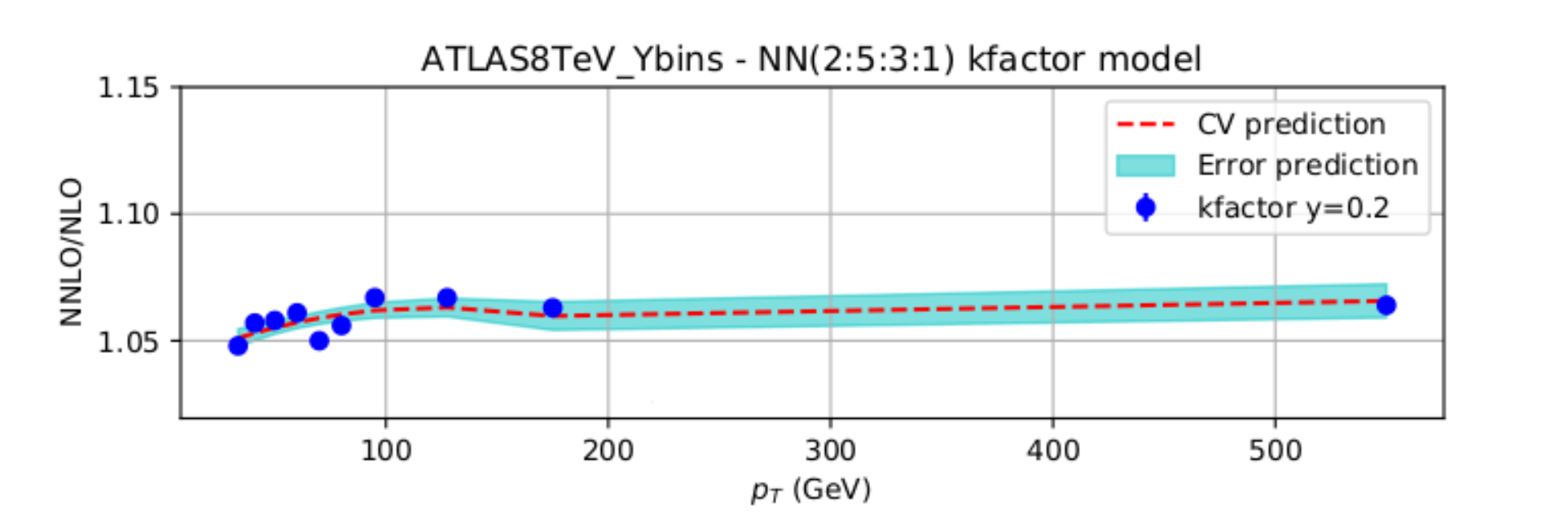}
  \caption{\small
    The NNLO/NLO cross-section ratio in  
 the central rapidity bin of the 8~TeV ATLAS $Z$ $p_T$
    distribution. The result of a fit and its associate uncertainty
    are also shown.
\label{fig:cf-zpt-fit}}
\end{center}
\end{figure}

Even the most accurate results for the NNLO/NLO correction factor
still display fluctuations, as shown in Fig.~\ref{fig:zptcf} where we
plot the NNLO/NLO cross-section ratio for the central rapidity bin of
the $8$~TeV ATLAS data. The points are shown together with their
nominal Monte Carlo integration uncertainty~\cite{Boughezal:2017nla}.
The point-to-point statistical fluctuation of the theoretical prediction appears
to be larger than the typical uncorrelated statistical uncertainty on
the ATLAS dataset, which is typically at the sub-percent or even permille
level. In order to check this, we have  fitted 
 an ensemble of neural networks to the
cross-section ratio, as a function of $p_T^Z$ for fixed rapidity. 
The fit has been
performed in each of the rapidity bins for the ATLAS and CMS data; 
more details are given in Ref.~\cite{Carrazza:2017bjw}.
The result of the fit
and its one-sigma 
uncertainty are  shown in Fig.~\ref{fig:cf-zpt-fit} for the central
rapidity bin of the ATLAS data.  

 The one-sigma uncertainty of the  fit, which is determined by the
 point-to-point 
fluctuation of the NNLO computation, is at the percent level, which is
  rather larger than the
statistical uncertainty of the data. Indeed, 
it is clear by inspection of
Figs.~\ref{fig:zptcf}-\ref{fig:cf-zpt-fit} that the point-to-point fluctuations of the NNLO/NLO ratio are much larger than those
of the data themselves (as seen in
Ref.s~\cite{Aad:2014xaa},\cite{Boughezal:2017nla}). We  conclude that
there is a 
residual theoretical uncertainty on the NNLO prediction which we
estimate to be of order of 1\% for all datasets. This conclusion has
been validated and cross-checked by repeating the fit with cuts or
different functional forms. We have therefore added an extra 1\% fully
uncorrelated theoretical uncertainty to this dataset (see also Ref.~\cite{Boughezal:2017nla}).

\subsection{Differential distributions and total
  cross-sections in $t\bar{t}$ production}
\label{sec:topdata}

Differential distributions for top pair production have been
included in NNPDF3.1 following the detailed study of Ref.~\cite{Czakon:2016olj}.
ATLAS and CMS have performed measurements of these distributions with a
variety of choices of kinematic variables, including the top quark rapidity
$y_t$, the rapidity of the top pair $y_{t\bar{t}}$, the transverse 
momentum of the top quark $p_T^t$, and the invariant mass of the top-antitop
system $m_{t\bar{t}}$. For ATLAS both absolute and normalized 
differential distributions are provided, whereas CMS only provides
normalized results.
Perturbative QCD corrections for all these distributions have been 
computed at NNLO~\cite{Czakon:2015owf,Czakon:2016dgf}.
In order to avoid double counting, only one distribution per
experiment can be included in the dataset, as the statistical correlations
between different distributions are not available.
The choice of differential distributions adopted in NNPDF3.1 follows the 
recommendation of Ref.~\cite{Czakon:2016olj}, where
a comprehensive study of the impact on the gluon PDF of 
various combinations of differential top pair distributions was performed. 
It was found that the normalized rapidity distributions have the
largest constraining power and lead to a good agreement between theory and data
for ATLAS and CMS.
The use of rapidity distributions has some further  advantages.
First, it reduces the risk of possible contamination by BSM effects.
For example, heavy resonances would be kinematically suppressed
in the rapidity distributions, but not in the tails of the $m_{t\bar{t}}$ and 
$p_T^t$ distributions.
Second, rapidity distributions exhibit a milder sensitivity upon variations 
of the value of $m_t$ than the $p_T^t$ and $m_{t\bar{t}}$ 
distributions~\cite{Czakon:2016vfr}.

We therefore include the $8$~TeV normalized
rapidity distributions in the lepton+jets final
state from ATLAS~\cite{Aad:2015mbv} and 
CMS~\cite{Khachatryan:2015oqa}, which 
correspond respectively to an integrated
luminosity of $20.3$ fb$^{-1}$ and $19.7$ fb$^{-1}$. 
We consider measurements in the full phase space, with observables 
reconstructed in terms of the top or top-pair kinematic variables, because
NNLO results are available only for stable top quarks.
We also include, again following Ref.\cite{Czakon:2016olj},
the most recent total cross-sections measurements at $7$, $8$ and $13$~TeV from
ATLAS~\cite{Aad:2014kva,Aaboud:2016pbd} and
CMS~\cite{Khachatryan:2016mqs,CMS:2016syx}.
They replace previous measurements from 
ATLAS~\cite{ATLAS:2012aa,ATLAS:2011xha,TheATLAScollaboration:2013dja}
and CMS~\cite{Chatrchyan:2013faa,Chatrchyan:2012bra,Chatrchyan:2012ria} 
included in NNPDF3.0. 

At NLO theoretical predictions have been generated with {\tt Sherpa}~\cite{Gleisberg:2008ta}, in a format 
compliant to {\tt APPLgrid}~\cite{Carli:2010rw}, using
the {\tt MCgrid} 
code~\cite{DelDebbio:2013kxa} and the {\tt Rivet}~\cite{Buckley:2010ar} 
analysis package, with {\tt OpenLoops}~\cite{Cascioli:2011va} for the NLO 
matrix elements. 
All calculations have been performed with large Monte Carlo integration 
statistics in order to ensure that residual numerical fluctuations are 
negligible.
Our results have been carefully benchmarked against those obtained from 
the code of~\cite{Czakon:2016dgf}.
Renormalization and factorization scales, $\mu_R$ and $\mu_F$
respectively, have 
been chosen based on the recommendation of Ref.~\cite{Czakon:2016dgf} as
\be
\label{eq:scale1}
\mu_R=\mu_F=\mu=H_T/4 \, , \qquad H_T \equiv \sqrt{m_t^2+\lp {p_T^t}\rp^2}
+ \sqrt{m_t^2+\lp {p_T^{\bar{t}}}\rp^2} \, ,
\ee
where $m_t=173.3$ GeV is the PDG world average for the
top-quark pole mass~\cite{Agashe:2014kda},
and $p_T^t$ ($p_T^{\bar{t}}$) is the top (anti-top) transverse momentum.
NLO theoretical predictions for normalized differential distributions have been obtained
by dividing their absolute counterparts by the cross-section
integrated over the kinematic range of the data.

\begin{figure}[t]
\begin{center}
  \includegraphics[scale=0.31,angle=270]{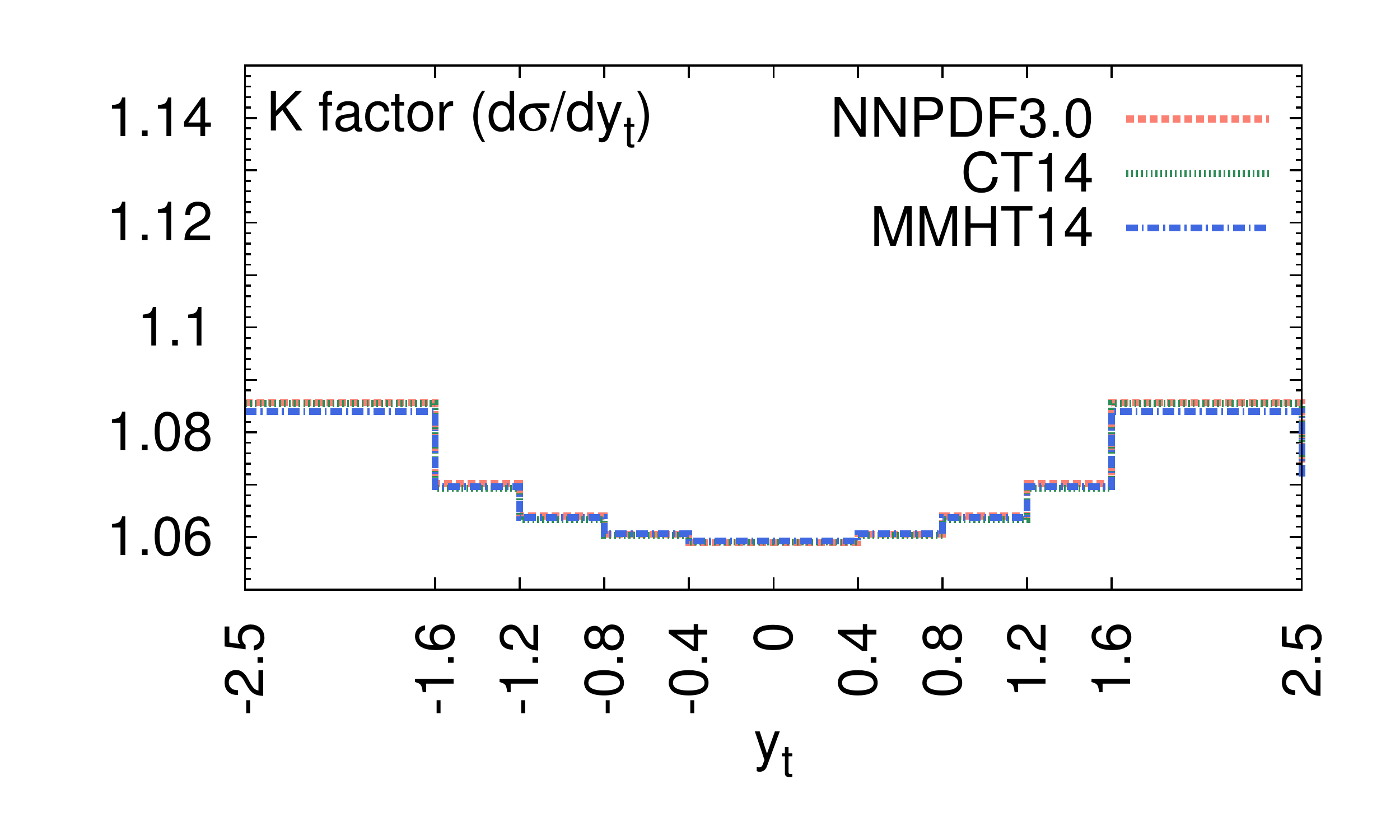}
  \includegraphics[scale=0.31,angle=270]{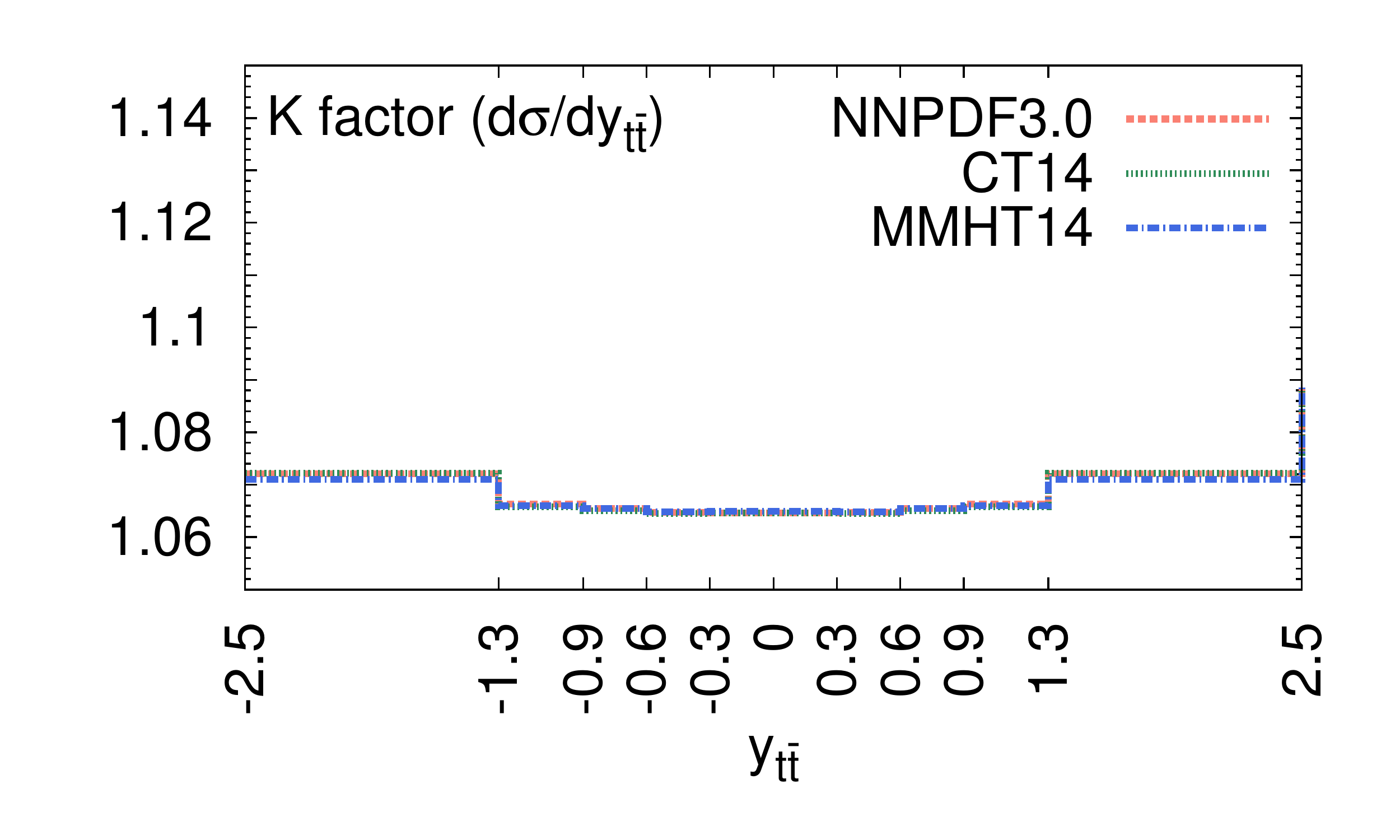}\\
  \caption{\small
    The NNLO/NLO cross-section ratio for the top quark rapidity $y_t$
    (left)
    and top-quark pair rapidity $y_{t\bar{t}}$ (right) corresponding
    to the 8~TeV ATLAS and CMS data.
     Results obtained with three different input PDF sets, NNPDF3.0,
     CT14, and MMHT14, are shown.
\label{fig:ttbar-cfact}}
\end{center}
\end{figure}

The NNLO correction factors have been computed separately
for the absolute  differential cross-sections and their normalizing total cross-sections. 
Differential cross-sections have been determined using the 
code of~\cite{Czakon:2016dgf}, with the scale choice Eq.~(\ref{eq:scale1}).
Results for the NNLO/NLO ratio are shown in
Fig.~\ref{fig:ttbar-cfact}, where it can be seen that the size of the
NNLO corrections is $6\%$ and $9\%$, actually smaller
than the data uncertainty, with a reasonably flat shape in the kinematic
region covered by the data.
We also show explicitly the dependence of the results on the PDF set
used in the calculation by using three different global PDF sets: it
is clear that this dependence is completely negligible.

Total cross-sections have been computed with the {\tt top++}
code~\cite{Czakon:2011xx} at NNLO+NNLL, and with fixed scales
$\mu_R=\mu_F=m_t$, following the recommendation of
Ref.~\cite{Czakon:2016dgf} which suggests that NNLO+NNLL resummed
cross-sections should be used in conjunction to NNLO differential
distributions if the latter are determined using a dynamical scale choice.

%

%

%
\newpage
\section{The NNPDF3.1 global analysis}

\label{sec:results}

We now present the results of the NNPDF3.1 global
analysis at LO, NLO and NNLO, and compare them with the previous
release NNPDF3.0 and  with other recent PDF sets.
Here we present results obtained using  the complete dataset
of Tab.~\ref{tab:completedataset}-\ref{tab:completedataset3}, discussed 
in Sect.~\ref{sec:expdata}. Studies of the impact of individual
measurements will be discussed along with PDF determinations from reduced
datasets in Sect.~\ref{sec:impactnewdata}. 

After a brief methodological summary, 
we discuss the  fit quality, and 
then examine individual PDFs and their uncertainties.
We compare NNPDF3.1 PDFs with NNPDF3.0 and with  CT14~\cite{Dulat:2015mca}, MMHT2014~\cite{Harland-Lang:2014zoa} 
and ABMP16~\cite{Alekhin:2017kpj}. 
We next examine the impact of independently parametrizing
charm, the principal methodological 
improvement in NNPDF3.1. Finally, we discuss theoretical
uncertainties, both related to QCD parameters 
and to missing higher order corrections to the
theory used for PDF determination.

In this Section all NLO and NNLO NNPDF3.1 results are produced using
the CMC~\cite{Carrazza:2015hva} optimized 100 replica Monte Carlo sets,
see Sect.~\ref{sec:delivery} below: despite only
including 100 replicas, these sets reproduce  the statistical features
of a set of at least about 400 replicas (see 
Sect.~\ref{sec:compression}).  We present here only a selection of
results: a more extensive set of results is available from a public
repository, see  Sect.~\ref{sec:delivery}.

\subsection{Methodology}
\label{sec:methodology}

NNPDF3.1 PDFs are determined with largely the same methodology as in NNPDF3.0:
the only significant change is that now charm is independently
parametrized. The PDF parametrization is identical to that
discussed in Sect.~3.2 of Ref.~\cite{Ball:2014uwa}, including the
treatment of preprocessing, but with the PDF
basis in Eq.~(3.4) of that reference now supplemented by an extra PDF
for charm, parametrized like all other PDFs (as per Eq.~(2) of
Ref.~\cite{Ball:2016neh}).
PDFs are parametrized at the scale $Q_0=1.65$~GeV whenever the charm PDF
is independently parametrized. For the purposes of comparison we also provide PDF sets constructed with perturbatively 
generated charm; 
in these sets, PDFs are parametrized at the scale $Q_0=1.0$~GeV. This ensures  
that the parametrization scale is always above the charm mass when
charm is independently parametrized, and below it when it is
perturbatively generated. 

As in Ref.~\cite{Ball:2014uwa} we use $\alpha_s(m_Z)=0.118$ as a
default throughout the paper, though determinations have also been
performed for several different values of
$\alpha_s$ (see Sect.~\ref{sec:delivery}).
Pole heavy quark masses are used throughout, with the main motivation
that for the inclusive observables used for PDF determination
$\overline{\rm MS}$ masses are inappropriate, since they distort the perturbative
expansion in the threshold region~\cite{Ball:2016qeg}. 
The default values of the heavy quark pole masses are  $m_c=1.51$~GeV for
  charm and $m_b=4.92$~GeV for bottom,
  following the recommendation of the Higgs cross-section working
  group~\cite{deFlorian:2016spz};
  PDF sets for different charm mass values, corresponding to
 the $\pm 1$-sigma uncertainty band from
  Ref.~\cite{deFlorian:2016spz}, are also provided, see
Sect.~\ref{sec:delivery}.

\subsection{Fit quality}
\label{sec:fquality}

\begin{table}[h]
\begin{center}
  \scriptsize
      \renewcommand{\arraystretch}{1.10}
\begin{tabular}{|l|c|c|c||c|c|}
\hline
  & \multicolumn{3}{c||}{NNPDF3.1  } & \multicolumn{2}{c|}{NNPDF3.0 } 
\\Dataset  &  NNLO  & NLO & LO  & NNLO  & NLO  \\
\hline
\hline
NMC    &    1.30     &   1.35   &     3.25       &   1.29   &   1.36  \\    
SLAC    &    0.75      &   1.17   &   3.35       &   0.66   &   1.08  \\    
BCDMS    &    1.21      &   1.17   &  2.20        &   1.31   &   1.21  \\    
CHORUS    &    1.11      &   1.06   &   1.16      &   1.11   &   1.14  \\    
NuTeV dimuon    &    0.82      &   0.87   &  4.75         &   0.69  &   0.61  \\    
\hline
HERA I+II inclusive    &    1.16      &   1.14   &  1.77         &   1.25   &   1.20  \\      
HERA $\sigma_c^{\rm NC}$    &    1.45      &   1.15 ()   &  1.21           &  [1.61]   &   [2.57]  \\    
HERA $F_2^b$    &    1.11      &   1.08   &   11.2           &   [1.13]   &   [1.12]  \\    
\hline
\hline
DY E866 $\sigma^d_{\rm DY}/\sigma^p_{\rm DY}$ &    0.41     &   0.40   &     1.06       &   0.47   &   0.53  \\    
DY E886 $\sigma^p$    &    1.43     &   1.05   &    0.81       &   1.69   &   1.17  \\    
DY E605  $\sigma^p$   &    1.21     &   0.97   &  0.66      &   1.09   &   0.87  \\    
\hline
CDF $Z$ rap    &    1.48     &   1.619   &    1.54         &   1.55   &   1.28  \\    
CDF Run II $k_t$ jets    &    0.87      &   0.84   &  1.07           &   0.82   &   0.95  \\    
\hline
D0 $Z$ rap    &    0.60      &   0.67   &    0.65         &   0.61   &   0.59  \\    
D0 $W\to e\nu$  asy   &    2.70      &   1.59   &   1.75         &   [2.68]   &   [4.58]  \\    
D0 $W\to \mu\nu$  asy    &    1.56      &   1.52   &    2.16          &   [2.02]   &   [1.43]  \\    
\hline
\hline
ATLAS total   &    {\bf 1.09}      &   {\bf 1.36}   &    {\bf 5.34}         &   {\bf 1.92 }  & {\bf 1.98}  \\    
ATLAS $W,Z$ 7 TeV 2010    &    0.96     &   1.04   &    2.38         &   1.42   &   1.39  \\    
ATLAS high-mass DY 7 TeV    &    1.54     &   1.88   &   4.05         &   1.60   &   2.17  \\    
ATLAS low-mass DY 2011    &    0.90      &   0.69   &   2.86         &   [0.94]   &   [0.81]  \\    
ATLAS $W,Z$ 7 TeV 2011    &    2.14     &   3.70   &    27.2       &   [8.44]   &   [7.6]  \\    
ATLAS jets 2010 7 TeV     &    0.94      &   0.92   &  1.22        &   1.12   &   1.07  \\    
ATLAS jets 2.76 TeV     &    1.03      &   1.03   &     1.50     &   1.31   &   1.32  \\    
ATLAS jets 2011 7 TeV     &    1.07     &   1.12   &   1.59       &   [1.03]   &   [1.12]  \\    
ATLAS $Z$ $p_T$ 8 TeV $(p_T^{ll},M_{ll})$       &    0.93    &   1.17   &   -         &   [1.05]  &  [1.28]  \\    
ATLAS $Z$ $p_T$ 8 TeV $(p_T^{ll},y_{ll})$    &    0.94      &   1.77   &   -         &   [1.19]   &   [2.49]  \\    
ATLAS $\sigma_{tt}^{\rm tot}$     &    0.86      &   1.92   &  53.2           &   0.67   &   1.07  \\    
ATLAS $t\bar{t}$ rap    &    1.45     &   1.31   &  1.99       &   [3.32]   &   [1.50]  \\
\hline
CMS total   &   {\bf  1.06}    &   {\bf 1.20 }  &    {\bf 2.13}        &   {\bf 1.19 }  &
{\bf 1.33}  \\    
CMS $W$ asy 840 pb     &    0.78     &   0.86   &    1.55       &   0.73   &   0.85  \\    
CMS $W$ asy 4.7 fb     &    1.75     &   1.77   &   3.16      &   1.75   &   1.82  \\    
CMS $W+c$ tot    &    -     &   0.54   &    16.5          &   -   &   0.93  \\    
CMS $W+c$ ratio    &    -      &   1.91   &    3.21         &   -   &   2.09  \\    
CMS Drell-Yan 2D 2011    &    1.27     &   1.23   &  2.15           &   1.20   &   1.19  \\    
CMS $W$ rap 8 TeV    &    1.01      &   0.70   &    4.32         &   [1.24]   &   [0.96]  \\    
CMS jets 7 TeV 2011     &    0.84      &   0.84   &  0.93            &   1.06   &   0.98  \\    
CMS jets 2.76 TeV    &    1.03      &   1.01   &    1.09          &   [1.22]   &   [1.18]  \\    
CMS $Z$ $p_T$ 8 TeV $(p_T^{ll},M_{ll})$    &    1.32    &   3.65   &   -         &   [1.59]   &   [3.86]  \\    
CMS $\sigma_{tt}^{\rm tot}$     &    0.20      &   0.59   &  53.4            &   0.56   &   0.10  \\    
CMS $t\bar{t}$ rap    &    0.94      &   0.96   &     1.32           &   [1.15]   &   [1.01]  \\
\hline 
LHCb total   &  {\bf  1.47}      &   {\bf 1.62}   &   {\bf 5.16}       &   {\bf 2.11}   &   {\bf 2.67}  \\
LHCb $Z$ 940 pb    &    1.49      &   1.27   &        2.51      &   1.29   &   0.91  \\    
LHCb $Z\to ee$ 2 fb    &    1.14      &   1.33   &    6.34         &   1.21   &   2.31  \\    
LHCb $W,Z \to \mu$ 7 TeV    &    1.76      &   1.60   &   4.70          &   [2.59]   &   [2.36]  \\    
LHCb $W,Z \to \mu$ 8 TeV    &    1.37     &   1.88   &   7.41            &   [2.40]   &   [3.74]  \\    
\hline
\hline
{\bf Total dataset}   &  {\bf  1.148}      & {\bf  1.168}   & {\bf  2.238}        & {\bf  1.284}   & {\bf  1.307}  \\    
\hline
\end{tabular}
\caption{\small
\label{tab:chi2tab_31-nlo-nnlo-30}
  The values of $\chi^2/N_{\rm dat}$ for the global fit and for 
all the datasets included in the NNPDF3.1 LO, NLO
  and NNLO PDF determinations. Values obtained using the NNPDF3.0 NLO and NNLO
  PDFs  are also shown: numbers in brackets correspond 
   to data not fitted in NNPDF3.0. Note that NNPDF3.0 values are
   produced using NNPDF3.1 theory settings, and are thus somewhat
   worse than those quoted in Ref.~\cite{Ball:2014uwa}.
}
\end{center}
\end{table}


In Table~\ref{tab:chi2tab_31-nlo-nnlo-30} we provide
values of $\chi^2/N_{\rm dat}$ both for the global fit and individually for all the datasets included in the
NNPDF3.1 LO, NLO
  and NNLO PDF determinations. These are compared to their NNPDF3.0 NLO and NNLO
  counterparts. The $\chi^2$
  is computed using the covariance matrix including all correlations,
  as published by the corresponding experiments.
 Inspection of this table
shows that the fit quality improves from LO to NLO to NNLO: not only
is there a significant improvement between LO and NLO, but there is also a marked improvement when going from NLO
to NNLO. It is interesting to note that this was not the case in
NNPDF3.0 where the fit quality at NNLO was in fact slightly worse than
at NLO (see Table~9 of Ref.~\cite{Ball:2014uwa}). This reflects the increased proportion of 
hadronic processes included in NNPDF3.1, for which NNLO corrections
are often substantial, and also, possibly, methodological improvements.

The overall fit quality with NNPDF3.1 is  rather better than that
obtained using NNPDF3.0 PDFs. 
Whereas this is clearly expected for LHC measurements which
were not included in NNPDF3.0, it is
interesting to note that the HERA
measurements which were already present in 3.0 (though in slightly different
uncombined form) are also better fitted. 
 The quality of the description with the
previous NNPDF3.0 PDFs is nevertheless quite acceptable for all the new data, 
indicating a general compatibility between NNPDF3.0 and NNPDF3.1. 
Note that NNPDF3.0 values in
Table~\ref{tab:chi2tab_31-nlo-nnlo-30} are computed using the NNPDF3.1
theory settings, thus in particular with different 
values of the heavy quark masses than those used in the NNPDF3.0 PDF
determination. Because of this, the NNPDF3.0 fit quality
shown shown in Table~9 of Ref.~\cite{Ball:2014uwa} is slightly better than 
that shown in Table~\ref{tab:chi2tab_31-nlo-nnlo-30}, yet even so 
the fit quality of NNPDF3.1 is better still. Specifically, 
concerning HERA data, the fit quality of
NNPDF3.0 with consistent theory settings can be read off Table~7 of
Ref.~\cite{Czakon:2016olj}: it corresponds to  $\chi^2/N_{\rm
  dat}=1.21$ thereby showing that indeed NNPDF3.1 provides a better
description. The reasons for this improvement will be discussed in
Sect.~\ref{sec:results-mc} below.

For many of the new LHC measurements, achieving a good description of the data
is only possible at NNLO. The
total $\chi^2/N_{\rm dat}$ for the ATLAS, CMS and
LHCb experiments is 1.09, 1.06 and 1.47 respectively at NNLO,
compared to 1.36, 1.20 and 1.62 at NLO. The datasets exhibiting
the largest improvement when going from NLO to NNLO are those with the
smallest experimental uncertainties. For example the
ATLAS $W,Z$ 2011 rapidity distributions
(from 3.70 to 2.14), the CMS 8 TeV $Z$ $p_T$ distributions
(from 3.65 to 1.32) and the LHCb 8 TeV $W,Z\to \mu$
rapidity distributions (from 1.88 to 1.37); in these experiments
uncorrelated statistical uncertainties are typically at the
sub-percent level.
It is likely that this trend will continue as LHC measurements become more precise.



\clearpage

\subsection{Parton distributions}
\label{sec:PDFcomparisons}

\begin{figure}[t]
\begin{center}
  \includegraphics[scale=0.75]{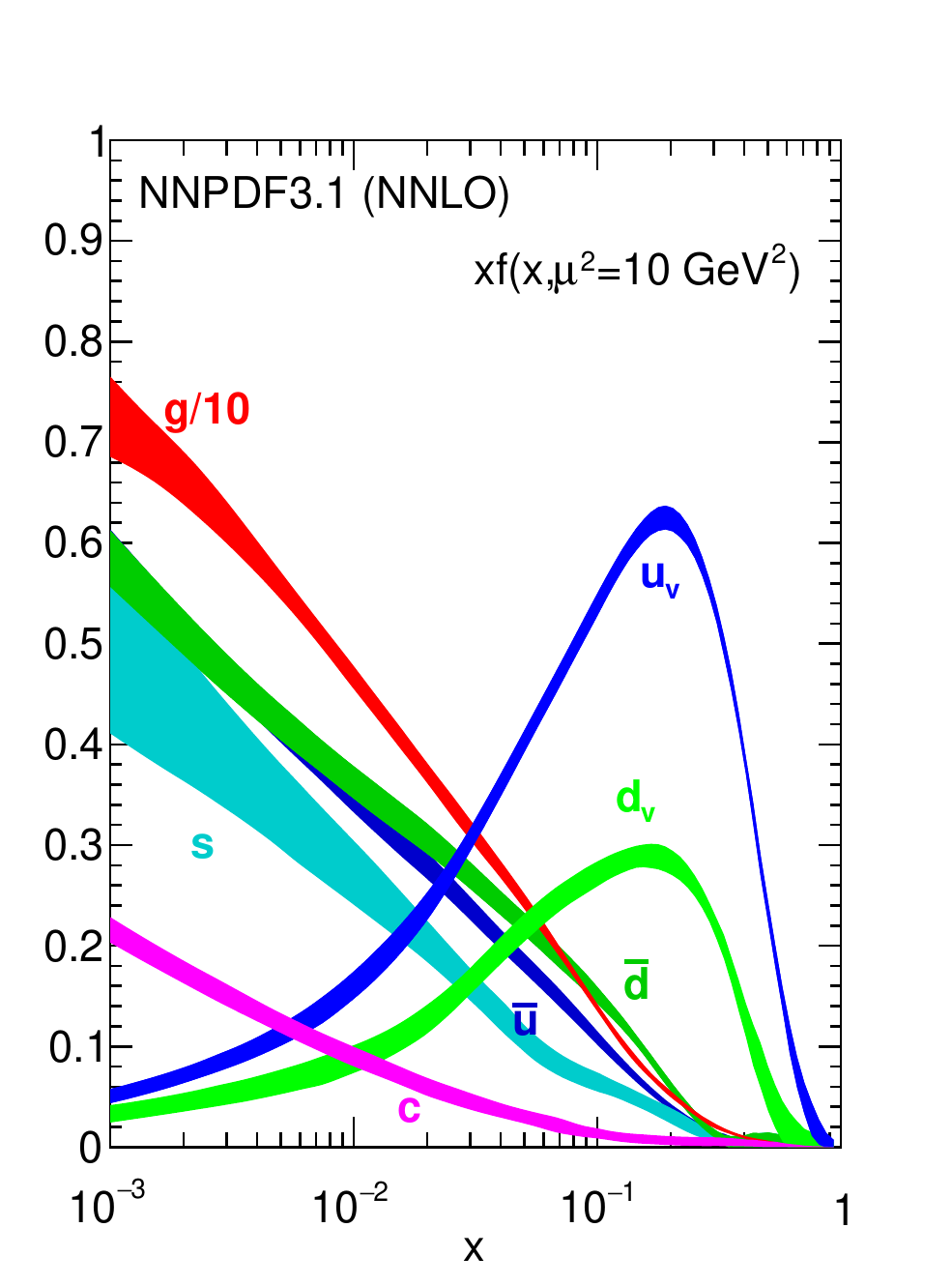}
  \includegraphics[scale=0.75]{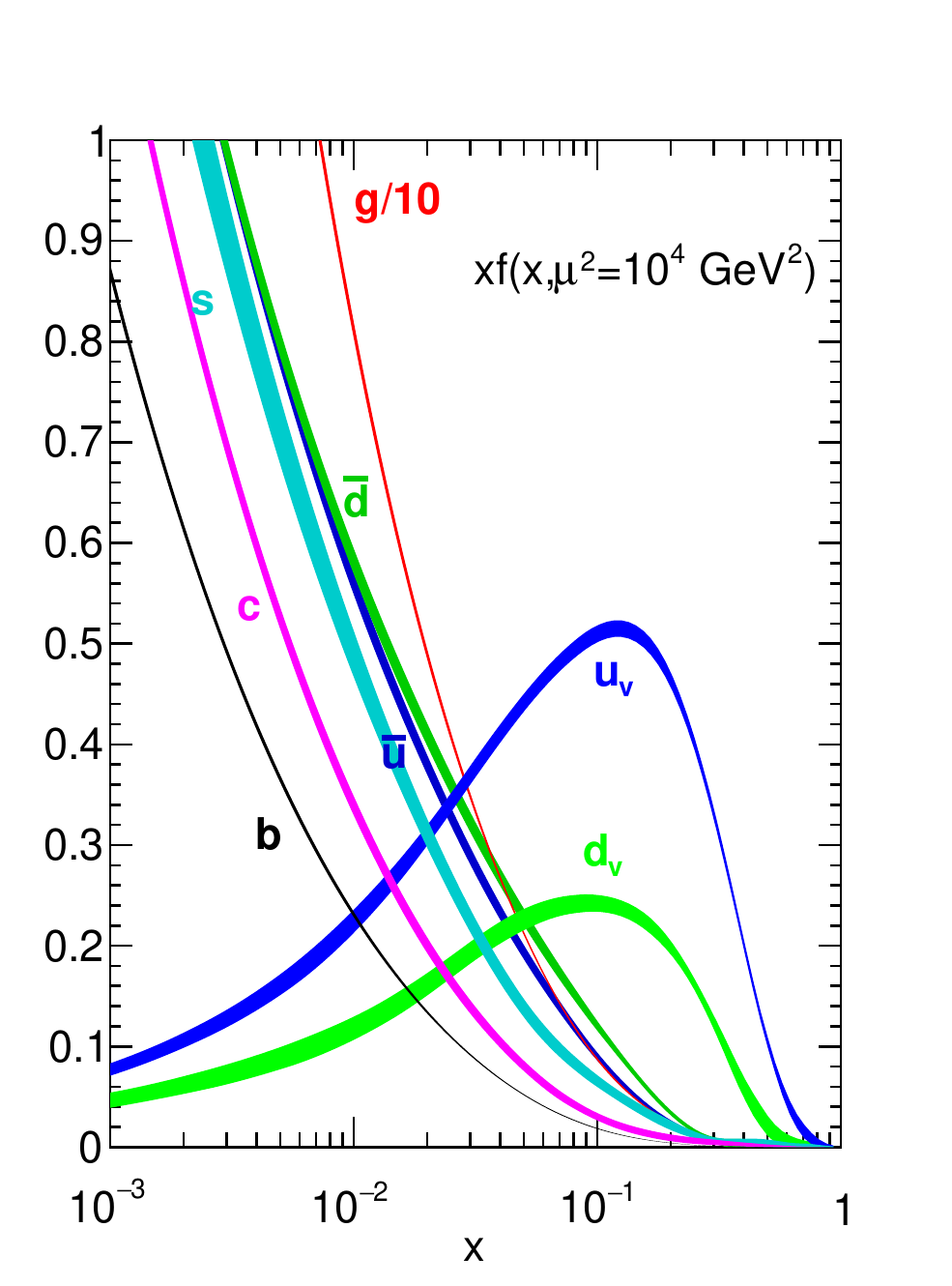}
  \caption{\small The NNPDF3.1 NNLO PDFs, evaluated
    at $\mu^2=10~{\rm GeV}^2$ (left) and $\mu^2=10^4~{\rm GeV}^2$ (right). 
    \label{fig:nnlopdfs}
  }
\end{center}
\end{figure}

We now inspect the baseline NNPDF3.1 parton distributions, and
compare them to NNPDF3.0 
and to  MMHT14~\cite{Harland-Lang:2014zoa},
CT14~\cite{Dulat:2015mca} and
ABMP16~\cite{Alekhin:2017kpj}.
The NNLO NNPDF3.1 PDFs are displayed in Fig.~\ref{fig:nnlopdfs}. It can be seen 
that although charm is now independently parametrized, it is still known more precisely than the strange PDF.
The most precisely determined PDF over most of the experimentally accessible range of $x$ is now 
the gluon, as will be discussed in more detail below.

\begin{figure}[t]
\begin{center}
  \includegraphics[scale=1]{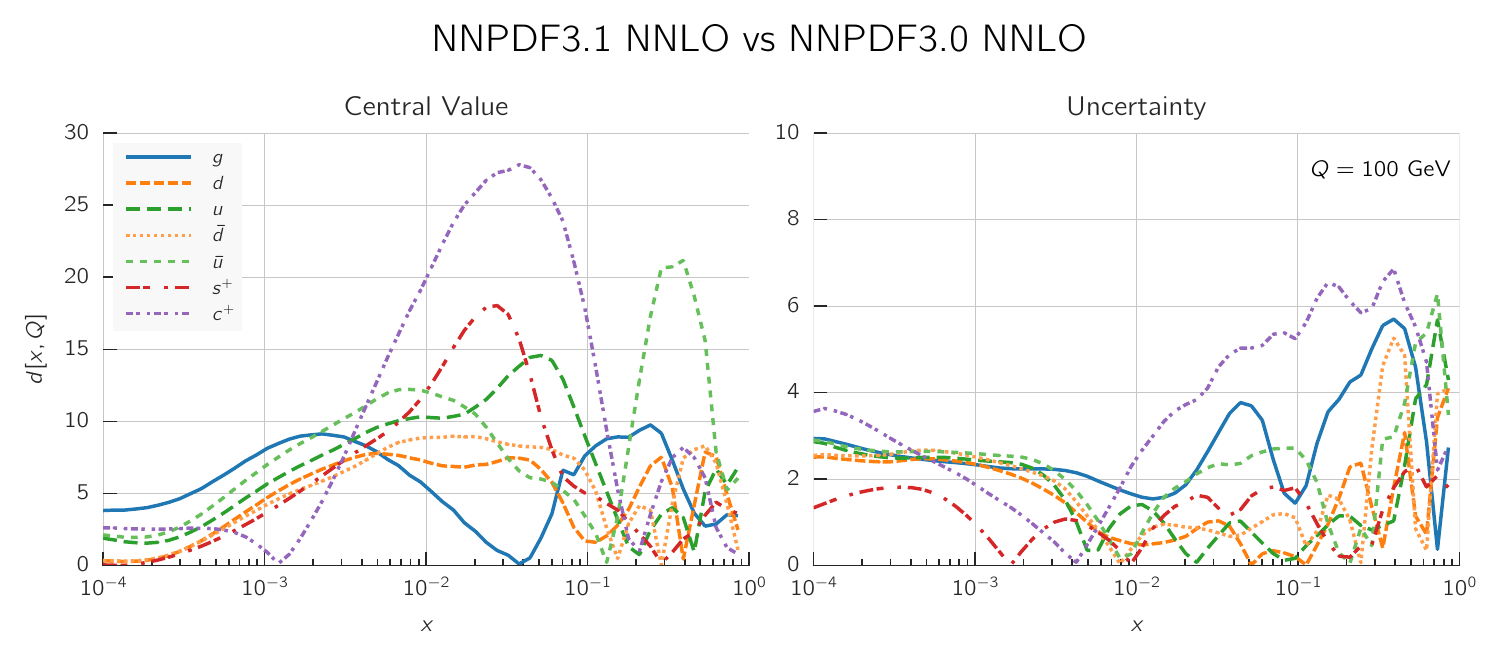}
  \caption{\small Distances between the central
    values (left) and the uncertainties (right) of
    the NNPDF3.0 and NNPDF3.1 NNLO PDF sets, evaluated
    at $Q=100$ GeV. Note the different in scale on the $y$ axis between
    the two plots.
    \label{fig:distances_31_vs_30}
  }
\end{center}
\end{figure}

In Fig.~\ref{fig:distances_31_vs_30} we show the  distance between the NNPDF3.1 and NNPDF3.0 PDFs.
According to  the definition of the distance given in Ref.~\cite{Ball:2010de}, $d\simeq 1$ corresponds to
statistically equivalent sets. Comparing two sets with $N_{\rm rep}=100$ replicas,
a distance of $d\simeq 10$ corresponds to a difference of one-sigma in units of the corresponding variance,
both for central values and for PDF uncertainties. For clarity only the distance between the total
strangeness distributions $s^+=s+\bar s$ is shown, rather than the strange and
antistrange separately.
We find important differences both at the level of central values and
of PDF errors for all flavors and in the entire range of $x$.
The largest distance is found for charm, which is independently
parametrized in NNPDF3.1,
while it was not in NNPDF3.0. Aside from this, the most significant
distances are seen in 
light quark distributions at large $x$ and strangeness at medium $x$. 

\begin{figure}[t]
  \begin{center}
      \includegraphics[scale=0.32]{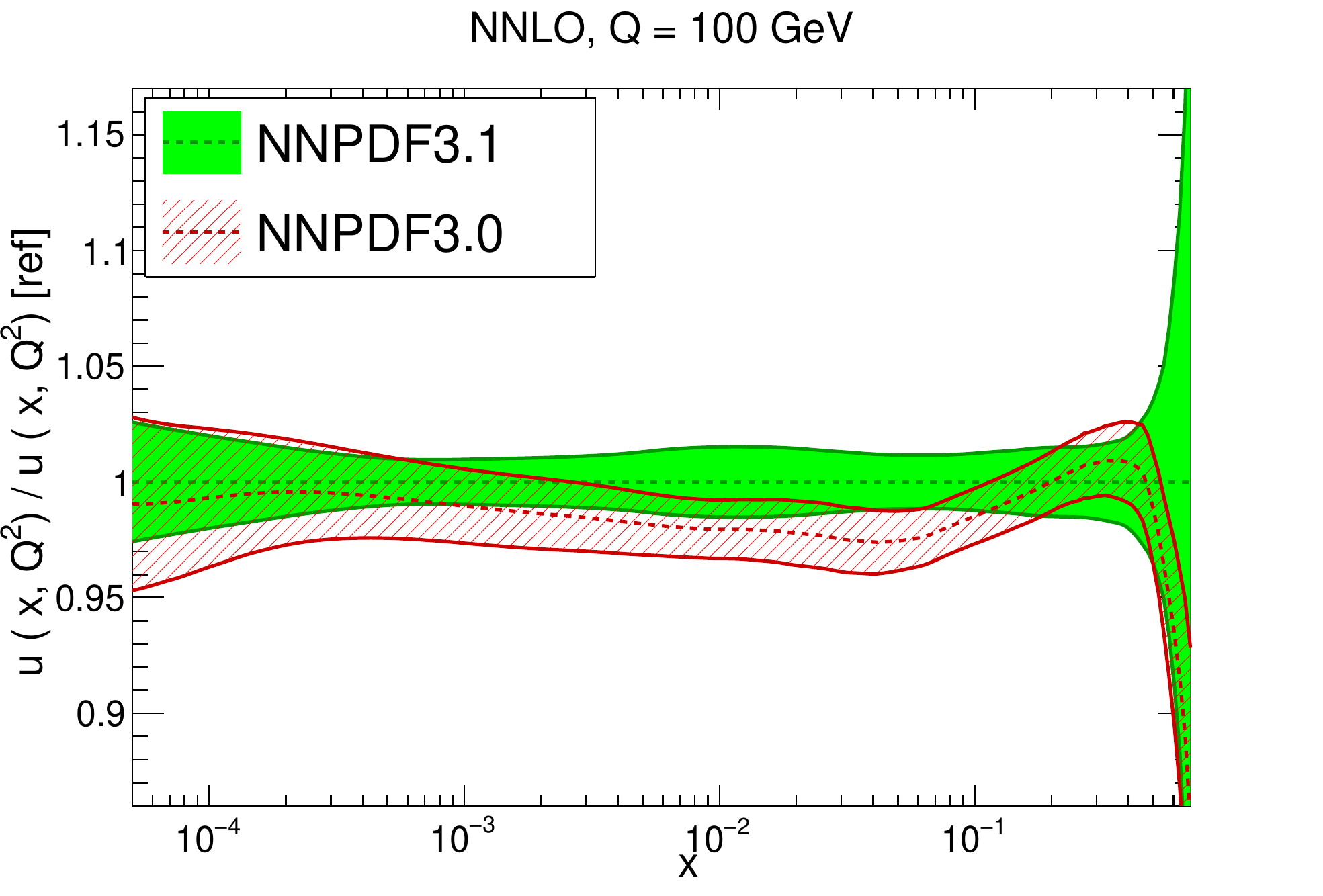}
    \includegraphics[scale=0.32]{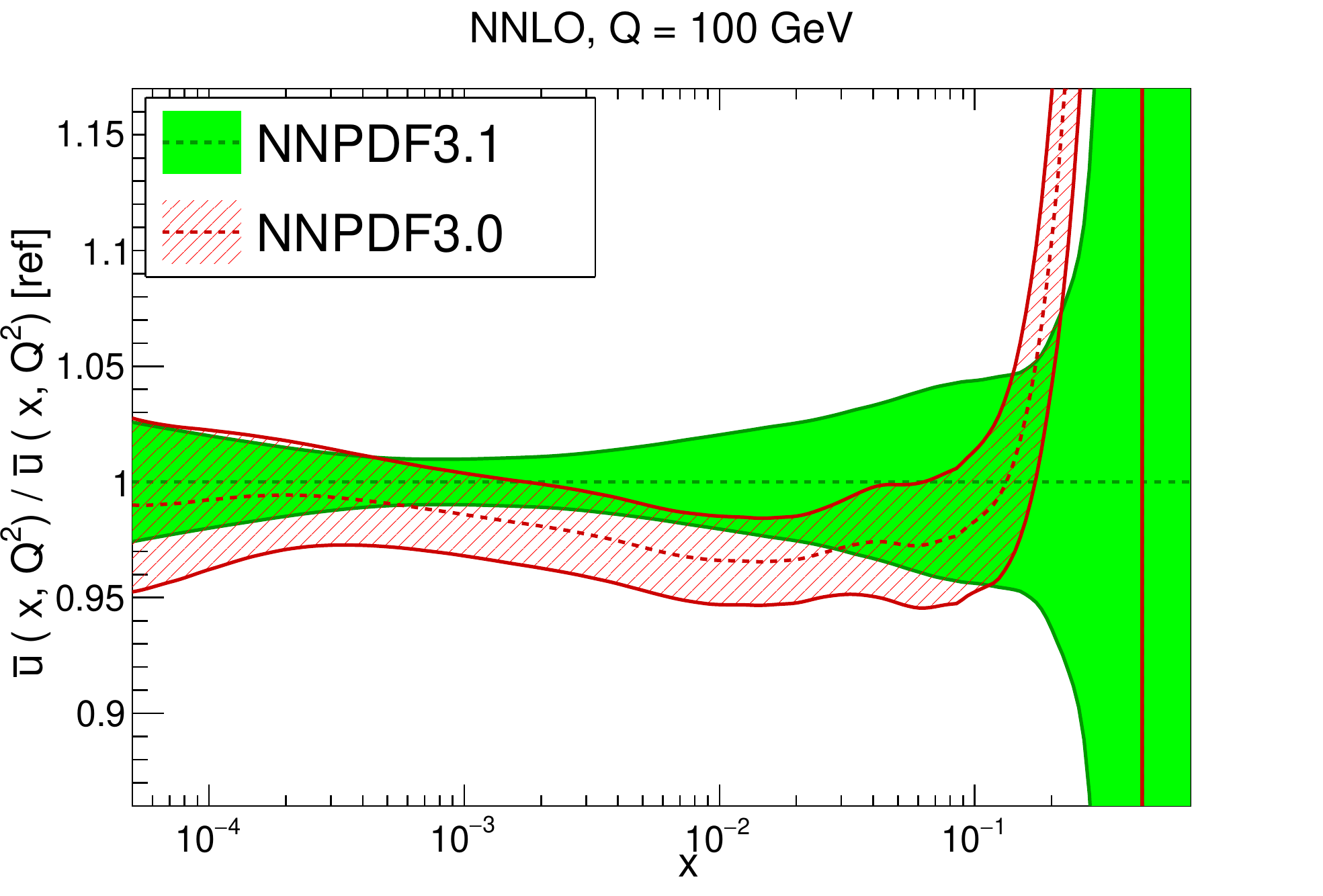}
    \includegraphics[scale=0.32]{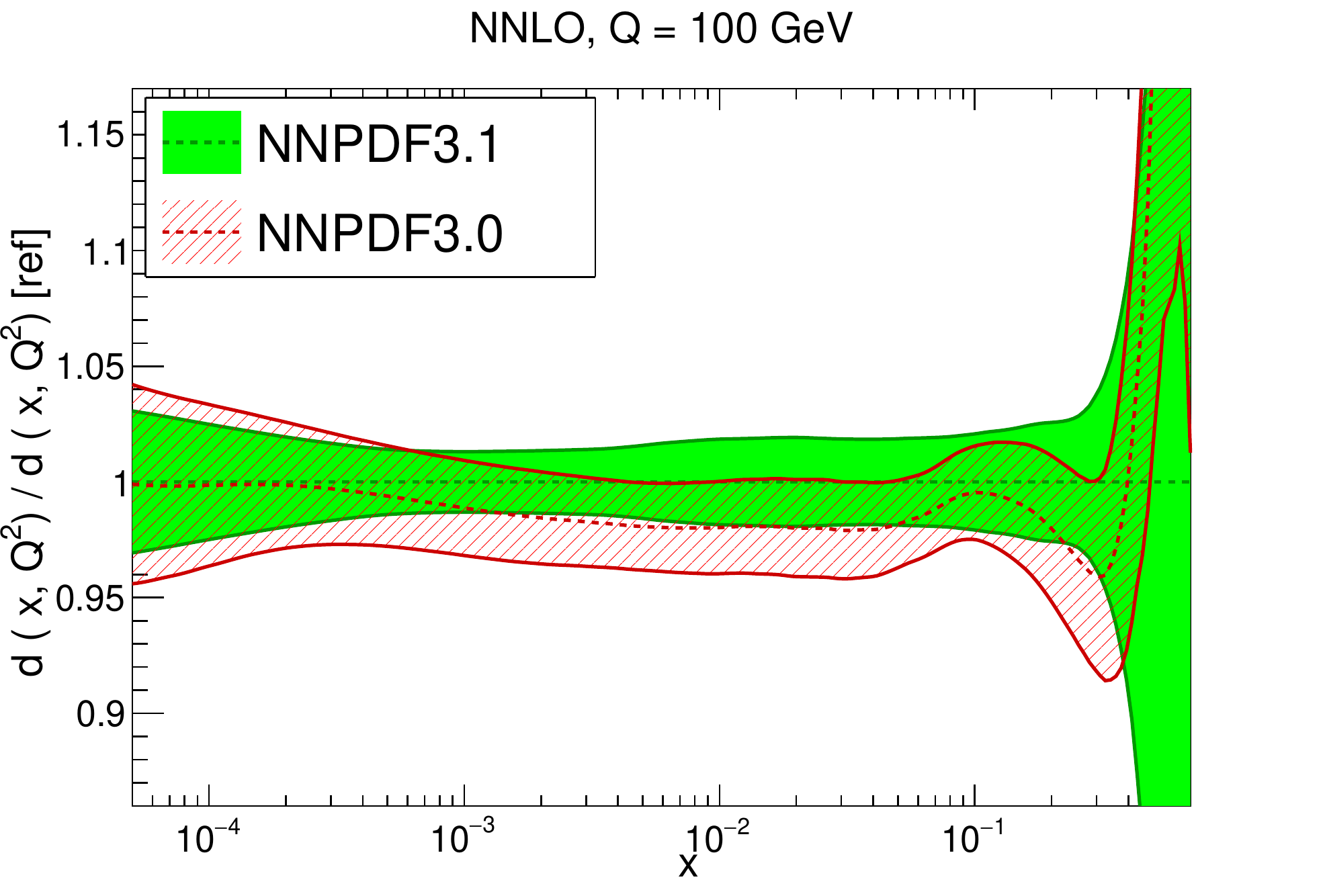}
    \includegraphics[scale=0.32]{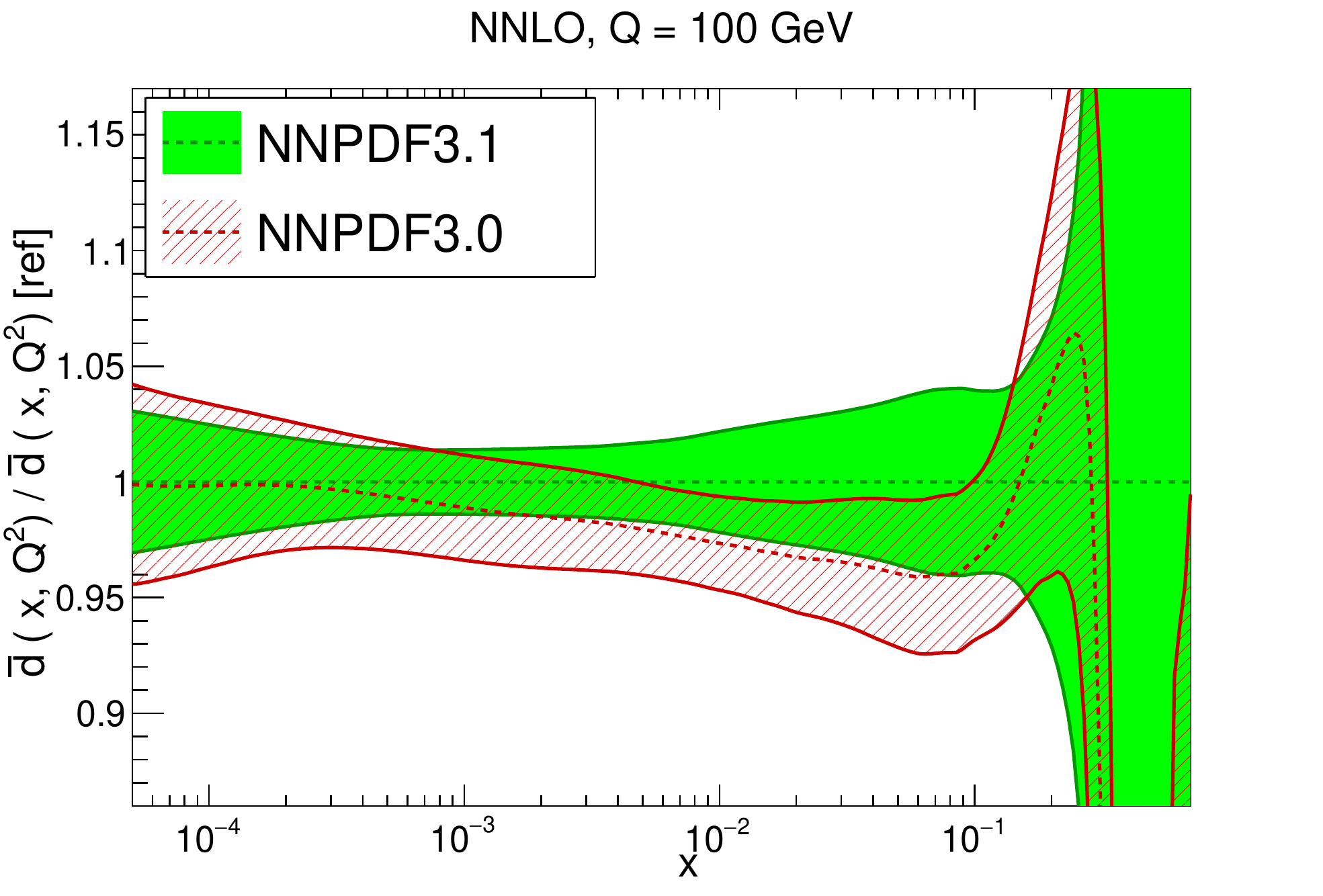}
  \includegraphics[scale=0.32]{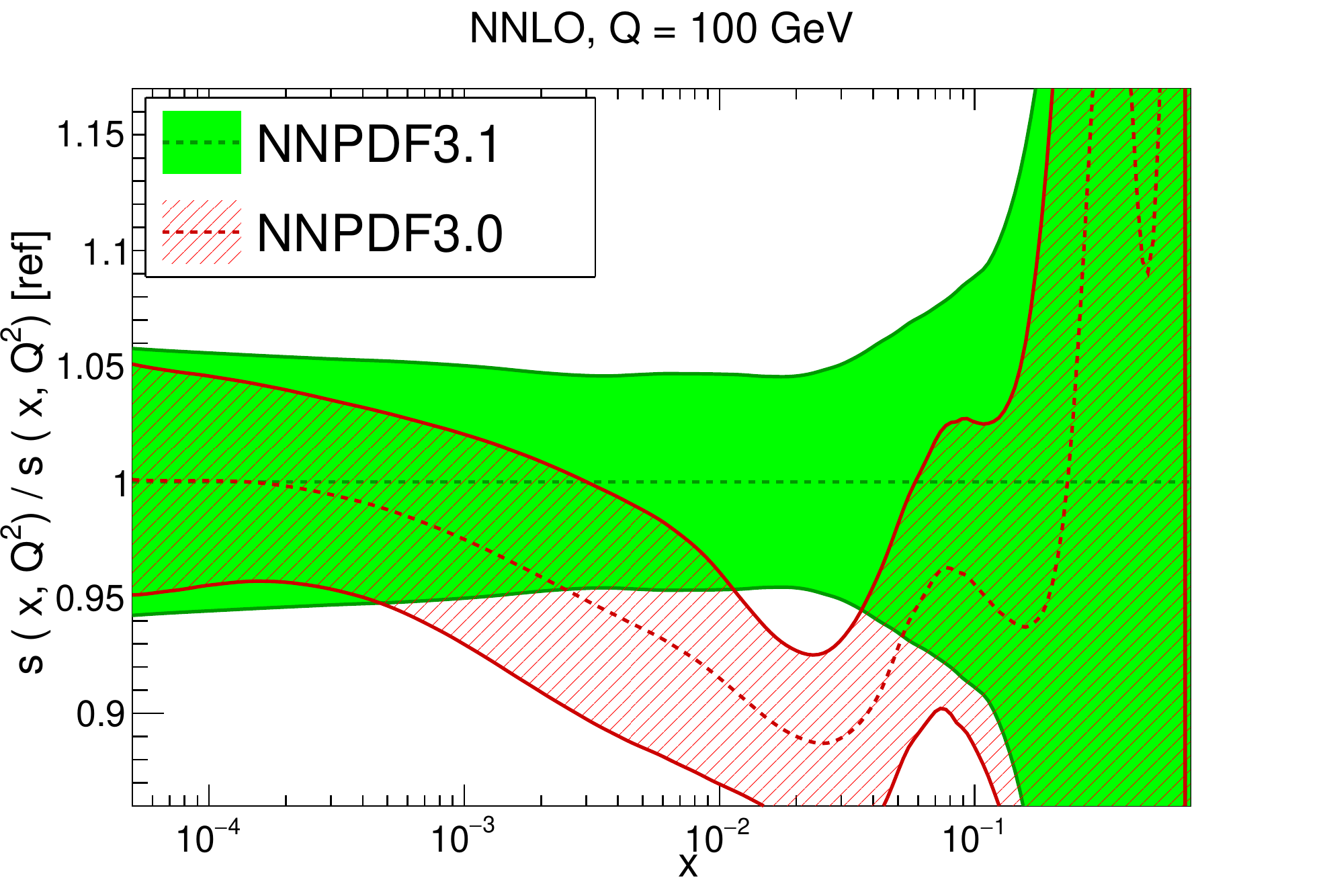}
  \includegraphics[scale=0.32]{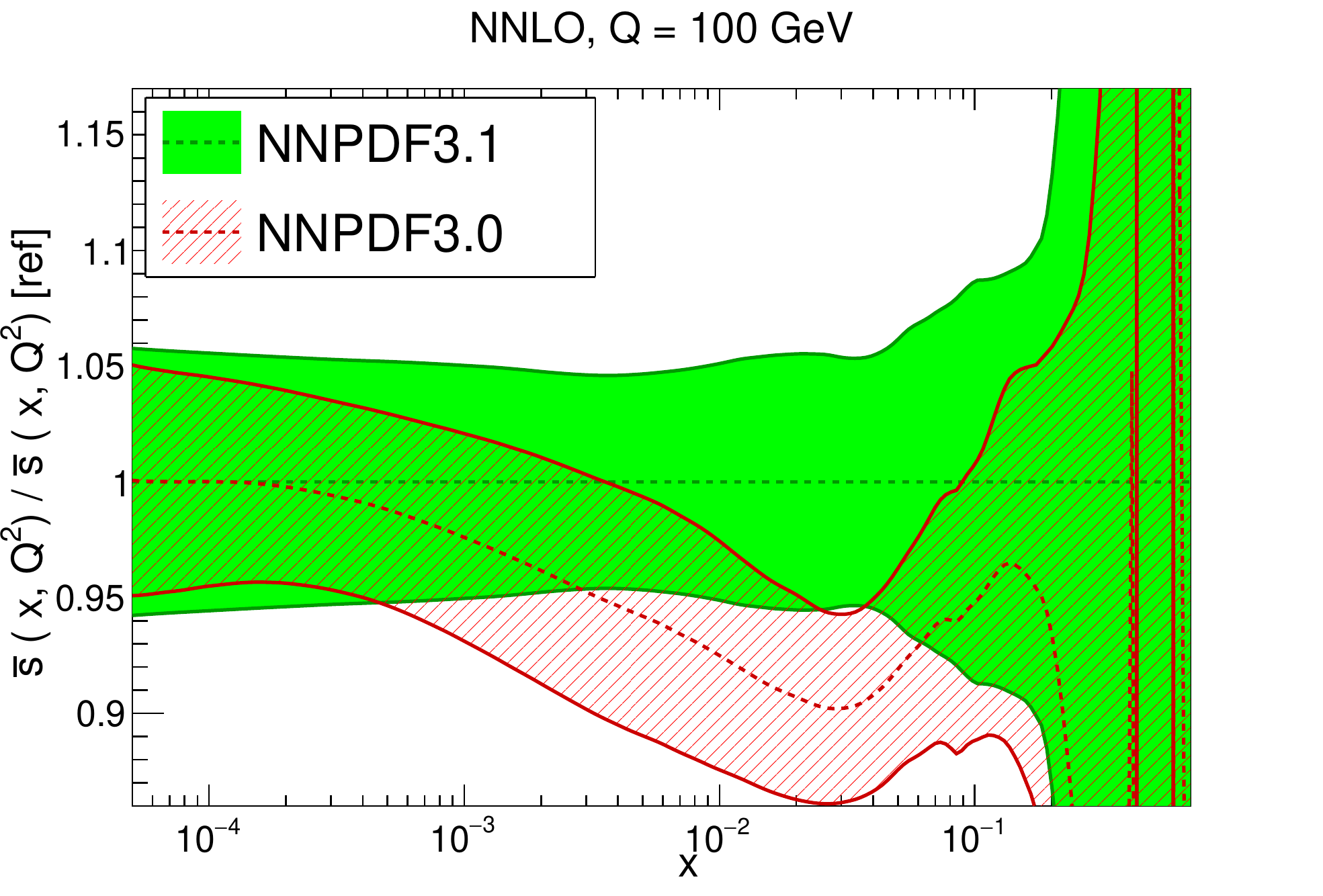}
  \includegraphics[scale=0.32]{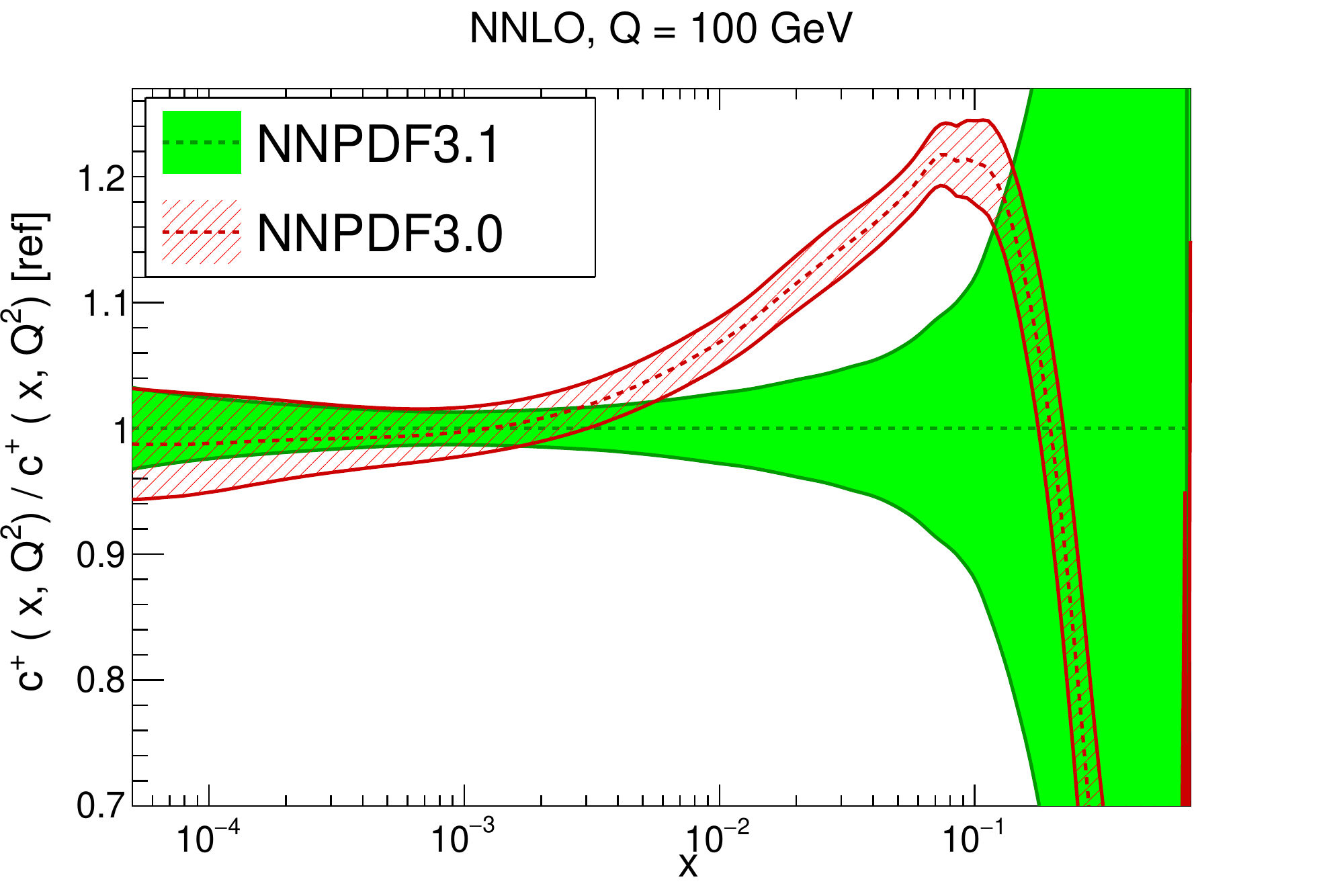}
  \includegraphics[scale=0.32]{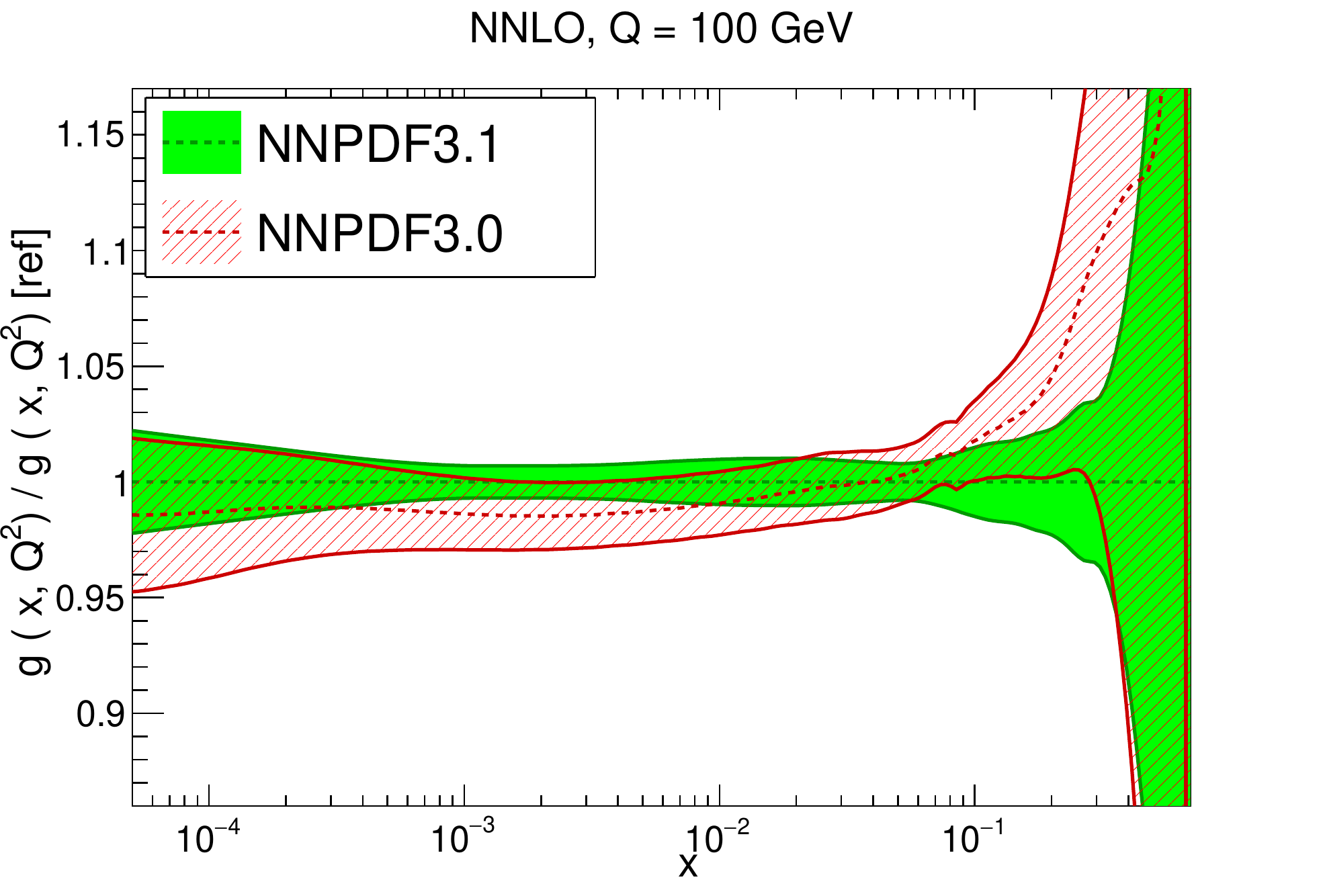}
    \caption{\small Comparison between NNPDF3.1 and NNPDF3.0 NNLO PDFs
      at $Q=100$ GeV. From top to bottom  up and antiup, down
      and antidown, strange and antistrange, charm and gluon are shown.
    \label{fig:31-nnlo-vs30}
  }
\end{center}
\end{figure}

In Fig.~\ref{fig:31-nnlo-vs30} we compare the full set of NNPDF3.1 NNLO PDFs with NNPDF3.0.
The NNPDF3.1 gluon is slightly larger than its NNPDF3.0 counterpart
in the $x\lsim 0.03$ region, while it becomes smaller at larger $x$,
with significantly reduced PDF errors.
The NNPDF3.1 light quarks and strangeness are larger than 3.0
at intermediate $x$, with the largest deviation seen for the strange
and antidown PDFs, while at both small and large $x$
there is good agreement between the two PDF determinations. The best-fit
charm PDF of NNPDF3.1 is significantly smaller in the intermediate-$x$ region 
compared to the perturbative charm of NNPDF3.0, while at larger $x$ it has 
significantly increased uncertainty.

\begin{figure}[t]
  \begin{center}
    \includegraphics[scale=0.32]{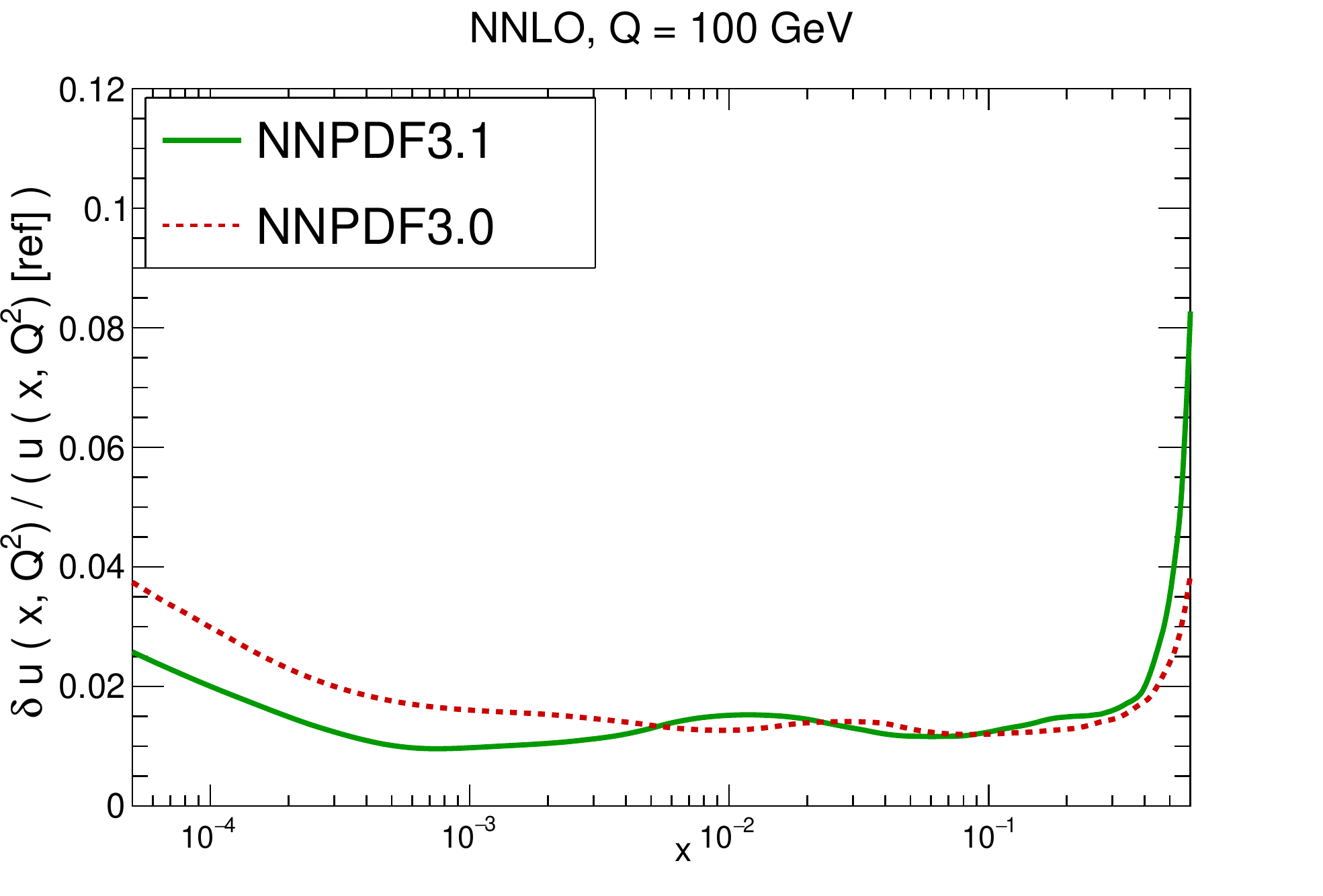}
    \includegraphics[scale=0.32]{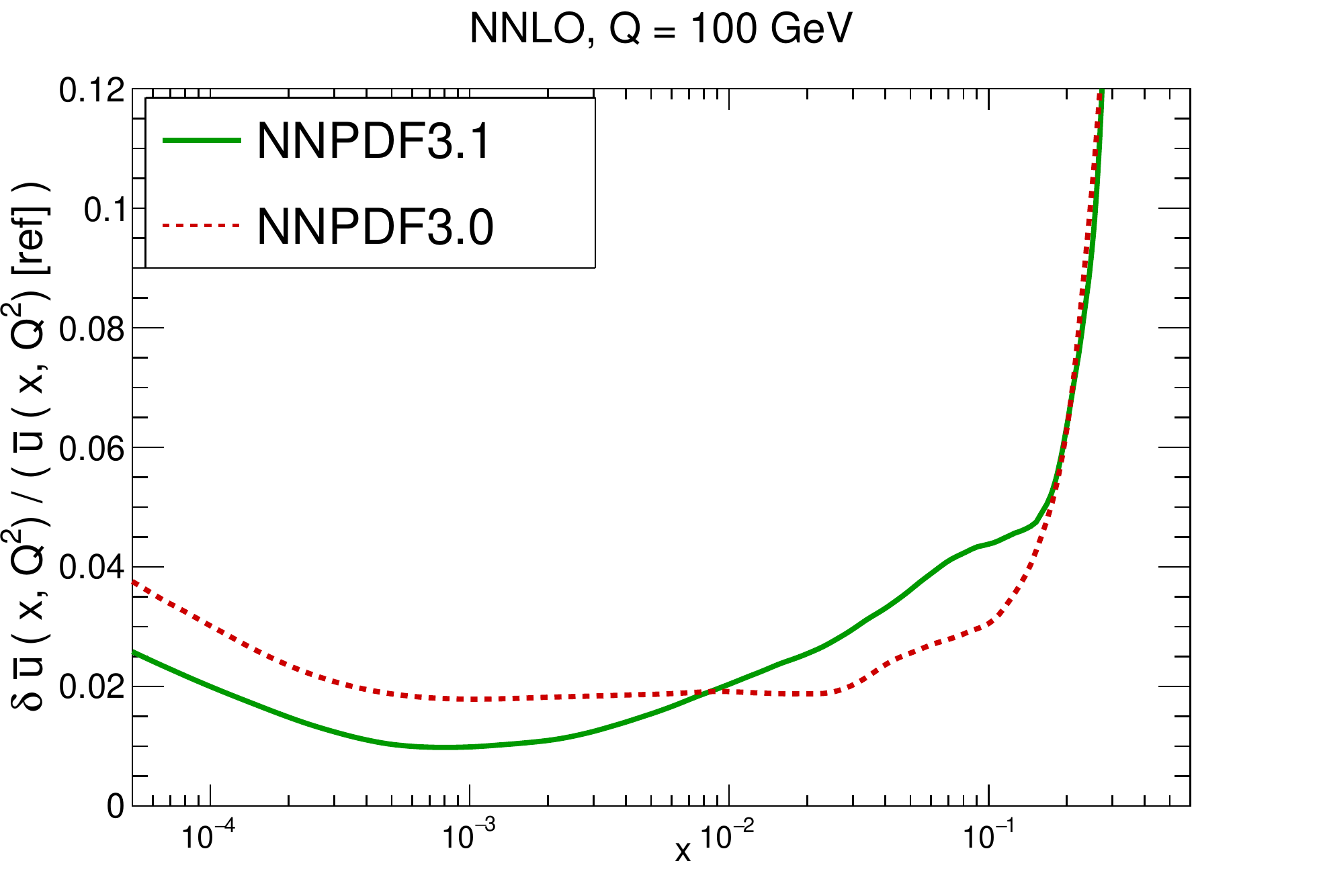}
    \includegraphics[scale=0.32]{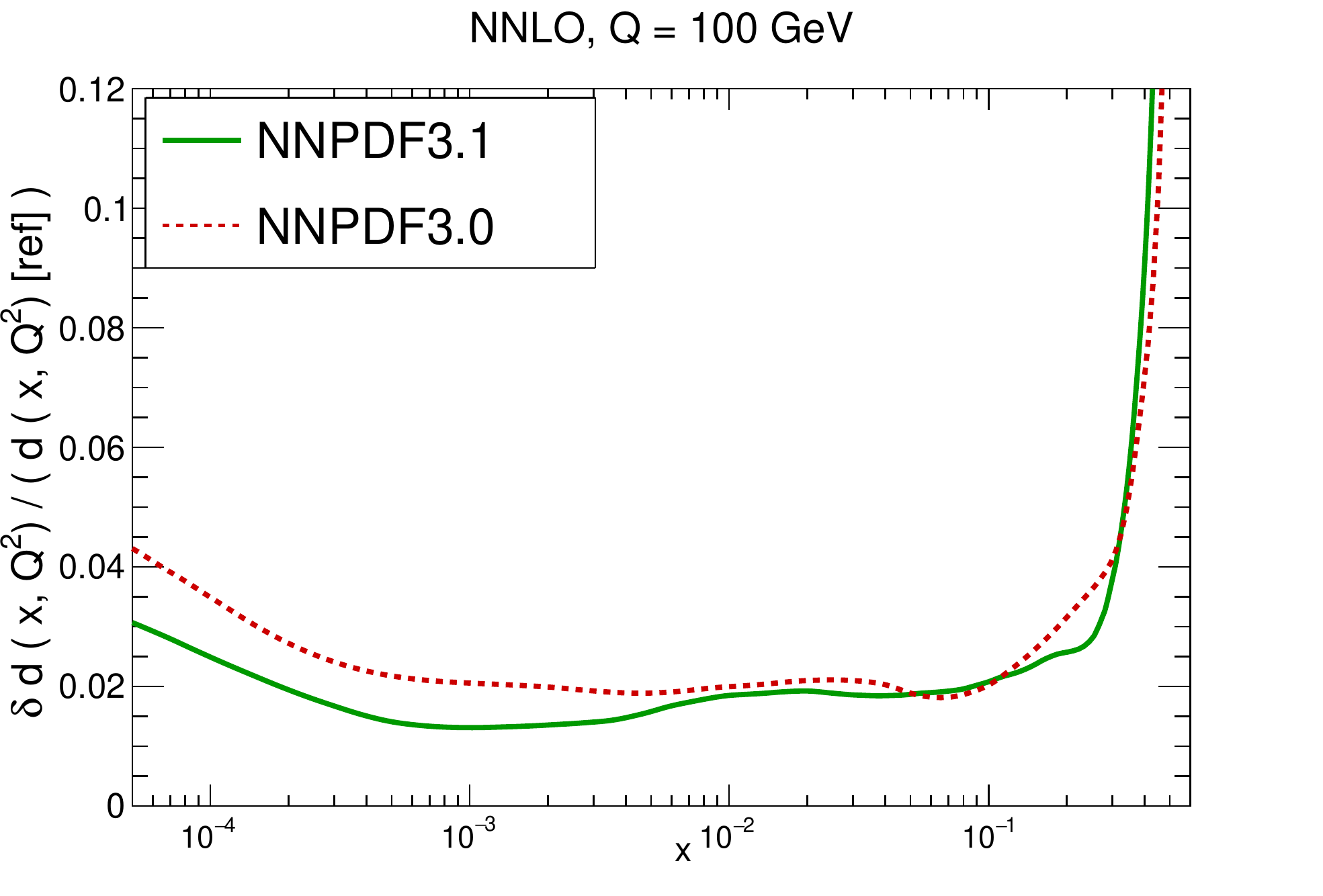}
    \includegraphics[scale=0.32]{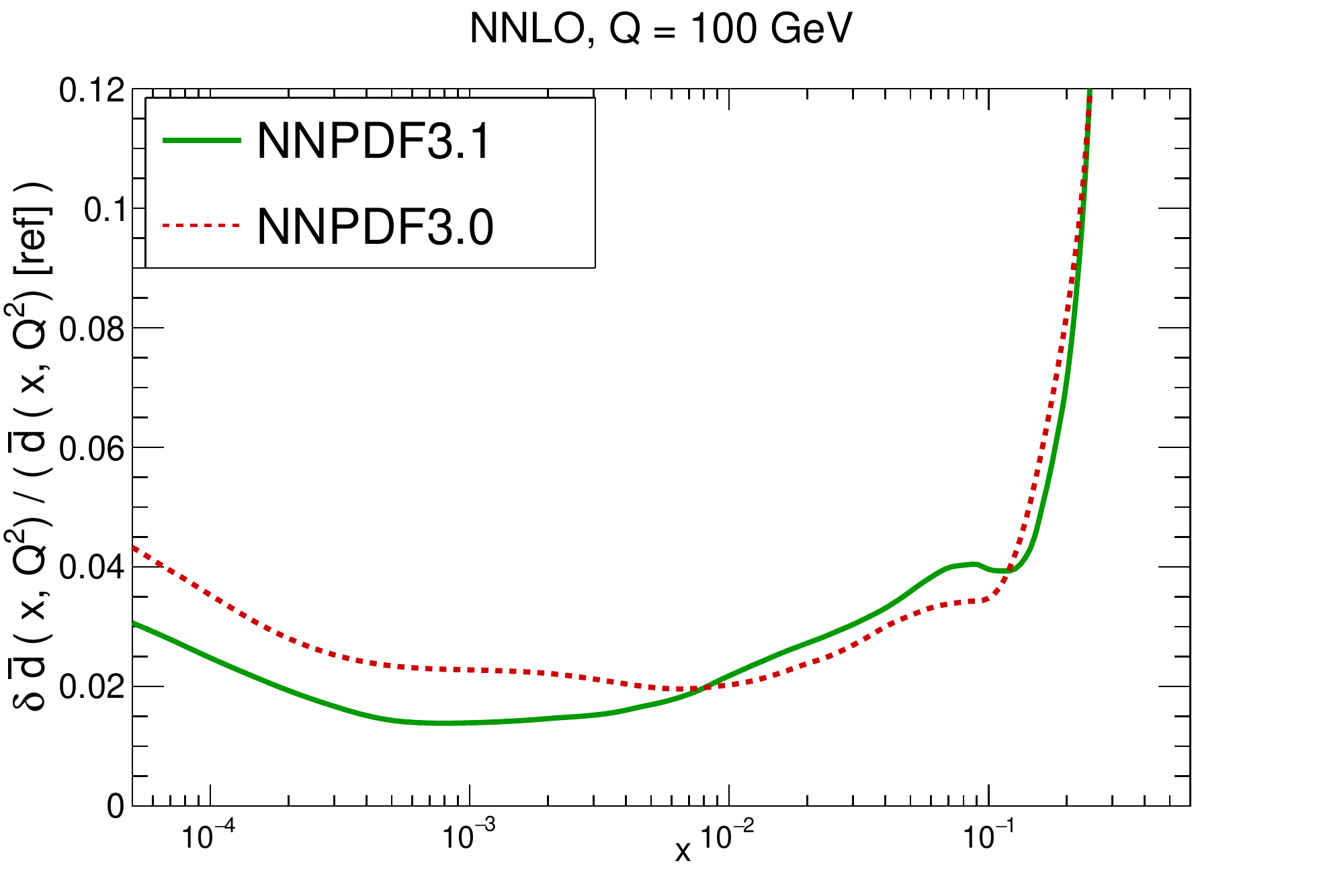}
  \includegraphics[scale=0.32]{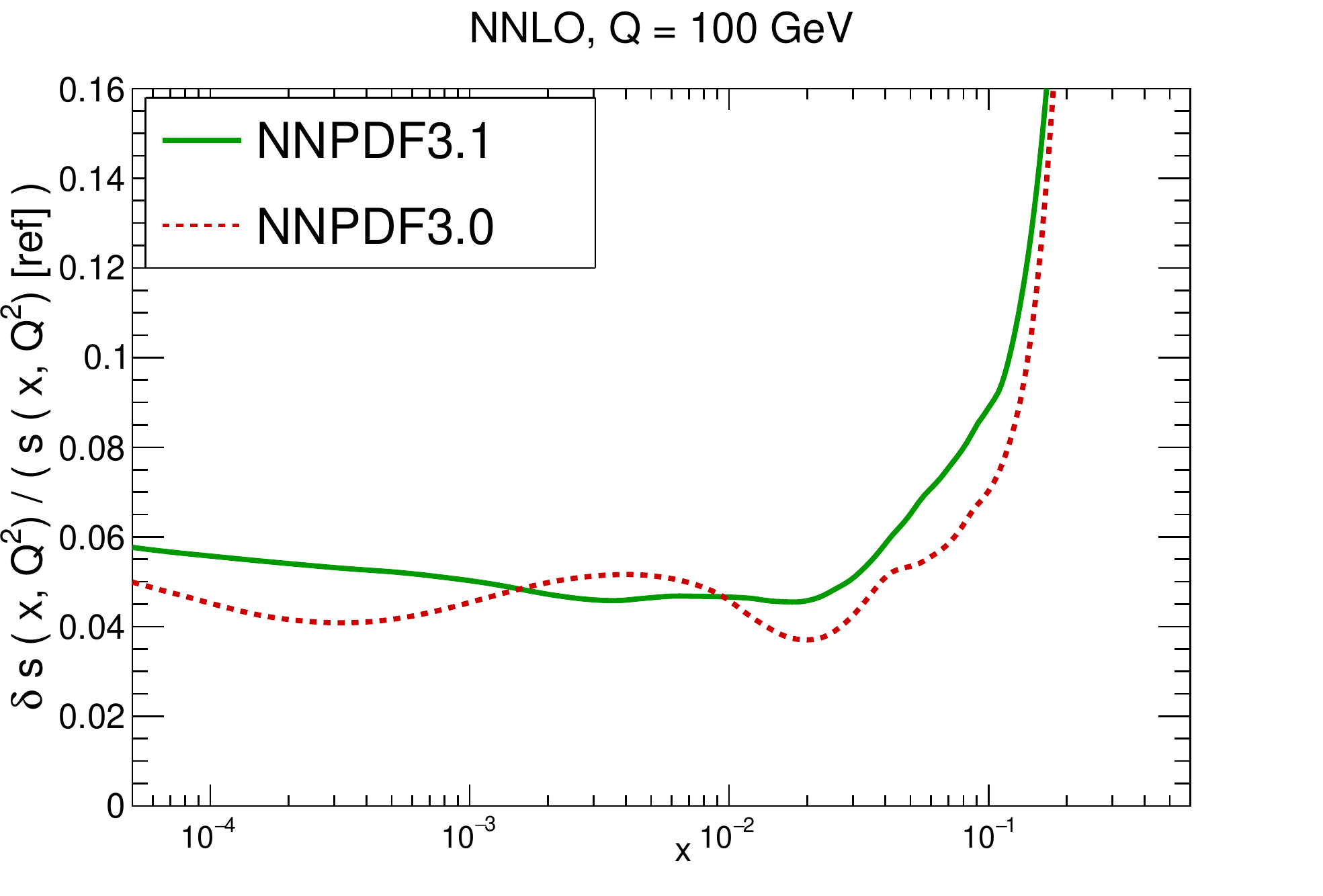}
  \includegraphics[scale=0.32]{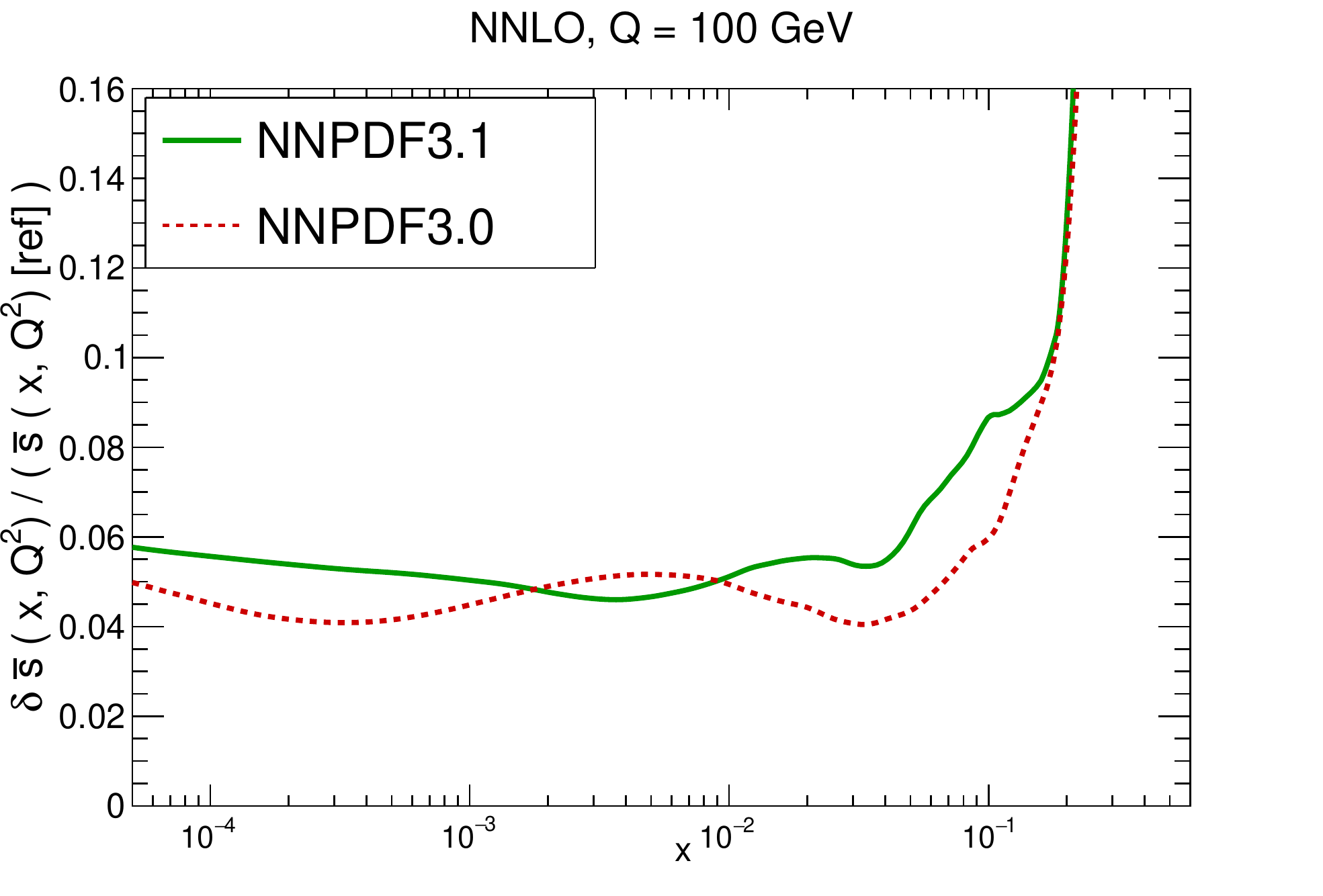}
  \includegraphics[scale=0.32]{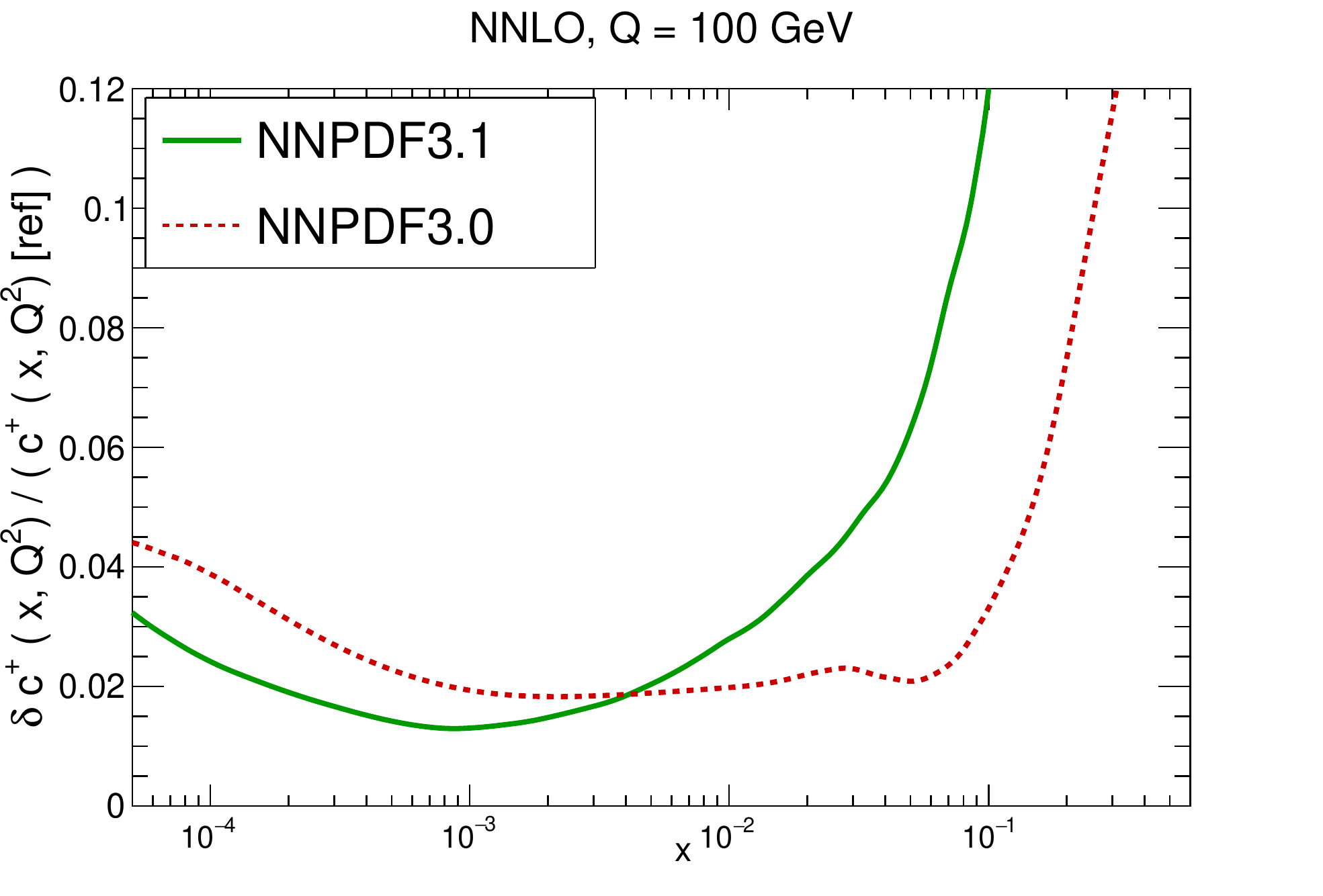}
  \includegraphics[scale=0.32]{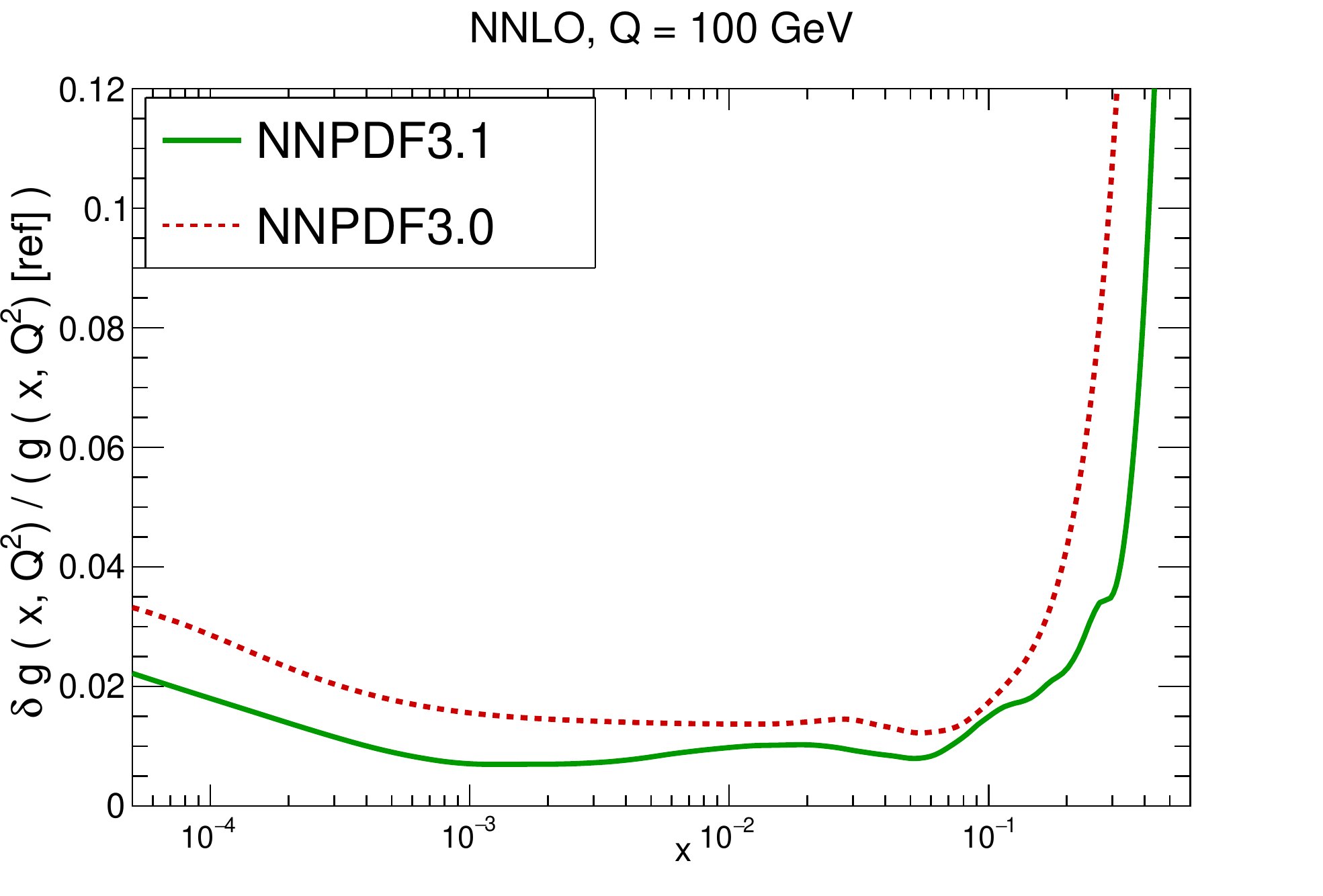}
  \caption{\small Comparison between NNPDF3.1 and NNPDF3.0 relative
    PDF uncertainties at $Q=100$; the PDFs are as in
    Fig.~\ref{fig:31-nnlo-vs30}. The uncertainties shown are all
    normalized to the NNPDF3.1 central value.
    \label{fig:ERR-31-nnlo-vs30}
  }
\end{center}
\end{figure}

A detailed comparison of the corresponding uncertainties is presented
in Fig.~\ref{fig:ERR-31-nnlo-vs30}, where we compare the relative
uncertainty on each PDF, defined as the ratio of
the one-sigma PDF uncertainty to the central value of the
NNPDF3.1 set. NNPDF3.1 uncertainties
are either comparable to those of NNPDF3.0, or are rather smaller. The only 
major exception to this is the charm PDF at intermediate and large $x$ for 
which uncertainties are substantially increased. On the other hand, the uncertainties 
in the gluon PDF are smaller in NNPDF3.1 over the entire range of $x$.
This is an important result, since one may have expected generally larger
uncertainties in NNPDF3.1 due to the inclusion of one additional
freely parametrized PDF. 
The fact that the only uncertainty which has enlarged significantly 
is that of the charm PDF suggests that not parametrizing charm may be a source of bias. 
The fact that central values change by a non-negligible amount, though
 compatible within uncertainties, while the uncertainties themselves are significantly
reduced, strongly suggests that NNPDF3.1 is more accurate than
 NNPDF3.0, as would be expected from the substantial amount of new data 
 included in the fit.
The effect of parametrizing charm on PDFs and their uncertainties will be
discussed in more detail in Sect.~\ref{sec:results-mc}, while the effects of 
the new data on both central values and uncertainties will be
discussed in Sect.~\ref{sec:disentangling}.

\begin{figure}[t]
  \begin{center}
 \includegraphics[scale=0.32]{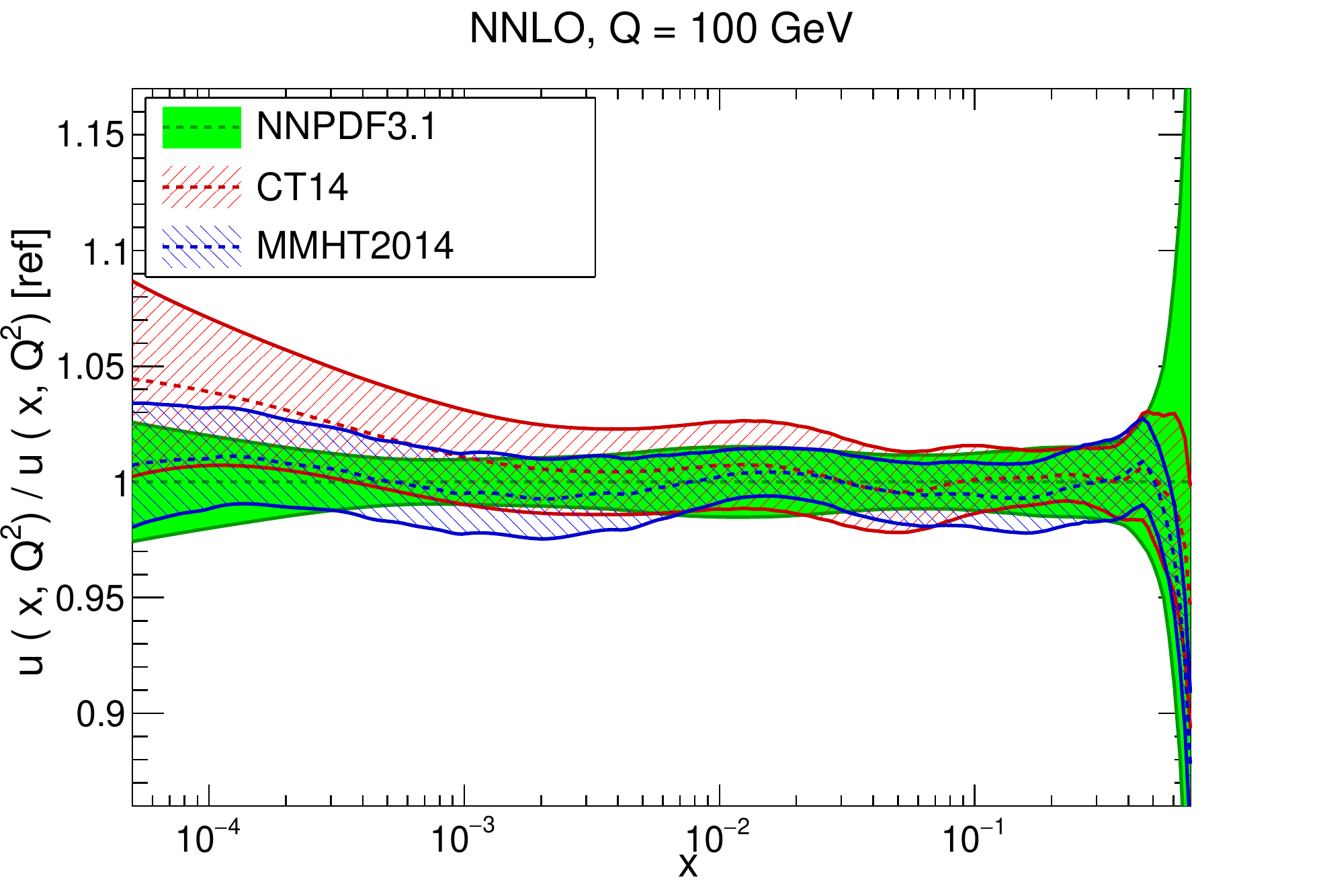}
    \includegraphics[scale=0.32]{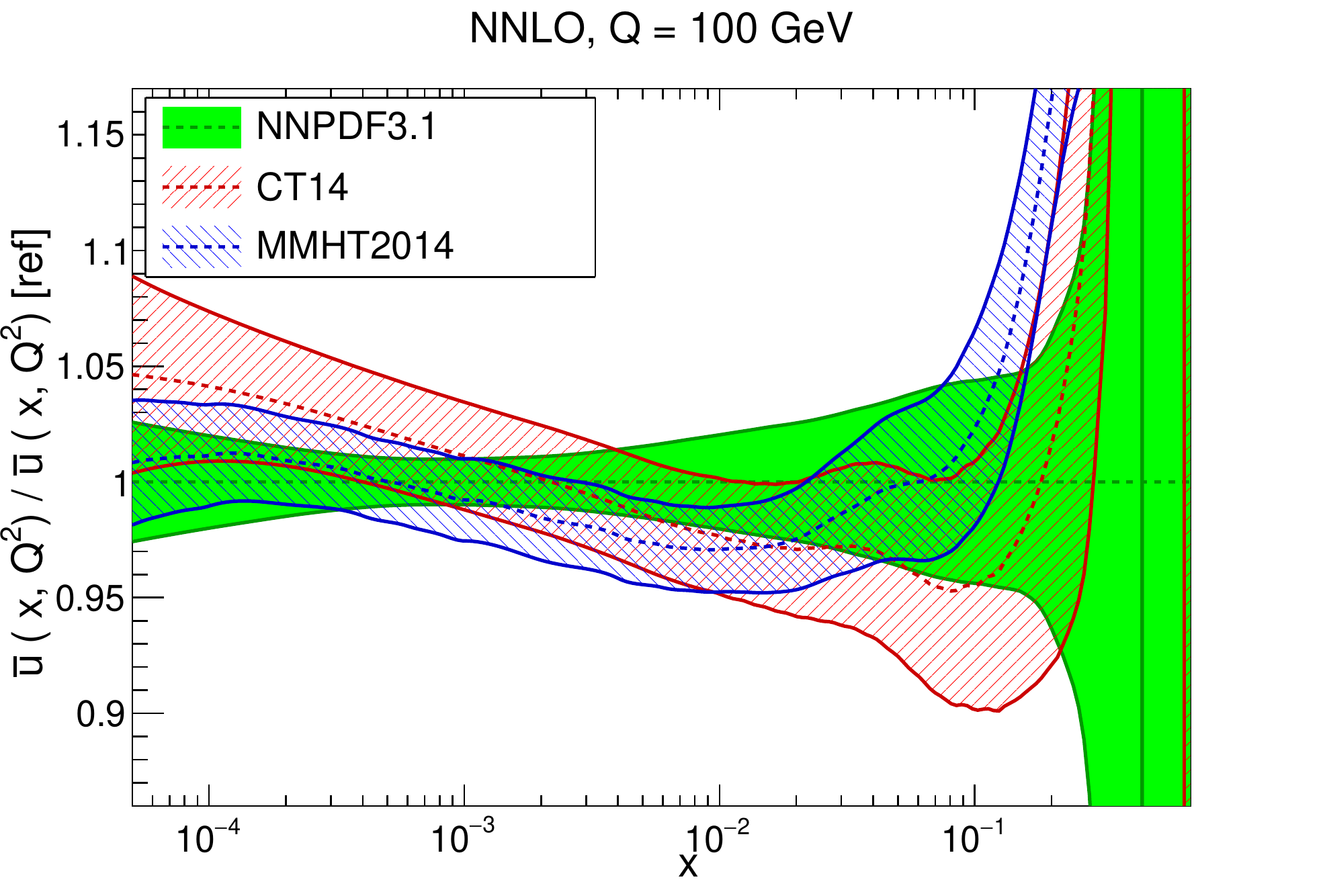}
    \includegraphics[scale=0.32]{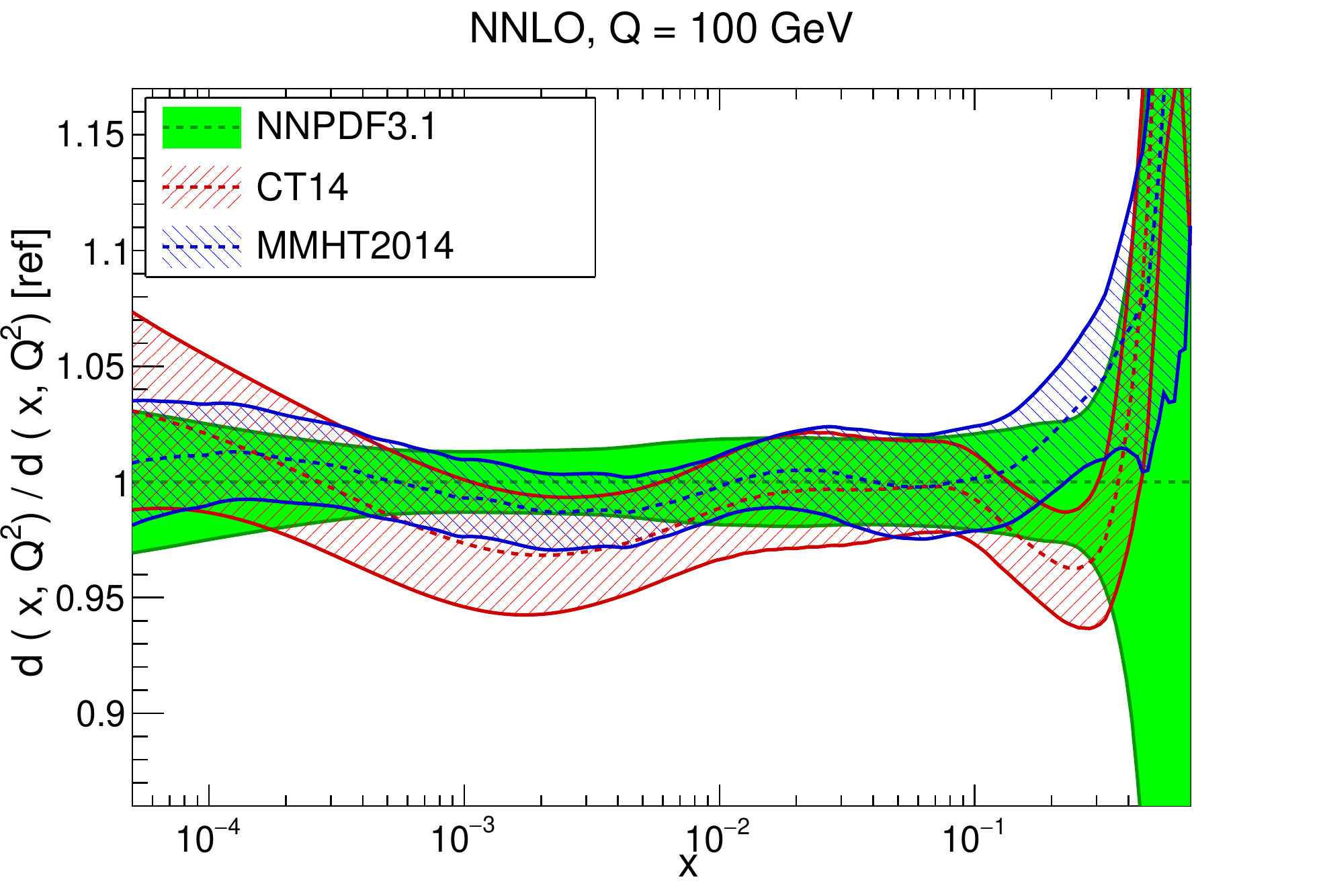}
    \includegraphics[scale=0.32]{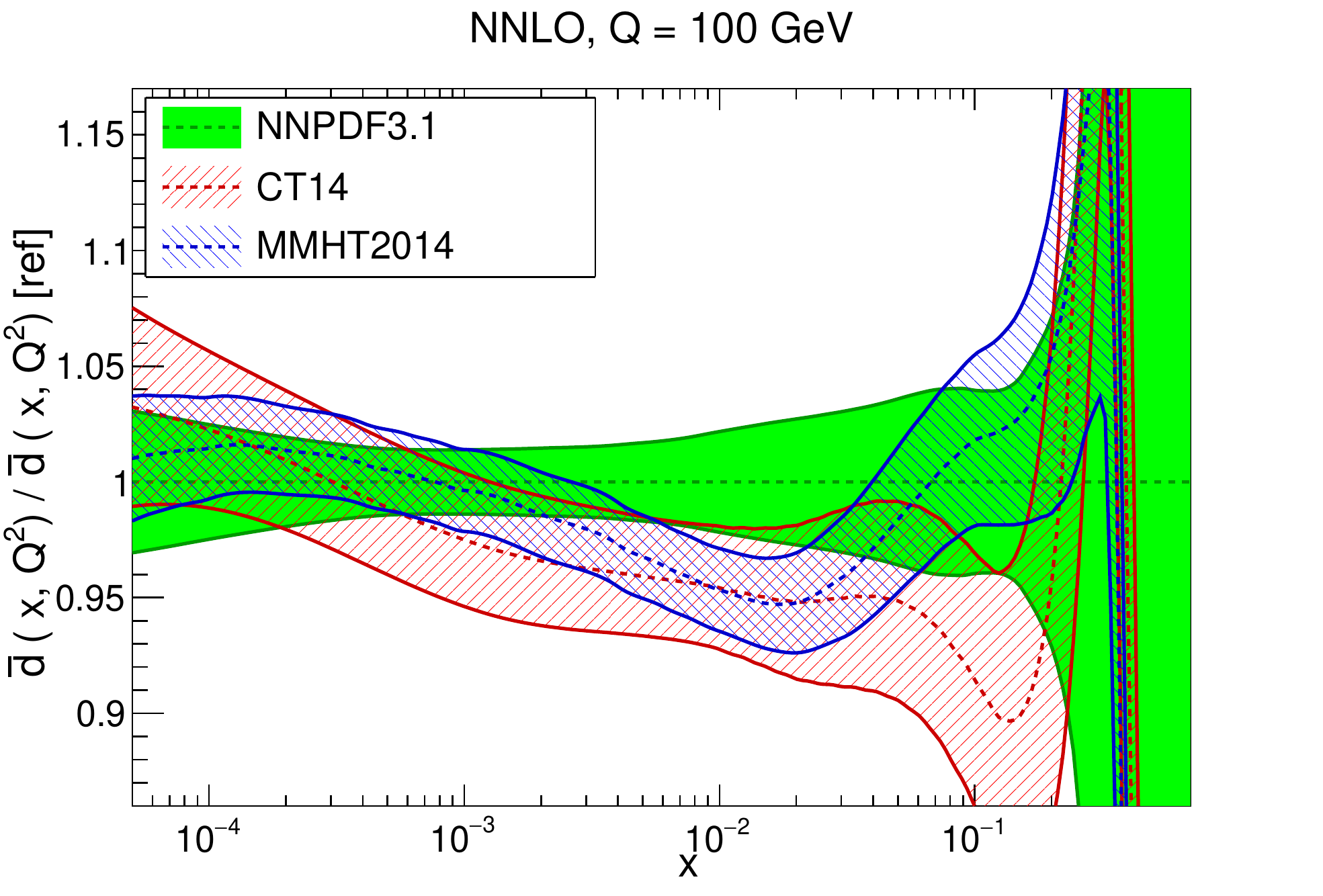}
  \includegraphics[scale=0.32]{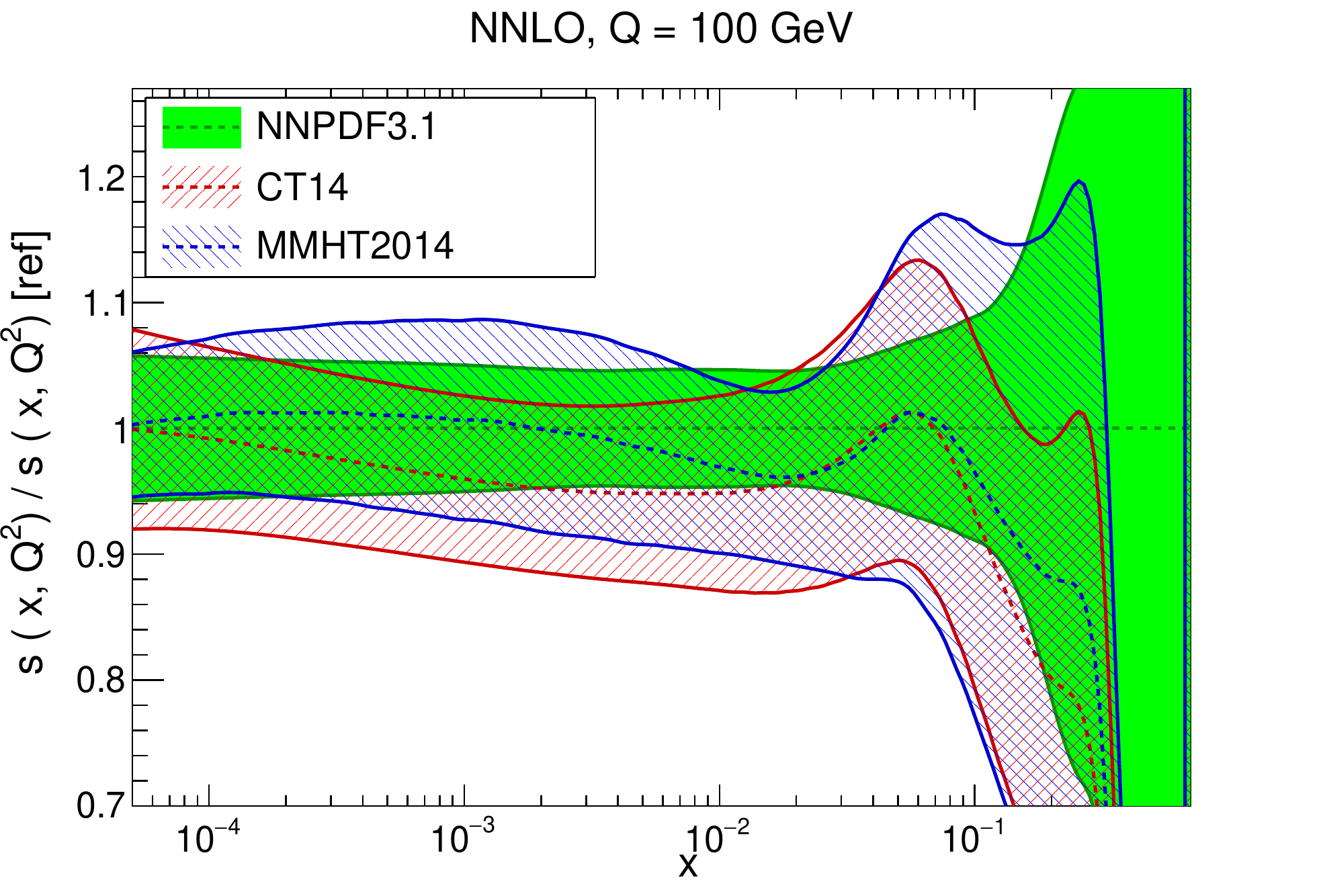}
  \includegraphics[scale=0.32]{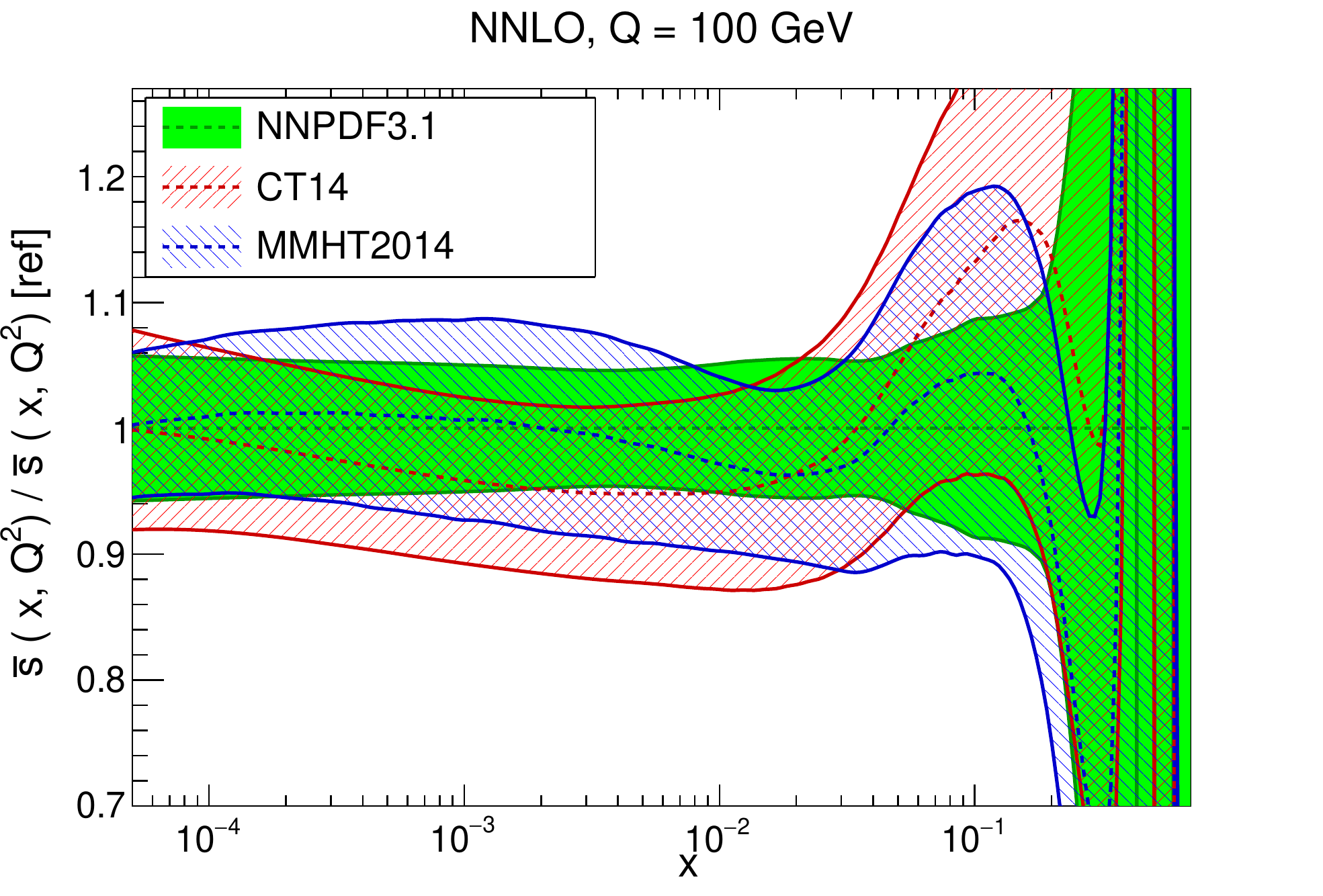}
  \includegraphics[scale=0.32]{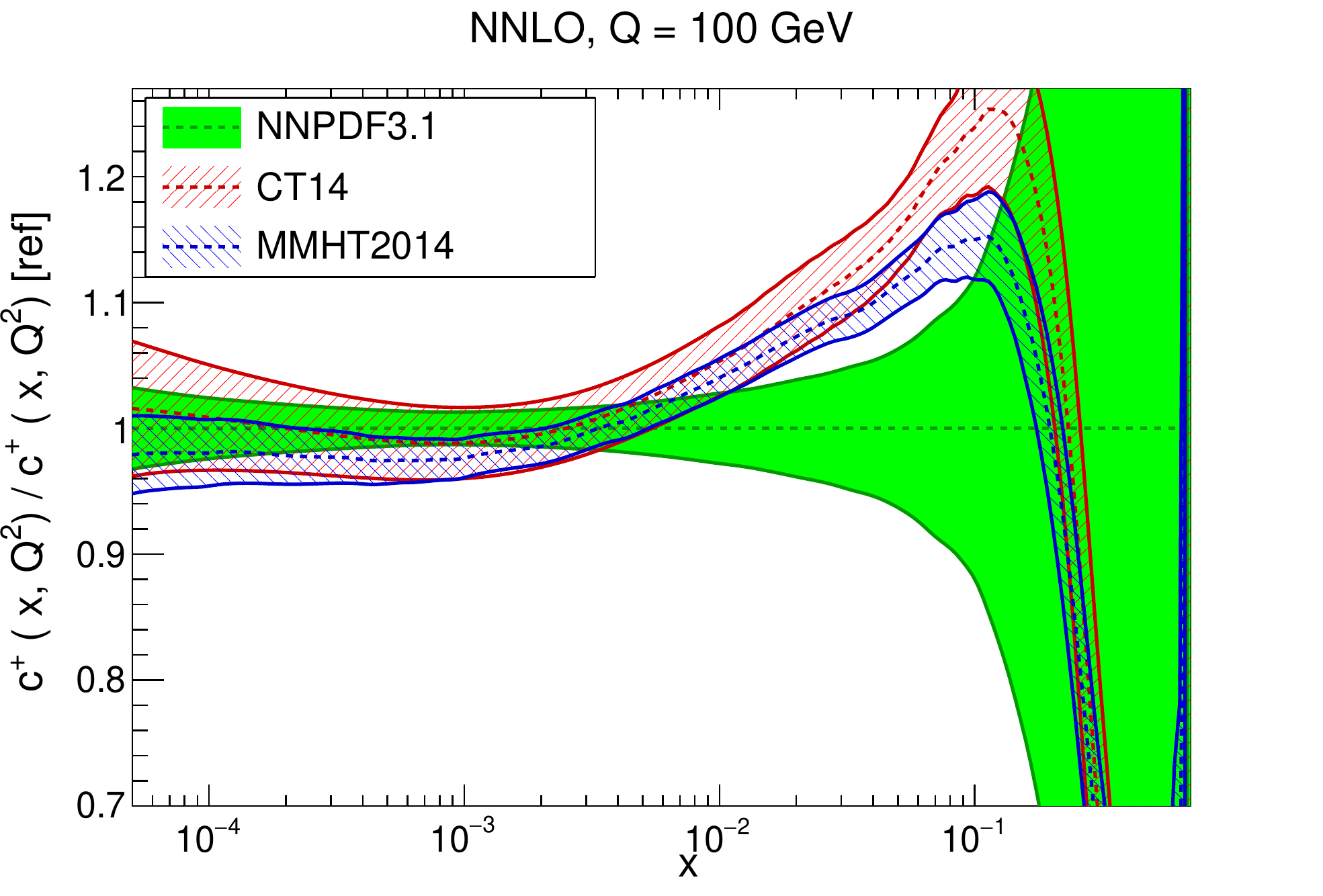}
  \includegraphics[scale=0.32]{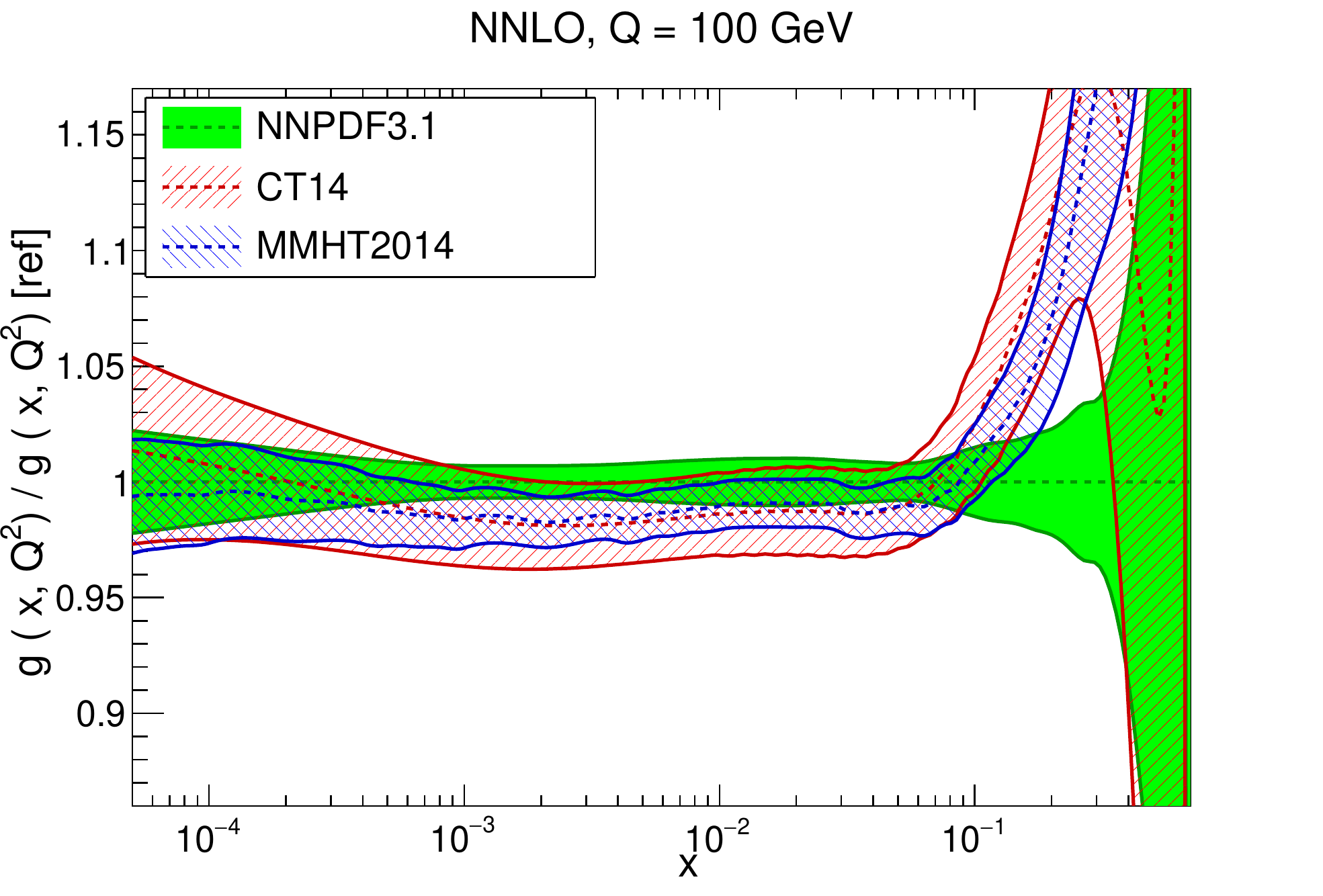}
  \caption{\small Comparison between NNPDF3.1, CT14 and MMHT2014 NNLO PDFs.
    The comparison is performed at $Q=100$ GeV, and results are shown
    normalized to the central value of NNPDF3.1; the PDFs are as in Fig.~\ref{fig:31-nnlo-vs30}.
    \label{fig:globalfits}
  }
\end{center}
\end{figure}

\begin{figure}[t]
  \begin{center}
 \includegraphics[scale=0.32]{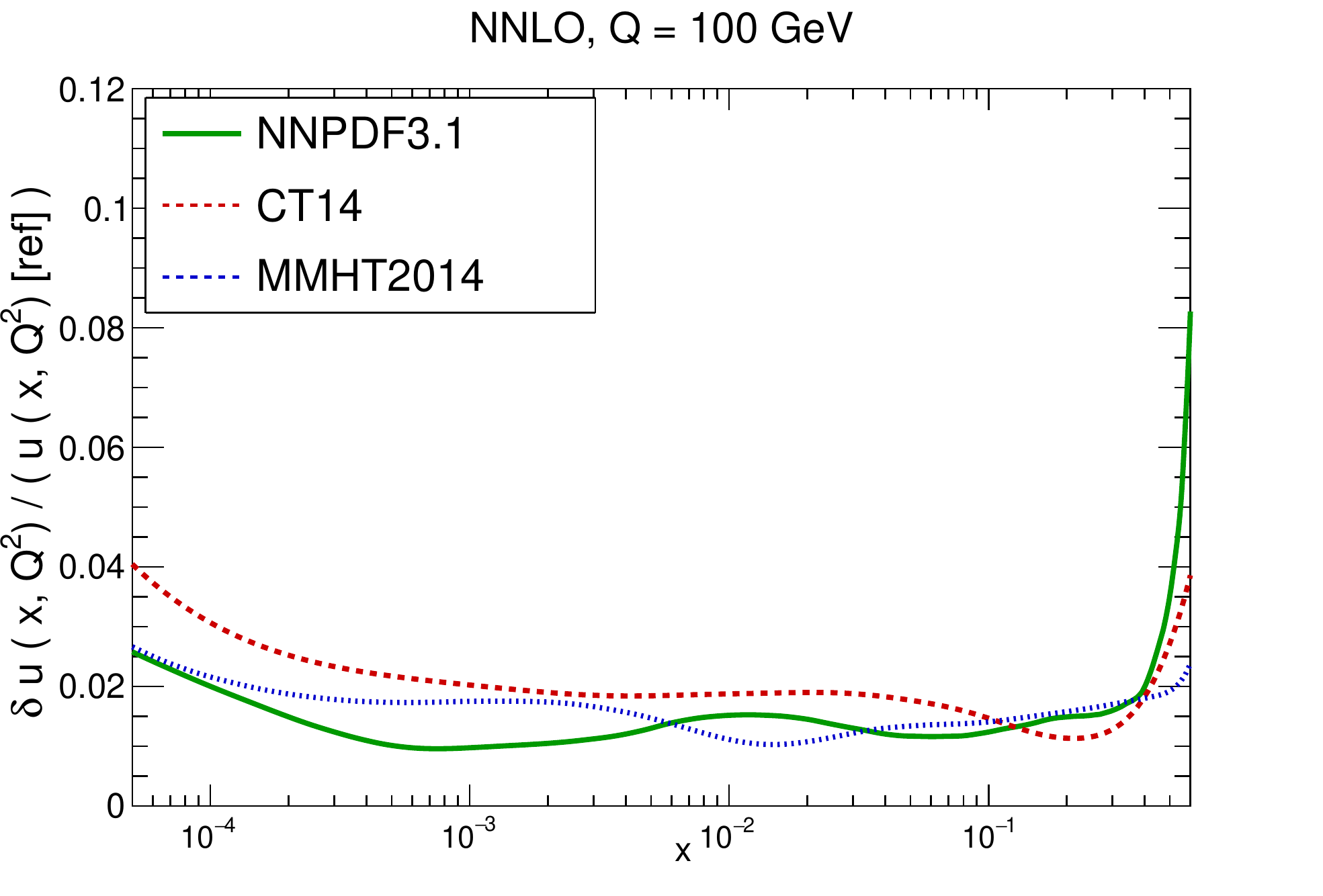}
    \includegraphics[scale=0.32]{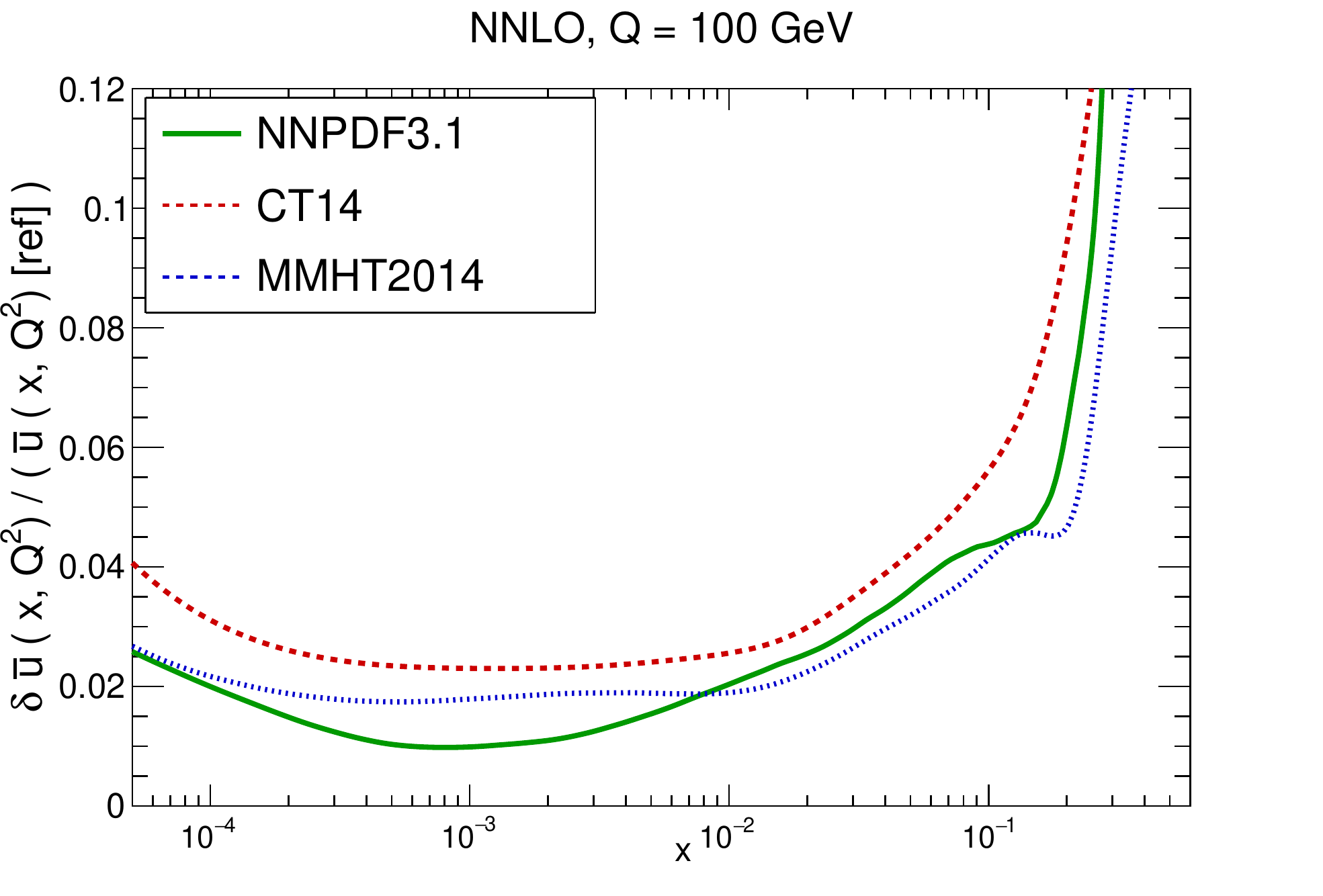}
    \includegraphics[scale=0.32]{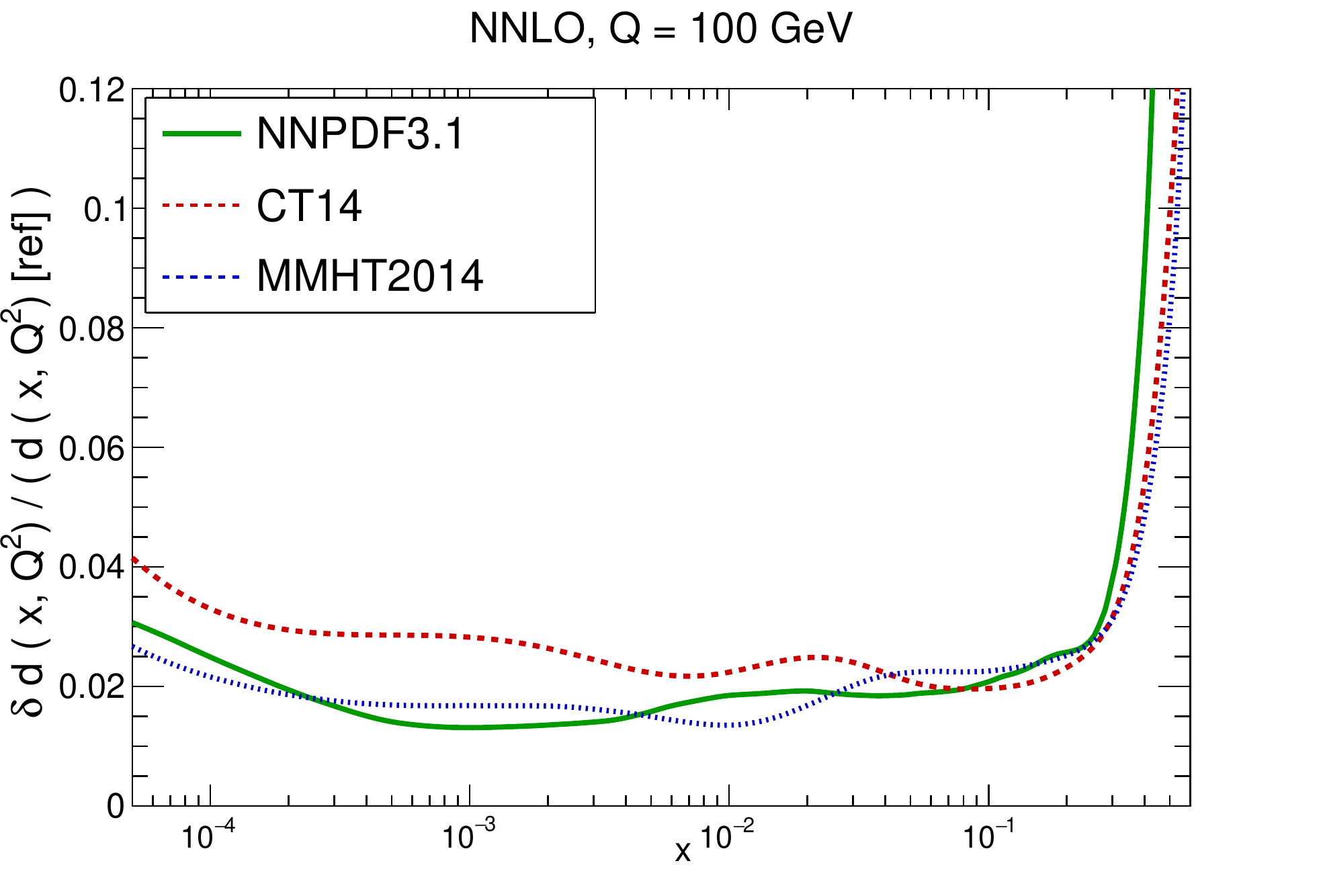}
    \includegraphics[scale=0.32]{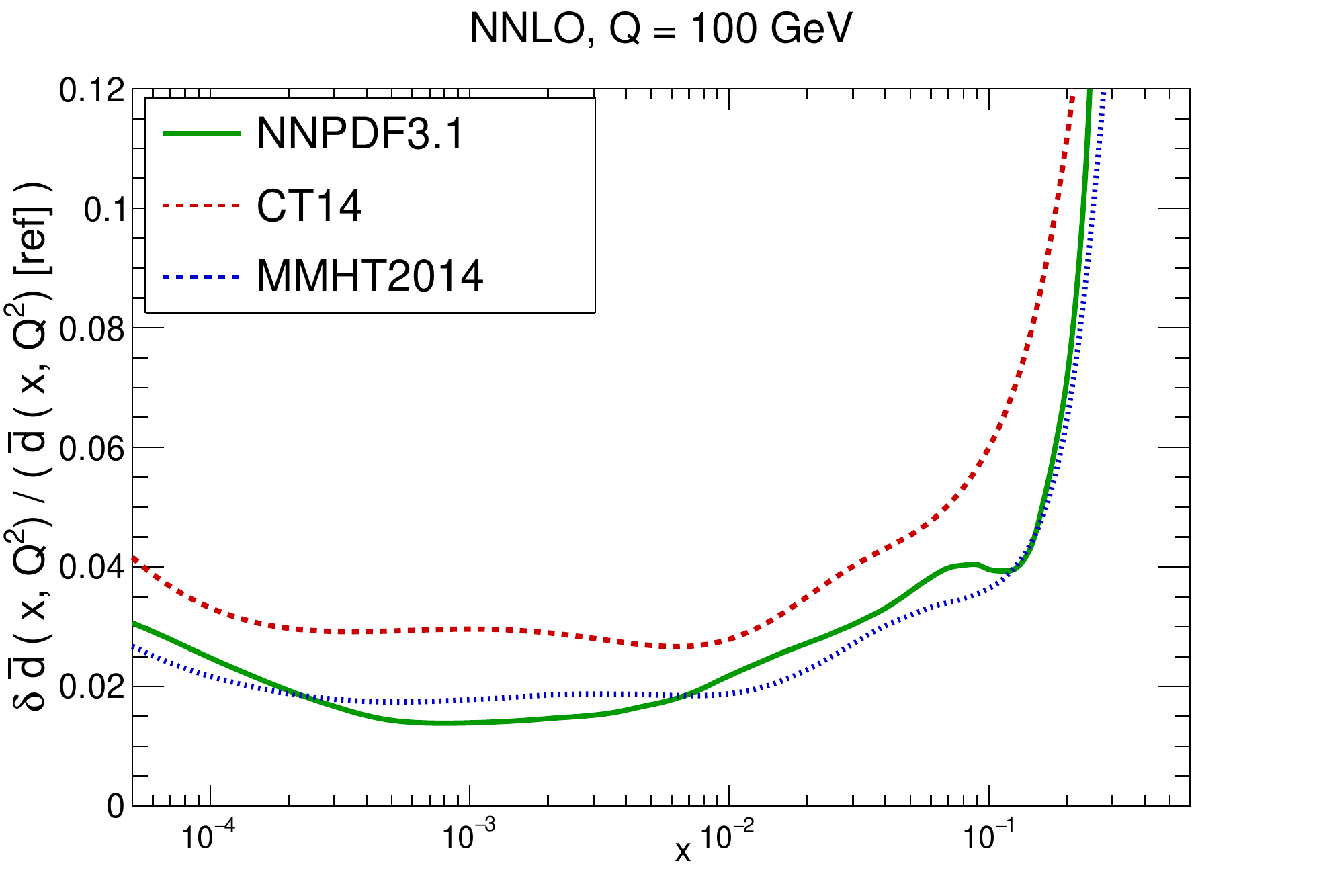}
  \includegraphics[scale=0.32]{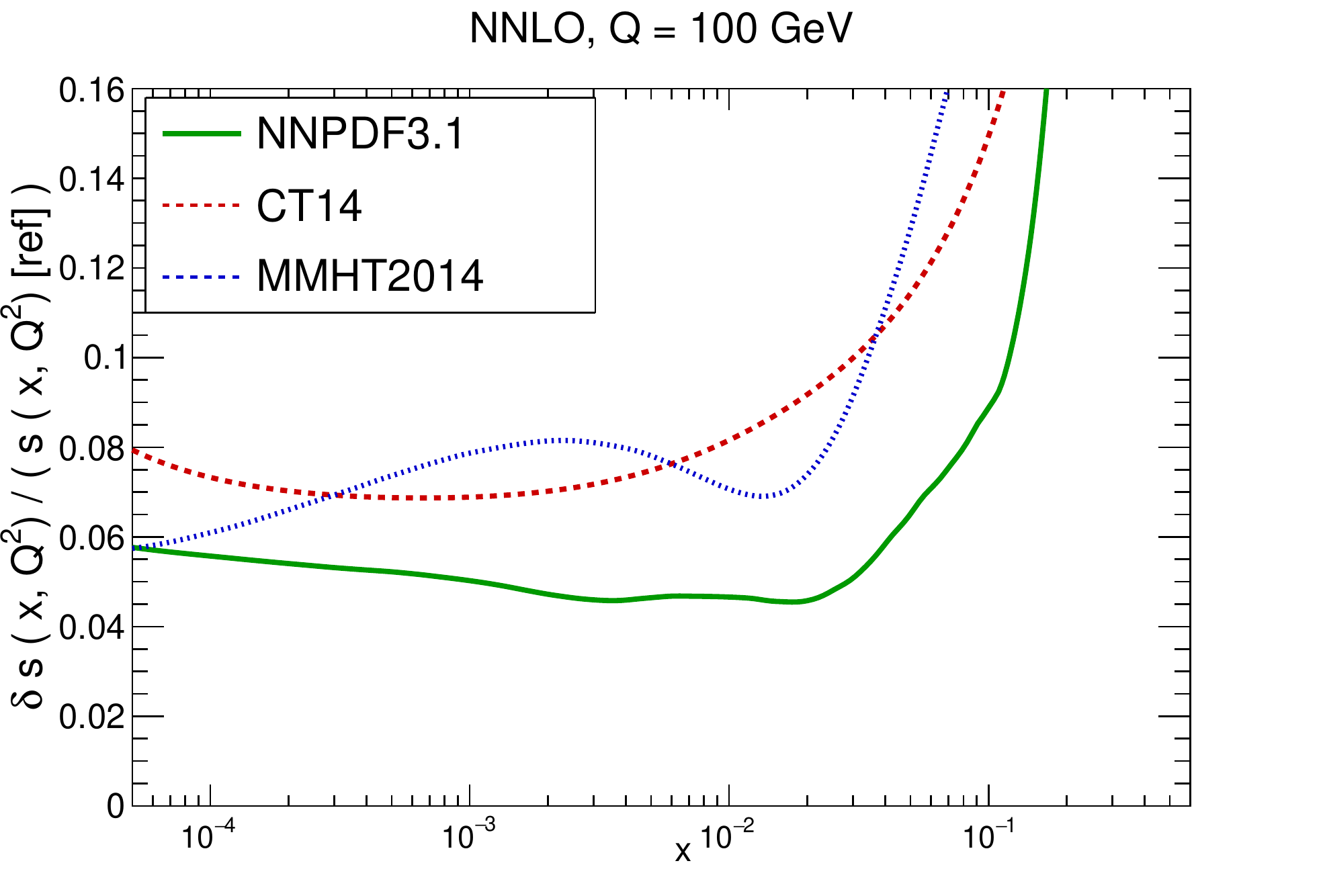}
  \includegraphics[scale=0.32]{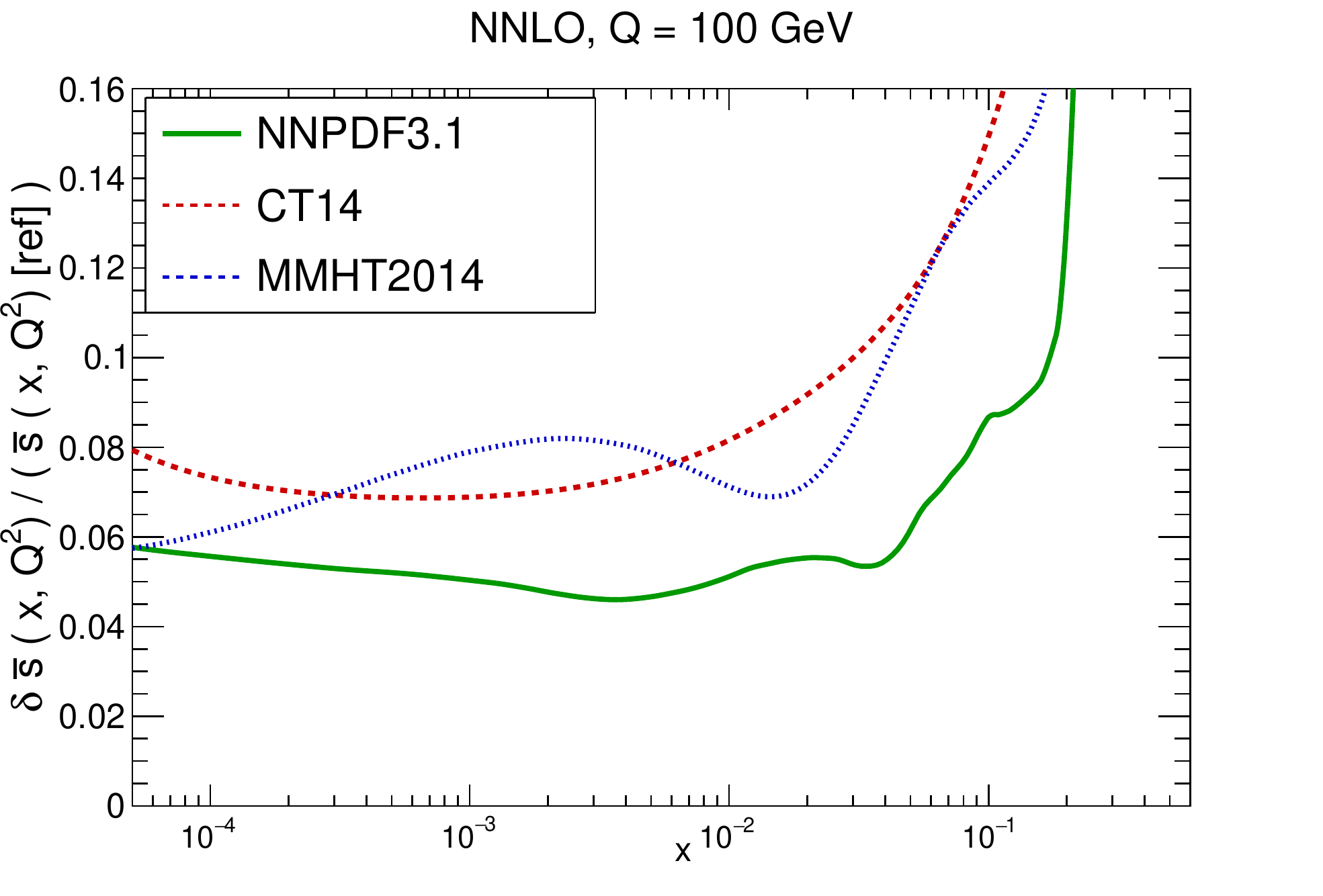}
  \includegraphics[scale=0.32]{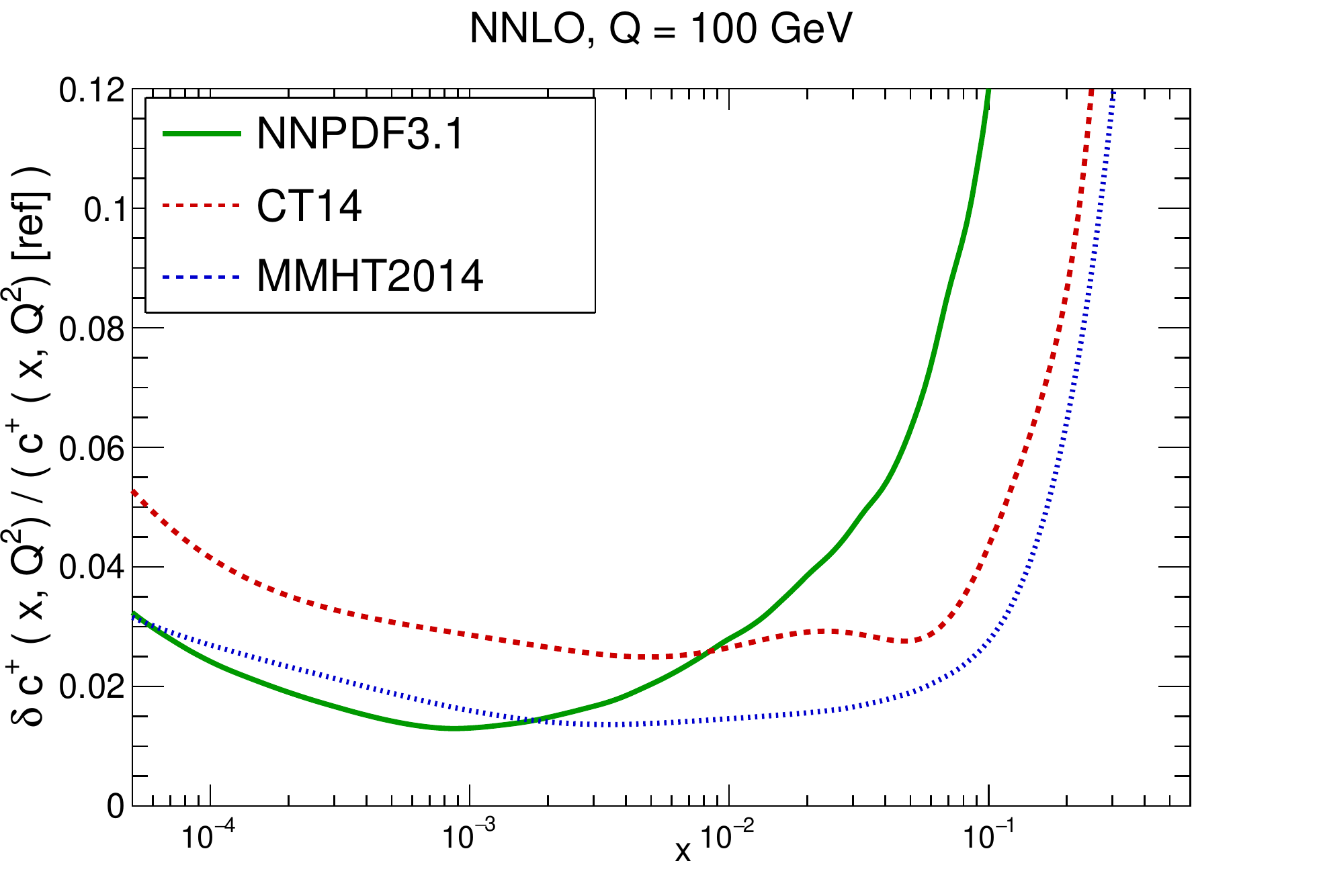}
  \includegraphics[scale=0.32]{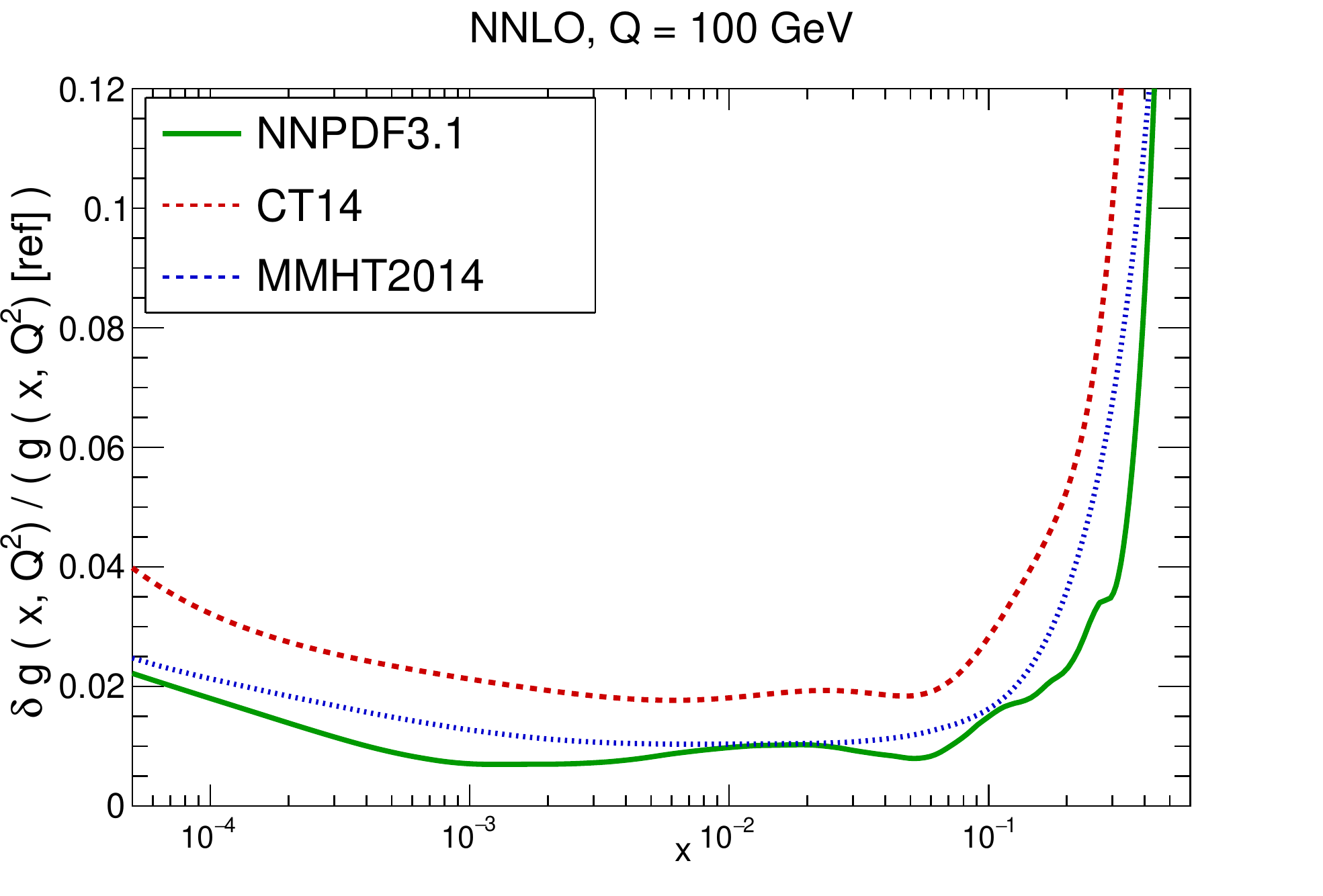}
  \caption{\small Comparison between NNPDF3.1, CT14 and MMHT2014 relative
    PDF uncertainties at $Q=100$; the PDFs are as in Fig.~\ref{fig:globalfits}.
    \label{fig:ERR-globalfits}
  }
\end{center}
\end{figure}

In Fig.~\ref{fig:globalfits} we compare the NNPDF3.1 PDFs to the other
global PDF sets included in the PDF4LHC15 combination along with NNPDF3.0, 
namely CT14 and MMHT14. This comparison is therefore indicative of the effect of
replacing NNPDF3.0 with NNPDF3.1 in the combination. 
The relative 
uncertainties in the three sets are compared in
Fig.~\ref{fig:ERR-globalfits}. Comparing Fig.~\ref{fig:globalfits} to
Fig.~\ref{fig:31-nnlo-vs30},  it is interesting to observe that  
several aspects of the pattern
of differences between NNPDF3.1 and the other global fits
are  similar to those between NNPDF3.1  and NNPDF3.0, and therefore
they are likely to have a similar origin. This is patricularly clear
for the charm and gluon.
The gluon in the region $x\lsim 0.03$, relevant for Higgs
production, is still in good 
    agreement between the three sets. However, now NNPDF3.1 is at the upper edge
    of the one-sigma range, i.e. the  NNPDF3.1 gluon
    in this region is enhanced.
    At large $x$ the NNPDF3.1 gluon is instead suppressed in comparison to
    MMHT14 and CT14. As we will show in Sects.~\ref{sec:results-mc}, \ref{sec:disentangling}
    the  enhancement is a consequence of parametrizing charm, while
    as we will show in 
    Sect.~\ref{sec:impacttop} the large-$x$ suppression is a direct
    consequence of including the 8~TeV top differential data. 
    The uncertainty in the NNPDF3.1 gluon PDF is now 
   noticeably smaller than that of either CT14 or MMHT14. 

For the quark PDFs, for up and down we find 
good agreement in the entire range of $x$. For the antidown PDF,
agreement is marginal, 
with NNPDF3.1 above MMHT14 and CT14 for $x\lsim 0.1$ and below them for larger
$x$. The strange fraction of the proton is larger in NNPDF3.1 than CT14 and 
MMHT14, and has rather smaller PDF uncertainties.
The best-fit NNPDF3.1 charm is suppressed at intermediate $x$ in comparison to the
perturbatively generated ones of CT14 and MMHT14, but has a much larger
uncertainty at large $x$ as would be expected, with the differences
clearly traceable  to the fact that in
   NNPDF3.1 charm  is freely parametrized.

\begin{figure}[t]
  \begin{center}
    \includegraphics[scale=0.32]{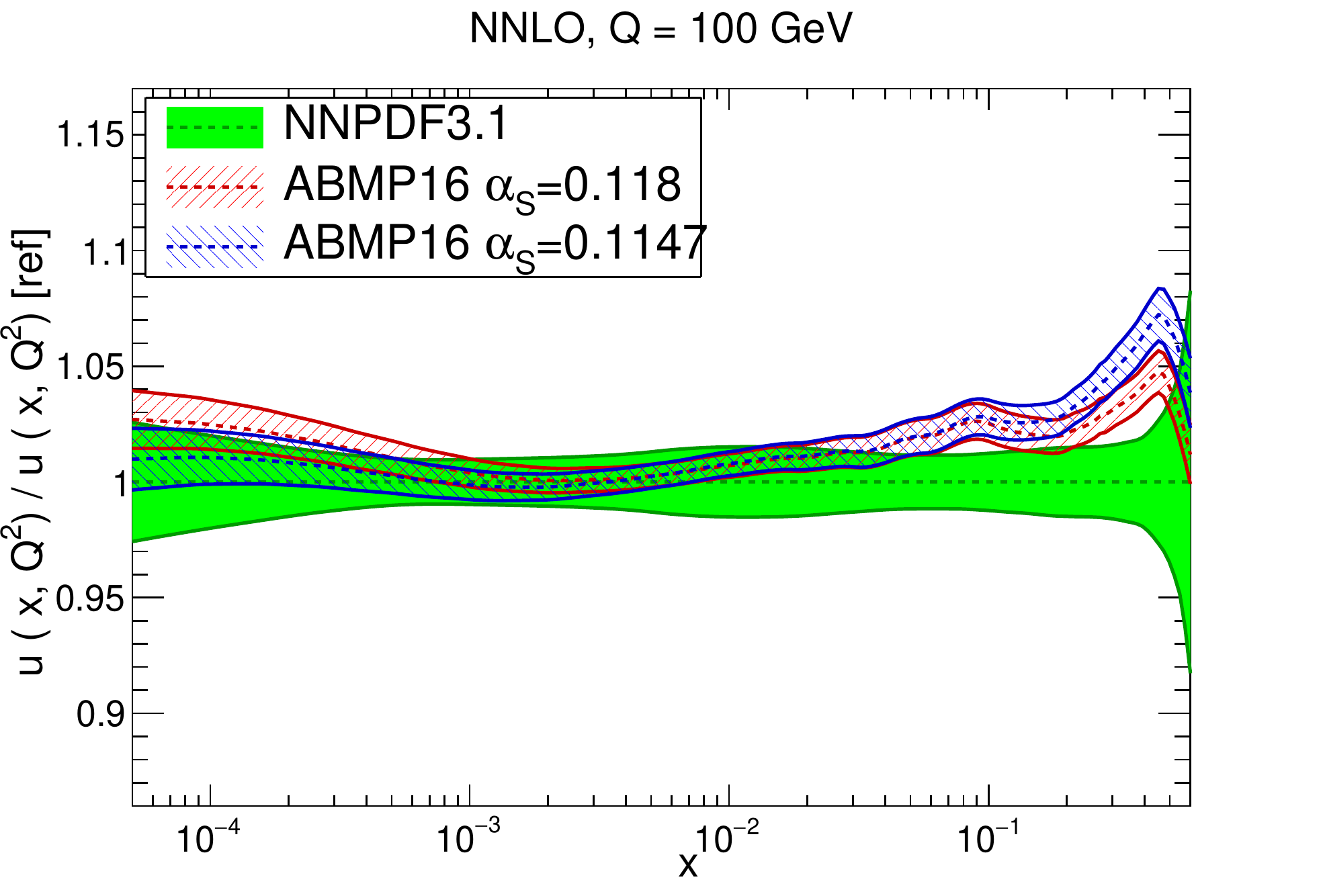}
    \includegraphics[scale=0.32]{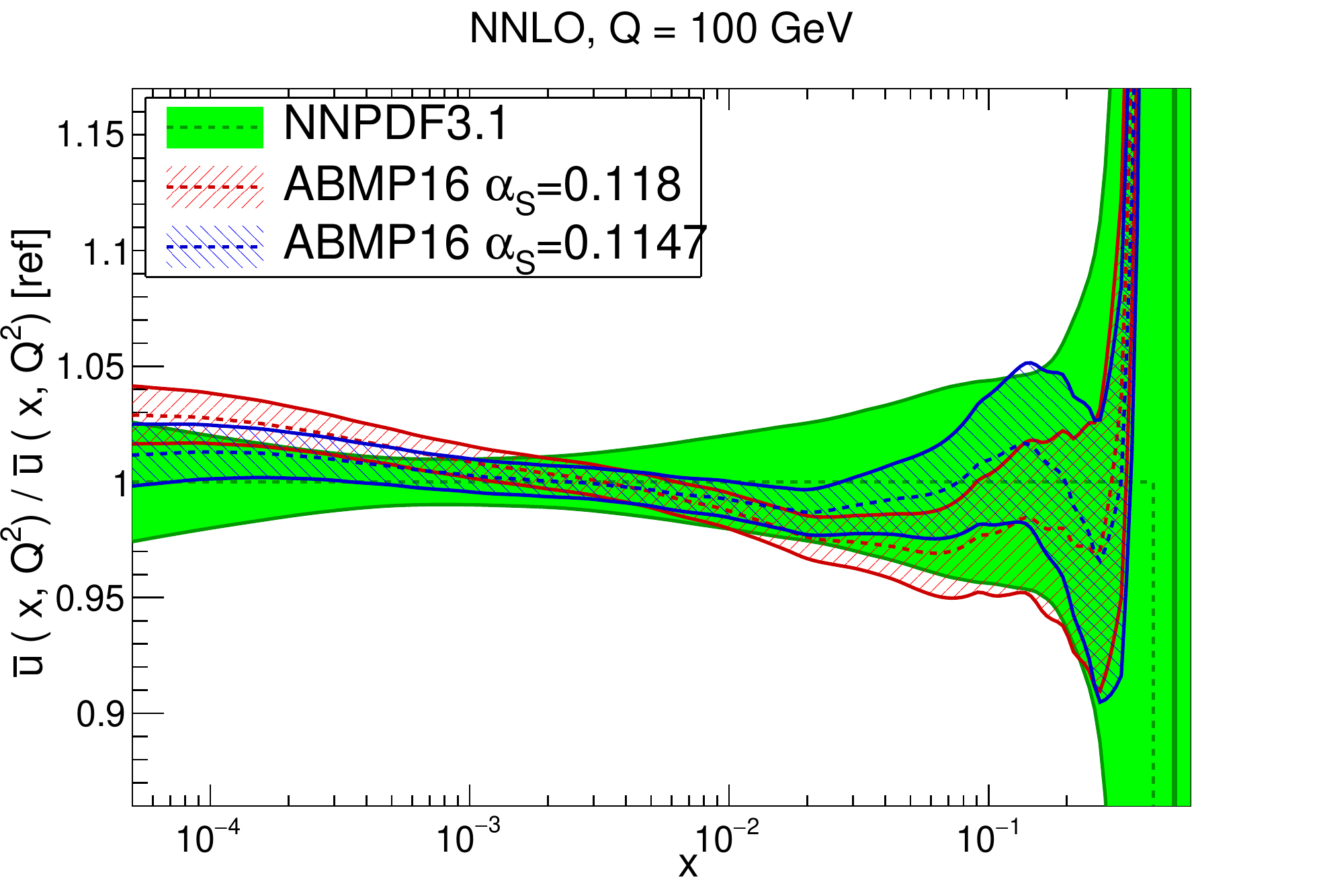}
    \includegraphics[scale=0.32]{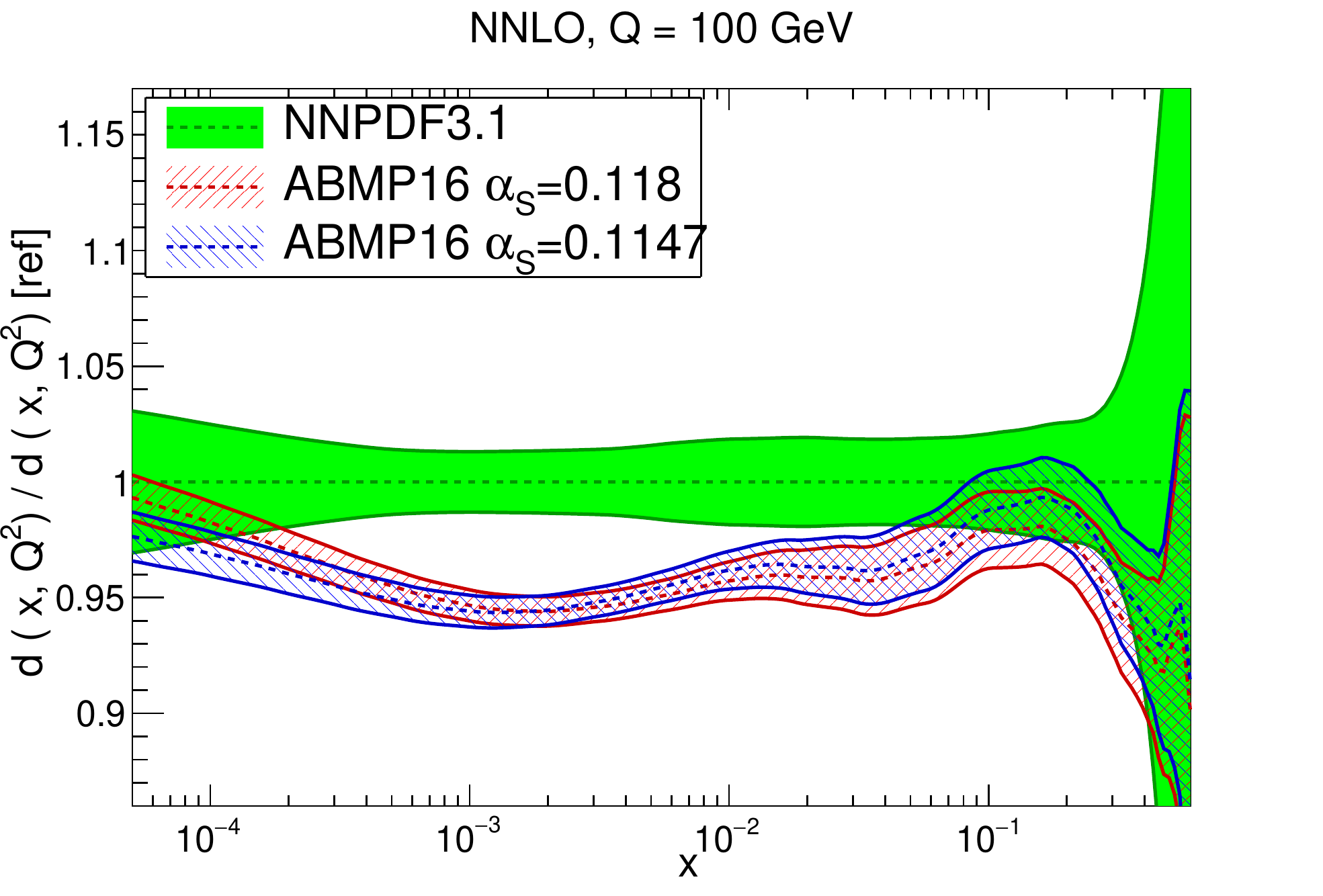}
    \includegraphics[scale=0.32]{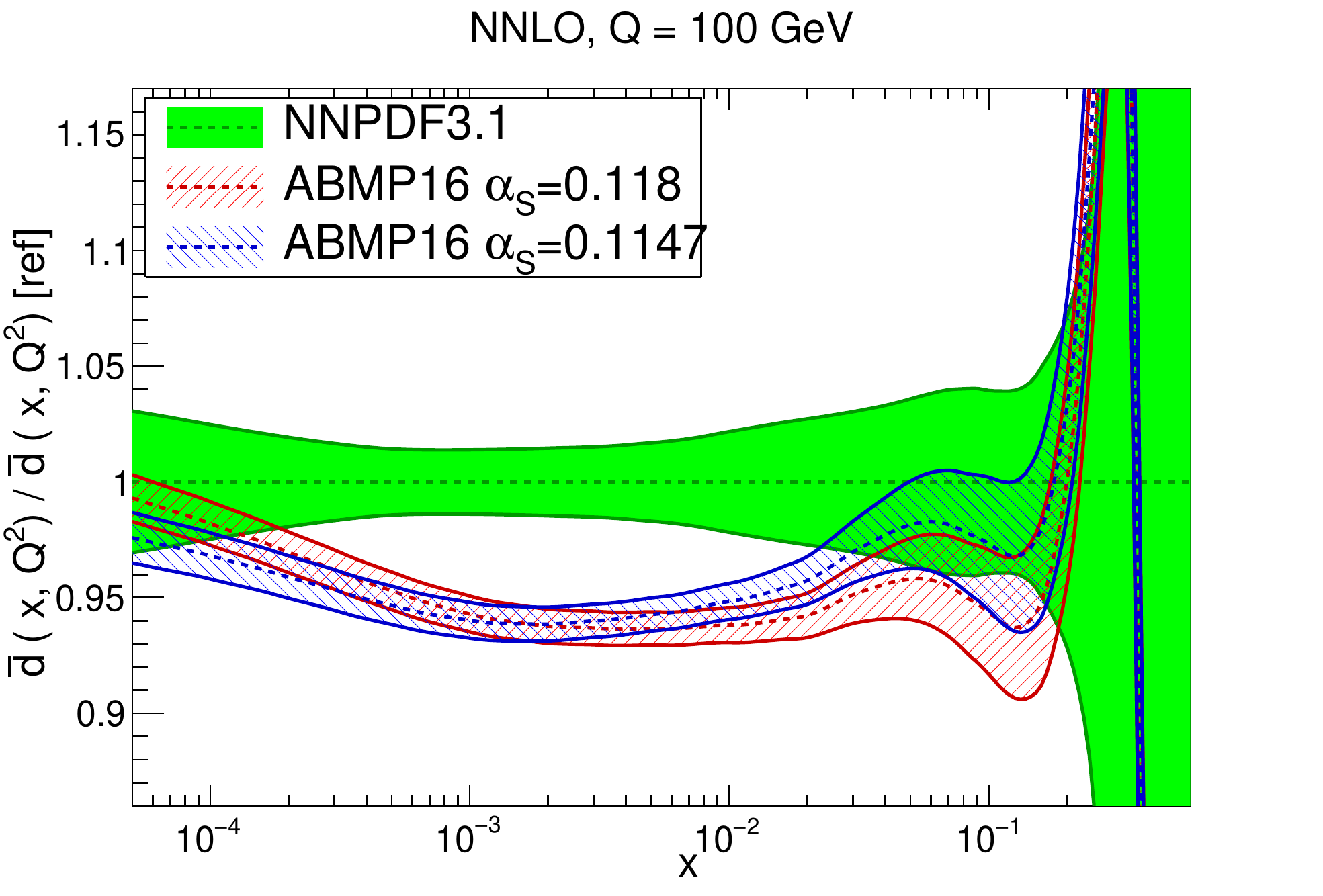}
  \includegraphics[scale=0.32]{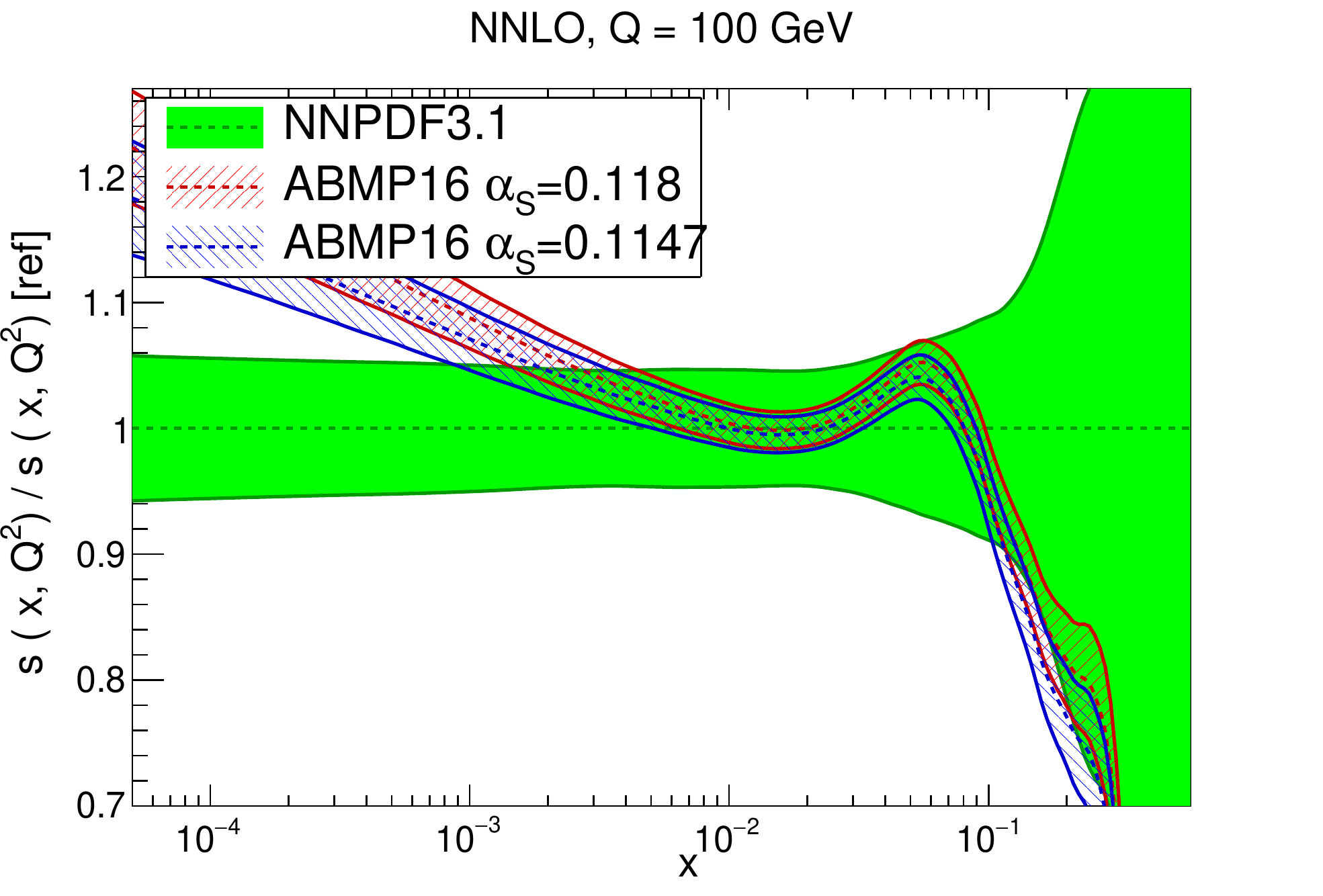}
  \includegraphics[scale=0.32]{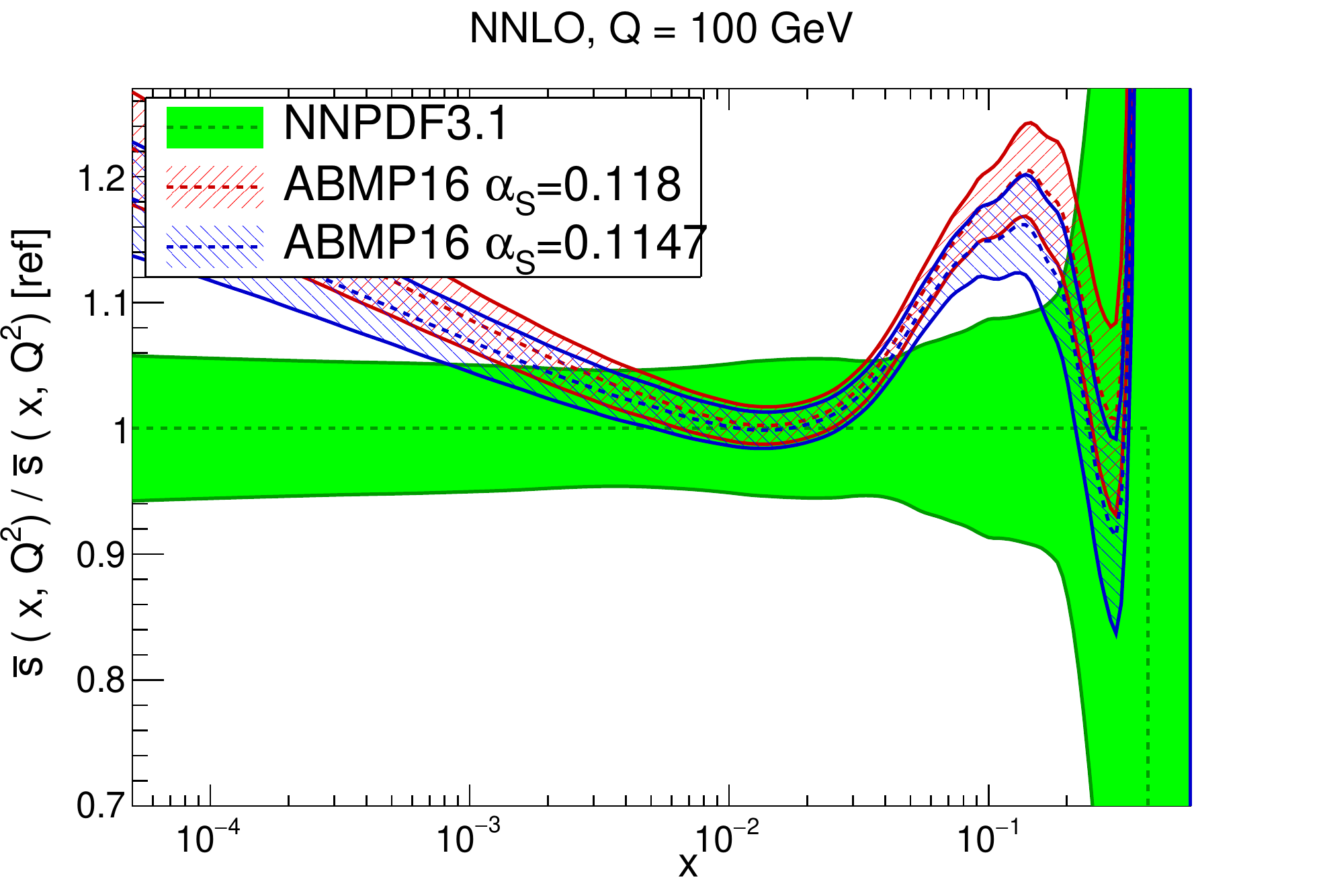}
  \includegraphics[scale=0.32]{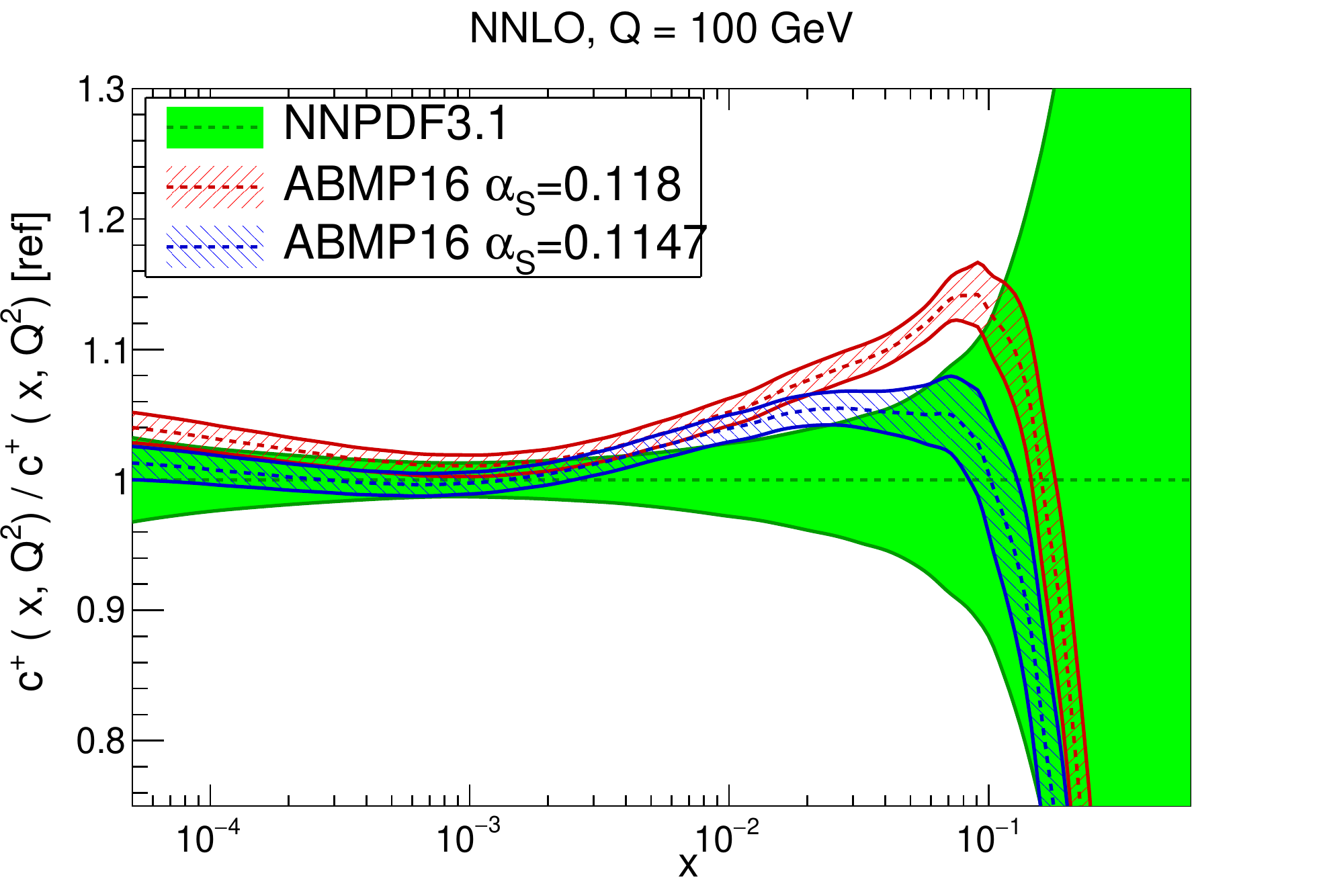}
  \includegraphics[scale=0.32]{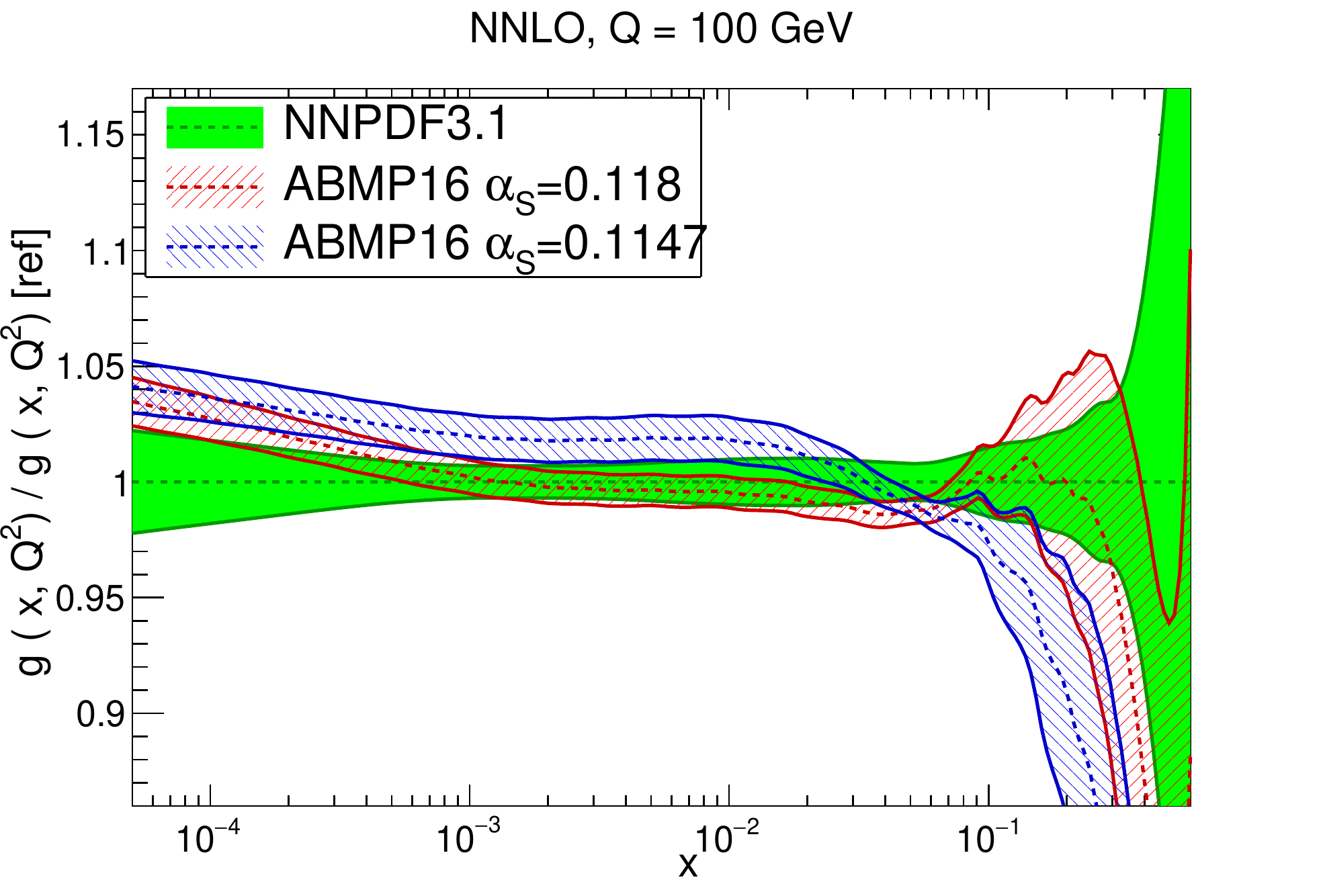}
  \caption{\small Same as Fig.~\ref{fig:globalfits} but now comparing
    to the ABMP16 NNLO $n_f=5$ sets both with their default
    $\alpha_s(m_Z)=0.1147$, and $\alpha_s(m_Z)=0.118$.
    \label{fig:abmp16}
  }
\end{center}
\end{figure}

Finally, in Fig.~\ref{fig:abmp16} we compare NNPDF3.1 
to the recent ABMP16 set. This set is released in
various fixed-flavor number schemes. Because we perform the comparison
at a scale $Q^2=10^4$ GeV$^2$, we choose the $n_f=5$ NNLO ABMP16 sets,
both with their default value $\alpha_s(m_Z)=0.1147$ and with
$\alpha_s(m_Z)=0.118$. 
When a common value of $\alpha_s(m_Z)=0.118$ is adopted, there
is generally reasonable agreement for the gluon PDF, except
at large $x$ where ABMP16 undershoots
NNPDF3.1.
Differences are larger in the case of light quarks:
the ABMP16 up distribution overshoots NNPDF3.1 at large $x$, while the down quark undershoots
in the whole $x$ range. Differences are largest for the strange PDF,
though comparing with Fig.~\ref{fig:globalfits} it is clear that ABMP16
differs by a similary large amount from MMHT14 and CT14.
 In general the ABPM16
sets have rather smaller uncertainties than NNPDF3.1. This
is especially striking for strangeness, where the difference in
uncertainty is particularly evident. This is to be contrasted
with the CT14 and MMHT14 sets, which have qualitatively similar uncertainties 
to NNPDF3.1 throughout the data region. 
The fact that the uncertainties for ABMP16 are so small can be traced
to their overly restrictive parametrization, and the fact that this set is produced using a
Hessian methodology, but unlike MMHT14 and CT14, with no tolerance
(see Refs.~\cite{Harland-Lang:2014zoa,Dulat:2015mca}).

\clearpage

\clearpage

\subsection{Methodological improvements: parametrizing charm}
\label{sec:results-mc}

The main methodological improvement in NNPDF3.1 over
NNPDF3.0 is the fact that the charm PDF is now parametrized  
in the same way as the light and strange quark PDFs.
To quantify the effect of this change, we have performed a repeat of the NNPDF3.1 analysis
but with charm treated as in all previous NNPDF PDF determinations, 
i.e., generated
entirely perturbatively through matching conditions implemented at NLO or NNLO. 
\begin{table}[h]
\begin{center}
  \scriptsize
      \renewcommand{\arraystretch}{1.10}
\begin{tabular}{|l|c|c||c|c|}
  \hline
  &  \multicolumn{2}{c||}{NNPDF3.1 pert. charm} &
  \multicolumn{2}{c|}{NNPDF3.1}\\
  \hline
Dataset  &    NNLO  & NLO 
& NNLO    & NLO  \\
\hline
\hline
NMC      &   1.38     &   1.38   &    1.30     &   1.35 \\    
SLAC      &   0.70      &   1.22  &    0.75      &   1.17   \\    
BCDMS       &   1.27      &   1.24  &    1.21      &   1.17   \\    
CHORUS       &   1.10      &   1.07    &    1.11      &   1.06  \\    
NuTeV dimuon       &   1.27      &   1.01  &    0.82      &   0.87  \\
\hline
HERA I+II inclusive       &   1.21      &   1.15&    1.16      &   1.14   \\
HERA $\sigma_c^{\rm NC}$       &   1.20 (1.42)    &   1.21 (1.35)
 &    1.45      &   1.15 (1.35)\\
HERA $F_2^b$       &   1.16      &   1.12
&    1.11      &   1.08  \\
\hline
\hline
DYE866 $\sigma^d_{\rm DY}/\sigma^p_{\rm DY}$
   &   0.46     &   0.48   &    0.41     &   0.40 \\
DYE886 $\sigma^p$       &   1.38     &   1.09 &    1.43     &   1.05    \\
DYE605  $\sigma^p$      &   1.05     &   0.83 &    1.21     &   0.97  \\
\hline
CDF $Z$ rap       &   1.44      &   1.46   &    1.48     &   1.62 \\
CDF Run II $k_t$ jets       &   0.86      &   0.86
&    0.87      &   0.84 \\
\hline
D0 $Z$ rap       &   0.60      &   0.64 &    0.60      &   0.67   \\
D0 $W\to e\nu$  asy      &   2.71      &   1.63
&    2.70      &   1.59    \\
D0 $W\to \mu\nu$  asy       &   1.42      &   1.38
&    1.56      &   1.52 \\
\hline
\hline
ATLAS total      &   {\bf 1.17}     &  {\bf 1.45}  &
{\bf 1.09}      &   {\bf 1.3}  \\    
  ATLAS $W,Z$ 7 TeV 2010      &   1.04      &   1.08 &   0.96     &   1.04  \\
  ATLAS high-mass DY 7 TeV      &   1.66     &   2.08
   &    1.54     &   1.88  \\
  ATLAS low-mass DY 2011      &   0.83      &   0.70
  &    0.90      &   0.69 \\
  ATLAS $W,Z$ 7 TeV 2011       &   2.74      &   4.29
   &    2.14     &   3.70  \\
ATLAS jets 2010 7 TeV       &   0.96      &   0.95  &    0.94      &   0.92  \\
ATLAS jets 2.76 TeV       &   1.06      &   1.13   &    1.03      &   1.03  \\
ATLAS jets 2011 7 TeV       &   1.11     &   1.14   &    1.07     &   1.12  \\
ATLAS $Z$ $p_T$ 8 TeV $(p_T^{ll},M_{ll})$          &   0.94     &   1.19
&    0.93    &   1.17 \\
ATLAS $Z$ $p_T$ 8 TeV $(p_T^{ll},y_{ll})$      &   0.96      &   1.84
 &    0.94      &   1.77 \\
ATLAS $\sigma_{tt}^{\rm tot}$       &   0.80      &   2.03  &    0.86      &   1.92  \\
ATLAS $t\bar{t}$ rap      &   1.39     &   1.18  &    1.45     &   1.31  \\
\hline
CMS total     &   {\bf 1.09}     &   {\bf 1.2}
&    {\bf  1.06}    &   {\bf 1.20 } \\
CMS $W$ asy 840 pb       &   0.69     &   0.80  &    0.78     &   0.86  \\
CMS $W$ asy 4.7 fb        &   1.75      &   1.76
 &    1.75     &   1.77\\
CMS $W+c$ tot      &   -     &   0.49  &    -     &   0.54 \\
CMS $W+c$ ratio       &   -     &   1.92   &    -      &   1.91 \\
CMS Drell-Yan 2D 2011       &   1.33      &   1.27
 &    1.27     &   1.23  \\
CMS $W$ rap 8 TeV       &   0.90      &   0.65 &    1.01      &   0.70    \\
CMS jets 7 TeV 2011       &   0.87      &   0.86
&    0.84      &   0.84  \\
CMS jets 2.76 TeV       &   1.06      &   1.05  &    1.03      &   1.01  \\
CMS $Z$ $p_T$ 8 TeV $(p_T^{ll},y_{ll})$      &   1.29      &   3.50
 &    1.32    &   3.65 \\
CMS $\sigma_{tt}^{\rm tot}$        &   0.21      &   0.67  &    0.20      &   0.59  \\
CMS $t\bar{t}$ rap       &   0.96      &   0.96
&    0.94      &   0.96 \\
\hline 
LHCb total      &   {\bf 1.48}      &  {\bf 1.77 }
&  {\bf  1.47}      &   {\bf 1.62}  \\
LHCb $Z$ 940 pb      &   1.31     &   1.08 &    1.49      &   1.27  \\
LHCb $Z\to ee$ 2 fb      &   1.47     &   1.66
 &    1.14      &   1.33 \\
LHCb $W,Z \to \mu$ 7 TeV       &   1.54      &   1.51
&    1.76      &   1.60 \\
LHCb $W,Z \to \mu$ 8 TeV       &   1.51     &   2.28
 &    1.37     &   1.88 \\
\hline
\hline
Total dataset \bf     &\bf   1.187    & \bf  1.197   &  \bf  1.148      & \bf  1.168 \\
\hline
\end{tabular}
\caption{\small
\label{tab:chi2tab_31-nlo-nnlo-30-pc} 
Same as Tab.~\ref{tab:chi2tab_31-nlo-nnlo-30}, but now comparing the
default NNPDF3.1 NNLO and NNLO sets to the  variant in which charm is
perturbatively generated. For HERA $\sigma_c^{\rm NC}$ the number in
parenthesis refer to the subset of data to which the NNLO FC cut of
Table.~\ref{eq:tablekincuts} is applied.
}
\end{center}
\end{table}


In Table~\ref{tab:chi2tab_31-nlo-nnlo-30-pc} we show the 
$\chi^2/N_{\rm dat}$ values when charm is  perturbatively
generated  at NLO and NNLO. Unsurprisingly the fit quality deteriorates when
charm is not independently parametrized PDF. 
This is what one would naively expect 
since perturbative charm imposes a constraint upon the fit, thereby reducing the number
of free parameters. 

 However, it is interesting to observe that 
the fit quality to the inclusive HERA data (1306 data points)
significantly deteriorates
 when going from NLO to NNLO with perturbative charm, whereas it
 remains stable when charm is independently parametrized. Concerning
 the charm structure function data, note that,
 as discussed in Sect.~\ref{sec:datadis} above, a further cut is applied to 
 the HERA $\sigma_c^{\rm NC}$ data at NNLO when charm is independently
 parametrized. In order to allow for a consistent comparison, in
 Table~\ref{tab:chi2tab_31-nlo-nnlo-30-pc} we 
 show in parenthesis the value of 
 $\chi^2/N_{\rm dat}$ computed for the 37 (out of 47) data points that survive
 this cut also for all other  cases. Hence, for this data the fit
 quality is simlar with perturbative and parametrized charm, and also
 similar at NLO and NNLO (slightly worse at NNLO, by an amount
 compatible with a statistical fluctuation).
The fact  that when  parametrizing charm 
 there no longer is a deterioration of fit quality when going
from NLO to NNLO suggests that this  
resolves a tension present at NNLO, with  perturbative charm, 
between HERA and hadron collider data. Likewise, a purely perturbative charm leads
to a substantial deterioration at NNLO for BCDMS, NMC and especially
for the NuTeV dimuon cross-sections. This can be traced to the fact that
independently parametrizing charm is essential to reconcile the HERA data with the
constraints on the strange content of the proton imposed by
the ATLAS $W,Z$ 2011 rapidity distributions. 

\begin{figure}[t]
  \begin{center}
    \includegraphics[scale=0.32]{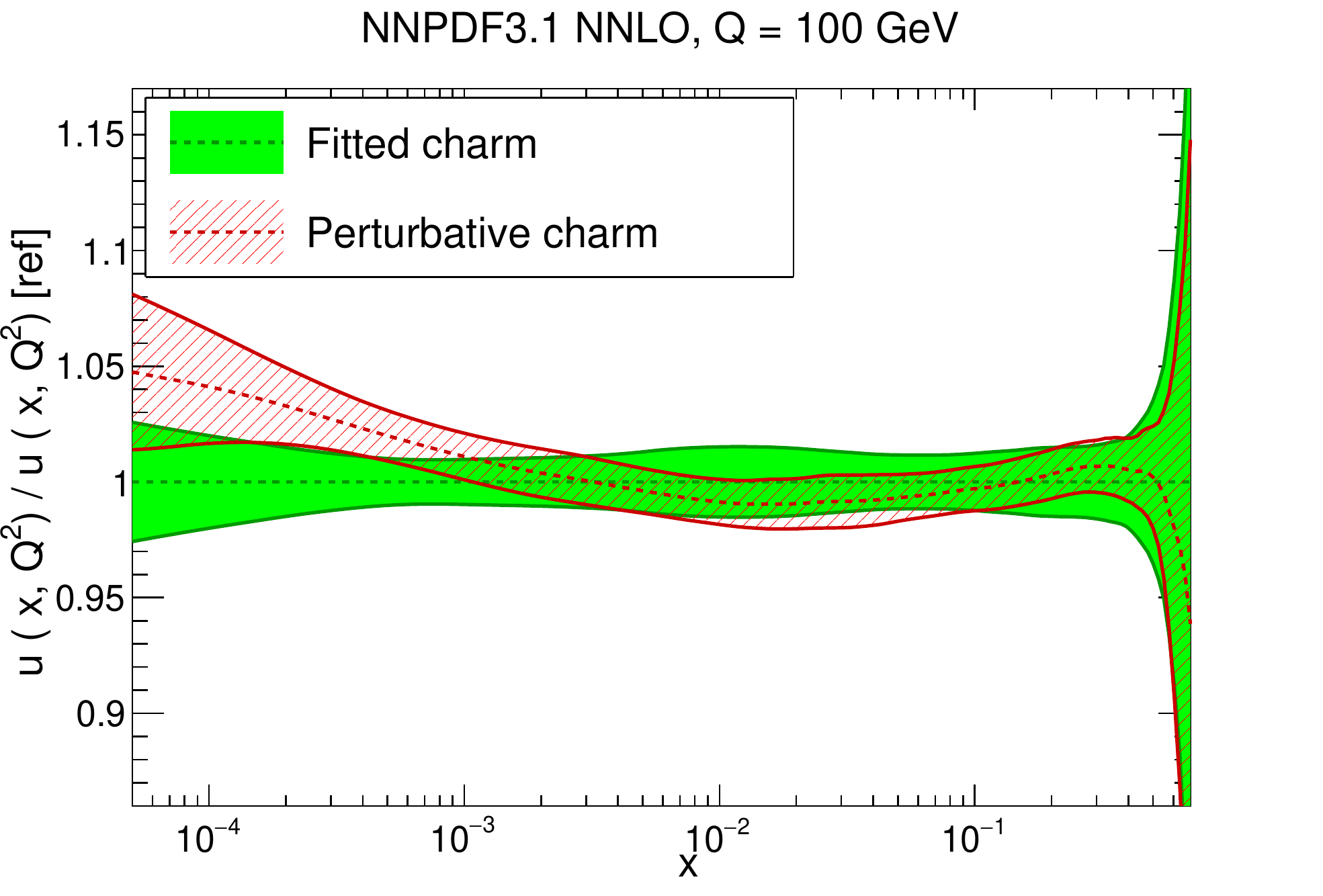}
    \includegraphics[scale=0.32]{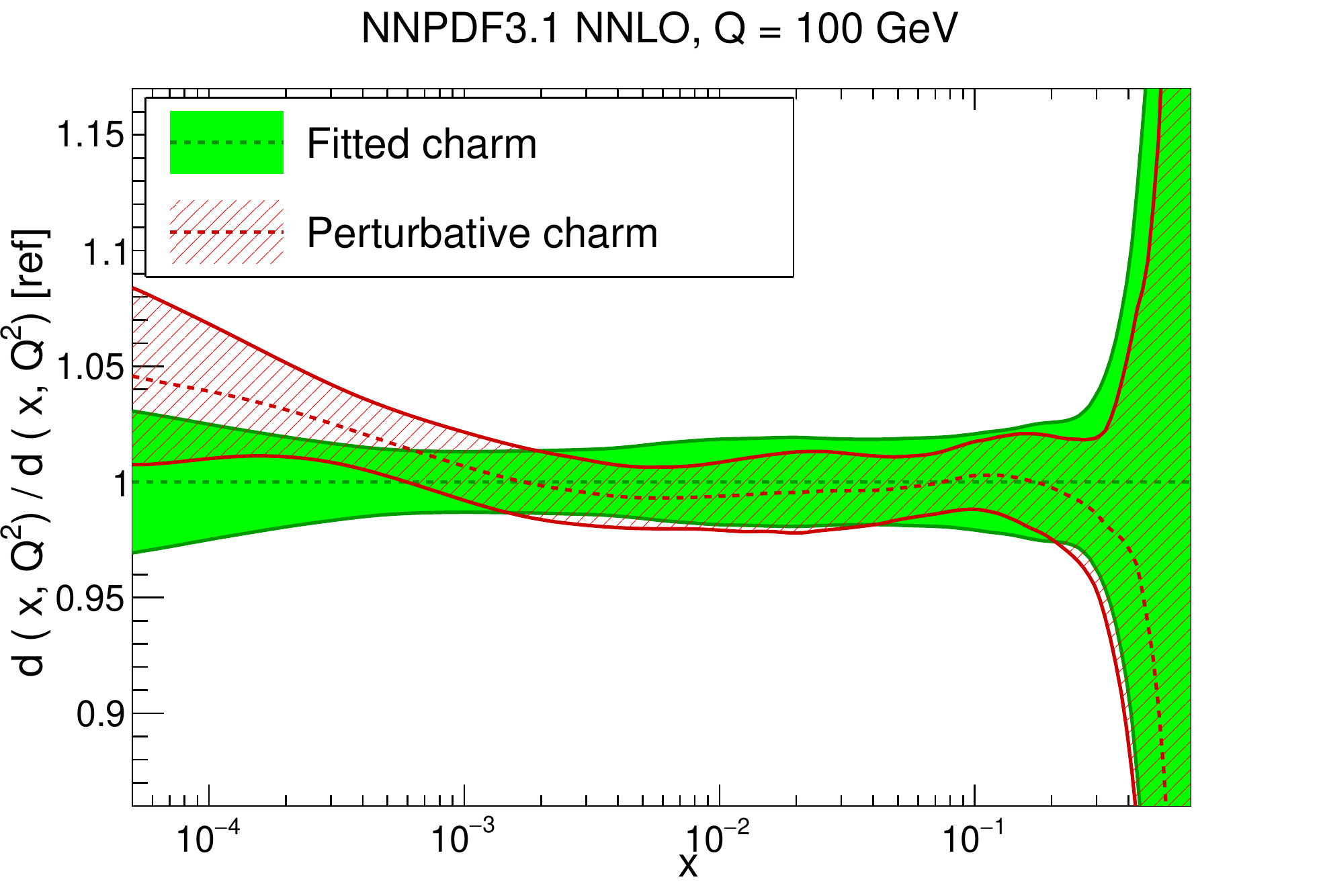}
    \includegraphics[scale=0.32]{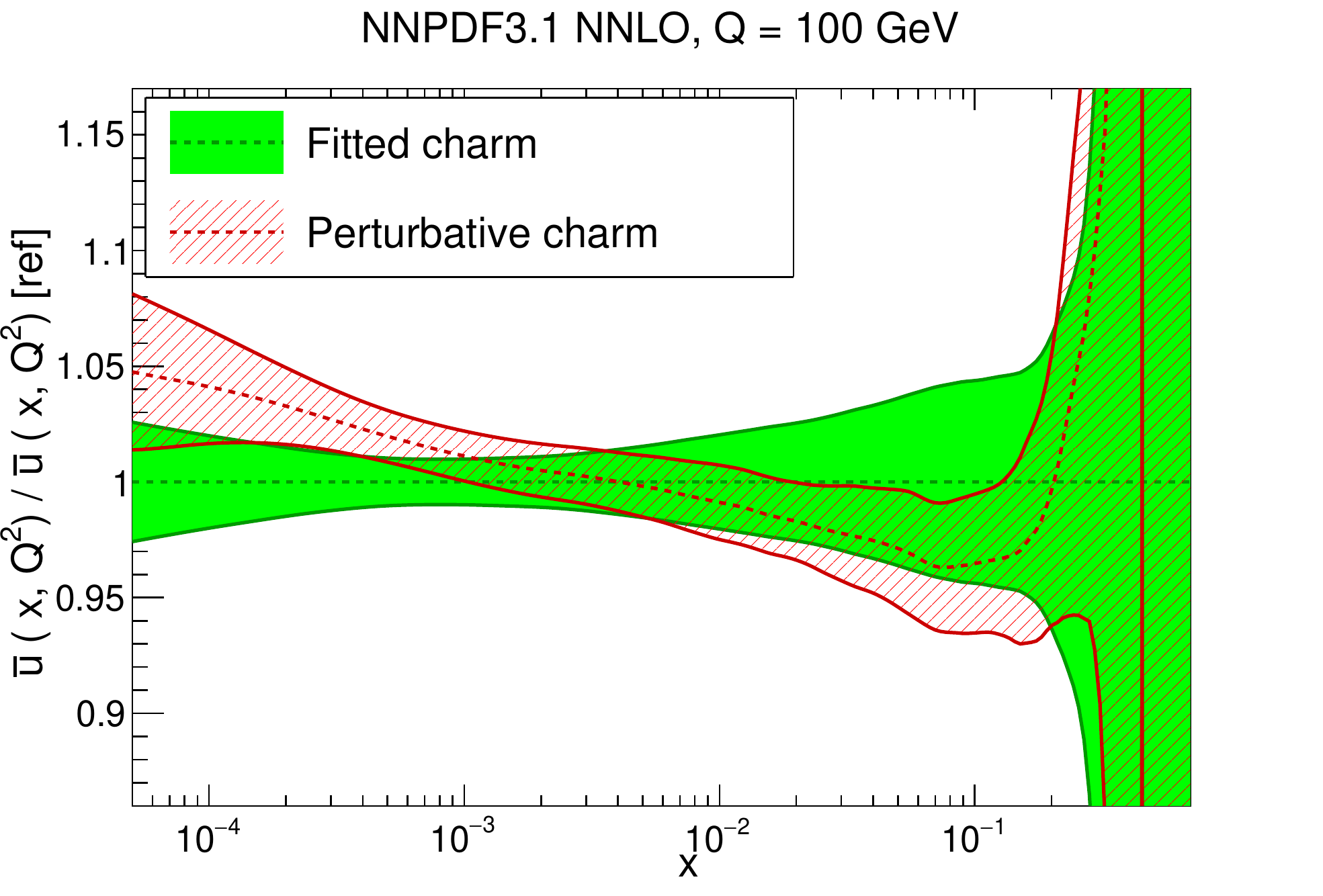}
    \includegraphics[scale=0.32]{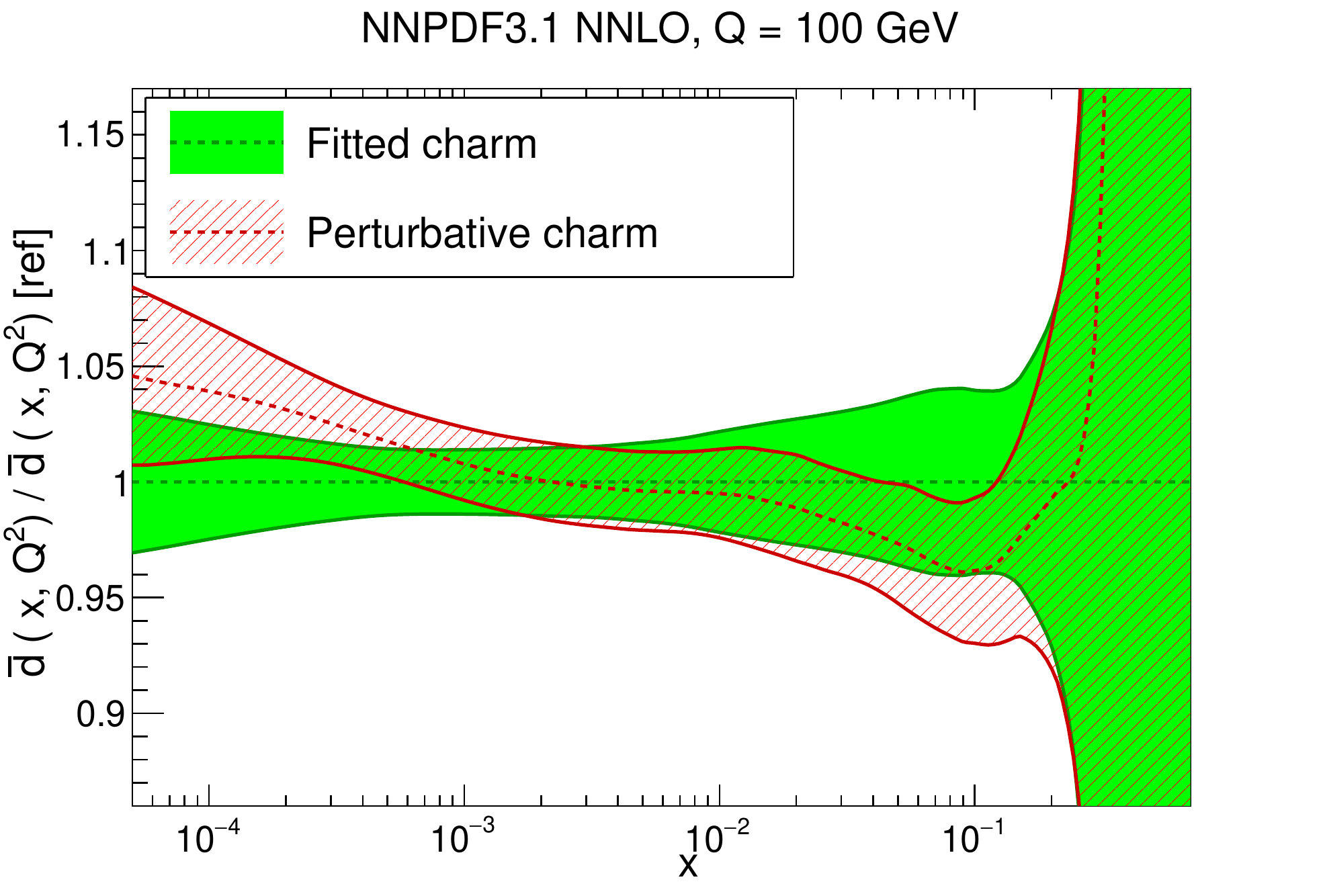}
    \includegraphics[scale=0.32]{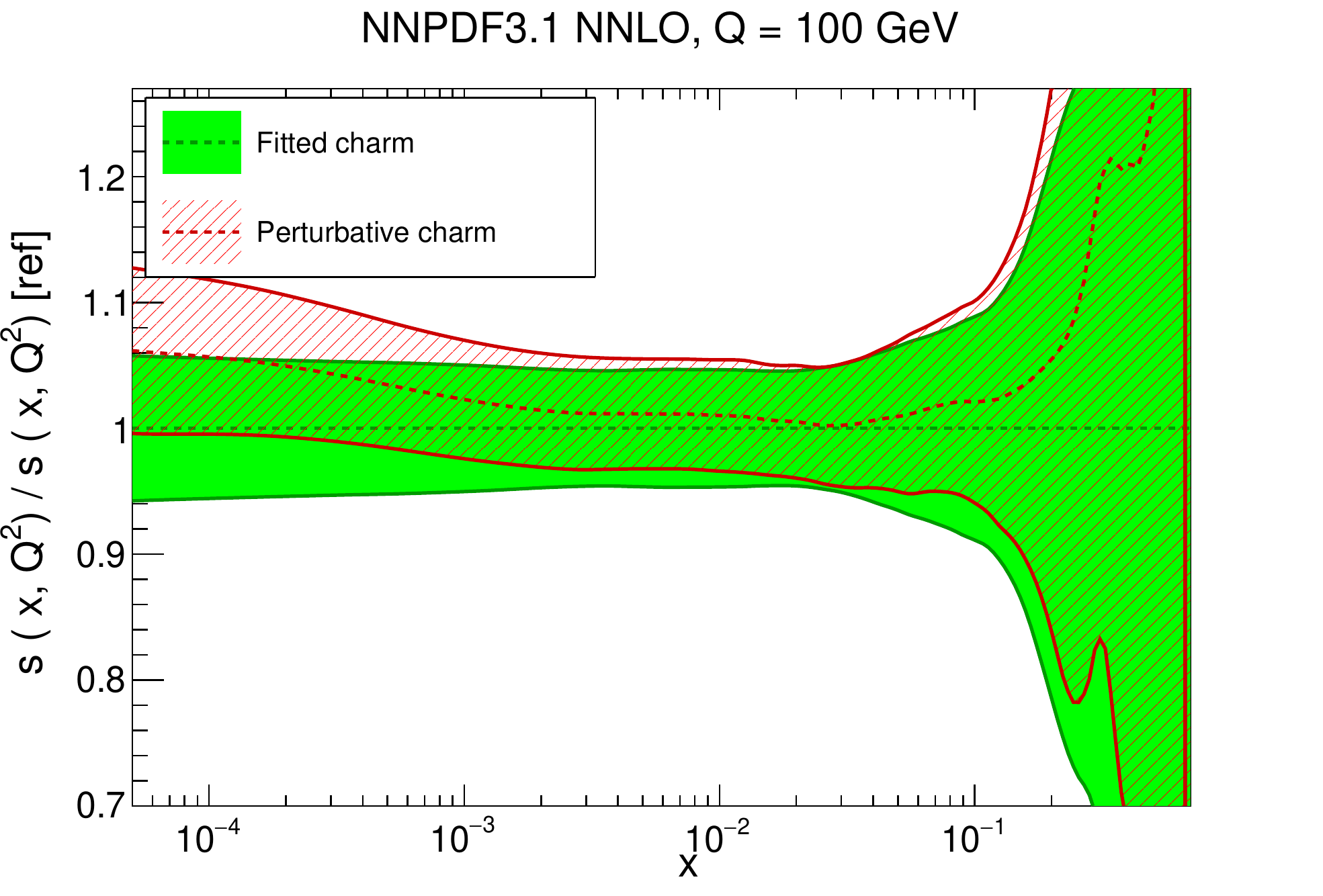}
    \includegraphics[scale=0.32]{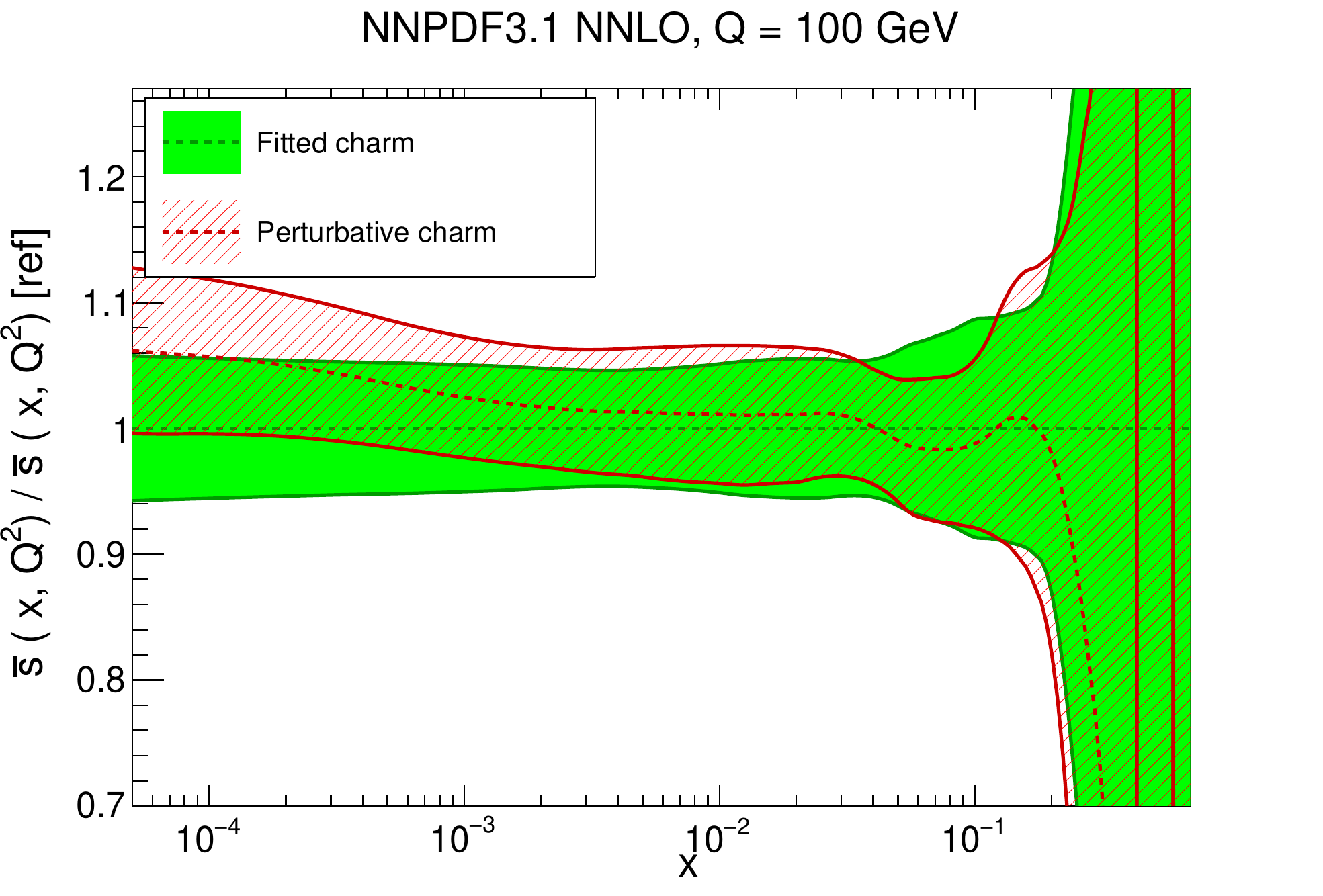}
    \includegraphics[scale=0.32]{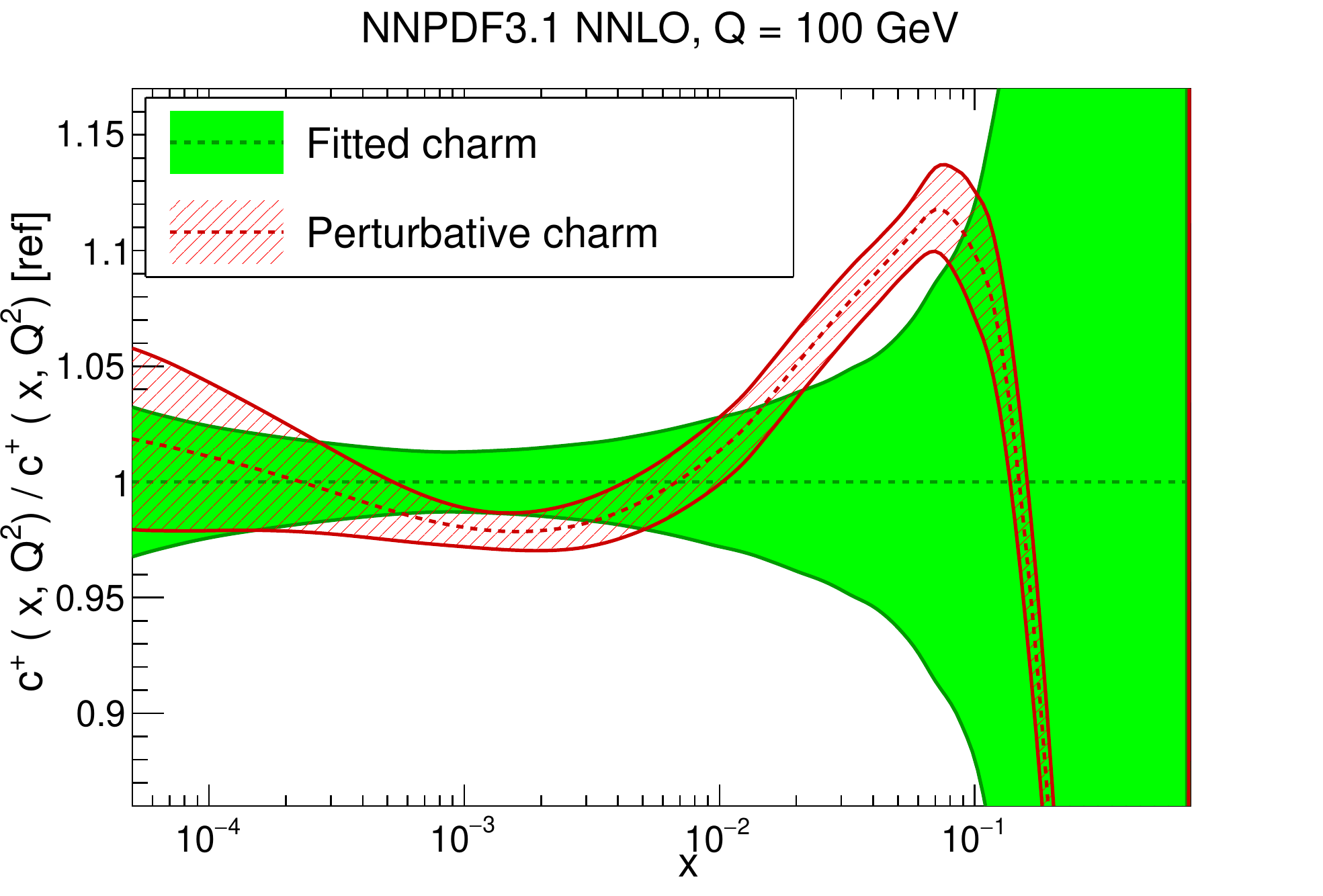}
    \includegraphics[scale=0.32]{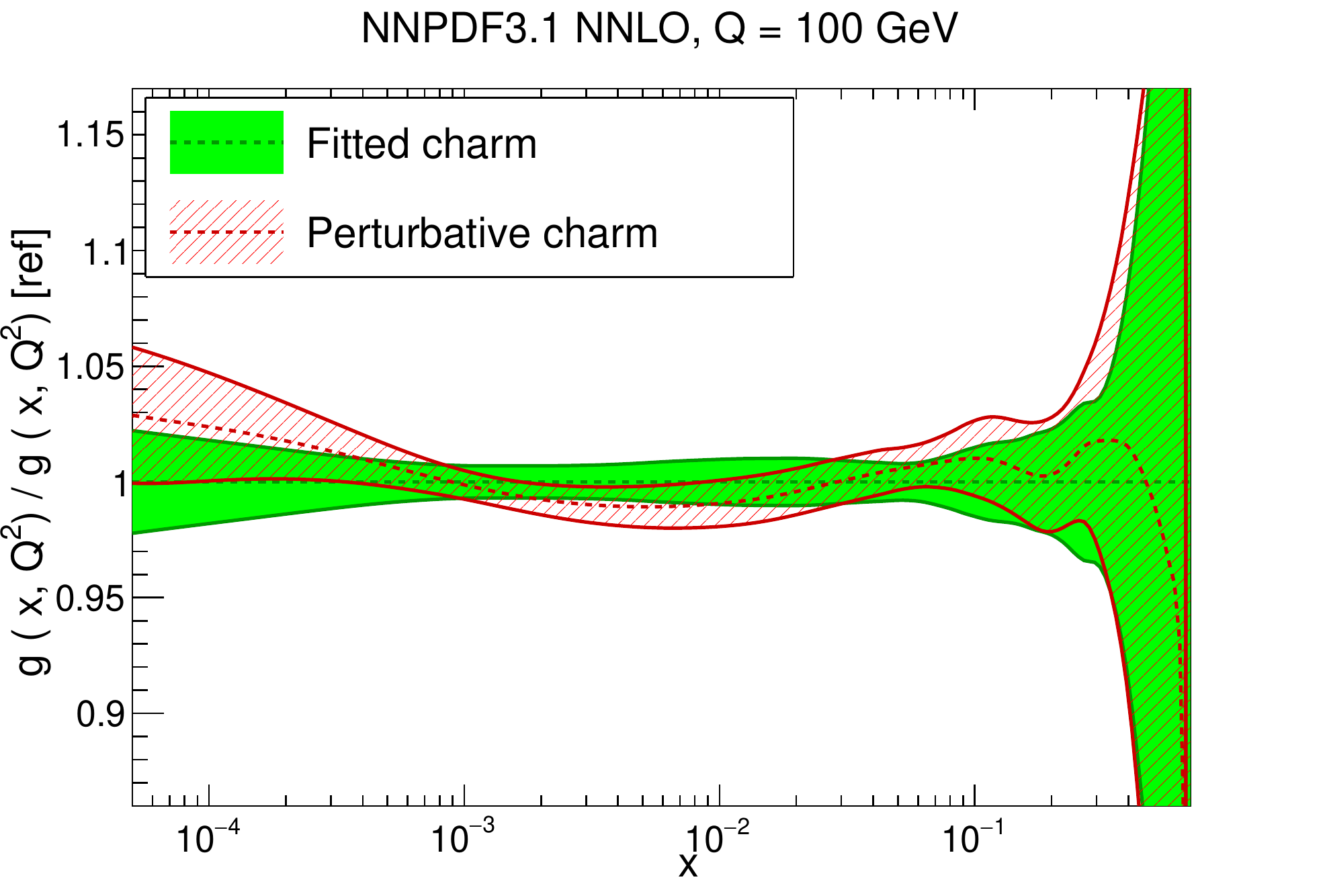}
  \caption{\small 
    Comparison of NNPDF3.1 NNLO PDFs to a variant 
in which charm is generated entirely perturbatively 
(and everything else is unchanged).
    \label{fig:31-nnlo-fitted-vs-pch}
  }
\end{center}
\end{figure}

In Fig.~\ref{fig:31-nnlo-fitted-vs-pch} we directly compare the PDFs
with parametrized and perturbative charm. The light quark PDFs and the
gluon are generally enhanced for $x\gsim0.003$ and reduced for smaller $x$
when charm is independently parametrized. The largest differences can be seen in the up quark, while
the strange and gluon distributions are more stable. The best-fit charm distribution has a
distinctly different shape and significantly larger uncertainty than
its perturbatively generated counterpart. As argued in
Ref.~\cite{Ball:2016neh} this shape might well be compatible with 
a charm PDF generated perturbatively at high perturbative orders. 

\begin{figure}[t]
  \begin{center}
    \includegraphics[scale=0.32]{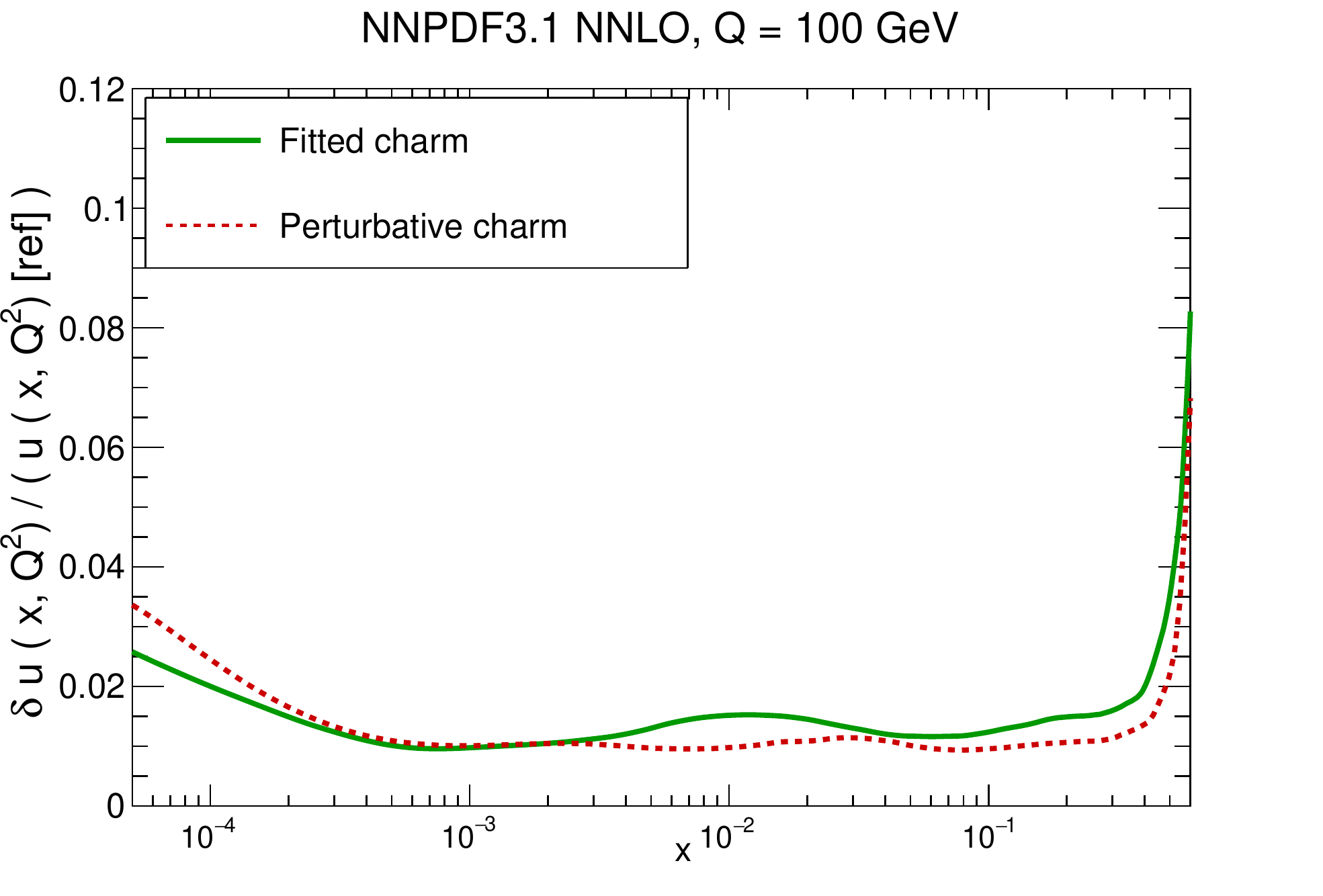}
    \includegraphics[scale=0.32]{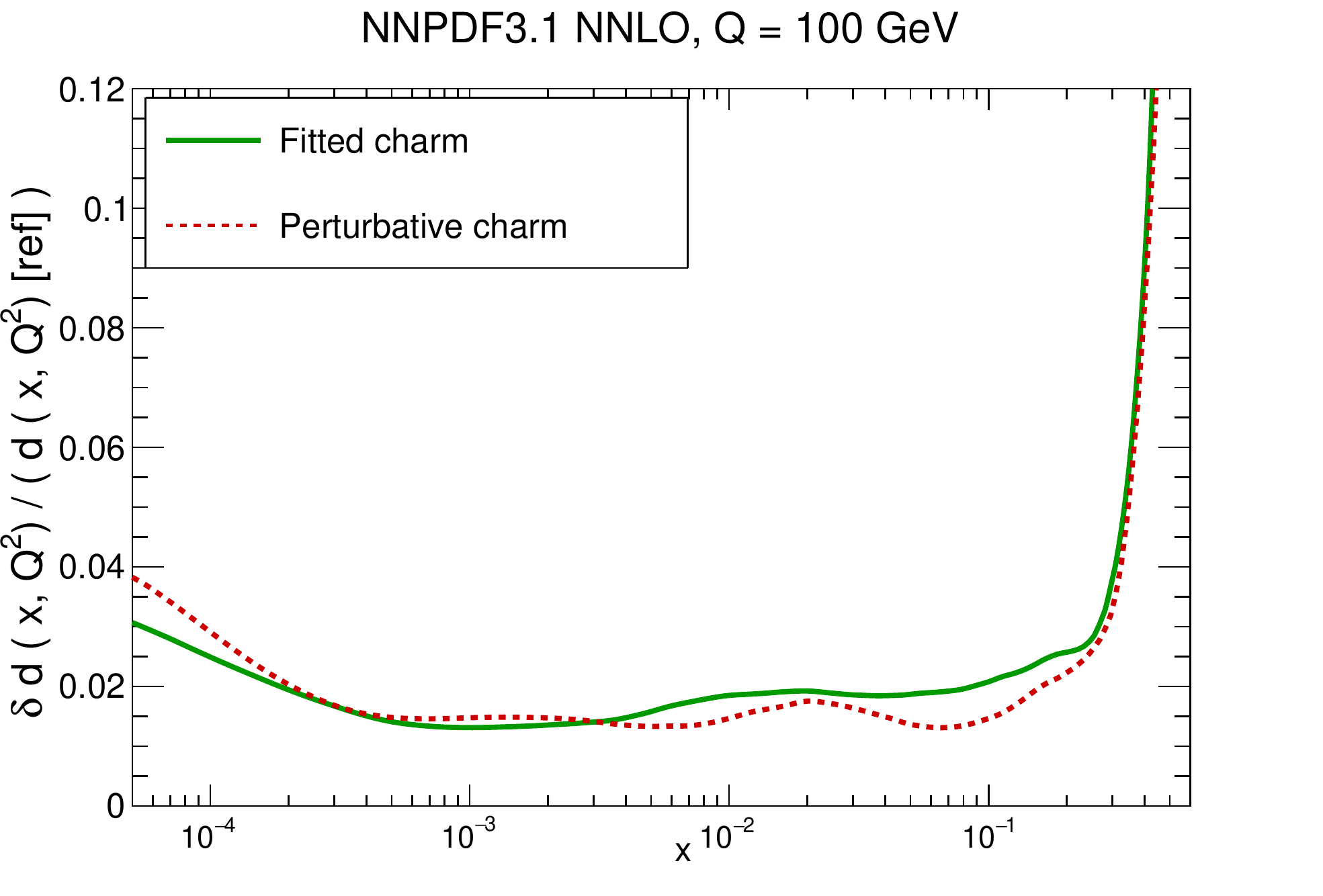}
    \includegraphics[scale=0.32]{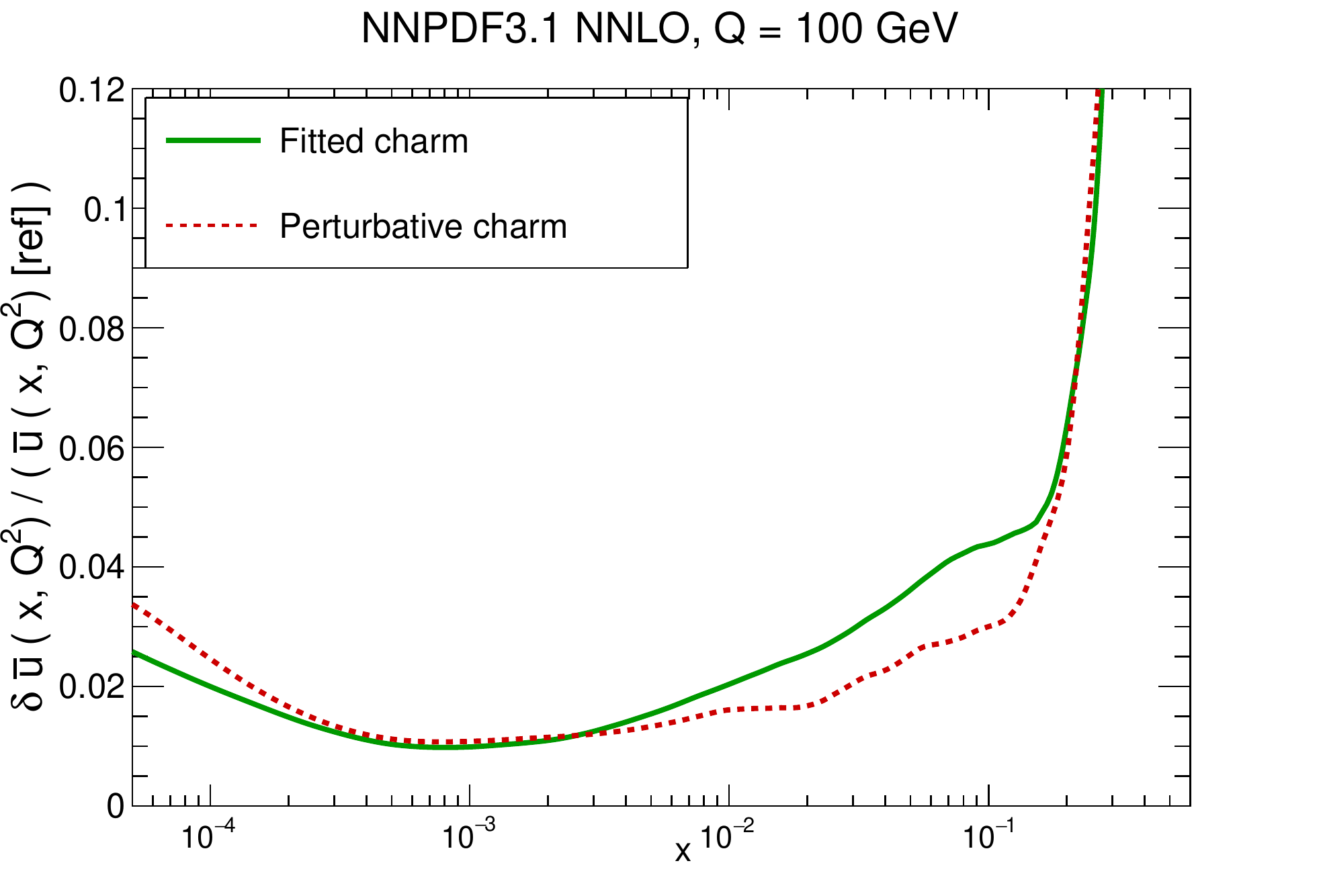}
    \includegraphics[scale=0.32]{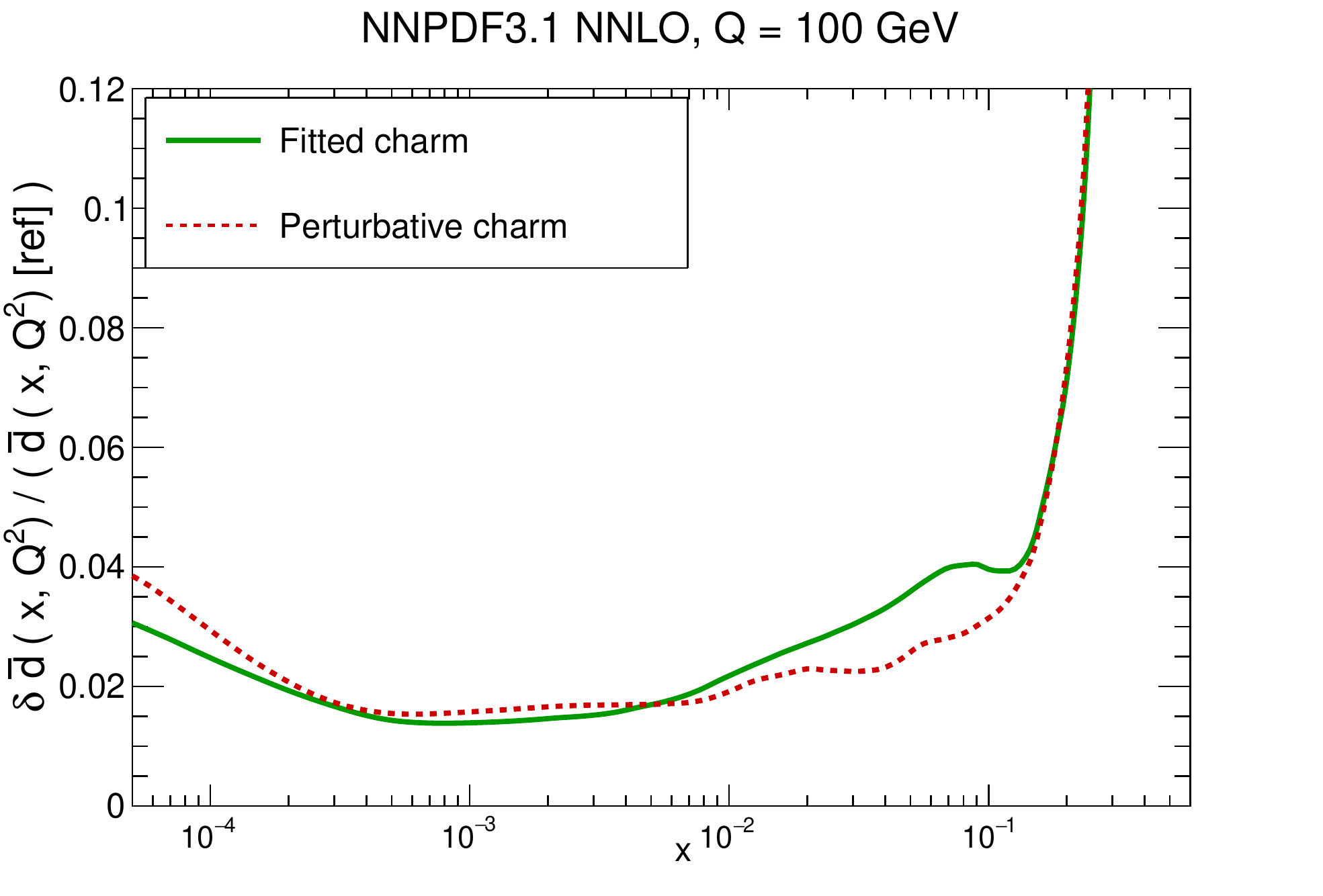}
    \includegraphics[scale=0.32]{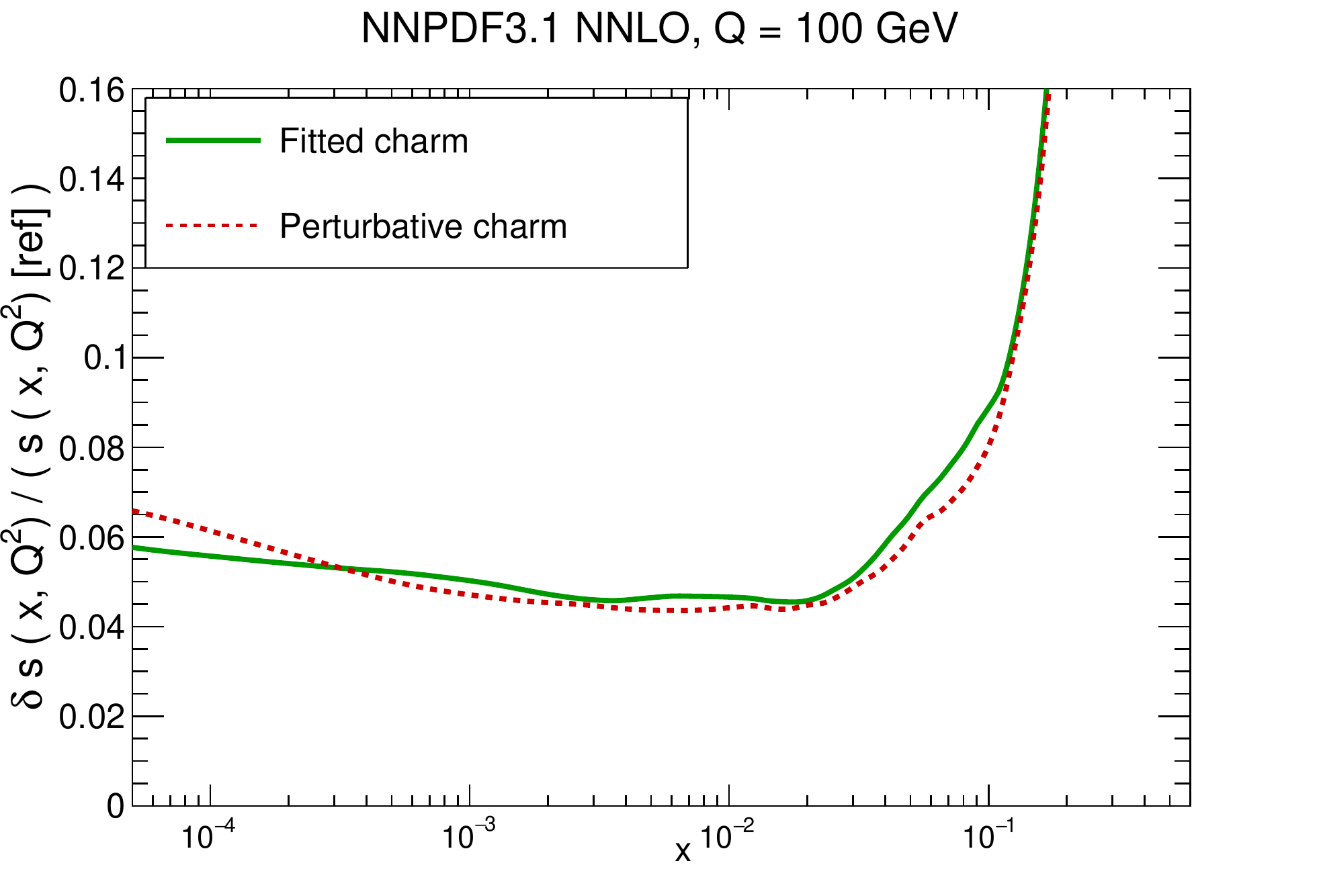}
    \includegraphics[scale=0.32]{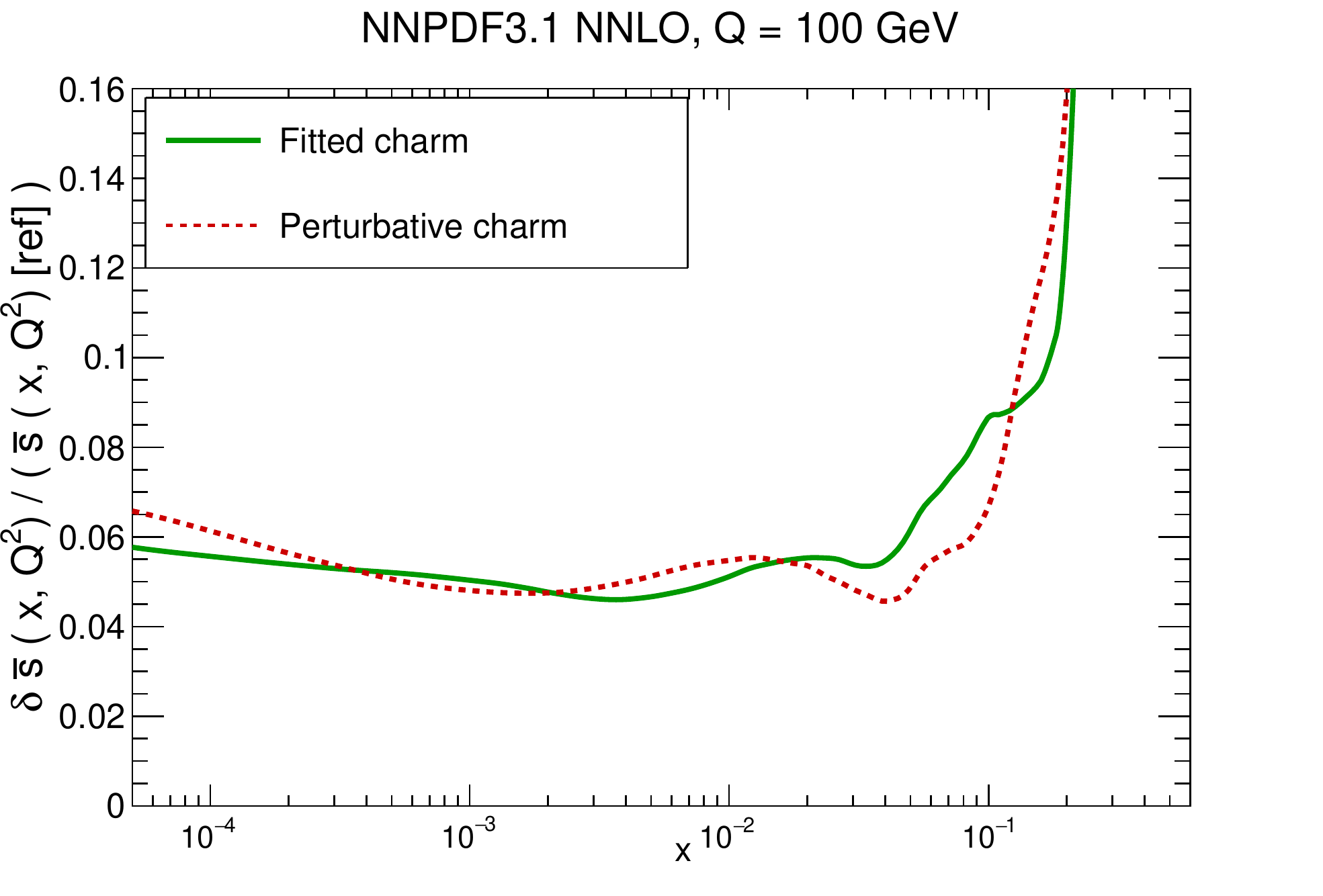}
    \includegraphics[scale=0.32]{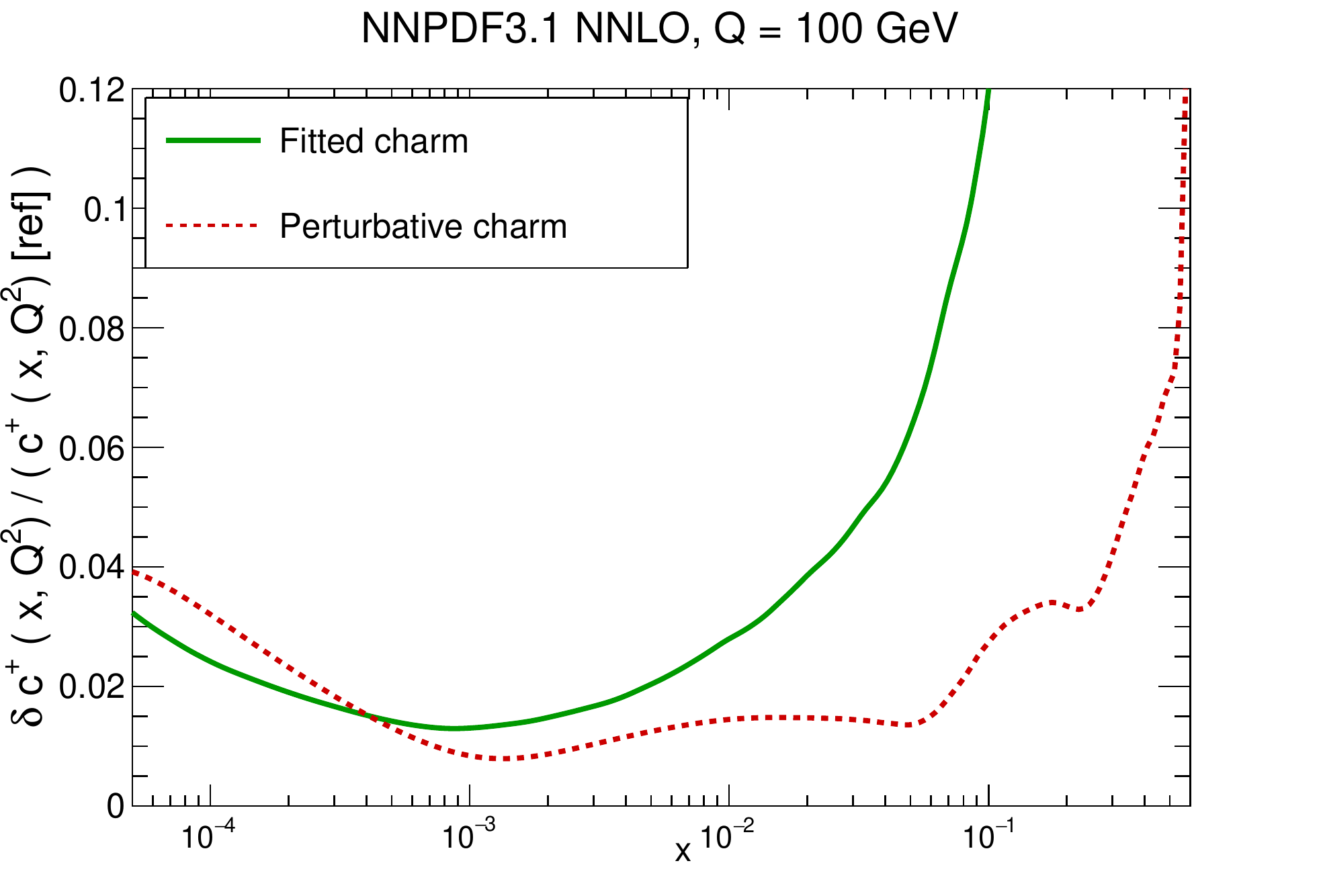}
    \includegraphics[scale=0.32]{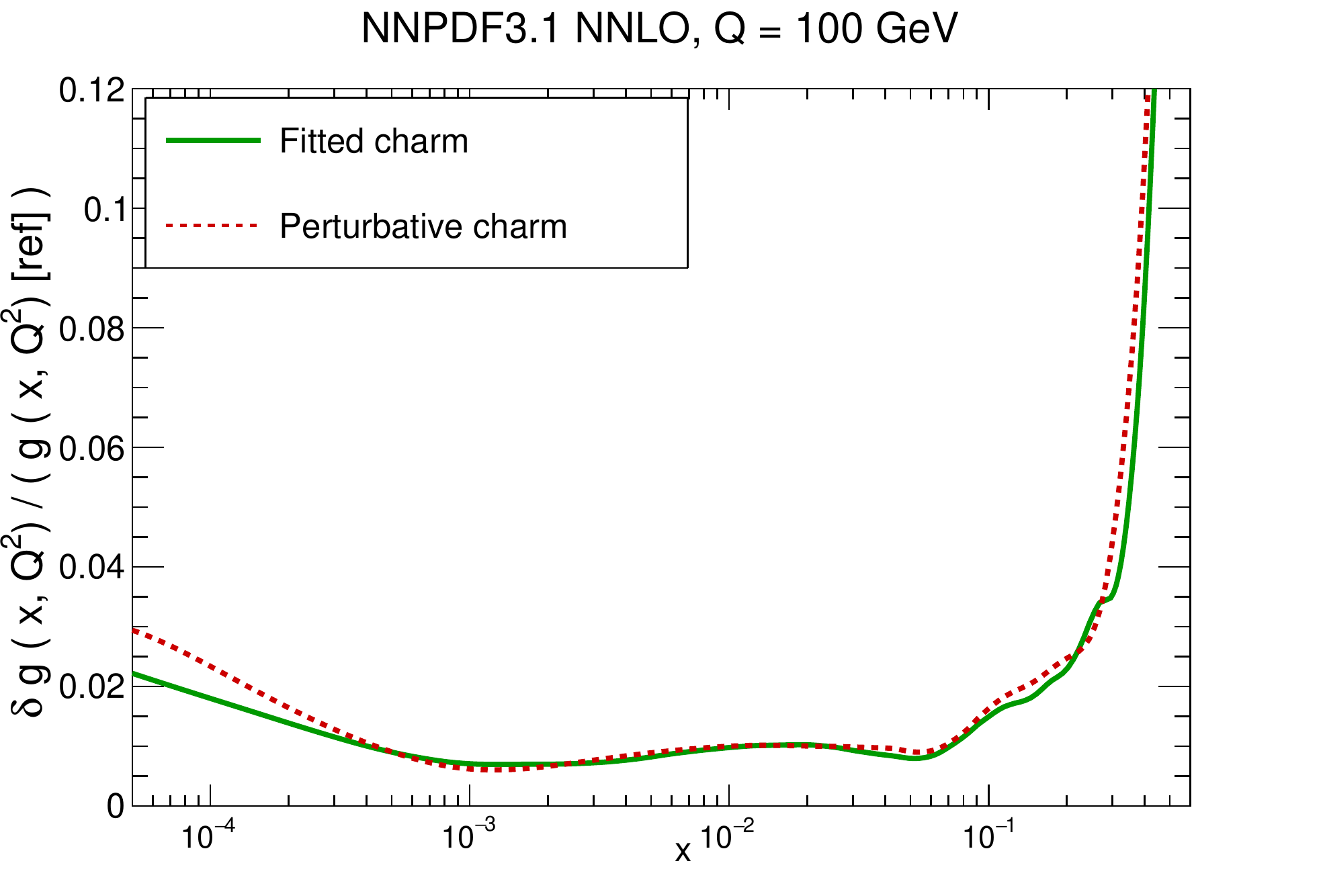}
  \caption{\small 
    Comparison of the fractional
    one-sigma PDF uncertainties in
    NNPDF3.1 NNLO with the corresponding version 
where charm is generated perturbatively
    (and everything else is unchanged).
    The PDF comparison plot was shown in
    Fig.~\ref{fig:31-nnlo-fitted-vs-pch}.
    \label{fig:31-nnlo-fitted-vs-pch-err}
  }
\end{center}
\end{figure}

In Fig.~\ref{fig:31-nnlo-fitted-vs-pch-err} we directly compare
PDF uncertainties. It is remarkable that the uncertainties other than
for charm are essentially unchanged when charm is independently parametrized,
with only a slight increase in sea quark PDF uncertainties for
$10^{-3}\lsim x \lsim 10^{-2}$. The uncertainty on the gluon is 
almost completely unaffected. The PDF uncertainty on charm when it is
independently parametrized  is in
line with that of other sea quark PDFs, while the uncertainty of the
perturbatively generated charm follows that of the gluon and is consequently 
much smaller.

\begin{figure}[t]
\begin{center}
  \includegraphics[scale=0.36]{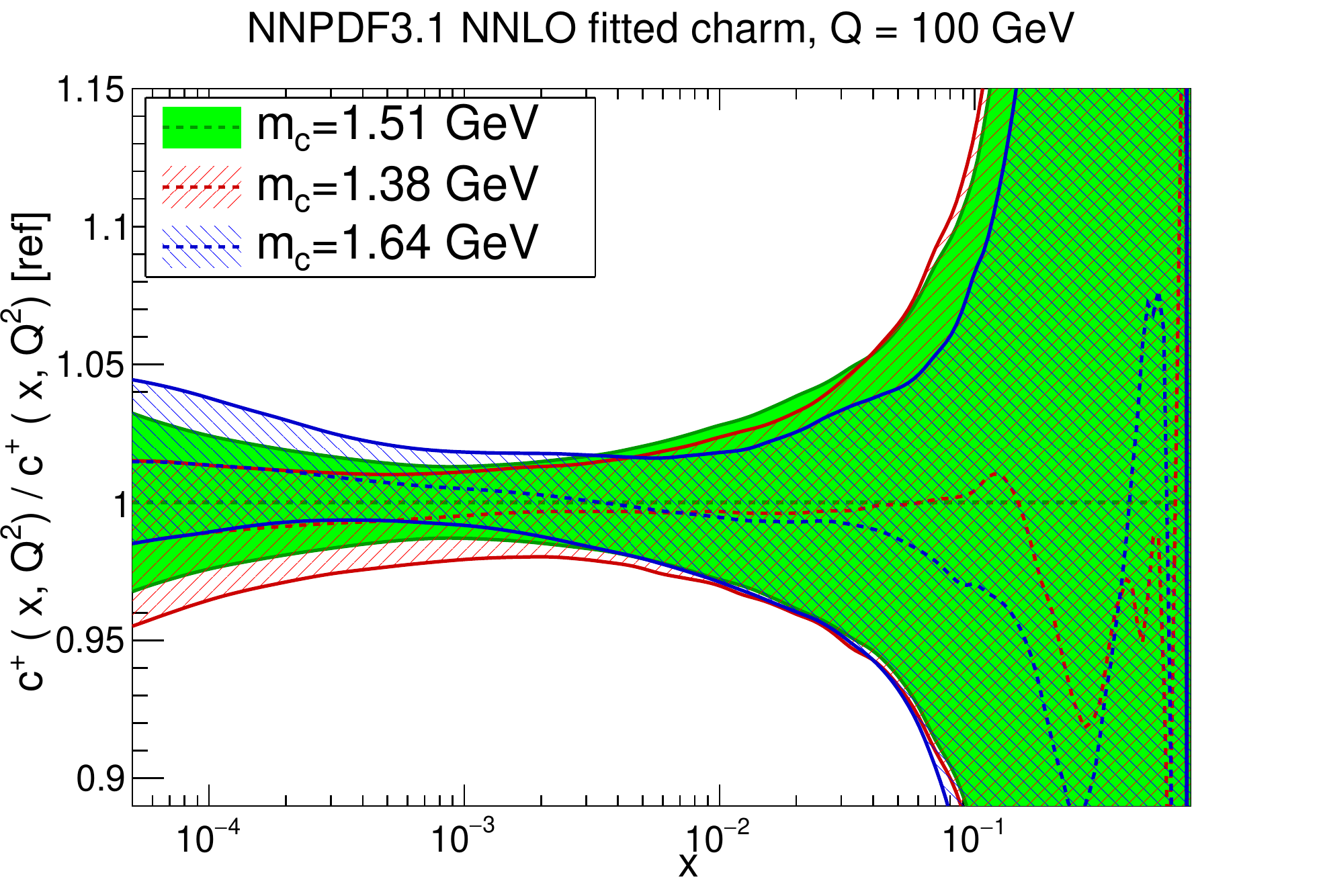}
  \includegraphics[scale=0.36]{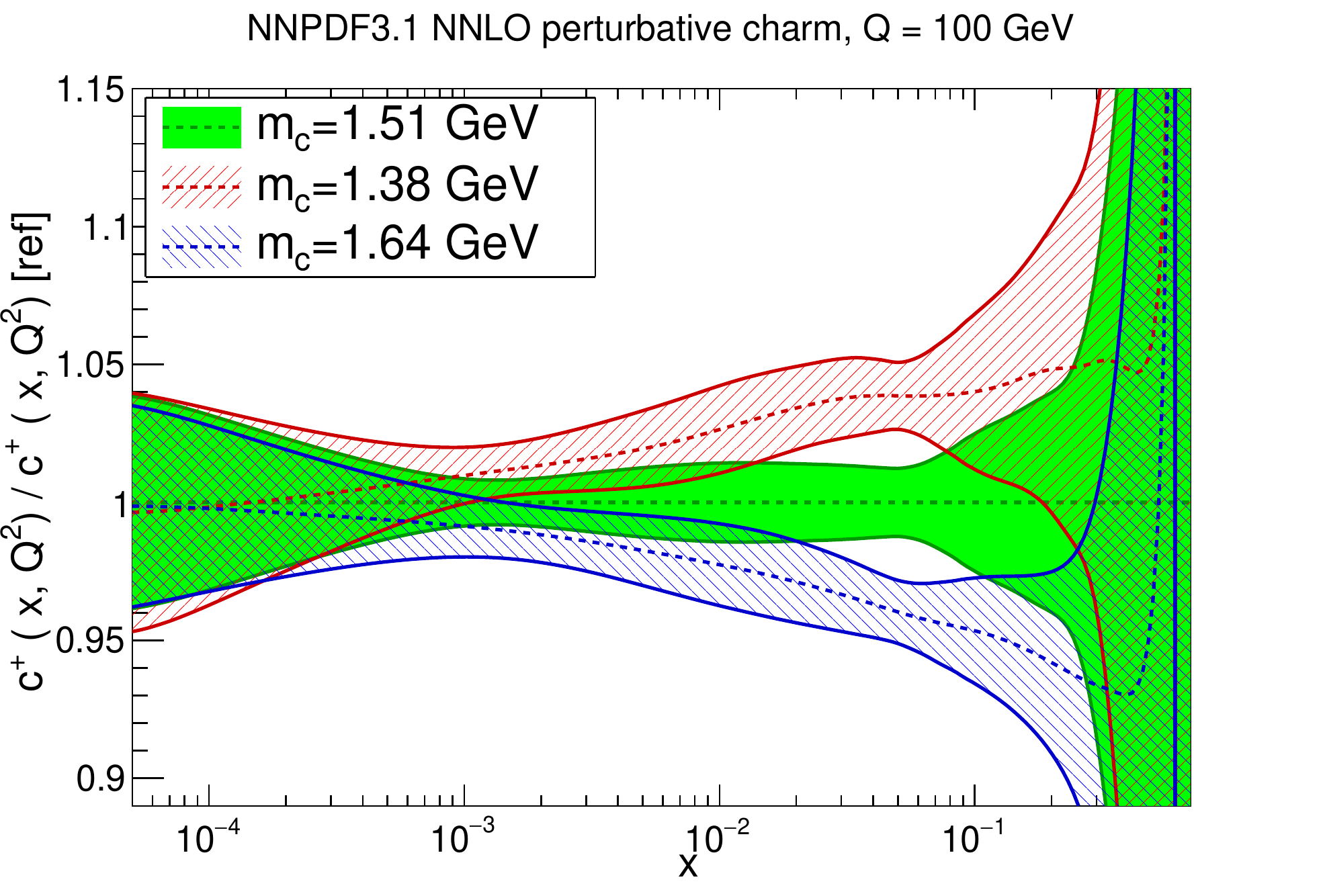}
  \includegraphics[scale=0.36]{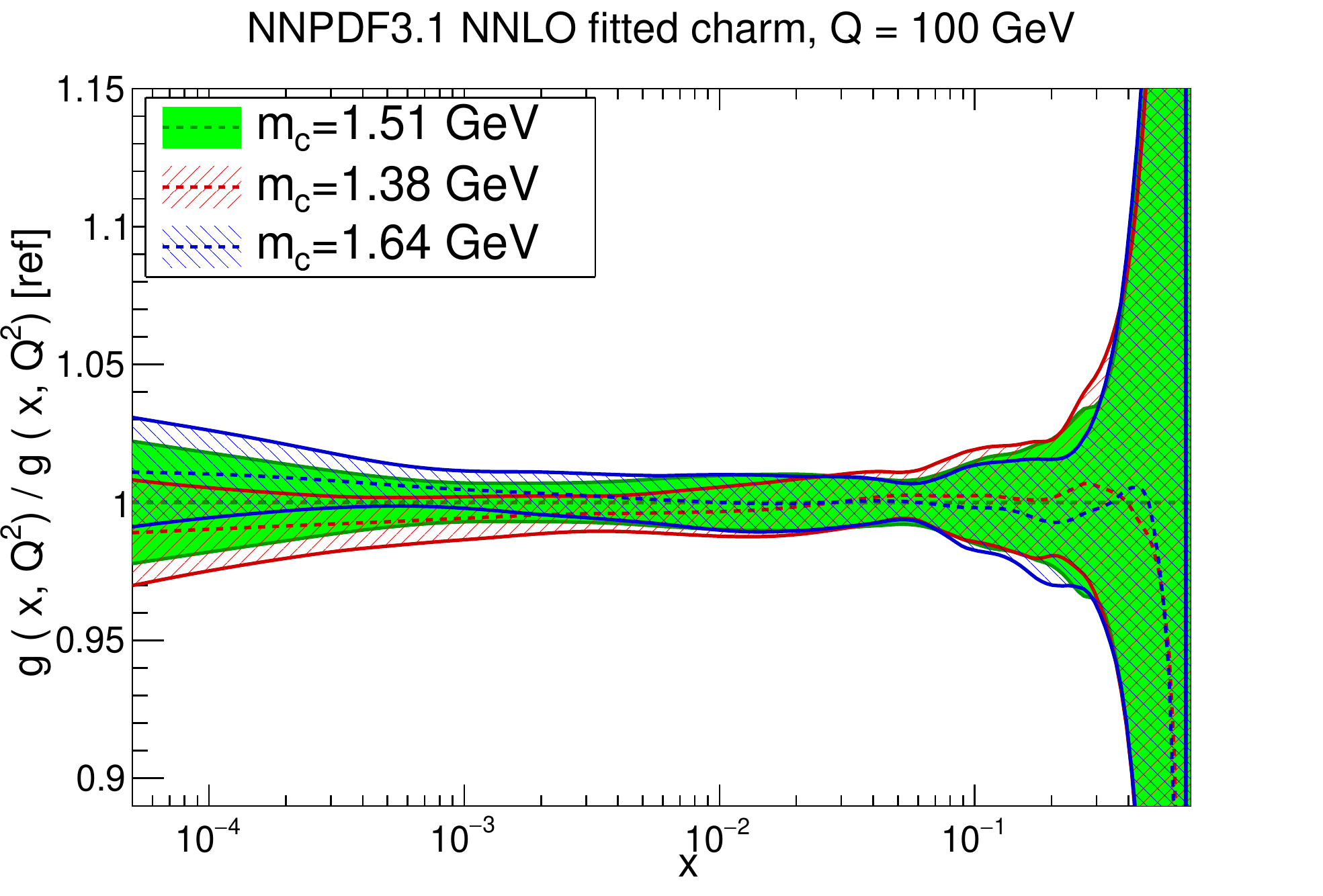}
  \includegraphics[scale=0.36]{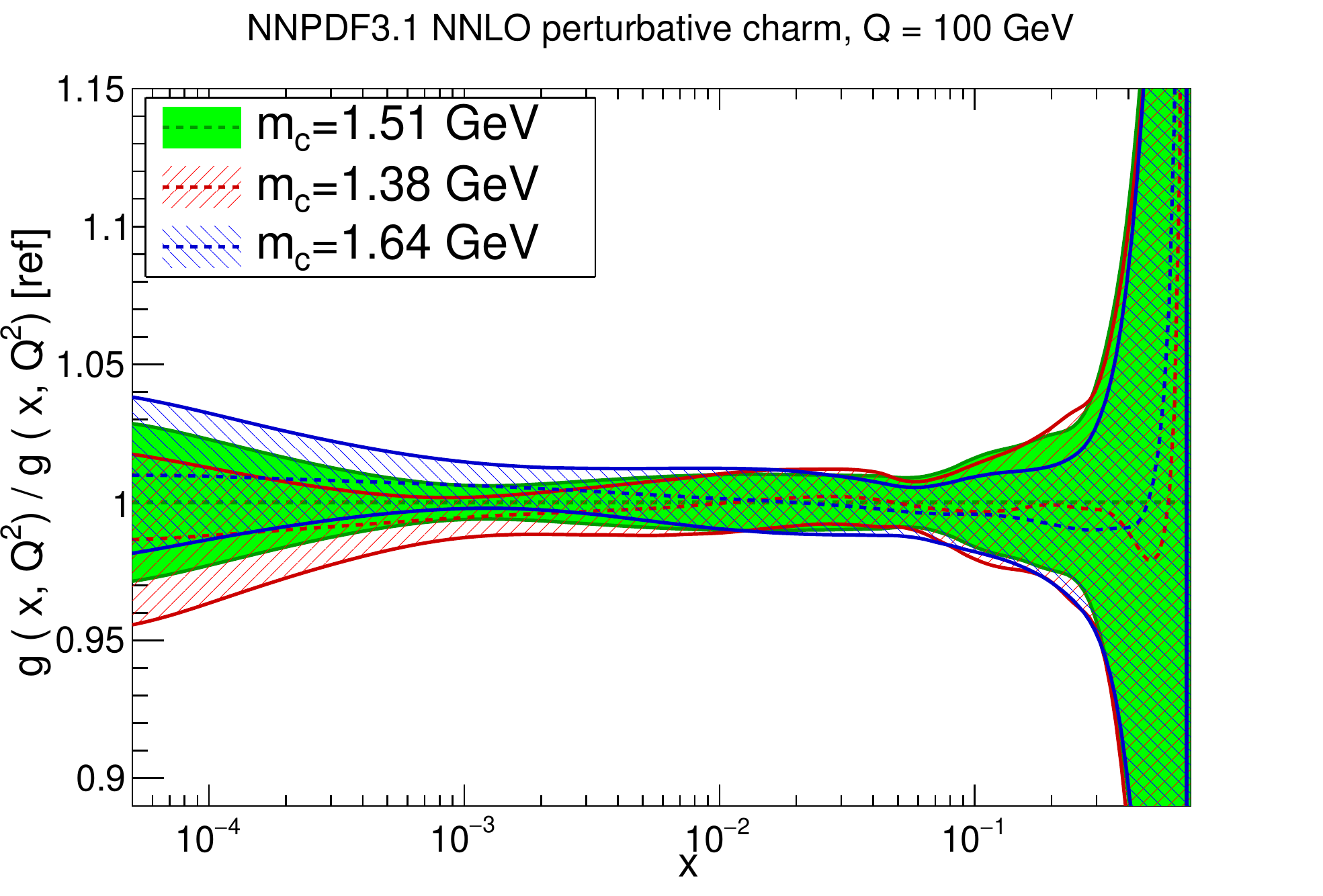}
  \includegraphics[scale=0.36]{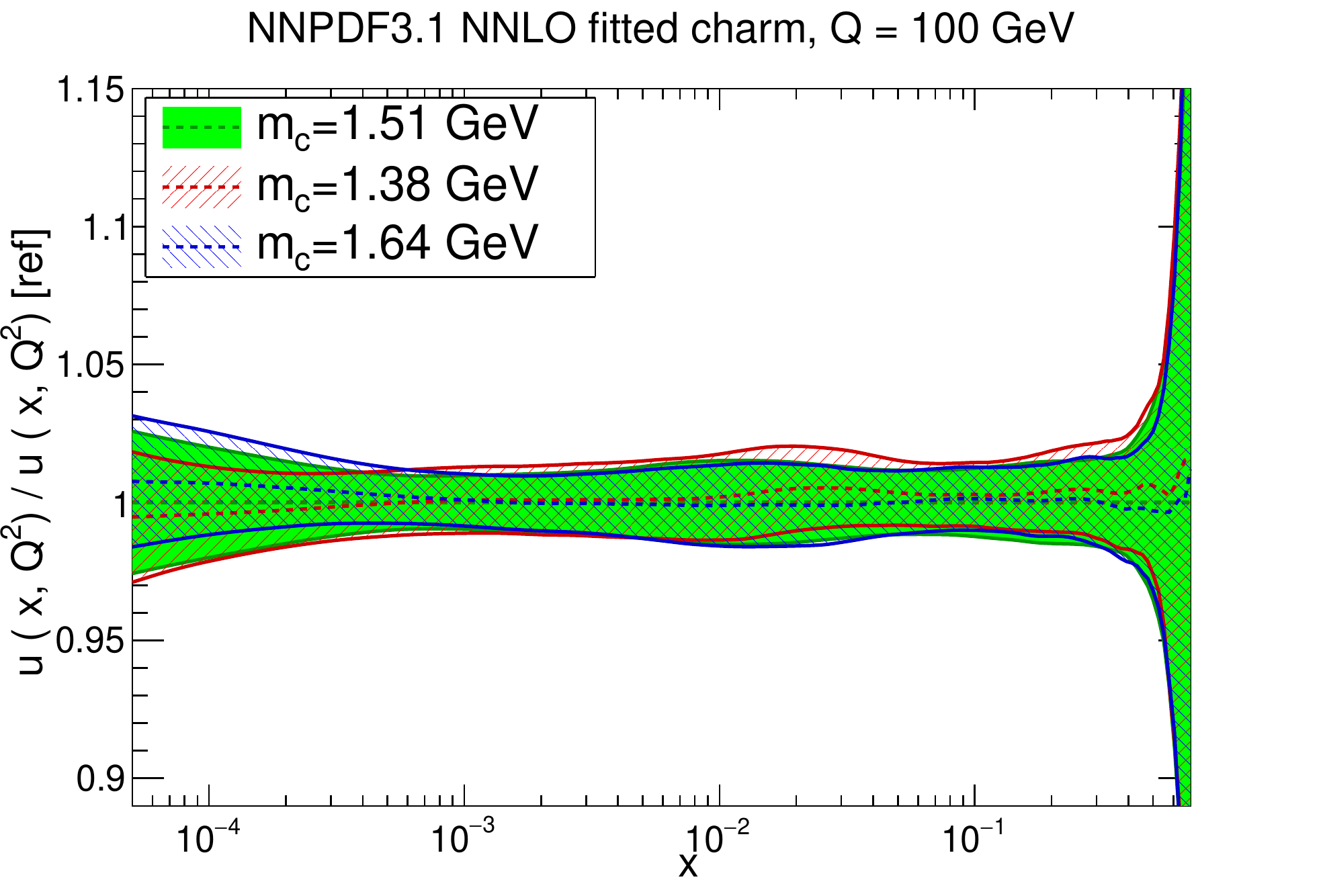}
  \includegraphics[scale=0.36]{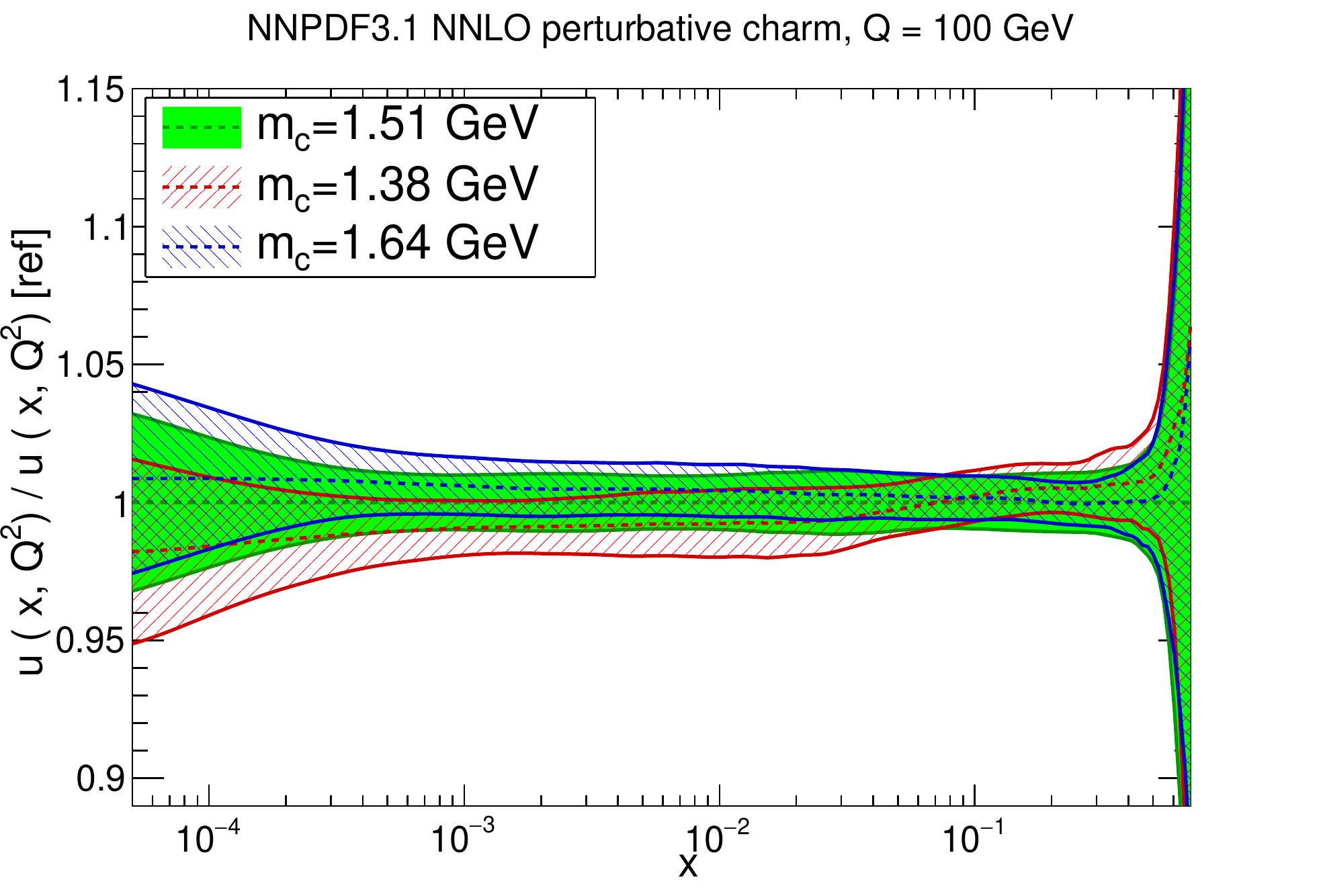}
  \includegraphics[scale=0.36]{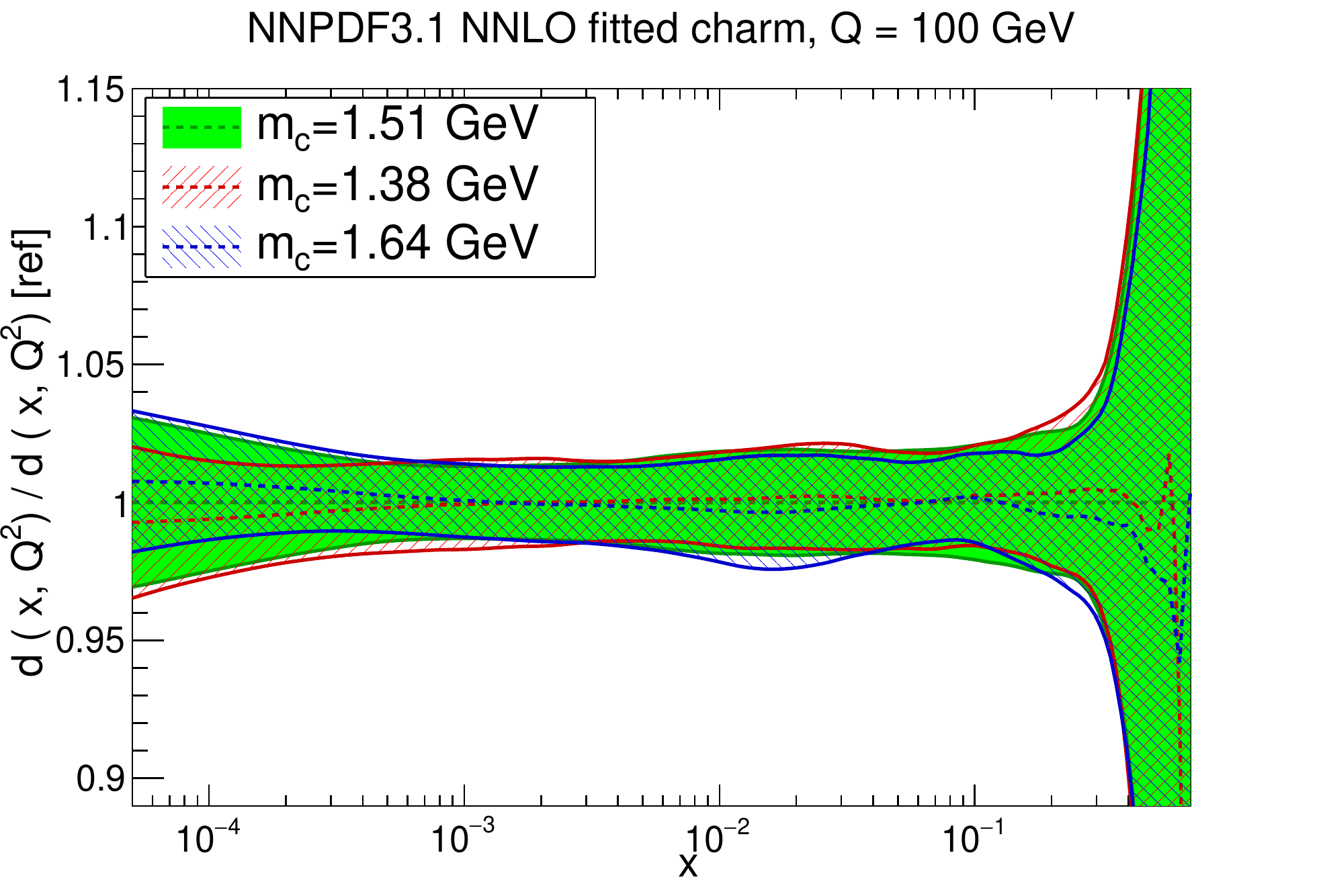}
  \includegraphics[scale=0.36]{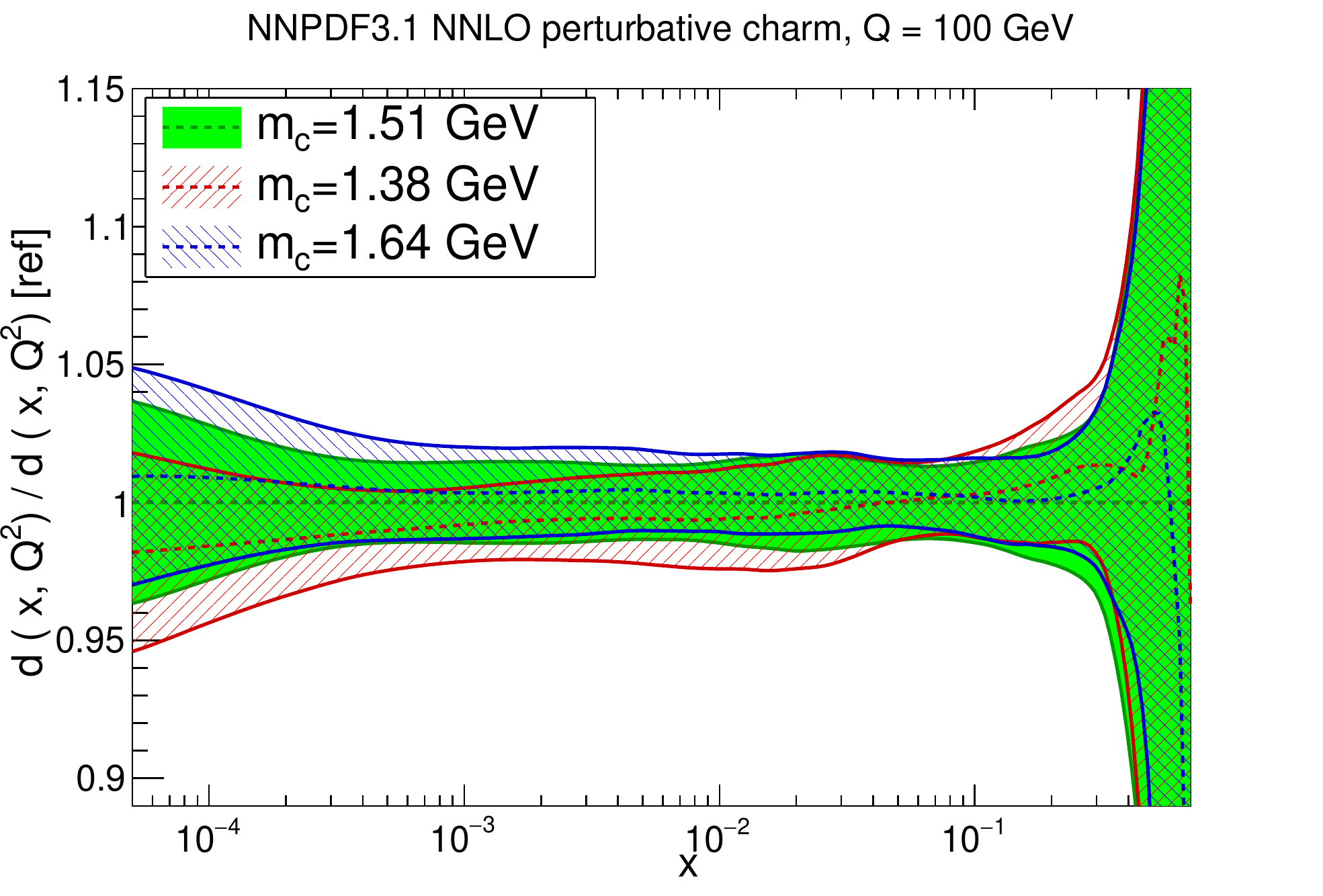}
  \caption{\small Dependence of the NNPDF3.1 NNLO PDFs on the charm
    mass. Results are shown both for parametrized charm  (left)
    and  perturbative charm (right), for  (from top to bottom) charm,
    gluon, up and down PDFs.
        \label{fig:nnlo-fc-mcdep-1}
  }
\end{center}
\end{figure}

A previous comparison of PDFs determined with parametrized or perturbative
charm was presented in Ref.~\cite{Ball:2016neh} and led to the
conclusion that parametrizing charm and determining it from the data
 greatly reduces the dependence on the
charm mass thereby reducing the overall PDF uncertainty when the
uncertainty due to the charm mass is kept into account. 
As mentioned, NNPDF3.1 PDFs are determined using heavy quark
pole mass values and uncertainties recommended by the
Higgs Cross-Section Working Group~\cite{deFlorian:2016spz}.
For charm, this corresponds to
$ m_c^{\rm pole}=1.51 \pm 0.13\,{\rm GeV}$.
In order to estimate the impact of this uncertainty, we have produced 
NNPDF3.1 NNLO sets with $m_c^{\rm pole}=1.38$~GeV and $m_c^{\rm pole}=1.64$~GeV.
Results are shown in Fig.~\ref{fig:nnlo-fc-mcdep-1} for some
representative PDFs, both for the default NNPDF3.1 and for the version
with perturbative charm. It is clear that the very strong
dependence of the charm PDF on $m_c$ which is found when
charm is perturbatively generated all but disappears when charm is
independently parametrized. While the gluon is always quite stable, the 
dependence of perturbatively generated charm on $m_c$ propagates to
the light quark distributions. These are therefore significantly
stabilized by parametrizing charm. Indeed, if charm is generated
perturbatively, the shift in up and down quark distribution upon one-sigma
variation of the charm mass is comparable to (though somewhat smaller
than) 
the PDF uncertainty.
When charm is independently parametrized 
this dependence is considerably reduced. 
With parametrized charm, collider
observables at high scales become essentially independent of the charm mass, 
in line with the  expectation from decoupling arguments.

\clearpage

\subsection{Theoretical uncertainties}
\label{sec:thunc}

PDF uncertainties on global PDF sets entering the PDF4LHC15
combination
 consist only of the uncertainty propagated
from experimental data and uncertainties due to the
methodology. These
can be controlled through closure testing. There are however further sources
of uncertainty due to the theory used in PDF determination, which we
briefly assess here. These can be divided into two main classes:
\begin{itemize}
\item  Missing higher order uncertainties (MHOU), arising due to the truncation of
  the QCD perturbative expansion at a given fixed order (LO, NLO or NNLO)
  in the theory used for PDF determination.
\item Parametric uncertainties, due to the uncertainties on the values
  of parameters of the theory used for PDF determination: the main ones
 are the values of $\alpha_s(m_Z)$ and
  of $m_c^{\rm pole}$.
\end{itemize}

\begin{figure}[t]
\begin{center}
  \includegraphics[scale=1]{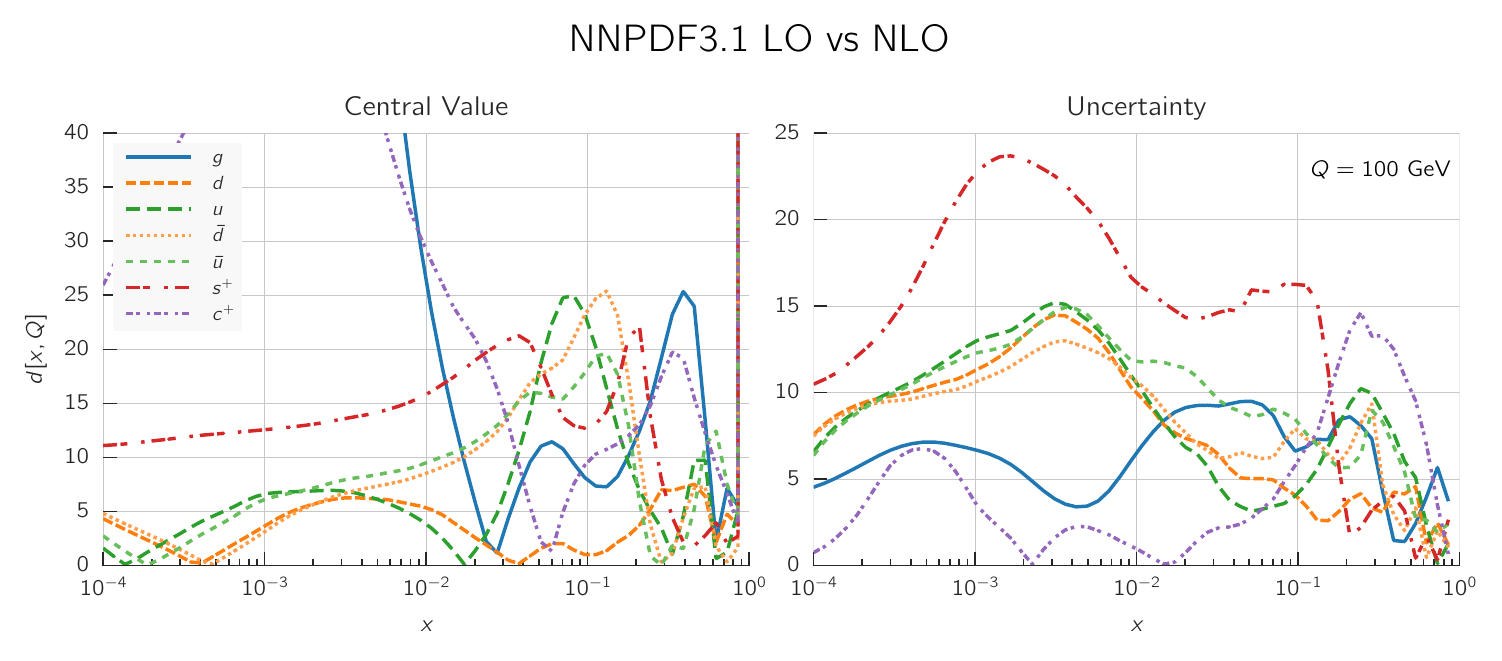}
  \includegraphics[scale=1]{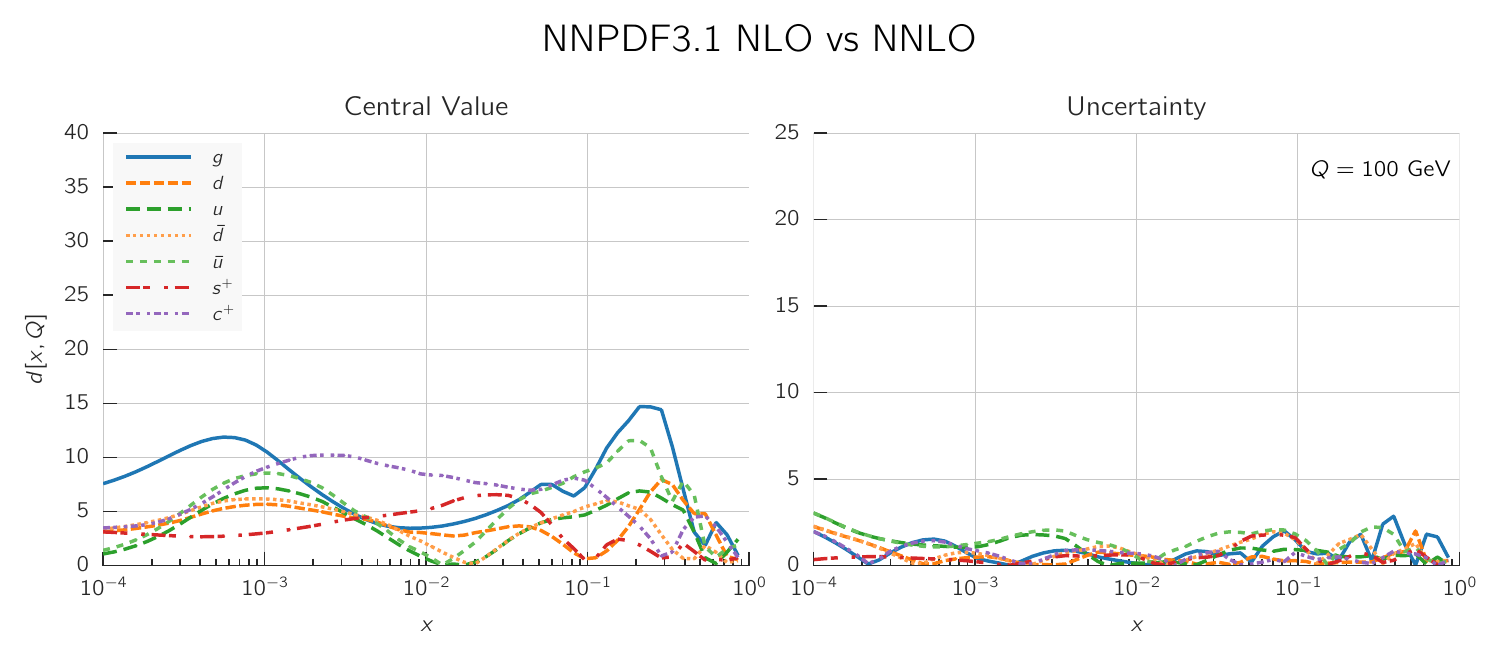}
  \caption{\small Distances between the
    LO and NLO (top) and the NLO and NNLO (bottom) NNPDF3.1
    NNLO PDFs
    at $Q=100$ GeV. Note the difference in scale on the $y$ axis between
    the two plots.
    \label{fig:distances_perturbativeorder}
  }
\end{center}
\end{figure}

\begin{figure}[t]
\begin{center}
  \includegraphics[scale=0.33]{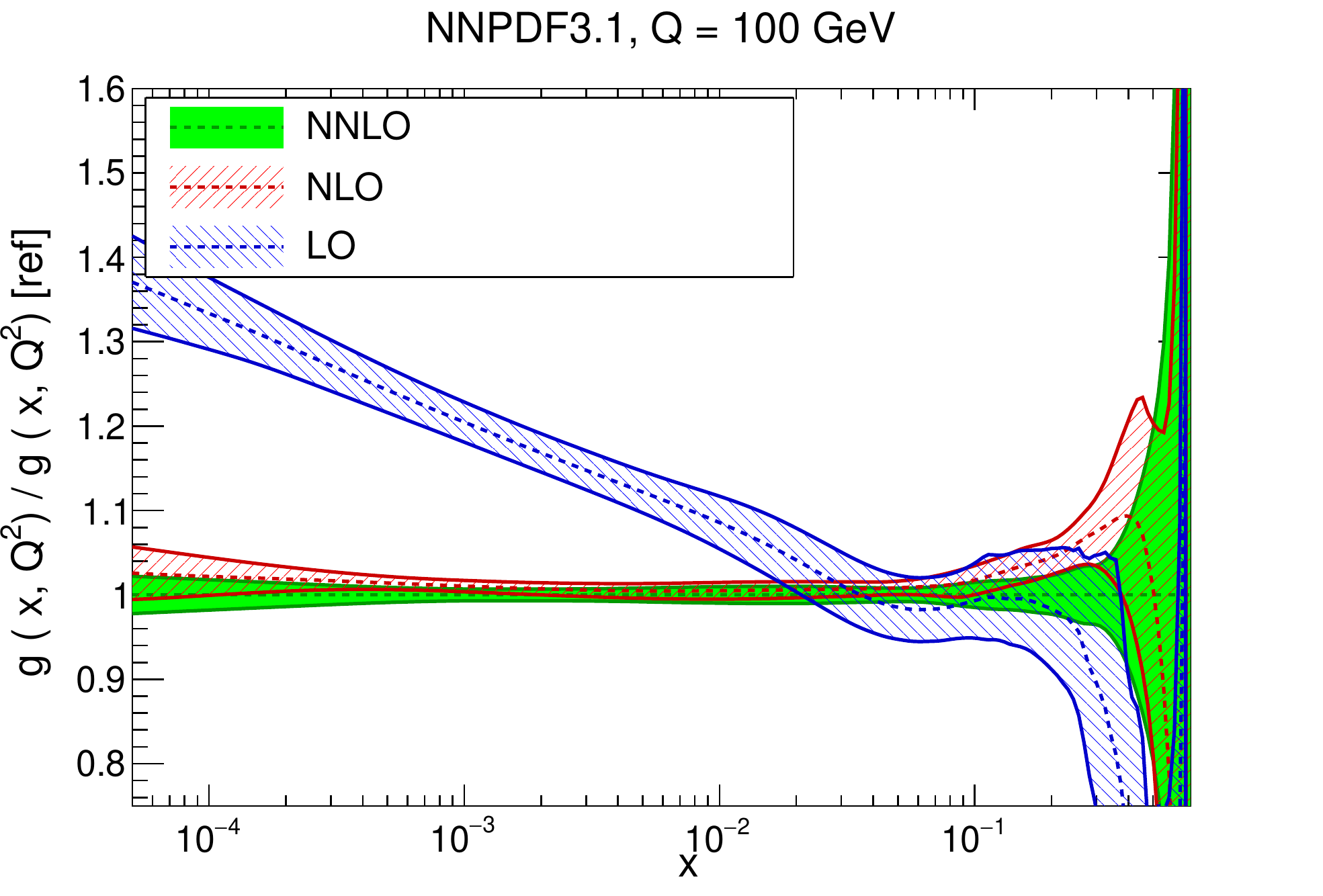}
  \includegraphics[scale=0.33]{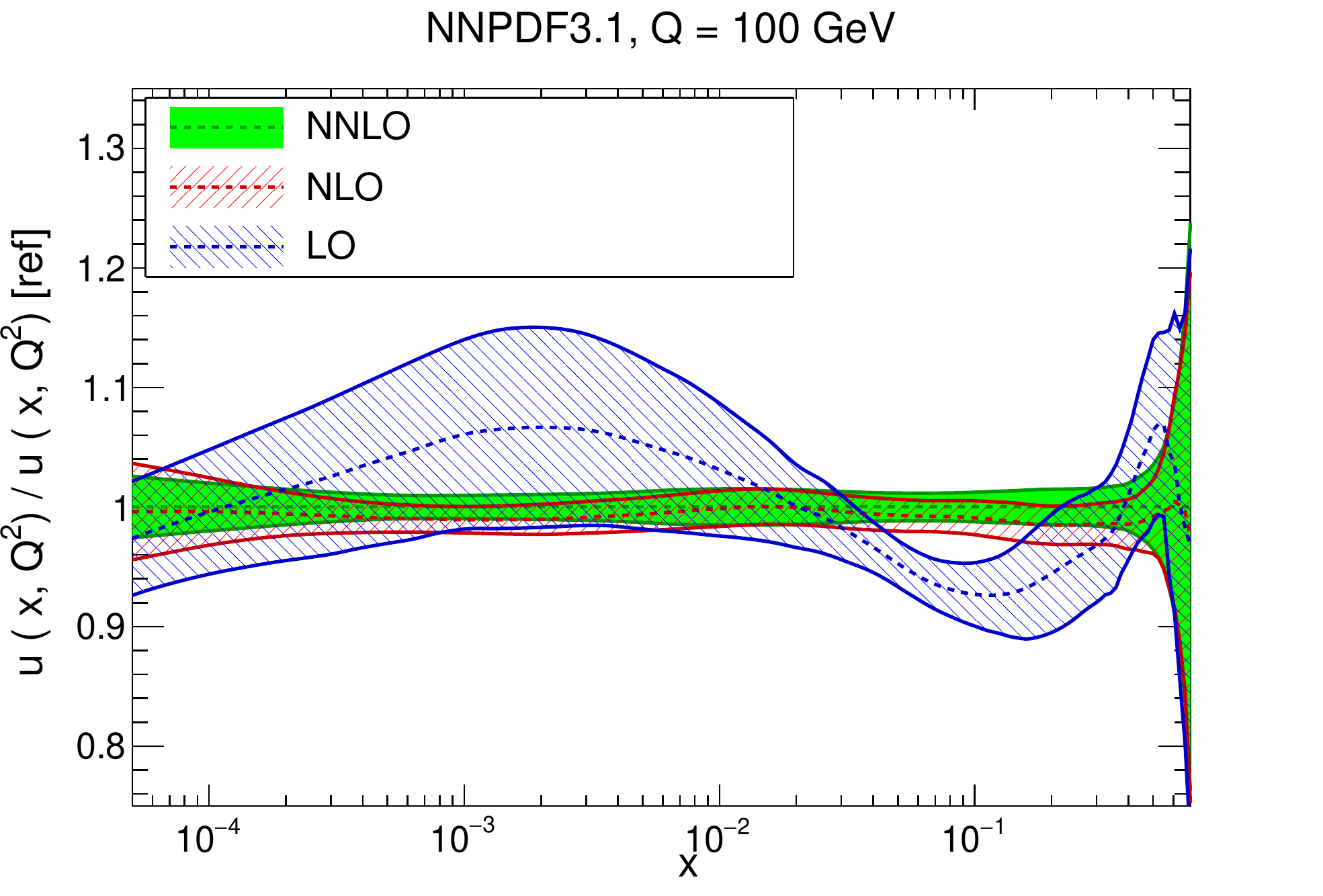}
  \includegraphics[scale=0.33]{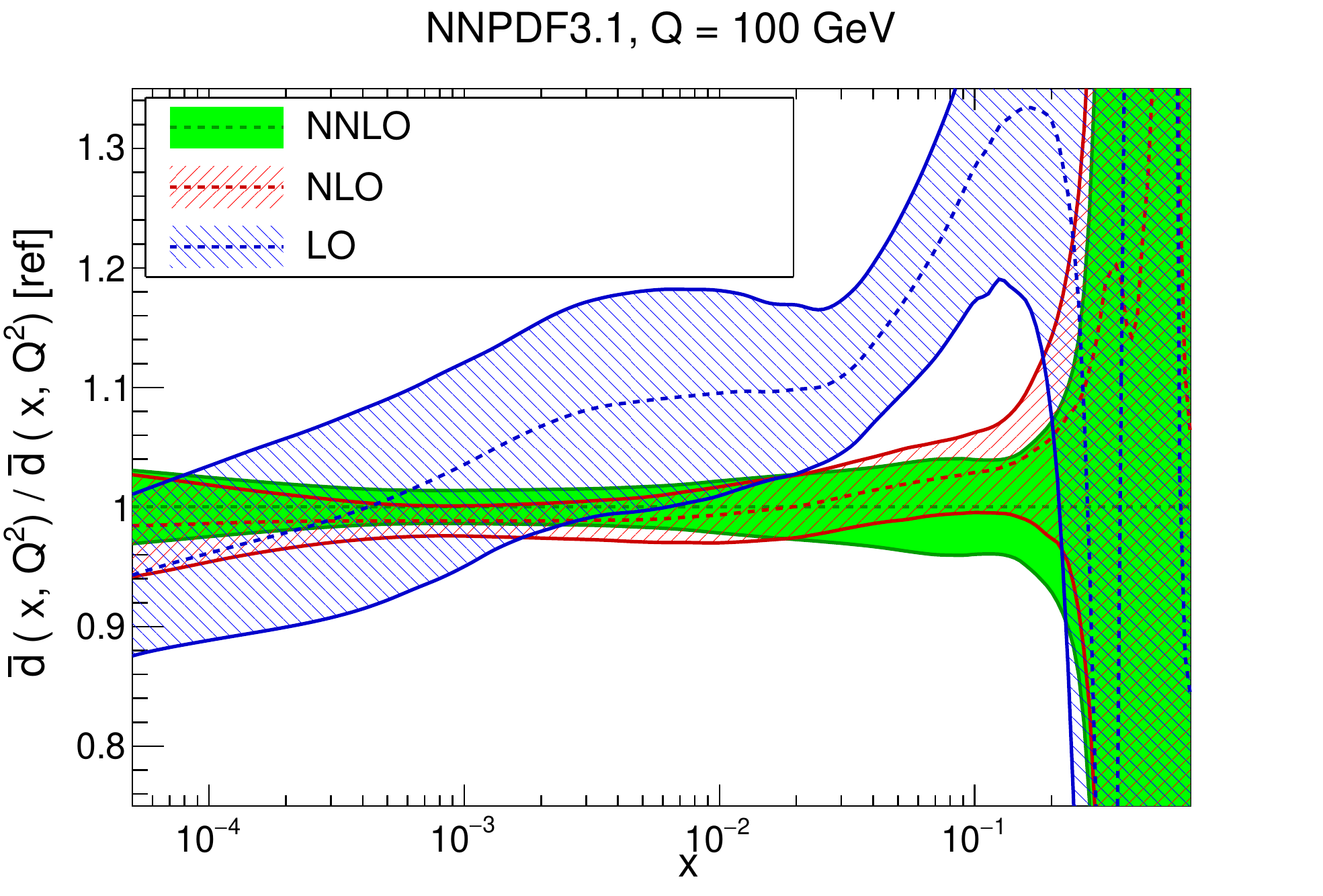}
  \includegraphics[scale=0.33]{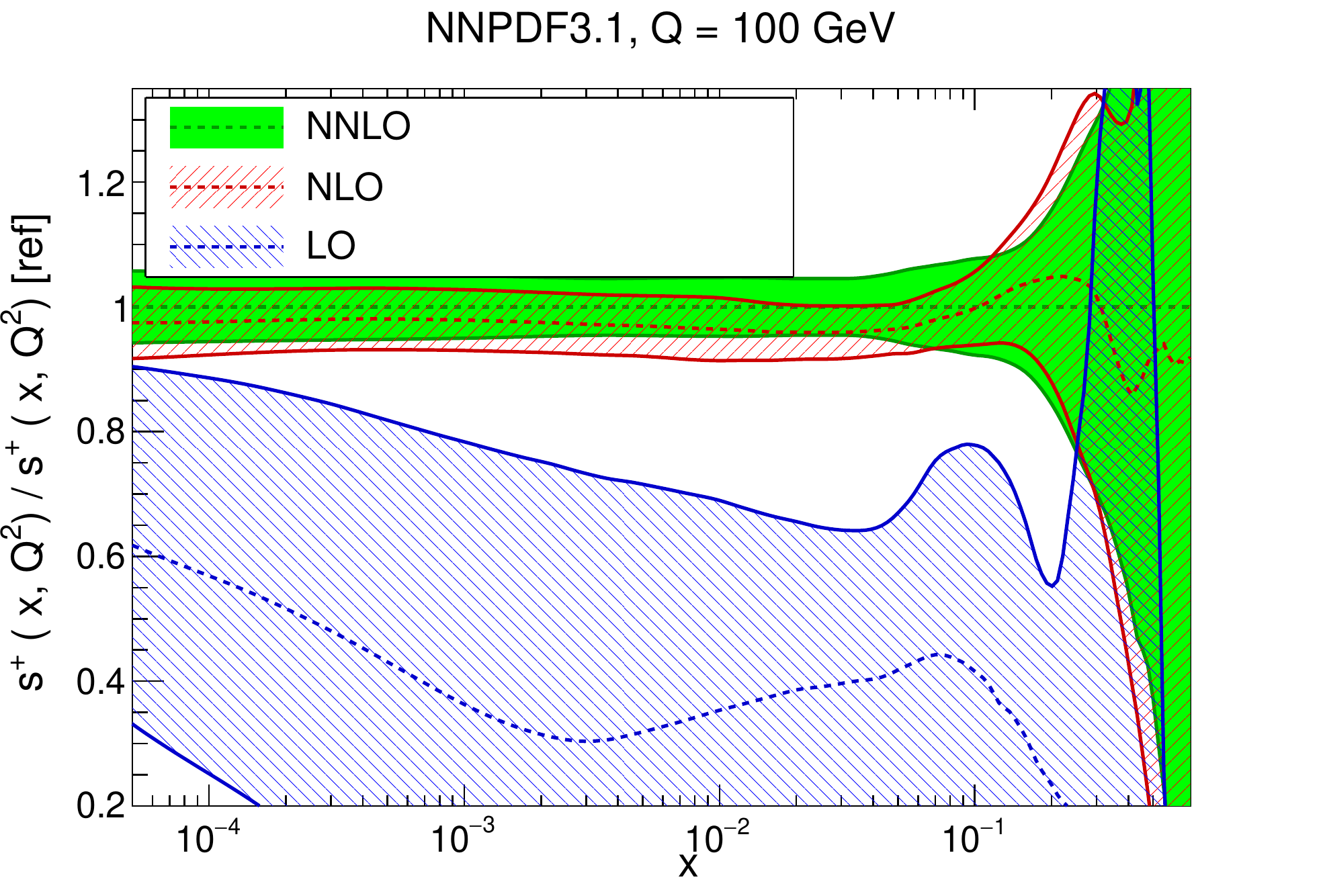}
  \caption{\small Comparison between some 
    of the LO, NLO and NNPDF3.1 NNLO PDFs: gluon and up (top),
    antidown and total strangeness (bottom). All results are shown at
    $Q=100$ GeV, normalized to the NNLO central value.}
    \label{fig:pdfcomp-pertorder}
\end{center}
\end{figure}

A full assessment of MHOU is an open problem, which we leave to future
investigations. For the time being, a first assessment can be obtained by
studying the perturbative stability of our results.
In Fig.~\ref{fig:distances_perturbativeorder} we show the distances at $Q=100$ GeV 
between all the PDFs in the
    LO and NLO sets, and in the NLO and NNLO sets. Some of the
 LO, NLO and NNLO PDFs are then compared directly in
 Fig.~\ref{fig:pdfcomp-pertorder}.  
Differences between the LO and NLO sets are very large, both for
central values and uncertainties, the latter being substantial at LO
due to the poor 
    fit quality. The shift in quark PDFs can be as large as two sigma
    ($d\simeq 20$), while the gluon at small $x$ is completely
    different between LO and NLO due to the fact that the singular small-$x$
    behaviour of the quark to gluon splittings only starts  at NLO, 
and due to the vanishing of gluon initiated  
    DIS and DY processes at LO.
On the other hand, when going from NLO to NNLO,
 PDF uncertainties are essentially unaffected. Central values are also 
 reasonably stable: the largest shifts, in the large-$x$ gluon and down quark and
 small-$x$ gluon, remain at or below the one-sigma level.

\begin{figure}[t]
\begin{center}
  \includegraphics[scale=0.36]{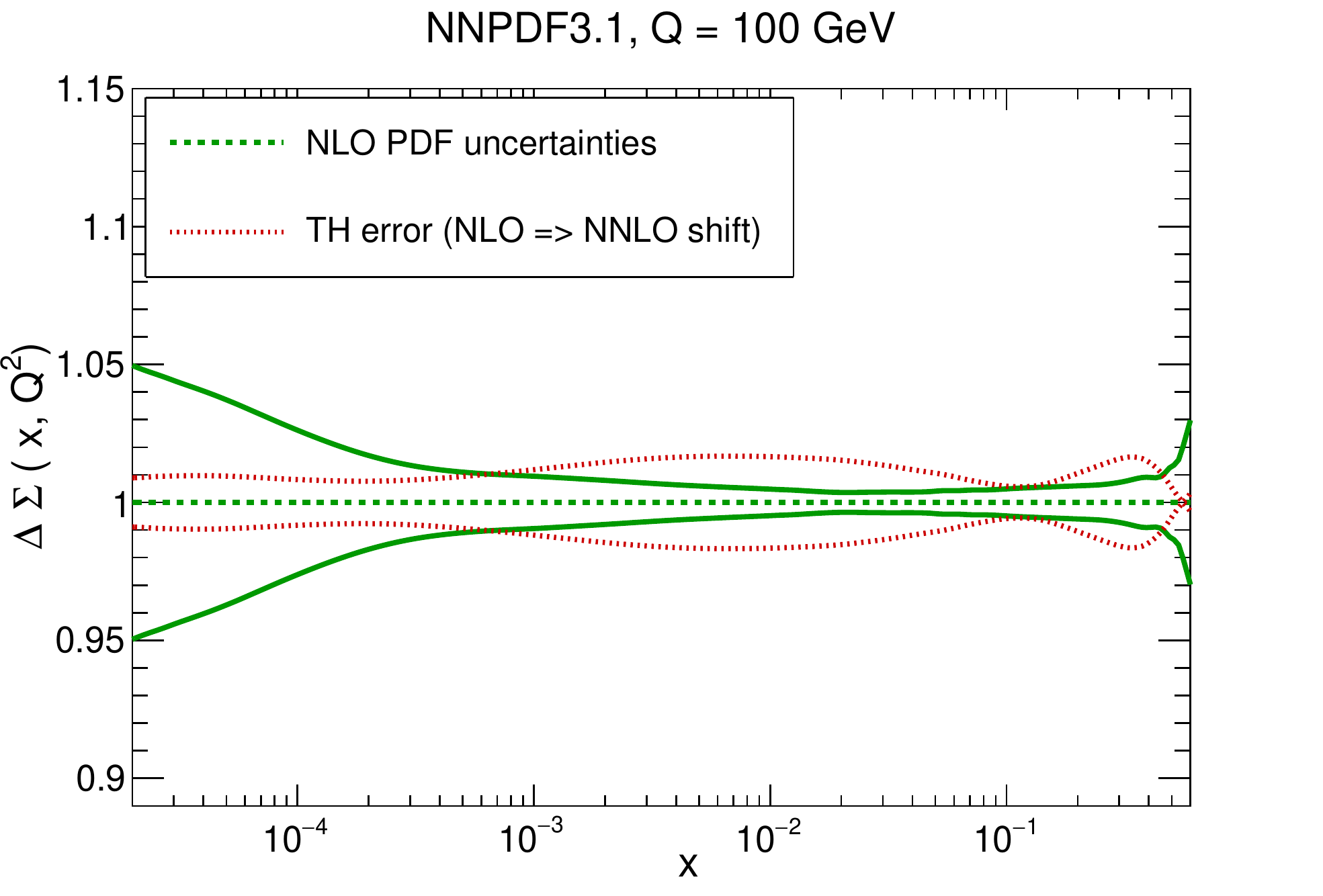}
  \includegraphics[scale=0.36]{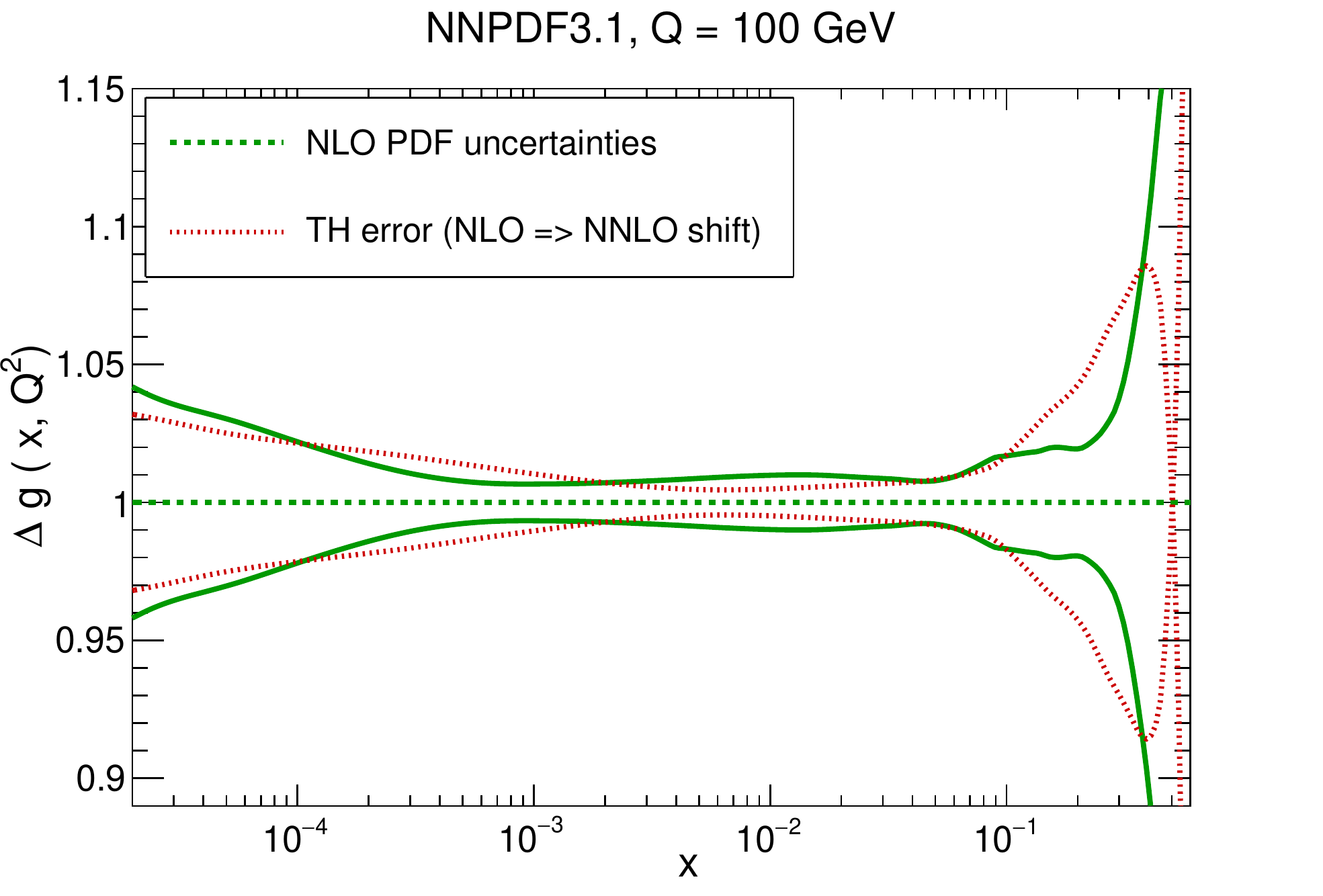}
  \includegraphics[scale=0.36]{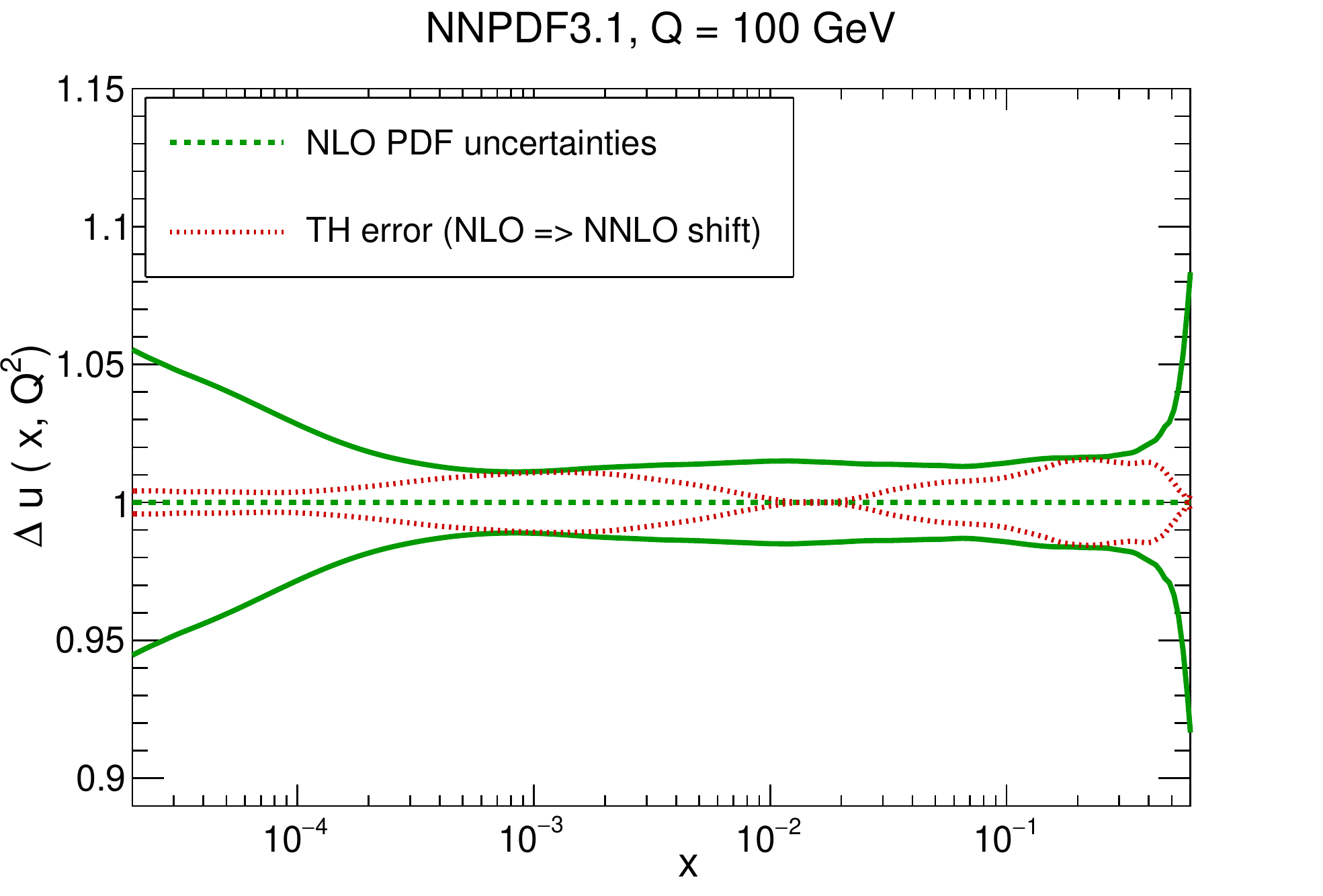}
  \includegraphics[scale=0.36]{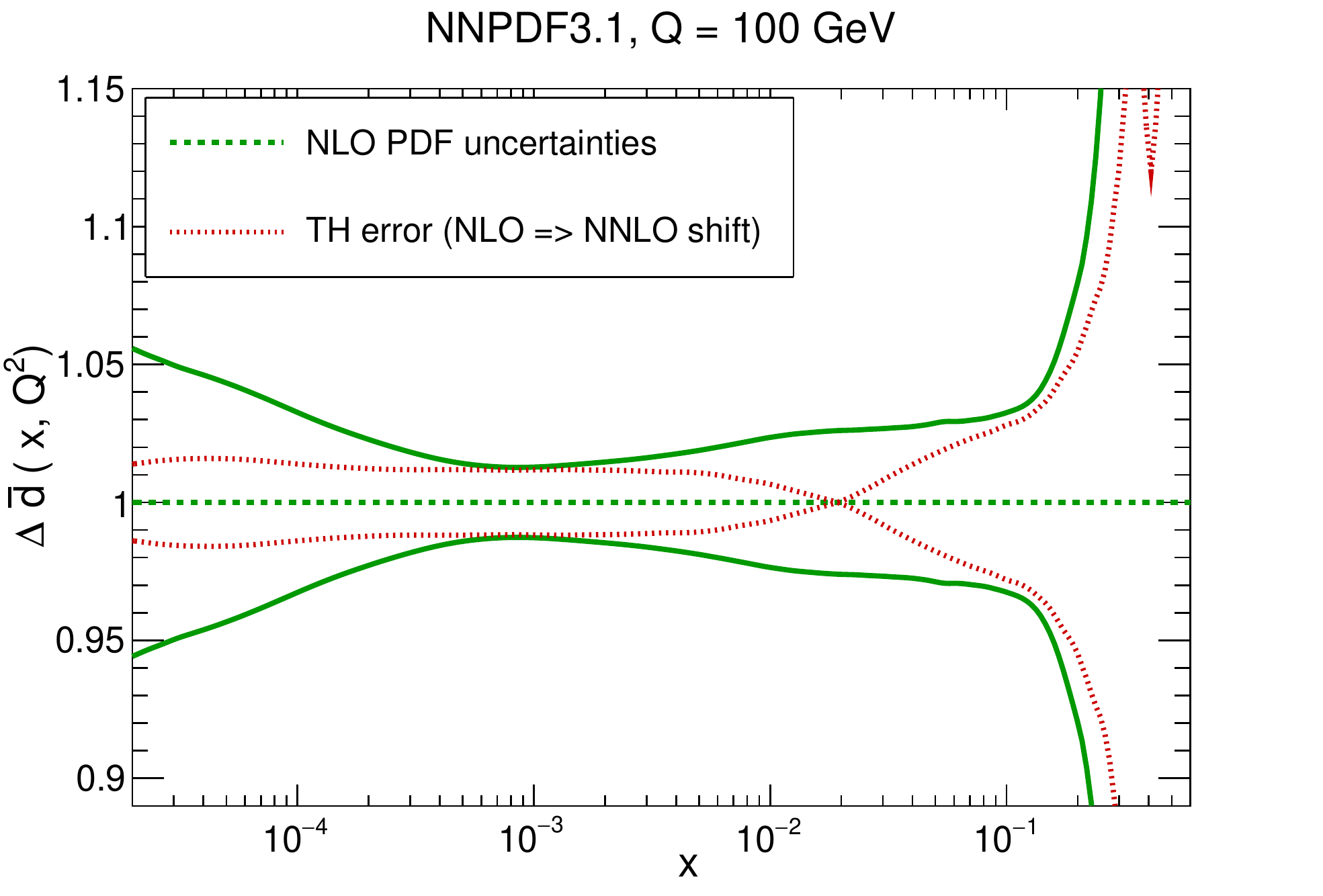}
  \caption{\small Comparison between the NLO PDF uncertainties and the
    shift between the NLO and NNLO PDFs. All results are shown as
    ratios to the NLO PDFs, for $Q=100$~GeV. The shift is
    symmetrized. We show results for the singlet, gluon (top); up and
    antidown (bottom) PDFs.
    \label{fig:THerror-shifts}
  }
\end{center}
\end{figure}

A quantitative estimate of the MHOU can be obtained
by computing the shift between the central values of the NLO and NNLO
NNPDF3.1 PDFs. The result is shown in Fig.~\ref{fig:THerror-shifts}
for some PDF combinations. In the plot, the shift has been
symmetrized, and is compared to the NLO standard PDF uncertainty.
In the quark singlet $\Sigma$ for $x\lesssim 10^{-3}$ the shift is
larger than the PDF uncertainty, while it is smaller for 
individual flavors (as illustrated by the two quark
distributions shown).  
This suggests that for individual quark flavors and the gluon at
NNLO, MHOU can be reasonably neglected at the current level of precision. 
However, for particular combinations (such as the singlet at small $x$) it is
unclear whether MHOU can be neglected even at NNLO, given that at NLO
they are larger than the PDF uncertainty.

%

We finally turn to parametric uncertainties. As we have discussed in
Sect.~\ref{sec:results-mc}, the dependence of PDFs upon the charm mass 
is almost entirely removed by parametrizing charm. The dependence on the $b$-quark 
mass is minor, except for the bottom PDFs
themselves~\cite{Ball:2011mu,Harland-Lang:2015qea}. Therefore, the only
significant residual parametric uncertainty is on the value of the strong
coupling. This uncertainty is routinely included along with the PDF
uncertainty; in order to do this consistently, one needs PDF sets
produced with different central values of $\alpha_s$ (see
e.g. Ref.~\cite{Butterworth:2015oua}).  We have 
determined  NNPDF3.1 NLO and NNLO  PDFs with $\alpha_s(m_Z)$
varied in the range
$0.108\le\alpha_s(m_Z)\le0.124$ (see Sect.~\ref{sec:delivery}).

\begin{figure}[t]
  \begin{center}
  \includegraphics[scale=0.36]{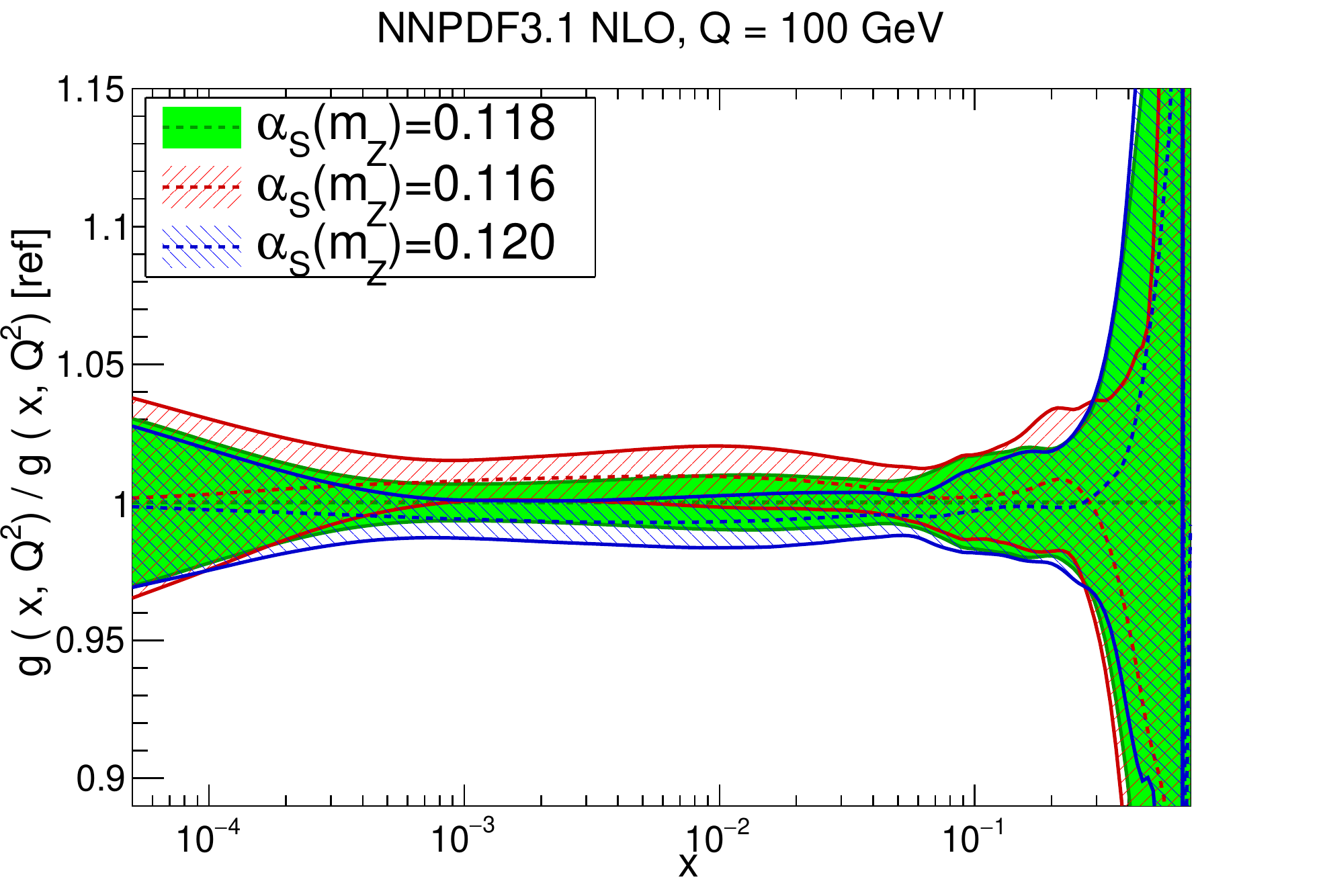}
  \includegraphics[scale=0.36]{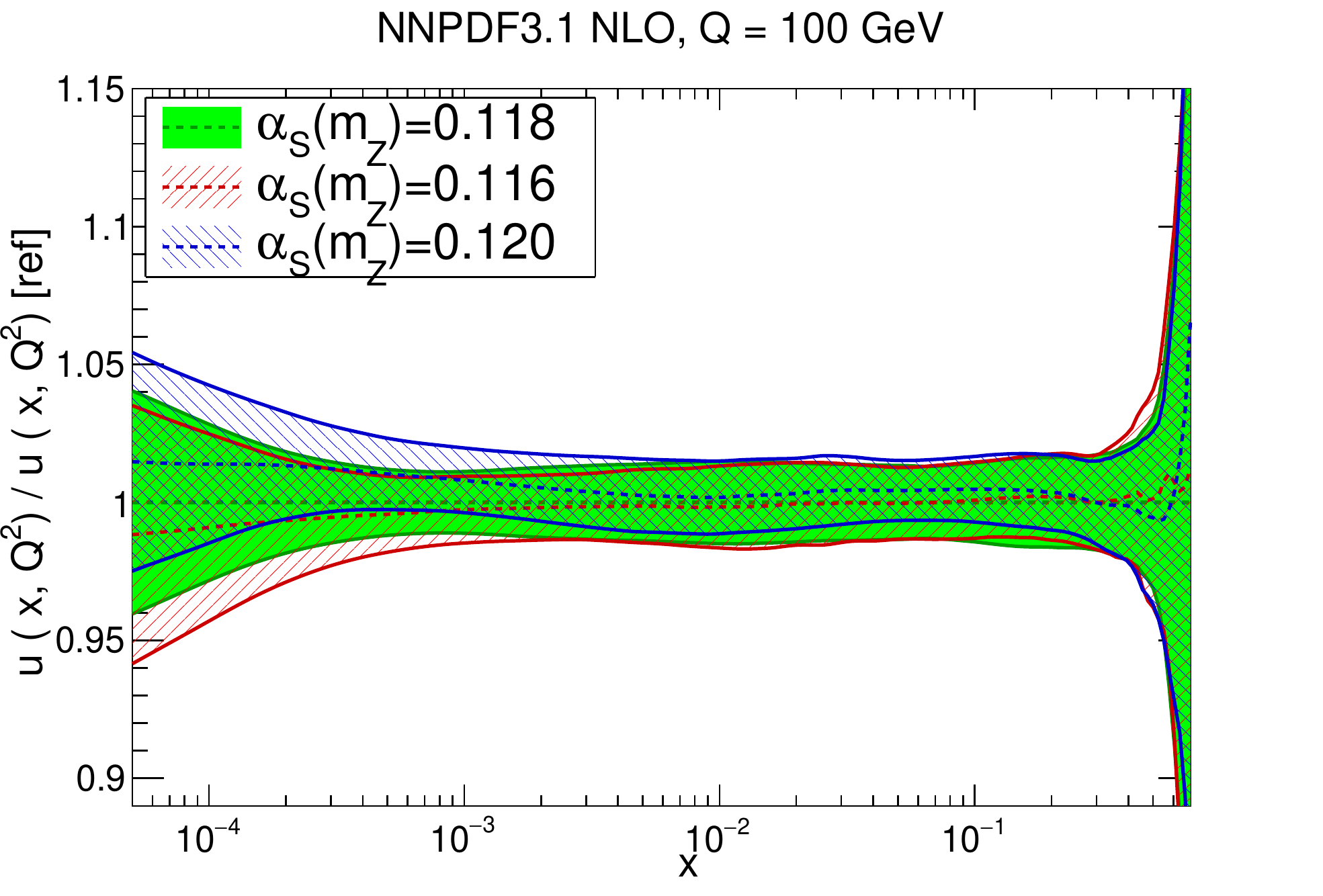}
  \includegraphics[scale=0.36]{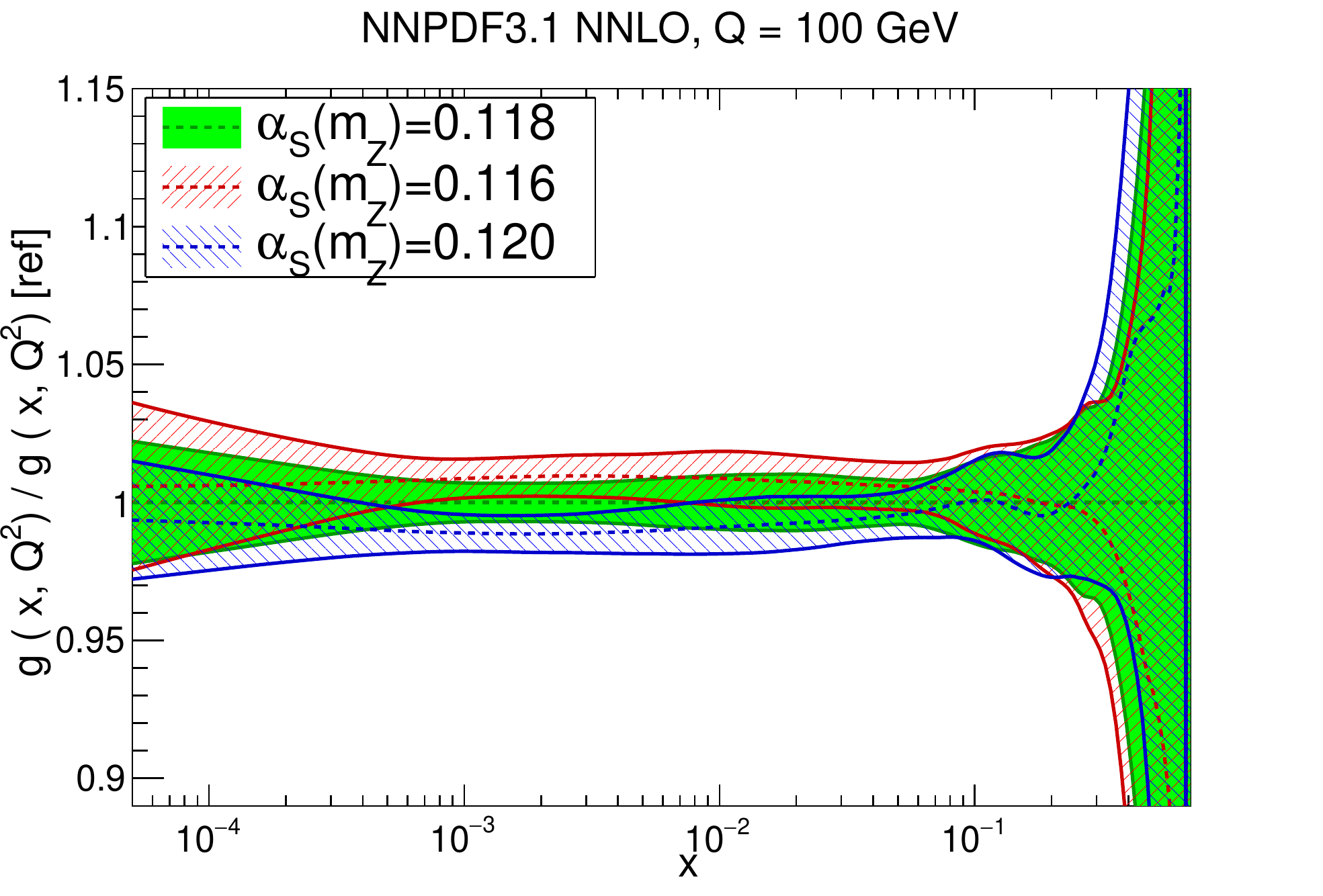}
  \includegraphics[scale=0.36]{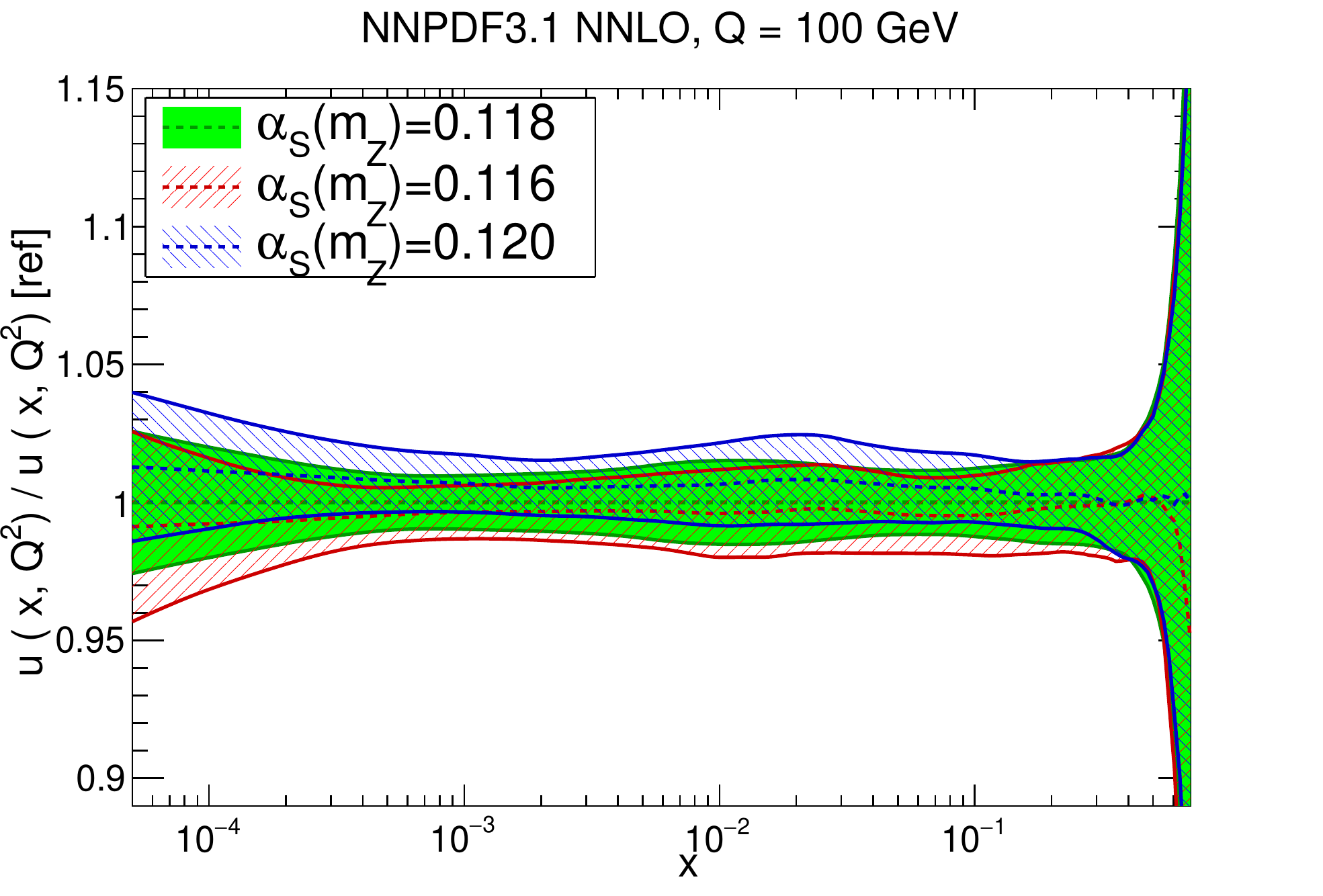}
  \caption{\small 
    \label{fig:PDFasvariations} Dependence of NNPDF3.1 NLO (top) and
    NNLO (bottom) PDFs on
    the value of $\alpha_s$. The gluon (left) and up quark (right) are
    shown at $Q=100$~GeV, normalized to the central value.
  }
\end{center}
\end{figure}

In Fig.~\ref{fig:PDFasvariations} we compare the up and gluon PDFs as
$\alpha_s(m_Z)$ is varied by $\Delta\alpha_s=\pm0.002$ about its
central value. As is well known,
the gluon is anti-correlated to $\alpha_s(m_Z)$ at small and medium $x$, but 
positively correlated to it at large $x$. The dependence on $\alpha_s$ is 
rather milder for quark PDFs, with positive correlation at small $x$, and very little
dependence altogether at large $x$.

\clearpage

\clearpage
\section{The impact of the new collider data}

\label{sec:impactnewdata}

We now study the dependence of the NNPDF3.1 PDF set upon 
the experimental information on which it is based. Firstly we disentangle
the effects of new data from the effects of methodological changes. Then 
we systematically quantify the impact on PDFs of each
new piece of experimental information added in
NNPDF3.1. Finally we discuss PDF determinations based on particular data subsets; 
PDFs determined only from collider data (i.e. excluding all fixed target data), 
 only from proton data (i.e. excluding all nuclear data), or excluding all LHC data.
As these PDF sets based on reduced dataset can also be useful for
specific phenomenological applications,  they are also made available
(see Section~\ref{sec:delivery} below). As in the previous Section, here
we will only present a selection of representative plots, the
interested reader is referred to a much larger set of plots available online 
as discussed in Section~\ref{sec:delivery}.

\subsection{Disentangling the effect of new data and methodology}
\label{sec:disentangling}

In Section~\ref{sec:results-mc} we have studied the impact of the main
methodological improvement introduced in NNPDF3.1, namely,
independently parametrizing the
charm PDF and determining it from the data. 
In order to completely disentangle the effect of data and methodology
we have performed a PDF determination using NNPDF3.1 methodology,
but the NNPDF3.0 dataset: specifically, we have removed from the
NNPDF3.1 dataset all the new data. There remain some small residual differences
between this restricted dataset and that of NNPDF3.0, specifically in 
some small differences in cuts and in the use of the combined HERA data
instead of the separate HERA-I and HERA-II sets. However, these differences
are expected to be minor~\cite{Rojo:2015nxa}.

\begin{figure}[t]
\begin{center}
  \includegraphics[scale=1]{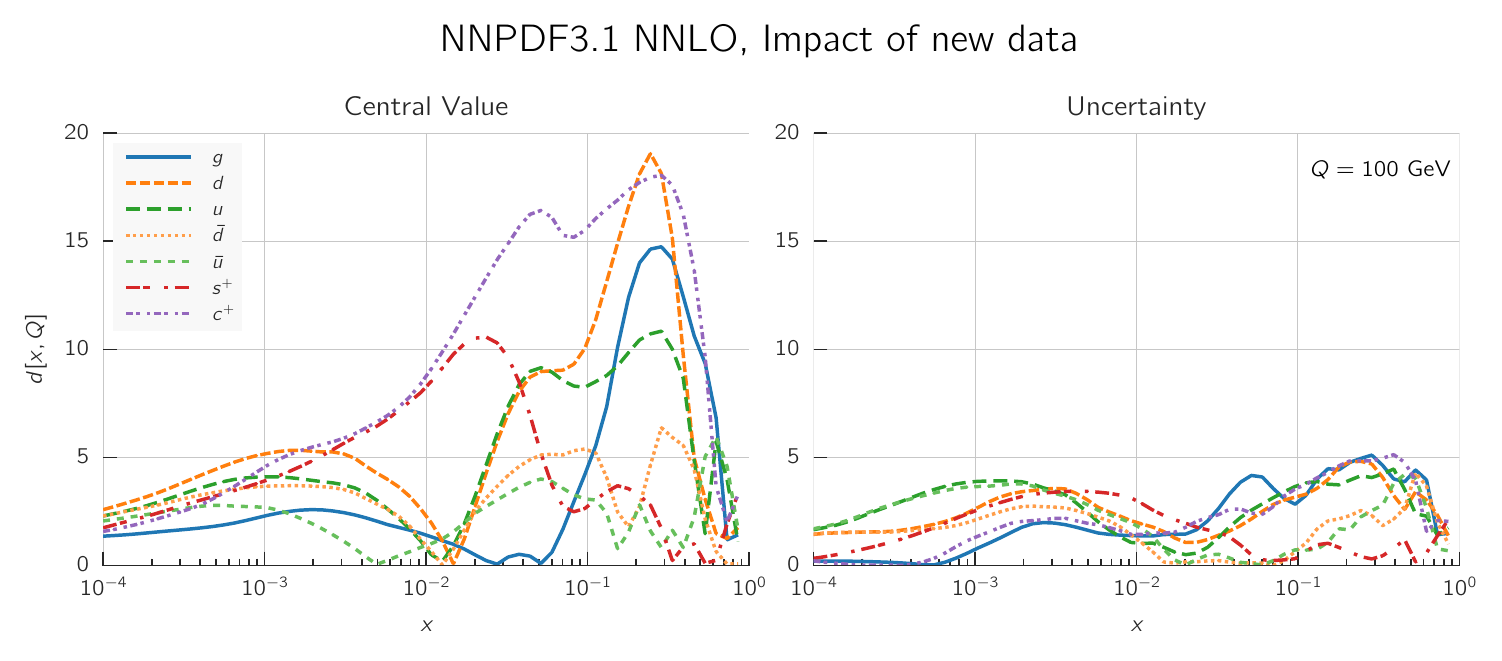}
  \caption{\small Same as Fig.~\ref{fig:distances_31_vs_30}, but now
    comparing the NNPDF3.1 NNLO
    global PDFs to PDFs determined using exactly the same methodology
    but with the NNPDF3.0 dataset.
    \label{fig:distances_nnpdf31_newdata}
  }
\end{center}
\end{figure}

\begin{figure}[t]
\begin{center}
  \includegraphics[scale=0.38]{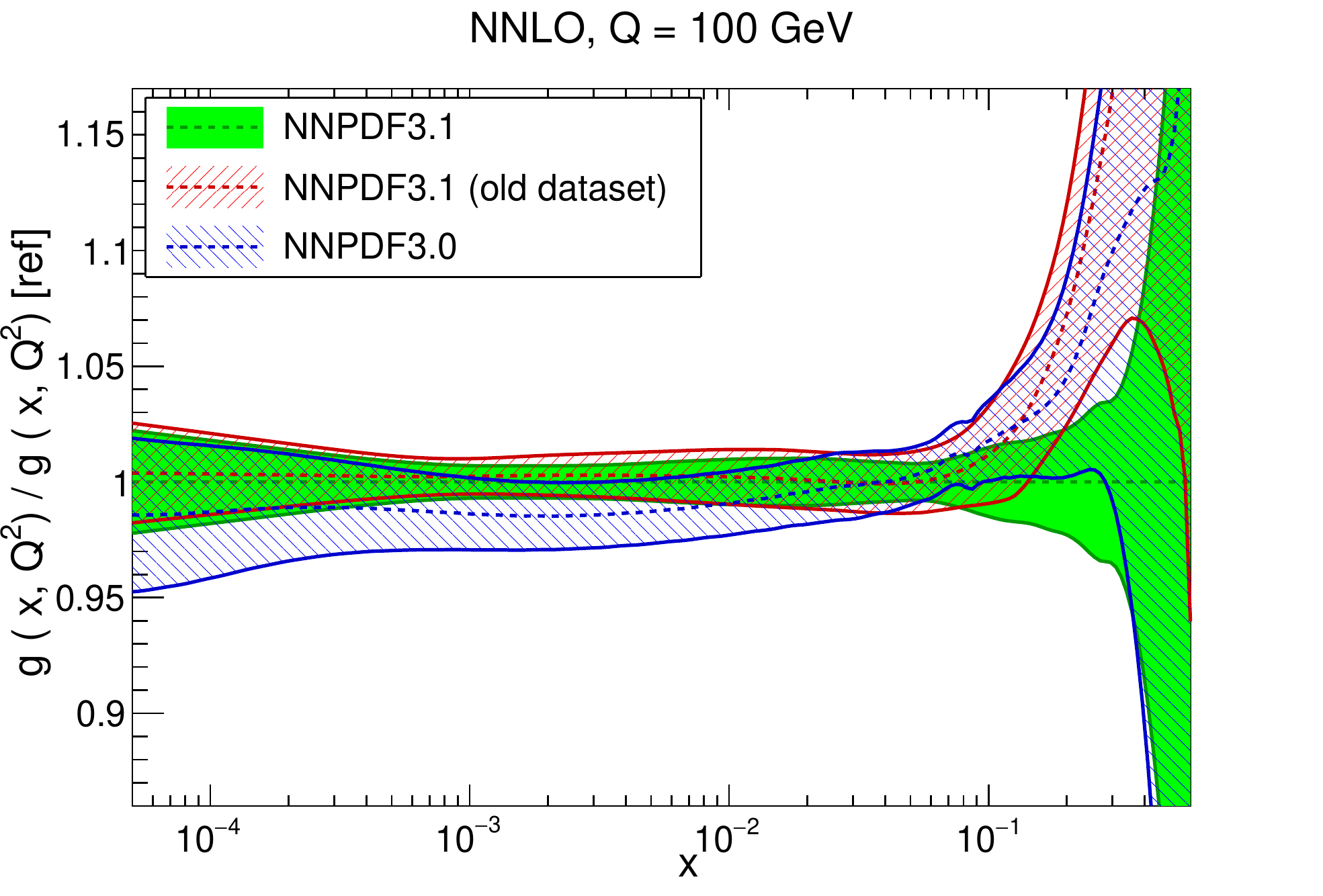}
  \includegraphics[scale=0.38]{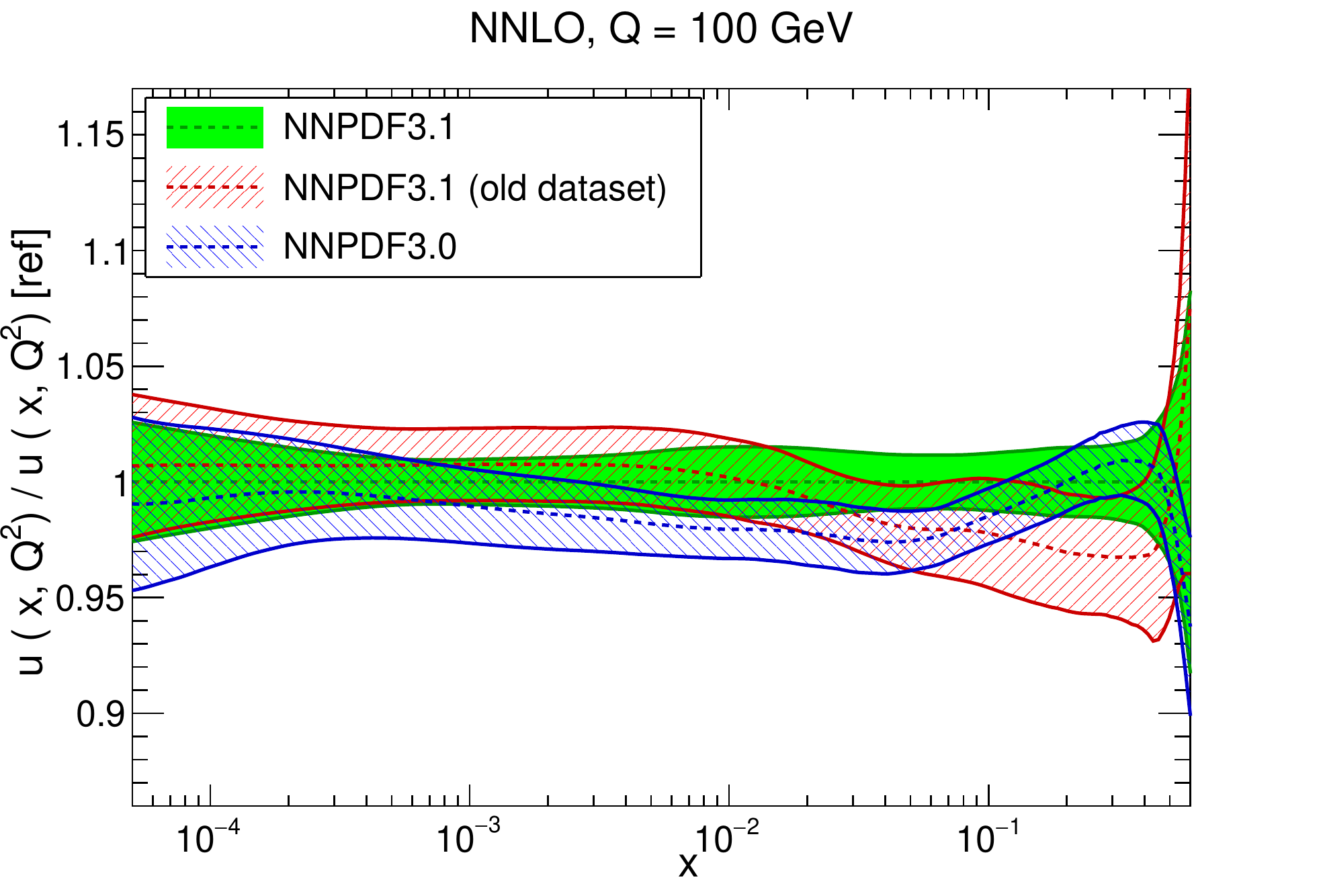}
  \includegraphics[scale=0.38]{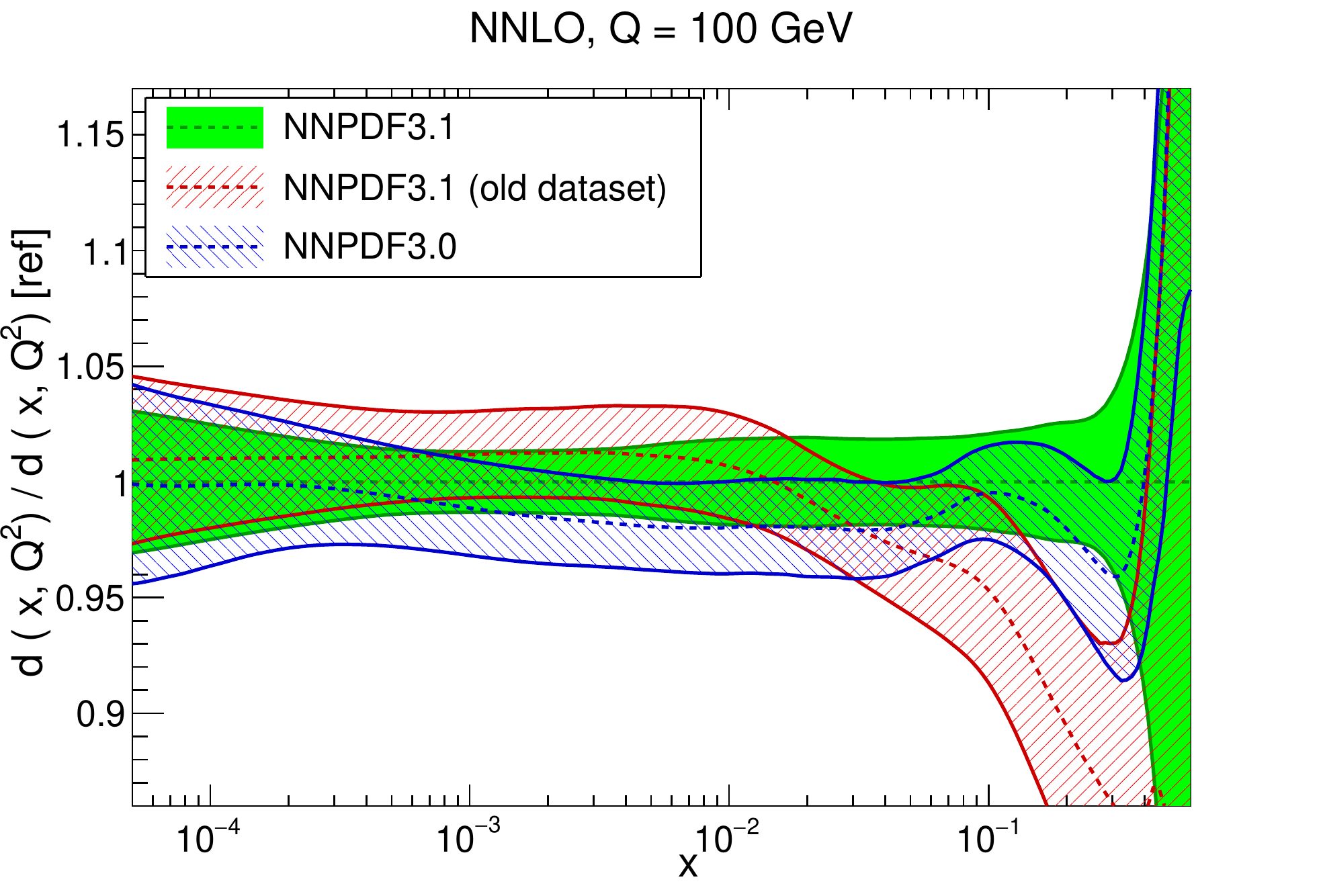}
  \includegraphics[scale=0.38]{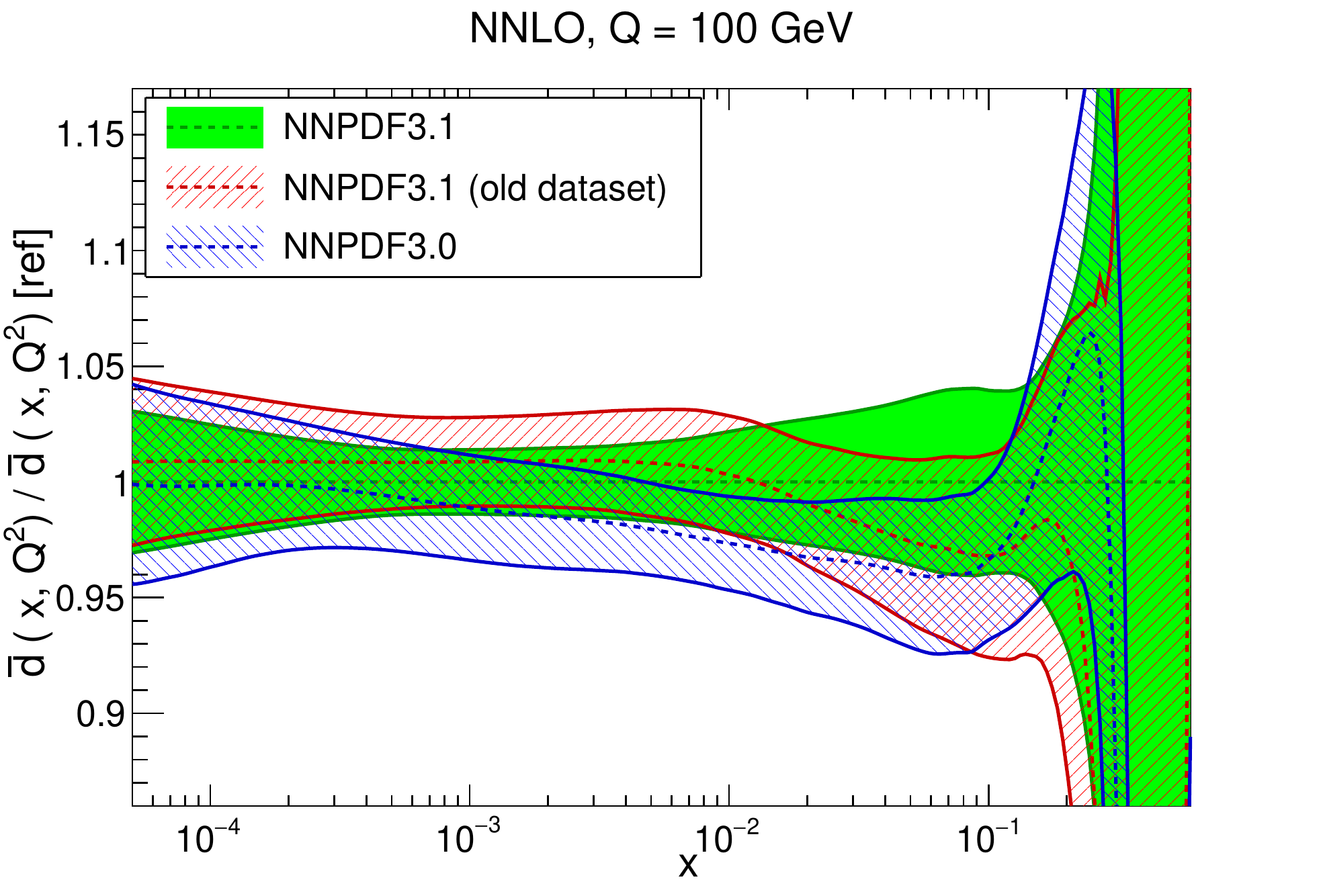}
   \includegraphics[scale=0.38]{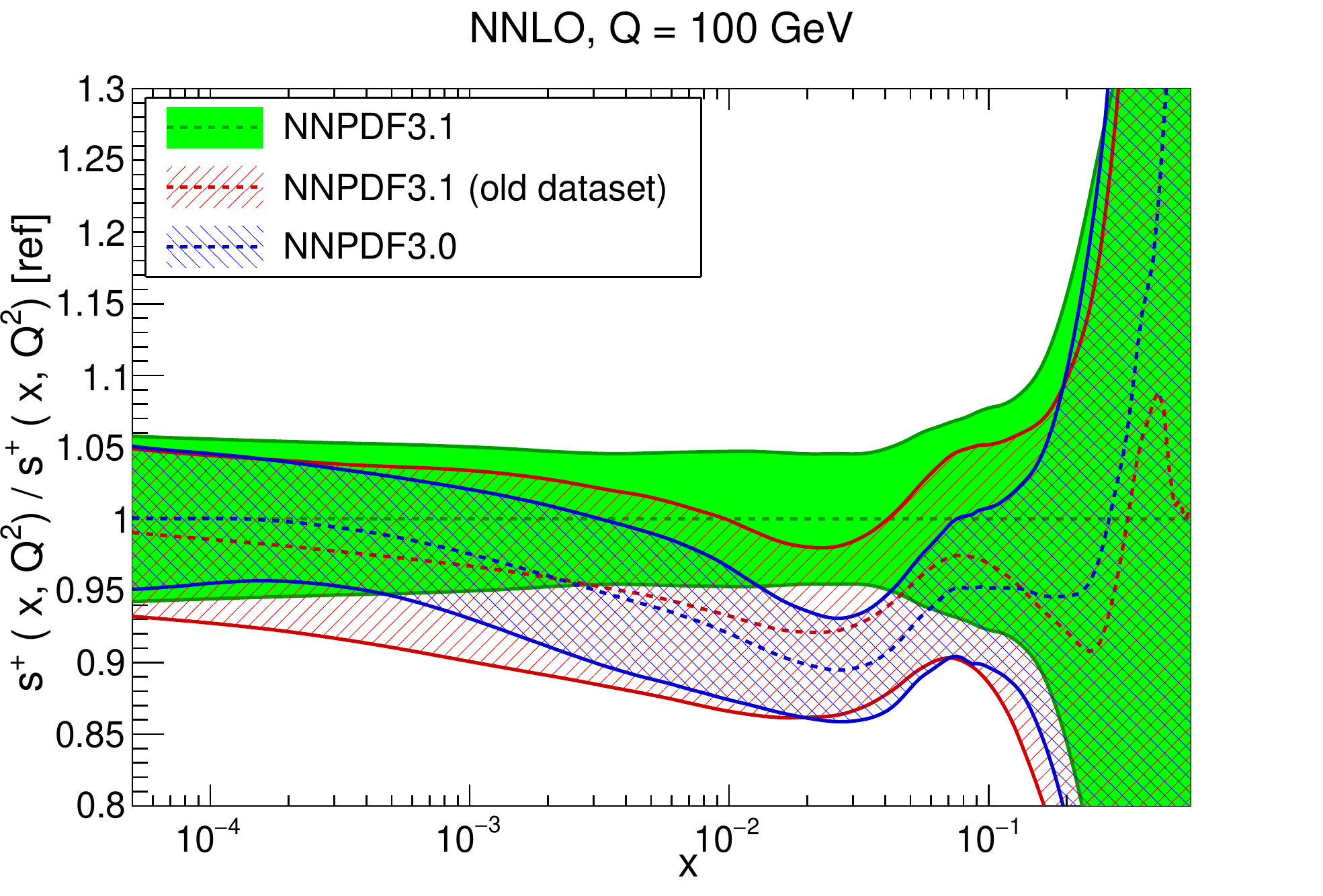}
   \includegraphics[scale=0.38]{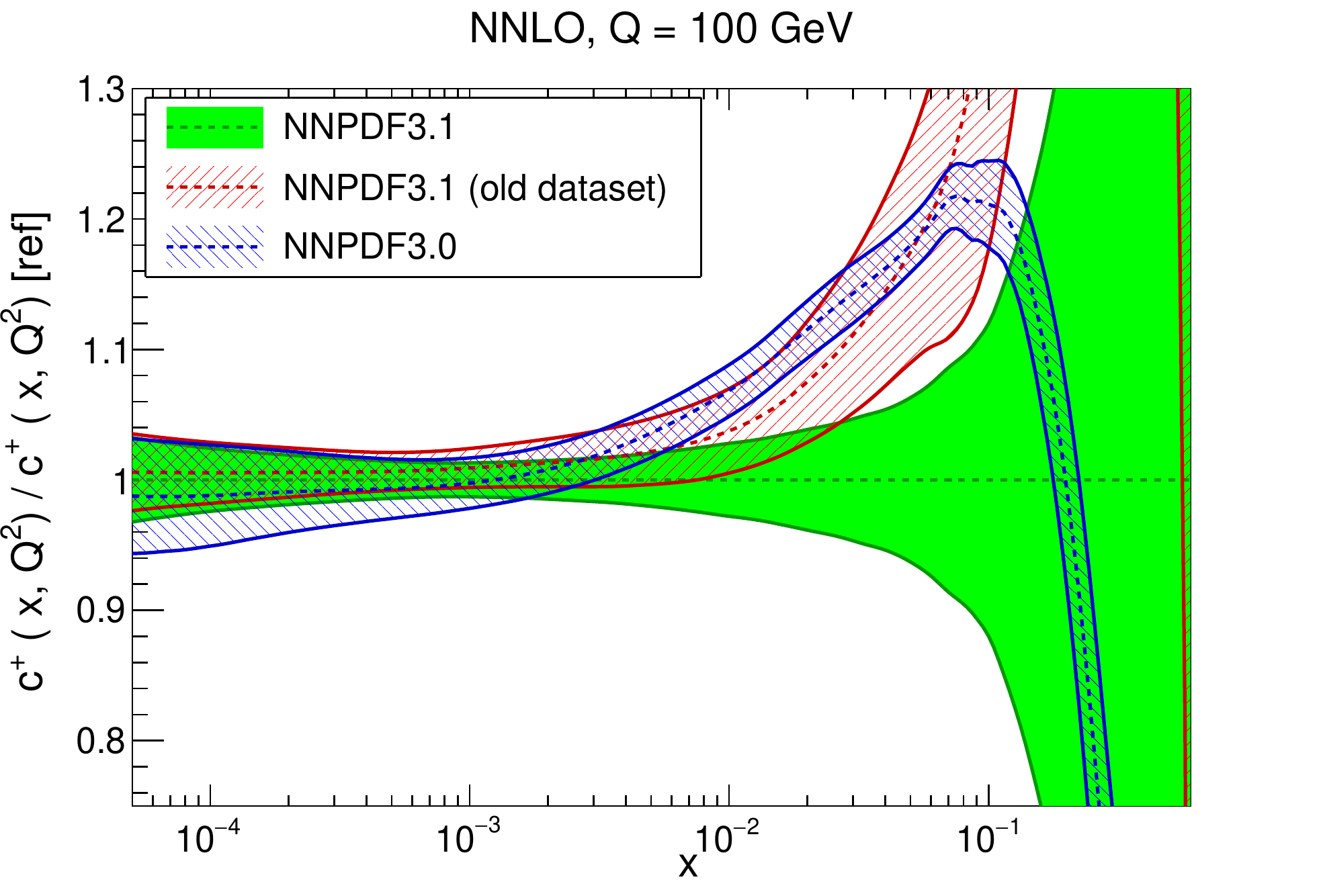}
   \caption{\small Same as Fig.~\ref{fig:31-nnlo-vs30}, but now also
     including PDFs determined using NNPDF3.1 methodology with the NNPDF3.0
     dataset. From left to right and from top to bottom the gluon, up,
     down, antidown,
     total strangeness and charm are shown.
    \label{fig:31-nnlo-old-vs-new}
  }
\end{center}
\end{figure}

In Fig.~\ref{fig:distances_nnpdf31_newdata} we show the
distances between the NNPDF3.1 NNLO PDF set,  and that 
based on the NNPDF3.0 dataset using the same methodology.
We see that the impact of the new data is mostly localized at
large $x$, for the up, down and charm quarks and the gluon,
and at medium $x$ for strangeness.
As far as uncertainties are concerned, we observe improvements of
up to half a sigma across a wide range in $x$ and for all PDF flavors.
In Fig.~\ref{fig:31-nnlo-old-vs-new} we compare some representative PDFs for 
NNPDF3.1, the set based on NNPDF3.0 data with NNPDF3.1 methodology, and the 
original NNPDF3.0. We
see that the overall effect of the new data and the new methodology are 
comparable, but that they act in different regions and for different PDFs.
     For instance, for the light quarks and the gluon the impact of the
     new methodology dominates for all $x\lsim 10^{-2}$, where it
     produces an enhancement, and specifically the enhancement of the
     gluon for $x\lsim 0.03$ which was discussed in Sect.~\ref{sec:PDFcomparisons}.
      At large $x$ instead the dominant effect is
     from the new data, which lead to a reduction of the gluon
     and an enhancement of the quarks. Whereas of course charm is very
     significantly affected by the change in methodology --- it was
     not independently parametrized
     in NNPDF3.0 --- for $x\gsim0.1$, the new data also have a big
     impact. In fact, while strangeness is mostly affected
     by the new data in the medium and small $x$ regions, charm and gluon are  
      most affected by them at large $x$.

\subsection{The transverse momentum of the $Z$ boson}
\label{sec:impactzpt}
The use of transverse momentum distributions has been advocated for a
long time (see e.g. Ref.~\cite{Forte:2013wc}) as a clean and powerful
constraint on PDFs, particularly the gluon.
As discussed in Section~\ref{sec:expdata}, it is now 
possible to include such data at NNLO thanks to the availability of the 
computation of this process up to NNLO QCD, along with precise data on $Z$ $p_T$ 
from ATLAS and CMS at 8 TeV. The impact of this dataset on PDFS  
has recently been studied in detail in Ref.~\cite{Boughezal:2017nla}.

\begin{figure}[t]
\begin{center}
  \includegraphics[scale=1]{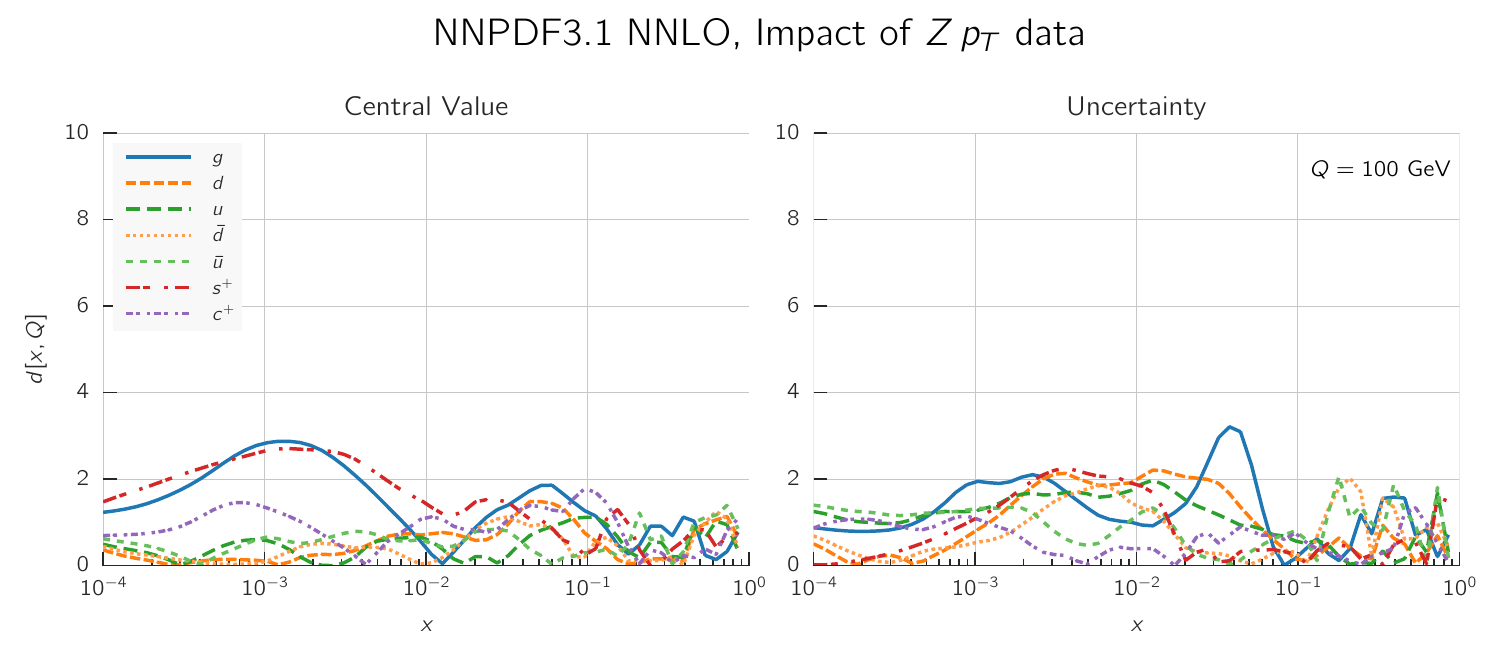}
  \caption{\small Same as Fig.~\ref{fig:distances_31_vs_30},  but now
    comparing the default NNPDF3.1 to a version of it with the  
8 TeV $Z$ $p_T$ data from ATLAS and CMS not included.
\label{fig:distances_noZpT}}
\end{center}
\end{figure}

\begin{figure}[t]
\begin{center}
  \includegraphics[scale=0.38]{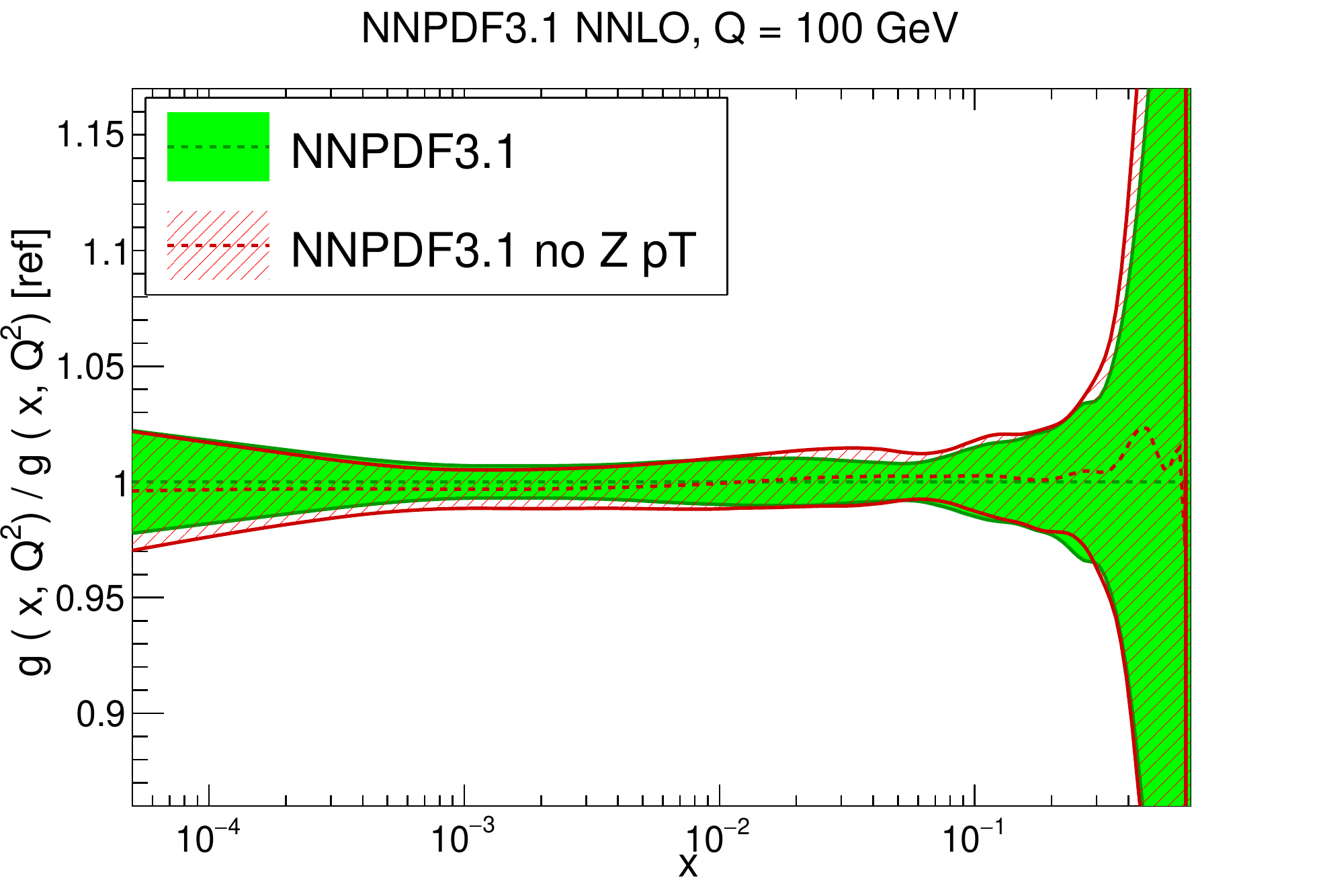}
  \includegraphics[scale=0.38]{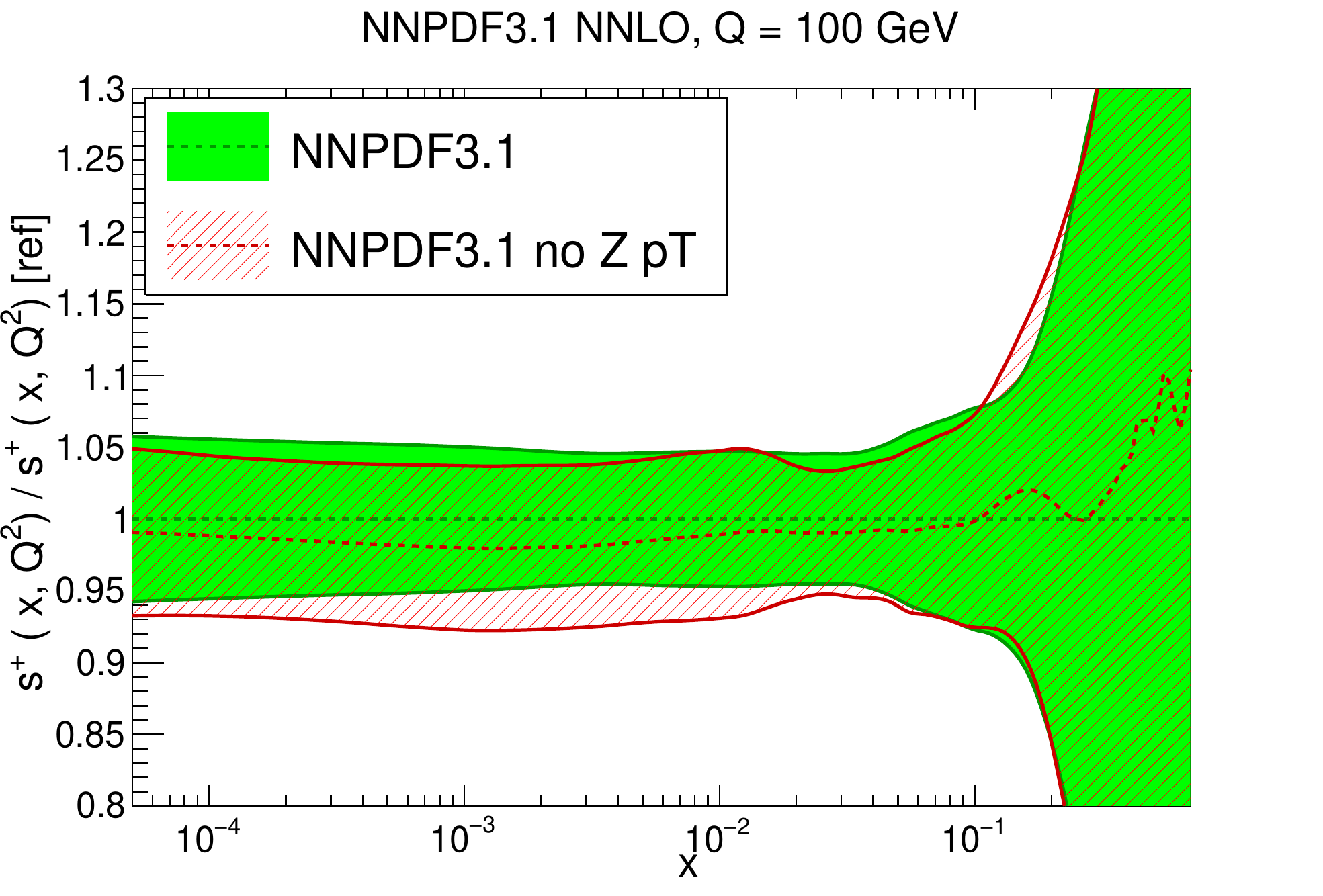}
  \includegraphics[scale=0.38]{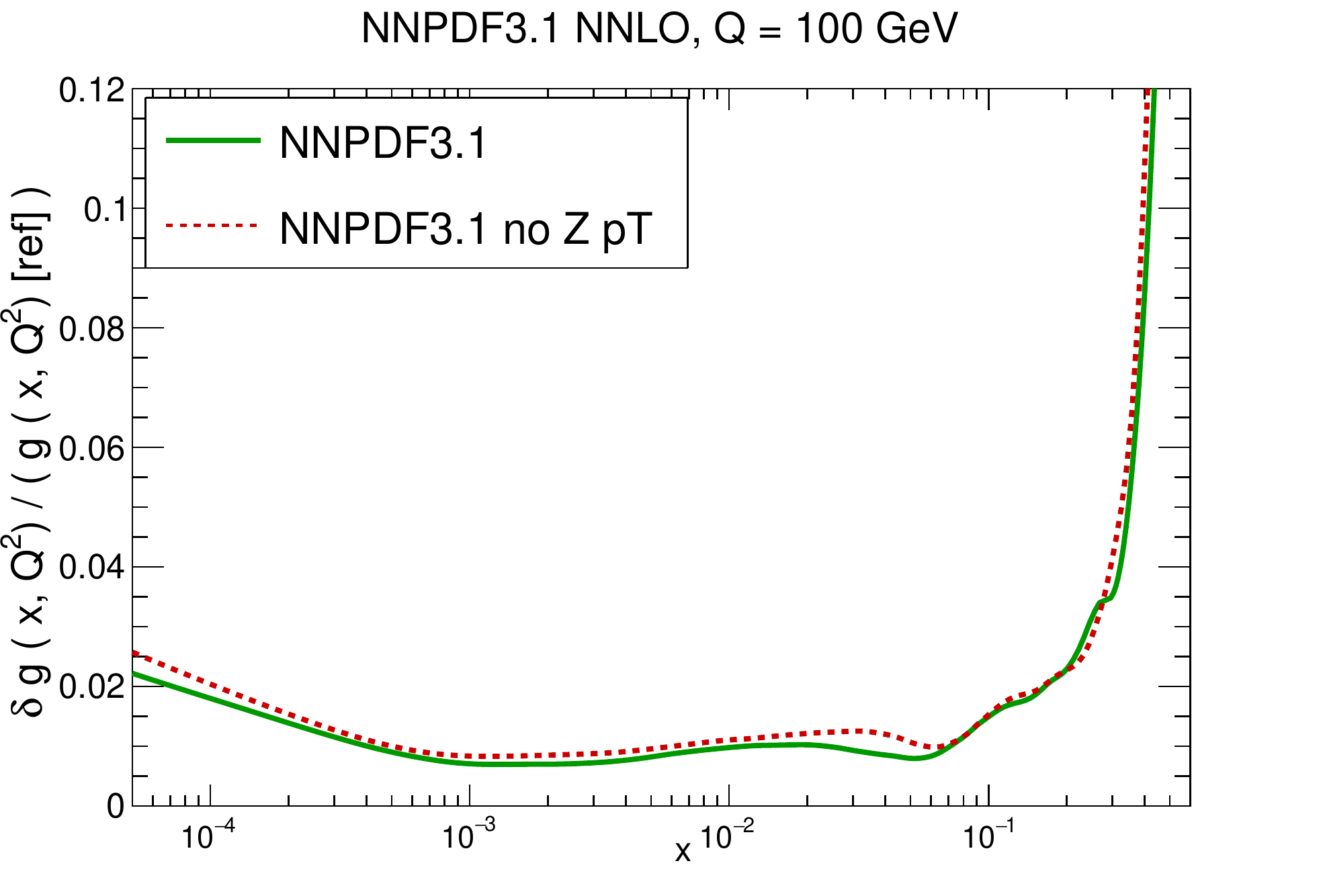}
  \includegraphics[scale=0.38]{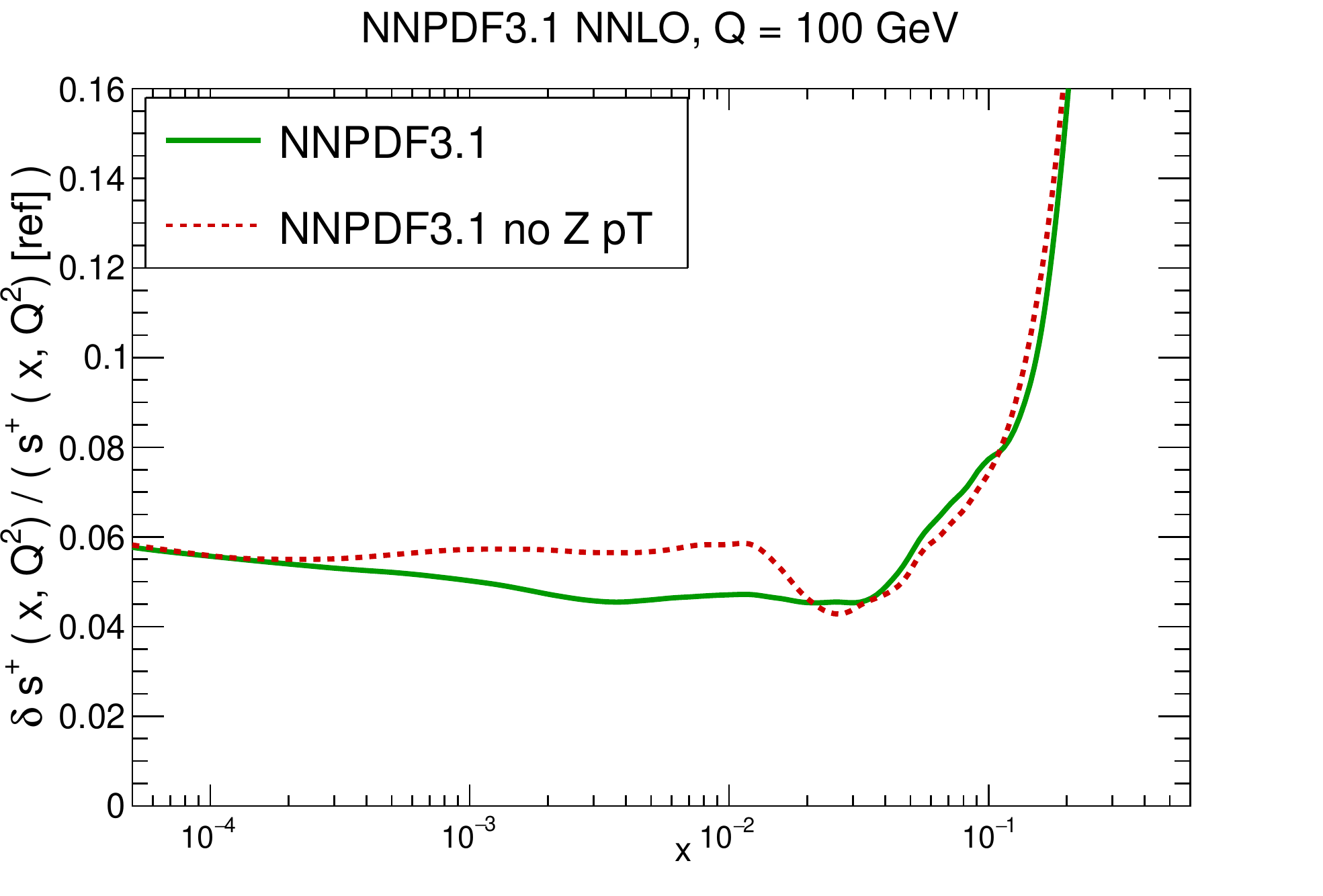}
  \caption{\small Same as Fig.~\ref{fig:31-nnlo-vs30} (top) and as
    Fig.~\ref{fig:ERR-31-nnlo-vs30} (bottom), but now
    comparing the default NNPDF3.1 to a version of it with the  
8 TeV $Z$ $p_T$ data from ATLAS and CMS not included. Results are
shown for the gluon (left) and total strangeness (right).
\label{fig:31-nnlo-ZpT}}
\end{center}
\vskip-0.5cm
\end{figure}

 NNPDF3.1 is the first global PDF determination to include this data. 
In order to assess the impact of this dataset, we have repeated the NNLO
determination, excluding all $Z$ $p_T$ data. 
In Fig.~\ref{fig:distances_noZpT} we show the
distances between this PDF set and the default: it is clear
that the effect on all PDFs is moderate, with changes below one third of a sigma. 
The largest differences are seen in the gluon, as
expected, and the strange distributions. The reason for this state of affairs can
be best understood by directly comparing PDFs and their uncertainties,
see Fig.~\ref{fig:31-nnlo-ZpT}. It is clear that central values move
very little while uncertainties are slightly reduced, therefore
demonstrating the excellent consistency of the constraint from these measurements
with the existing dataset. The $Z$ $p_T$  dataset therefore reinforces the reliability of our gluon
determination. It also reduces somewhat the uncertainty on the total strangeness.
While in Ref.~\cite{Boughezal:2017nla} this dataset was found to have a rather
stronger impact than shown here, it should be noted that this was the
case when
determining PDFs from  the NNPDF3.0 dataset (less
 the jet data). In NNPDF3.1 more data are added, specifically top pair
 differential distributions:  a smaller impact
 of the $Z$ $p_T$ dataset when added to a wider prior  is not unexpected.

In addition to the 8 TeV measurements from ATLAS and CMS there also exists 
a measurement of the normalized distribution at 7 TeV from ATLAS. The inclusion
of this dataset is problematic because the covariance matrix for a
normalized distribution depends on the cuts imposed on the dataset,
and only the covariance matrix for the full dataset is available. This
issue was studied in detail in
Ref.~\cite{Boughezal:2017nla}. Furthermore, this dataset is
superseded by the more precise 8~TeV measurement. Therefore, it
has not been included in the NNPDF3.1 dataset. However, we have
studied its potential impact by including it in a dedicated PDF determination, 
with its nominal published covariance matrix unmodified
despite the cuts.

\begin{figure}[t]
\begin{center}
  \includegraphics[scale=1]{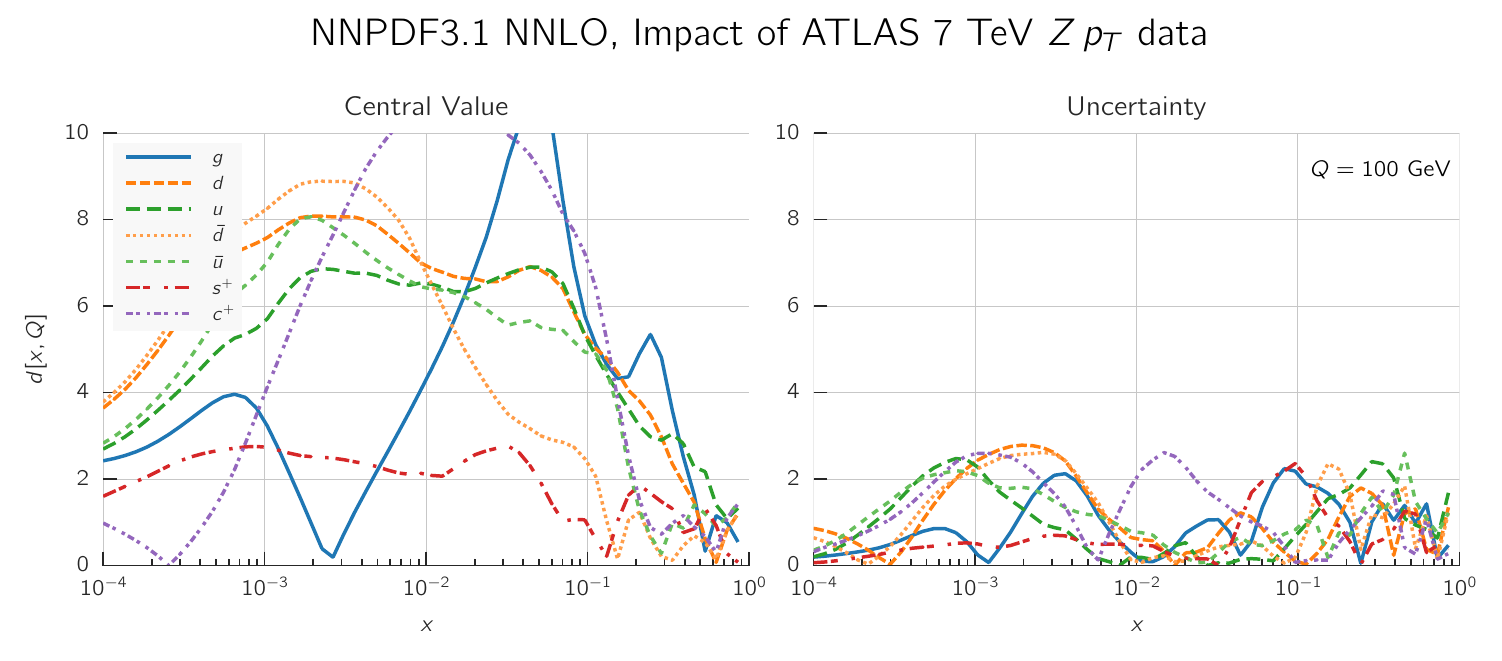}
  \caption{\small Same as Fig.~\ref{fig:distances_31_vs_30},  but now
    comparing the default NNPDF3.1 to a version of it with  
the 7~TeV $Z$ $p_T$ ATLAS data also included.
 \label{fig:distances_effect_Zpt7}}
\end{center}
\end{figure}

\begin{figure}[t]
\begin{center}
  \includegraphics[scale=0.38]{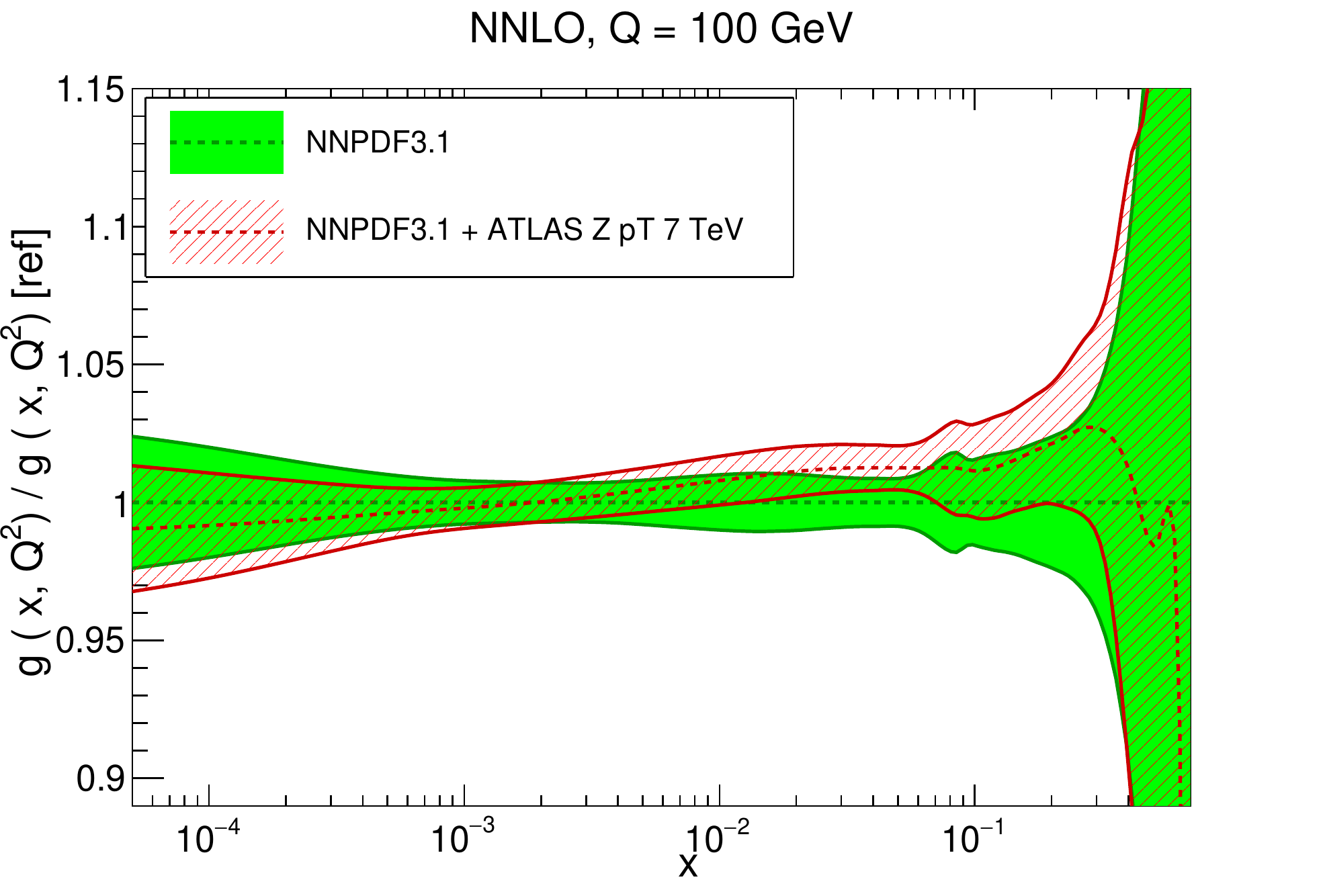}
  \includegraphics[scale=0.38]{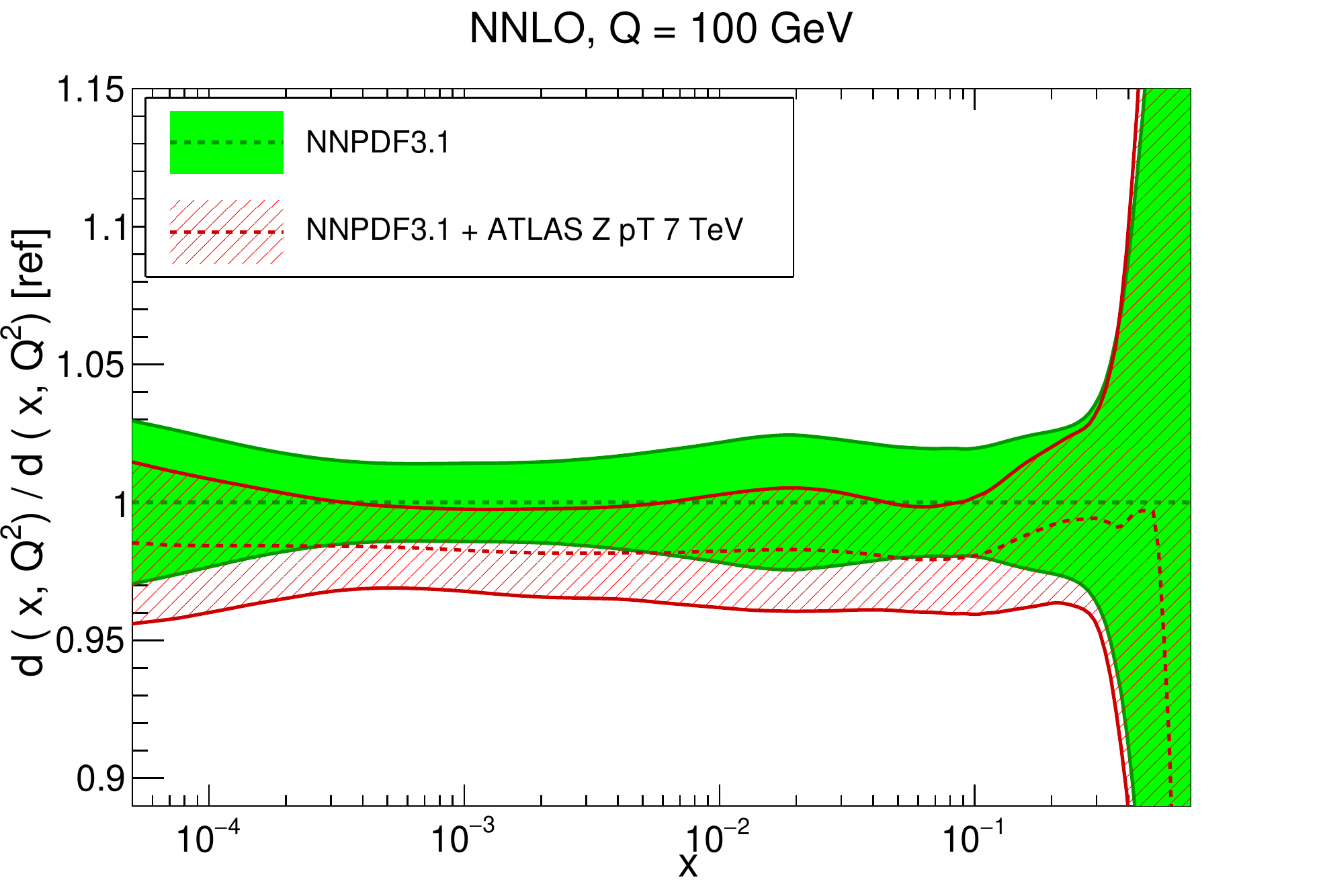}
  \caption{\small 
Same as Fig.~\ref{fig:31-nnlo-vs30}  but now
    comparing the default NNPDF3.1 to a version of it  with  
the 7~TeV $Z$ $p_T$ ATLAS data also included. Results are
shown for the gluon (left) and down quark (right).
\label{fig:31-nnlo-AZPT7TEV}}
\end{center}
\end{figure}

In Table~\ref{tab:NNLOZpTchi2} we provide
 the $\chi^2/N_{\rm dat}$ values for all the LHC
    $Z$ $p_T$ measurements, for both the NNPDF3.1 NNLO baseline 
(including  the ATLAS and CMS $Z$ $p_T$ 8 TeV data), and also from 
the determination also including the ATLAS $Z$ $p_T$ 7 TeV data.
    In the first column, the value of the $\chi^2/N_{\rm dat}$ for the 7 TeV data
    is in parenthesis to indicate that, unlike all other values, 
   it is a prediction
    and not the outcome of a fit. It is clear that
    the ATLAS $Z$ $p_T$ 7~TeV dataset is very poorly reproduced by the default NNPDF3.1
    set, and even after its inclusion in the dataset it cannot be
    accommodated. In fact, its inclusion is accompanied by a
    deterioration in the fit quality to ATLAS 8~TeV data, which are
    more accurate and supersede them. Furthermore, 
    there are also indications of tension between this dataset and the ATLAS
    $W$/$Z$ rapidity 
    distributions, whose total $\chi^2$ deteriorates by
    $12$ units (with 46 datapoints). 
The distances between these two PDF sets, displayed 
in Fig.~\ref{fig:distances_effect_Zpt7}, show that 
the gluon and quarks are shifted by almost one sigma by the
inclusion of the ATLAS $Z$ $p_T$ 7~TeV data. This is explicitly
shown in Fig.~\ref{fig:31-nnlo-AZPT7TEV} for the gluon and down
quark. It is apparent that uncertainties are however almost unchanged by the
inclusion of this dataset.

\begin{table}[b!]
  \centering
  \small
  \begin{tabular}{|c|c|c|}
    \hline
   &         NNPDF3.1 NNLO   &  + ATLAS $Z$ $p_T$ 7 TeV data \\
    \hline
    \hline
ATLAS $Z$ $p_T$ 7 TeV $(p_T^{ll},y_{ll})$    &     [6.78]           &   3.40    \\
ATLAS $Z$ $p_T$ 8 TeV $(p_T^{ll},M_{ll})$    &       0.93         &    0.98   \\
ATLAS $Z$ $p_T$ 8 TeV $(p_T^{ll},y_{ll})$     &    0.93            &   1.17    \\
CMS $Z$ $p_T$ 8 TeV $(p_T^{ll},M_{ll})$     &      1.32          &   1.33    \\
\hline
  \end{tabular}
  \caption{\small
    The values of $\chi^2/N_{\rm dat}$ for the LHC
    $Z$ $p_T$ data using the NNPDF3.1 NNLO PDF set, and for a new PDF
    determination which
    also includes the ATLAS $Z$ $p_T$ 7 TeV data.
    \label{tab:NNLOZpTchi2}
  }
\end{table}

While we cannot say how a better treatment of the
covariance matrix would affect the results, we must conclude that
within our current level of understanding, 
inclusion of the ATLAS 7~TeV $Z$ $p_T$ dataset would have a significant
impact on PDFs, without an improvement in precision, and with signs of
tension between this dataset and both the remaining $Z$ $p_T$ datasets, and
other $W$ and $Z$ production data. Therefore its inclusion in the global dataset
does not appear to be justified.

\subsection{Differential distributions for top  pair production}
\label{sec:impacttop}

\begin{figure}[t]
\begin{center}
  \includegraphics[scale=1]{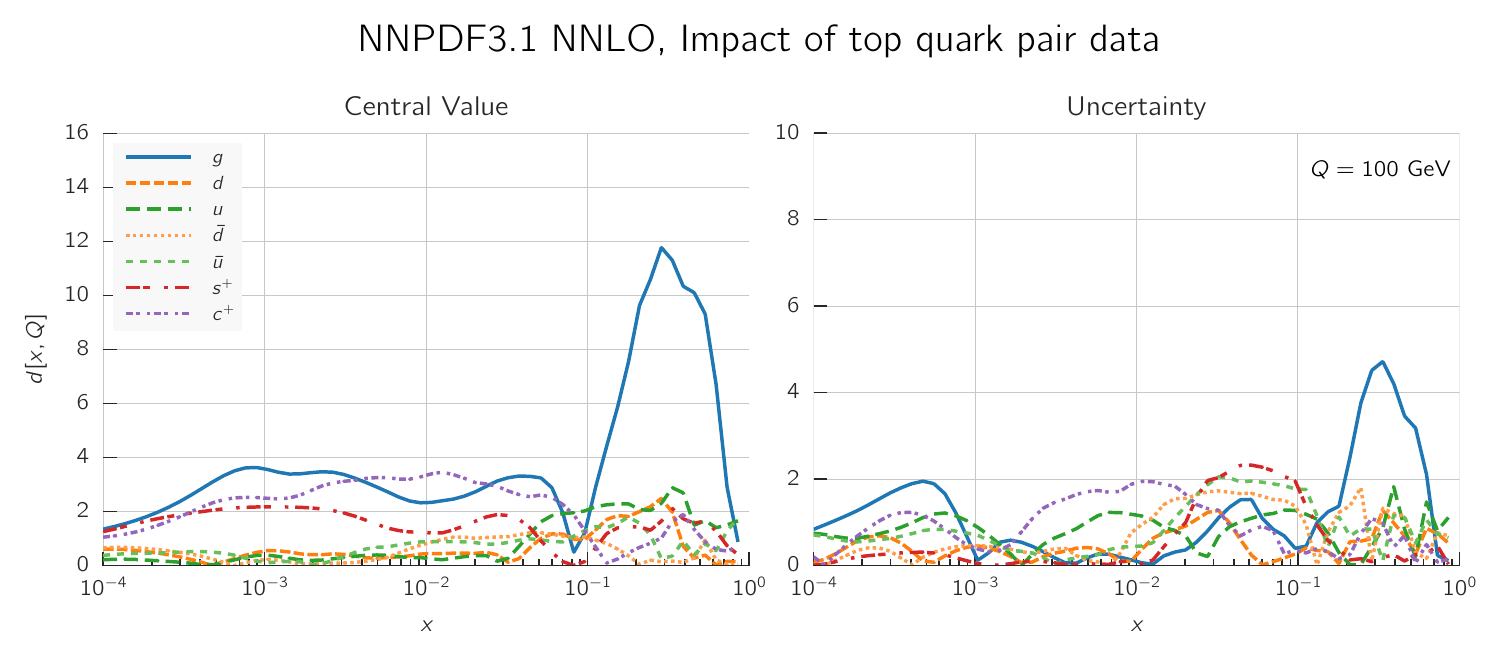}
  \caption{\small 
Same as Fig.~\ref{fig:distances_noZpT} but now excluding all top data (total cross-sections and differential
distributions). Note the different scale on the $y$ axis in the left plot.
    \label{fig:distances_notop}
  }
\end{center}
\end{figure}

The impact of differential top pair production on PDFs
and the optimal selection of top datasets has been discussed
extensively in Ref.~\cite{Czakon:2016olj}.
Here we briefly study the impact of the top data on NNPDF3.1 by
comparing with PDFs determined removing the top data from the dataset.
In Fig.~\ref{fig:distances_notop} we show the distances between these
PDF sets. Large differences can be seen in the gluon central value and uncertainty
for $x\gsim 0.1$: 
these data constrain the gluon for values as large as $x\simeq 0.6$~\cite{Czakon:2016olj}, a region in which
constraints from other processes are not available. The effect on other PDFs is moderate, with the largest
impact seen on charm at small $x$.

\begin{figure}[t]
\begin{center}
  \includegraphics[scale=0.38]{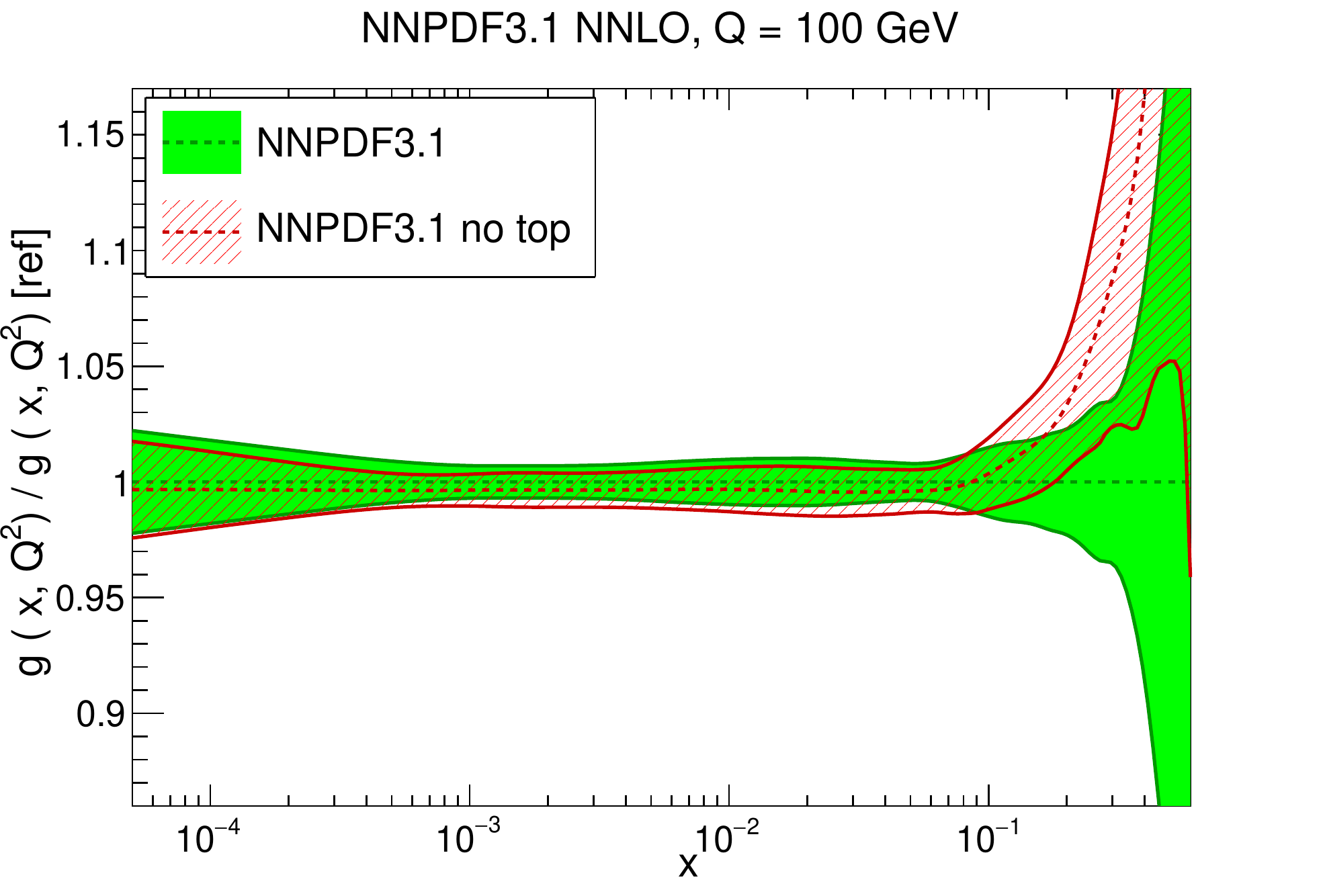}
  \includegraphics[scale=0.38]{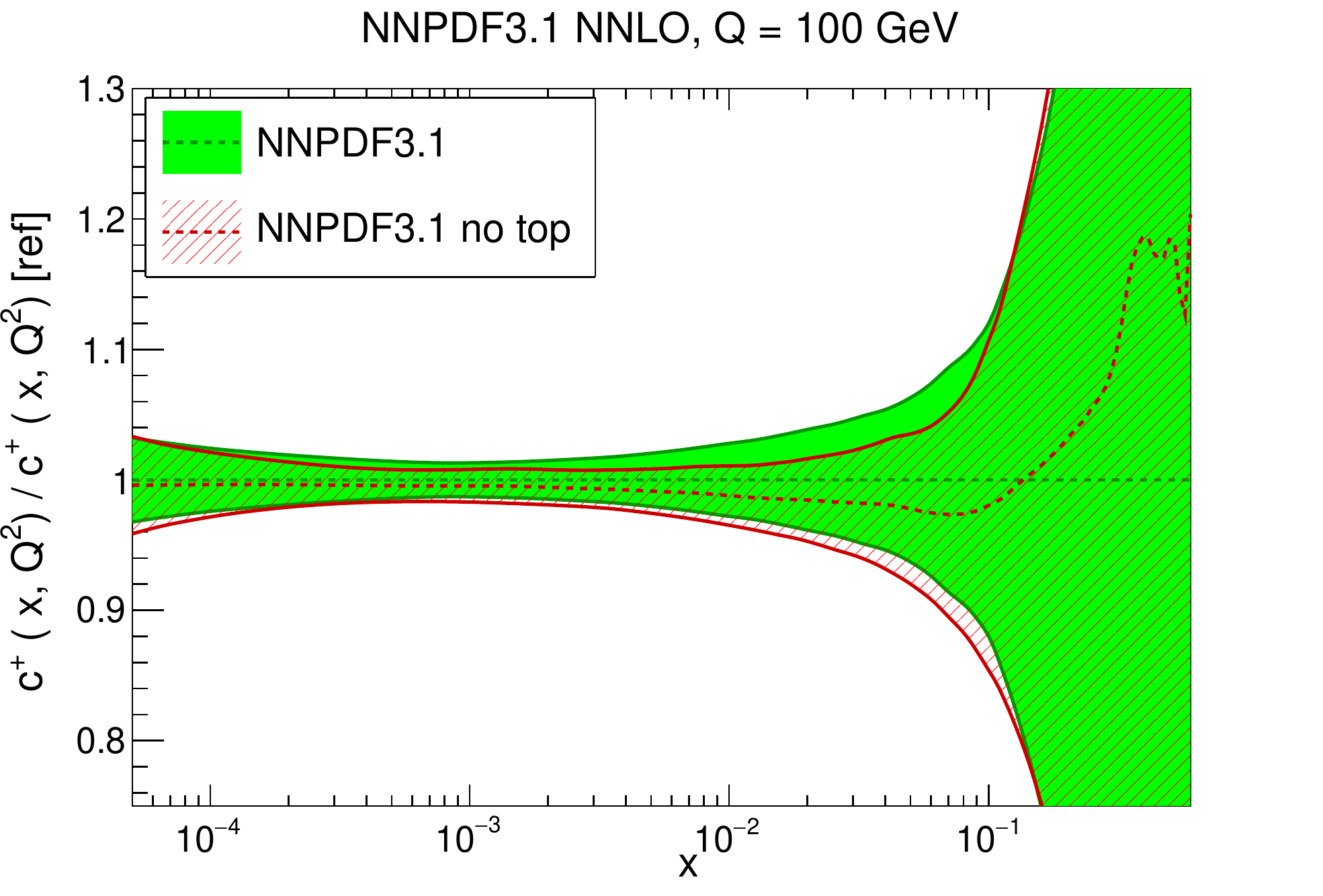}
  \includegraphics[scale=0.38]{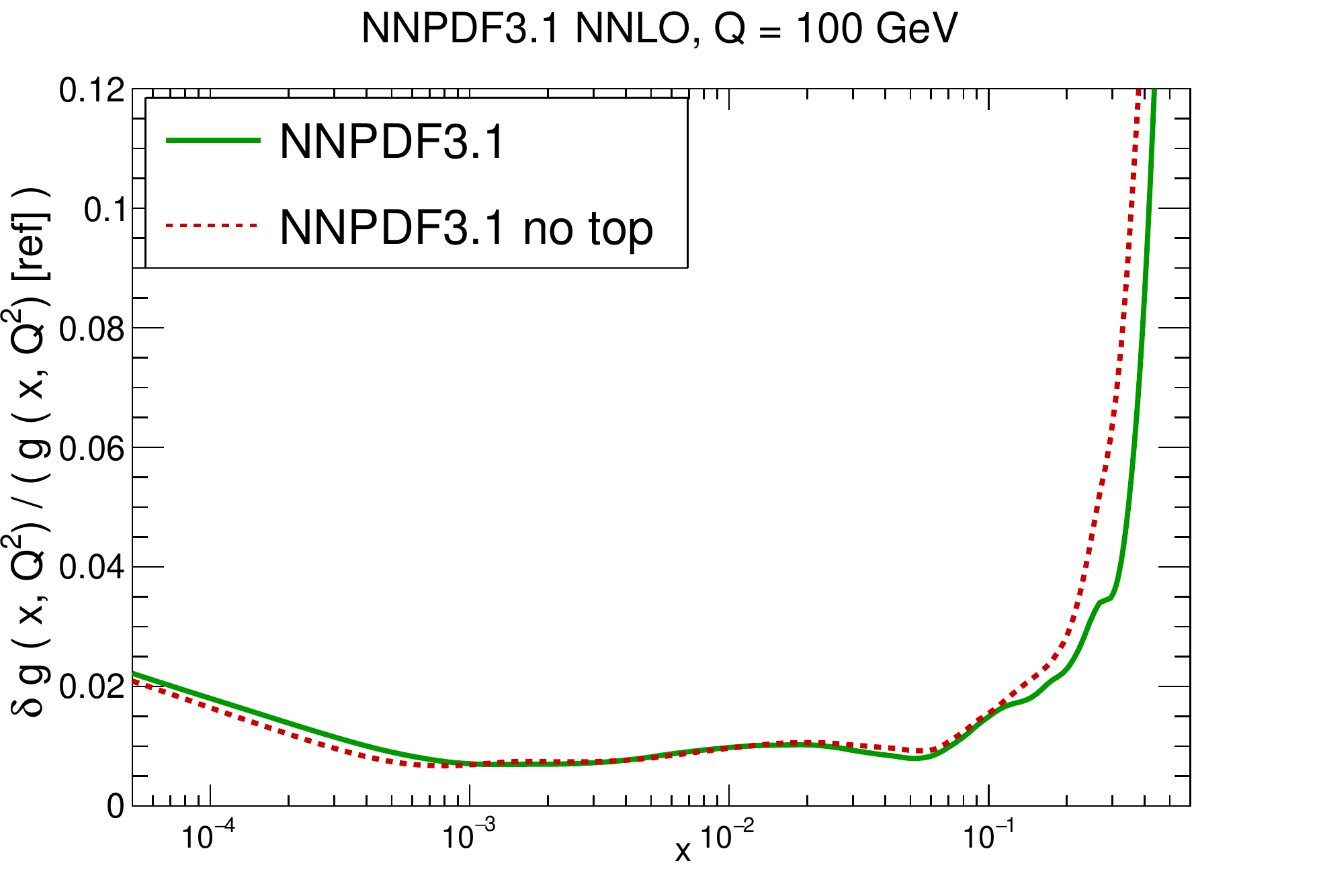}
  \includegraphics[scale=0.38]{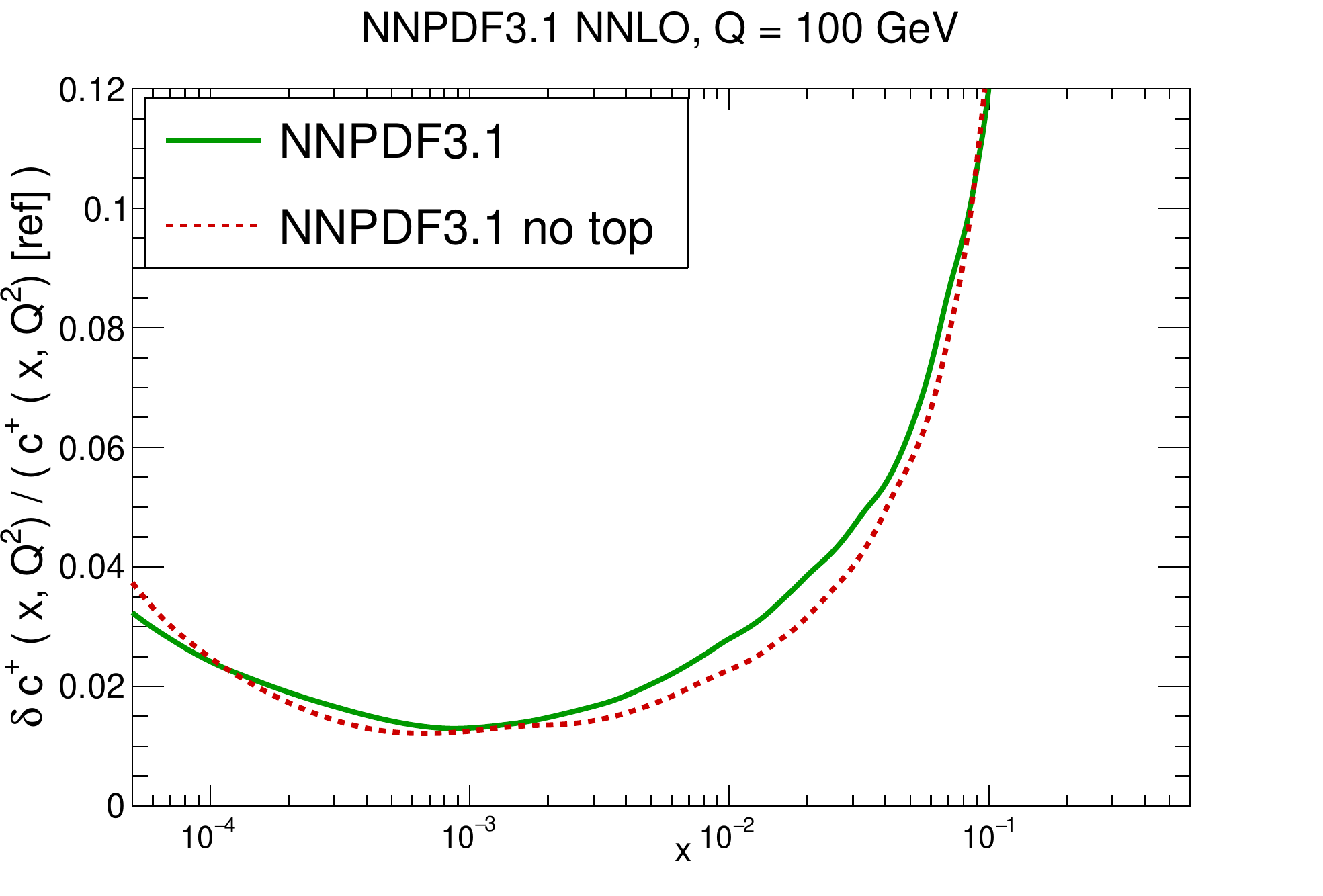}
  \caption{\small 
Same as Fig.~\ref{fig:31-nnlo-ZpT} but now excluding all top data (total cross-sections and differential
distributions).
Results are shown for the gluon (left) and charm (right), the PDFs above and their uncertainties below.
\label{fig:31-nnlo-top}}
\end{center}
\end{figure}

The differences between the two PDF sets are demonstrated in Fig.~\ref{fig:31-nnlo-top}, where the gluon and the
charm quark are shown. There is a substantial reduction in the uncertainty of the large $x$ gluon, 
with the central value without top data being considerably higher than 
 the narrow error band of the result when top is included. This
 suggests a significant increase in the precision of the gluon determination
 due to the top data. For the large $x$ gluon the differences between NNPDF3.1 and
NNPDF3.0 seen in Figs.~\ref{fig:31-nnlo-vs30}-\ref{fig:31-nnlo-old-vs-new} are 
therefore partly driven by the top data.
The impact on quark PDFs is marginal, as can be seen in the case of charm.

As already mentioned in Sect.~\ref{sec:topdata}, it has been shown in
Ref.~\cite{Czakon:2016vfr} that the sensitivity of the rapidity
distribution on the top mass is minimal. In fact, in 
Ref.~\cite{Czakon:2016olj} it was shown 
that if the top mass is varied by~1~GeV,
 NLO theoretical predictions for the normalized rapidity distributions
at the LHC~8~TeV 
vary by 0.6\% at most in the kinematic range covered by the data,
which is much less than the uncertainty  on the data, or the size of
the  NNLO corrections. This strongly suggests that our results are essentially
independent of the value of the top mass.

\subsection{Inclusive jet production}
\label{sec:jetdata}

While jet data have been used for PDF determination for a long time,
their full NNLO treatment is only becoming possible now, thanks to the recent
completion of the relevant computation~\cite{Currie:2016bfm,Currie:2017ctp}.
However,
as discussed in Section~\ref{sec:datajets}, NNLO
 corrections are not yet available for
all datasets included in NNPDF3.1. Consequently in the default
NNPDF3.1 PDF determination, jets have been included using NNLO PDF
      evolution and NLO matrix elements supplemented by an extra
      theory uncertainty determined through scale variation. Here we
      assess generally the effect of jet data, and in particular the
      possible impact of this approximation.

\begin{figure}[t]
\begin{center}
  \includegraphics[scale=1]{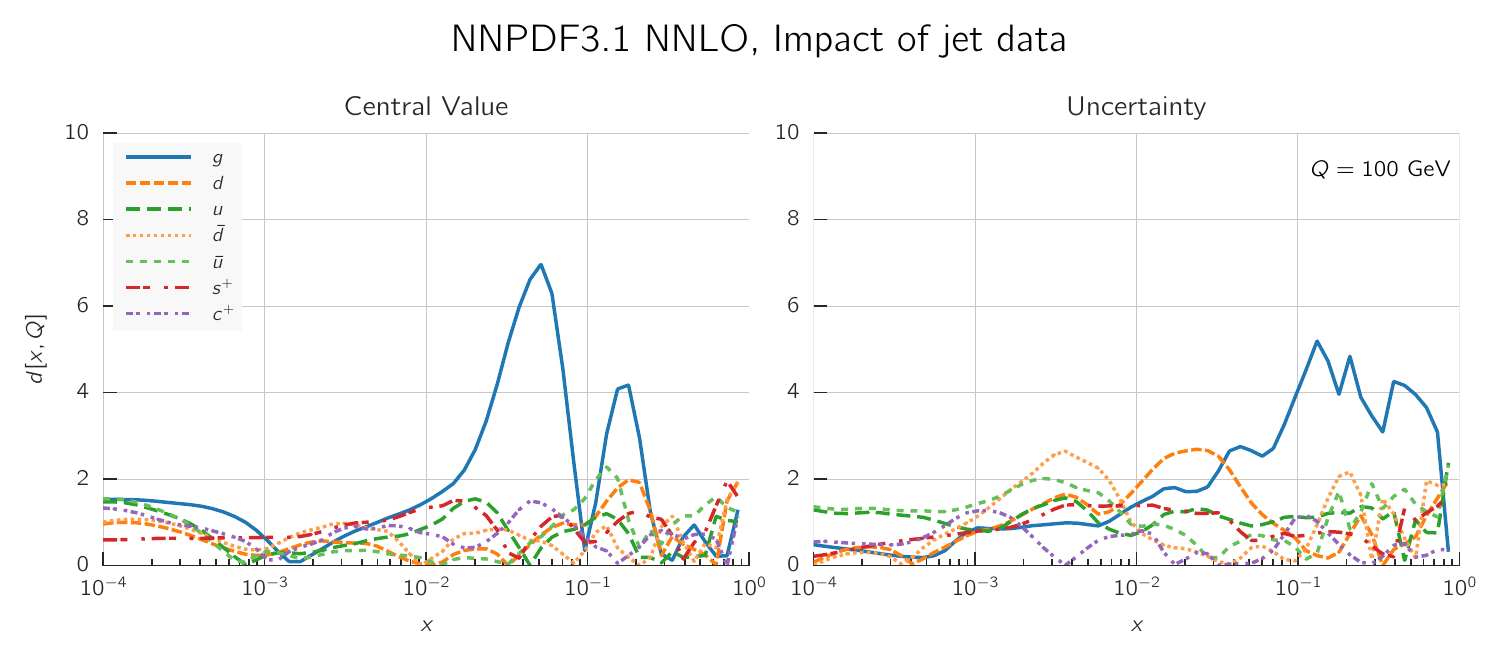}
  \caption{\small  Same as Fig.~\ref{fig:distances_noZpT} but now excluding all jet data.
    \label{fig:distances_nojets}
  }
\end{center}
\end{figure}

\begin{figure}[t]
\begin{center}
  \includegraphics[scale=0.38]{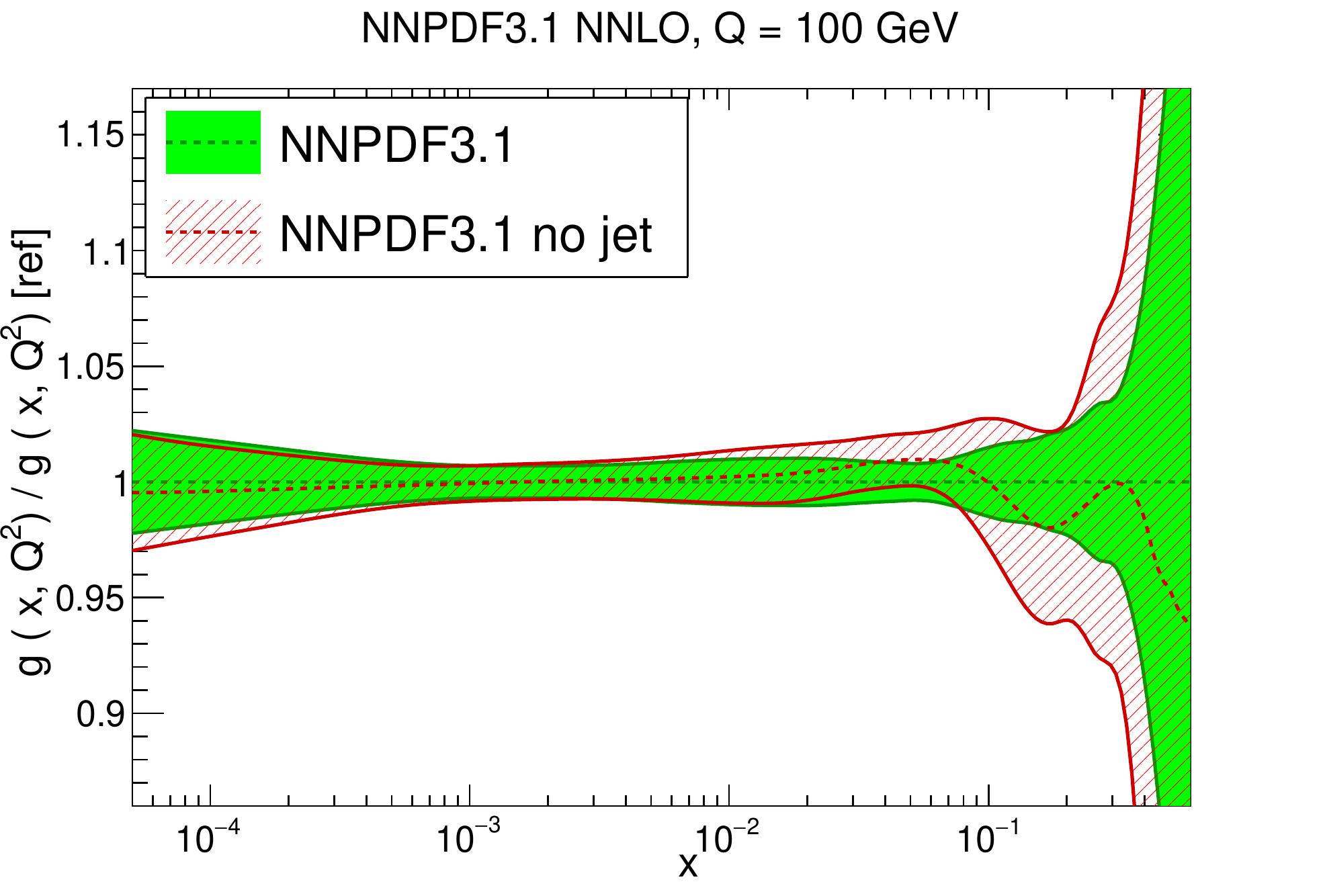}
  \includegraphics[scale=0.38]{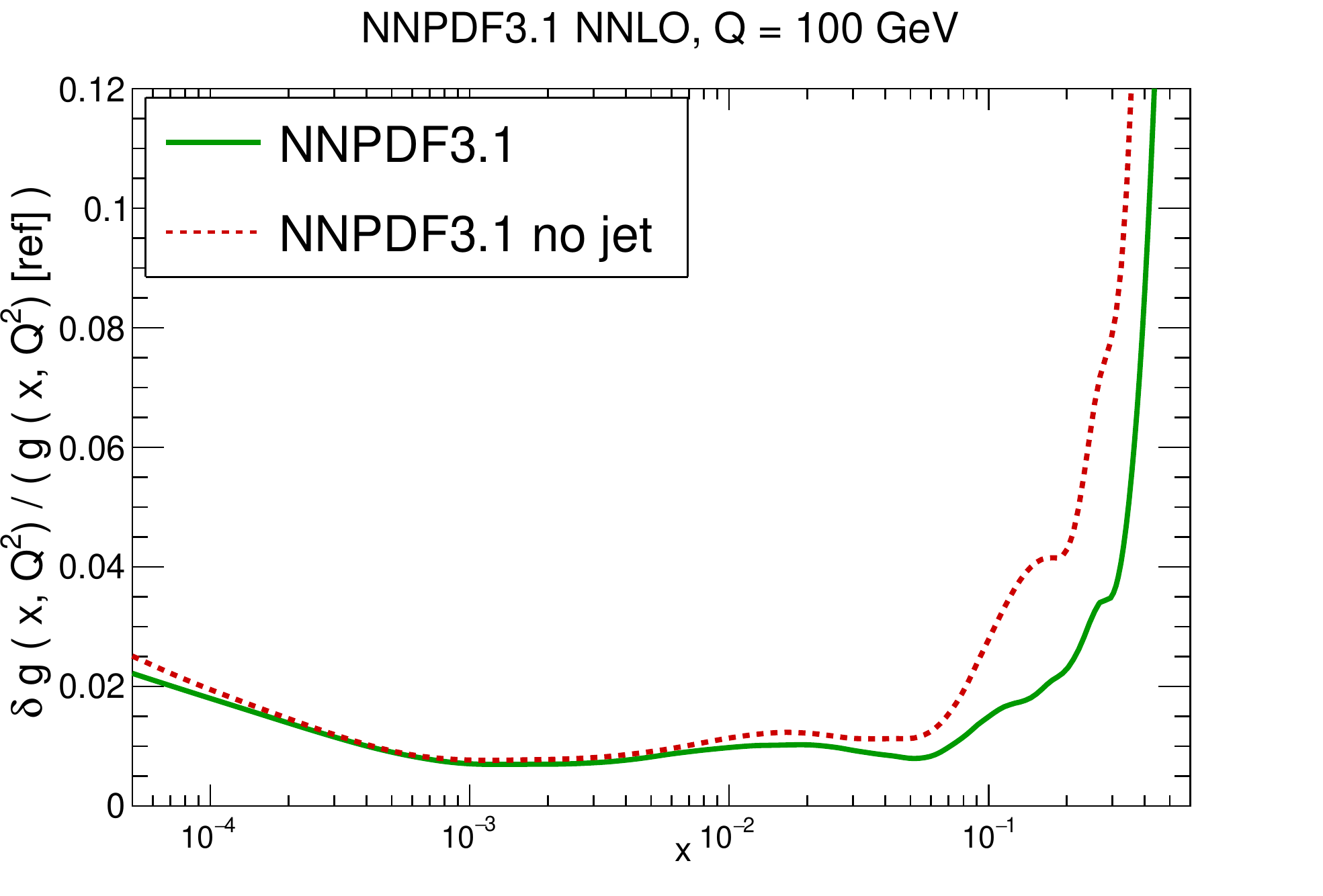}
    \caption{\small Comparison between the default NNPDF3.1 NNLO PDFs
      an alternative determination in which all jet data have been
      removed: the gluon (left) and the percentage uncertainty on it
      (right) are shown
\label{fig:jetdata-nnlo}}
\end{center}
\end{figure}

To this end, we first repeat the NNPDF3.1 determination
but excluding jet data. The distances between these PDFs and the
default are shown
in Fig.~\ref{fig:distances_nojets}. It is clear that 
jet data have a moderate and very localized impact, on the gluon
in the region $0.1\lsim x\lsim 0.6$, at most at the half-sigma level,
with essentially no impact on other PDFs. The changes in all other PDFs
are compatible with a statistical fluctuation.
A direct comparison of the gluon PDFs and their uncertainties in
Fig.~\ref{fig:jetdata-nnlo} confirms this. The uncertainty on the gluon is reduced by up to a factor of 
two by the jet data in this region, with the central value of the gluon within the narrower 
uncertainty band of the default set.

It is interesting to observe that in NNPDF3.0 the impact of the jet
data was rather more significant, with uncertainties being reduced by
a large factor for all $x\gsim 0.1$. In NNPDF3.1 the gluon at large $x$ is 
strongly constrained by the top data, as discussed in
Section~\ref{sec:impacttop}. Specifically, 
the addition of the jet data leaves the gluon unchanged in this
region, see Fig.~\ref{fig:jetdata-nnlo}, but addition of the top data produces a significant shift, as
seen in Fig.~\ref{fig:31-nnlo-top}. This 
suggests excellent compatibility between the jet and top data, with
the large $x$ gluon now mostly determined by the top data. This also
explains the insensitivity to the NNLO correction to jet production,
to be discussed shortly.

\begin{figure}[t]
\begin{center}
  \includegraphics[scale=0.38]{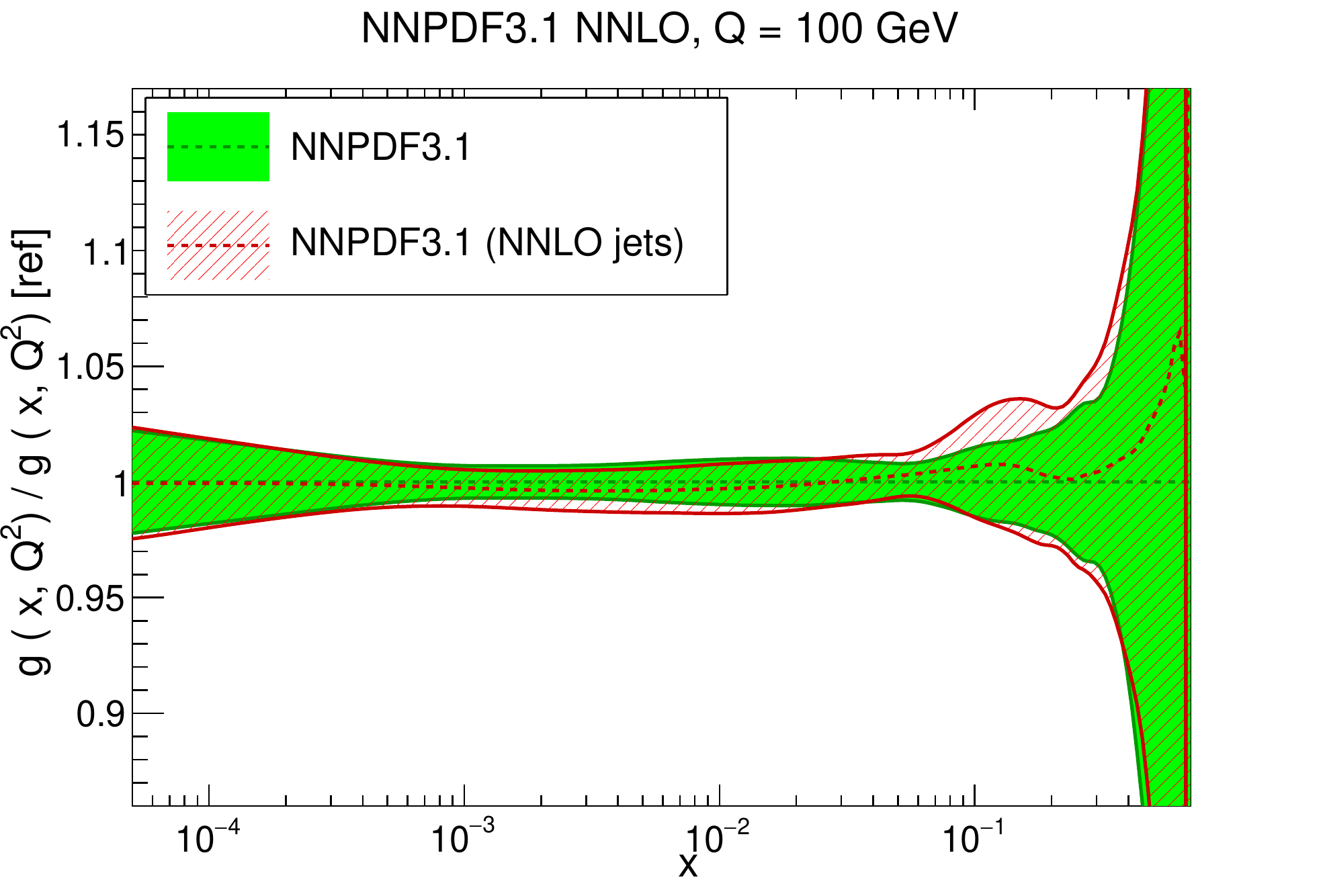}
  \includegraphics[scale=0.38]{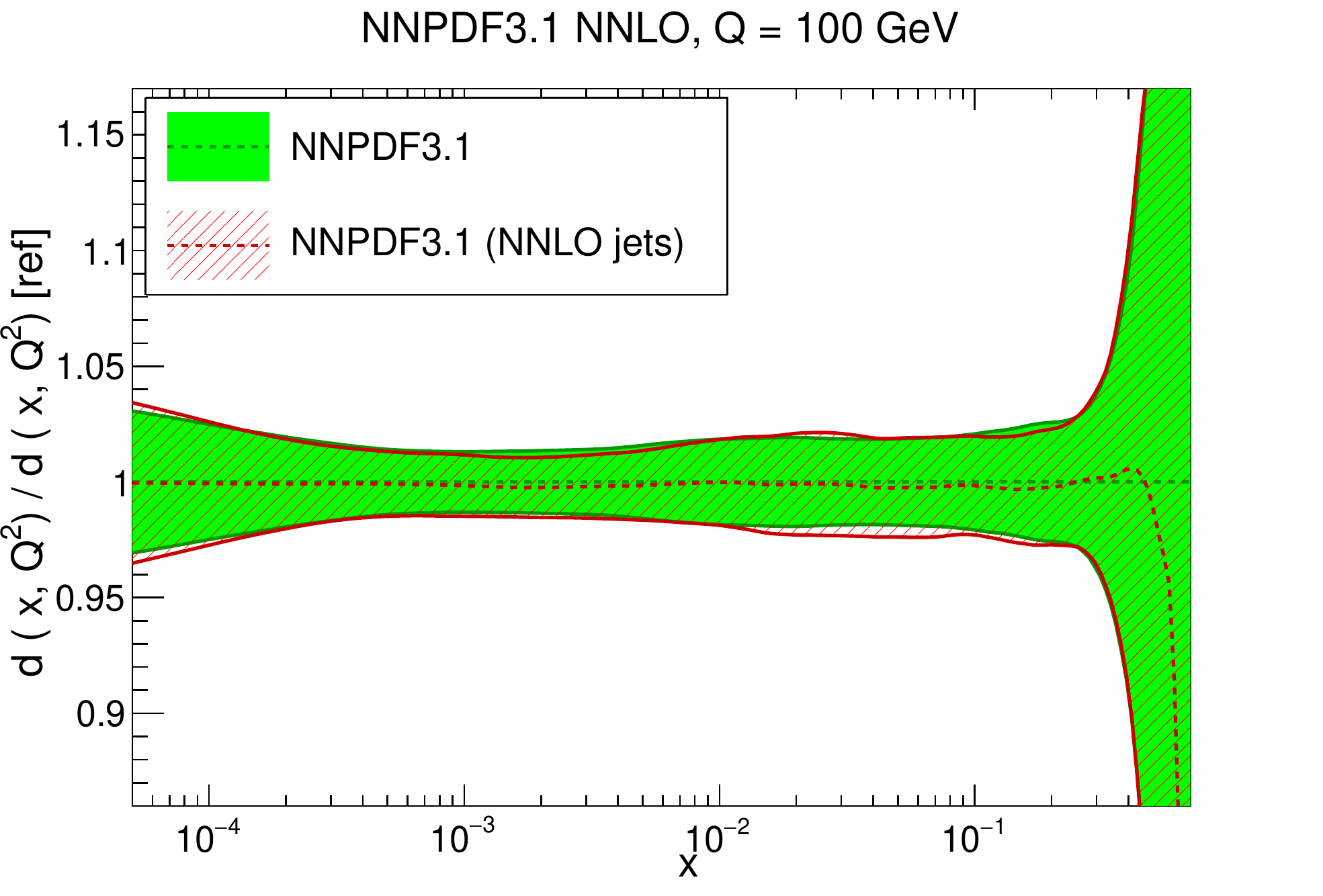}
    \caption{\small 
Same as Fig.~\ref{fig:31-nnlo-vs30}  but now
comparing the default NNPDF3.1 NNLO PDFs
      to an alternative determination in which 
ATLAS and CMS 7~TeV jet data have been included
      using exact NNLO theory. The gluon (left) and down
      (right) PDFs are shown.
\label{fig:jetdataex-nnlo}}
\end{center}
\end{figure}

Despite there reduced impact, the jet data still
play a non-negligible role. One may therefore worry about the reliability of the
theoretical treatment, based on NLO matrix elements with theory
uncertainties. In order to assess this, we have repeated the PDF
determination but now
using full NNLO theory in the case of the 2011 
7~TeV LHC jet data where it is available.
      The other jet datasets, namely the CDF Run II $k_T$ jets, the
      ATLAS and CMS $\sqrt{s}=2.76$ TeV datasets, and the ATLAS 2010
      7~TeV dataset, are treated as in
      the baseline. Essentially no change in PDFs is found, as
      illustrated in Fig.~\ref{fig:jetdataex-nnlo} where the gluon
      and down PDFs are shown.
      Such a result is consistent with the percent level NNLO corrections
      found when using our choice of the jet $p_T$ as the central scale, shown in 
      Fig.~\ref{sec:datajets}. 

Also, as mentioned in
      Sect.~\ref{sec:datajets}, only the central rapidity bin of the
      ATLAS 2011 7~TeV data has been included, because we have found
      that, while a good description can be achieved if each of the
      rapidity bins is included in turn, or if the uncertainties are
      decorrelated between rapidity bins, it is impossible to achieve
      a good description of all rapidity bins with correlations
      included. One may therefore wonder whether the inclusion of
      other rapidity bins would lead to different results for the
      PDFs, despite the fact that they have less PDF
      sensitivity~\cite{Rojo:2014kta}.  In order to check this, we
      have compared to the data the prediction for all of the ATLAS 2011 7~TeV
      data using the default NNPDF3.1 set, and  determined the
      $\chi^2$ for each rapidity bin separately. For the five rapidity
      bins which have not been included, from central to forward, we
      find $\chi^2/N_{\rm dat}=1.27,\>0.95,\>1.06,\>0.97,\>0.73$, with
      respectively $N_{\rm dat}=29,\>26,\>23,\>19,\>12$, to be
      compared to the value  $\chi^2/N_{\rm dat}=1.06$ for $N_{\rm
        dat}=31$ of Tab~\ref{tab:NNLOjetschi2} for the central
      rapidity bin which is  included. We 
      conclude that all rapidity bins are well reproduced, and thus
      none of them can have a significant pull on PDFs that might
      change the result if they were included.

\begin{figure}[t]
\begin{center}
  \includegraphics[width=0.49\textwidth]{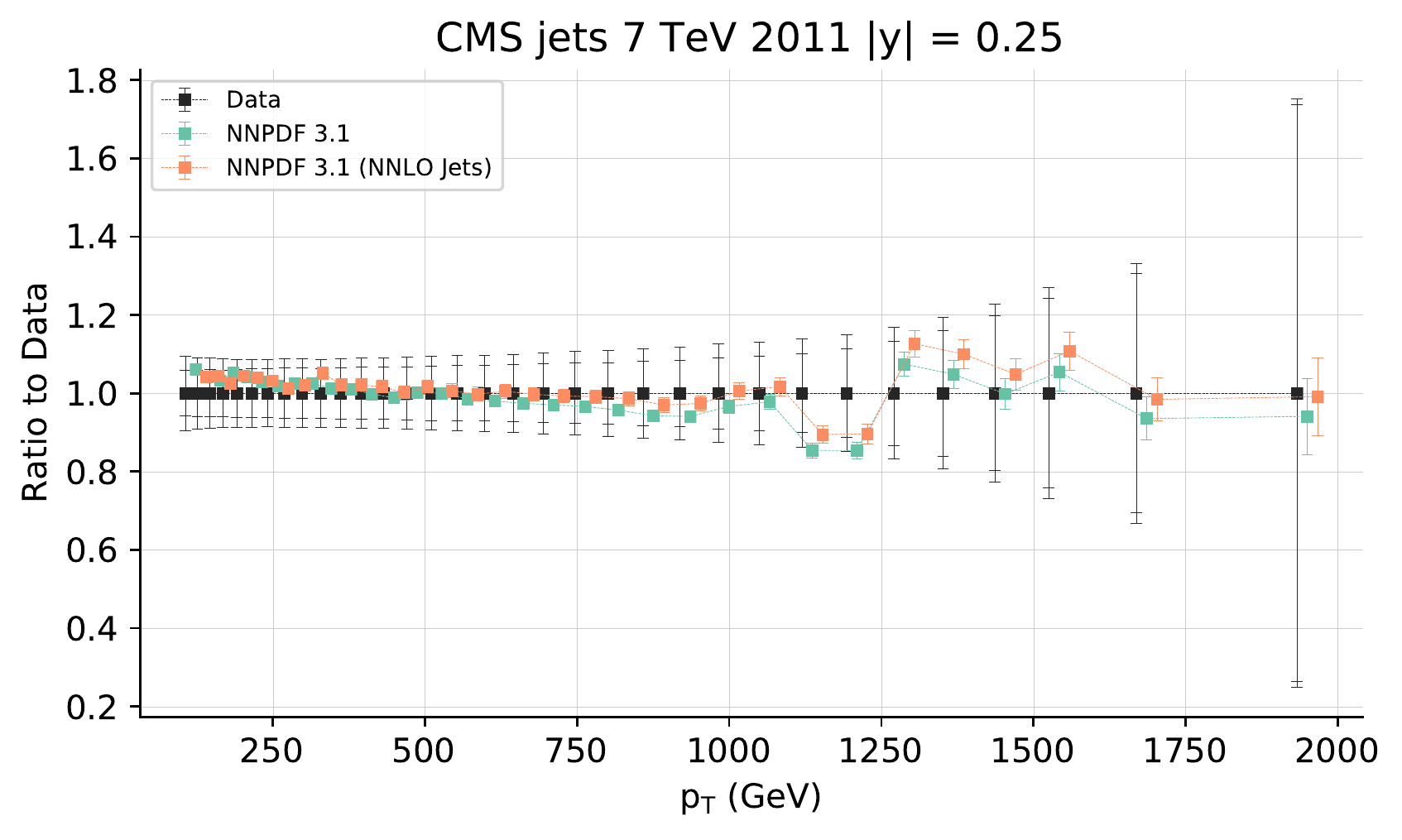}
  \includegraphics[width=0.49\textwidth]{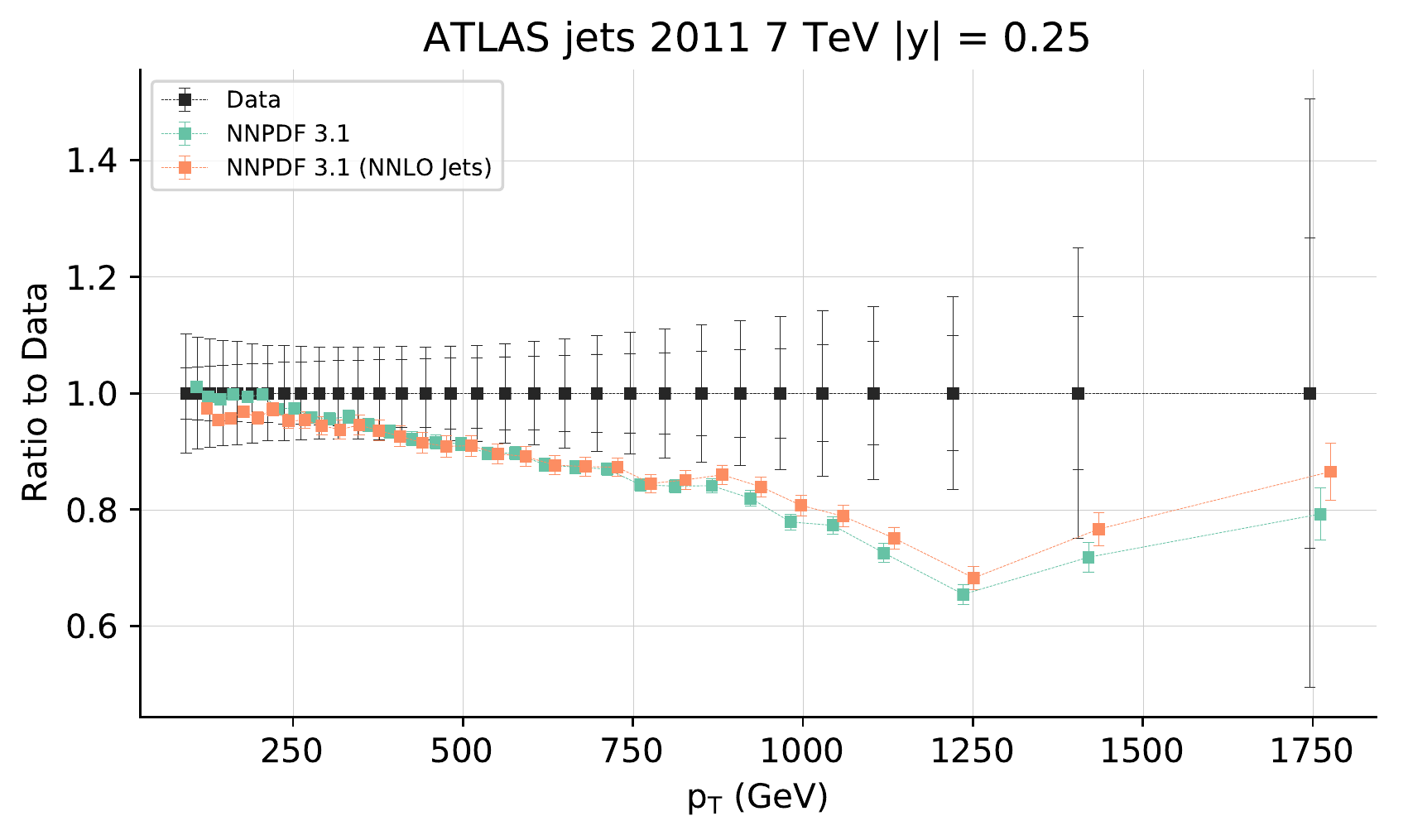}
  \caption{\small Comparison between CMS (left)
    and ATLAS (right) one-jet inclusive data 
    at 7~TeV from 2011, and best-fit results obtained using
NLO theory supplemented by scale uncertainties or exact NNLO
theory. The uncertainties shown on the best-fit prediction is the PDF
uncertainty, while that on the data is the diagonal (outer error bar)
and the scale uncertainty on the NLO prediction (inner error bar).
\label{fig:jetdata-nnlo-datath}}
\end{center}
\end{figure}

In order to understand better the impact of the NNLO corrections and
their effect on the PDFs, in 
Fig.~\ref{fig:jetdata-nnlo-datath} we compare the
best-fit prediction to 
the 7~TeV  2011 CMS 
    and ATLAS data for the two PDF sets compared in
    Figs.~\ref{fig:distances_nojets}-\ref{fig:jetdataex-nnlo}.
The corresponding values of $\chi^2/N_{\rm dat}$ are collected in
Table~\ref{tab:NNLOjetschi2}, both for these and all other jet data.
First of all, one should note that for all experiments for which an exact NNLO
 computation is not yet available, listed on the top part of
Table~\ref{tab:NNLOjetschi2}, the $\chi^2$ values obtained with these two PDF
sets are almost identical. Because the predictions for these experiments 
are computed using the same theory (NLO with extra scale uncertainty),
this shows that the change in PDFs is very small.
For the experiments for which NNLO theory is available, $\chi^2$
values also change very little. A comparison of the data to theory (see
Fig.~\ref{fig:jetdata-nnlo-datath}) shows that the NLO and NNLO
predictions are, as expected, quite close. Nevertheless, the NNLO
prediction is in slightly better agreement with the data, and this is
reflected by the
better value of the NNLO $\chi^2$ shown in
Table~\ref{tab:NNLOjetschi2}, despite the
the extra scale uncertainty (shown as an inner error bar on the data
in Fig.~\ref{fig:jetdata-nnlo-datath}) that is added to
the NLO prediction only.

As a final check, we have repeated the PDF determination but with a cut excluding
large $p_T$ jet data, for which the NNLO corrections are larger
(compare Fig.~\ref{fig:jetcf}). Specifically, we have only kept 7~TeV
jet data with $p_T\le 240$~GeV,  2.76~TeV
jet data with $p_T\le 95$~GeV,  1.96~TeV
jet data with $p_T\le 68$~GeV. We found that the ensuing PDFs are
statistically equivalent (distances of order one-two) to those from
the default set.  
We conclude that the impact of the approximate treatment of jet data
on the default NNPDF3.1 set is very small.

\begin{table}[H]
  \centering
  \begin{tabular}{|c|c|c|}
    \hline
   &         NNPDF3.1   &  exact NNLO  \\
    \hline
    \hline
CDF Run II $k_t$ jets     &          0.84      &      0.85 \\
ATLAS jets 2.76 TeV   & 1.05       &     1.03 \\
CMS jets 2.76 TeV  & 1.04        &    1.02 \\
ATLAS jets 2010 7 TeV  & 0.96     &       0.95 \\
\hline
ATLAS jets 2011 7 TeV &  1.06      &      0.91 \\
CMS jets 7 TeV 2011 7 TeV &   0.84   &         0.79 \\
\hline
  \end{tabular}
  \caption{\small
    The values of $\chi^2/N_{\rm dat}$ for all the jet datasets obtained
    using either of the two PDF sets compared in
    Fig.~\ref{fig:jetdataex-nnlo}, and in each case the theory used in
    the corresponding PDF determination. For the datasets in the top part of the
    table the exact NNLO computations are not yet available and NLO theory
    with scale uncertainty is used throughout, while for those in the
    bottom part of the table NNLO theory is used for the right column.
    \label{tab:NNLOjetschi2}
  }
\end{table}

\clearpage

\subsection{Electroweak boson production in the forward region}
\label{sec:lhcbimpact}

\begin{figure}[t]
\begin{center}
  \includegraphics[scale=1]{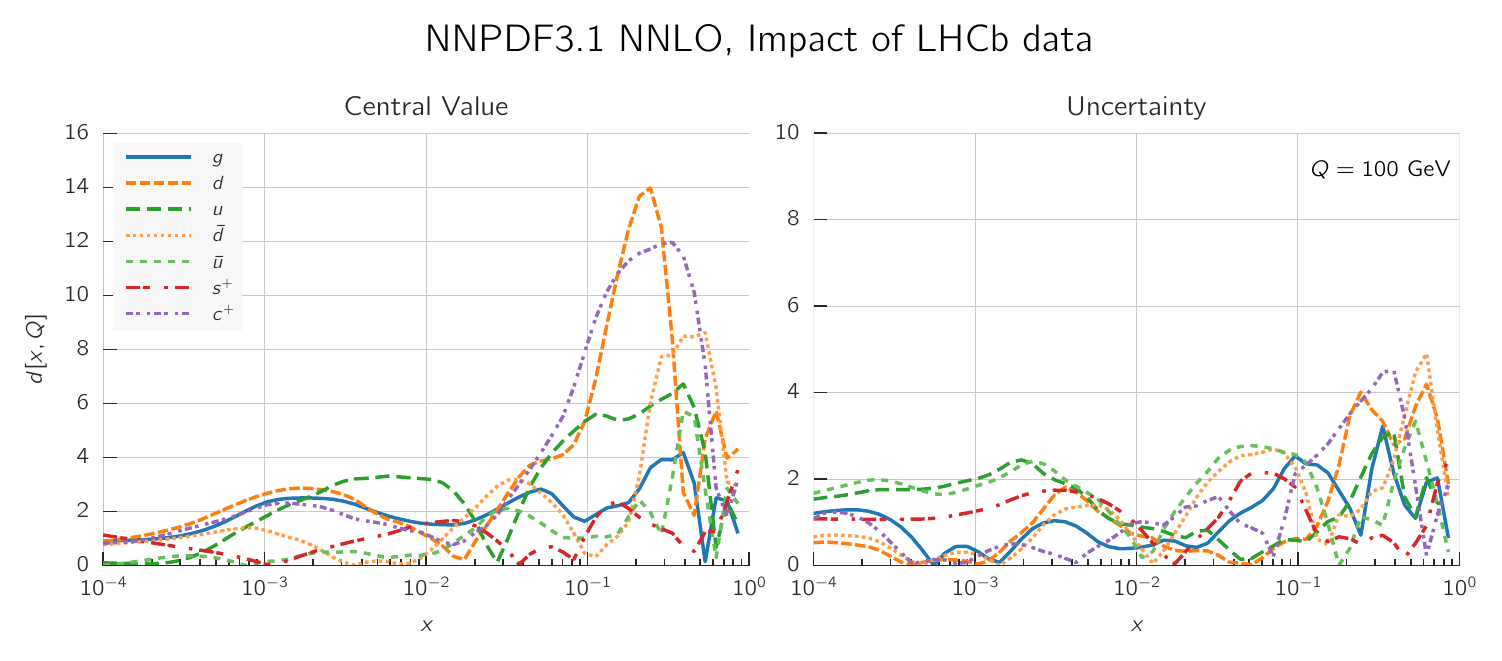}
  \caption{\small Same as Fig.~\ref{fig:distances_noZpT} but now
    excluding all LHCb data. Note the different scale on the $y$ axis 
in the left plot.
    \label{fig:distances_noLHCb}
  }
\end{center}
\end{figure}

Electroweak production data from the LHCb experiment open up a new
kinematic region and therefore provide new constraints on flavor separation
at large and small $x$. While LHCb data were included in NNPDF3.0, 
the release of the legacy Run I LHCb measurements at 7 TeV and 8 TeV, which include all
correlations between $W$ and $Z$ data, greatly increases their utility in
PDF determination.
In Fig.~\ref{fig:distances_noLHCb} distances are shown
between the NNPDF3.1 NNLO default and PDFs determined excluding all
LHCb data.
The impact is significant for all quark PDFs, especially in the
valence region:
hence this data has a substantial impact on flavor
separation, most notably at large $x$.

\begin{figure}[t]
\begin{center}
  \includegraphics[scale=0.34]{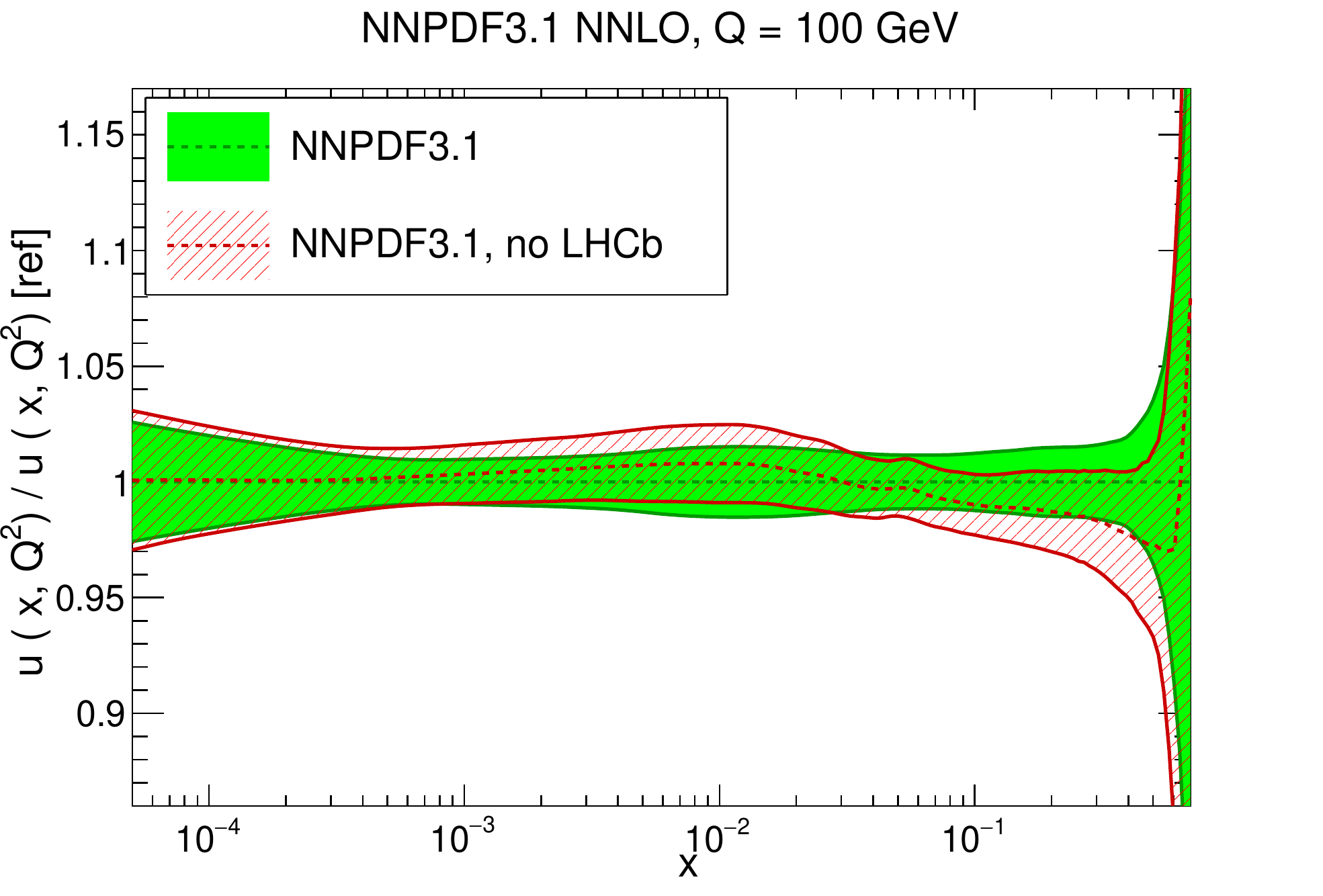}
  \includegraphics[scale=0.34]{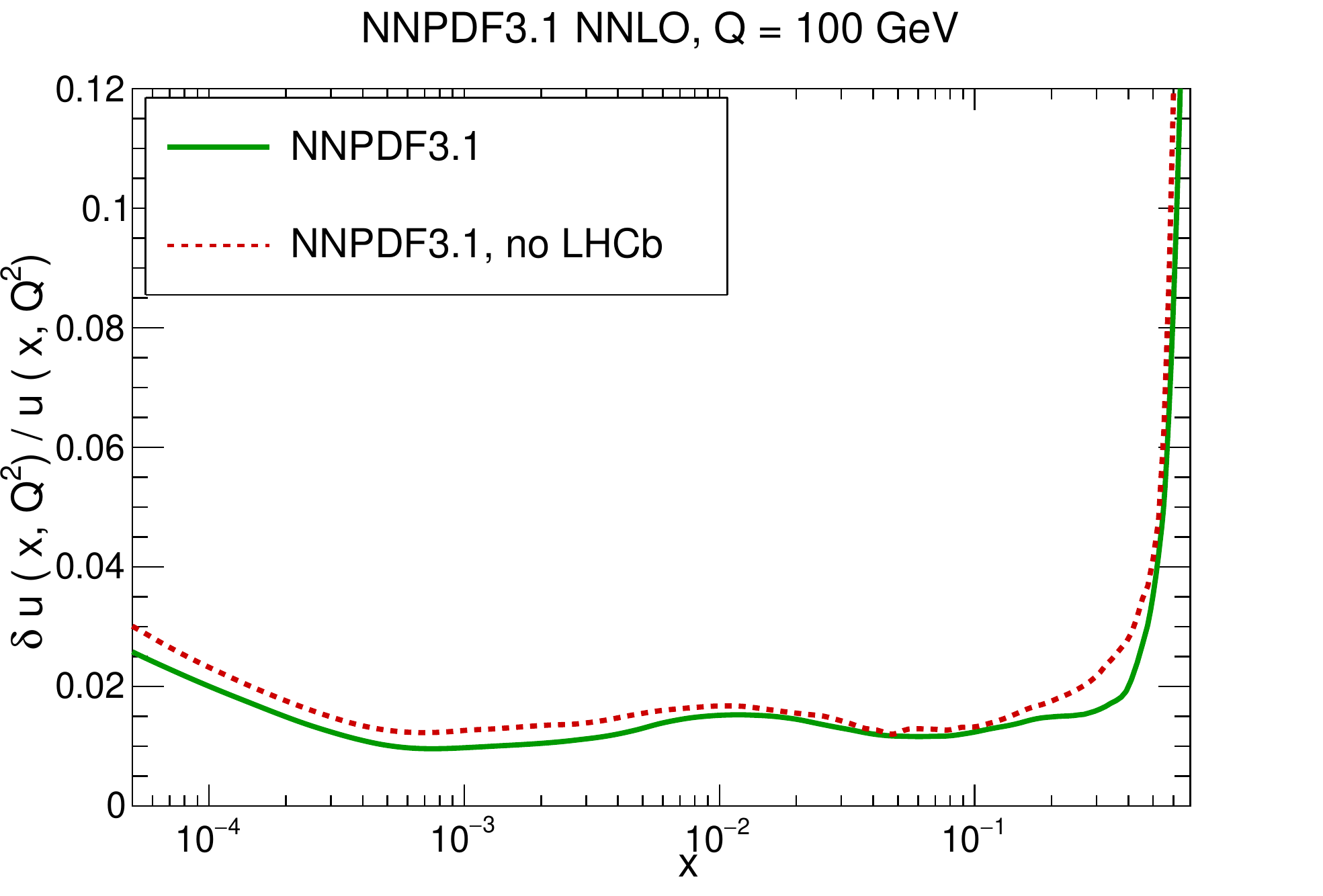}
  \includegraphics[scale=0.34]{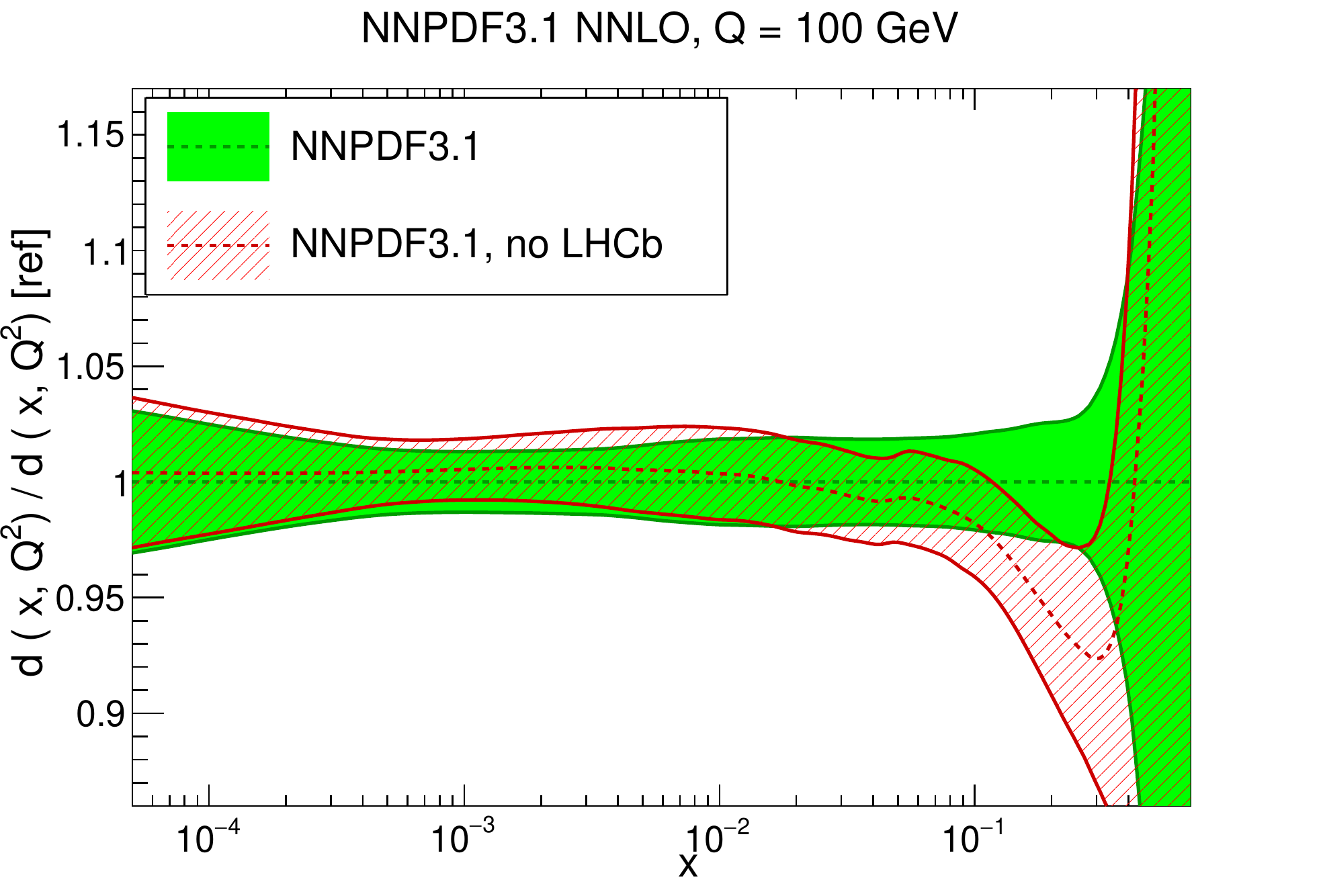}
  \includegraphics[scale=0.34]{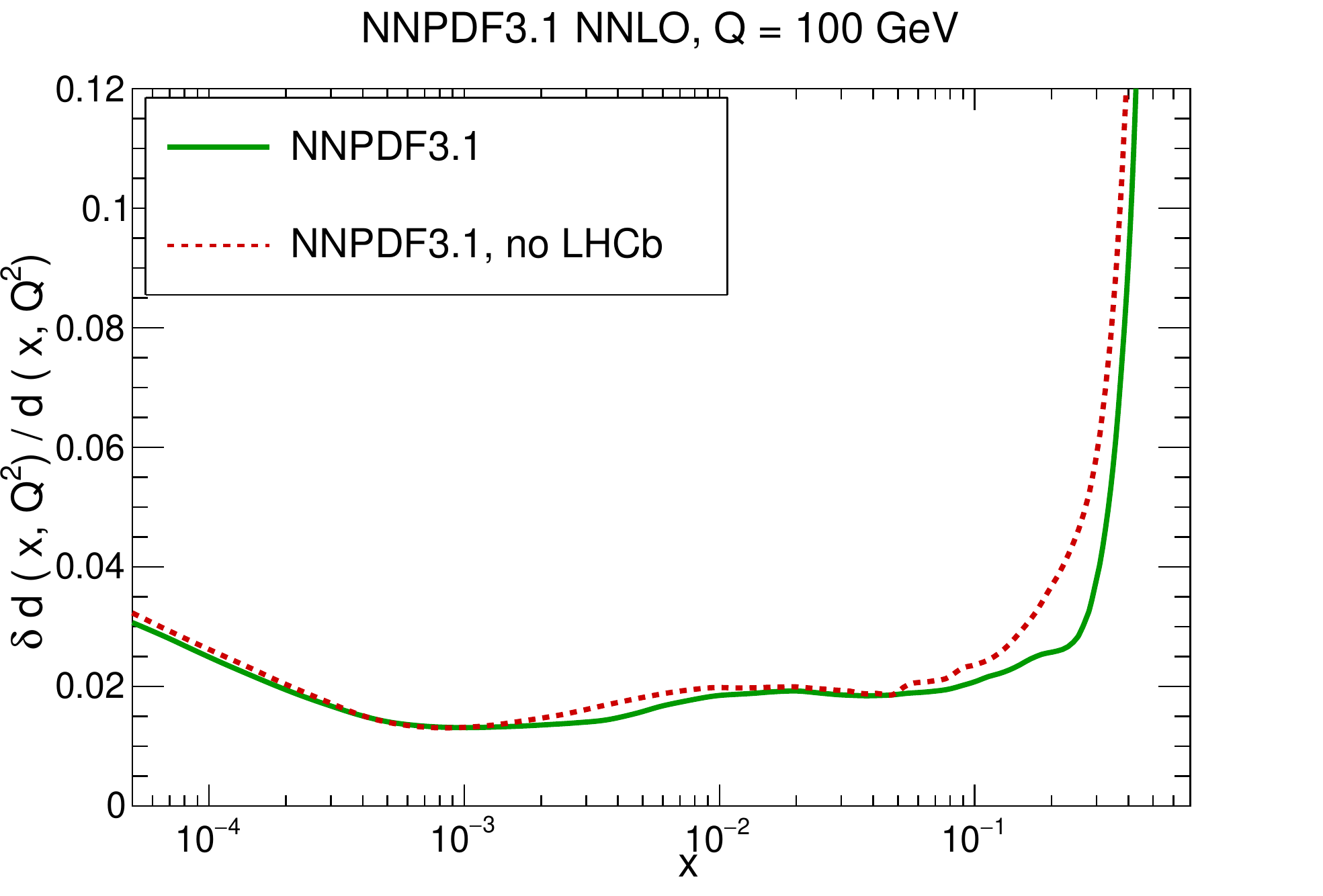}
  \includegraphics[scale=0.34]{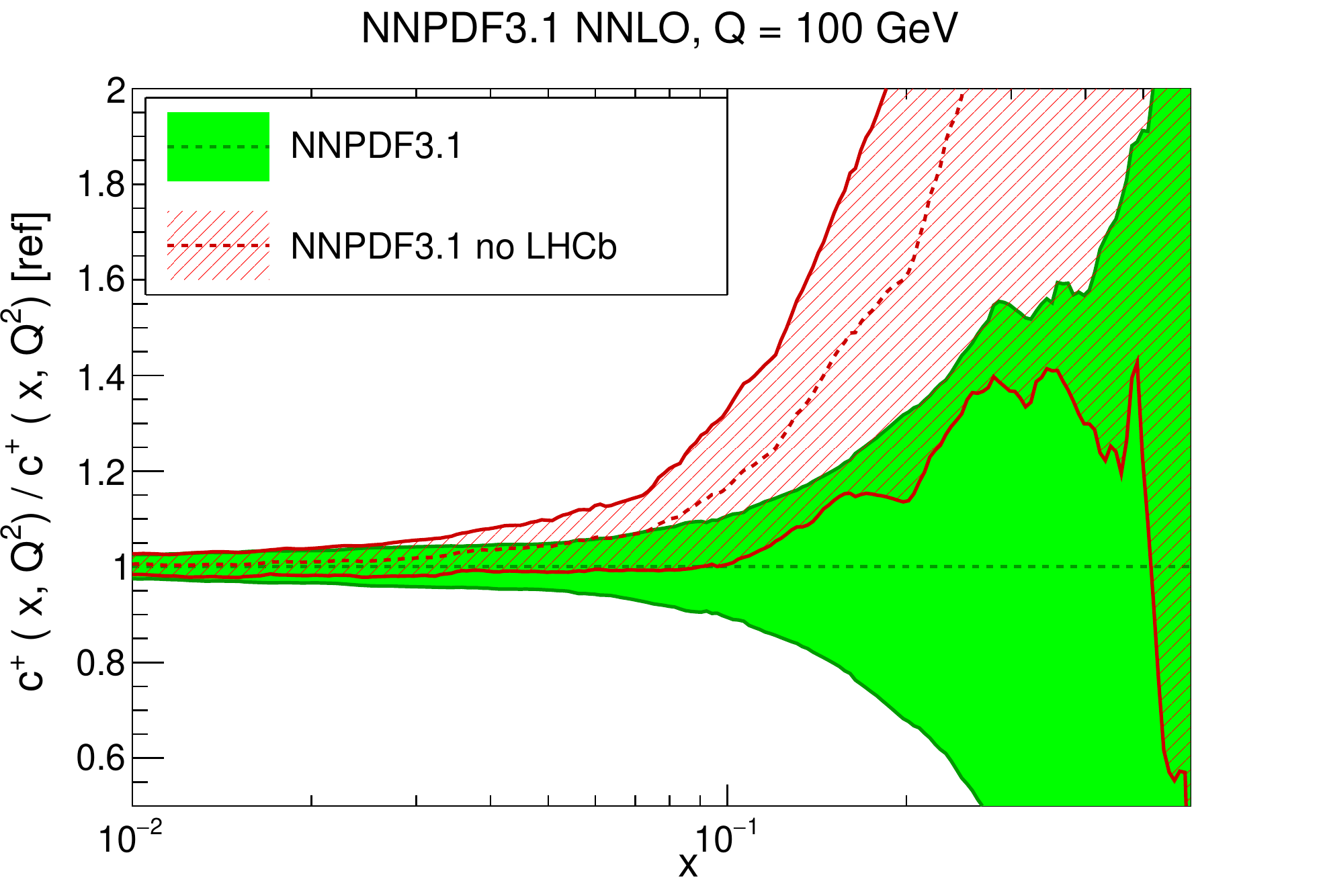}
  \includegraphics[scale=0.34]{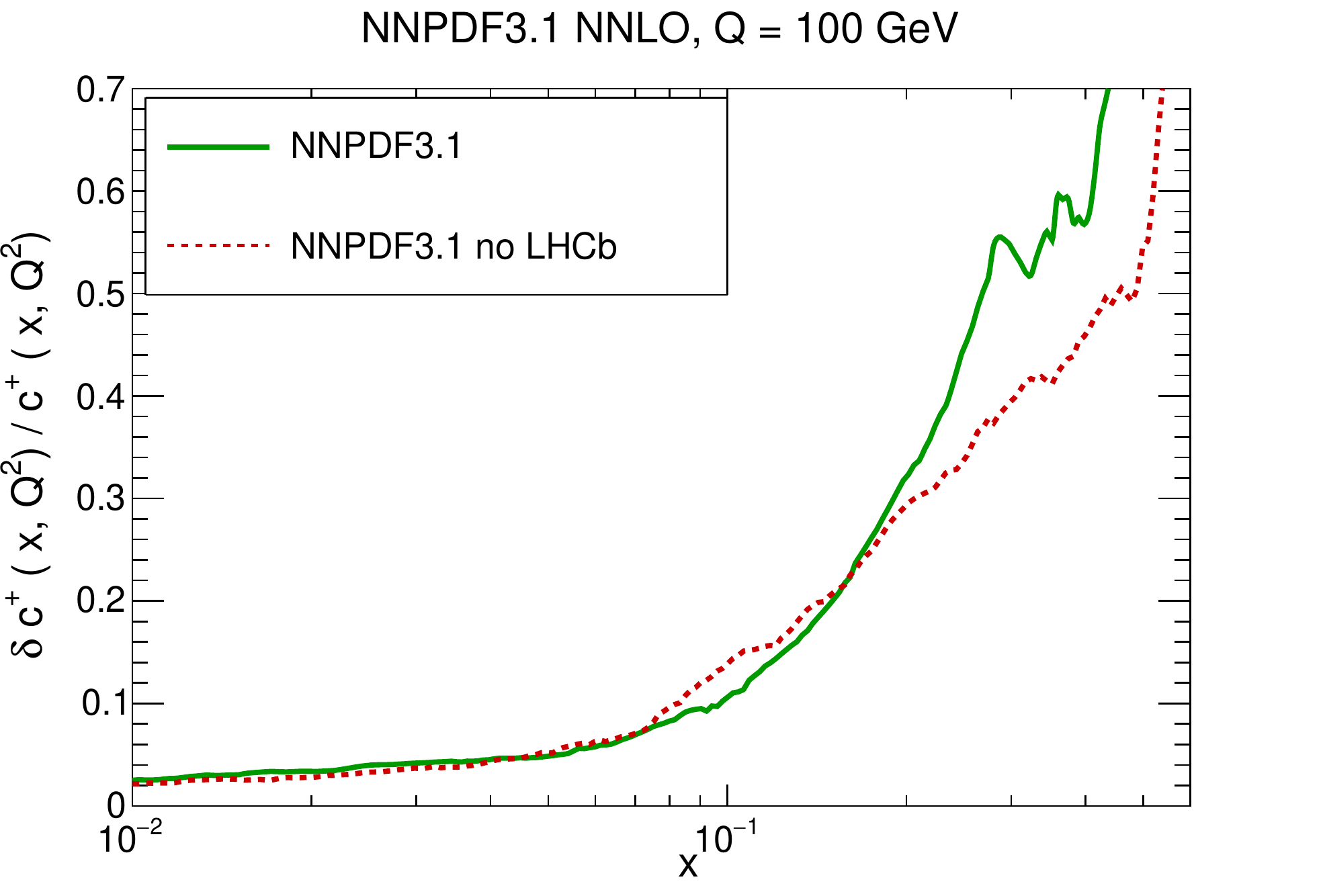}
  \caption{\small 
Same as Fig.~\ref{fig:31-nnlo-ZpT} but now excluding all LHCb data.
Results are presented, from top to bottow, for  the up, down and charm
PDFs. Both PDFs (left) and uncertainties (right) are shown.
\label{fig:lhcbdata-nnlo}}
\end{center}
\end{figure}
This is explicitly demonstrated 
for the up, down and charm PDFs in
Fig.~\ref{fig:lhcbdata-nnlo}, where the percentage PDF 
uncertainties are 
also shown. The LHCb data play a
significant role in the data-driven large-$x$ enhancement of light quark
PDFs discussed in Section~\ref{sec:disentangling} (see
Fig.~\ref{fig:31-nnlo-old-vs-new}), and is largely responsible for
the sizable impact of new data on charm for $x\gsim0.1$, where they
significantly reduce the uncertainty.
 Effects are more marked at medium
and large $x$, peaking at around $x\simeq 0.3$: in this region the PDF
uncertainty is also substantially reduced; the reduction in
uncertainty is especially marked for the down PDF.

\begin{figure}[t]
\begin{center}
  \includegraphics[scale=0.45]{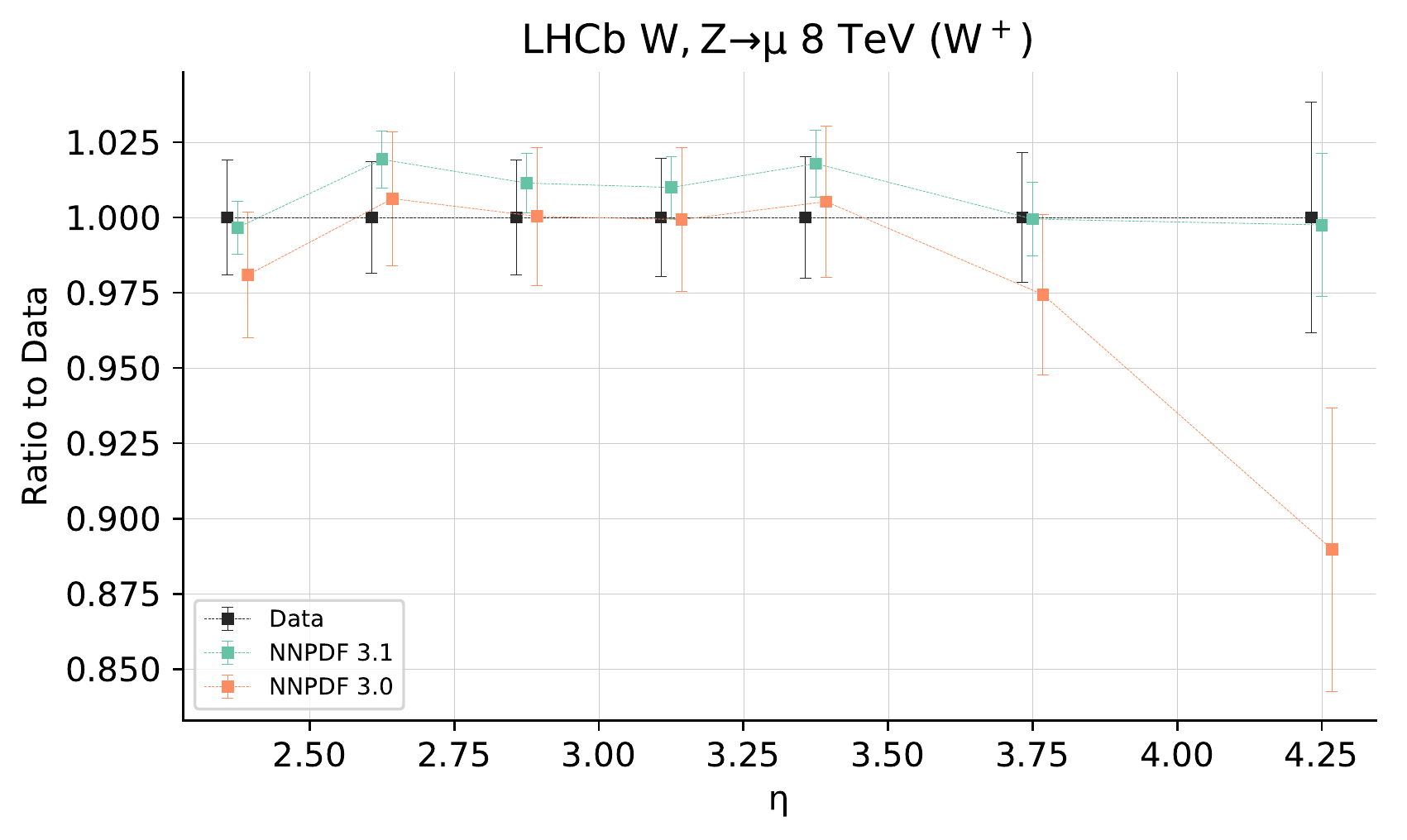}
  \includegraphics[scale=0.45]{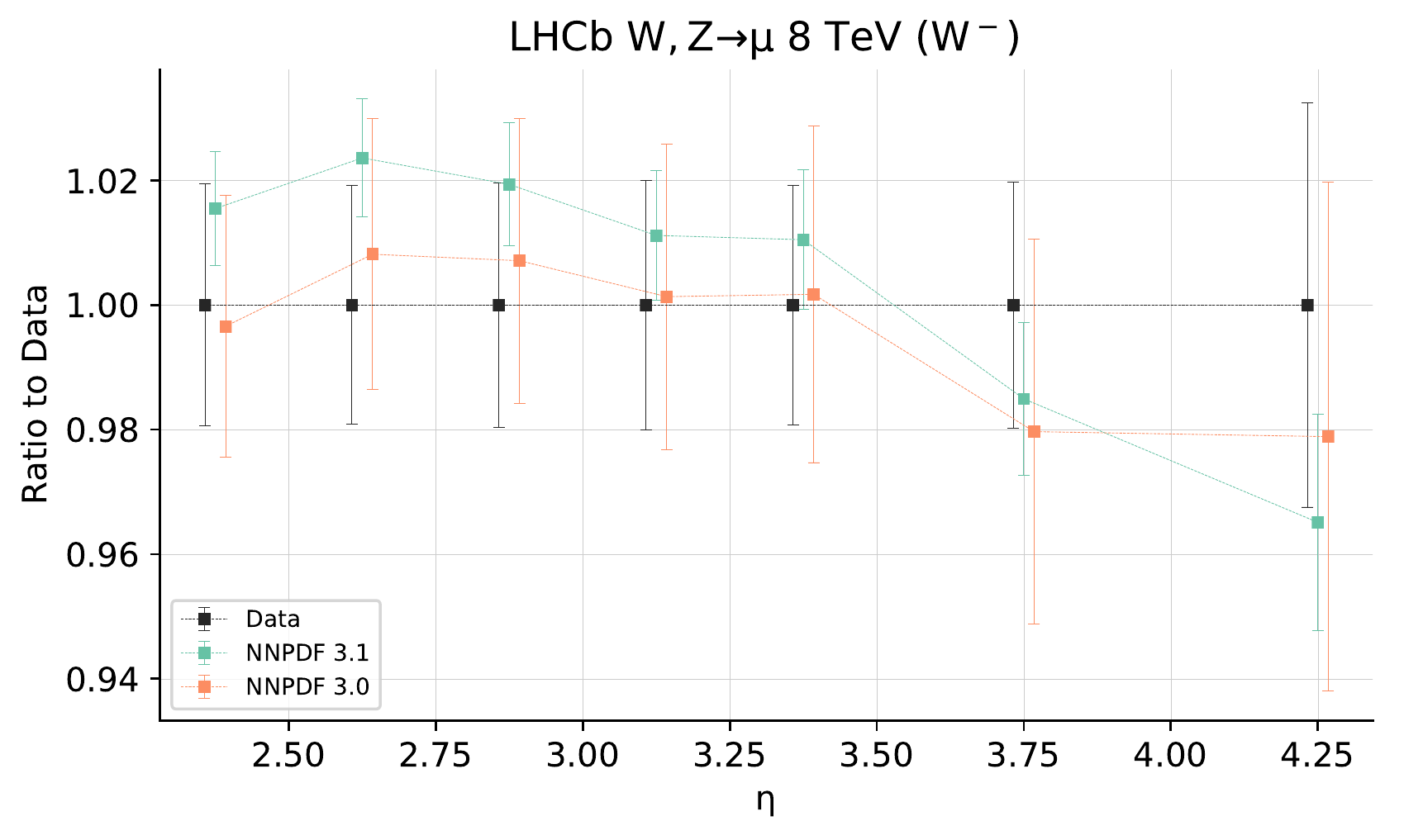}
  \caption{\small Comparison between 8~TeV LHCb muon $W^+$ (left) and
    $W^-$ (right) production data to NNLO predictions obtained using
    NNPDF3.1 and NNPDF3.0. The uncertainties shown are the diagonal
experimental uncertainty for the data, and the PDF uncertainty for
the best-fit prediction. 
\label{fig:lhcbdata-datatheory}}
\end{center}
\end{figure}

In order to see the impact of the LHCb data directly, in
Fig.~\ref{fig:lhcbdata-datatheory} we compare 
the 8~TeV LHCb muon $W^+$ and $W^-$  data
to predictions obtained using NNPDF3.0 and NNPDF3.1. The improvement is
clear, particularly for large rapidities. There is also a noticeable
reduction in PDF 
uncertainty on the prediction.

\begin{figure}[h]
\begin{center}
  \includegraphics[scale=1]{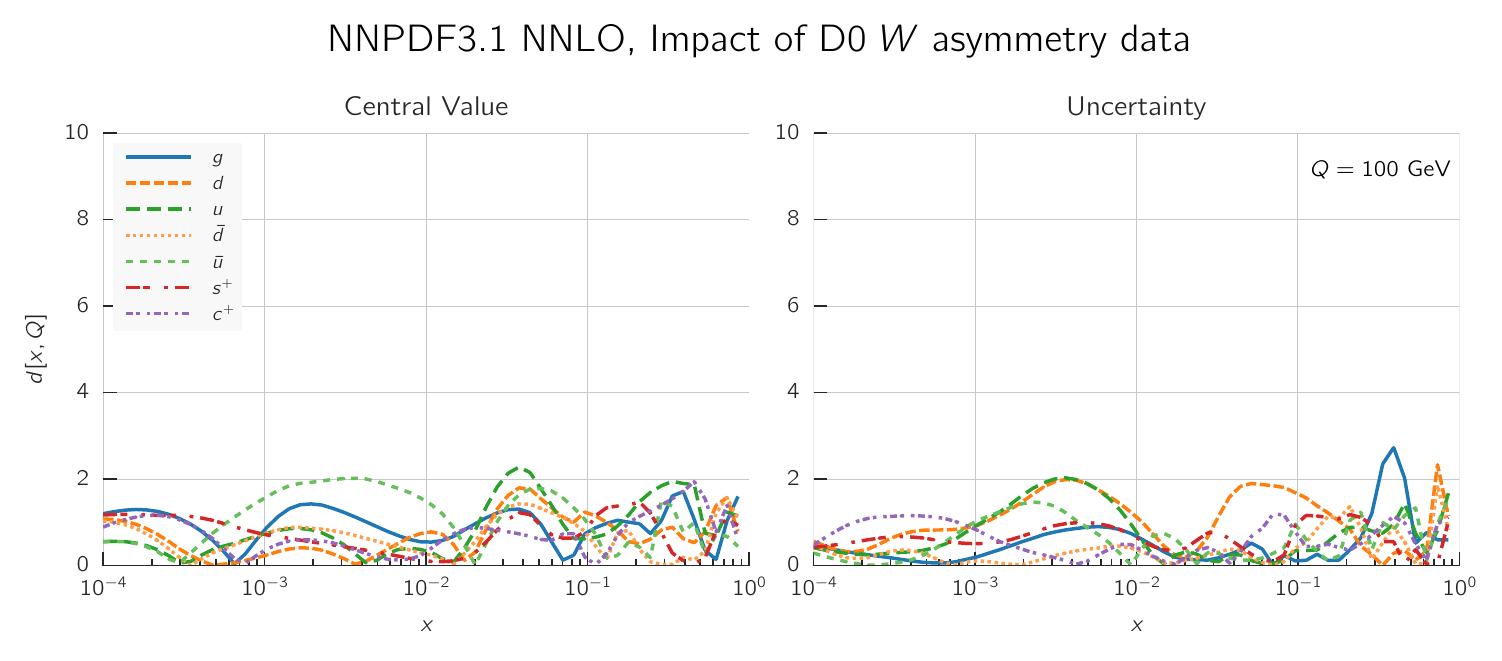}
  \caption{\small
Same as Fig.~\ref{fig:distances_noZpT} but now excluding 
 D0 $W$ asymmetry data.
    \label{fig:distances_nod0wasy}
  }
\end{center}
\end{figure}

\subsection{$W$ asymmetries from the Tevatron}
\label{sec:legtevdata}

\begin{figure}[h]
\begin{center}
  \includegraphics[scale=0.38]{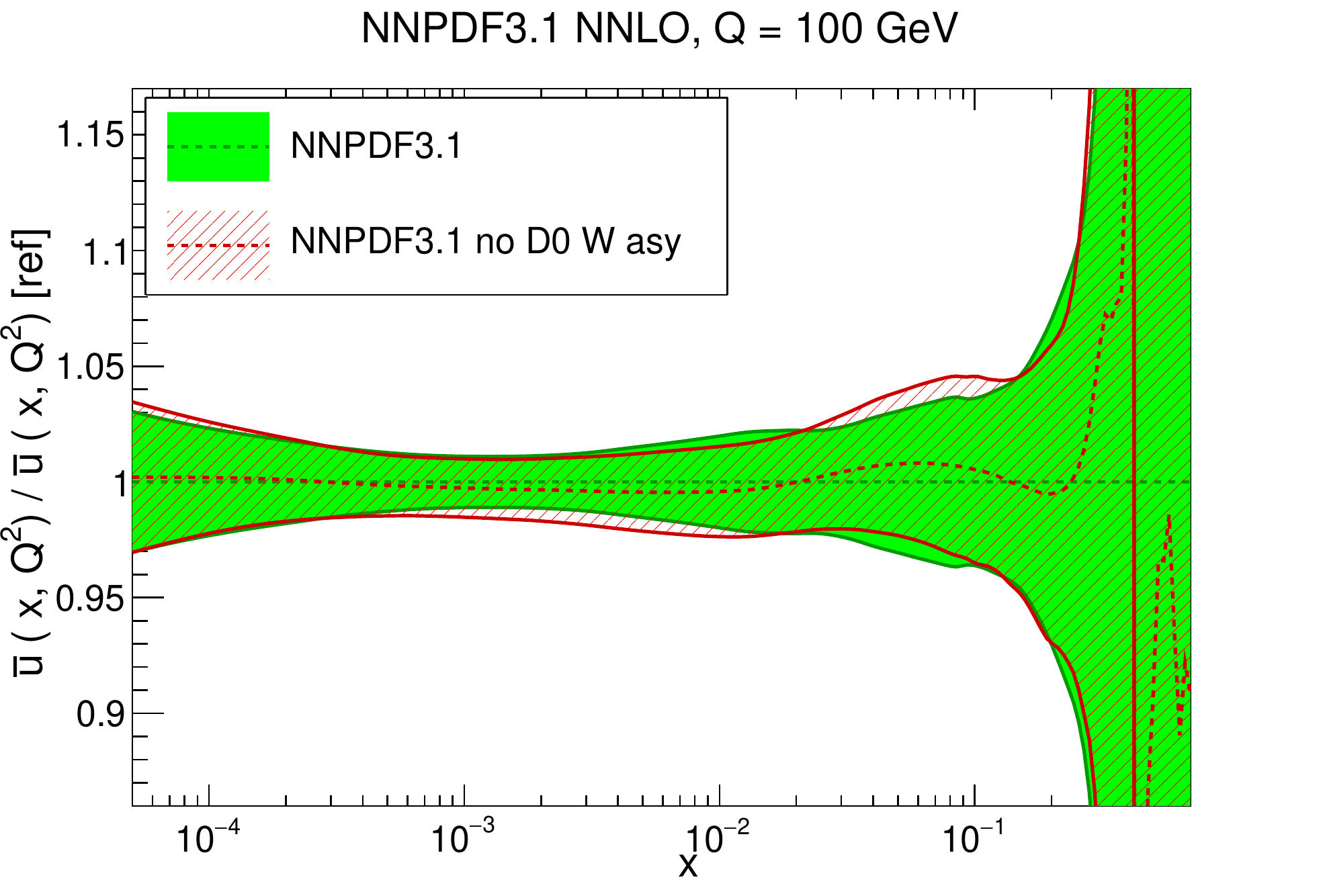}
  \includegraphics[scale=0.38]{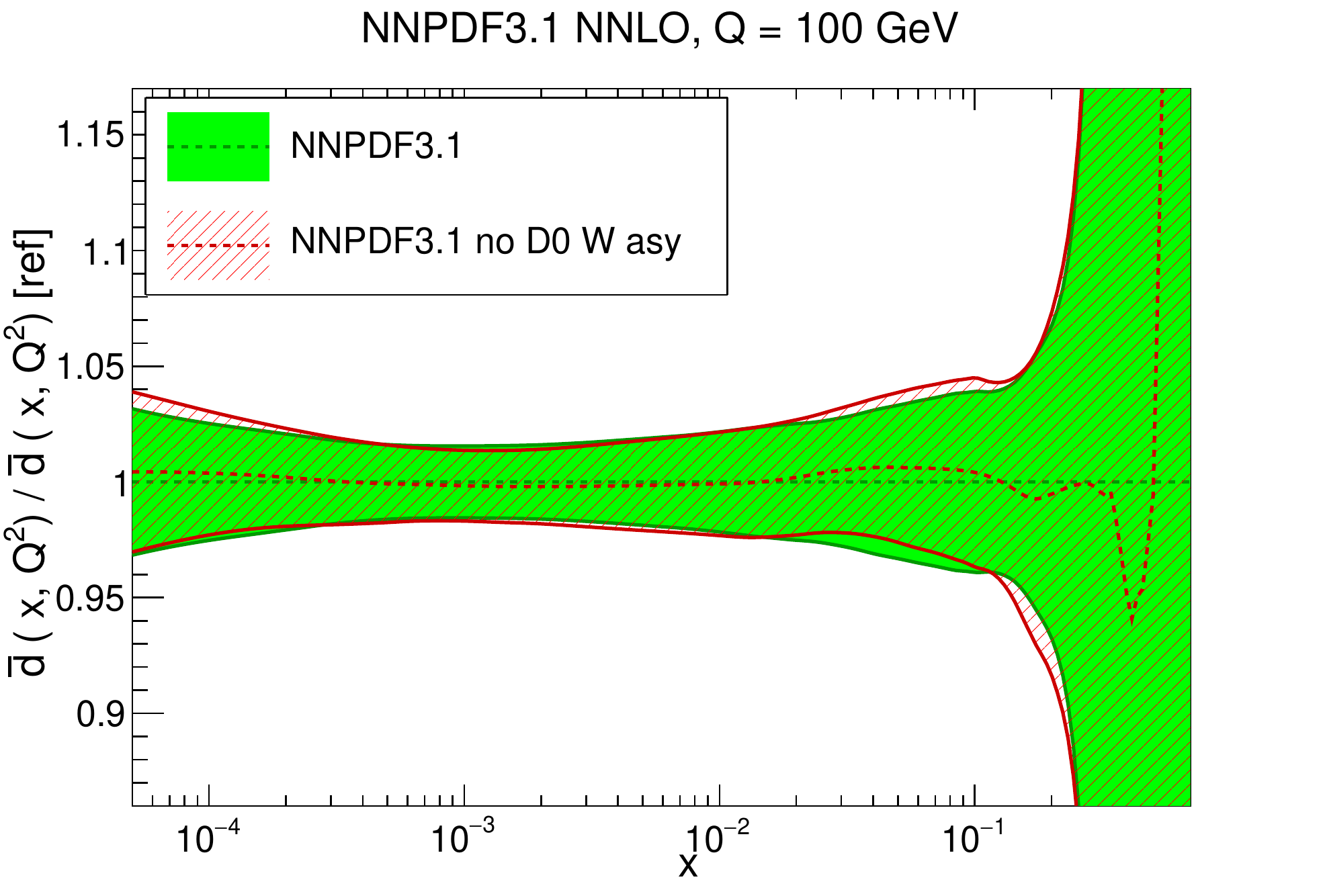}
  \caption{\small 
Same as Fig.~\ref{fig:31-nnlo-ZpT} but now excluding D0 $W$
asymmetries.
The antiup (left) and antidown (right) PDFs are shown.
    \label{fig:d0wasydata-nnlo}
  }
\end{center}
\end{figure}

$W$ production data from the Tevatron have for many years been the
leading 
source of 
information on quark flavor decomposition. The final legacy 
D0 $W$ asymmetry measurements in the electron and muon
channels are included in NNPDF3.1, superseding all previous data.
In Fig.~\ref{fig:distances_nod0wasy} we perform a distance comparison between the default NNPDF3.1 and PDFs determined
excluding this dataset. Distances are generally small, an observation confirmed by direct PDF comparison in  Fig.~\ref{fig:d0wasydata-nnlo}.
However, we have seen in Tab.~\ref{tab:chi2tab_31-nlo-nnlo-30} that
the fit quality for this dataset is rather better with NNPDF3.1 than
with the  previous NNPDF3.0. The moderate impact of this dataset 
is due to its excellent consistency with the abundant LHC data, which are now driving flavor separation. 
This
data thus provides further evidence for the reliability
 of the flavor separation in NNPDF3.1.

\subsection{The ATLAS $W,Z$ production data and strangeness}
\label{sec:atlaswz}

\begin{figure}[t]
\begin{center}
  \includegraphics[scale=1]{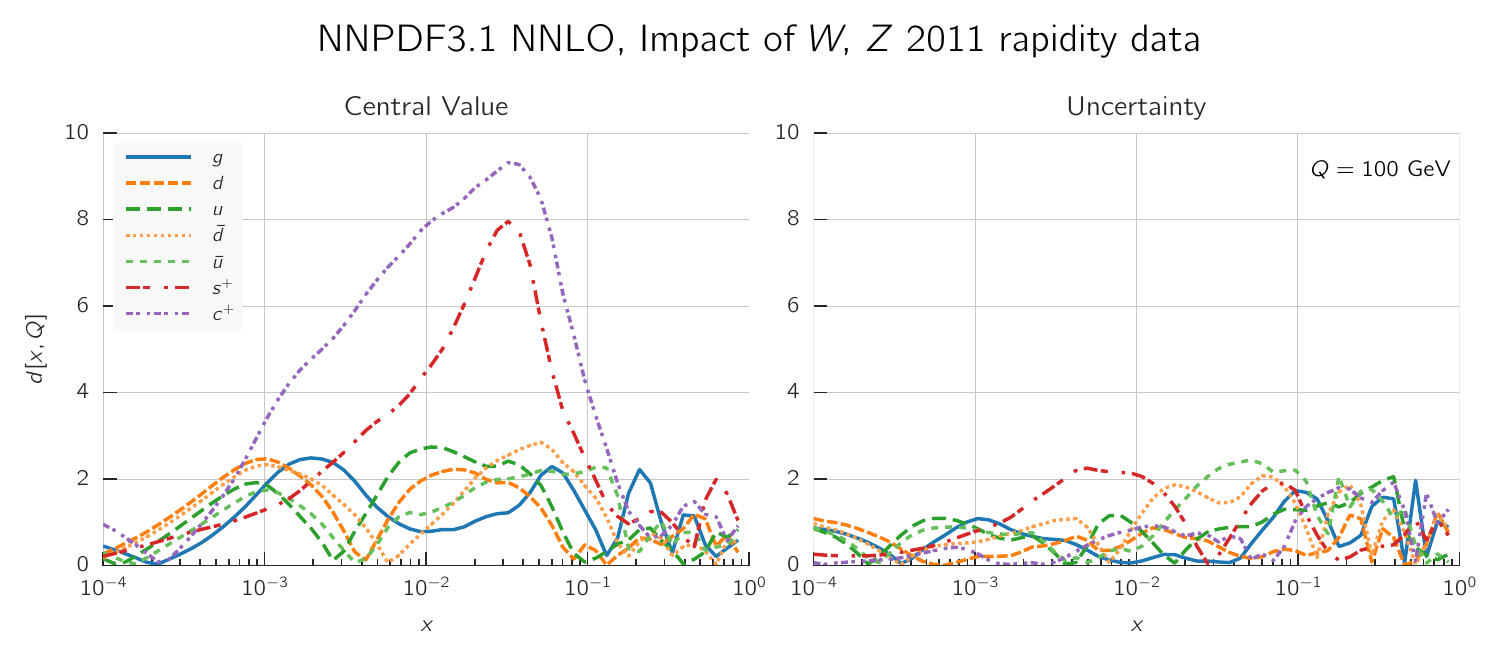}
  \caption{\small
Same as Fig.~\ref{fig:distances_noZpT} but now excluding 
2011 ATLAS $W,Z$ rapidity
    distributions.
    \label{fig:distances_atlaswz11rap}
  }
\end{center}
\end{figure}

ATLAS $W$ and $Z$ production data were already included in NNPDF3.0,
but recent measurements based on the 
2011 dataset~\cite{Aaboud:2016btc} have much smaller statistical
uncertainties. This dataset, like the previous ATLAS measurement,
 has been claimed to have a large impact on
strangeness. This is borne out by the plot,
Fig.~\ref{fig:distances_atlaswz11rap}, of the distance between the
default NNPDF3.1 and a version from which this dataset has been
excluded. Indeed the largest effect --- almost at the one sigma level on central
values --- is seen on the strange and charm PDFs, with a rather smaller
impact on all other PDFs.

\begin{figure}[t]
\begin{center}
  \includegraphics[scale=0.38]{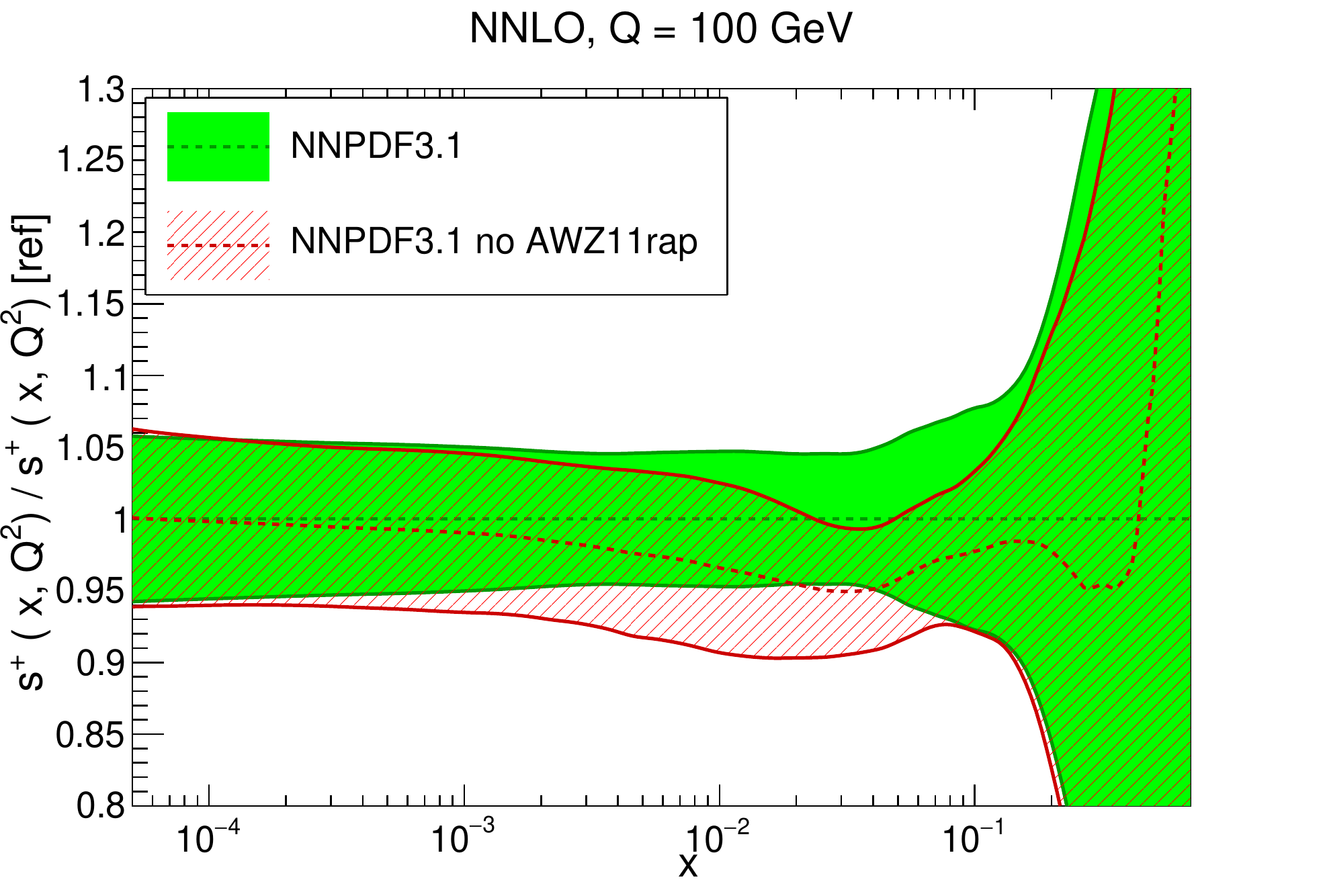}
  \includegraphics[scale=0.38]{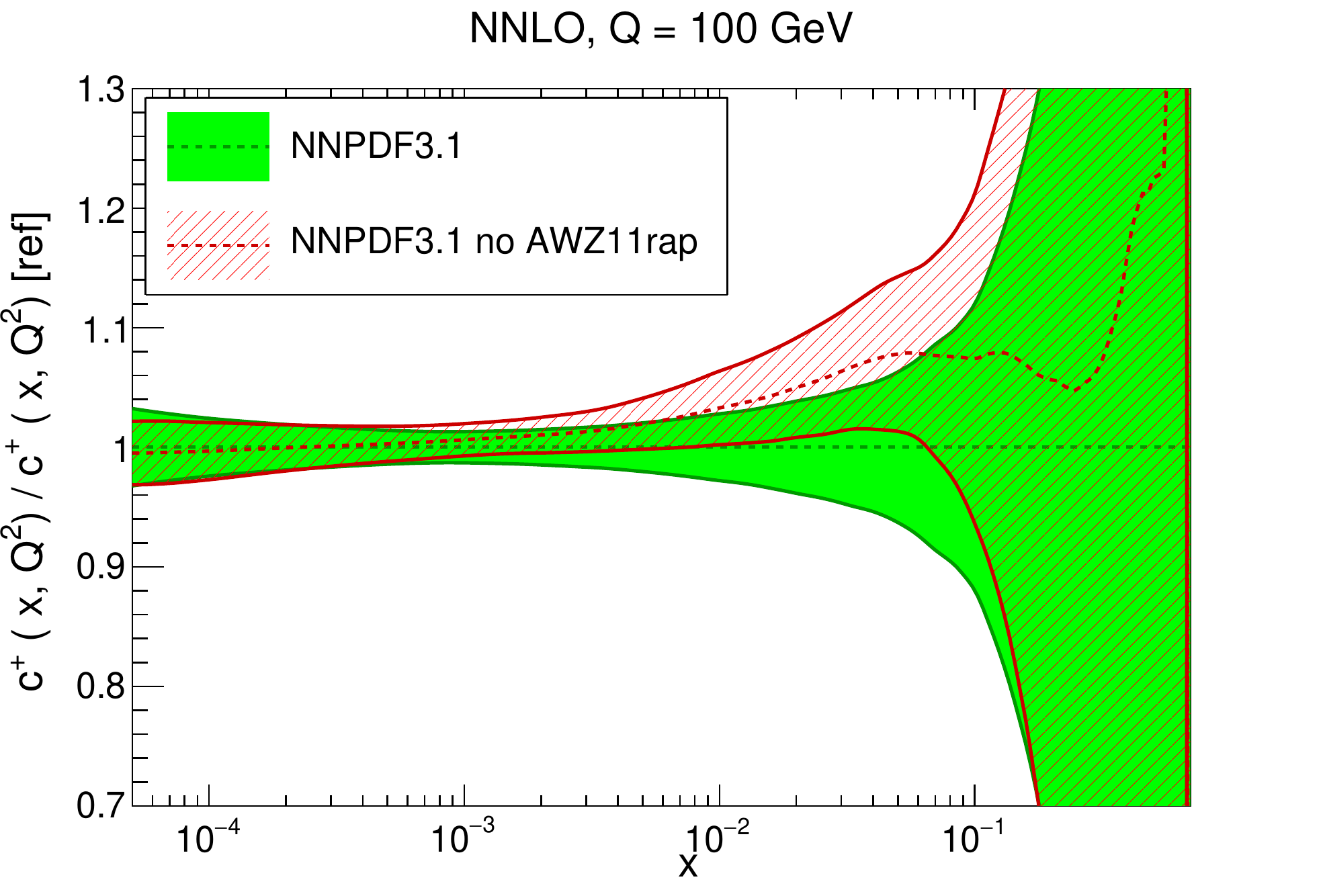}
  \caption{\small 
Same as Fig.~\ref{fig:31-nnlo-ZpT} but now 
excluding 
2011 ATLAS $W,Z$ rapidity
    distributions.
The total strange (left) and charm (right) PDFs are shown.
    \label{fig:pdfs-noAWZrap11}
  }
\end{center}
\end{figure}

A direct comparison of the strange and charm PDFs in
Fig.~\ref{fig:pdfs-noAWZrap11} shows that strangeness is significantly
enhanced in the medium/small $x$ region 
by the inclusion of the ATLAS data, while charm is
suppressed. As discussed in Section~\ref{sec:results-mc} and shown in
Fig.~\ref{fig:31-nnlo-fitted-vs-pch}, this suppression of charm cannot
be accommodated when charm is perturbatively generated, and therefore
parametrizing charm is important in order to be able to reconcile the ATLAS
data with the global dataset which in this $x$ range is severely constrained by
HERA data. The strange and charm content of the proton will be
discussed in detail in Sections~\ref{sec:strangeness} and \ref{sec:phenocharm} below.

\begin{figure}[t]
\begin{center}
  \includegraphics[scale=0.42]{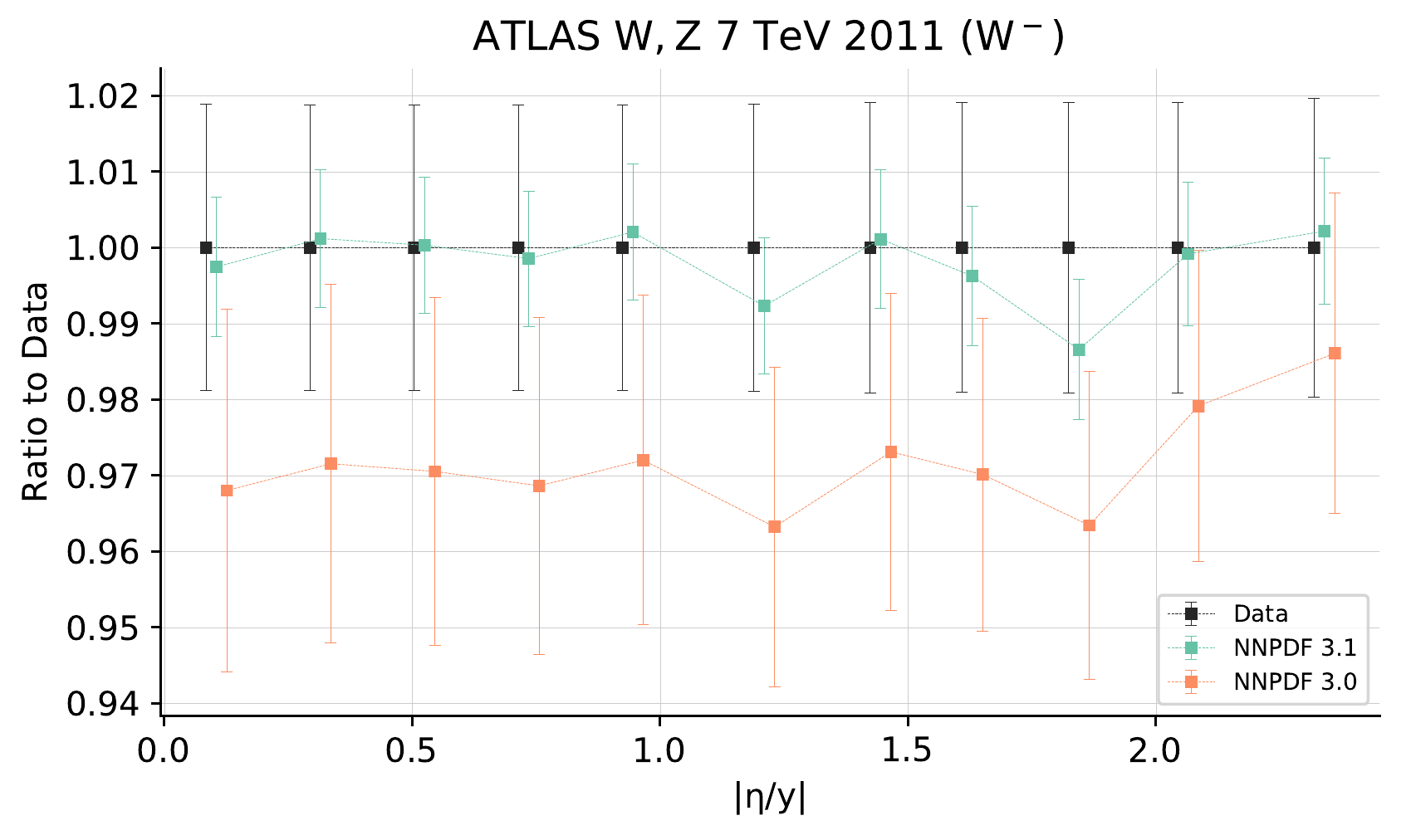}
  \includegraphics[scale=0.42]{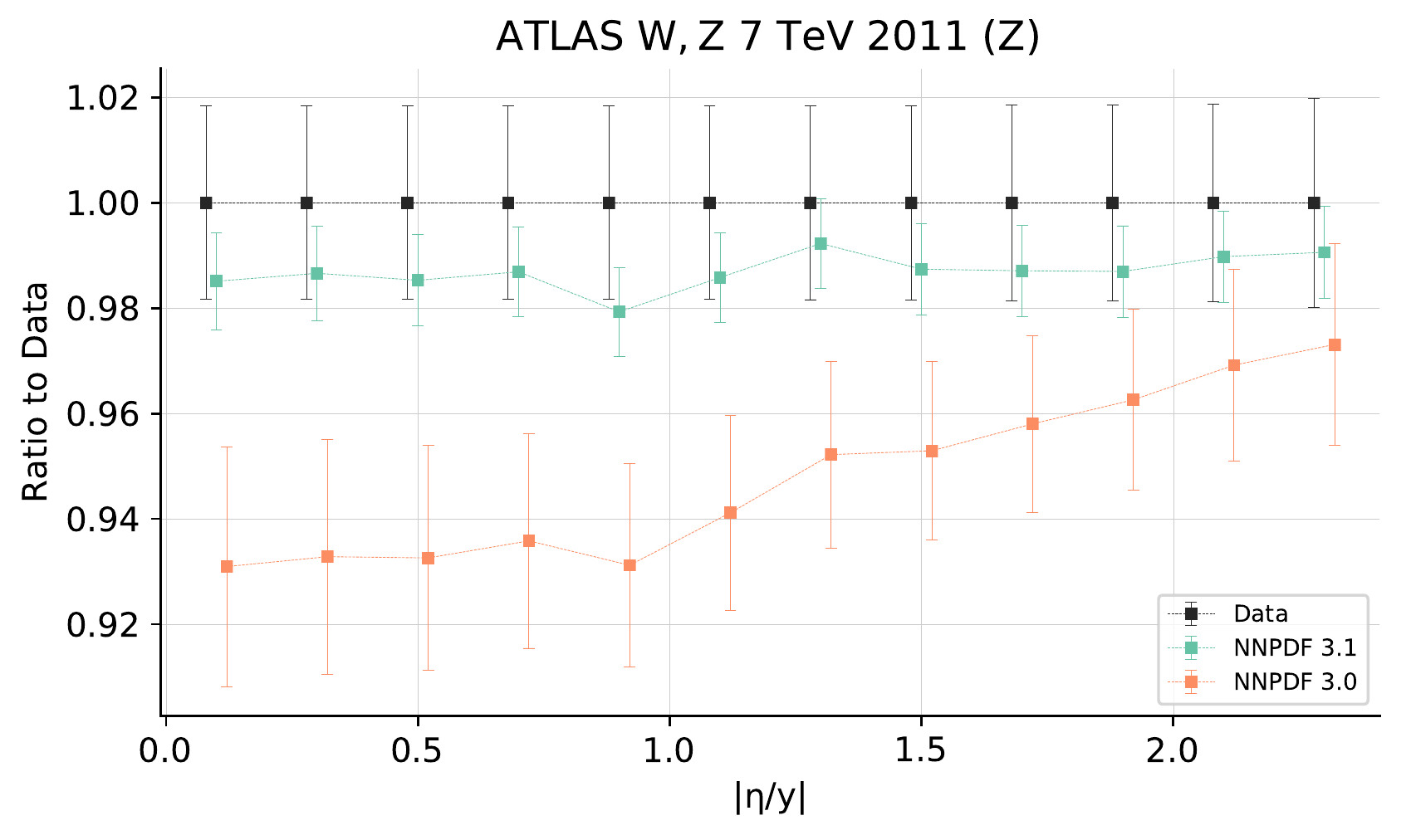}
  \caption{\small Comparison between the 2011 ATLAS 7~TeV  $W^-$
    (left) and $Z$ (right)
    data to NNLO predictions obtained using NNPDF3.1 and NNPDF3.0; $W$
    production data 
     are plotted versus the pseudorapidity of the forward lepton
     $\eta_l$, while $Z$ production data are plotted vs the dilepton
     rapidity $y_{ll}$. 
 \label{fig:atlaswz11-datatheory}}
\end{center}
\end{figure}

\begin{figure}[t]
\begin{center}
  \includegraphics[scale=0.42]{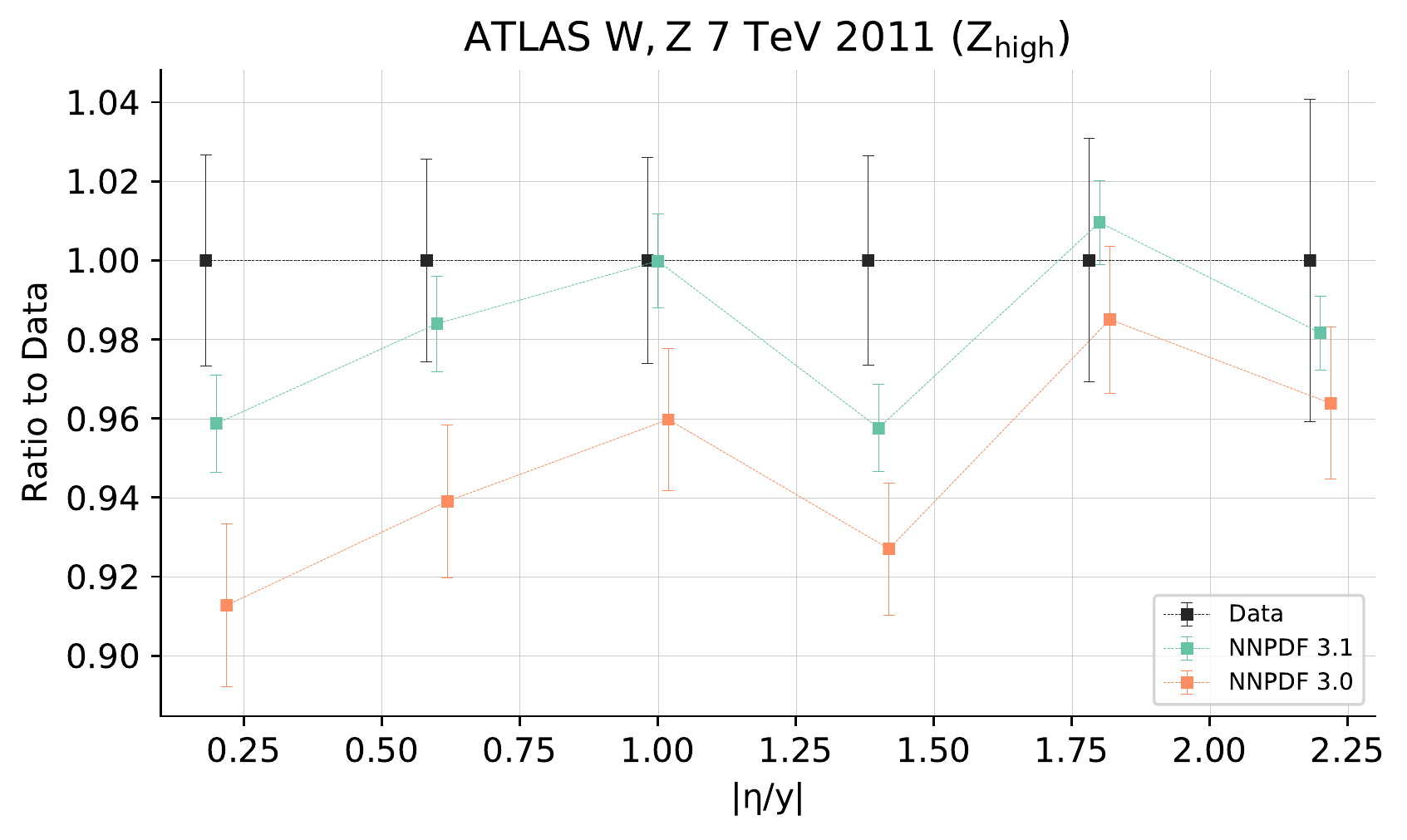}
  \includegraphics[scale=0.42]{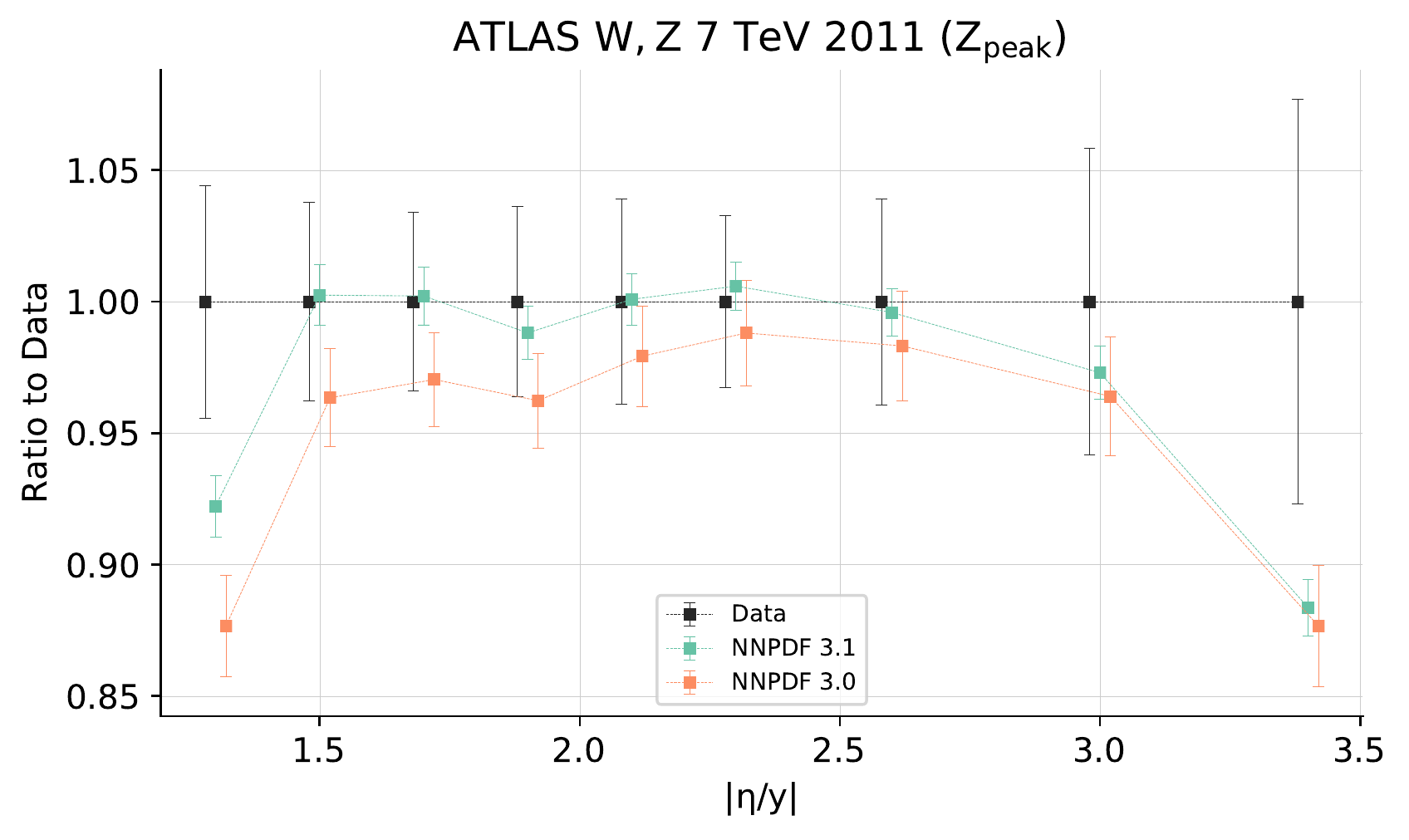}
  \caption{\small Same as Fig.~\ref{fig:atlaswz11-datatheory} but now
    for two of the four data bins which have not been included in the
    NNPDF3.1 determination: high-mass $Z$ production at central rapidity (left)
    and on-shell $Z$ production at forward rapidity (right).
 \label{fig:atlaswz11-datatheorypred}}
\end{center}
\end{figure}

NNPDF3.1 achieves a good description of the data, as illustrated in 
Fig.~\ref{fig:atlaswz11-datatheory} where the NNLO prediction obtained
using NNPDF3.0 is also shown. It is clear that the agreement is
greatly improved, as demonstrated in the $\chi^2$ values shown
in Tab.~\ref{tab:chi2tab_31-nlo-nnlo-30}. It is also interesting to 
note the significant reduction in PDF uncertainties. As mentioned in
Sect.~\ref{data:inclusiveWZ}, these data have been included only
partially in the NNPDF3.1 determination: specifically $Z$ production
data off peak  or at forward rapidity have not been included. We have
checked however that the description of these data in NNPDF3.1 is
equally good, and similarly improved in comparison to NNPDF3.0,
with $\chi^2/N{\rm dat}$ of order unity. A comparison of these data
for two of the four bins which have not been included to
the NNPDF3.1 and NNPDF3.0 is shown in Fig.~\ref{fig:atlaswz11-datatheorypred}.

\subsection{The CMS 8~TeV double-differential Drell-Yan distributions}
\label{sec:cms8tevimpact}

Like ATLAS, CMS has also published updated electroweak boson
production data. The NNPDF3.0 PDF determination already included 
double-differential (in rapidity and invariant mass) Drell-Yan data
at 7 TeV from the CMS 2011 dataset~\cite{Chatrchyan:2013tia}.
An updated version of the same measurement at 8~TeV based on 2012
data was presented in Ref.~\cite{CMS:2014jea}, including both the absolute 
cross-sections and the ratio of 8~TeV and 7~TeV measurements.

\begin{figure}[t]
\begin{center}
  \includegraphics[scale=1]{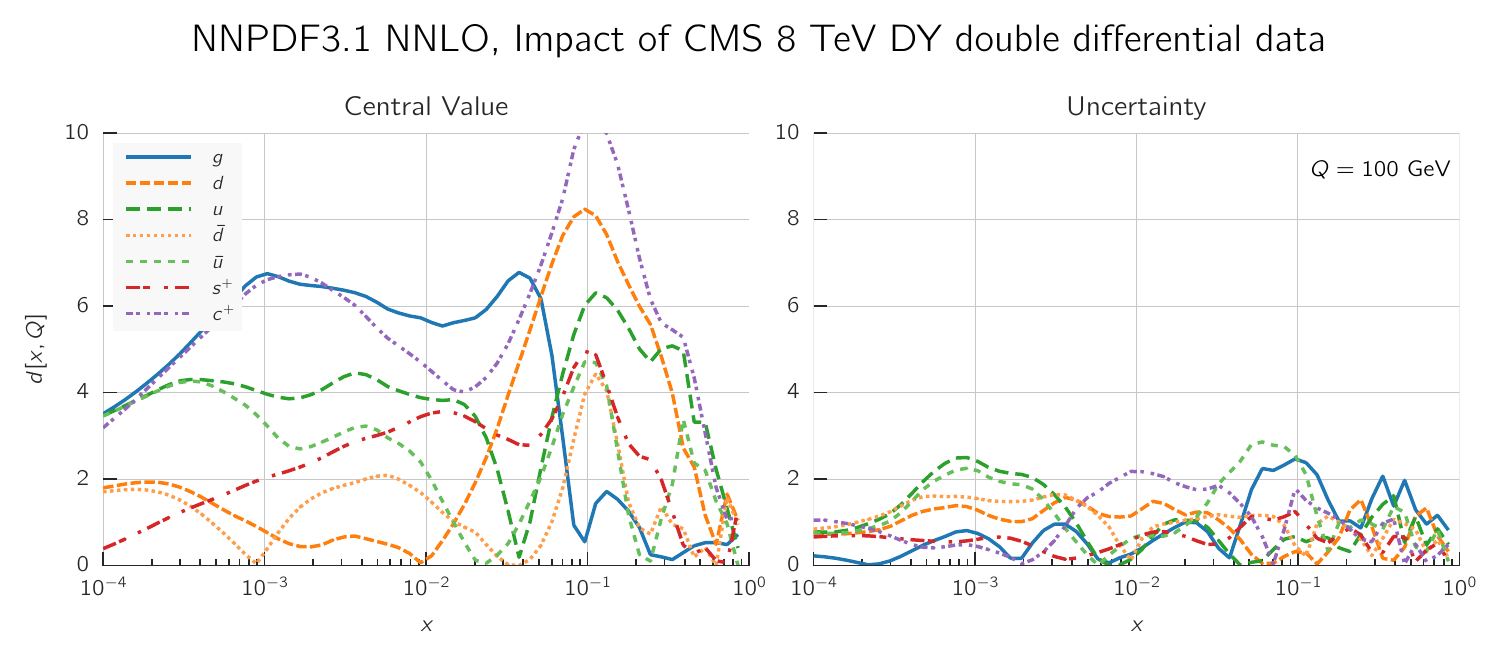}
  \caption{\small
Same as Fig.~\ref{fig:distances_31_vs_30}  but now
    comparing the default NNPDF3.1 to a version of it  with  
the 8~TeV CMS double-differential Drell-Yan  data also included.
    \label{fig:distances_cmsdy2d12}
  }
\end{center}
\end{figure}

This data has very small uncorrelated systematic
uncertainties. Unfortunately, only the full covariance
matrix, with no breakdown of individual correlated systematics, has been made 
available. The combination of these two facts makes it impossible to include 
this experiment in the NNPDF3.1 dataset, as we now explain.
In Fig.~\ref{fig:distances_cmsdy2d12} we show
the distances between the NNPDF3.1 NNLO PDF set and a modified version of it
where this dataset has been included.
While the impact on uncertainties is moderate, clearly this dataset has a significant 
impact at the level of central-values on all PDFs for
almost all $x$ values, with a particularly important impact on 
the medium/small $x$ gluon. 
This is somewhat surprising, given that Drell-Yan
production only provides an indirect handle on the gluon PDF.

\begin{figure}[t]
\begin{center}
  \includegraphics[scale=0.38]{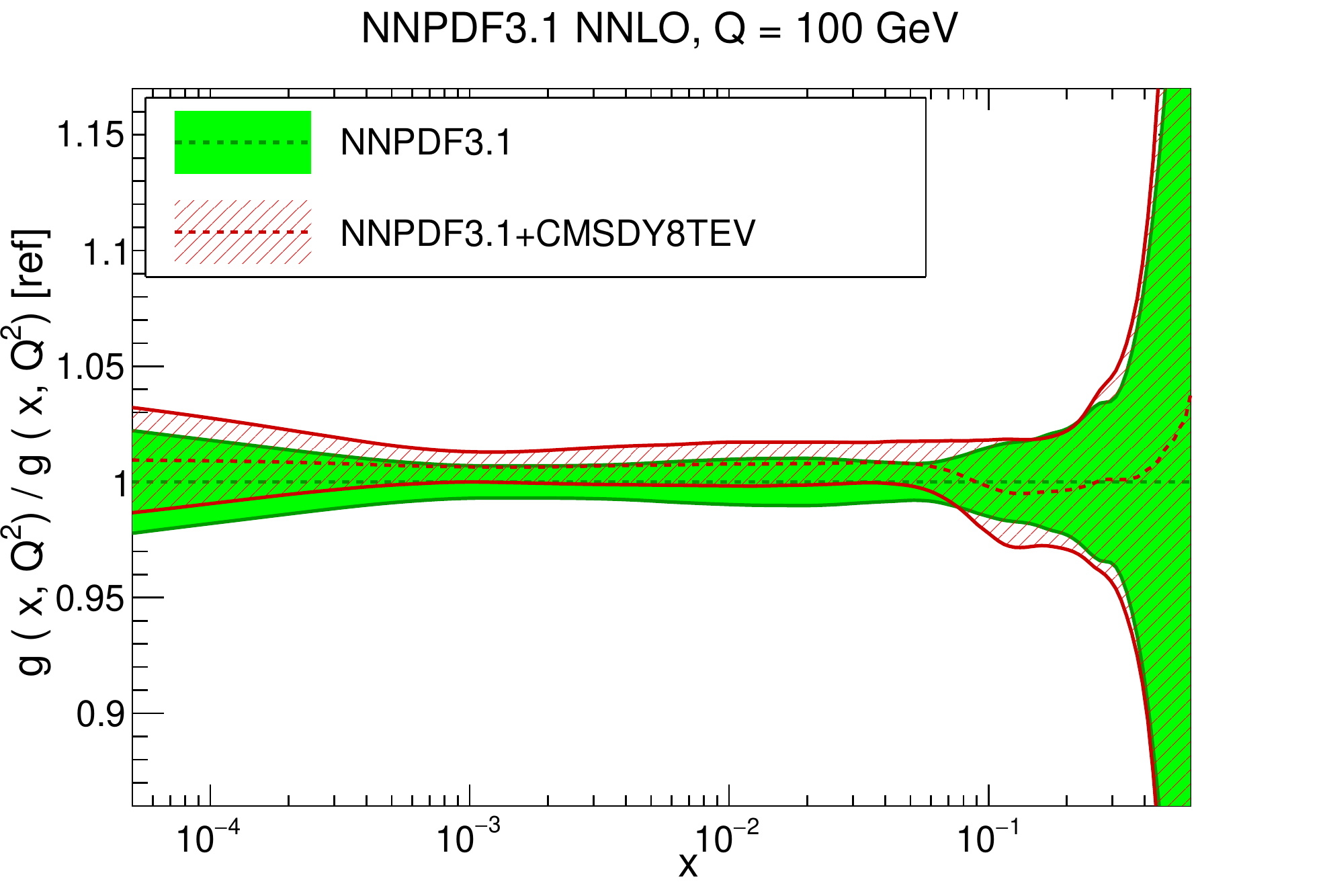}
  \includegraphics[scale=0.38]{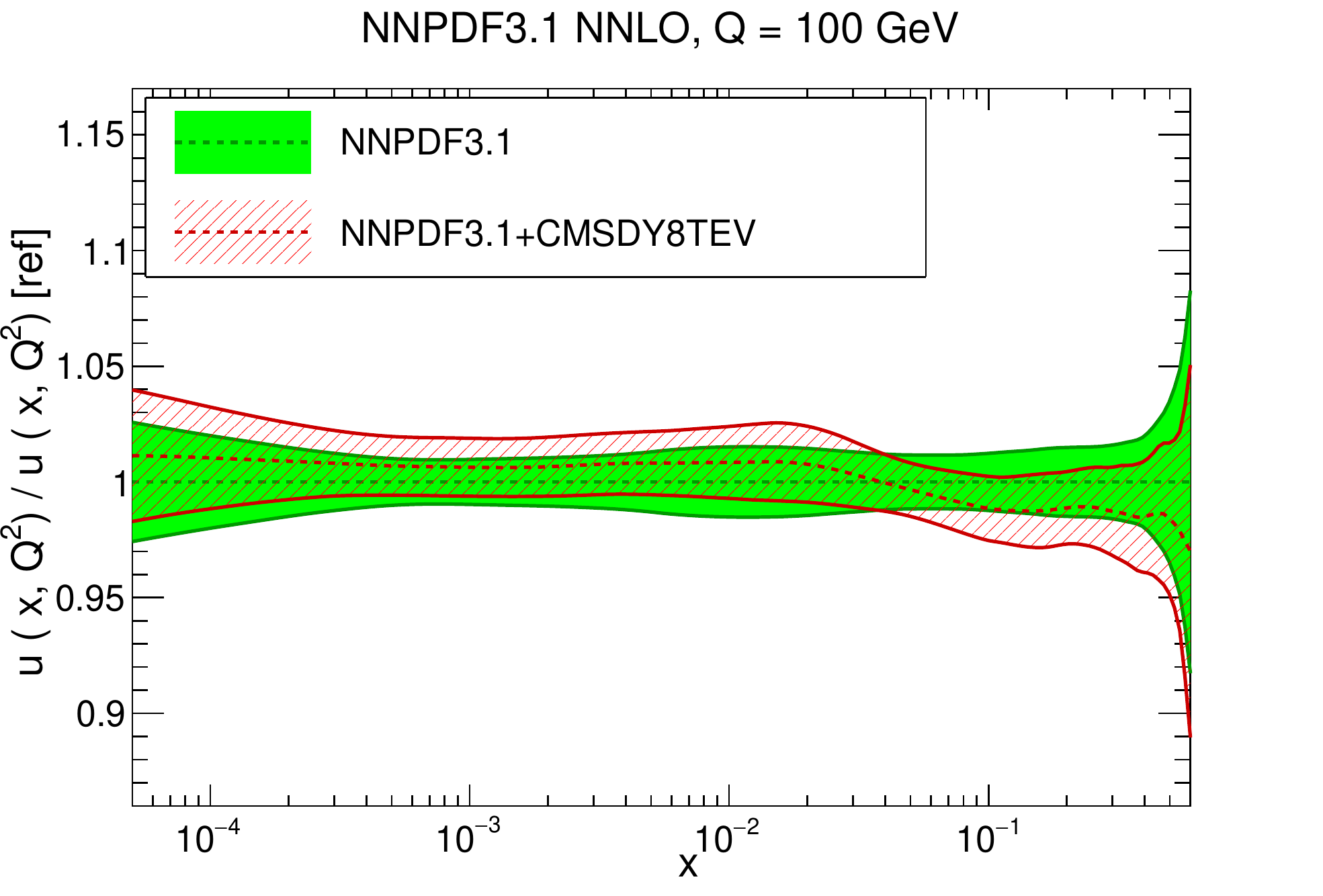}
  \includegraphics[scale=0.38]{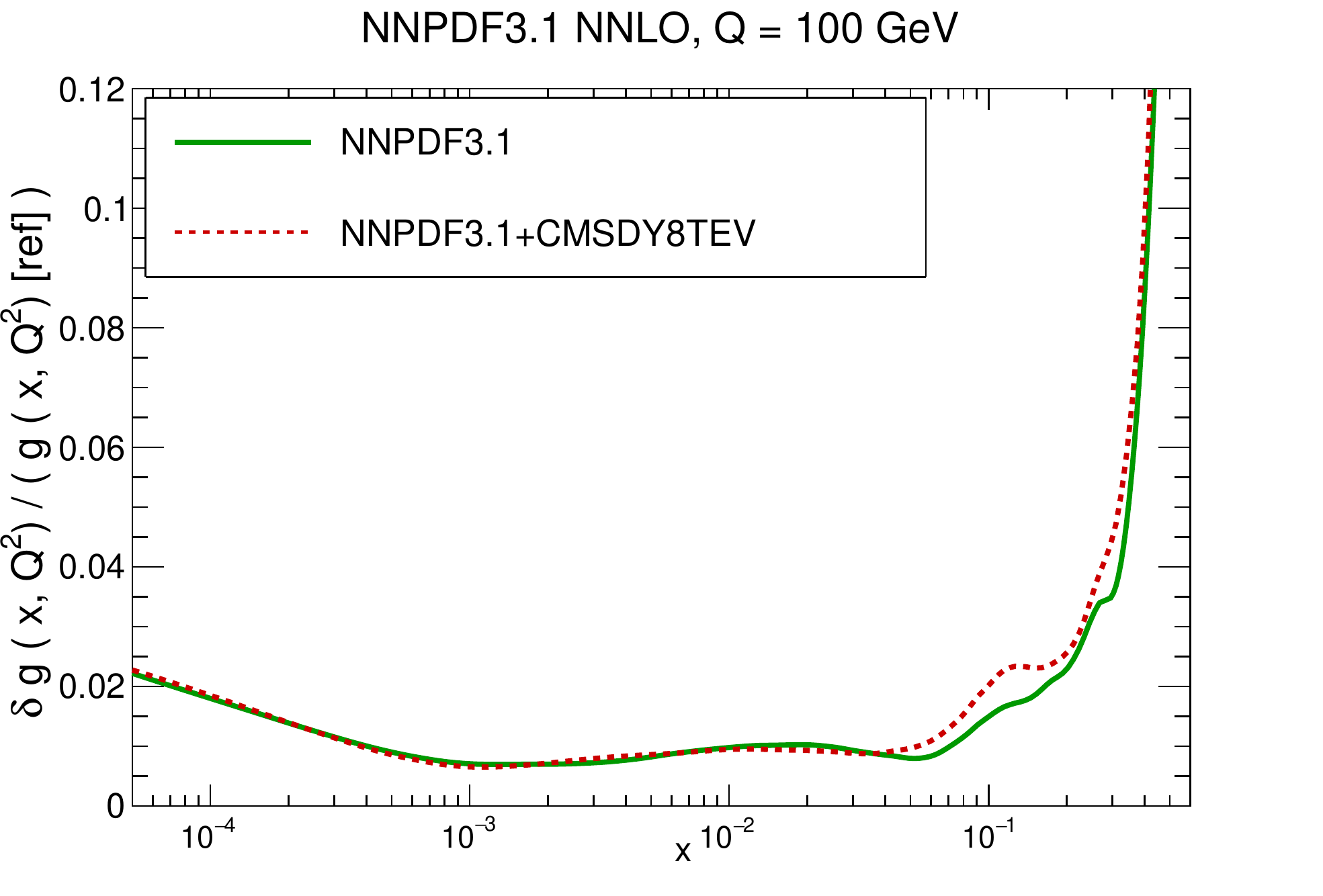}
  \includegraphics[scale=0.38]{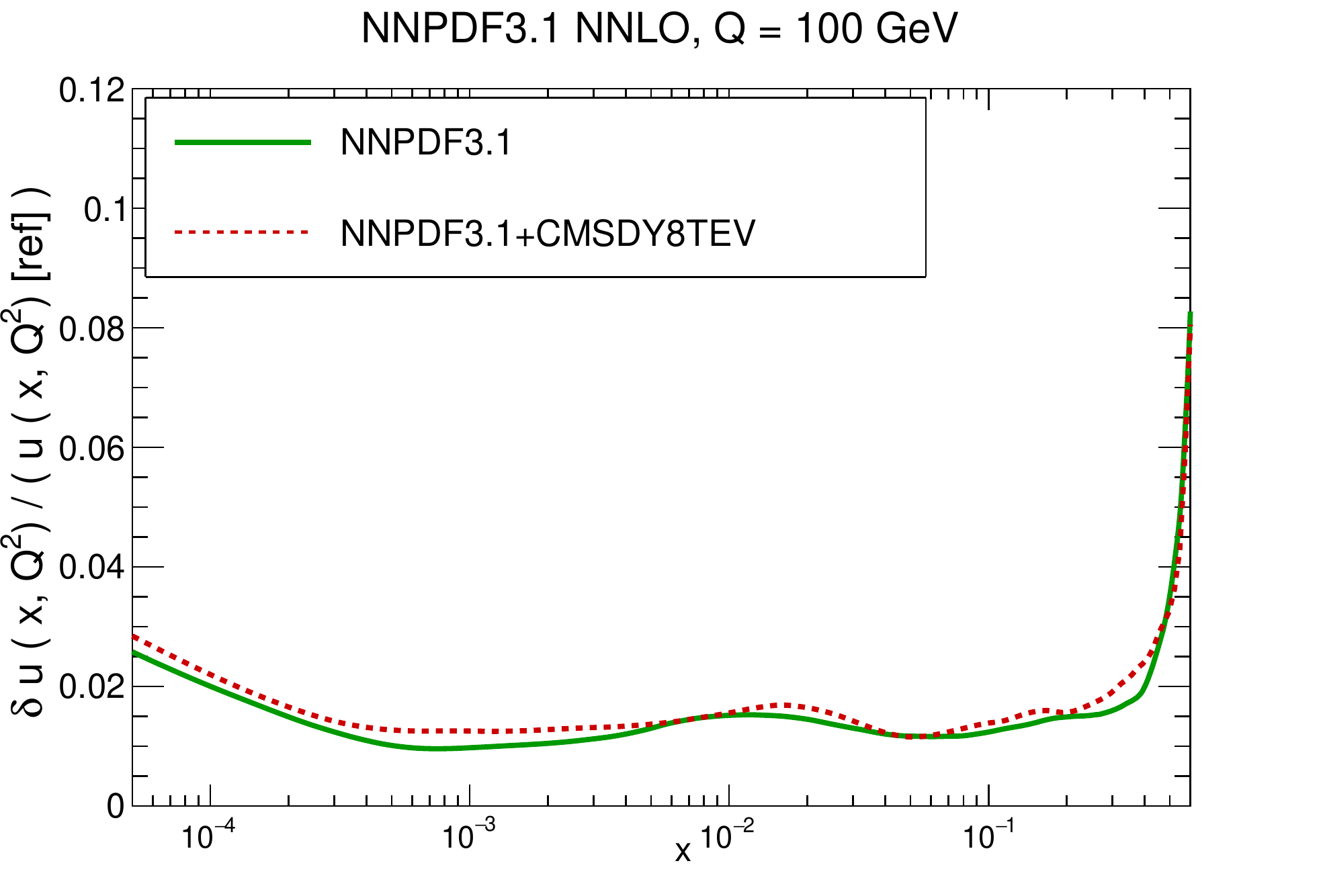}
  \caption{\small 
Same as Fig.~\ref{fig:31-nnlo-vs30}  (top) but now
    comparing the default NNPDF3.1 to a version of it  with  
the 8~TeV CMS double-differential Drell-Yan  data also included. The
corresponding 
percentage uncertainties are also shown (bottom). Results are
shown for the gluon (left) and up quark (right).
    \label{fig:pdfs-cmsdy2d12}
  }
\end{center}
\end{figure}

A direct comparison of PDFs and their uncertainties 
in Fig.~\ref{fig:pdfs-cmsdy2d12} shows that these data induce an
upwards shift by up to one sigma of the gluon for $x\lsim 0.1$, and
a downward shift of the light quark PDFs for $x\gsim 0.1$, 
by a comparable amount. This, however, is not accompanied by a
reduction of PDF uncertainties, which increase a little, as
also shown in Fig.~\ref{fig:pdfs-cmsdy2d12}. 

Furthermore, while the fit quality of the 8~TeV CMS
double-differential Drell-Yan data remains poor after their inclusion in the fit,
with a value of $\chi^2/N_{\rm dat}=2.88$, there is a
certain deterioration in fit quality of all other experiments. Indeed,
the total $\chi^2$ to all the other data
deteriorates by $\Delta \chi^2=11.5$.
A more detailed inspection shows
that the most marked deterioration is seen in the HERA combined
inclusive DIS data, with $\Delta
\chi^2=19.7$. 
This means that there is tension between the CMS data and
the rest of the global dataset, and more specifically tension with the HERA data,
which are most sensitive to the small $x$ gluon.

We must conclude that this experiment appears to be inconsistent with
the global dataset, and particularly with the data 
with which we have the least reasons to doubt, namely the combined HERA data
and their determination of the gluon. In the absence of more detailed
information on the covariance matrix it is not possible to further
investigate the matter, and the 8~TeV CMS
double-differential Drell-Yan data have consequently not been included in the global dataset.

\subsection{The EMC $F_2^c$ data and intrinsic charm.}
\label{sec:emc}

The advantages of introducing an independently parametrized charm PDF
were advocated in Ref.~\cite{Ball:2016neh}, where a first global PDF
determination including charm was presented, based on the NNPDF3.0 methodology and
dataset. The default NNPDF3.0 dataset was supplemented by charm
deep-inelastic structure function data from the EMC
collaboration~\cite{Aubert:1982tt}. This dataset is quite old, but it
remains the only measurement of the charm structure function in the
large $x$ 
region. With the wider NNPDF3.1 dataset the EMC dataset is no longer
quite so indispensable, specifically in view of phenomenology at the
LHC:
it has thus been omitted from the default NNPDF3.1 
determination as doubts have been raised about its reliability. 

\begin{figure}[t]
\begin{center}
  \includegraphics[scale=1]{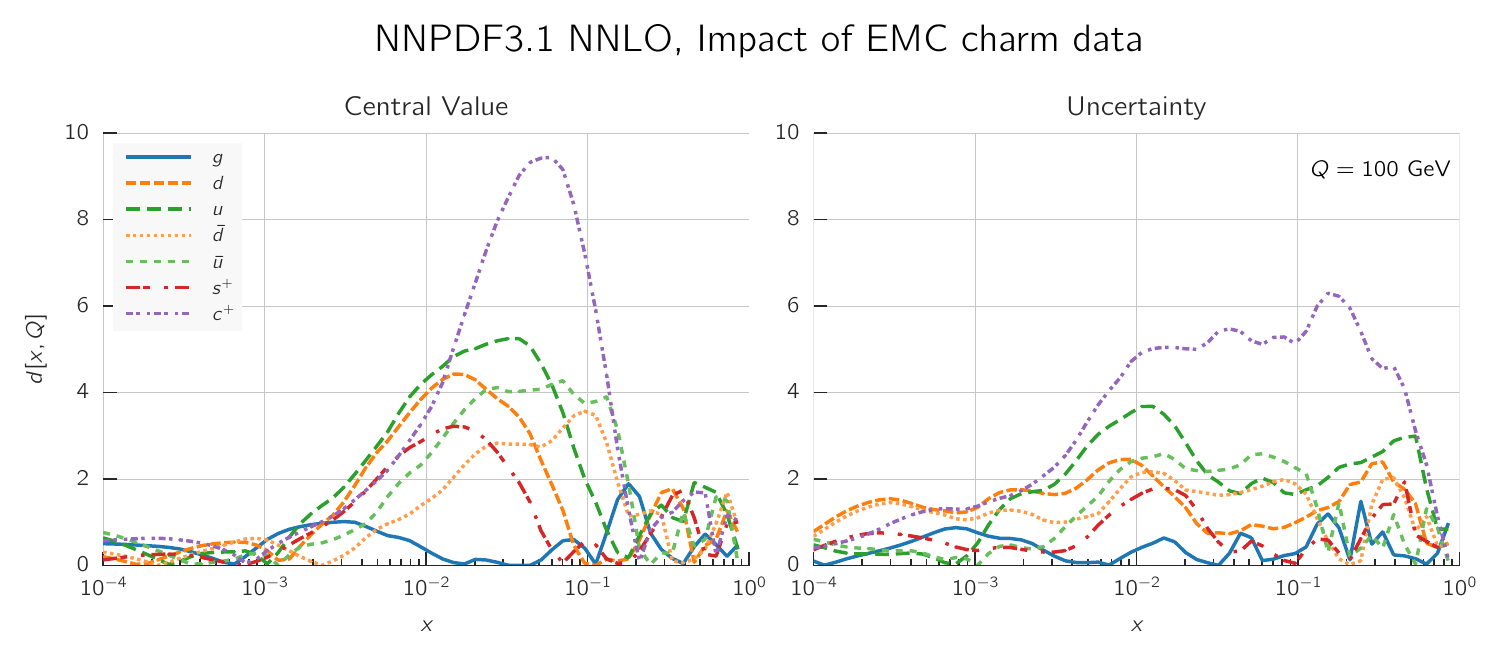}
  \caption{\small Same as Fig.~\ref{fig:distances_effect_Zpt7},  but now
    comparing the default NNPDF3.1 to a version of it with  
the EMC $F_2^c$ dataset also included.
    \label{fig:distances_emc}
  }
\end{center}
\end{figure}

\begin{figure}[t]
\begin{center}
  \includegraphics[scale=0.38]{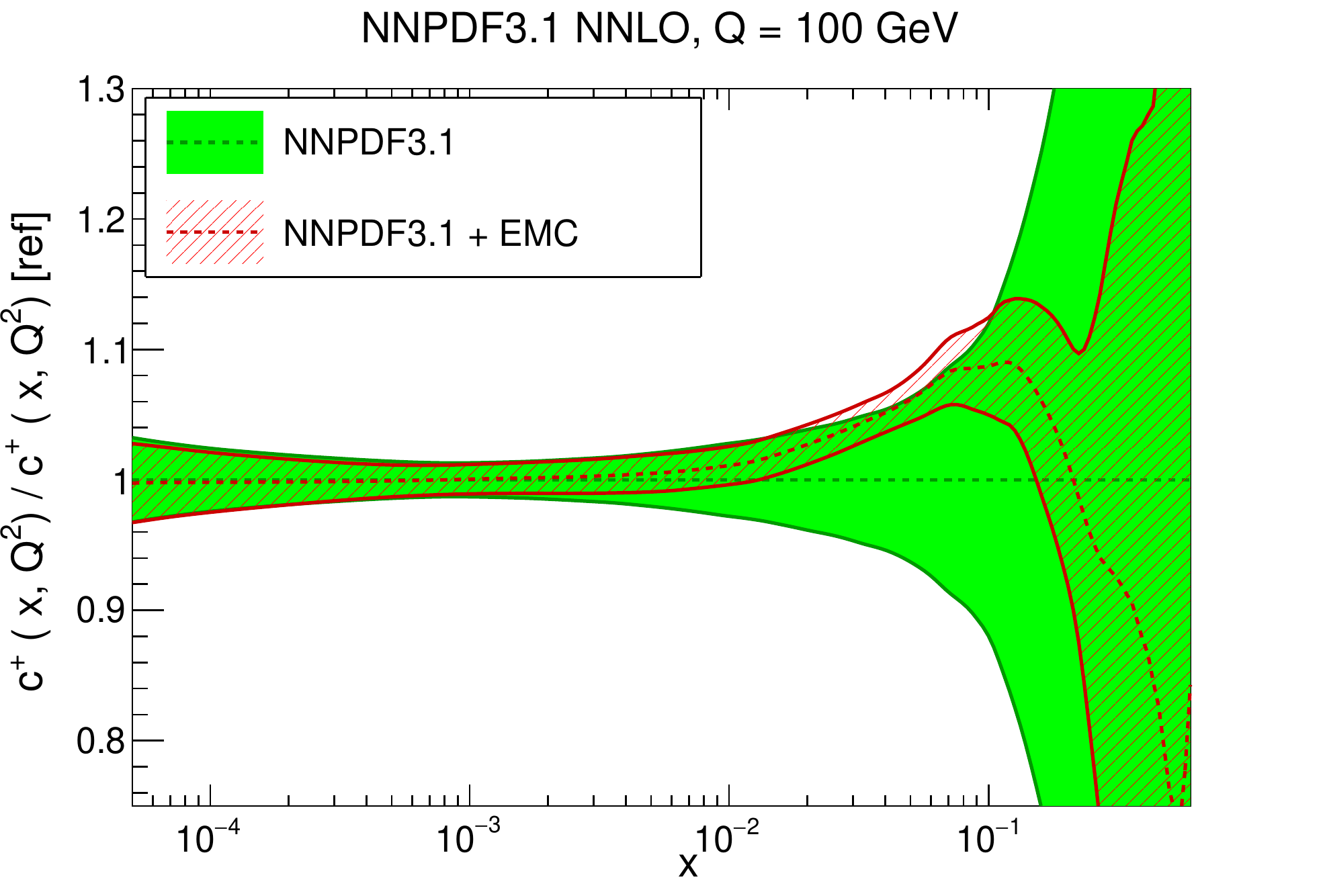}
  \includegraphics[scale=0.38]{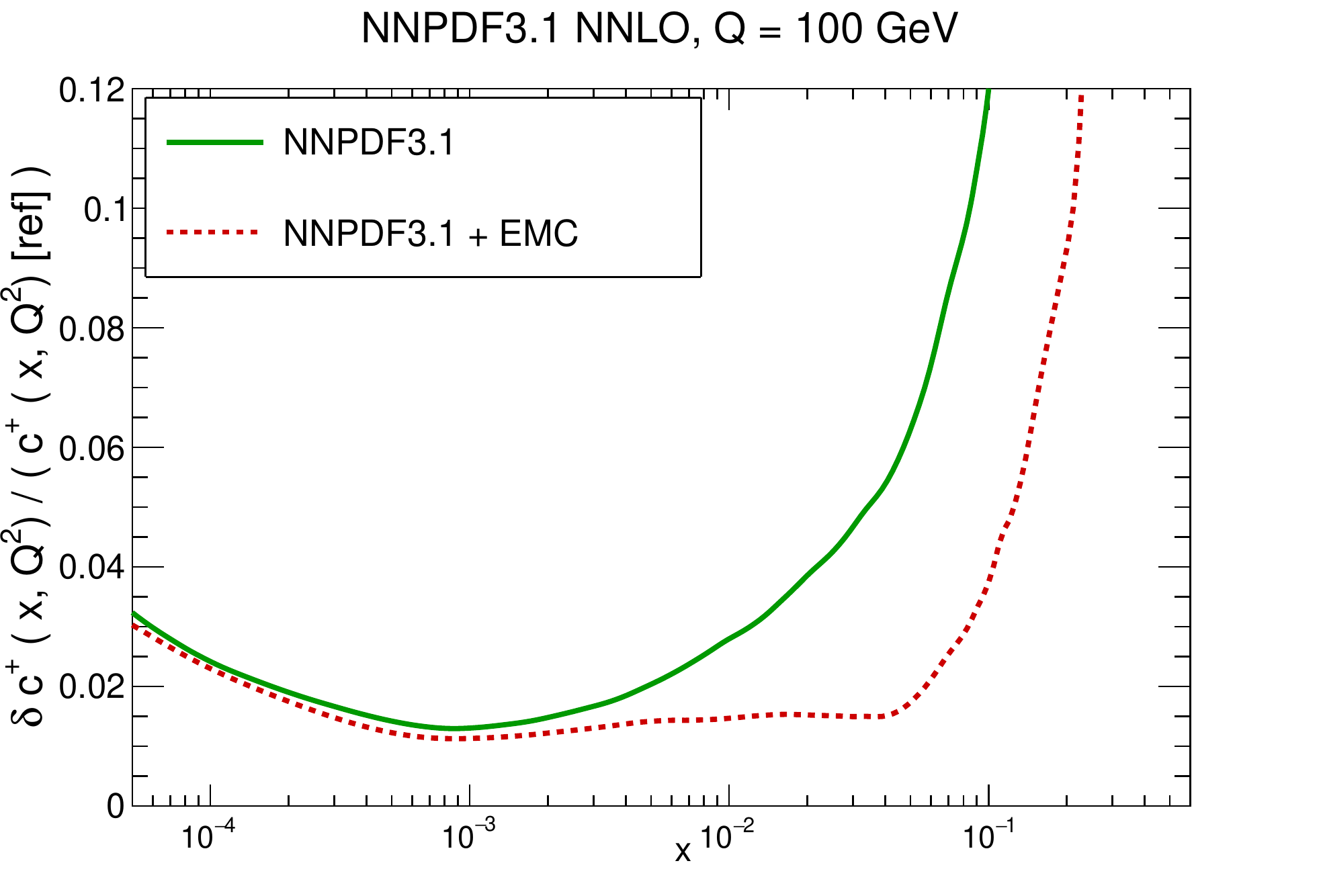}
  \includegraphics[scale=0.38]{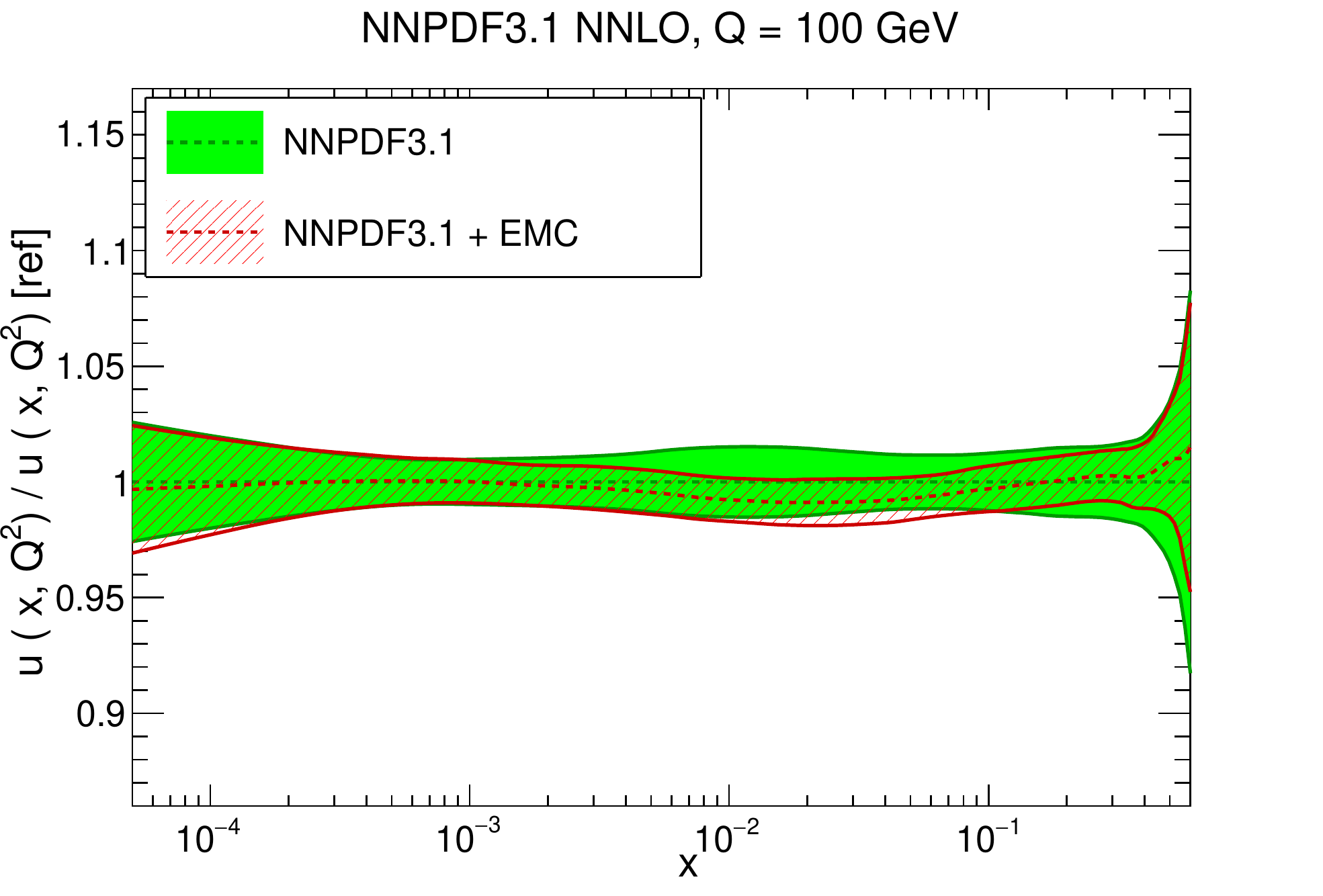}
  \includegraphics[scale=0.38]{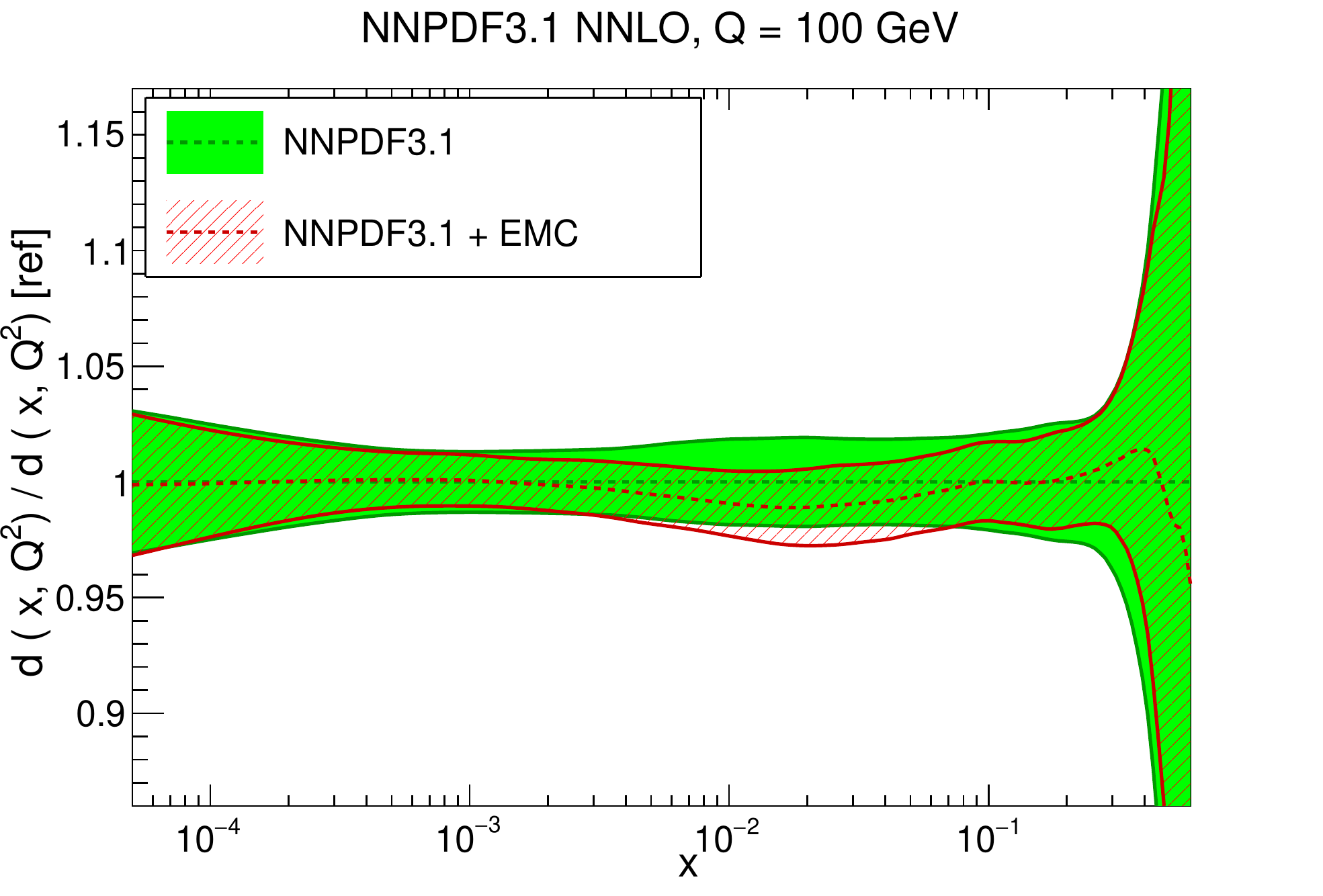}
  \caption{\small 
Same as Fig.~\ref{fig:31-nnlo-AZPT7TEV} but now
    comparing the default NNPDF3.1 to a version of it with  
the EMC $F_2^c$ dataset also included. 
Results are
shown for the charm (top left),
up (bottom left) and
down (bottom right) PDFs. The  relative PDF
uncertainty on charm is also shown (top right).
\label{fig:31-nnlo-emc}}
\end{center}
\end{figure}

However, a number of checks performed in
Refs.~\cite{Ball:2016neh,Rottoli:2016lsg},
such as variations of kinematical cuts and
systematic uncertainties, do not suggest any serious compatibility issues, and 
rather confirmed this dataset as being as reliable as the other older 
datasets with fixed nuclear targets routinely included in global PDF determinations.
Therefore it is interesting to revisit the issue of the impact of
this dataset within the context of 
NNPDF3.1. To this purpose we have produced a modified version of the global
NNPDF3.1  NNLO analysis in which the EMC dataset~\cite{Aubert:1982tt} is added to
the default NNPDF3.1 dataset. In Fig.~\ref{fig:distances_emc} 
the distances between this PDF set and the default are shown. 
The EMC dataset has a non-negligible impact on charm, at the one
sigma level, and also to a
lesser extent, at the half-sigma level,
on all light quarks, with only the gluon left essentially unaffected. 
The bulk of the effect is localized in the region   $0.01\lsim x\lsim 0.3$.
The PDFs are directly compared in Fig.~\ref{fig:31-nnlo-emc}.
The EMC data lead to an increase in the charm distribution
towards the upper edge of its error band in the default PDF set for
$0.02\lsim x\lsim 0.2$, while reducing the uncertainty
on it by a sizable factor. The light quark PDFs are correspondingly
slightly suppressed, and their uncertainties also reduced a little.

The values of $\chi^2/N_{\rm dat}$ for the
    deep-inelastic scattering experiments and for the total
    dataset before and after inclusion of the EMC data in the
    NNPDF3.1 dataset are shown in Table~\ref{tab:EMCchi2}. The fit quality of the 
    EMC charm dataset is greatly improved by its
    inclusion, without any significant change of the fit quality
    for any other DIS data: the hadron collider
    data are even less sensitive. The 
    inclusion of EMC charm data therefore appears to give a more accurate charm determination, 
    with no cost elsewhere, and so usage
    of this PDF set is recommended when precise charm PDFs at large $x$ are
    required. The phenomenological implications of the charm PDF will be
    discussed in Section~\ref{sec:phenocharm} below.

\begin{table}[H]
  \centering
  \small
  \begin{tabular}{|c|c|c|}
    \hline
   &         NNPDF3.1 NNLO   &  + EMC charm data \\
    \hline
    \hline
NMC    &    1.30     &  1.29    \\
SLAC    &    0.75      &  0.76   \\
BCDMS    &    1.21      &  1.24  \\
CHORUS    &    1.11      & 1.10   \\
NuTeV dimuon    &    0.82      &   0.88    \\
\hline
HERA I+II inclusive    &    1.16      &    1.16       \\
HERA $\sigma_c^{\rm NC}$    &    1.45      &  1.42    \\
HERA $F_2^b$    &    1.11      &   1.11   \\
EMC $F_2^c$     &   [4.8]   &   0.93 \\
\hline
\hline
{\bf Total dataset}   &  {\bf  1.148}      &  {\bf  1.145}   \\
\hline
  \end{tabular}
  \caption{\small The values of $\chi^2/N_{\rm dat}$ for the
    deep-inelastic scattering experiments, as well as for the total
    dataset, for the
    NNPDF3.1 NNLO PDF set and for a new PDF determination which
    also includes the EMC charm structure function data.
    \label{tab:EMCchi2}
  }
\end{table}

\clearpage

\subsection{The impact of LHC data}
\label{sec:nolhc}

\begin{figure}[t]
\begin{center}
  \includegraphics[scale=1]{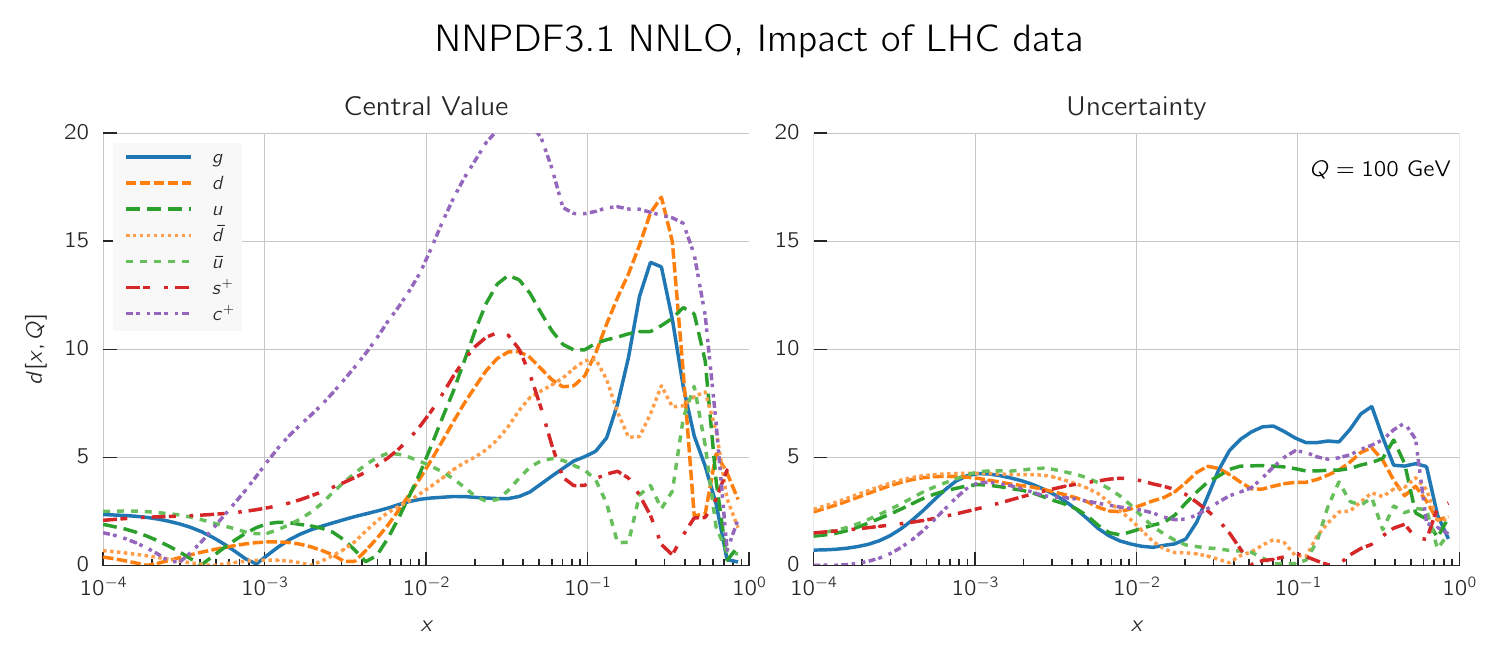}
  \caption{\small Same as Fig.~\ref{fig:distances_noZpT} but now
    excluding all LHC data. 
    \label{fig:distances_lhc}
  }
\end{center}
\end{figure}

We have seen that the LHC data have a significant impact on various
PDFs. Both in order to precisely gauge this impact, and in view of
possible applications in which usage of PDFs without LHC data is required,
we have produced a PDF set in which all LHC data are excluded from
the NNPDF3.1 dataset. The distance between the ensuing PDF set and the
default are shown in Fig.~\ref{fig:distances_lhc}. 

\begin{figure}[t]
\begin{center}
  \includegraphics[scale=0.38]{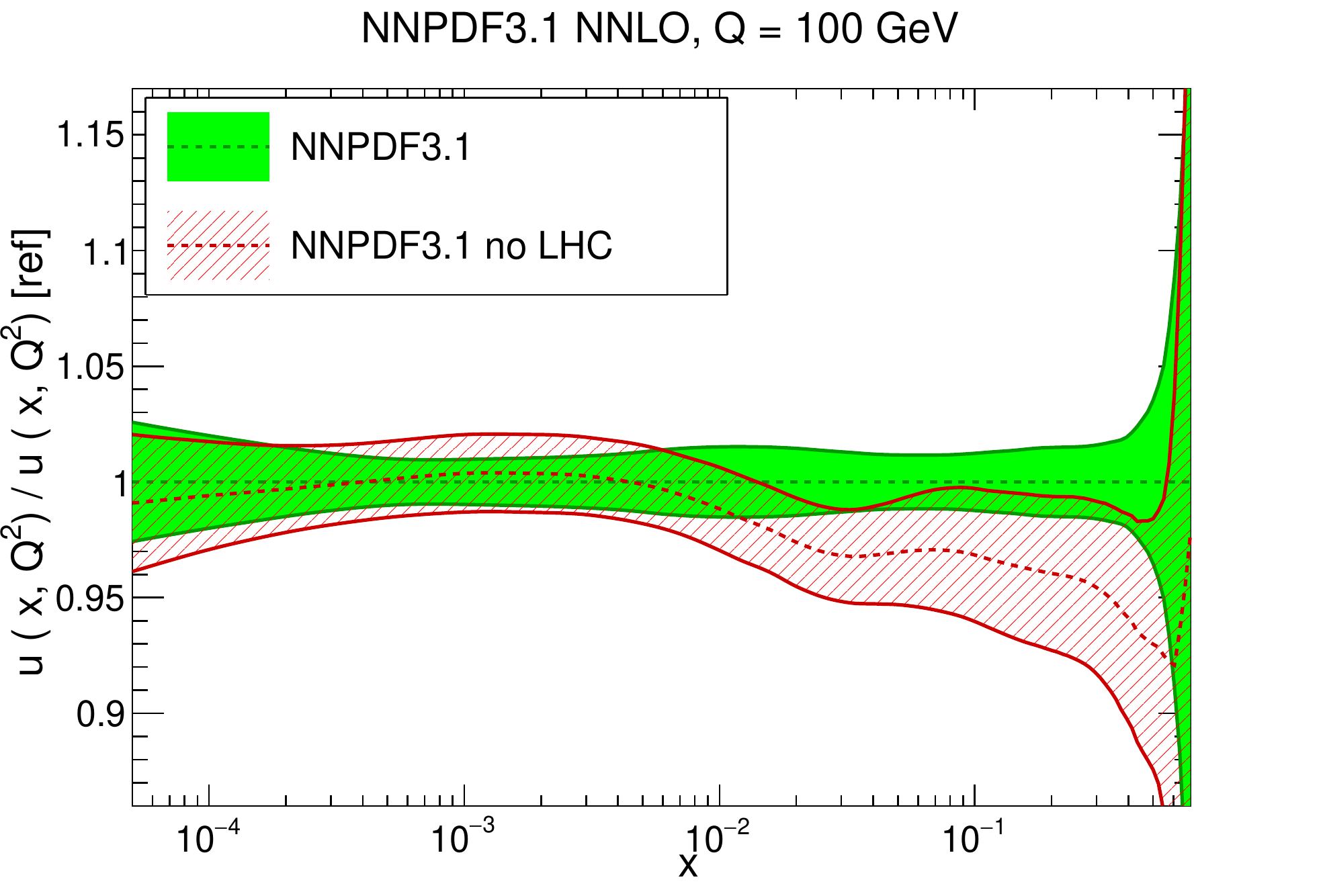}
  \includegraphics[scale=0.38]{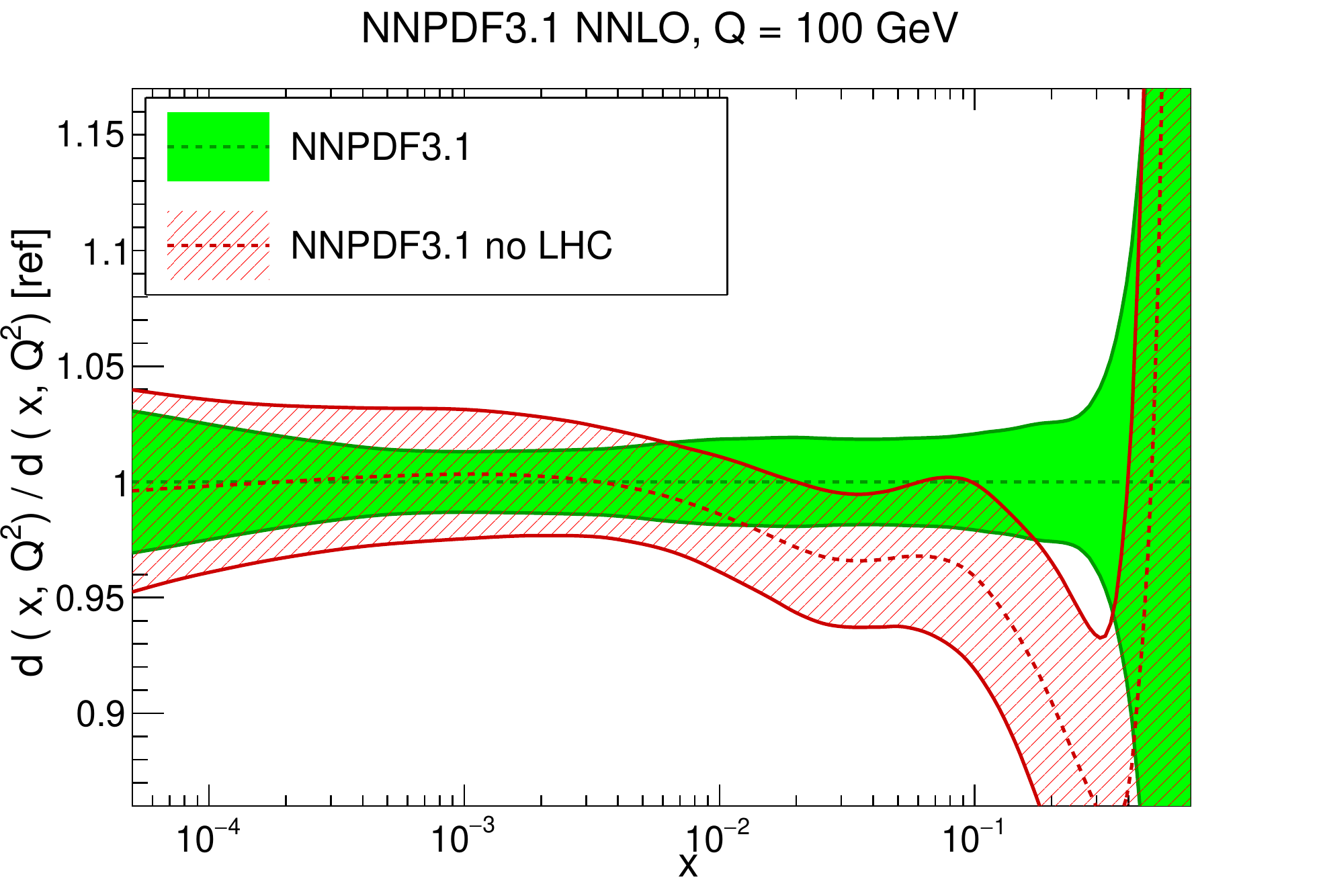}
  \includegraphics[scale=0.38]{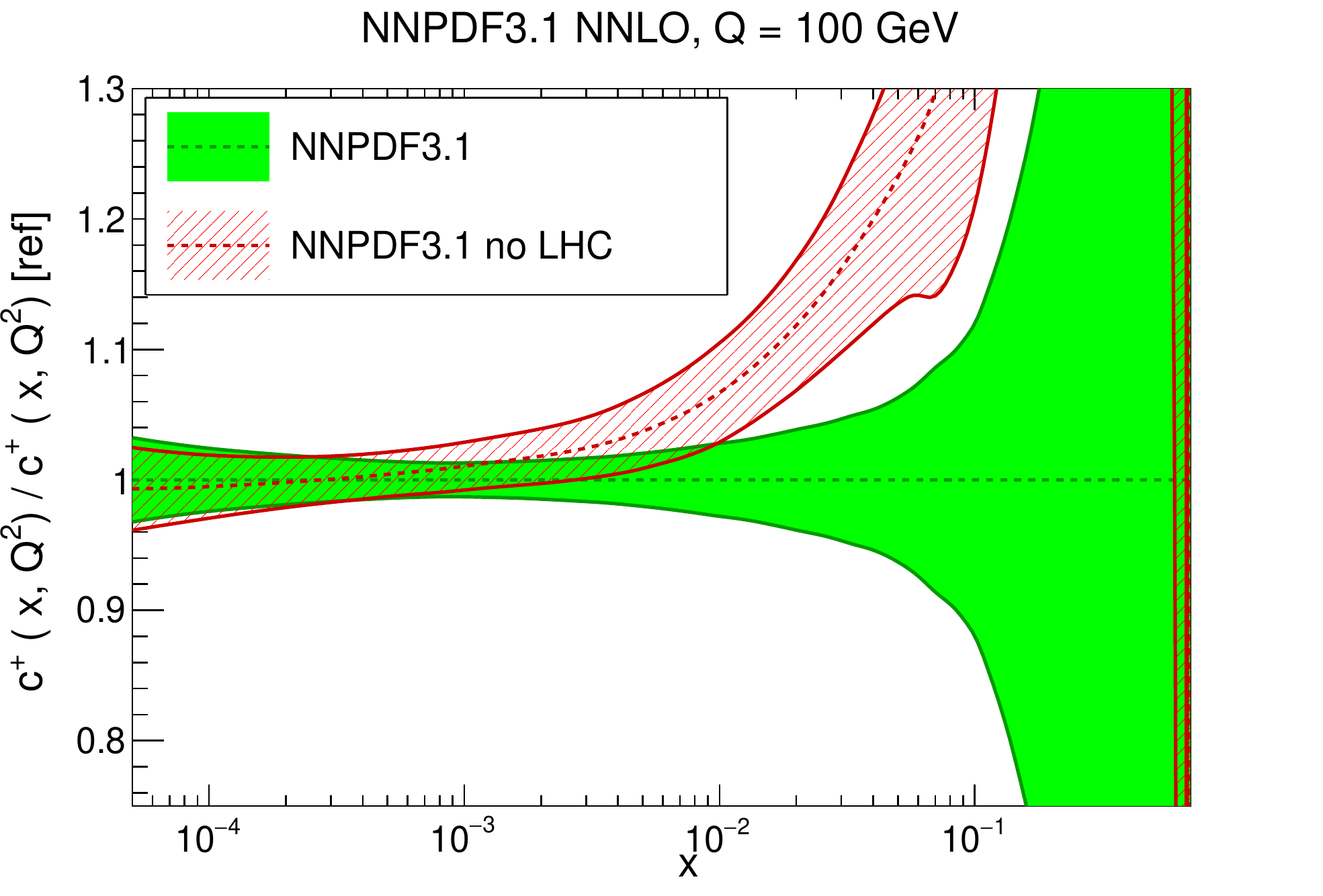}
  \includegraphics[scale=0.38]{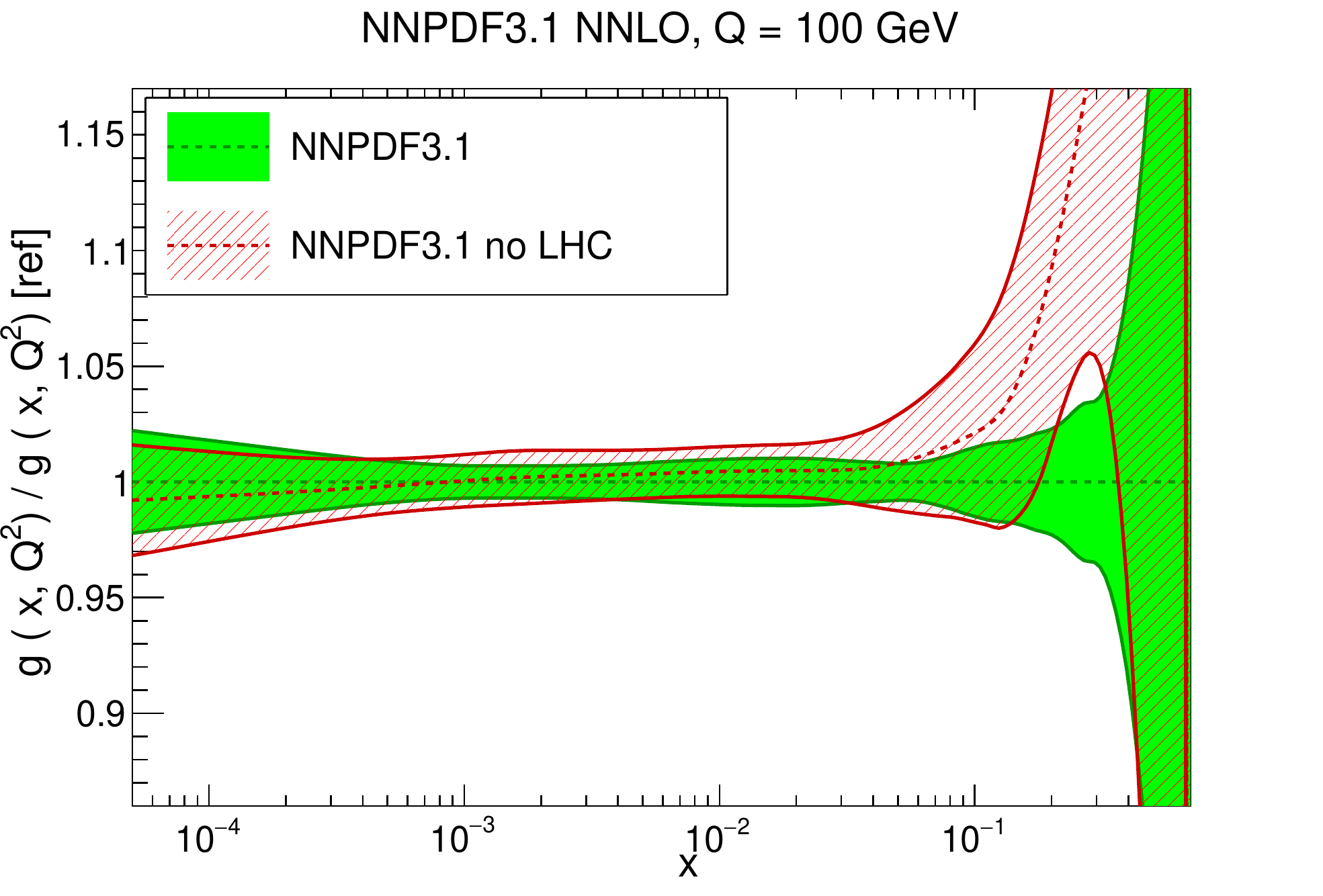}
  \caption{\small 
Same as Fig.~\ref{fig:31-nnlo-ZpT} 
but now excluding all LHC data. Results are
shown for the up (top left), down (top right), charm (bottom left) and
gluon (bottom right) PDFs.
\label{fig:31-nnlo-lhc}}
\end{center}
\end{figure}

The cumulative effect of the data which were discussed in
Sects.~\ref{sec:impactzpt}---\ref{sec:lhcbimpact} and \ref{sec:atlaswz}
is considerable. Most PDFs are affected at the one-sigma level and in some
cases (such as the down and charm quarks) at up to the two-sigma level. 
This is confirmed by direct comparison of the PDFs, see
Fig.~\ref{fig:31-nnlo-lhc}. The difference between the two fits
appears to be mostly driven by the CHORUS, BCDMS and fixed-target
Drell-Yan data, whose
$\chi^2$ improves respectively by 84, 32  and 38 units when removing the
LHC data. Other datsets display much more smaller differences,
typically compatible with statistical fluctuations, and in some cases (such as
for the SLAC data) the fit quality is actually somewhat better in the
global fit.

On the other hand, it is clear that the
shifts between PDFs without LHC data and those including them are 
compatible with the respective PDF uncertainties, and that the
uncertainties on the PDFs determined without LHC data are not so large
as to render them useless for phenomenology. We conclude that  the
default set remains considerably more accurate and should be used for
precision phenomenology. However, the use of 
PDFs determined without some or all the LHC
data may be mandatory in searches for new physics, in order to make
sure that possible new physics effects are not reabsorbed in the
PDFs. In such circumstances, we conclude that even though the uncertainty in
the PDFs without LHC data is not competitive, the level of 
deterioration is not so great as to make searches for new physics
altogether impossible.

\subsection{Nuclear targets and nuclear corrections.}
\label{sec:nonucl}
The NNPDF3.1 dataset includes several measurements taken upon nuclear
targets. DIS data from the SLAC, BCDMS and NMC experiments along with
the E886 fixed-target Drell-Yan data involve measurements of deuterium.
All neutrino data and the fixed-target E605 Drell-Yan data, are obtained with heavy
nuclear targets. All of these data were already included in previous
PDF determinations, including NNPDF3.0. The impact of nuclear
corrections was studied in 
Ref.~\cite{Ball:2014uwa} and found to be under control. 
However, the much wider dataset might now permit the removal of 
these data from the global dataset: whereas removing data inevitably
entails some loss of precision, this might be more
than compensated by the increase in accuracy due to the complete elimination 
of any dependence on uncertain nuclear corrections.

\begin{figure}[t]
\begin{center}
  \includegraphics[scale=1]{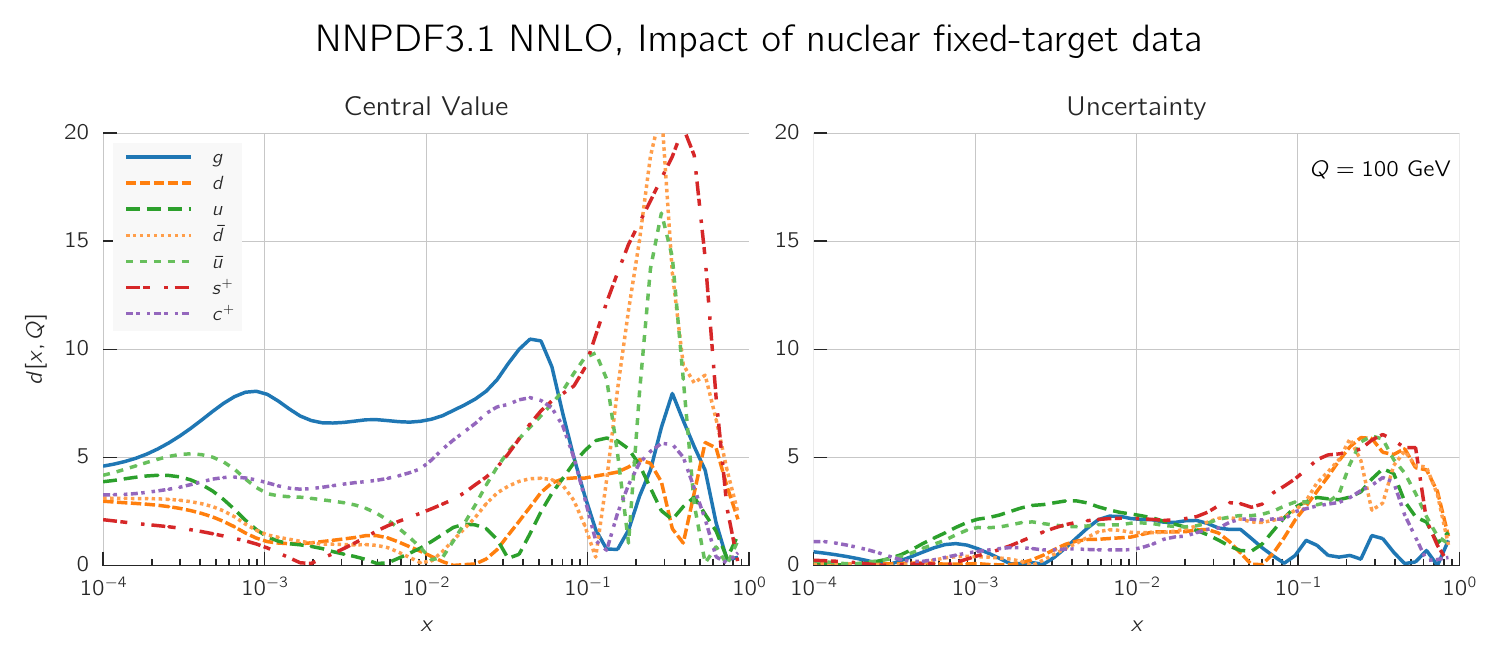}\\
   \includegraphics[scale=1]{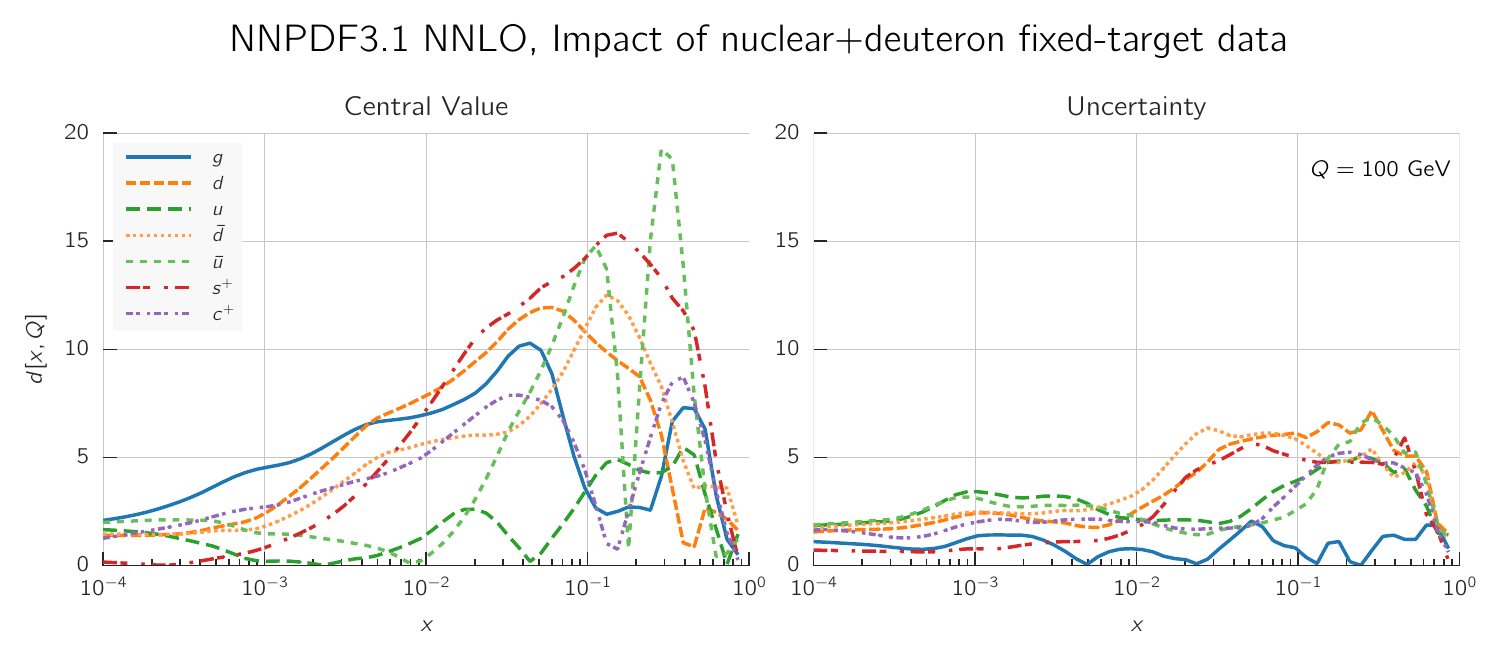}
  \caption{\small Same as Fig.~\ref{fig:distances_noZpT} but now
    excluding all data with heavy nuclear targets, but keeping
    deuterium data (top) or excluding all data with any nuclear target
    and only keeping proton data (bottom) 
    \label{fig:distances_proton}
  }
\end{center}
\end{figure}

In order to assess this, we performed two additional PDF
determinations 
with the NNPDF3.1
methodology. Firstly, by removing all heavy nuclear target data but keeping deuterium
data, and secondly removing all nuclear data and only keeping proton data.
The distances between the default and these two PDF sets are shown in
Fig.~\ref{fig:distances_proton}. At large $x$ the impact of
nuclear target data is significant, at the one to two sigma level,
mostly on the flavor separation of the sea. The deuterium data also have a
significant impact, particularly in the intermediate $x$ range.

\begin{figure}[t]
\begin{center}
  \includegraphics[scale=0.35]{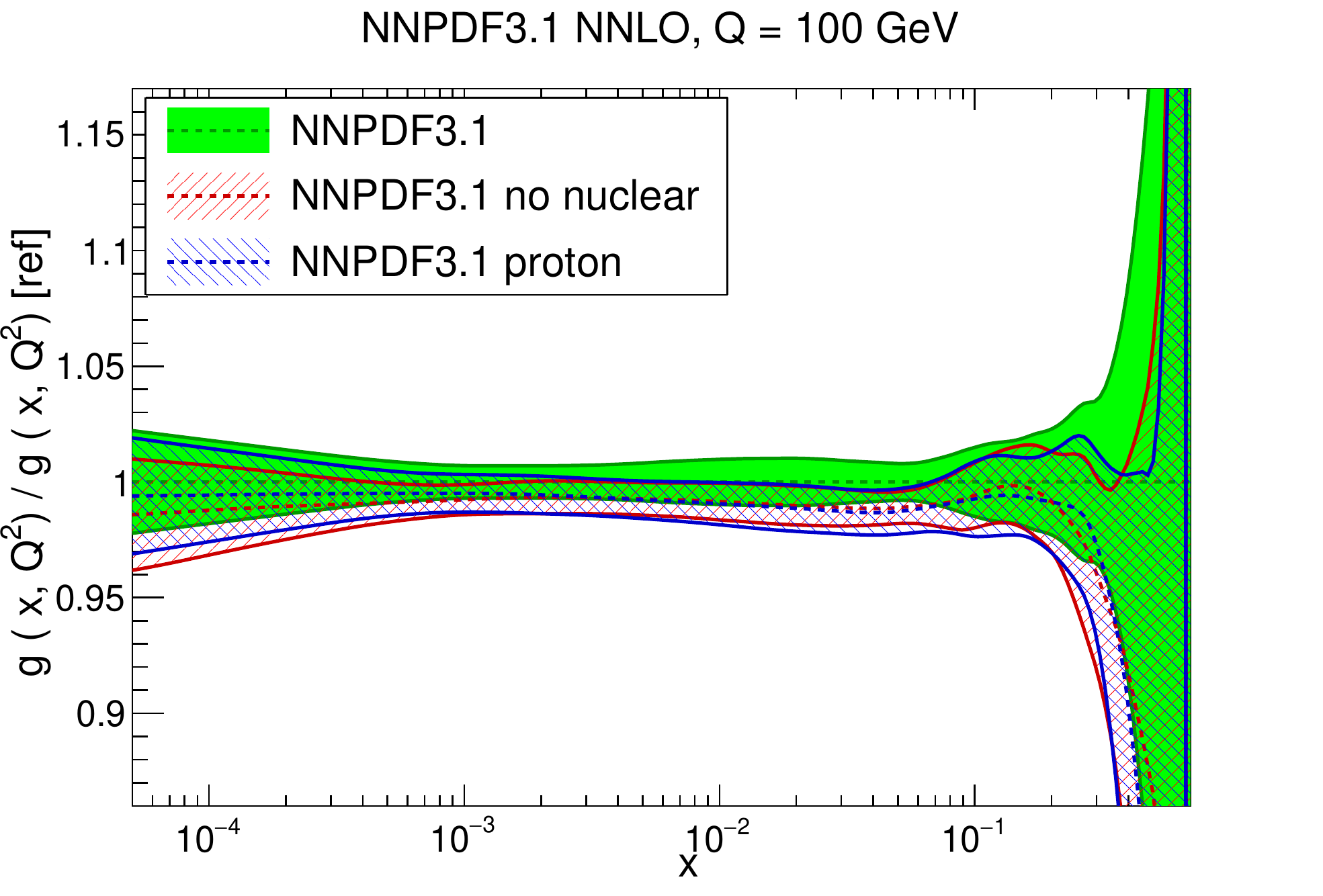}
  \includegraphics[scale=0.35]{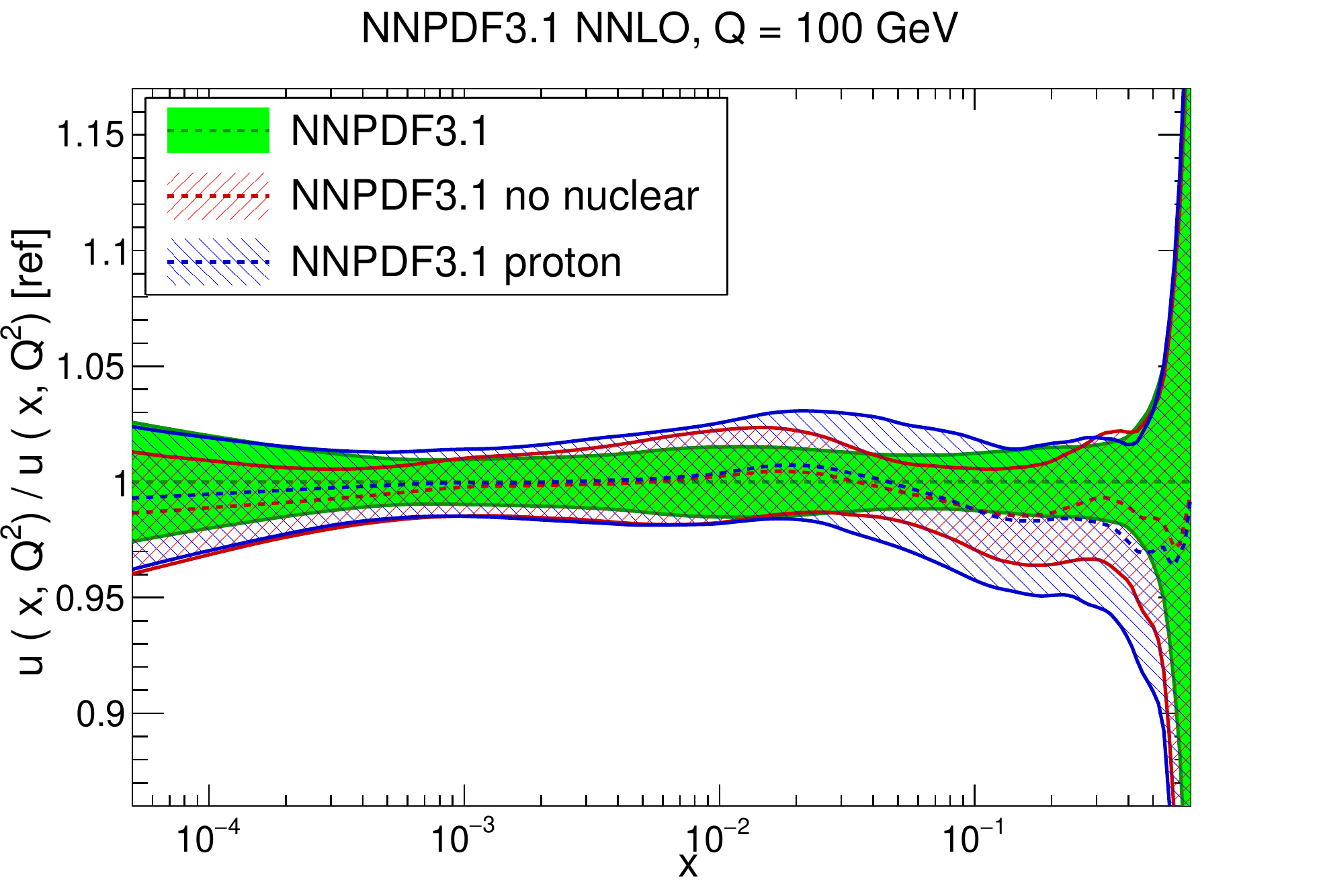}
  \includegraphics[scale=0.35]{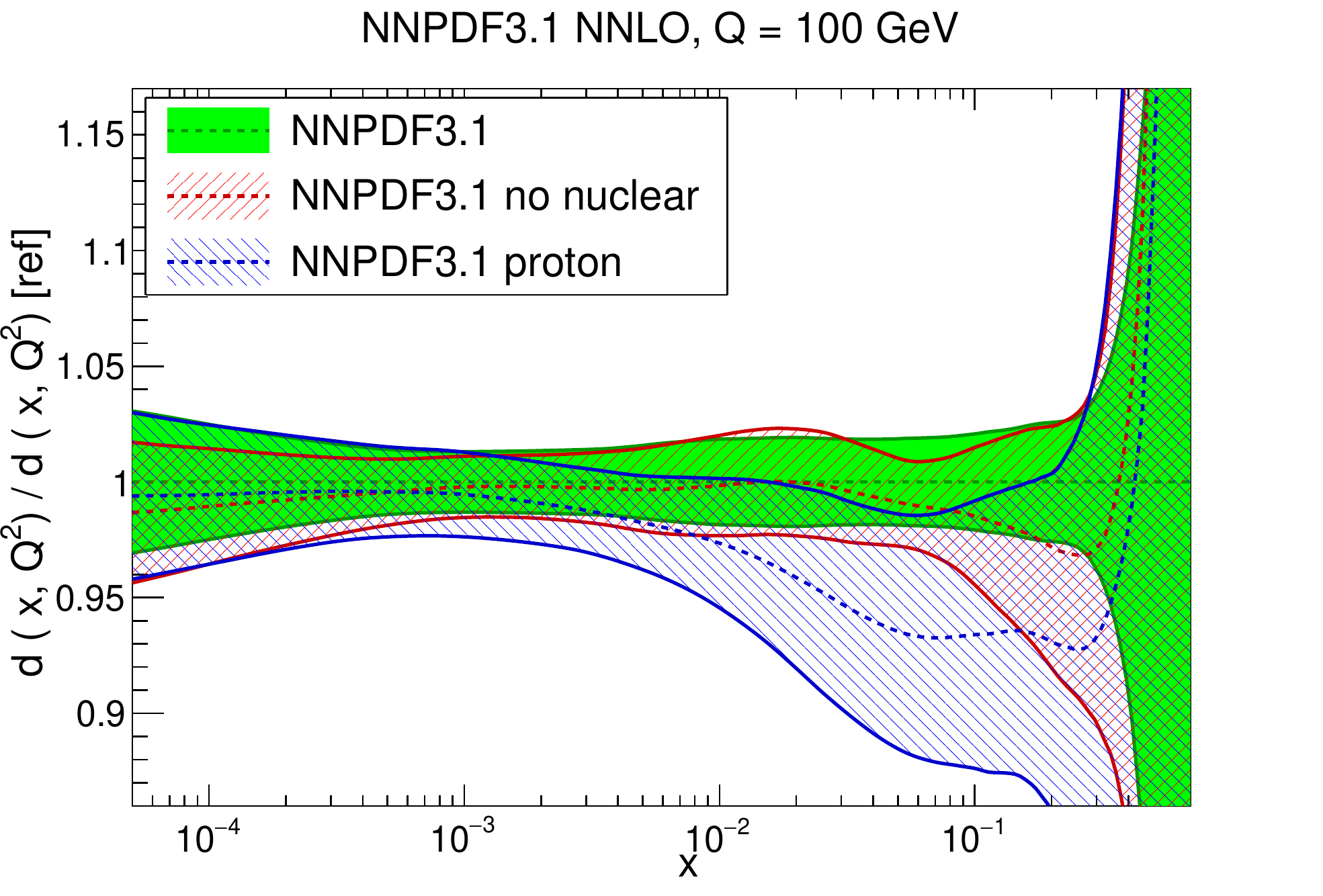}
  \includegraphics[scale=0.35]{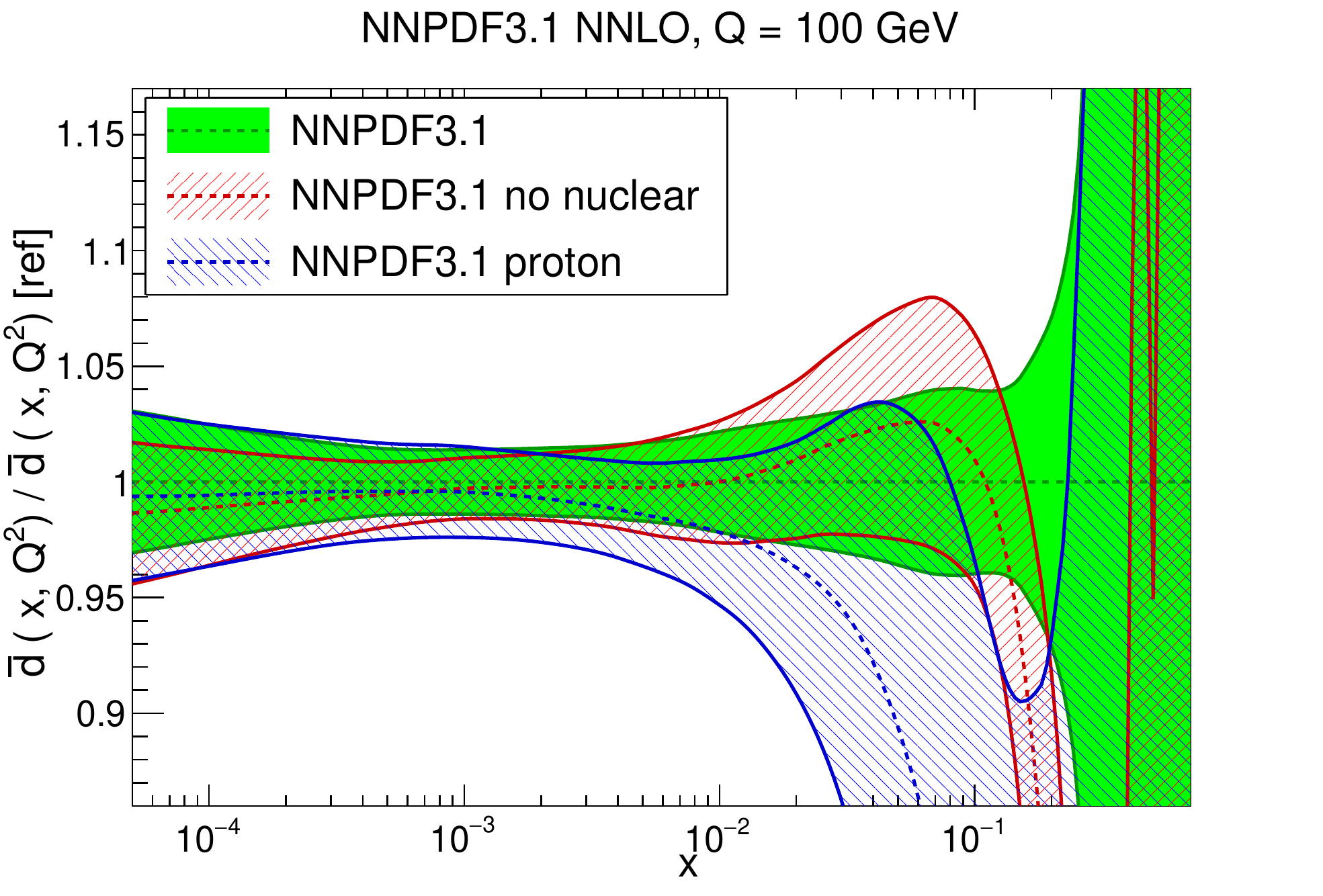}
  \caption{\small 
Same as Fig.~\ref{fig:31-nnlo-ZpT} 
but now excluding all data with heavy nuclear targets, but keeping
    deuterium data, or excluding all data with any nuclear target
    and only keeping proton data. Results are
shown for the gluon (top left), up (top right), down (bottom left) and
antidown (bottom right).
\label{fig:31-nnlo-proton}}
\end{center}
\end{figure}

A direct comparison of PDFs, in Fig.~\ref{fig:31-nnlo-proton}, and 
their uncertainties, in Fig.~\ref{fig:ERR-31-nnlo-datasetvar}, shows
that indeed PDFs determined with no heavy nuclear target data are
reasonably compatible with the global set, though with rather larger
uncertainties, especially for strangeness. Indeed,
best-fit results without heavy nuclear targets, or
even without deuterium data, are all compatible within their respective
uncertainties, which is consistent with the previous conclusion that
the absence of nuclear corrections for these data does not lead
to significant bias at the level of current PDF uncertainties. 
On the other hand, PDFs determined with only
proton data  while compatible to within one sigma with the
global set within their larger uncertainties, show a substantial loss of
precision. This is particularly notable for down quarks, due to the
importance of 
deuterium data in pinning down the isospin triplet PDF combinations. 

Because  deuterium data have a significant impact on the fit, one may
worry that nuclear corrections to the deuterium data are now no longer
negligible, at the accuracy of the present PDF determination. In order
to investigate this issue in greater detail, we have performed a
variant of the NNPDF3.1 NNLO default PDF determination
in which all deuterium data are corrected using the same
nuclear corrections as used by MMHT14 (specifically, Eqs.~(9,10) of
Ref.~\cite{Harland-Lang:2014zoa}).

\begin{figure}[t]
\begin{center}
  \includegraphics[scale=1]{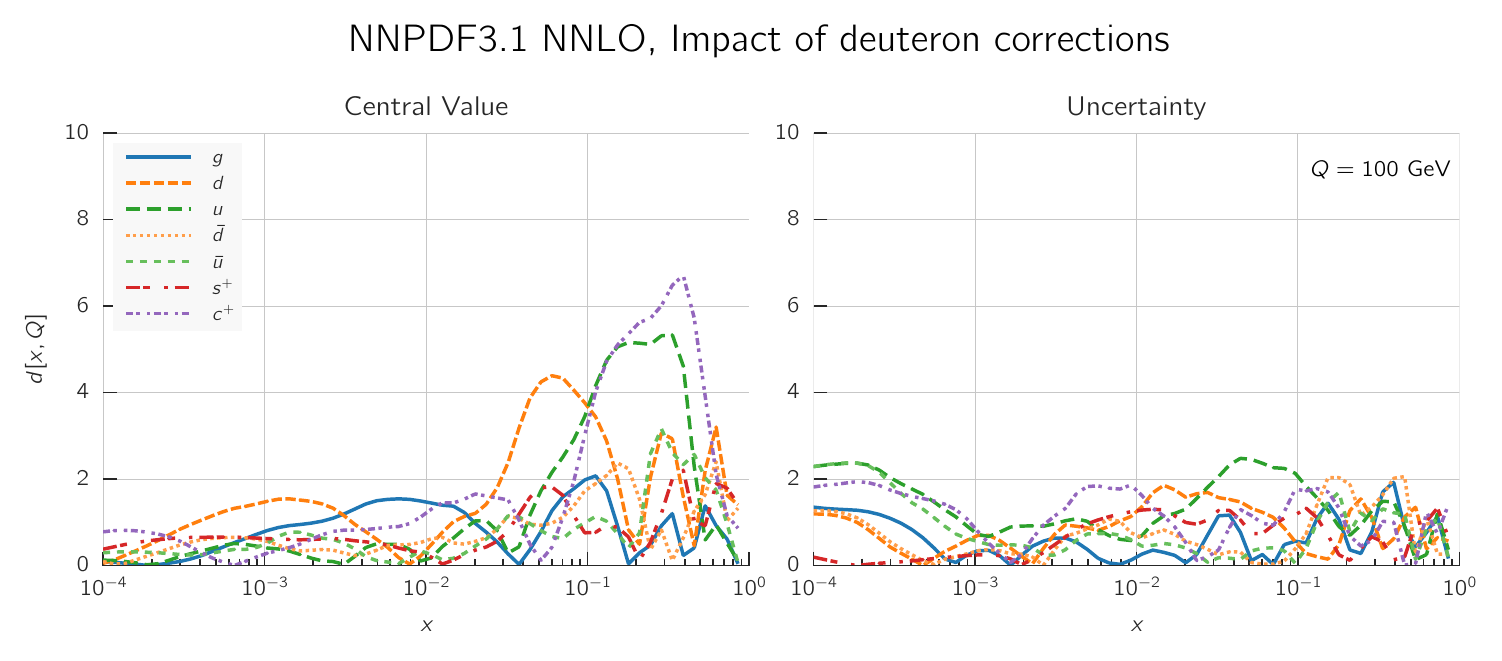}
  \caption{\small Same as Fig.~\ref{fig:distances_noZpT} but now
comparing the default NNPDF3.1 to a version in which all
deuterium data have been corrected using the nuclear corrections from
Ref.~\cite{Harland-Lang:2014zoa}.
    \label{fig:distances_nuclcorr}
  }
\end{center}
\end{figure}

\begin{figure}[t]
\begin{center}
  \includegraphics[scale=0.35]{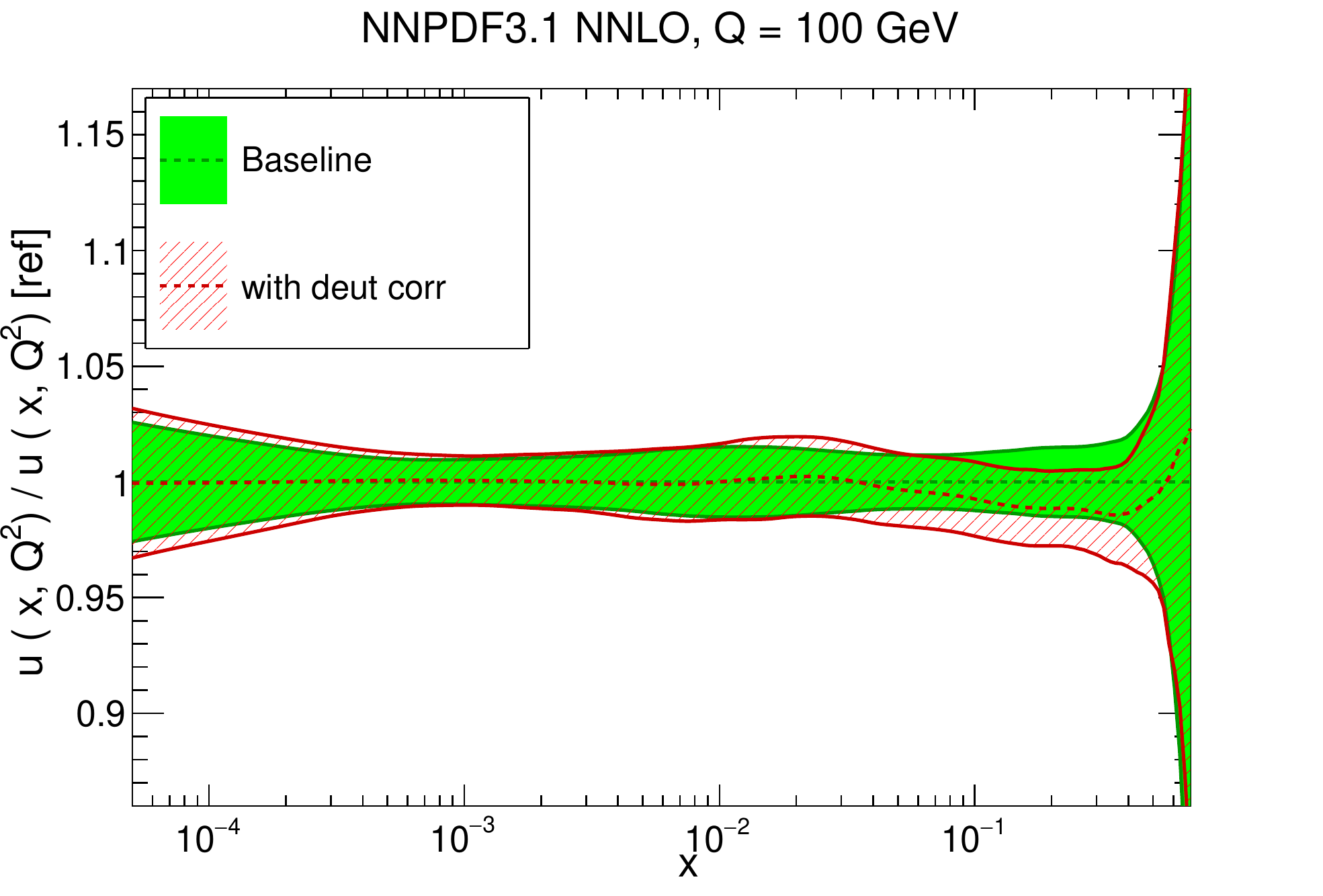}
  \includegraphics[scale=0.35]{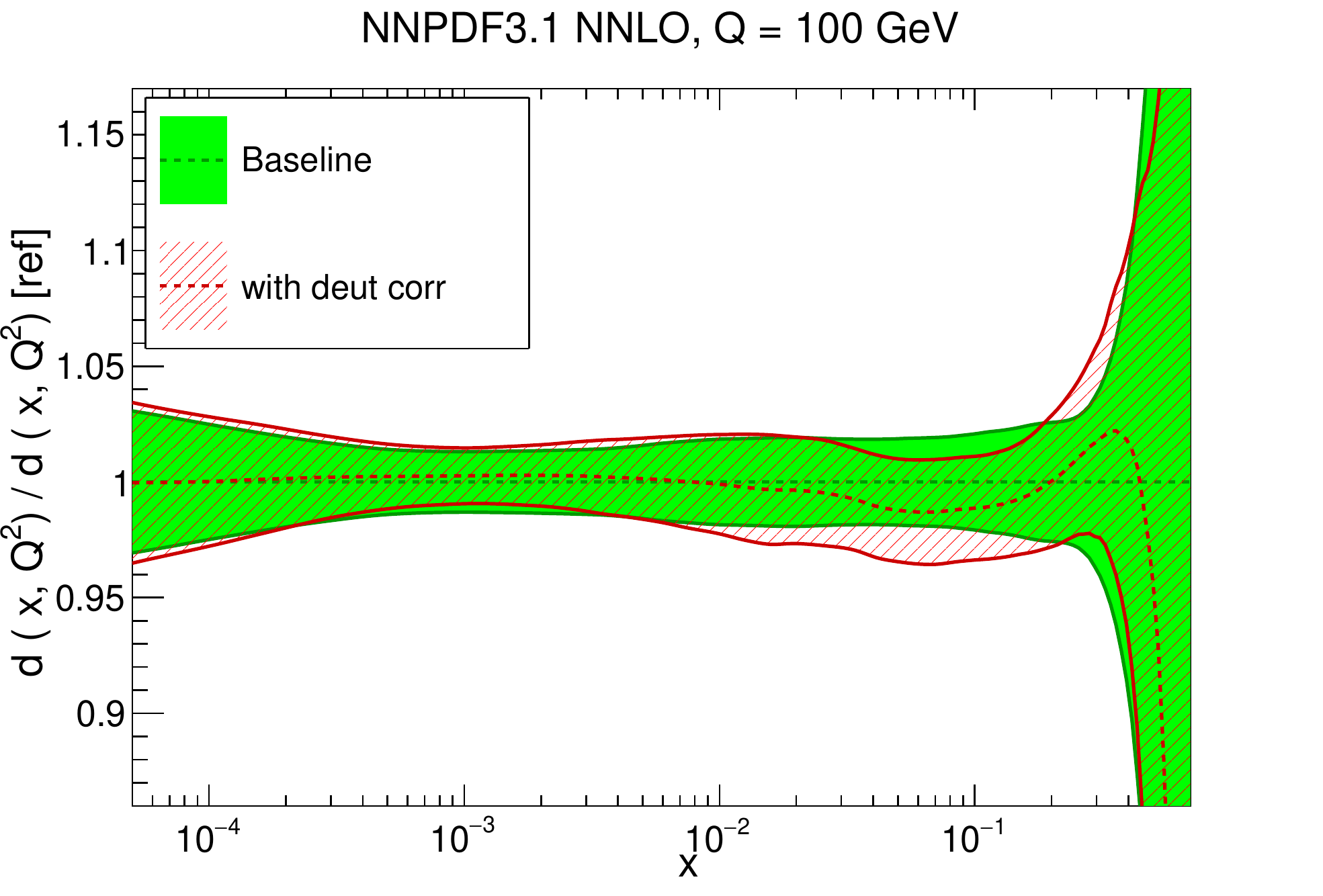}
  \includegraphics[scale=0.35]{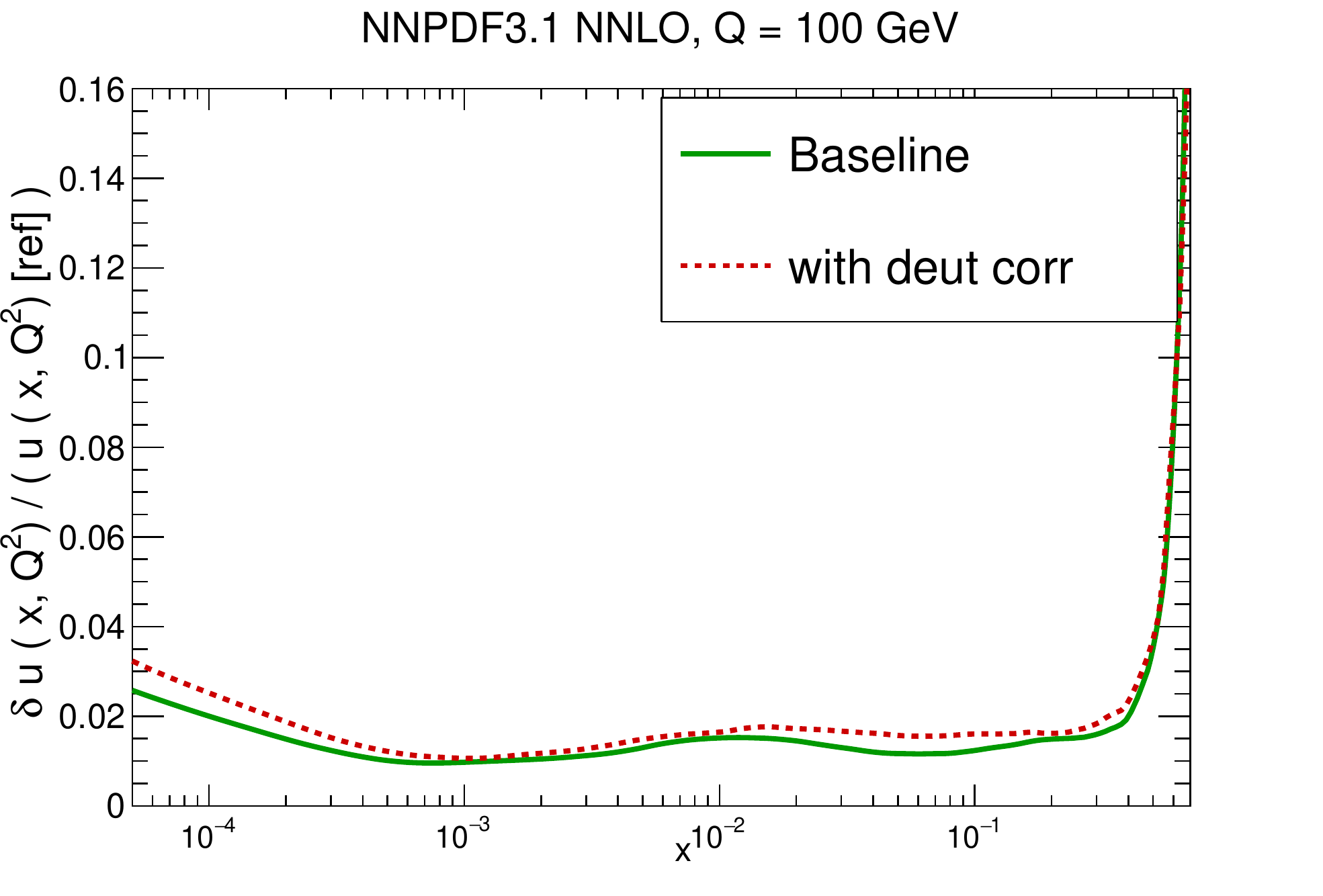}
  \includegraphics[scale=0.35]{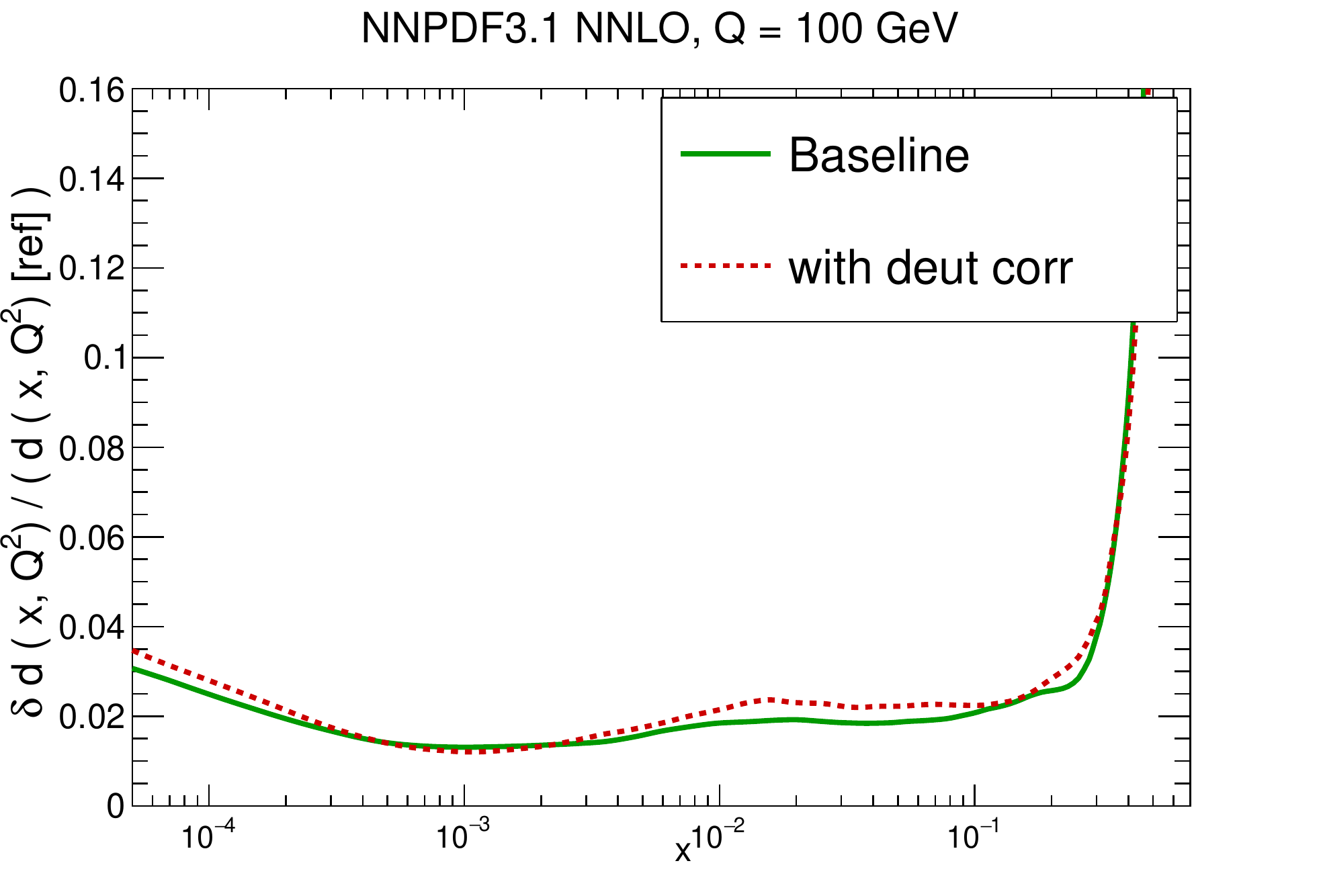}
  \caption{\small 
Same as Fig.~\ref{fig:31-nnlo-ZpT} 
but now comparing the default NNPDF3.1 to a version in which all
deuterium data have been corrected using the nuclear corrections from
Ref.~\cite{Harland-Lang:2014zoa}. Results are shown for the up (left=
and down (right) PDFs. The uncertainties are also shown (botton row).
\label{fig:31-nnlo-nuclcorr}}
\end{center}
\end{figure}
In terms of fit quality we find
that the inclusion of nuclear corrections leads to a
slight deterioration in the quality of the fit, with a value 
of $\chi^2/N_{\rm  dat}=1.156$, to be compared to the defaut 
$\chi^2/N_{\rm  dat}=1.148$ (see Table~\ref{tab:chi2tab_31-nlo-nnlo-30}).  
In
particular 
we find that for the NMC, SLAC, and
BCDMS data the values of $\chi^2/N_{\rm  dat}$ with (without) nuclear
corrections are respectively 0.94(0.95), 0.71(0.70), and 1.11(1.11).
Therefore, the addition of deuterium corrections has no significant
impact on the fit quality to these data. 

The distances between PDFs determined including deuterium corrections
and the default are shown in Fig.~\ref{fig:distances_nuclcorr}. They
are seen to be moderate and always below the half-sigma level, and confined mostly to the up and down PDFs, as expected. These PDFs  
are  shown in Fig.~\ref{fig:31-nnlo-nuclcorr}, which confirms the
moderate effect of the deuterium correction.
It should be noticed that the PDF uncertainty, also
shown in Fig.~\ref{fig:31-nnlo-nuclcorr}, is somewhat increased when
the deuterium corrections are included. The relative shift for other
PDFs are yet smaller since they are affected by larger uncertainties,
which are also somewhat increased by the inclusion of the nuclear corrections.


In view of the theoretical uncertainty involved in estimating nuclear
corrections, and bearing in mind that we see no evidence of an
improvement in fit quality while we note a slight increase in PDF
uncertainties when including deuterium corrections using the model of
Ref.~\cite{Harland-Lang:2014zoa}, we conclude that the impact of
deuterium corrections on the NNPDF3.1 results is sufficiently small that they
may be safely ignored even within the current high precision of PDF
determination. Nevertheless, more detailed dedicated studies of
nuclear corrections, also in relation to the construction of nuclear
PDF sets, may well be worth pursuing in future studies.

In conclusion, for the time being it is still appears advantageous to retain
nuclear target data in the global dataset for general-purpose PDF
determination. However, if very high accuracy is required (such as,
for instance, in the determination of standard model parameters) it
might be preferable to use PDF sets from which all data with nuclear
targets have been omitted. 

\begin{figure}[t]
  \begin{center}
    \includegraphics[scale=0.32]{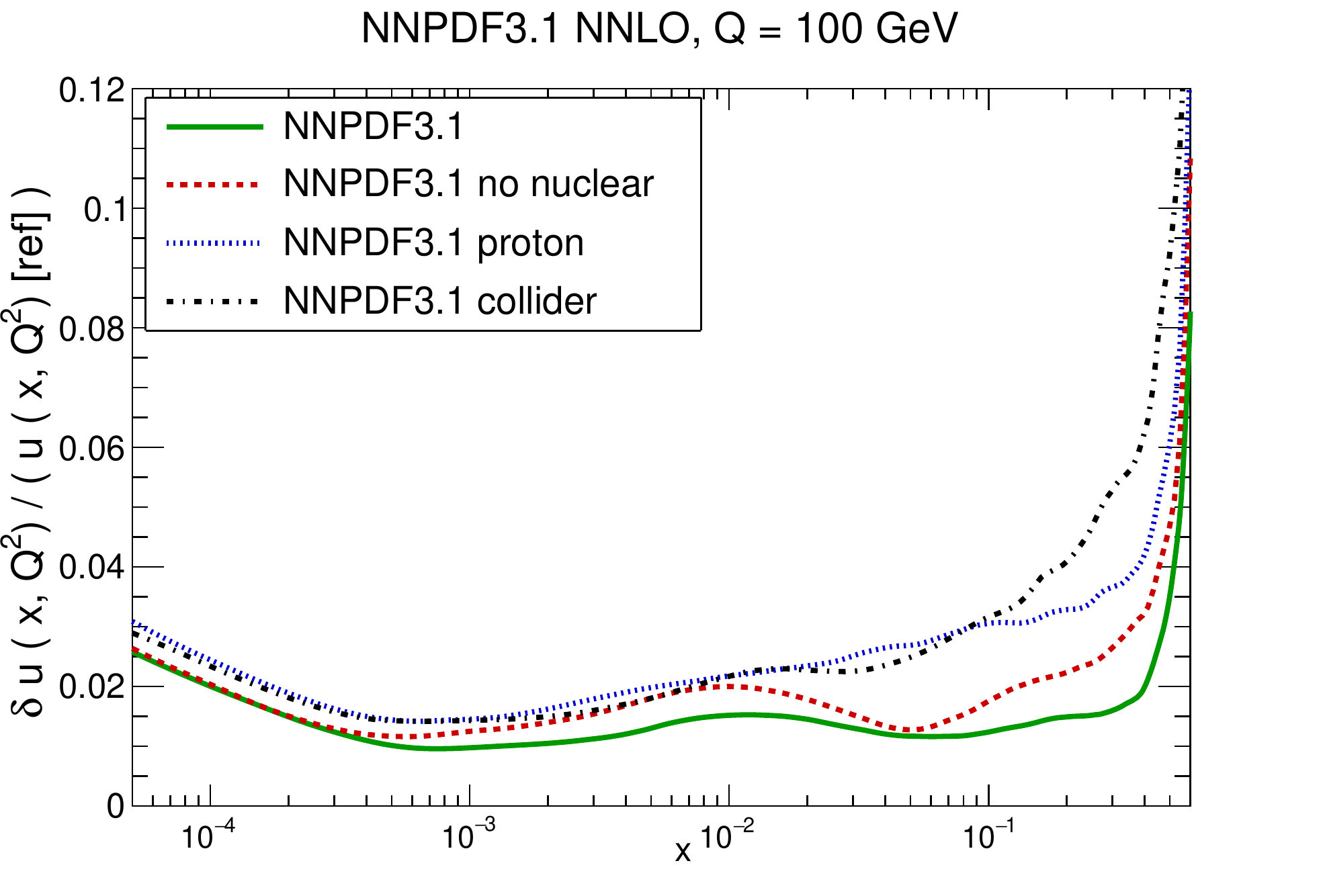}
    \includegraphics[scale=0.32]{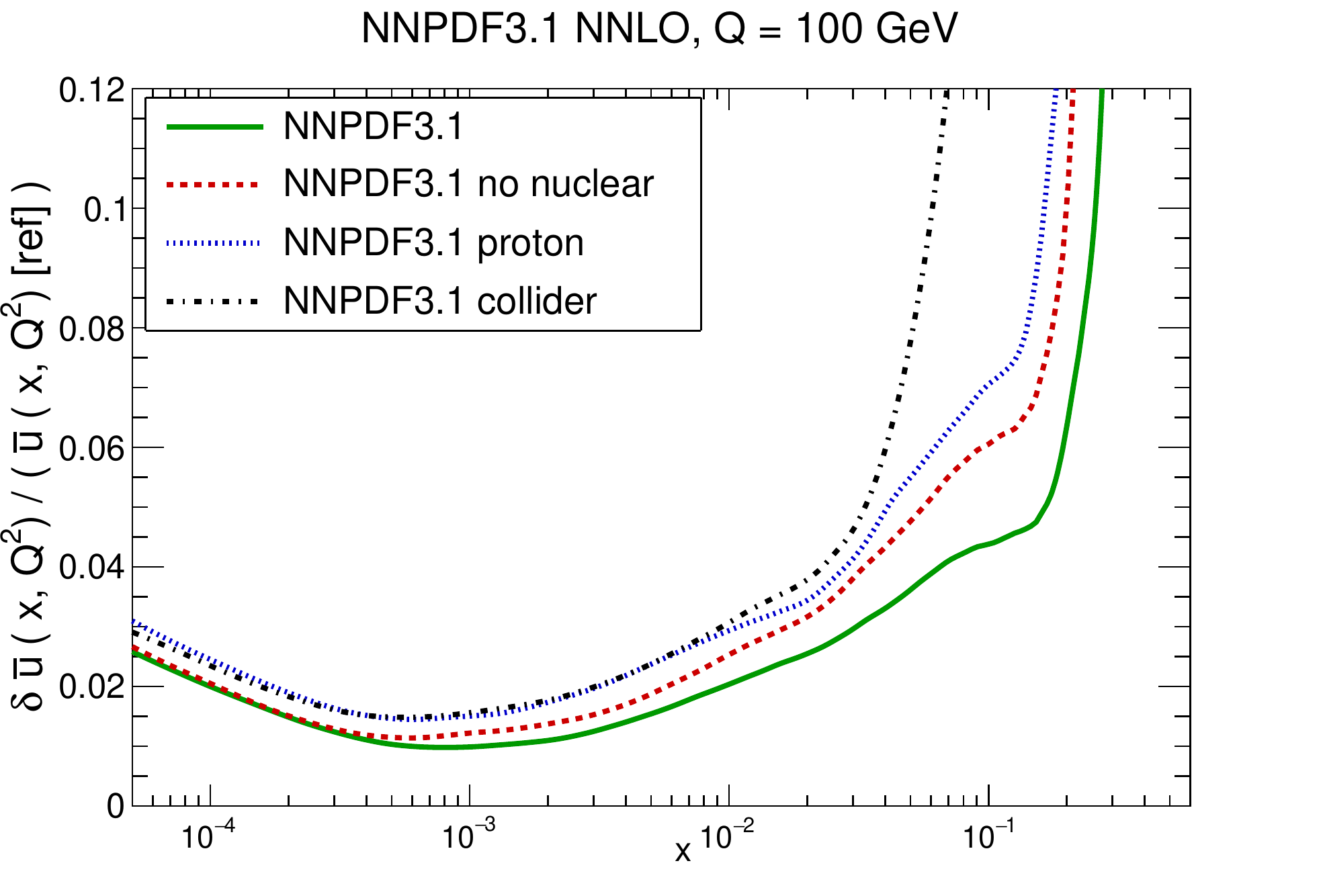}
    \includegraphics[scale=0.32]{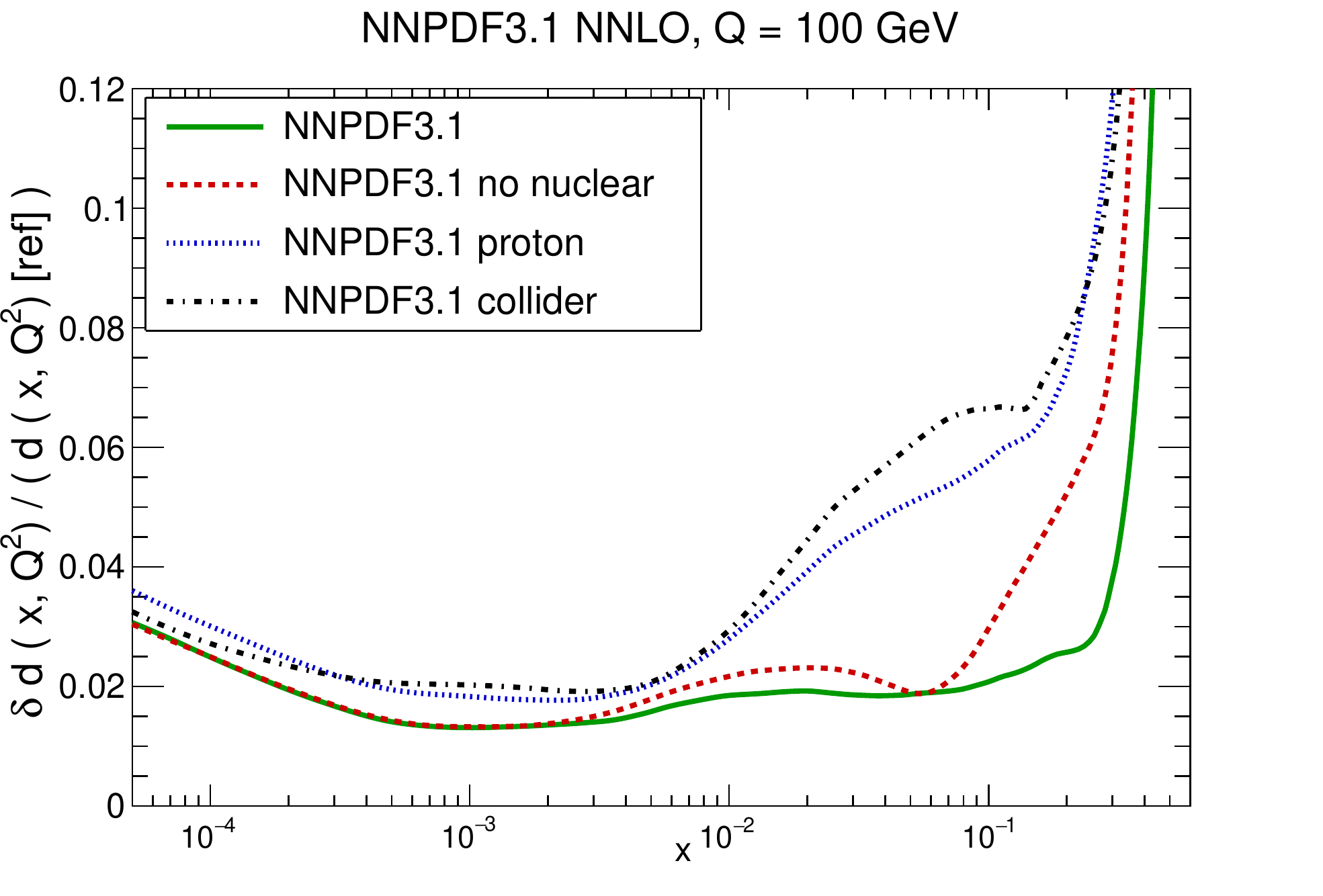}
    \includegraphics[scale=0.32]{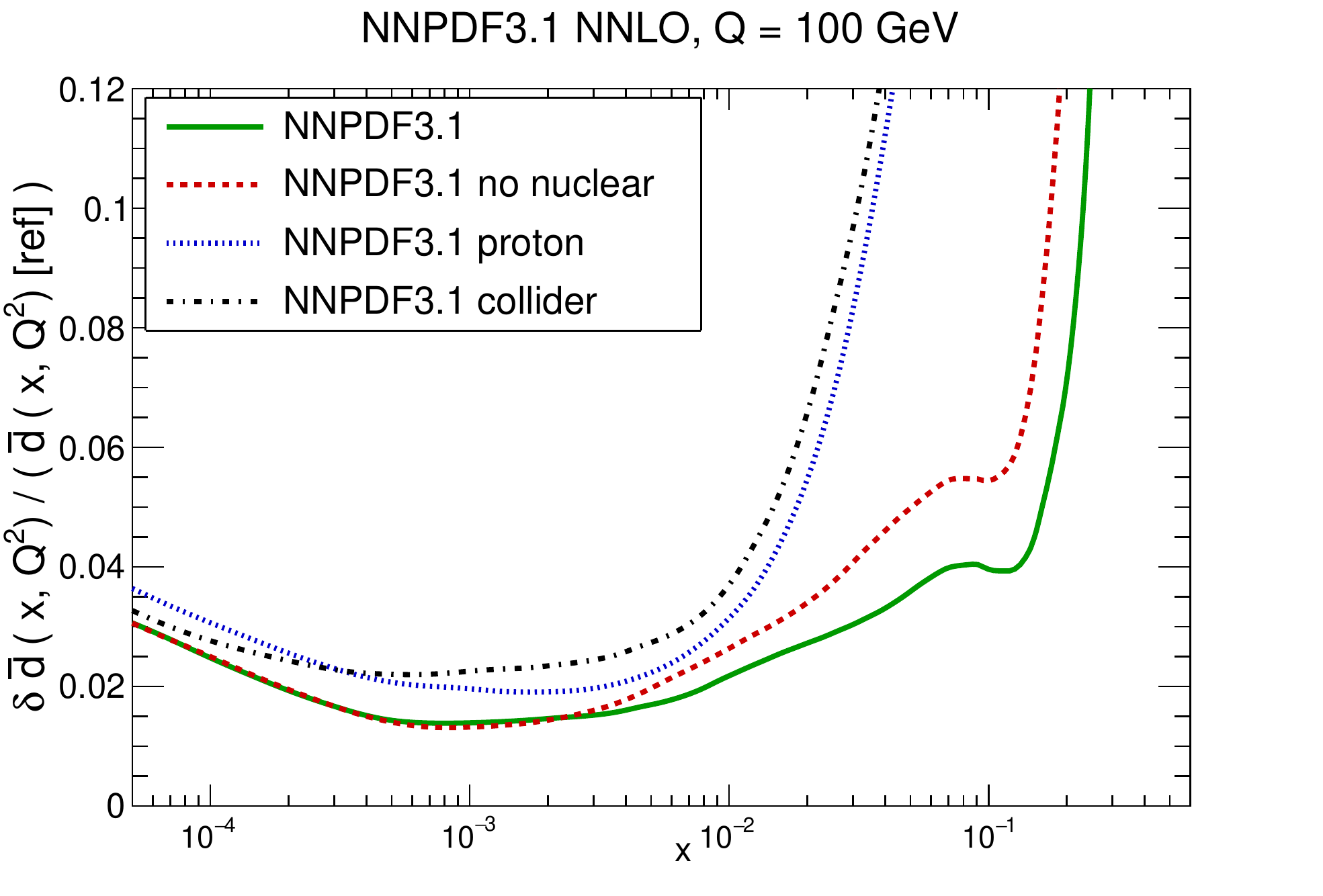}
  \includegraphics[scale=0.32]{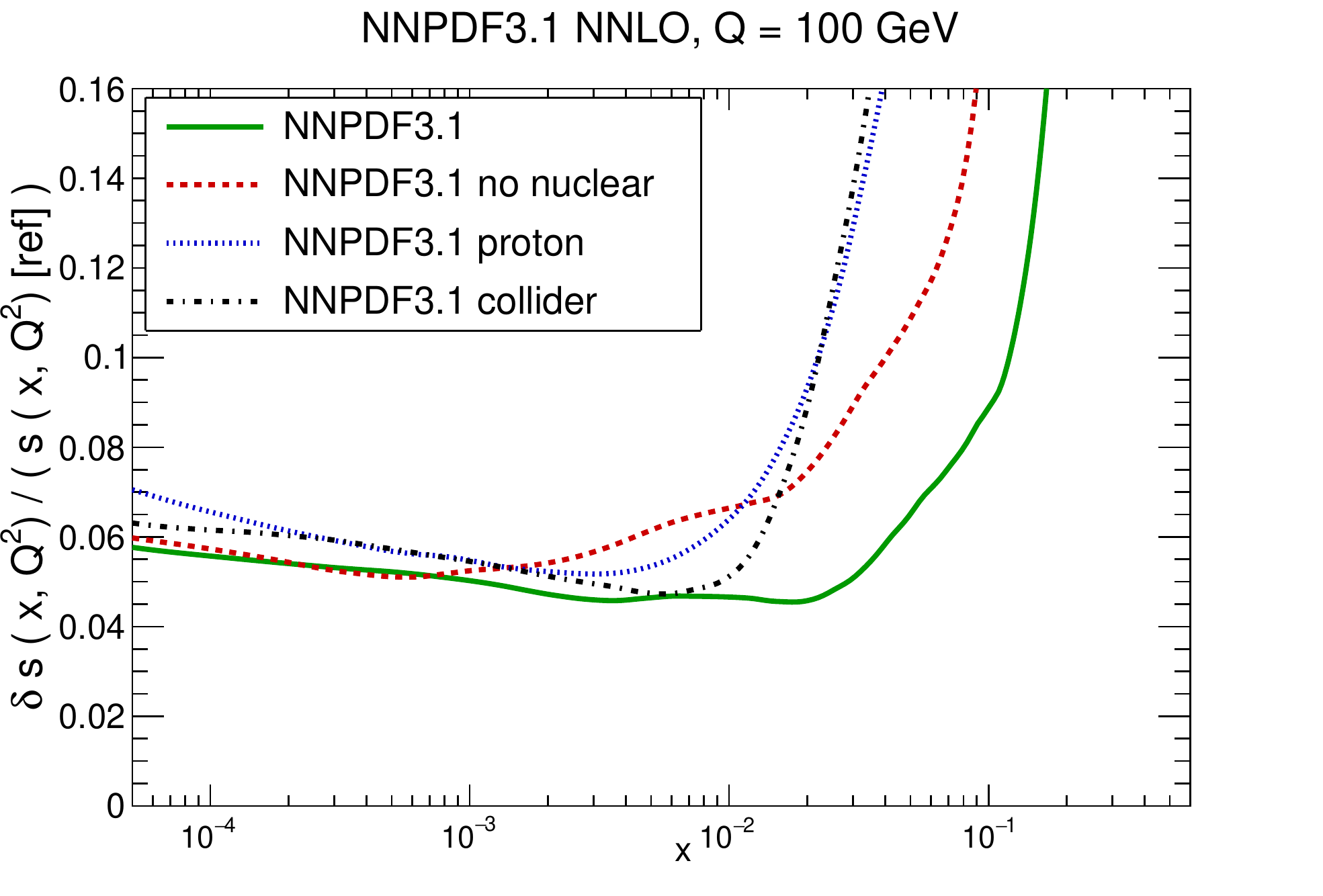}
  \includegraphics[scale=0.32]{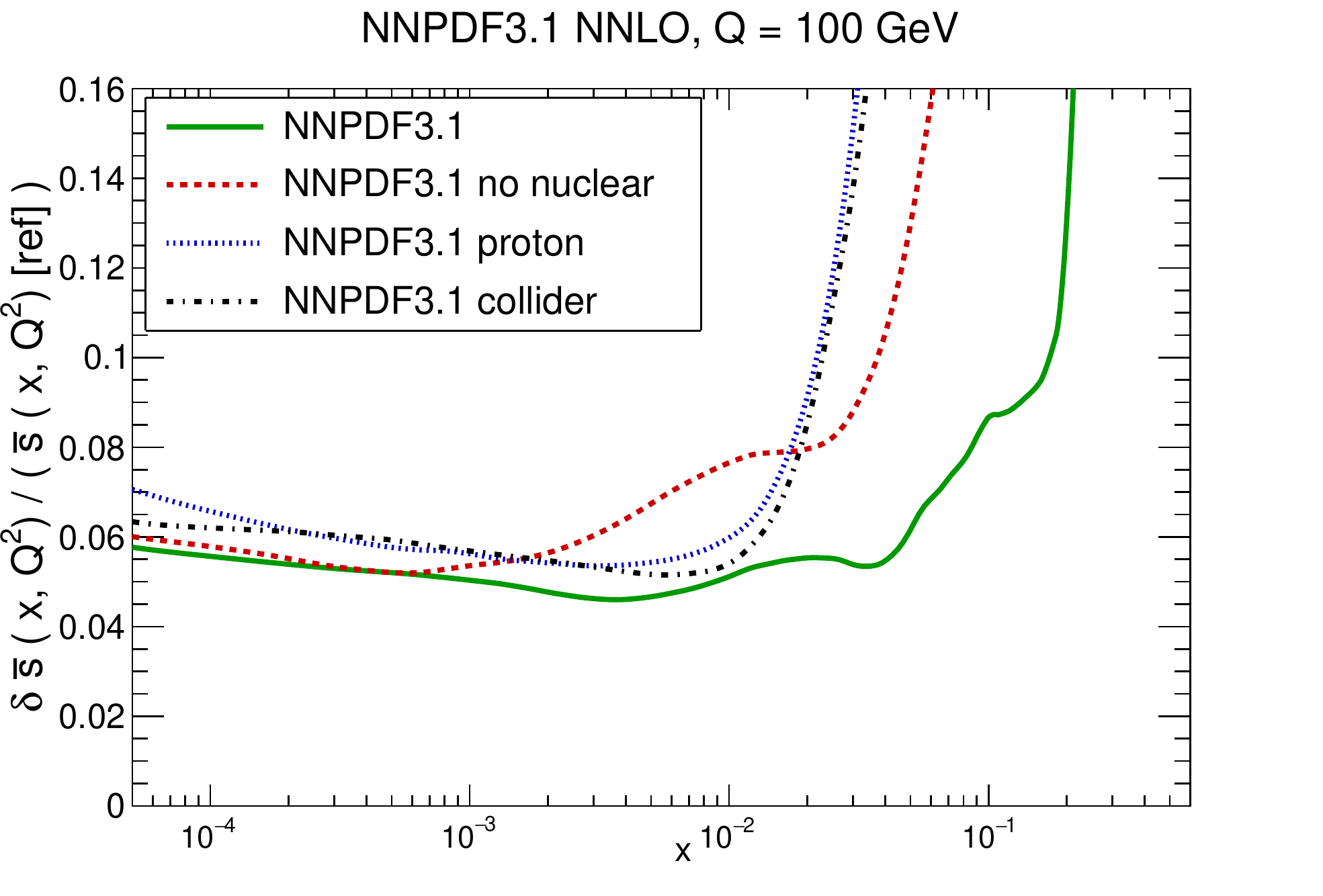}
  \includegraphics[scale=0.32]{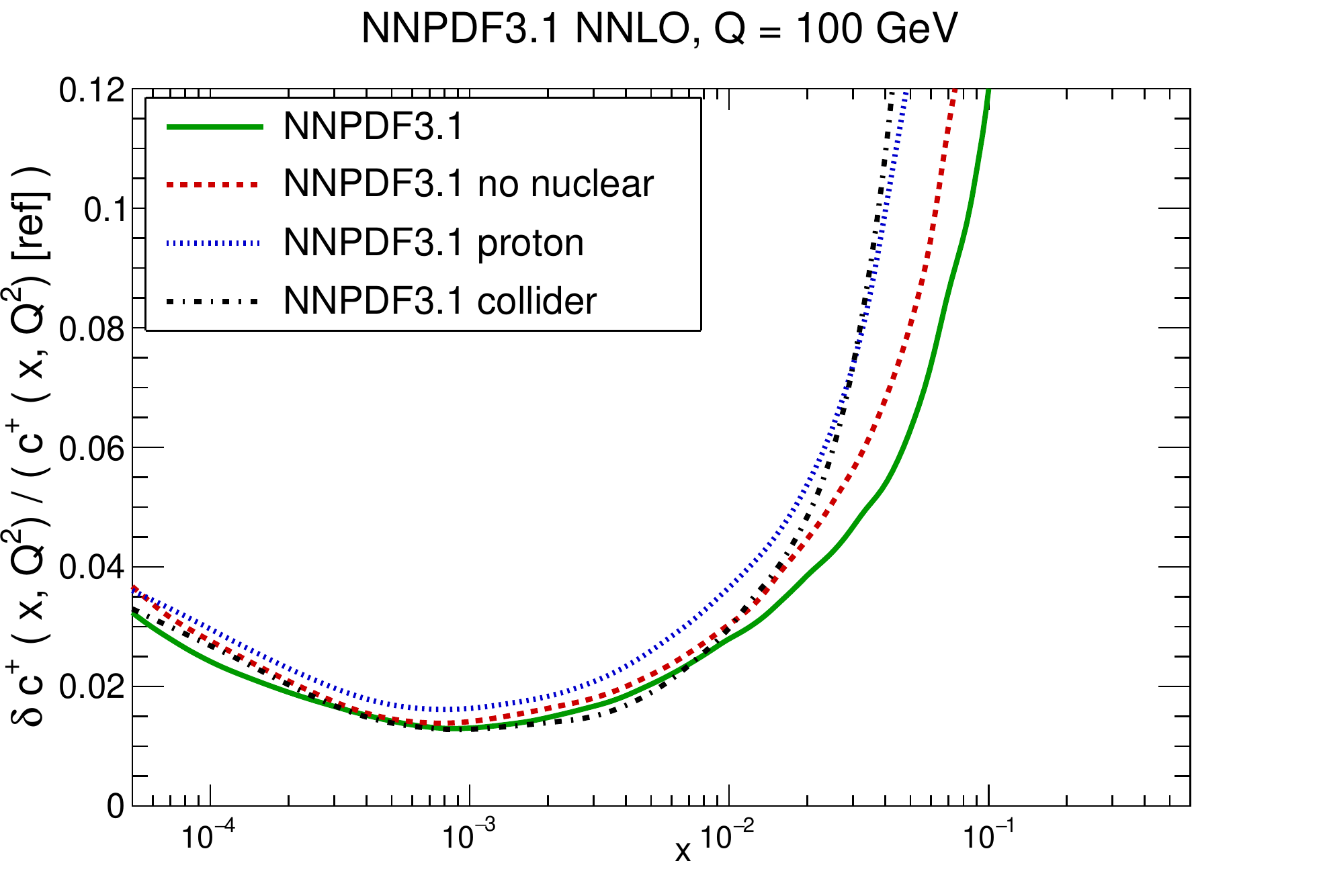}
  \includegraphics[scale=0.32]{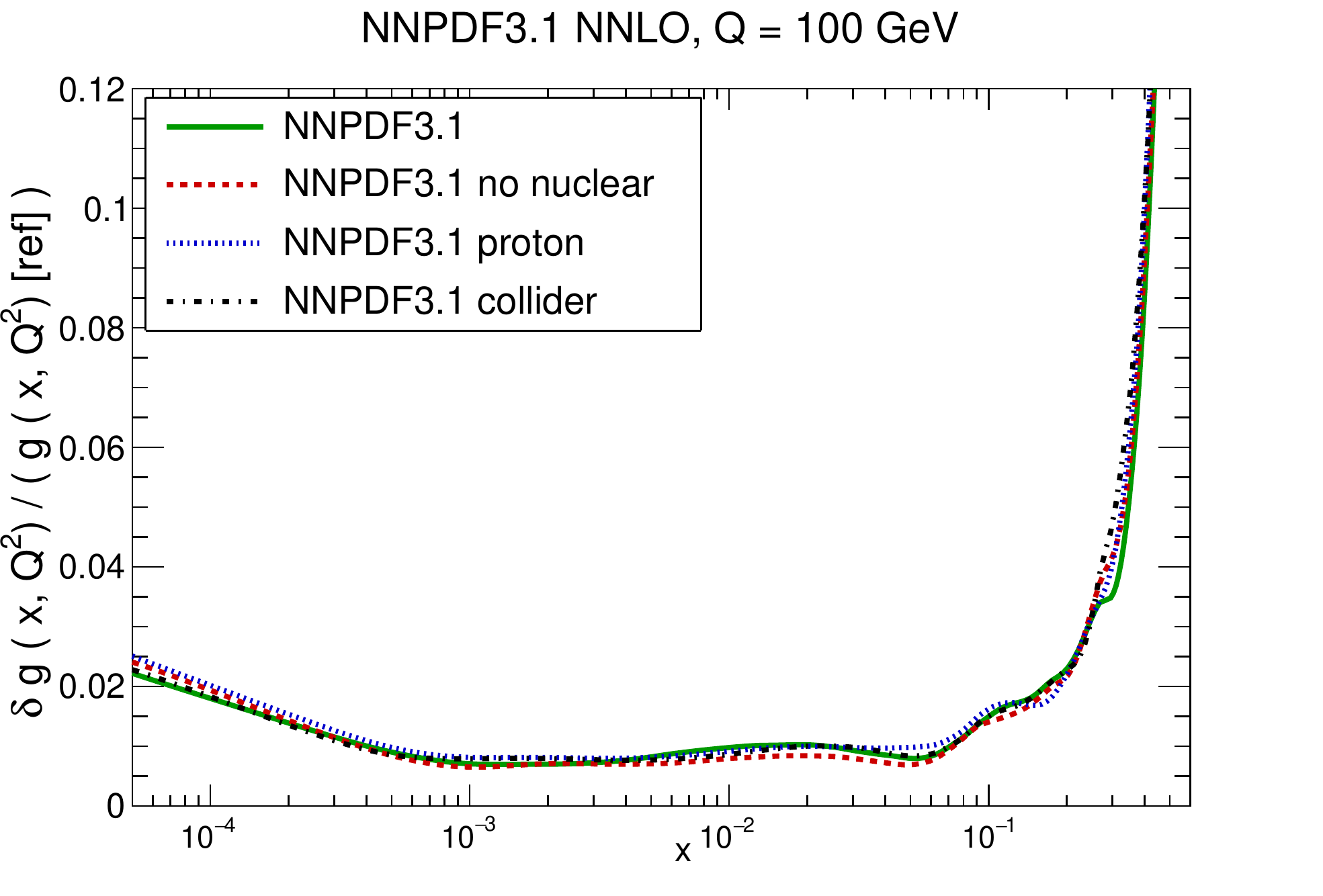}
  \caption{\small Comparison of the relative
    PDF uncertainties at $Q=100$ between the NNPDF3.1 and the no heavy nuclei, 
proton--only and collider--only PDF determinations. The uncertainties shown are all
    normalized to the NNPDF3.1 central value.
    \label{fig:ERR-31-nnlo-datasetvar}
  }
\end{center}
\end{figure}

\subsection{Collider-only parton distributions}
\label{sec:collideronly}

A yet more conservative option to that discussed in the previous
Section is to retain only collider data from HERA, the
Tevatron and the LHC. The motivation for this suggestion,
first presented in the NNPDF2.3 study~\cite{Ball:2012cx}, is that
this excludes data taken at low
scales, which may be subject to potentially large perturbative and non-perturbative
corrections. Furthermore, data taken on nuclear targets, and all of the older datasets 
are eliminated, thereby leading to a more reliable set of PDFs.
However, previous collider-only PDFs had very large uncertainties, due to the 
limited collider dataset then available.

\begin{figure}[t]
\begin{center}
  \includegraphics[scale=1]{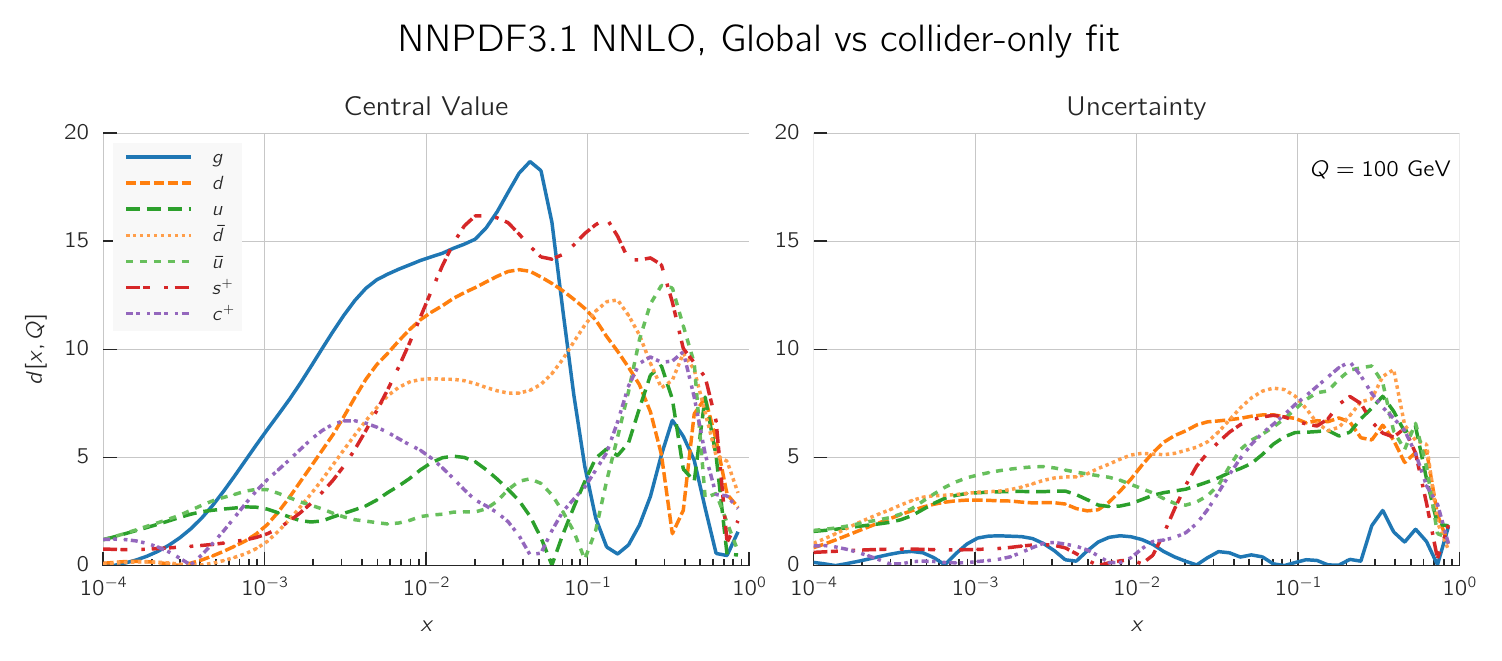}
  \caption{\small 
Same as Fig.~\ref{fig:distances_noZpT} but now
 only keeping collider data.
    \label{fig:distances_collider}
  }
\end{center}
\end{figure}

In order to re-assess the situation with the current, much wider LHC dataset,
we have repeated a collider-only PDF determination. This amounts to
repeating the proton only PDF determination 
described in the previous Section, but now with the proton fixed target data also removed. 
The distances between the ensuing
PDFs and the default NNPDF3.1 are shown in
Fig.~\ref{fig:distances_collider}. Comparing to
Fig.~\ref{fig:distances_proton} we notice
that distances are similar for most PDFs, the main exception
being the gluon, which in the intermediate $x$ region has now shifted 
by almost two sigma. 

\begin{figure}[t]
\begin{center}
  \includegraphics[scale=0.38]{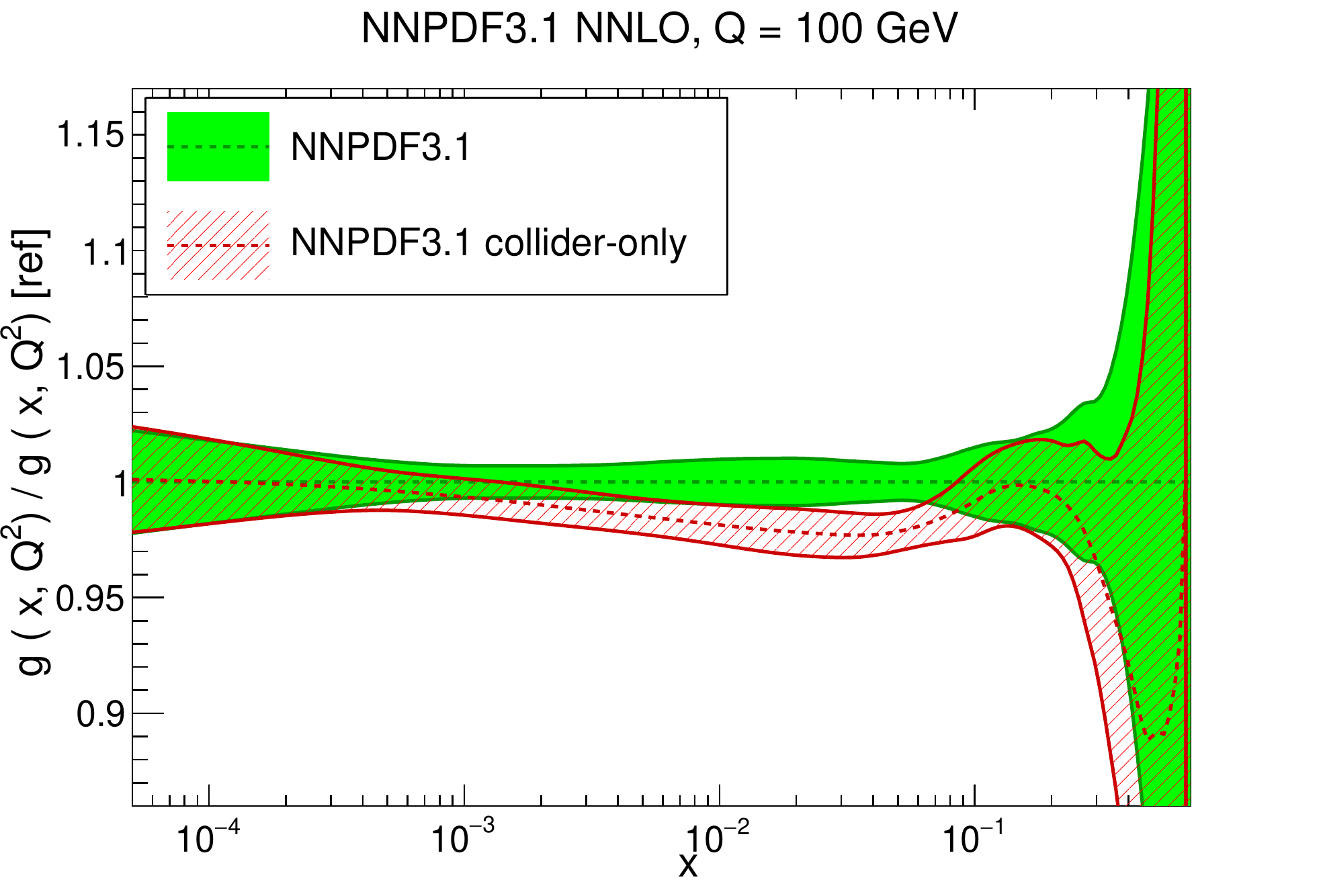}
  \includegraphics[scale=0.38]{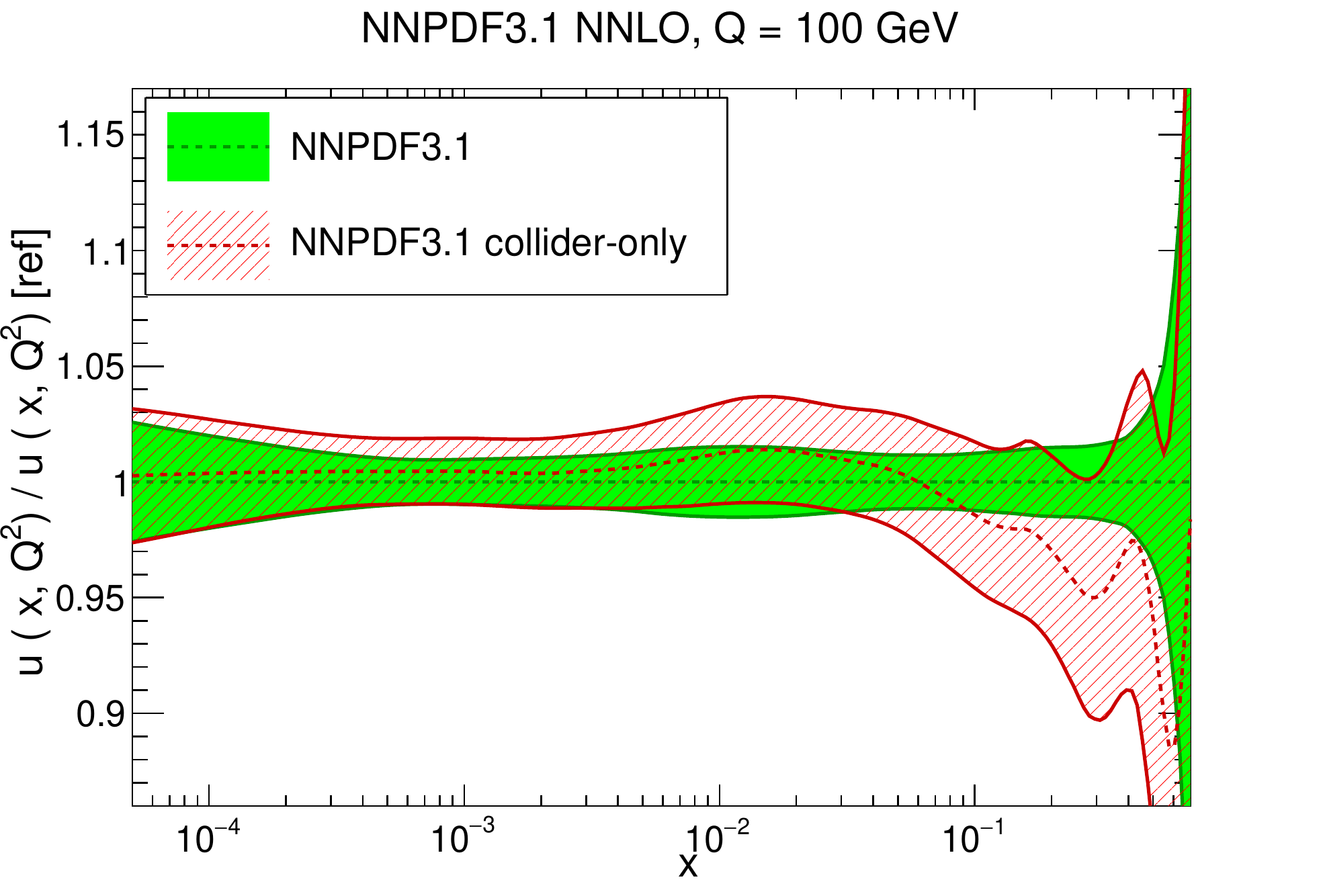}
  \includegraphics[scale=0.38]{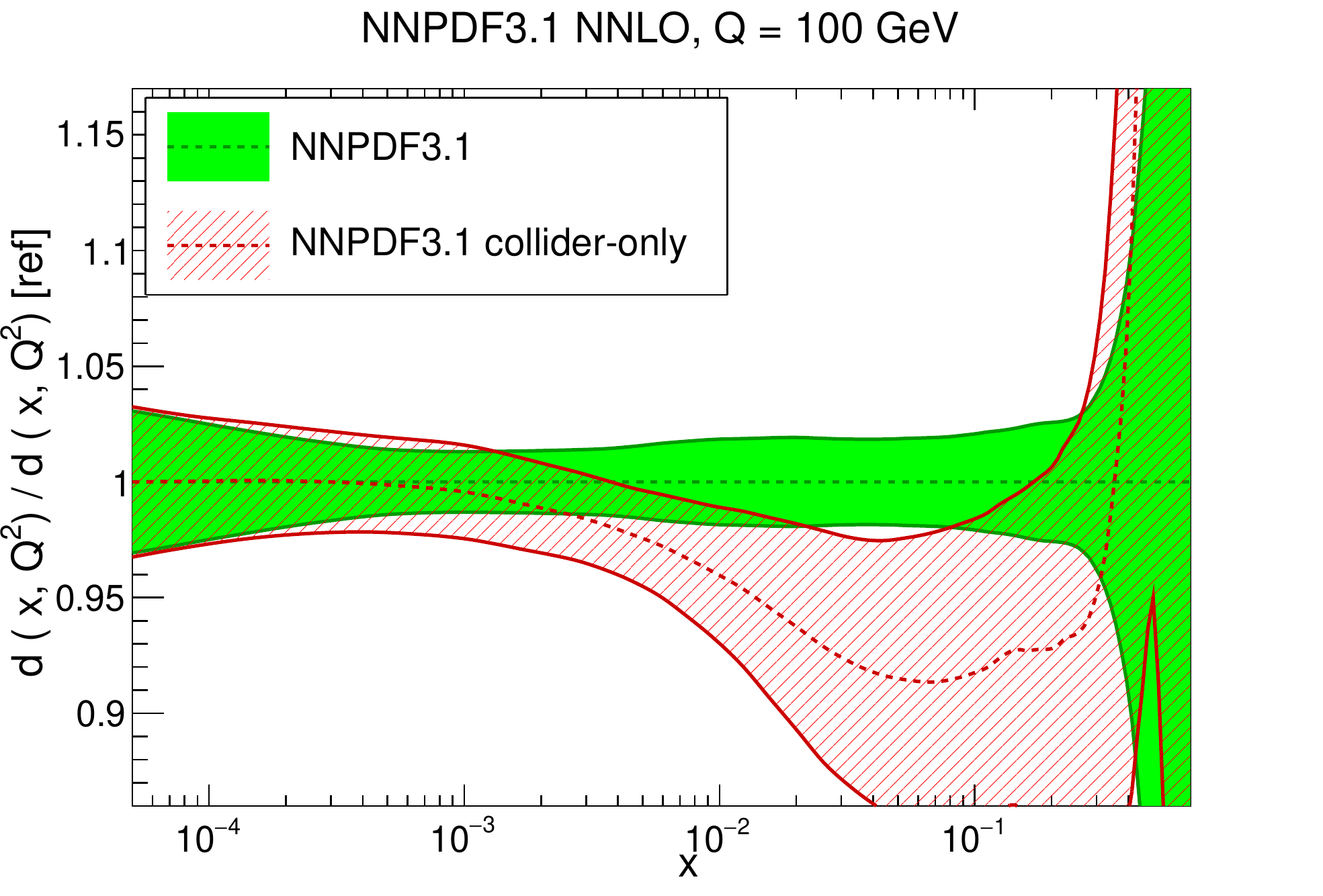}
  \includegraphics[scale=0.38]{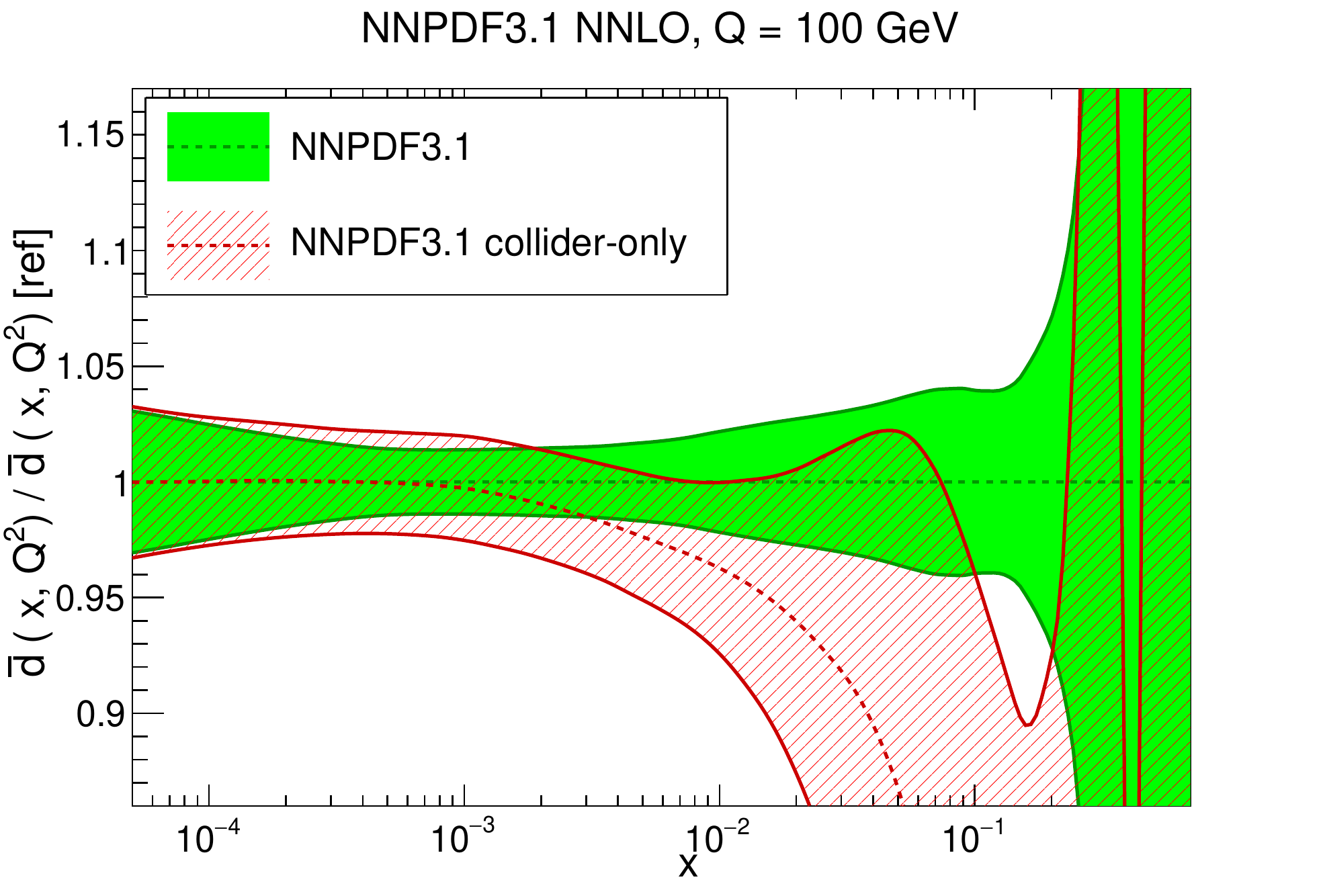}
  \caption{\small 
Same as Fig.~\ref{fig:31-nnlo-ZpT} 
but now only keeping collider data.
Results are
shown for the gluon (top left), up (top right), down (bottom left) and
antidown (bottom right).
    \label{fig:pdfs-collideronly}
  }
\end{center}
\end{figure}

This is confirmed by a direct comparison of PDFs in
Fig.~\ref{fig:pdfs-collideronly} and their uncertainties in Fig.~\ref{fig:ERR-31-nnlo-datasetvar}. 
The valence quarks (especially up) are reasonably stable, but the sea now is quite unstable upon 
removal of all the fixed target data, visibly more than in the proton only set
Fig.~\ref{fig:31-nnlo-proton}, with in particular a substantial increase in the 
uncertainty of the anti-up quark distribution at large $x$. Furthermore, the gluon, which in
Fig.~\ref{fig:31-nnlo-proton} was quite stable in the proton-only PDF
set, now undergoes a significant downward shift at intermediate $x$, even though its 
uncertainty is not substantially increased. 

We conclude that, despite impressive improvements due to recent LHC measurements, 
a collider-only PDF determination is still not very useful for general phenomenological applications.

\clearpage
\section{Implications for phenomenology}
\label{sec:pheno}

We now present some initial studies
of the phenomenological implications of NNPDF3.1 PDFs. 
Firstly, we summarize the status of PDF uncertainties building
upon the discussion in the previous Section; we discuss the status of 
PDF uncertainties, and then we focus on
the strange and charm PDF. We then discuss PDF luminosities which are
the primary input to hadron collider processes, and predictions for
LHC processes, specifically $W$, $Z$ and Higgs production at
the LHC. 
As elsewhere, only a selection of results is presented here, with
a much larger set is  available online 
as discussed in Section~\ref{sec:delivery}.

\subsection{Improvements in PDF uncertainties}

\label{sec:phenopdfs}

\begin{figure}[t!]
\begin{center}
  \includegraphics[scale=0.38]{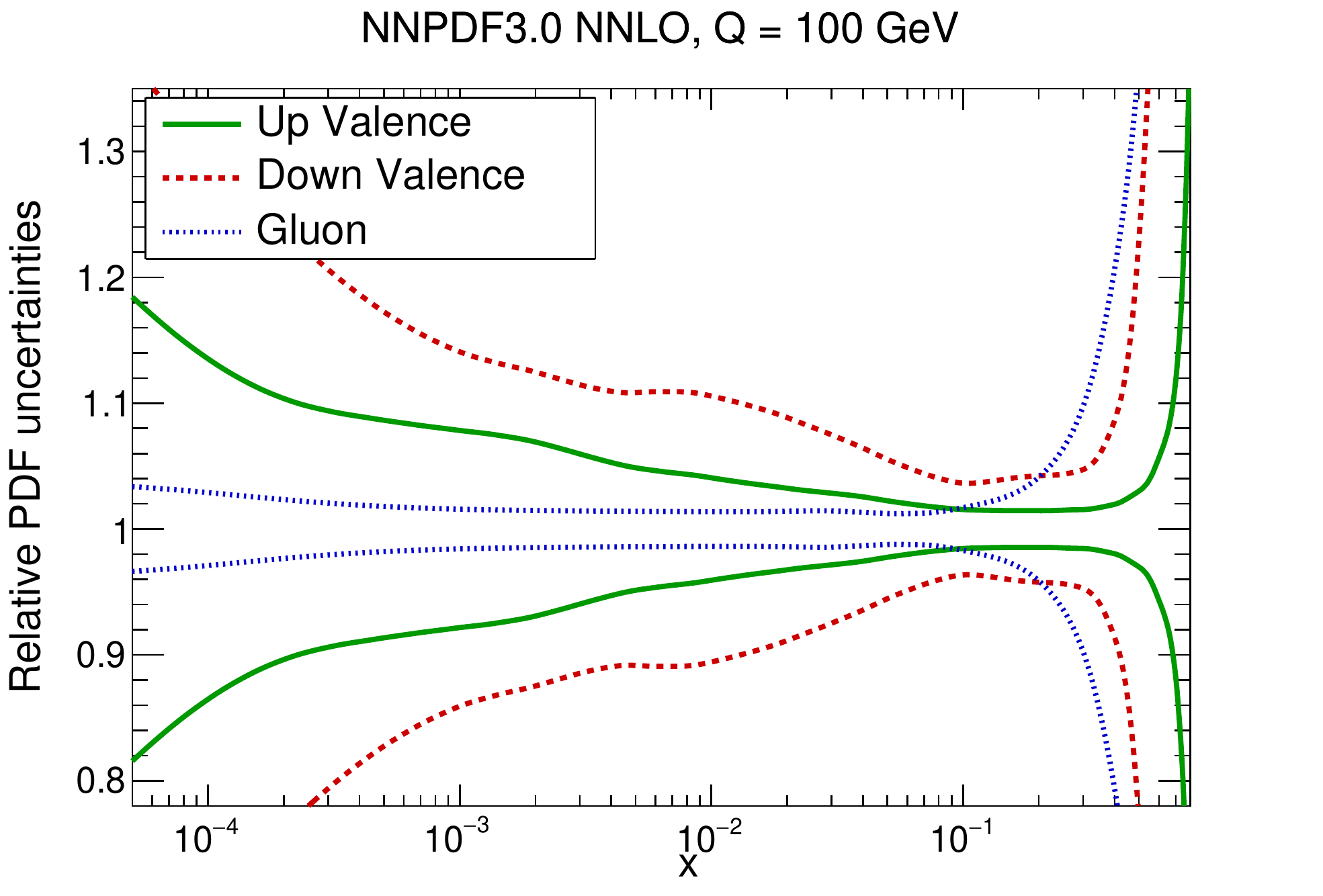}
  \includegraphics[scale=0.38]{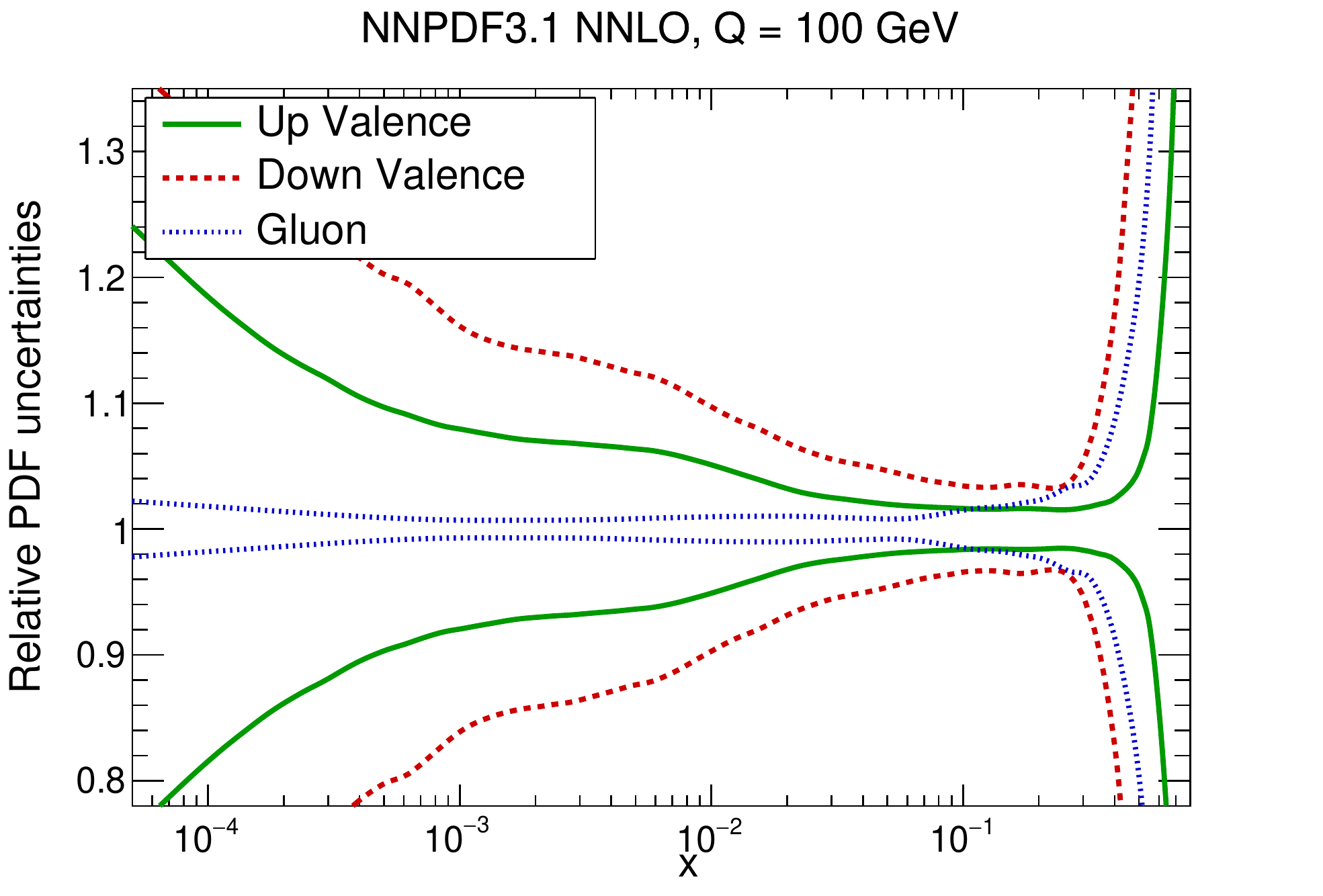}
  \includegraphics[scale=0.38]{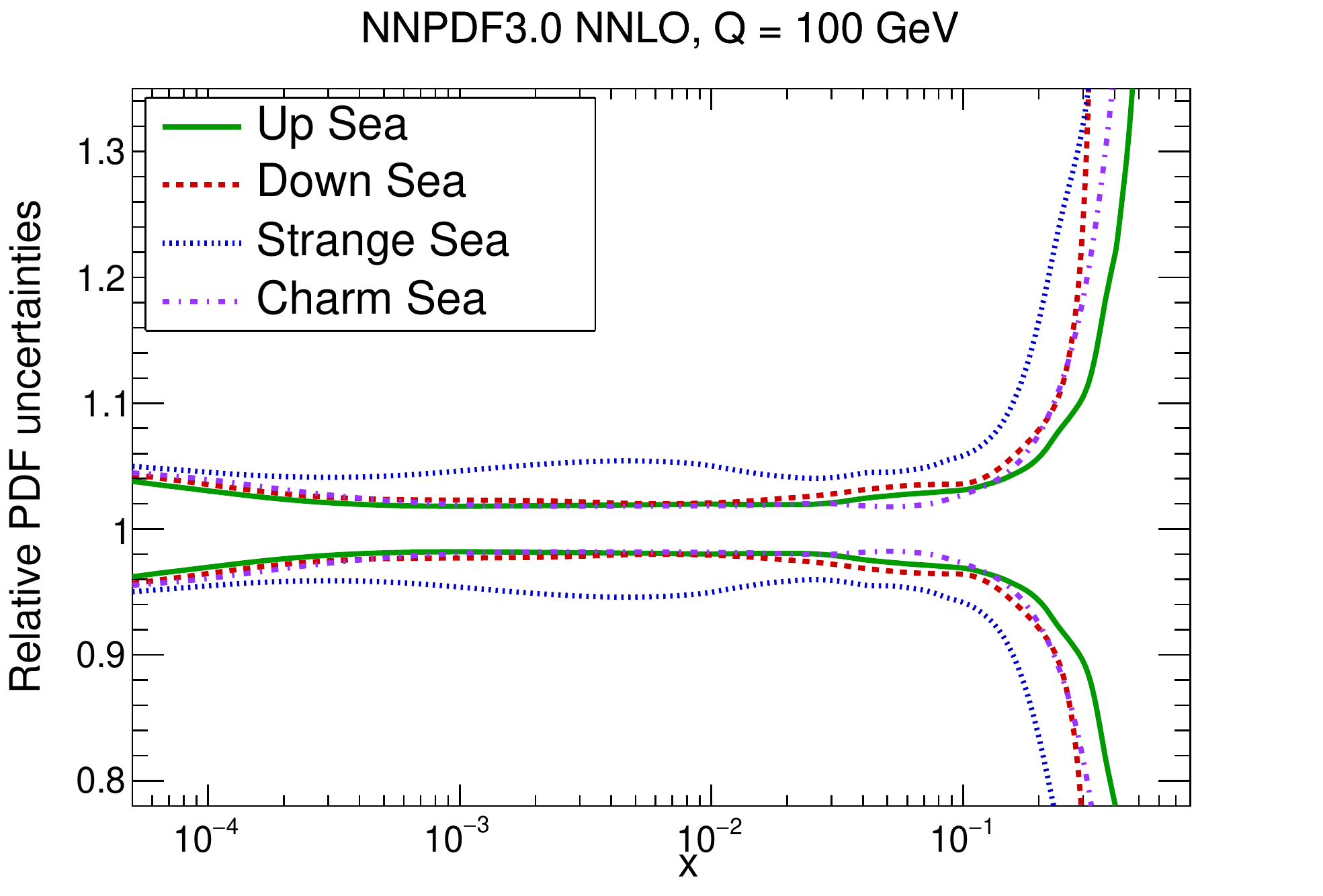}
  \includegraphics[scale=0.38]{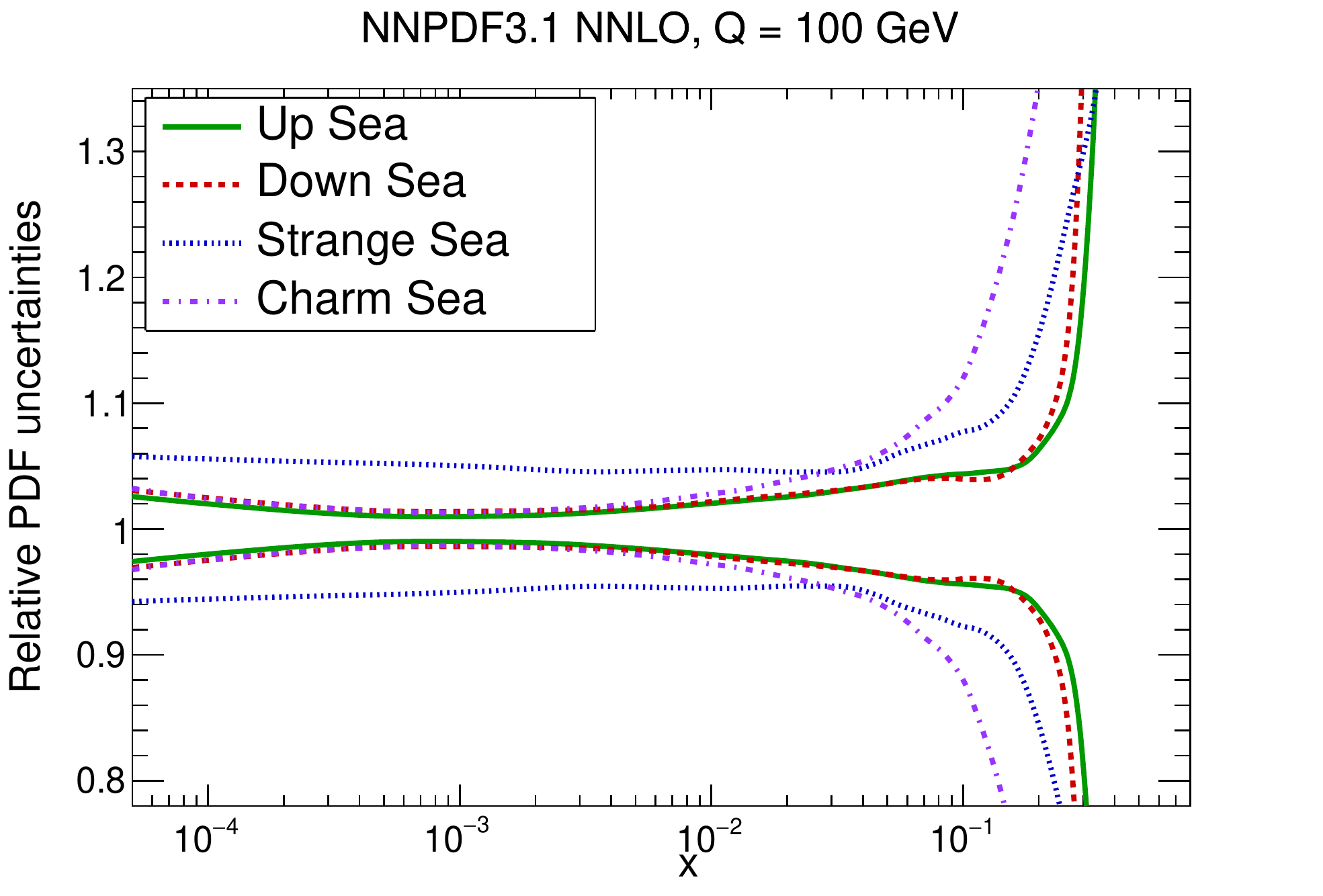}
  \includegraphics[scale=0.38]{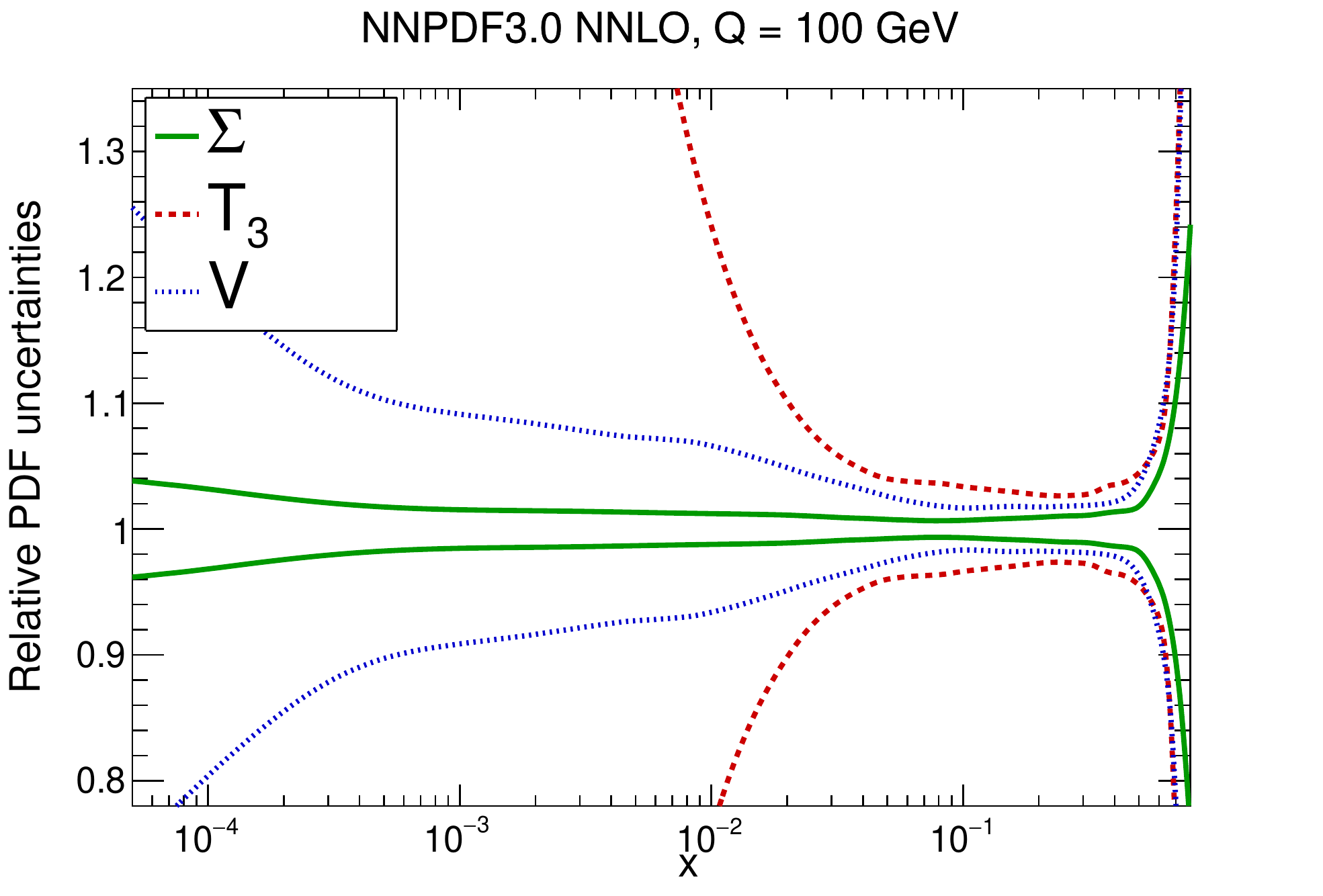}
  \includegraphics[scale=0.38]{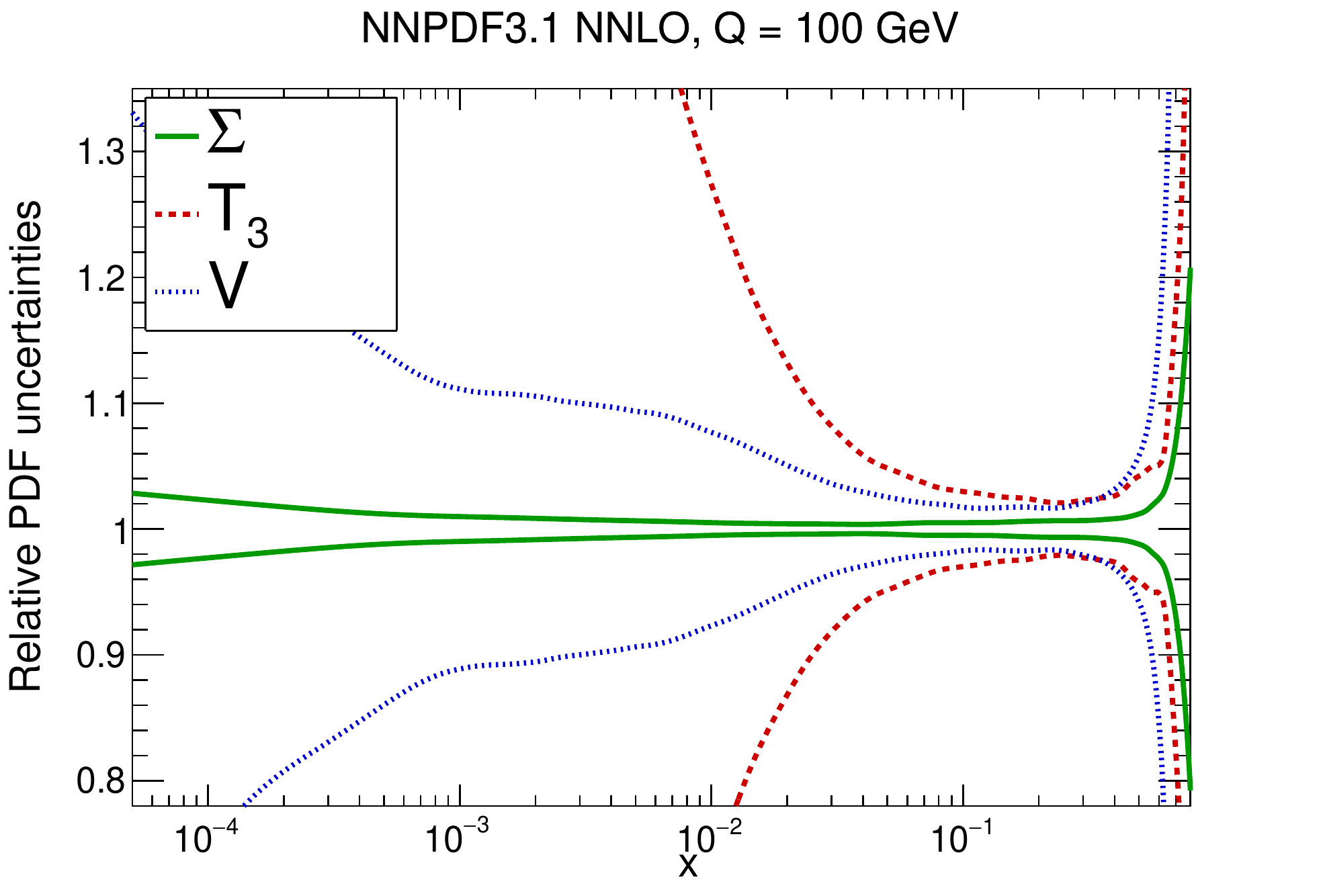}
  \caption{\small 
Comparison of relative uncertainties on NNPDF3.0 (left) and NNPDF3.1
  (right) NNLO PDFs, normalized to the NNPDF3.1
  NNLO central value. The two light quark valence PDFs and the gluon
   are shown  (top) along with all individual sea PDFs (center) and the
   singlet, valence and isospin triplet combinations (bottom).
\label{fig:pdfunc31vs30}}
\end{center}
\end{figure}

After discussing in Section~\ref{sec:impactnewdata} the impact on
NNPDF3.1 PDFs of each
individual new piece of data and after separating off the effect of the
new methodology, we can now study the combined effect of all the new data 
by comparing PDF uncertainties in NNPDF3.0 and NNPDF3.1. This is done in
Fig.~\ref{fig:pdfunc31vs30}, where we compare relative PDF uncertainties
(all computed with a common normalization) on PDFs in the NNPDF3.0 and
NNPDF3.1 sets, shown as valence (i.e. $q-\bar q$), and sea 
(i.e. $\bar q$) for up and down  and $q+\bar q$ for strange and
charm. Results are shown both for individual PDF flavors, and for the
singlet, valence, and triplet combinations defined in Eq.~(3.4) of
Ref.~\cite{Ball:2016neh}). 

The most visible effect is the very considerable reduction in gluon
uncertainty, which is now at the percent level for almost all $x$. As
discussed in  Section~\ref{sec:impactnewdata}, this is due to the
combination of many mutually consistent constraints on the gluon from
DIS (especially at HERA), $Z$ transverse momentum distributions, jet
production, and top pair production, which taken together cover a
very wide kinematic range. The singlet quark combination, which mixes
with the gluon, shows a comparable improvement for all $x\lsim 0.1$,
but less marked at large $x$.

Interestingly, for quark PDFs the pattern of uncertainties is
different in the flavor basis versus  the ``evolution'' basis as given in Eq.~(3.4) of
Ref.~\cite{Ball:2016neh}). Specifically, 
the aforementioned reduction in uncertainty on the singlet combination
is not seen in any of the 
light quark valence distributions, which generally have comparable
uncertainties in NNPDF3.1 and NNPDF3.0. This is due to the fact that
flavor separation is somewhat more uncertain in NNPDF3.1, due to the
fact that charm is now independently parametrized. This is compensated
by the availability of more experimental information (in particular
LHCb and ATLAS data), but not at small and very large $x$. Indeed, the
valence and triplet distributions  have generally somewhat larger
uncertainties  in
NNPDF3.1 than in NNPDF3.0, except for $x\sim0.1$.

This  pattern is also seen in  sea uncertainties. At small
$x\lsim 10^{-2}$ they are all comparable, and all (except for strangeness) 
rather smaller than
the corresponding uncertainties in NNPDF3.0, since they are driven by
the mixing of the singlet and the gluon through perturbative
evolution at small $x$~\cite{das}. Even the charm PDF, now
independently parametrized, has
a smaller uncertainty. In the large-$x$ region, instead, the less
accurate knowledge of flavor separation kicks in, and  relative uncertainties 
are larger.
Clearly, in NNPDF3.0
charm had an unnaturally small uncertainty, since at large $x$ 
perturbatively-generated charm is tied to the gluon. In NNPDF3.1, instead, the
hierarchy of uncertainties on sea PDFs in the valence region is what
one would expect, with up and down known most accurately, and strange
and charm affected by increasingly large uncertainties. The
uncertainty in NNPDF3.1 on the up and especially down sea components is a little 
increased, but 
still comparable to NNPDF3.0, while the uncertainty in strangeness is
stable and that on 
charm significantly increased, as it should be given that large-$x$
charm is largely unconstrained by data. In this respect, it is
interesting to observe that a more accurate determination of charm
and other sea PDFs at large $x$ can be achieved through the inclusion
of the EMC dataset as discussed in Section~\ref{sec:emc} above. Future LHC
data on processes such as $Z+c$ production may confirm the reliability
of the EMC dataset.

\subsection{The strange PDF}
\label{sec:strangeness}

Whereas there is broad consensus on the size, i.e. the central value,
of up and down PDFs, for which there is good agreement between existing
determinations within their small uncertainty, the size of the strange
PDF has been the object of some controversy, which we revisit here in
view of NNPDF3.1 results.
Specifically, the strange fraction of proton quark sea, defined as
\be
\label{eq:rs}
R_s (x,Q^2)=\frac{ s(x,Q^2)+\bar{s}(x,Q^2) }{ \bar{u}(x,Q^2)+\bar{d}(x,Q^2) } \, .
\ee
and the corresponding ratio of momentum fractions
\be
\label{eq:rsint}
K_s (Q^2)=\frac{\int_0^1 dx\, x \lp s(x,Q^2)+\bar{s}(x,Q^2)\rp }{ 
\int_0^1 dx\, x\lp \bar{u}(x,Q^2) + \bar{d}(x,Q^2)\rp }  \, ,
\ee
have been traditionally assumed to be significantly smaller than one,
and in PDF sets produced before the strange PDF could be extracted
from the data, such as e.g. NNPDF1.0~\cite{Ball:2008by}, it was often
assumed that $R_s\sim\frac{1}{2}$, for all $x$, and thus also 
$K_s\sim\frac{1}{2}$. This level of strangeness suppression is indeed found in
many recent global PDF sets, in which the strongest handle on the strange
PDF is provided by deep-inelastic neutrino inclusive $F_2$ and  charm
$F_2^c$ (``dimuon'') data. 

This was challenged in Ref.~\cite{Aad:2012sb} where, on the
basis of ATLAS $W$ and $Z$ production data, combined with HERA DIS
data, 
it was argued instead that,
in the measured region, the strange fraction $R_s$ is of order one. In
Refs.~\cite{Ball:2012cx,Ball:2015oha}, respectively based on the
NNPDF2.3, and NNPDF3.0 global analyses, both of which
includes the data of Ref.~\cite{Aad:2012sb}, it was concluded that
whereas the ATLAS data do favor a larger strange PDF, they have a moderate
impact on the global PDF determination 
due to large uncertainties, and also, that if
the strange PDF is only determined from HERA and ATLAS data, the
central value is consistent with the conclusion of
Ref.~\cite{Aad:2012sb}, but the uncertainty is large enough to lead to
agreement with the suppressed strangeness of the global PDF sets to within
one sigma. In Ref.~\cite{Ball:2015oha} it was also shown that the CMS
$W+c$ production data~\cite{Chatrchyan:2013uja}, which were included
there for the first time and which are also included in NNPDF3.1, 
though only in the NLO
determination because of lack of knowledge of the NNLO corrections,
have a negligible impact due to their large uncertainties. 

As we discussed in Section~\ref{sec:atlaswz},  ATLAS $W$ and $Z$
production data have been supplemented by the rather more accurate
dataset of Ref.~\cite{Aaboud:2016btc}, also claimed to 
favor enhanced strangeness. Indeed, we have seen in
Section~\ref{sec:atlaswz} that strangeness is significantly enhanced by
the inclusion of these data, and also, in Section~\ref{sec:results-mc},
that this enhancement  can be accommodated in the global PDF
determination 
thanks to the independently parametrized  charm PDF, which is a new
feature to NNPDF3.1.
It is thus interesting to re-asses strangeness in NNPDF3.1, by
comparing theoretically motivated choices of dataset: we will thus compare
to the
previous NNPDF3.0 results for strangeness obtained using the default
NNPDF3.1, the 
collider-only PDF set of Section~\ref{sec:collideronly}, which can be
considered to be theoretically more reliable, and a PDF set which we
have constructed by using NNPDF3.1 methodology, but only including all
HERA inclusive structure function data from
Tab.~\ref{tab:completedataset} and the ATLAS data of
Ref.~\cite{Aaboud:2016btc}. Because inclusive DIS data alone cannot
determine separately strangeness~\cite{Forte:2010dt} this is then a
determination of strangeness which fully relies on the ATLAS data.

\begin{figure}[t]
\begin{center}
  \includegraphics[scale=0.35]{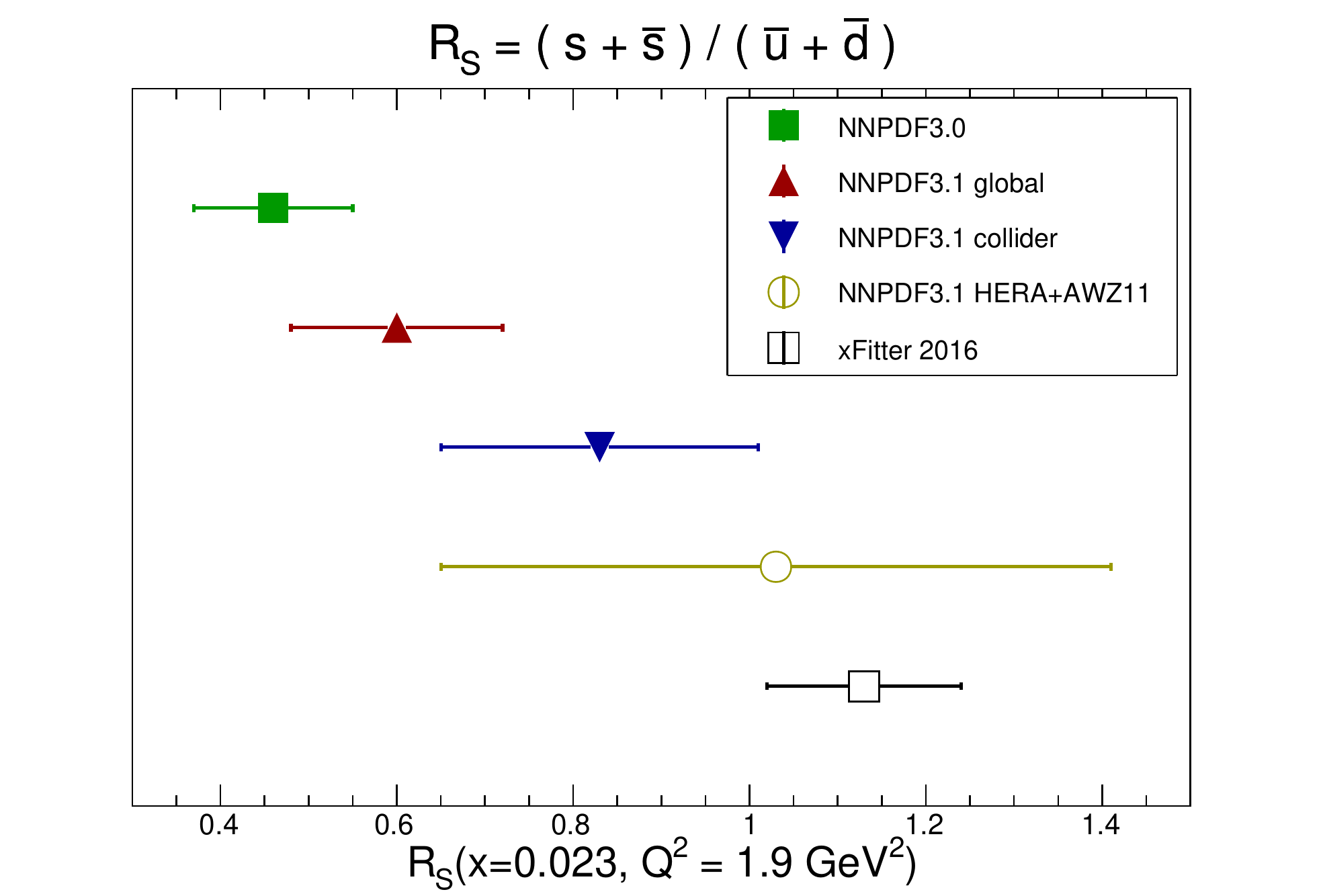}
  \includegraphics[scale=0.35]{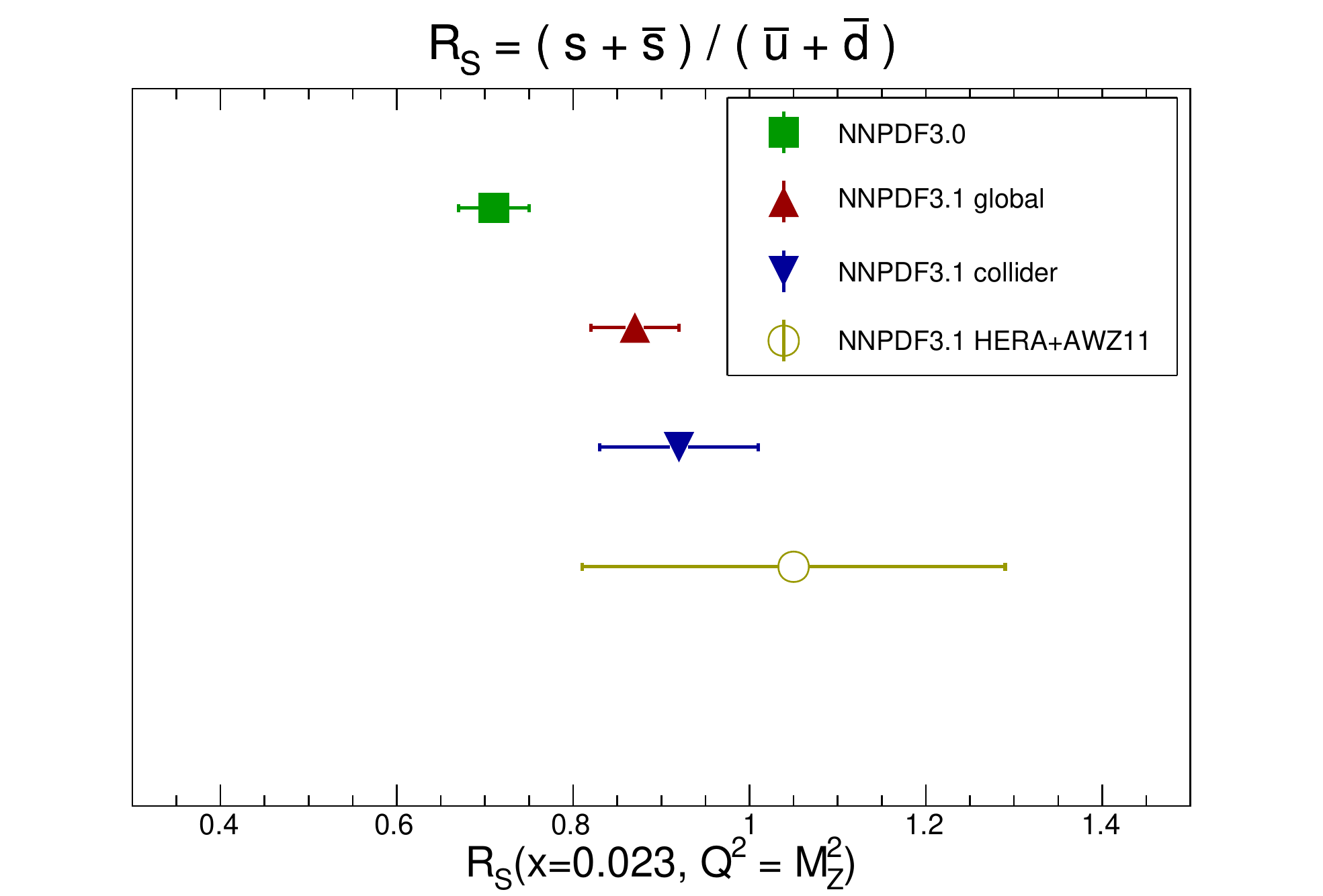}
  \caption{\small Graphical representation of the results
    of Table~\ref{tab:rs}.
\label{fig:xRsplot}}
\end{center}
\end{figure}

In Table~\ref{tab:rs}  we show NNLO
results, obtained using these different PDF sets,
 for $R_s(x,Q)$ Eq.~\ref{eq:rs} at
 $Q=1.38~{\rm GeV}$ (thus below charm threshold) and $Q=m_Z$ and
$x=0.023$, an $x$ value chosen by ATLAS in order to maximize
sensitivity. Results are also compared to that of
Ref.~\cite{Aaboud:2016btc}.
A graphical representation of the table is in Fig.~\ref{fig:xRsplot}.

\begin{table}[t]
  \begin{center}
    \small
        \renewcommand{\arraystretch}{1.2}
\begin{tabular}{|l|c|c|}
\hline
\centering PDF set &  $R_s(0.023,1.38~\mathrm{GeV})$ & $R_s(0.023,M_{Z})$ \\
\hline
\hline
NNPDF3.0    & 0.45$\pm$0.09    & 0.71$\pm$0.04 \\
NNPDF3.1     & 0.59$\pm$0.12  & 0.77$\pm$0.05 \\
NNPDF3.1 collider-only   & 0.82$\pm$0.18  & 0.92$\pm$0.09 \\
NNPDF3.1 HERA + ATLAS $W,Z$   & 1.03$\pm$0.38  & 1.05$\pm$0.240\\	
\hline
{\tt xFitter} HERA + ATLAS $W,Z$ (Ref.~\cite{Aaboud:2016btc}) 	& $1.13
\,{}^{+0.11}_{-0.11}$& - \\
\hline
\end{tabular}
\end{center}
\caption{\small \label{tab:rs} 
  The strangeness fraction $R_s(x,Q)$  Eq.~(\ref{eq:rs}) at $x=0.023$,
  at a low scale and a high scale. We show results obtained using  
NNPDF3.0, and NNPDF3.1 baseline, collider-only and  HERA+ATLAS $W,Z$
sets, compared to the {\tt xFitter}  ATLAS value  Ref.~\cite{Aaboud:2016btc}.
}
\end{table}

First, comparison of  the NNPDF3.1 HERA+ATLAS $W,Z$ result with that of
Ref.~\cite{Aaboud:2016btc}, based on the same data, shows agreement at
the one-sigma level, with a similar central value and a
greatly increased 
uncertainty, about four times larger, most likely because of the more
flexible parametrization and because of independently parametrizing charm.
Second, strangeness in NNPDF3.1 is rather larger than in NNPDF3.0: as
we have shown in Sects.~\ref{sec:results-mc},\ref{sec:atlaswz} this is
largely due to the effect of the
ATLAS~$W,Z$~2011 data, combined with determining  charm from the data:
indeed, it is clear
from Fig.~\ref{fig:31-nnlo-old-vs-new} that the new data and new
methodology both lead to strange enhancement, with the former effect
dominant but the latter not negligilbe. 
This enhancement is more marked in the collider-only PDF
set, which leads to a value which is very close to that coming from
the ATLAS data. This suggests some tension between 
strangeness preferred by collider data and the rest of the dataset,
i.e., most likely, neutrino data.

\begin{figure}[t]
\begin{center}
  \includegraphics[scale=0.37]{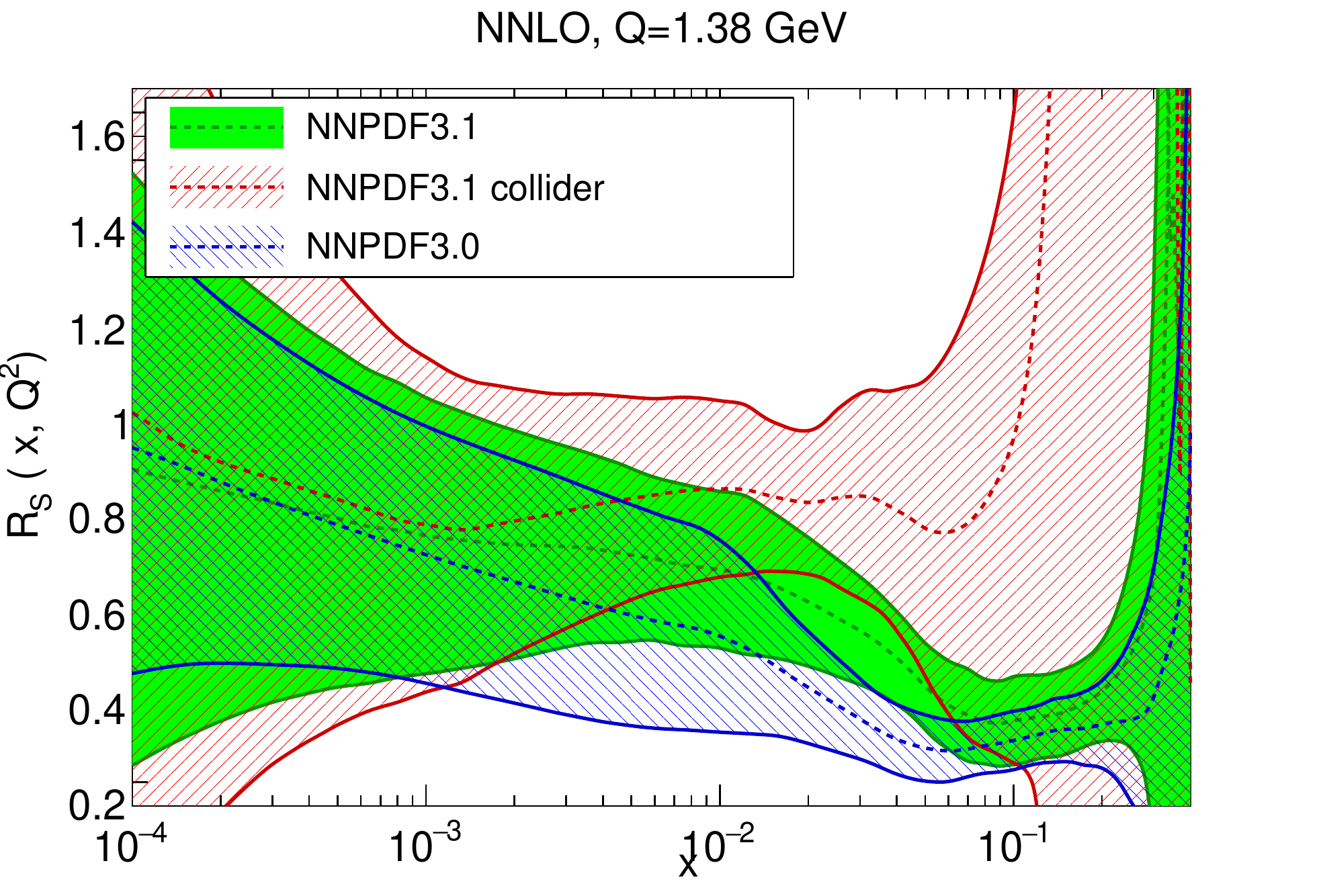}
  \includegraphics[scale=0.37]{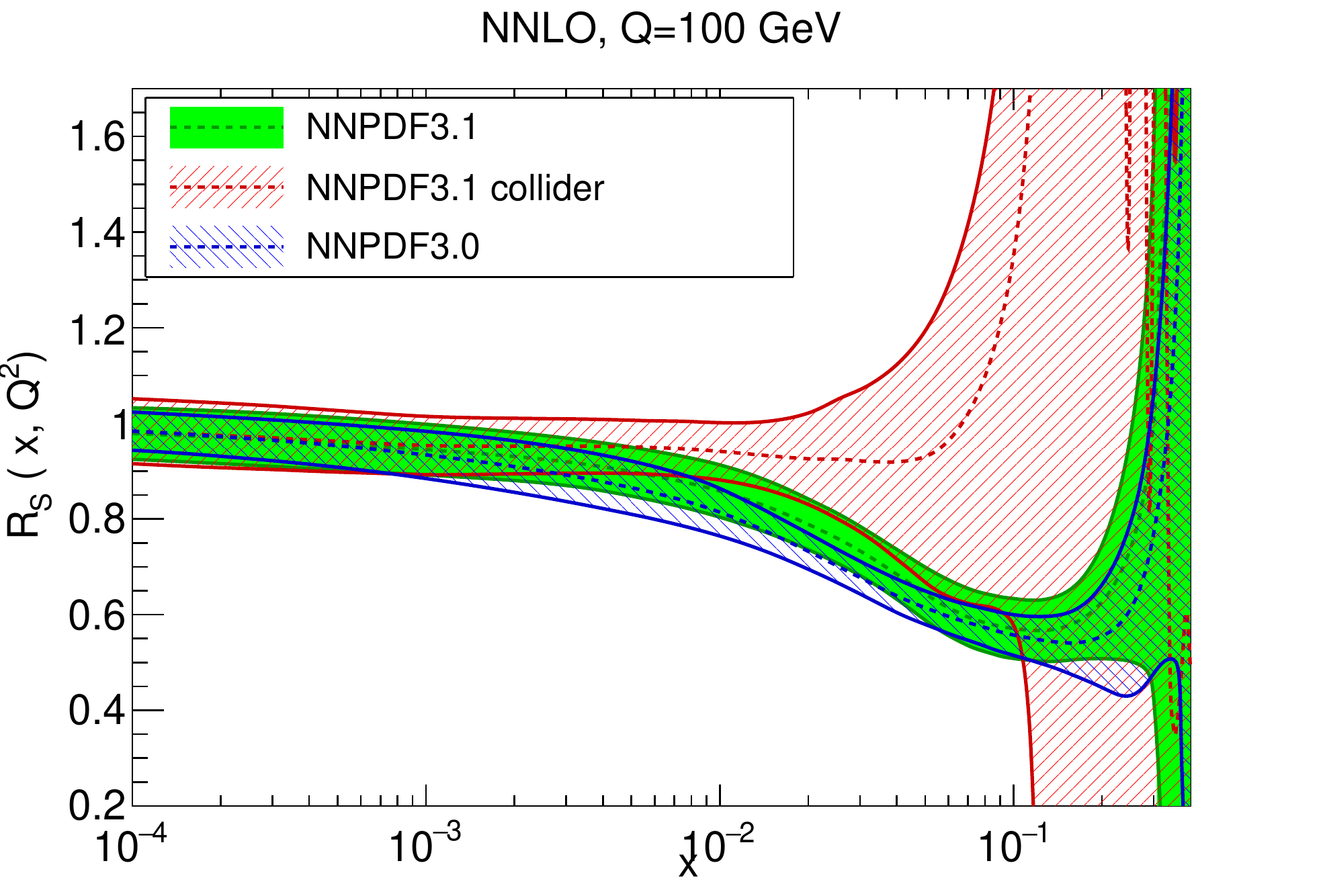}
   \includegraphics[scale=0.37]{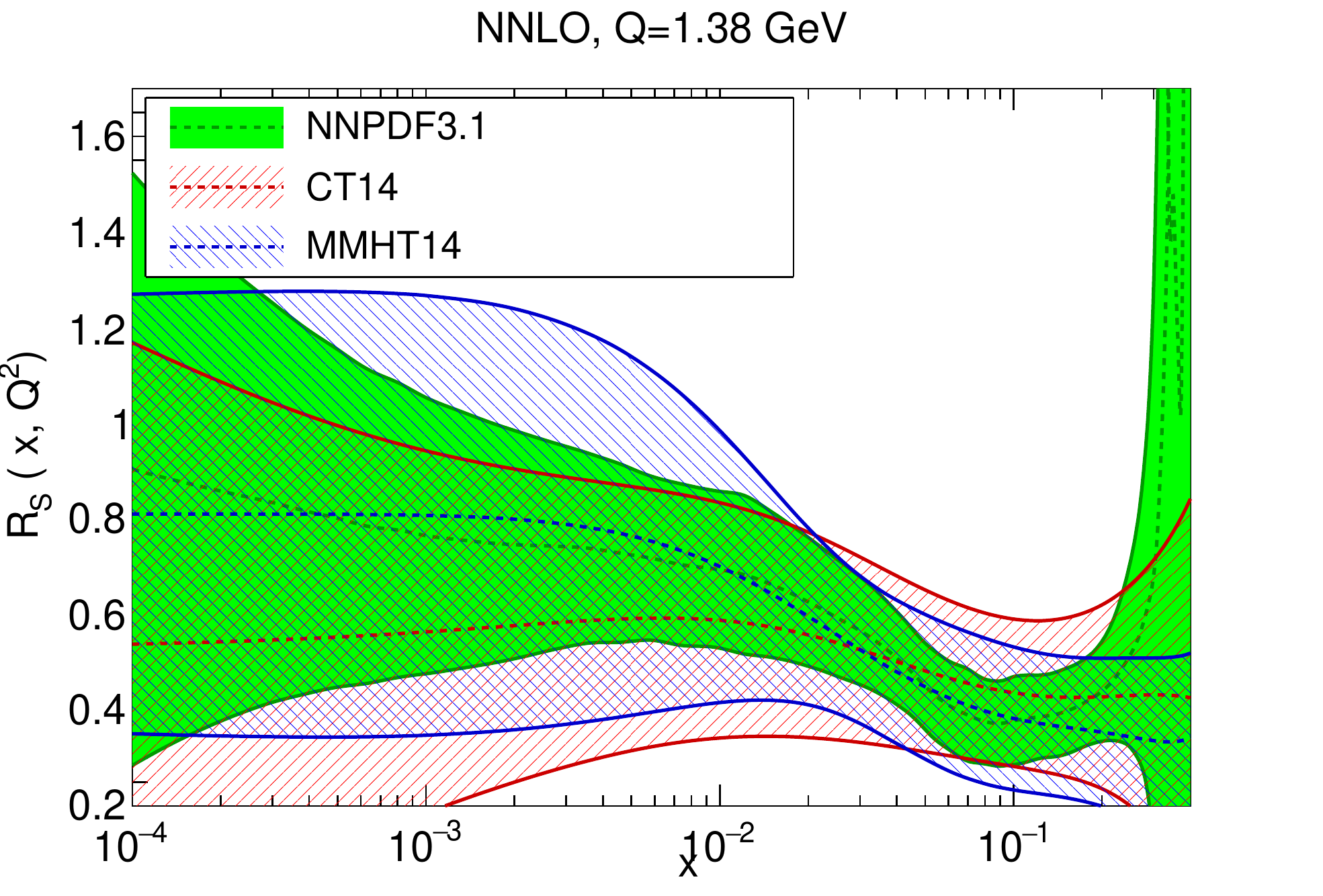}
  \includegraphics[scale=0.37]{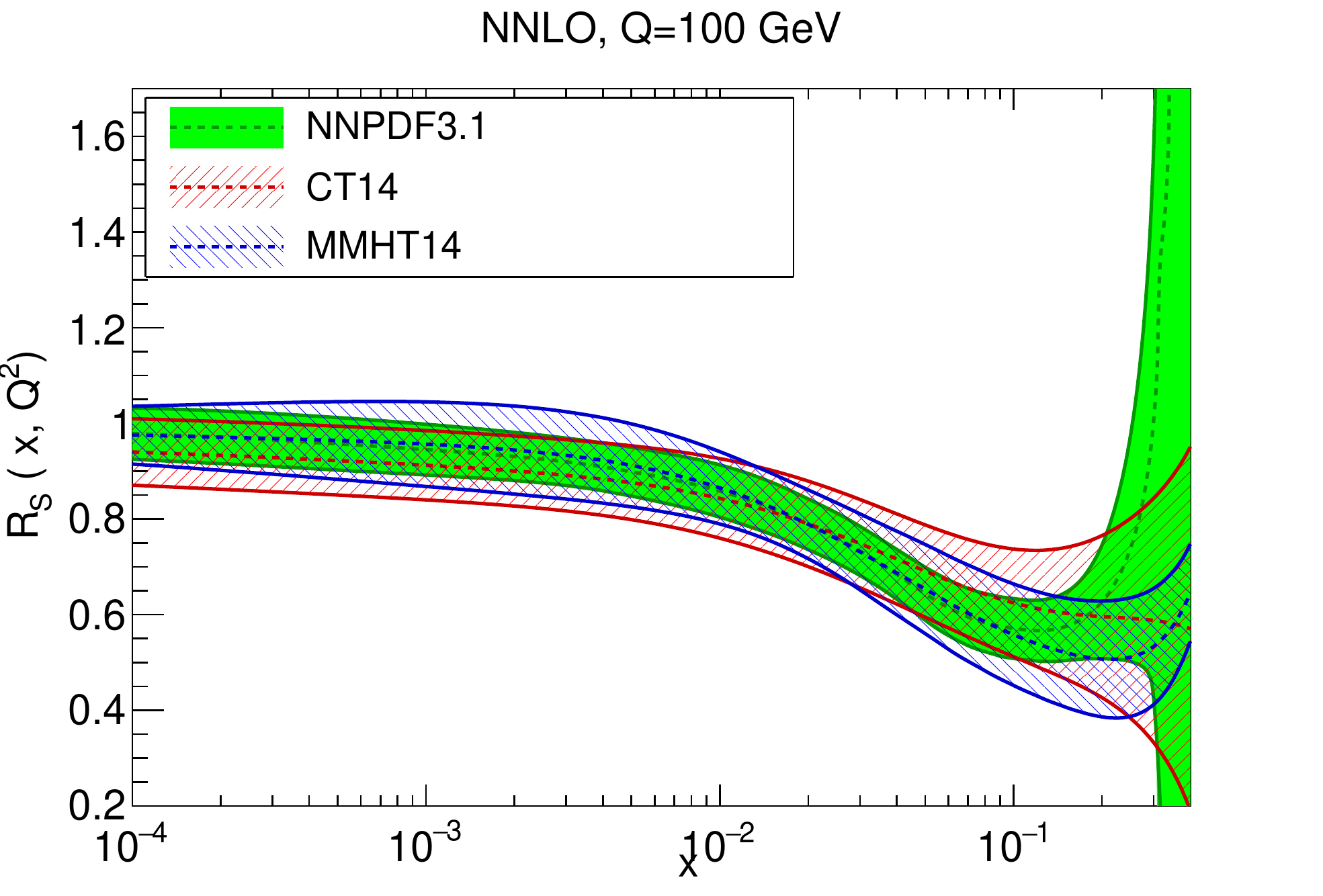}
  \caption{\small The strangeness ratio $R_s(x,Q)$ Eq.~(\ref{eq:rs})
    as a function of 
    $x$ for two values of $Q$, $Q=1.38$ GeV (left)
    and $Q=m_Z$ (right).
     Results are shown comparing NNPDF3.1 to  NNPDF3.1 and the
     collider-only NNPDF3.1 (top), and to CT14 and MMHT (bottom).
\label{fig:xRs-abs-atlaswz2011}}
\end{center}
\end{figure}

It is interesting to repeat this analysis for the full $x$ range. 
This is done in Fig.~\ref{fig:xRs-abs-atlaswz2011}, where
$R_s(x,Q)$ Eq.~(\ref{eq:rs}) is plotted as a function
of $x$ again at low and high scales, now only including NNPDF3.0, and
the default and collider-only versions of NNPDF3.1. It is clear that
in the collider-only PDF set strangeness is largely unconstrained   at
large $x$, whereas the global fit is constrained by neutrino data to
have a suppressed value $R_s\sim 0.5$. At lower $x$ we see the tension
between this and the constraint
from the collider data, which prefer a larger value.

In Fig.~\ref{fig:xRs-abs-atlaswz2011} we also compare the strangeness ratio $R_s(x,Q)$ of
NNPDF3.1 with that of CT14 and MMHT14.
We find that there is good consistency in the entire range of $x$,
while the PDF errors
in NNPDF3.1 are typically smaller than those of the other two sets,
especially at  large scales.
It is also interesting to note how in NNPDF3.1 the PDF uncertainties
in the ratio $R_s$ blow up at very large $x$, reflecting the lack
of direct information on strangeness in that kinematic region.

\begin{table}[t]
  \begin{center}
        \renewcommand{\arraystretch}{1.2}
\begin{tabular}{|l|c|c|}
\hline
\centering PDF set &  $K_s(Q=1.38~\mathrm{GeV})$ & $K_s(Q=M_{\rm Z})$ \\
\hline
\hline
NNPDF3.0   &  $0.45 \pm 0.07$  & $0.72 \pm 0.04$ \\
NNPDF3.1    & $0.53 \pm 0.07$  &  $ 0.75 \pm 0.04 $  \\
NNPDF3.1 collider-only   & $3.4 \pm 2.5$  & $1.5 \pm 0.6$  \\
NNPDF3.1 HERA + ATLAS $W,Z$   & $-1.0 \pm 7.0$  & $2.8 \pm 1.7$ \\
\hline
\end{tabular}
\end{center}
\caption{\small \label{tab:rsint} 
  The strangeness momentum fraction Eq.~(\ref{eq:rsint}) 
  at a low scale and a high scale. We show results obtained using  
NNPDF3.0, and NNPDF3.1 baseline, collider-only and  HERA+ATLAS $W,Z$
PDF sets.}
\end{table}

We now turn to  the strange momentum fraction $K_s(Q^2)$
Eq.~(\ref{eq:rsint}); values for the same PDF sets and scales are
shown in Table~\ref{tab:rsint}. Results are quite similar to those
found from the analysis of  Table~\ref{tab:rs}.
For the NNPDF3.1 collider-only and especially the HERA + ATLAS $W,Z$ fits, the
central value of $K_s$ is  
unphysical, with a huge uncertainty; essentially, all one can say is
that the strange momentum fraction $K_s$ is completely uncertain. 
This shows rather dramatically that the relatively precise values 
in Table~\ref{tab:rs} only hold in a rather narrow $x$ range.
It will be interesting to see whether more LHC data, possibly leading
to a competitive collider-only fit,  will confirm
strangeness enhancement and allow for an accurate determination of
strangeness in a wider range of $x$.

\subsection{The charm content of the proton revisited}
\label{sec:phenocharm}

The charm content of the proton determined by fitting the charm PDF
was quantified within the
NNPDF global analysis framework in~\cite{Ball:2016neh}, where results
obtained when  charm is independently parametrized, or perturbatively
generated , were compared for the first time.
The analysis there was performed at NLO only, and the dataset was very similar to that
of the NNPDF3.0 fit.
We now re-examine the fitted charm PDF at NNLO in perturbative QCD, and in the 
context of the inclusion of the new datasets,
in particular top, and LHCb and ATLAS electroweak boson production,
which sizably affects and
constrain the charm PDF. 

Indeed, we have seen in
Section~\ref{sec:disentangling}, in particular
Fig.~\ref{fig:31-nnlo-old-vs-new}, that the new data added in
NNPDF3.1 considerably reduces the charm PDF uncertainty, but also affects its
central value, which changes by more than one sigma at large $x$.
Also, in Ref.~\cite{Ball:2016neh} charm at large $x$ was mostly
constrained by the EMC data which we discussed in Section~\ref{sec:emc}
and which are not included in the default NNPDF3.1 PDF
determination. As we have seen in Section~\ref{sec:emc} these data still
have a significant impact on charm, hence a re-assessment of charm
determination is in order both when this dataset is included and when it
is not. We therefore now compare results obtained using the
default NNPDF3.1 NNLO set, the modified version of that in which charm
is generated perturbatively as discussed in
Section~\ref{sec:results-mc}, and the modified set in which the EMC data are added
to the NNPDF3.1 dataset, as discussed in Section~\ref{sec:emc}.

\begin{figure}[t]
\begin{center}
  \includegraphics[scale=0.35]{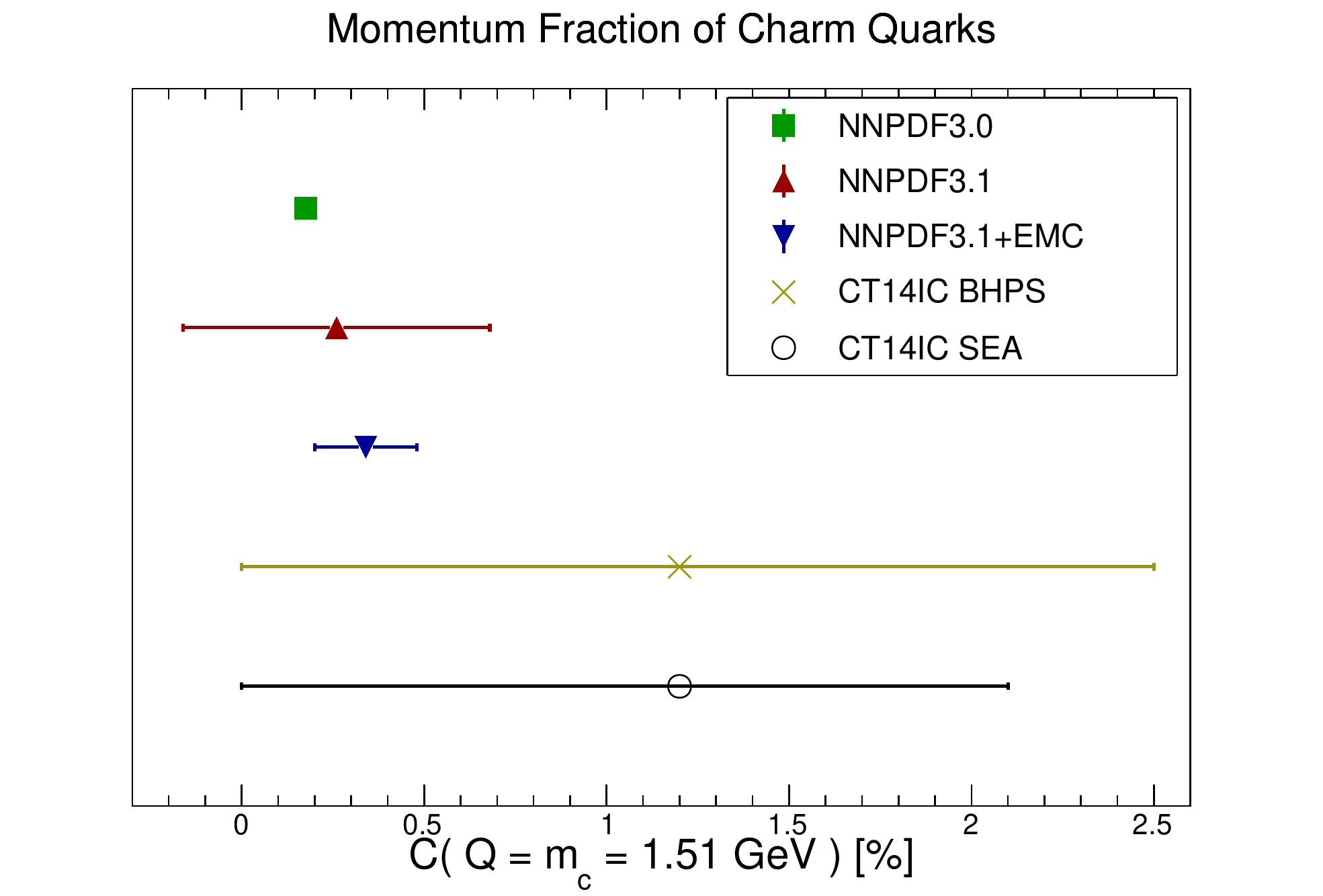}
  \includegraphics[scale=0.35]{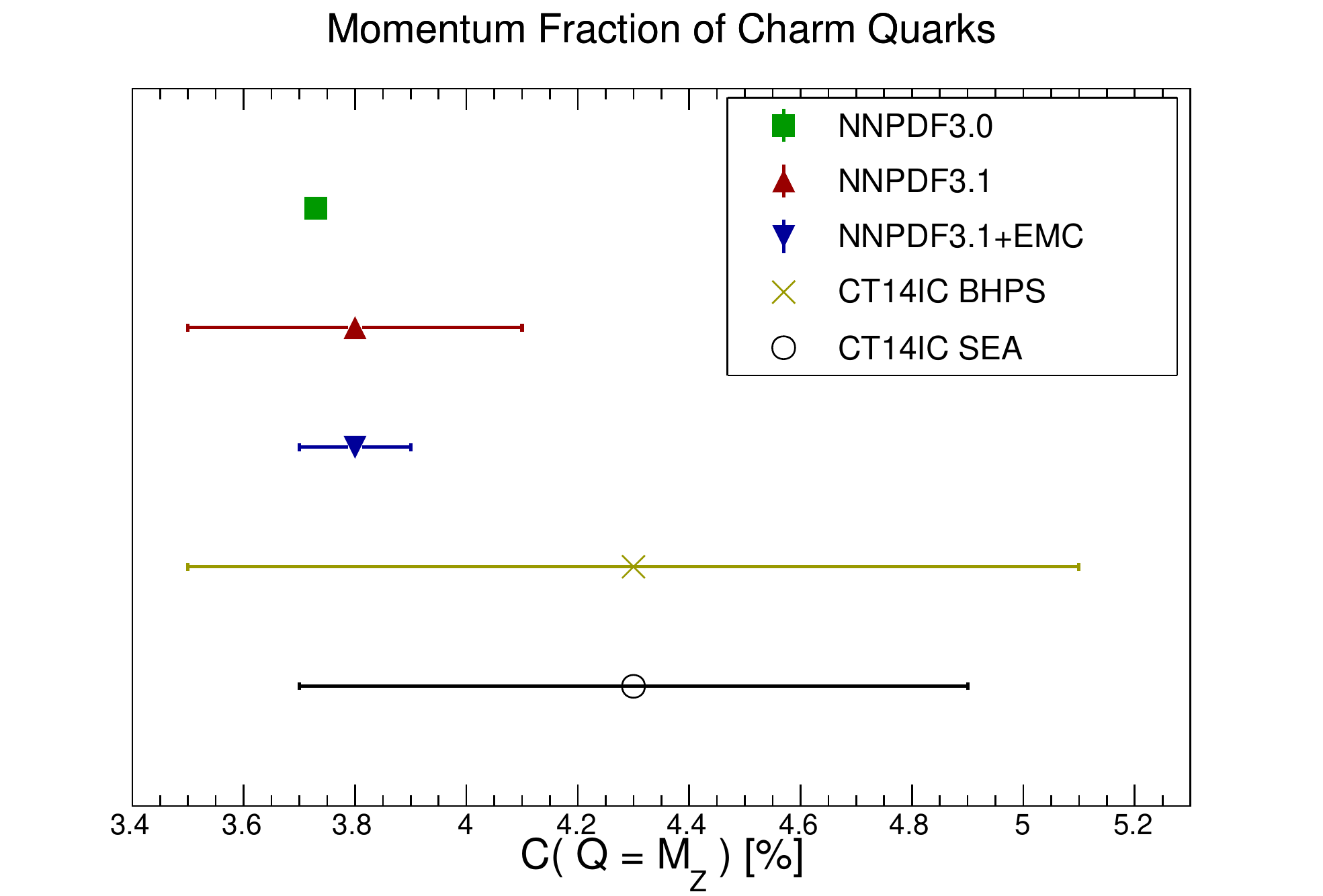}
  \caption{\small Graphical representation of the results
    for $C(Q^2)$ from Table~\ref{tab:momsr} 
    and $Q=m_c$ GeV (left) and $Q=m_Z$ (right). Model estimates from
    Ref.~\cite{Dulat:2013hea} are also shown.
\label{fig:xCplot}}
\end{center}
\end{figure}

In Table~\ref{tab:momsr} we show the charm momentum fraction, defined as
\be
\label{eq:charmmomfrac}
C(Q^2) \equiv \int_0^1\, dx\,\lp xc(x,Q^2)+x\bar{c}(x,Q^2)\rp \ ,
\ee
for two values of $Q^2$, at the charm threshold $Q=m_c$, and at
$Q=m_Z$, computed from using these PDF sets.
A graphical representation of the results from 
Table~\ref{tab:momsr} is shown in Fig.~\ref{fig:xCplot}.
There is a very large difference, by two orders of
magnitude, between the uncertainty on the momentum fraction, according
to whether charm is independently parametrized, or perturbatively generated.
However, the central values
agree with each other
within the large uncertainty determination when charm is parametrized, with
the corresponding central value only slightly larger than that when
charm is  perturbative
(though hugely different on the scale of the uncertainty on the
perturbatively generated result).
Upon adding the EMC data the
uncertainty is reduced by about a factor of three, and the central
value somewhat increased, consistently with the effect of this dataset on
the charm PDF discussed in Section~\ref{sec:emc}.

\begin{figure}[t]
\begin{center}
  \includegraphics[scale=0.30]{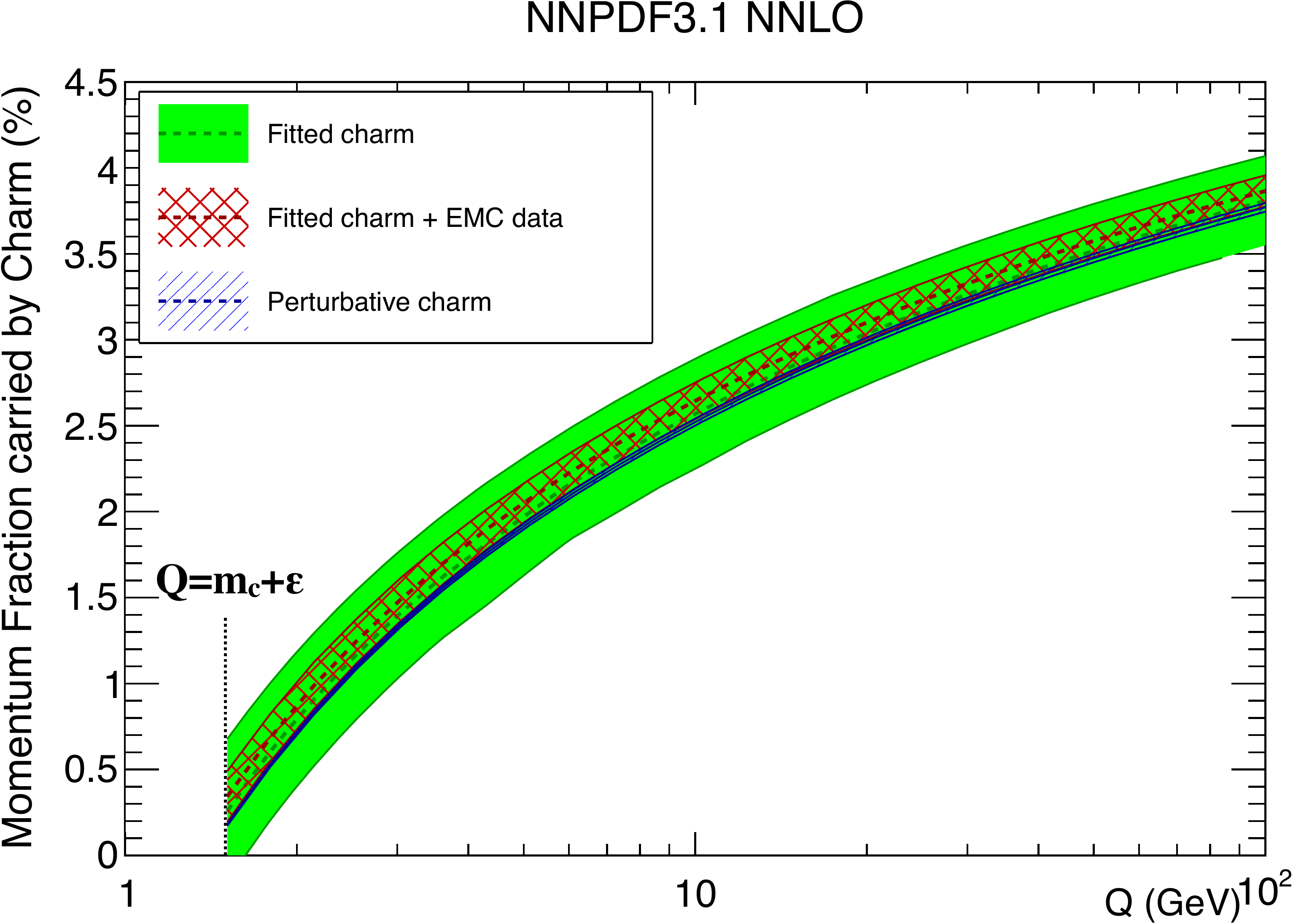}
  \caption{\small The charm momentum fraction of Tab.~\ref{tab:momsr}
    plotted  as a function of  scale $Q$.
\label{fig:momfrac}}
\end{center}
\end{figure}

We can interpret the  difference between the total momentum fraction
when charm is independently parametrized and determined from the data, 
and that when charm is perturbatively generated, as the
``intrinsic'' (i.e. non-perturbative) charm momentum fraction.
Including EMC data when charm is parametrized,
at  $Q=m_c$  we find that it is given by
$C(m_c)_{\rm FC}-C(m_c)_{\rm PC} =  \left(0.16 \pm 0.14\right)$\%: this provides evidence
for a small intrinsic charm component in the proton at the
one $\sigma$ level, somewhat improving the estimates of
Ref.~\cite{Ball:2016neh}, but 
still with a considerable degree of uncertainty.
Our estimate for the non-perturbative charm component of the proton
is considerably smaller than those allowed in the CT14IC model analysis~\cite{Dulat:2013hea}, 
also shown in Fig.~\ref{fig:xCplot}. However, it is larger than expected 
in Ref.~\cite{Jimenez-Delgado:2014zga}, where an upper bound of $0.5$\% at the 
four-$\sigma$ level is claimed. Both these analyses have difficulty fitting the EMC charm 
structure function data, due to an overly restrictive functional form
for the charm PDF. At high scale all estimates of $C(Q)$ are dominated by the perturbative
growth of the charm PDF at small $x$, but the one-$\sigma$ excess in the fit 
with EMC data persists, though as an ever diminishing fraction of the whole. 
This is demonstrated very clearly in Fig.~\ref{fig:momfrac}, where we plot the dependence 
of the charm momentum fraction $C(Q^2)$ on the scale $Q$.

\begin{table}[t]
  \center
   \renewcommand{\arraystretch}{1.4}
  \begin{tabular}{|l|c|c|}
    \hline
    PDF set & $C(m_c)$ & $C(m_Z)$ \\
    \hline
    \hline
    NNPDF3.1           & $\lp 0.26 \pm 0.42\rp $\%  &  $\lp 3.8 \pm 0.3\rp $\%\\
    NNPDF3.1 with pert. charm        & $\lp 0.176 \pm 0.005\rp $\%  &  $\lp 3.73 \pm 0.02\rp $\% \\
    NNPDF3.1 with EMC data  &   $\lp 0.34 \pm 0.14\rp $\% & $\lp 3.8  \pm 0.1\rp $\% \\
          \hline
  \end{tabular}
  \caption{\small \label{tab:momsr}
    The charm momentum fraction  $C(Q^2)$, Eq.~(\ref{eq:charmmomfrac}),
    just above the charm threshold $Q=m_c$~GeV and at
    $Q=m_Z$. Results are shown for NNPDF3.1, and its modified versions
    in which EMC data are added to the dataset,  or charm is not fitted. 
  }
  \end{table}

\begin{figure}[t]
  \begin{center}
    \includegraphics[scale=0.35]{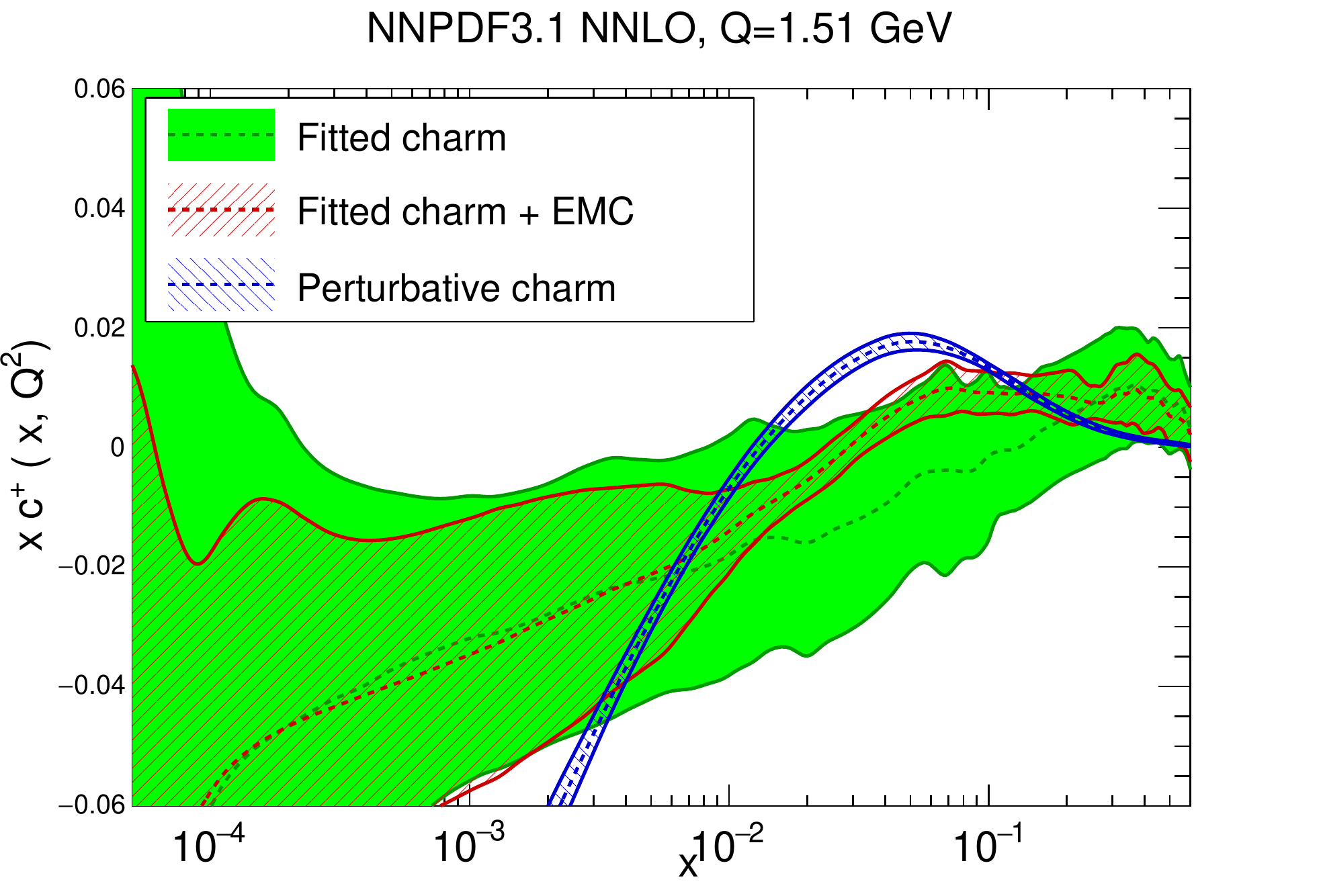}
  \includegraphics[scale=0.35]{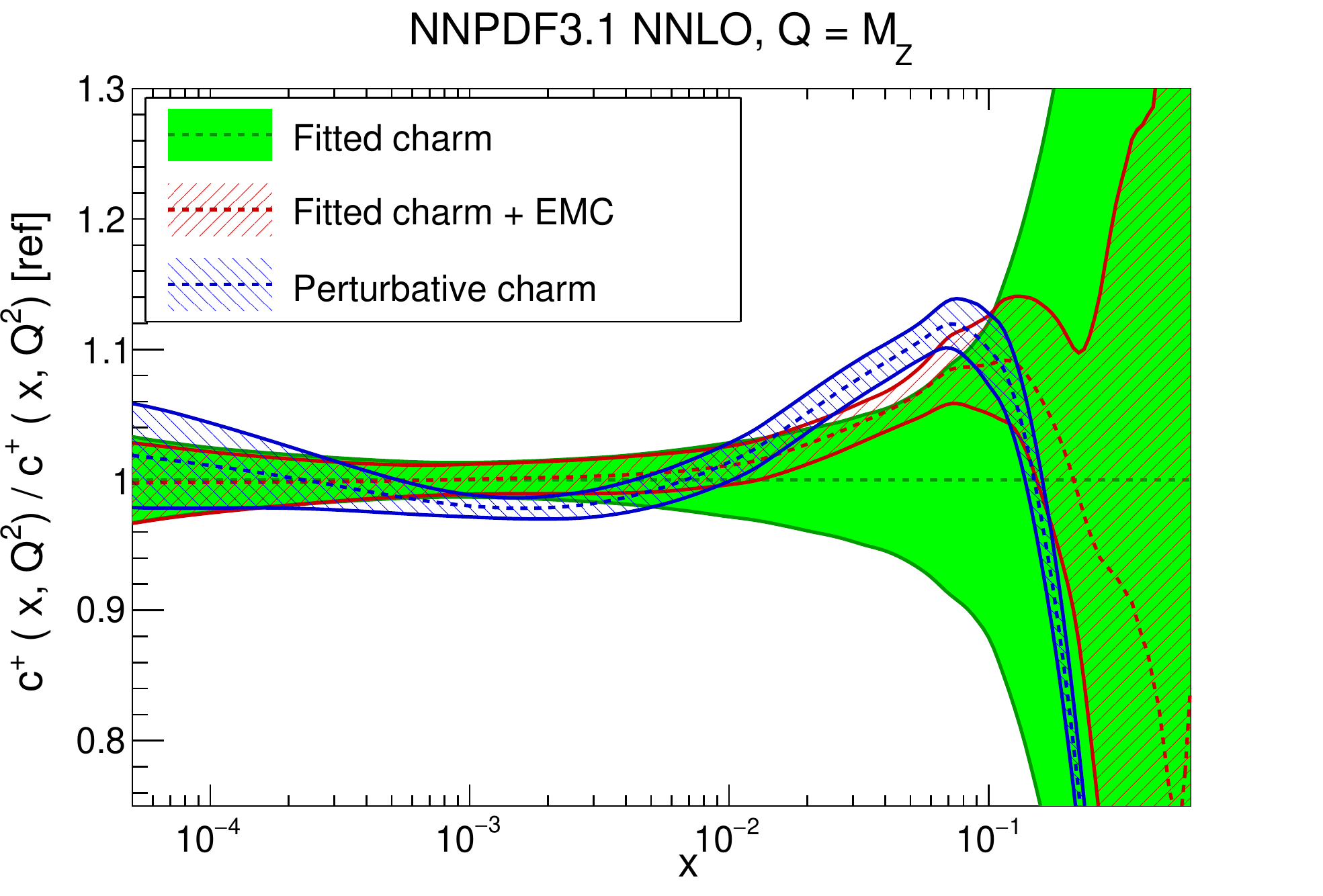}
  \includegraphics[scale=0.35]{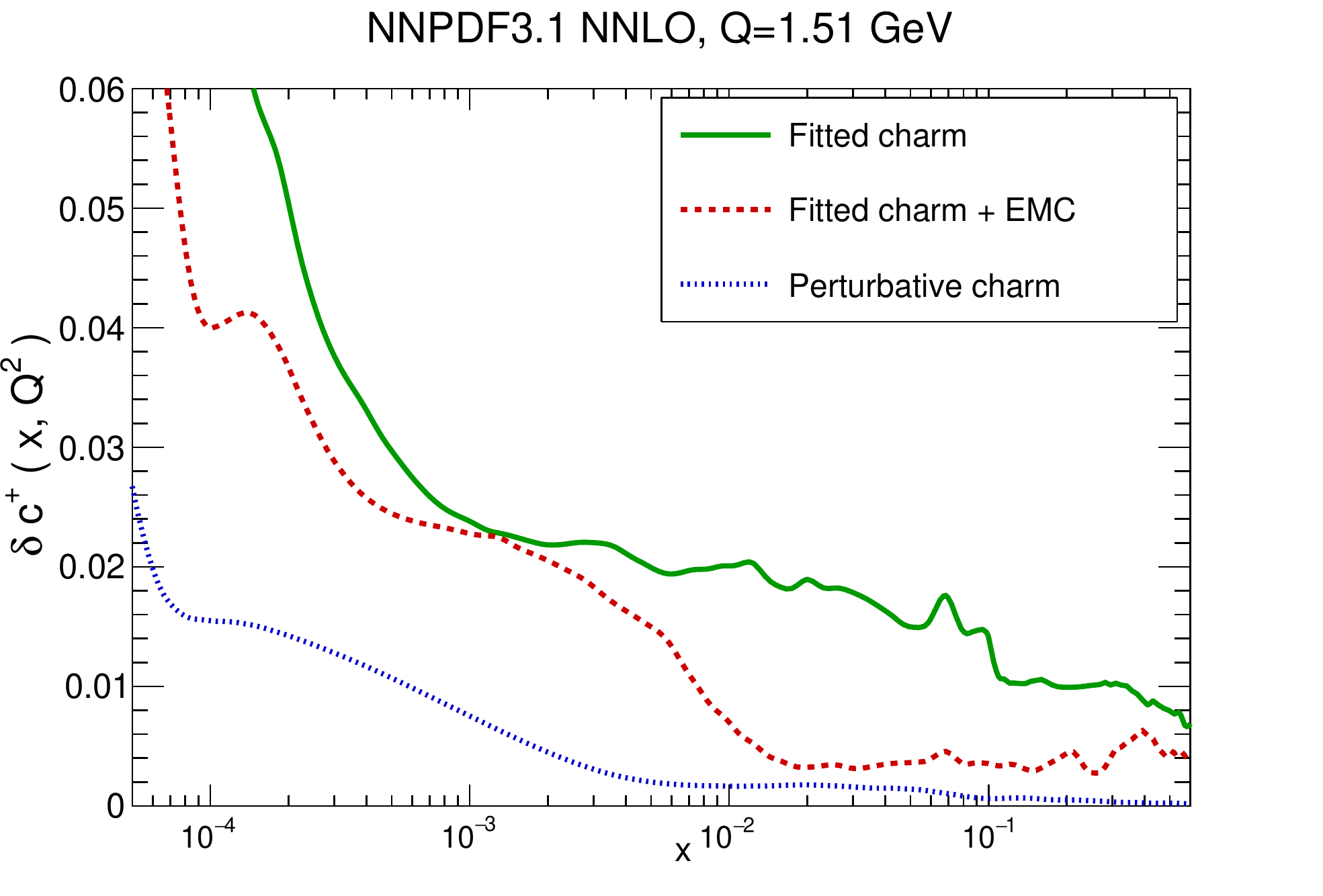}
  \includegraphics[scale=0.35]{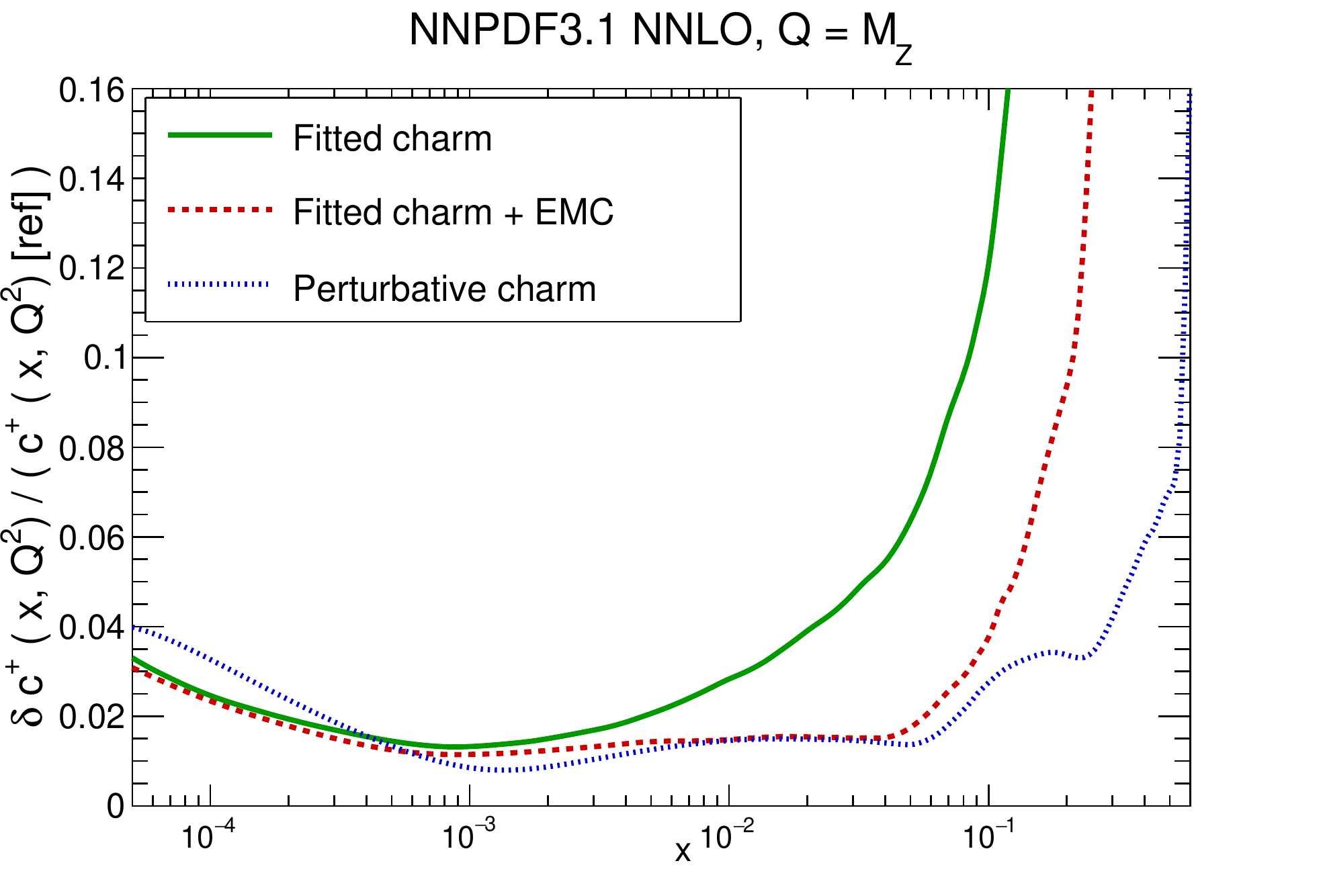}
    \caption{\small Comparison of the charm PDF at the scale and for
      the PDF sets of Tab.~\ref{tab:momsr}. Both the PDF (top) and the
      relative uncertainty (bottom) are shown.
\label{fig:PDFlargexcharm}}
\end{center}
\end{figure}

The origin of these values of the charm momentum fraction can be
understood by directly  comparing the charm PDF, which is done
in Fig.~\ref{fig:PDFlargexcharm}, again 
 just above threshold $Q=m_c$ and at $Q=m_Z$, in the latter case
as a ratio to the baseline NNPDF3.1 result.
The agreement of the charm  momentum fraction when it is
perturbatively generated or when it is parametrized and determined
from the data is related to the fact that, when parametrized, 
the best-fit charm
has qualitatively the same shape as charm generated perturbatively at
NNLO, as observed in Ref.~\cite{Ball:2016neh} and
Section~\ref{sec:results-mc} above. 
However, at threshold $Q=m_c$ GeV the best-fit charm
is larger than the perturbative component at large $x$, $x\gsim 0.2$,
albeit with large uncertainties, that are somewhat reduced when the EMC
dataset is added, without a significant change in shape. 
Upon addition of the EMC data, in the
medium-small  $10^{-2}\lsim x\lsim 10^{-1}$ the PDF is pushed at the upper edge
of the uncertainty band before addition, with considerably reduced
uncertainty. The unrealistically small uncertainty on the
perturbatively generated charm PDF is apparent, and also the reduction
in uncertainty due to the EMC data for all $x\gsim 10^{-3}$, already
discussed in Section~\ref{sec:emc}.

\begin{figure}[t]
\begin{center}
  \includegraphics[scale=0.37]{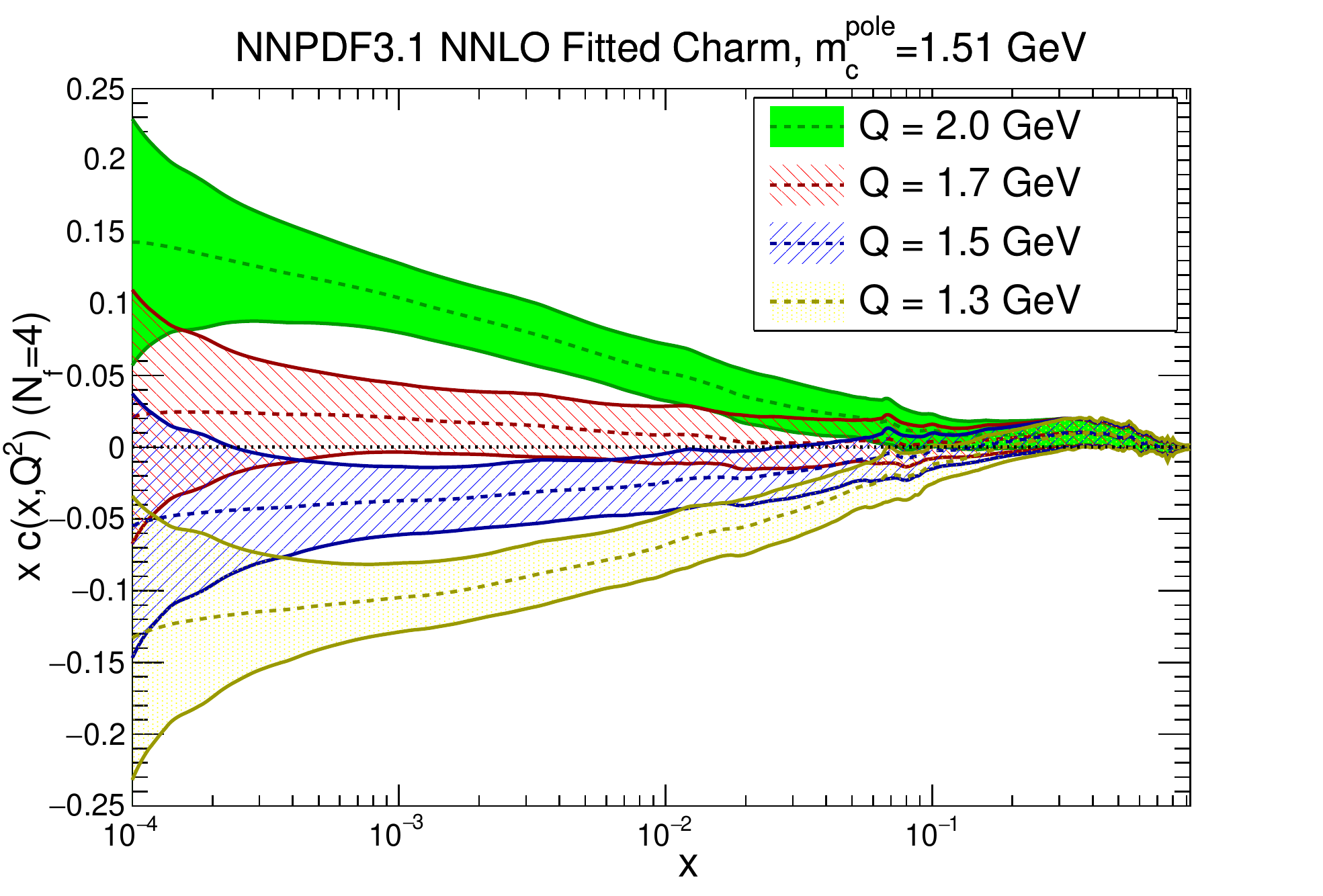}
  \includegraphics[scale=0.37]{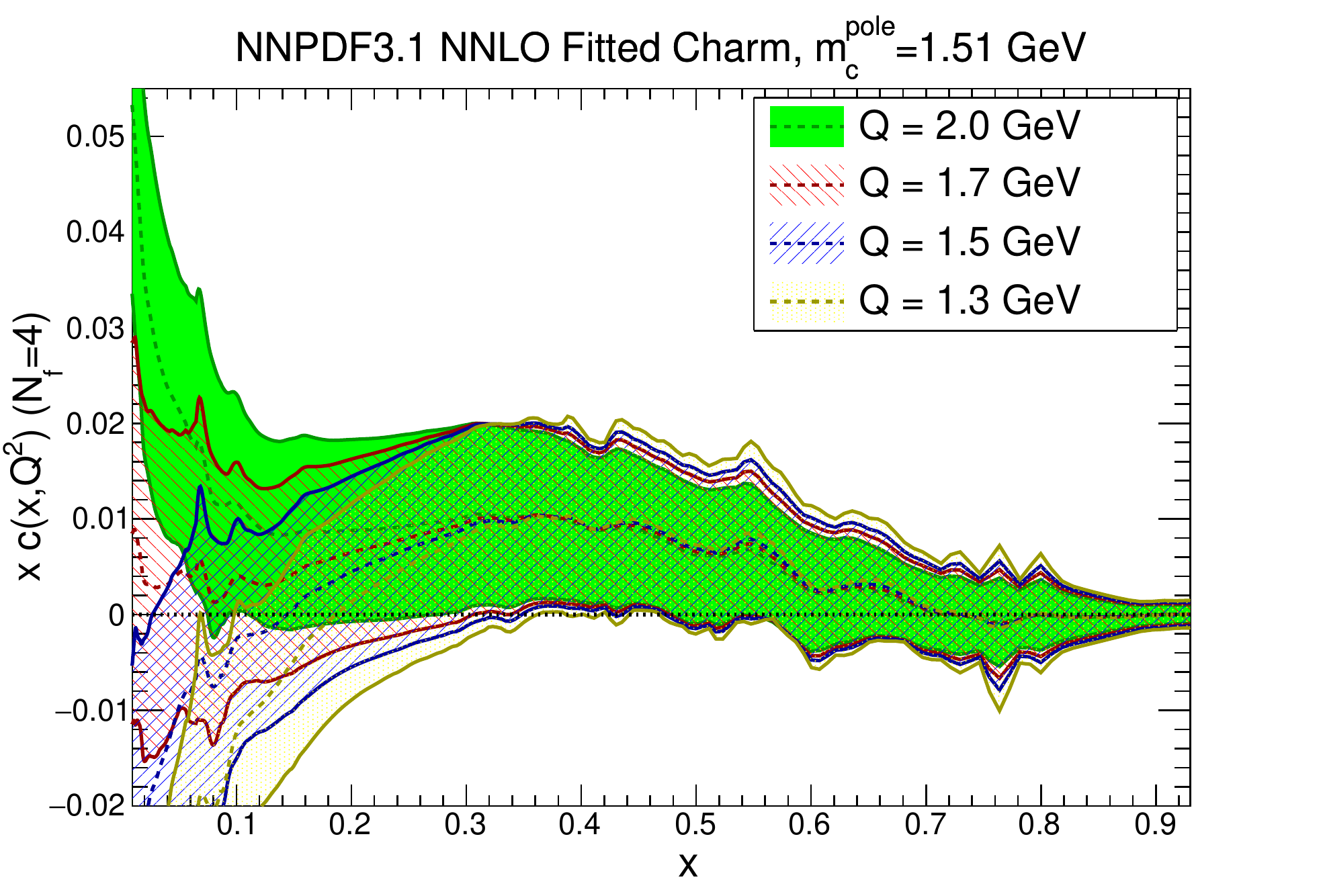}
  \includegraphics[scale=0.37]{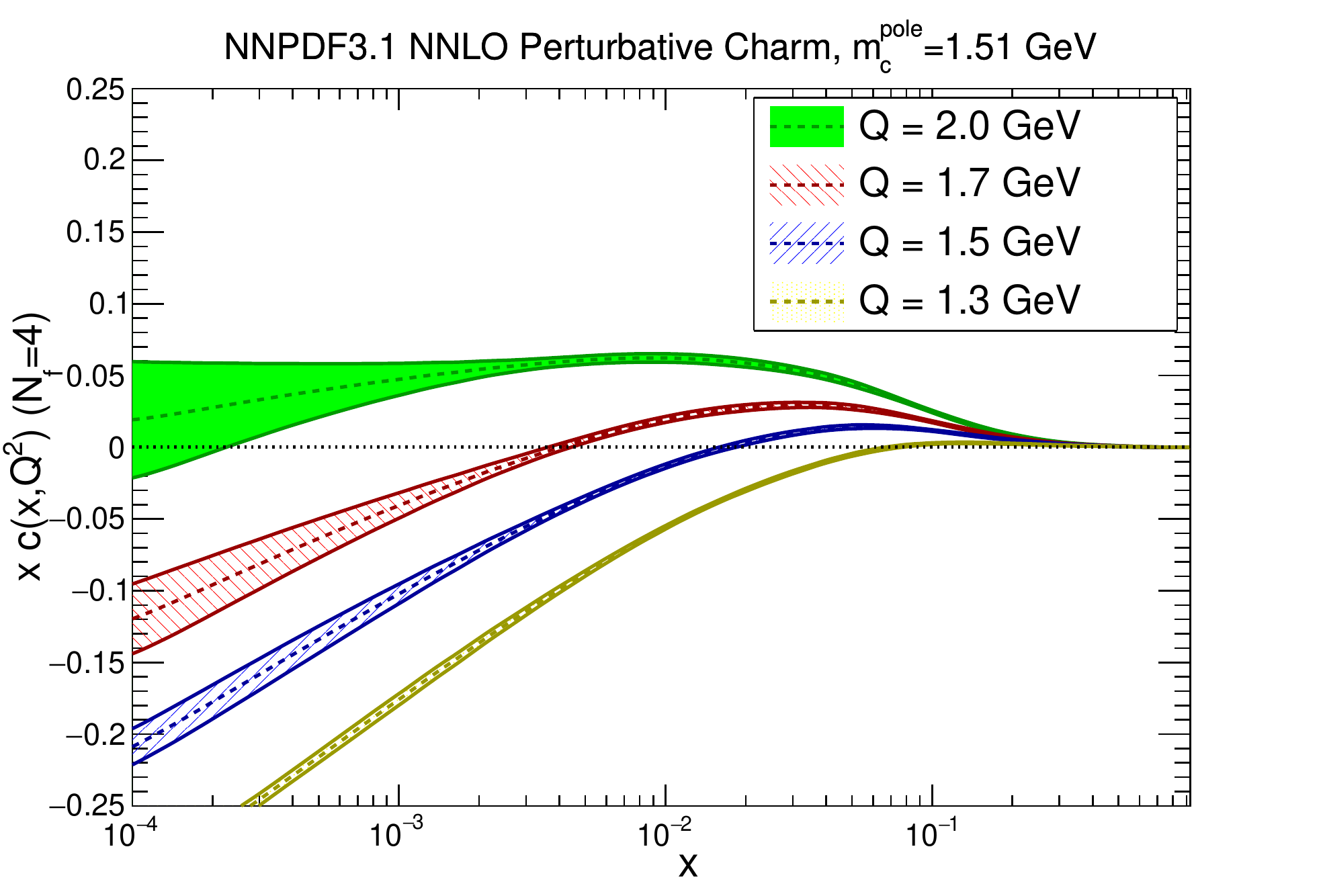}
  \includegraphics[scale=0.37]{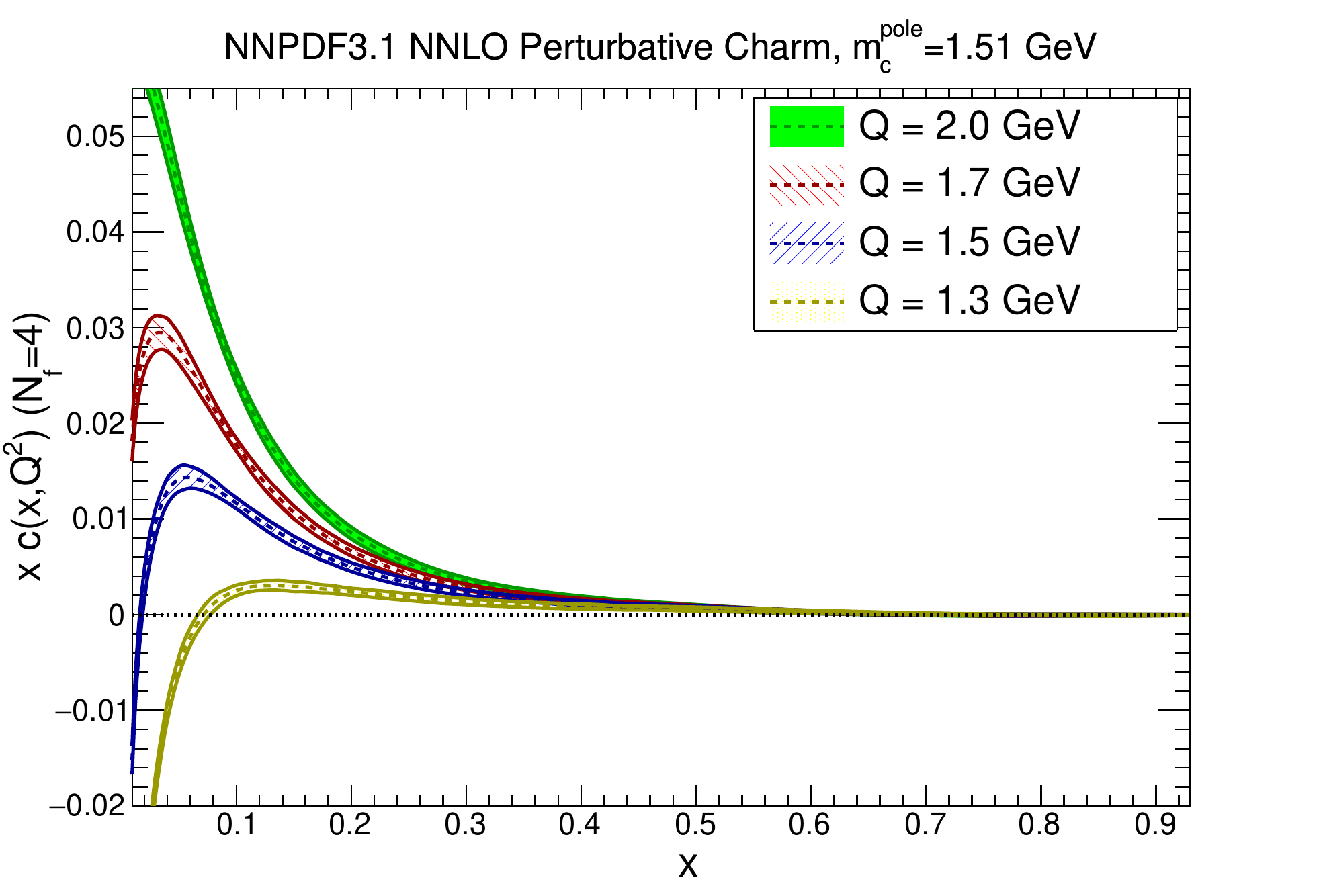}
  \caption{\small The charm PDF in the $n_f=4$ scheme at small $x$
    (left) and large $x$ (right plot) for different values of $Q$,
    in the NNPDF3.1 NNLO PDF set (top) and when assuming that charm is
    perturbatively generated (bottom).
\label{fig:charmPDFnf4}}
\end{center}
\end{figure}

The origin of the differences between the charm PDF when
perturbatively generated, or parametrized
and determined from the data,
and a possible decomposition of the latter into an ``intrinsic''
(non-perturbative) and a perturbative component can be understood by
studying their scale dependence close to threshold, in analogy to a
similar analysis presented in Ref.~\cite{Ball:2016neh}.
This is done 
in Fig.~\ref{fig:charmPDFnf4}, where the charm PDF is shown
(in the $n_f=4$ scheme) as a function of $x$ both when charm is
parametrized and when it is perturbatively generated.
On the one hand, the large $x\gsim0.1$ component of the
 charm PDF is essentially scale independent: perturbative
charm vanishes identically in this region, so the fitted result in
this region may
    be interpreted as being of non-perturbative origin, i.e. ``intrinsic''.

On the other hand, for smaller $x$ the charm PDF depends
strongly on scale. When perturbatively generated, it  is sizable and
positive already 
at $Q\sim 2$~GeV for all $x\gsim 10^{-3}$, while at threshold it 
becomes large and negative for all $x\lsim10^{-2}$, possibly because
of large unresummed small-$x$ contributions. The best-fit parametrized
 charm PDF, within its larger uncertainty, is rather flatter and 
smaller in modulus  
essentially for all $x\lsim 10^{-1}$, 
except at the scale-dependent point at which
perturbative charm changes sign. This difference in shape between
fitted charm and perturbative charm for all $x\lsim 0.1$ is rather
larger than the uncertainty on either  perturbative or  fitted charm. 
This observation seems to support the conclusion
that the assumption that charm is perturbatively generated might be a
source of bias.

The updated NNPDF3.1 analysis is consistent with the conclusion of our 
earlier charm studies~\cite{Ball:2016neh}, namely 
that a non-perturbative charm component in the proton is consistent with
global data, though the new high-precision collider data that we include now
sets more stringent bounds on how large this component can be.
In particular, once the EMC charm structure function dataset is added,
we see from Table~\ref{tab:momsr} that a non-perturbative charm
momentum fraction of $\simeq 0.5\%$ represents a 
deviation from our best fit value of around two to three sigma. Thus models of 
intrinsic charm which carry as much as 1\% of the proton's momentum 
are strongly disfavored by currently available data.

\subsection{Parton luminosities}
\label{sec:lumis}

\begin{figure}[t]
\begin{center}
  \includegraphics[scale=0.38]{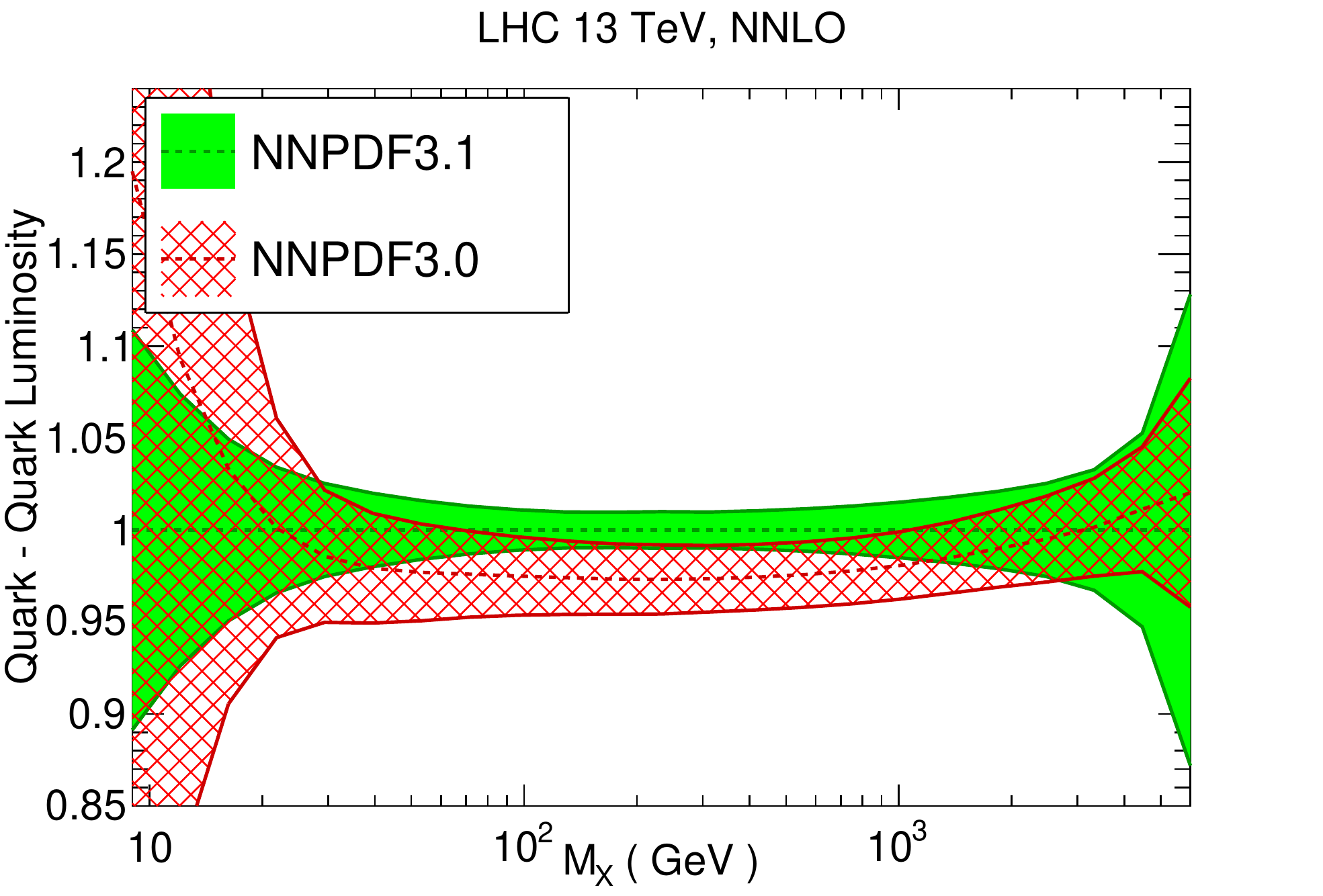}
  \includegraphics[scale=0.38]{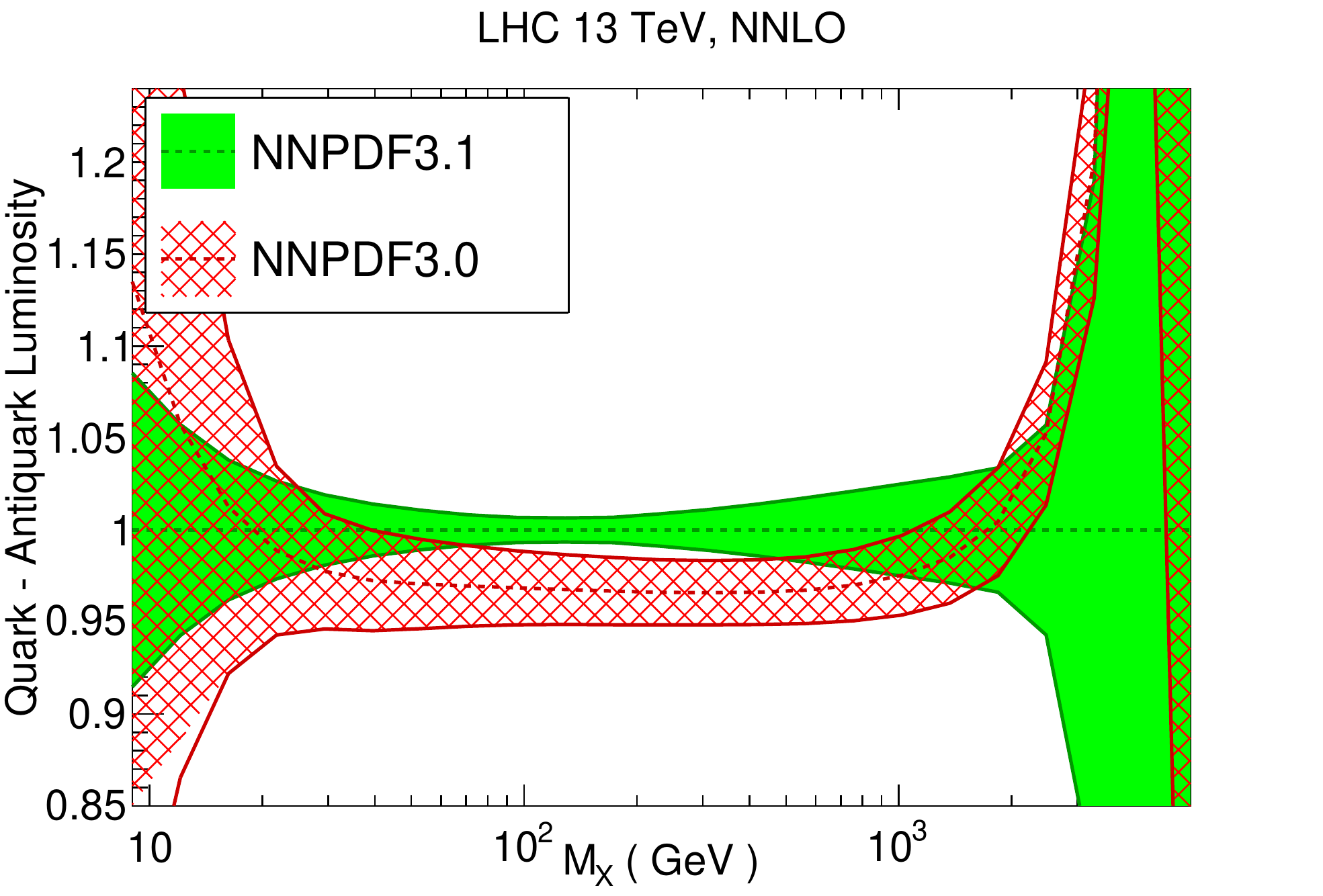}
  \includegraphics[scale=0.38]{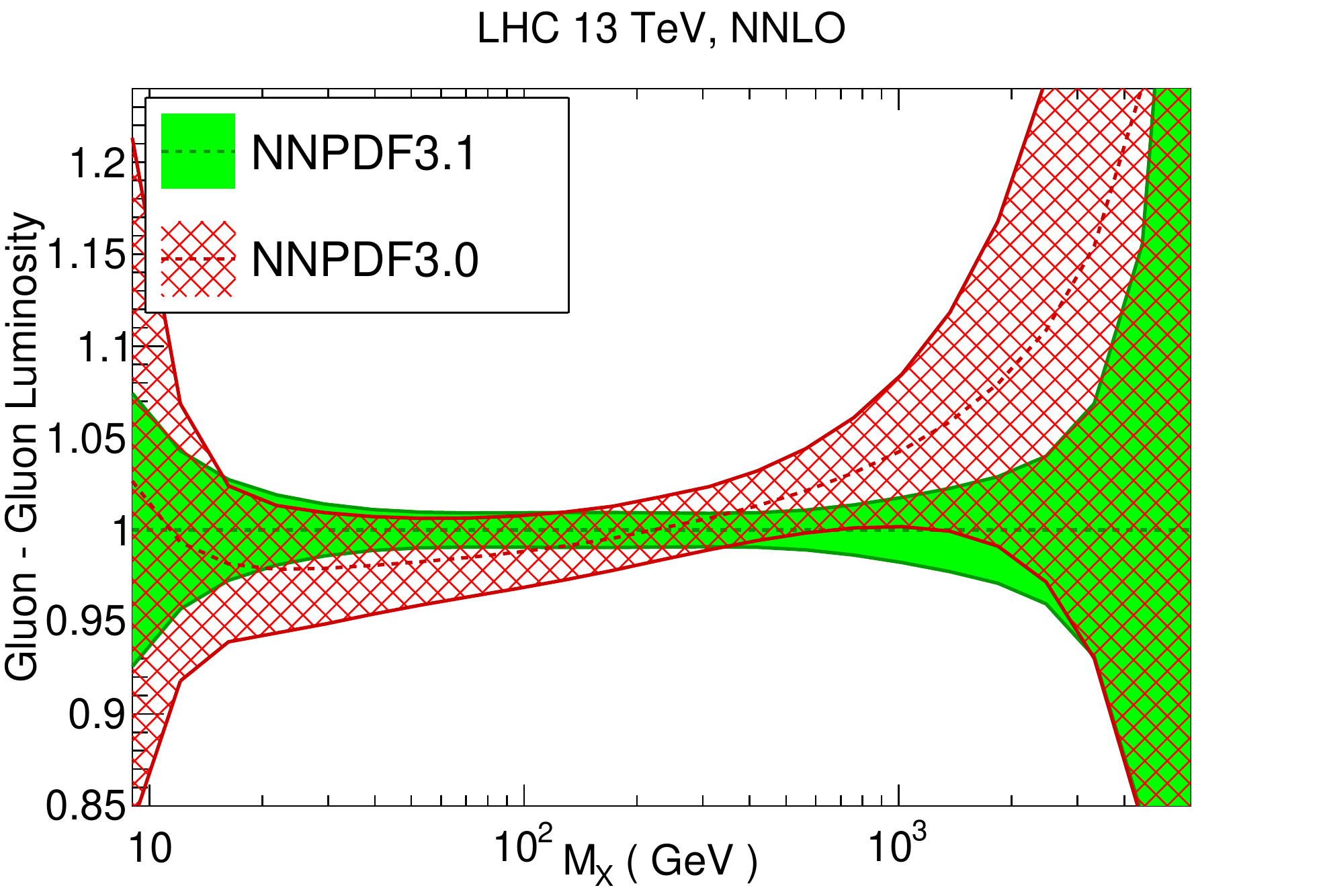}
   \includegraphics[scale=0.38]{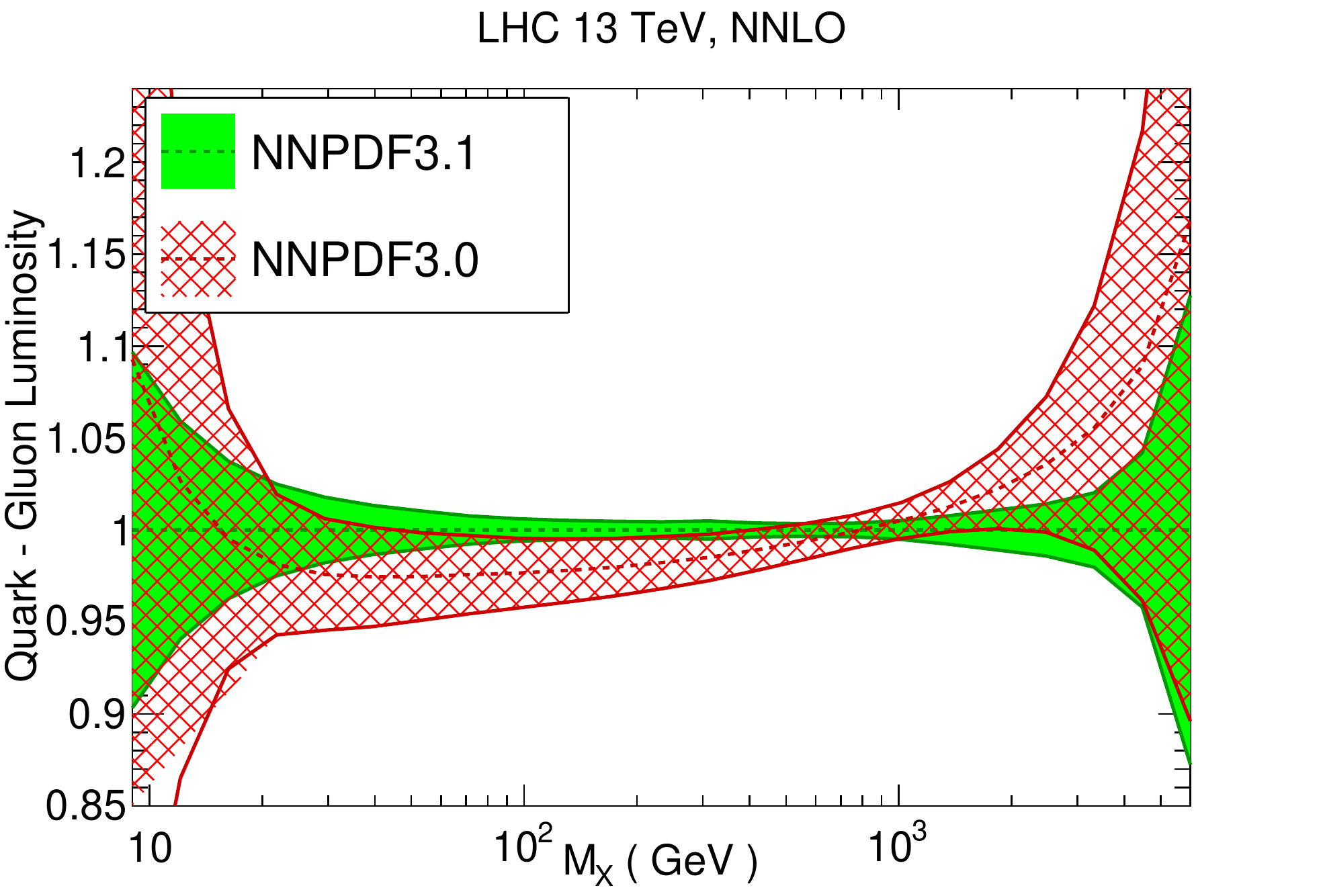}
   \caption{\small Comparison of parton luminosities with the NNPDF3.0
     and NNPDF3.1 NNLO PDF sets for the LHC~$13$ TeV. From left to
     right and from top to bottom 
 quark-antiquark, quark-quark, gluon-gluon
     and quark-gluon PDF luminosities are shown.     %
     Results are shown normalized to the central value of NNPDF3.1.
\label{fig:lumi-31-vs-30}}
\end{center}
\end{figure}

\begin{figure}[h]
\begin{center}
   \includegraphics[scale=0.40]{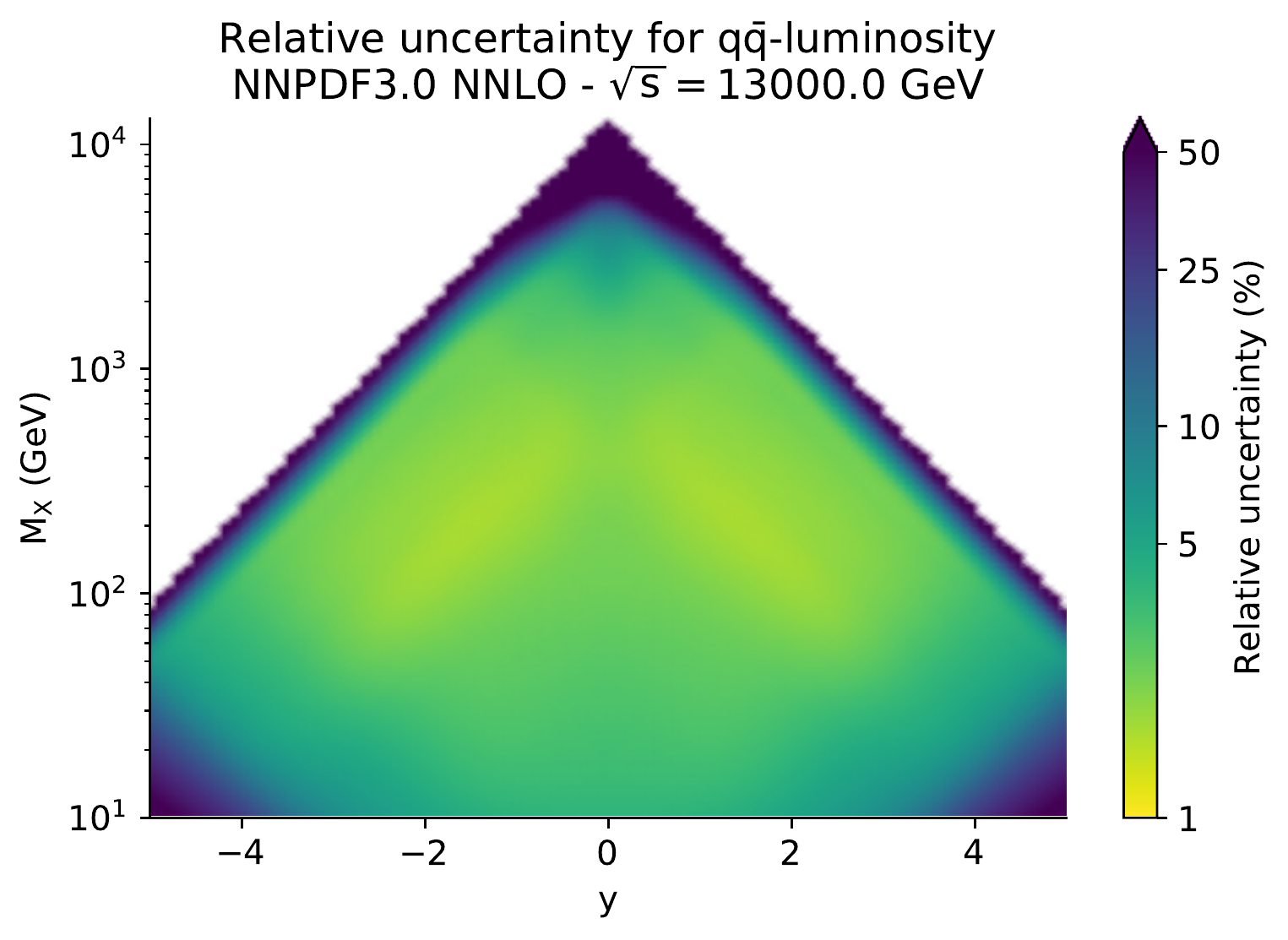}   
   \includegraphics[scale=0.40]{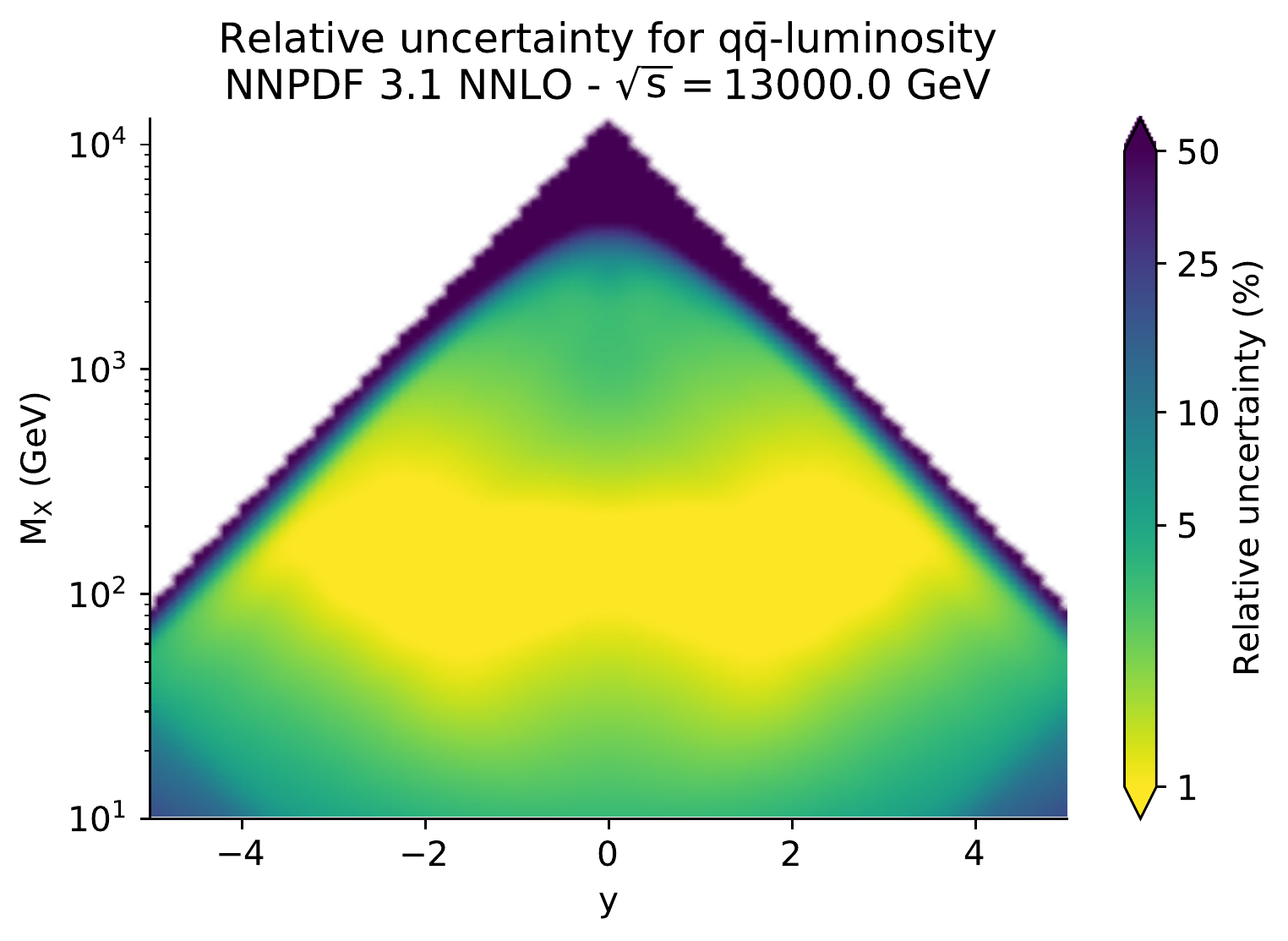}   
\includegraphics[scale=0.40]{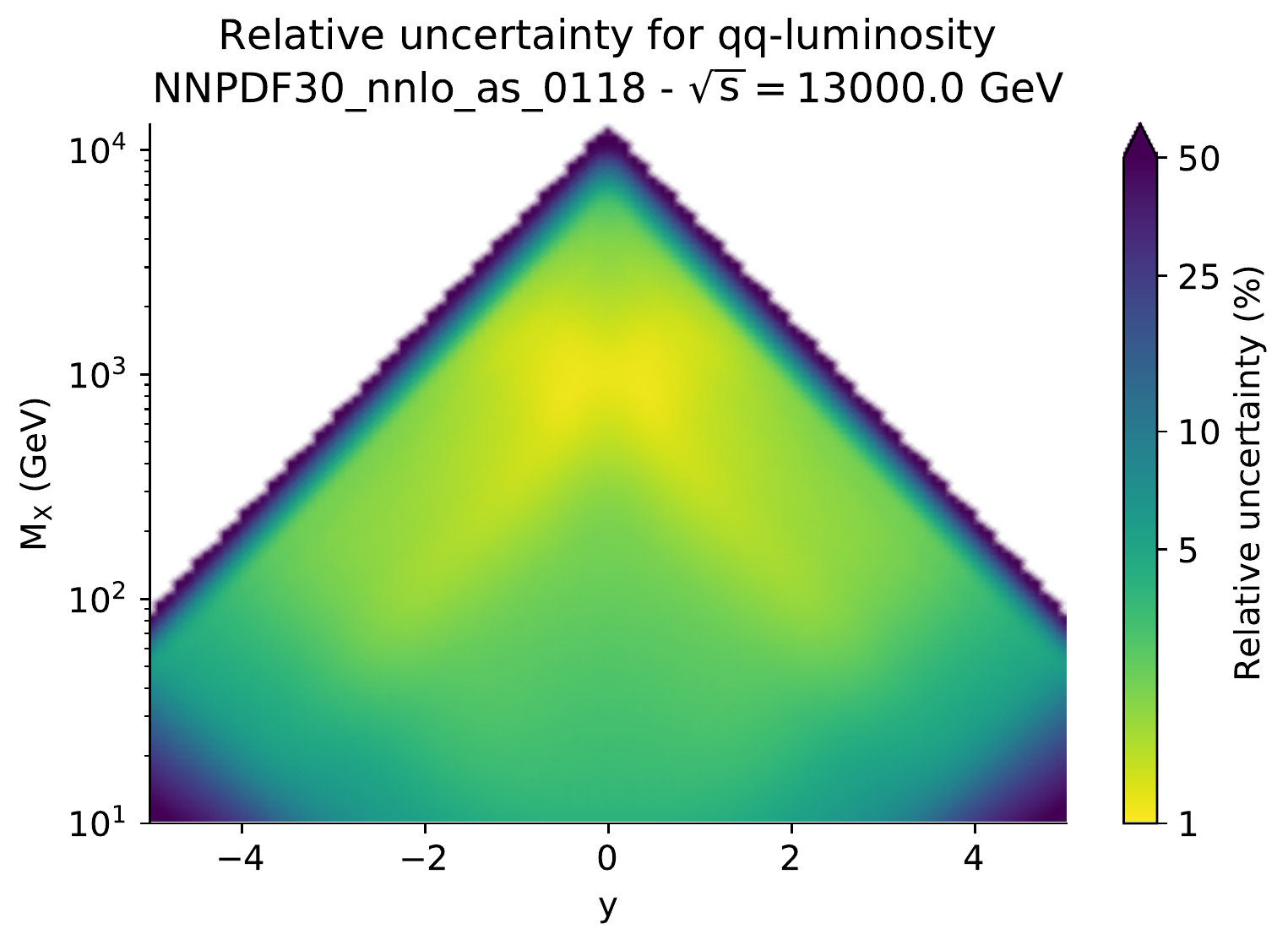}
\includegraphics[scale=0.40]{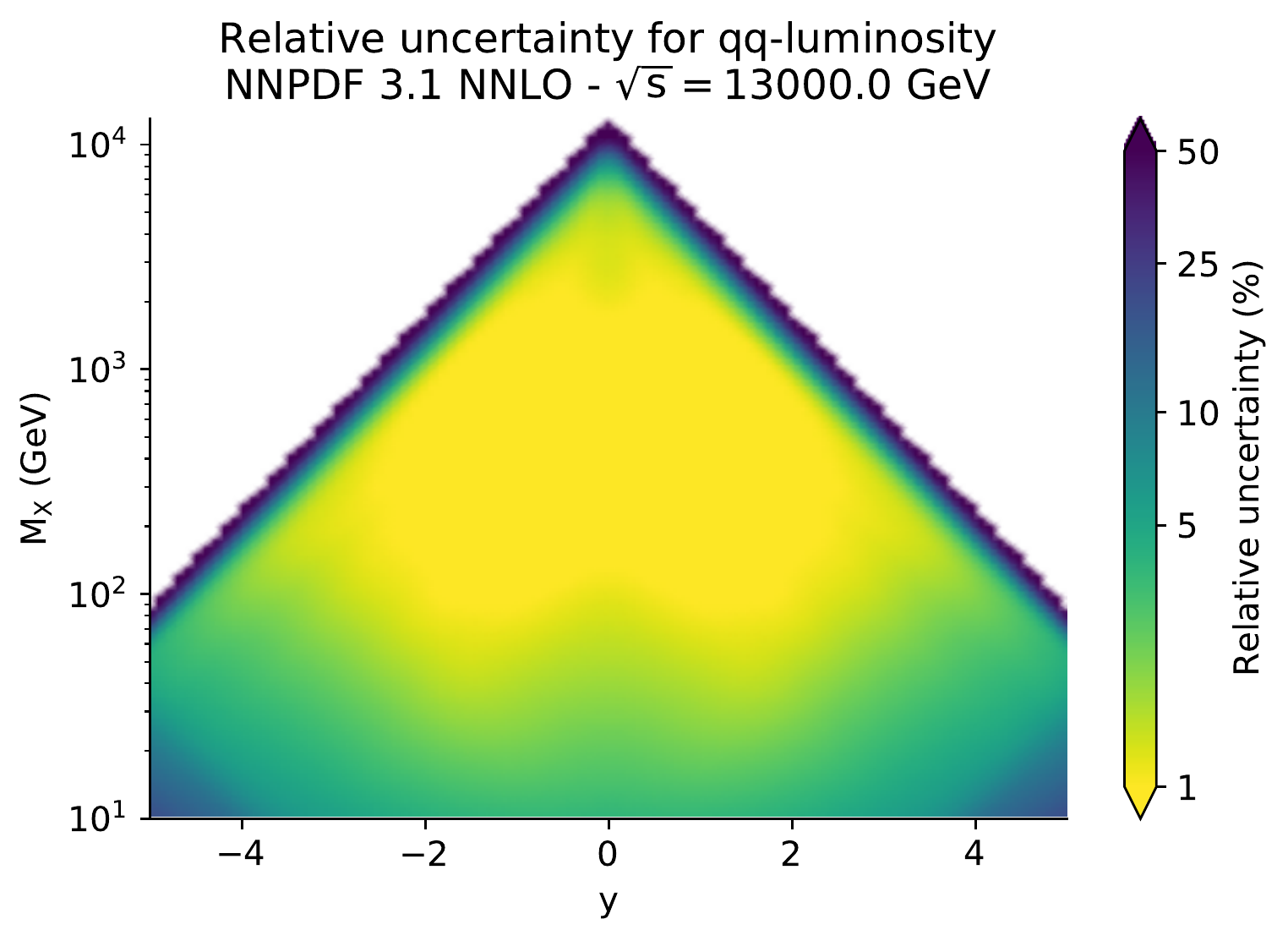}
\includegraphics[scale=0.40]{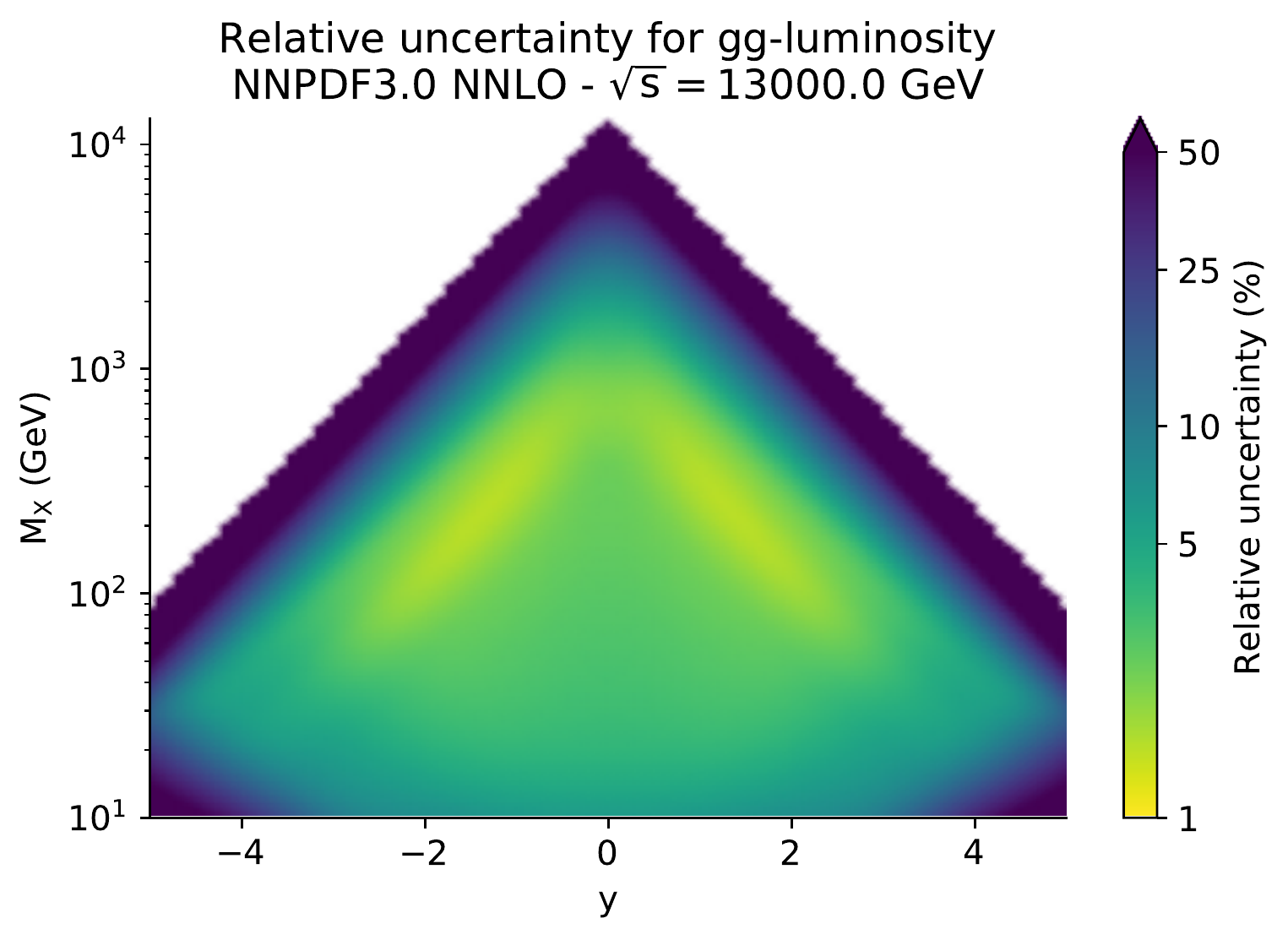}
\includegraphics[scale=0.40]{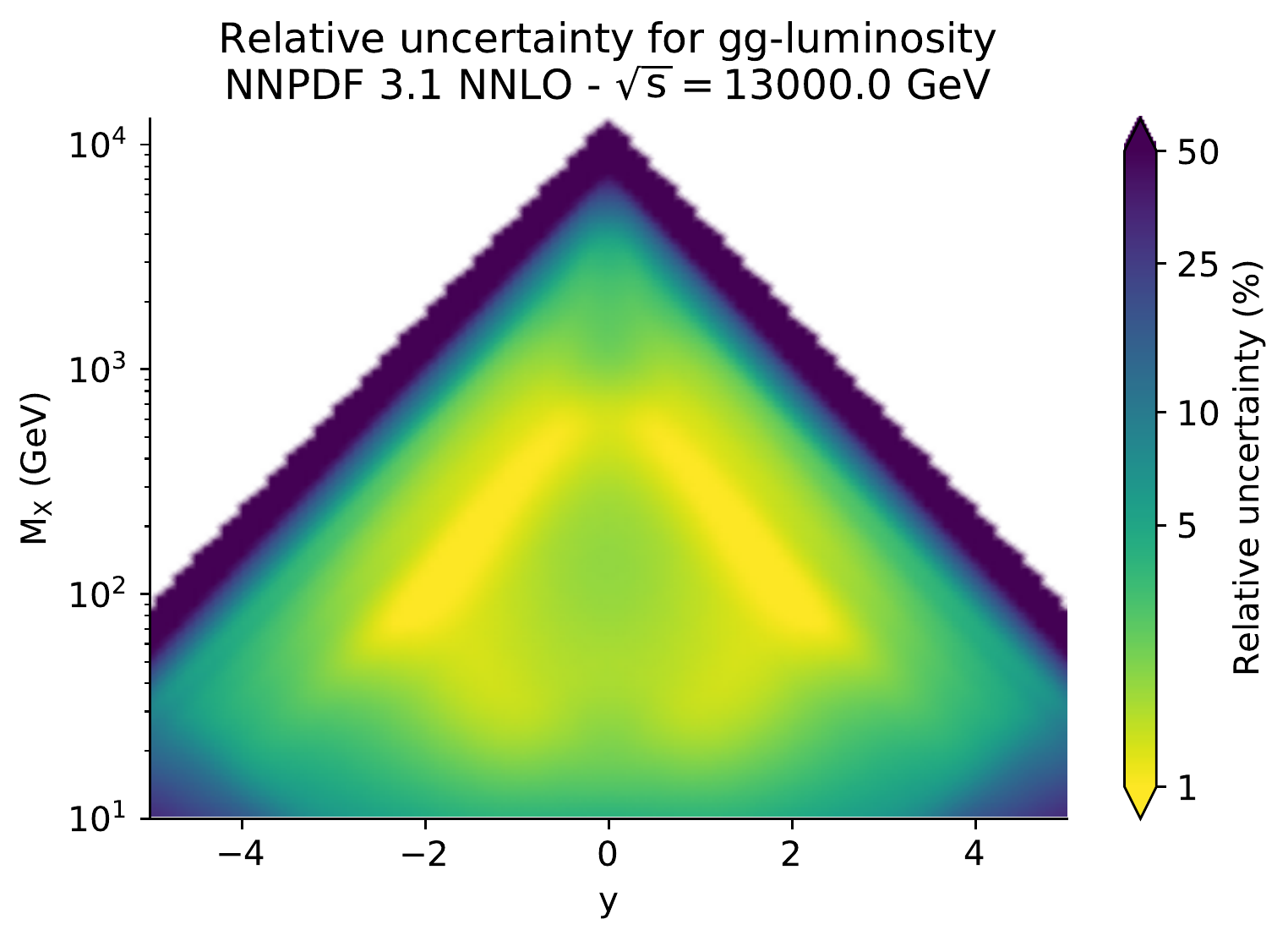}
  \includegraphics[scale=0.40]{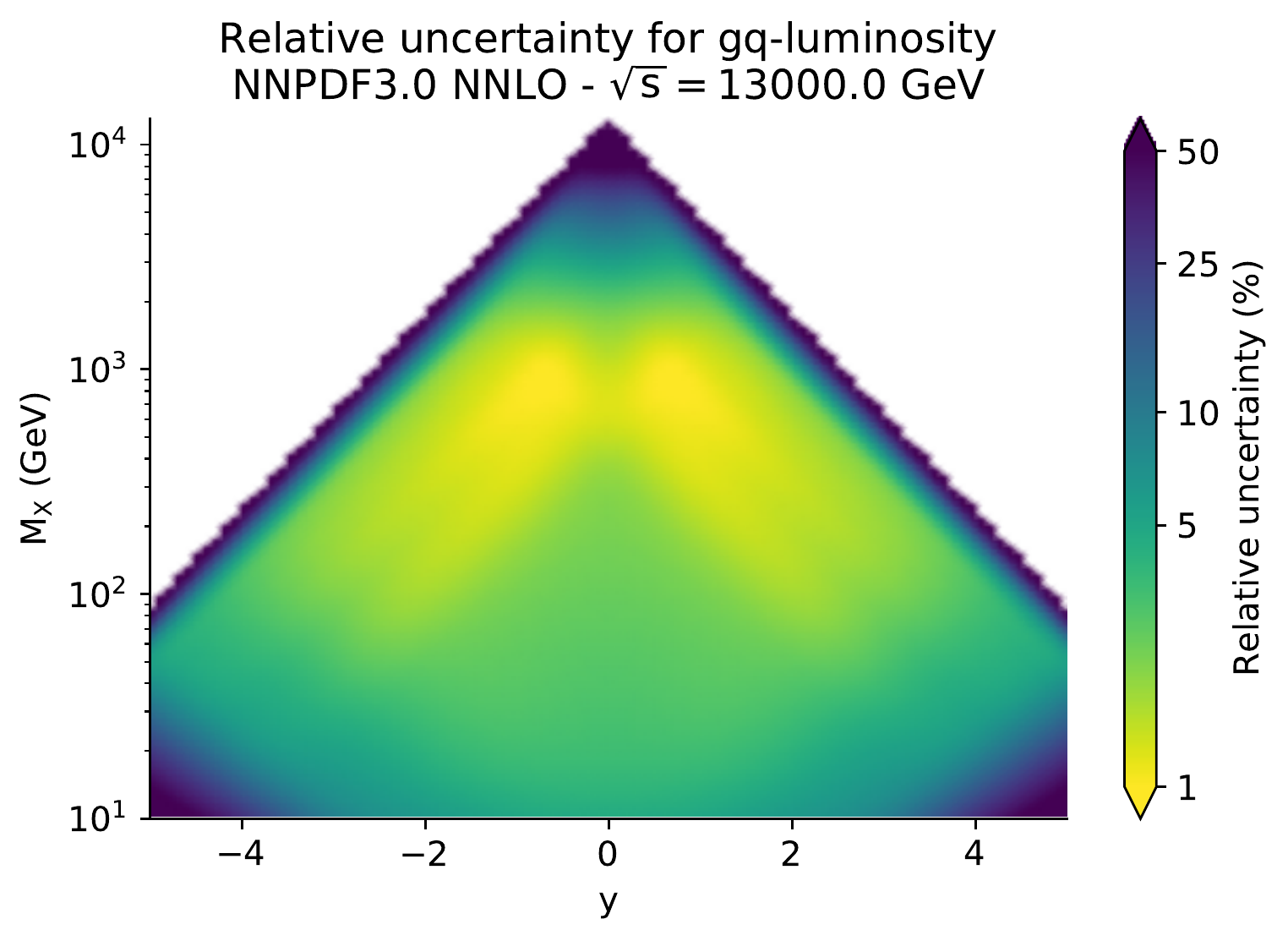}
  \includegraphics[scale=0.40]{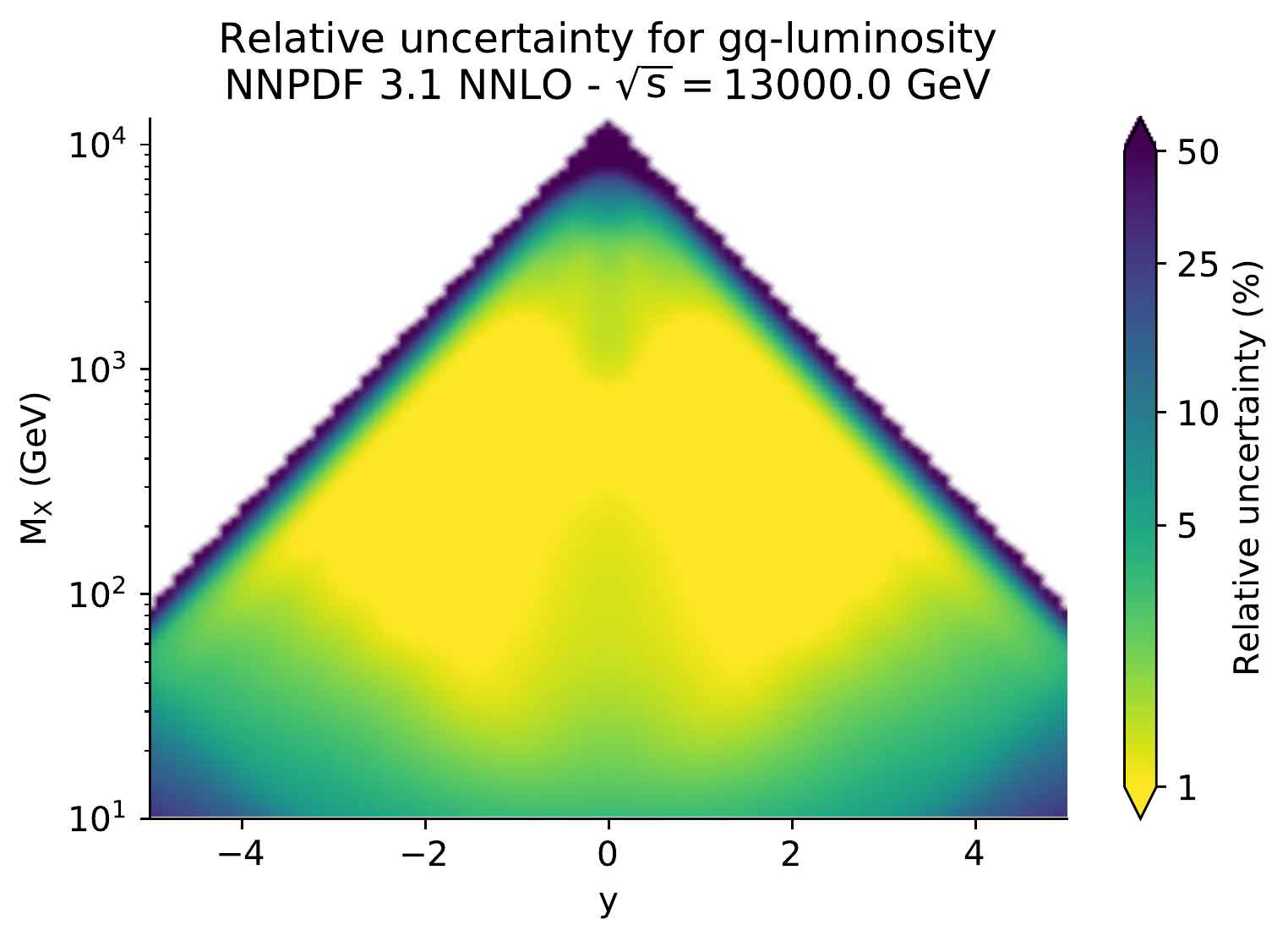}
 \includegraphics[scale=0.40]{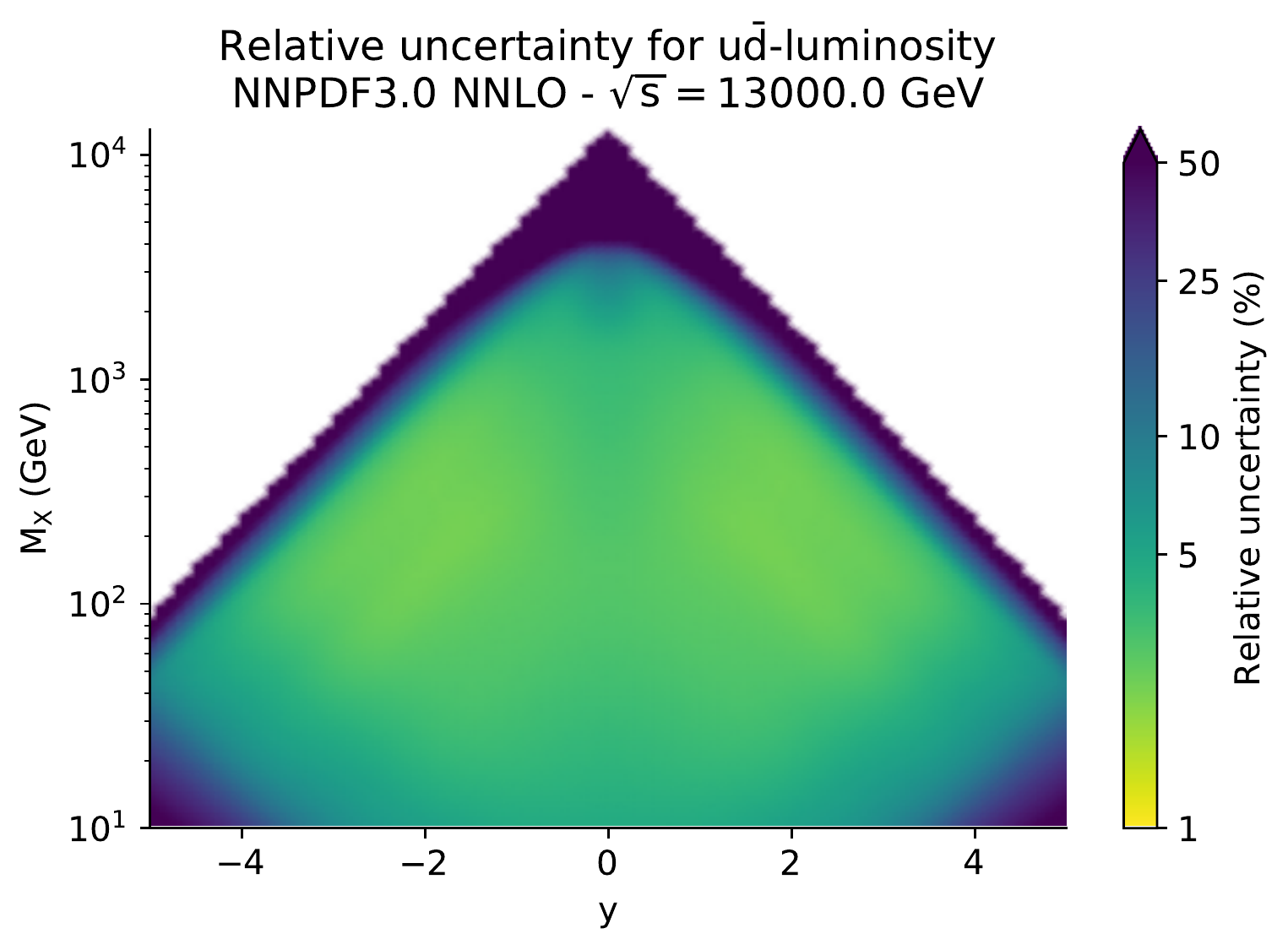}
 \includegraphics[scale=0.40]{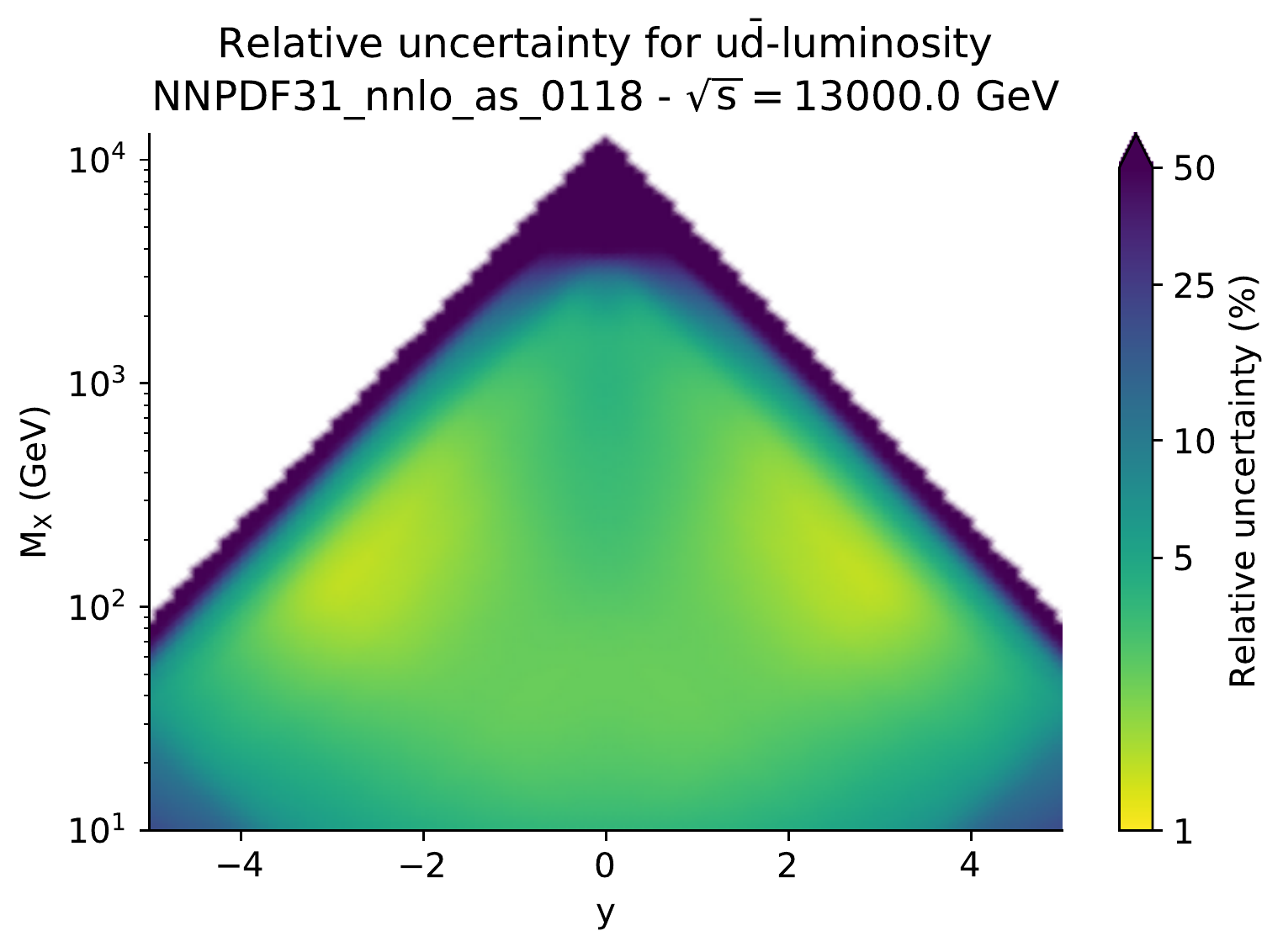}
   \caption{\small The relative uncertainty on the luminosities of
     Fig.~\ref{fig:lumi-31-vs-30}, plotted as  a function of
     the invariant mass $M_X$ and the rapidity $y$ of the final
     state; the left plots show results for NNPDF3.0 and the right
     plots for NNPDF3.1 (upper four rows). The bottom row shows
     results for the up-antidown luminosity.
\label{fig:lumi-nnpdf31-2d}}
\end{center}
\end{figure}

After analytically discussing the phenomenological implication of
individual PDFs and their uncertainties we now turn to parton
luminosities (defined as in Ref.~\cite{Mangano:2016jyj}) which drive
hadron collider processes. 
Parton luminosities from the NNPDF3.0 and NNPDF3.1 NNLO sets for the
LHC~$13$~TeV are compared
in   Fig.~\ref{fig:lumi-31-vs-30}, and their
uncertainties are displayed in  Fig.~\ref{fig:lumi-nnpdf31-2d} as a
two-dimensional contour plot as a function of the invariant mass $M_y$
and rapidity $y$ of the final state, all normalized to the NNPDF3.1
central value. We show results for the quark-quark, quark-antiquark,
  gluon-gluon  and gluon-quark luminosities, relevant for
the measurement of final states which do not couple to individual
flavors (such as $Z$ or Higgs). In the uncertainty plot we also show
for reference the up-antidown luminosity, relevant e.g. for $W^+$
production. 

Two features of this comparison are apparent. First, quark
luminosities are generally larger for all invariant masses, while the
gluon luminosity is somewhat enhanced for smaller invariant and
somewhat suppressed for  larger invariant masses in
NNPDF3.1 in comparison to NNPDF3.0.
The size of the shift in the quark sector is of order of one sigma or
sometimes even larger, while for the gluon is generally rather below
the one-sigma level. This of course reflects the pattern seen in
Section~\ref{sec:PDFcomparisons} for PDFs,
see in particular Fig.~\ref{fig:31-nnlo-vs30}.
Secondly, uncertainties are greatly reduced in NNPDF3.1 in comparison to
NNPDF3.0. This reduction is impressive and apparent in the plots of
Fig.~\ref{fig:lumi-nnpdf31-2d}, where it is clear that while
uncertainties were typically of order 5\% in most of phase space for
NNPDF3.0, they are now of the order of 1-2\% in a wide central
rapidities range
$|y|\lsim 2$ and for final state masses $100$~GeV$\lsim M_x\lsim 1$~TeV.
This is a direct consequence of  the reduction in uncertainties on both the
gluon and quark singlet PDF discussed in Section~\ref{sec:phenopdfs}
above. Indeed, luminosities which are sensitive to the flavor
decomposition, such as the up-antidown luminosity, also shown in
Fig.~\ref{fig:lumi-nnpdf31-2d}, do not display a significant reduction
in uncertainties when going to NNPDF3.0 to NNPDF3.1.

\begin{figure}[t]
\begin{center}
  \includegraphics[scale=0.38]{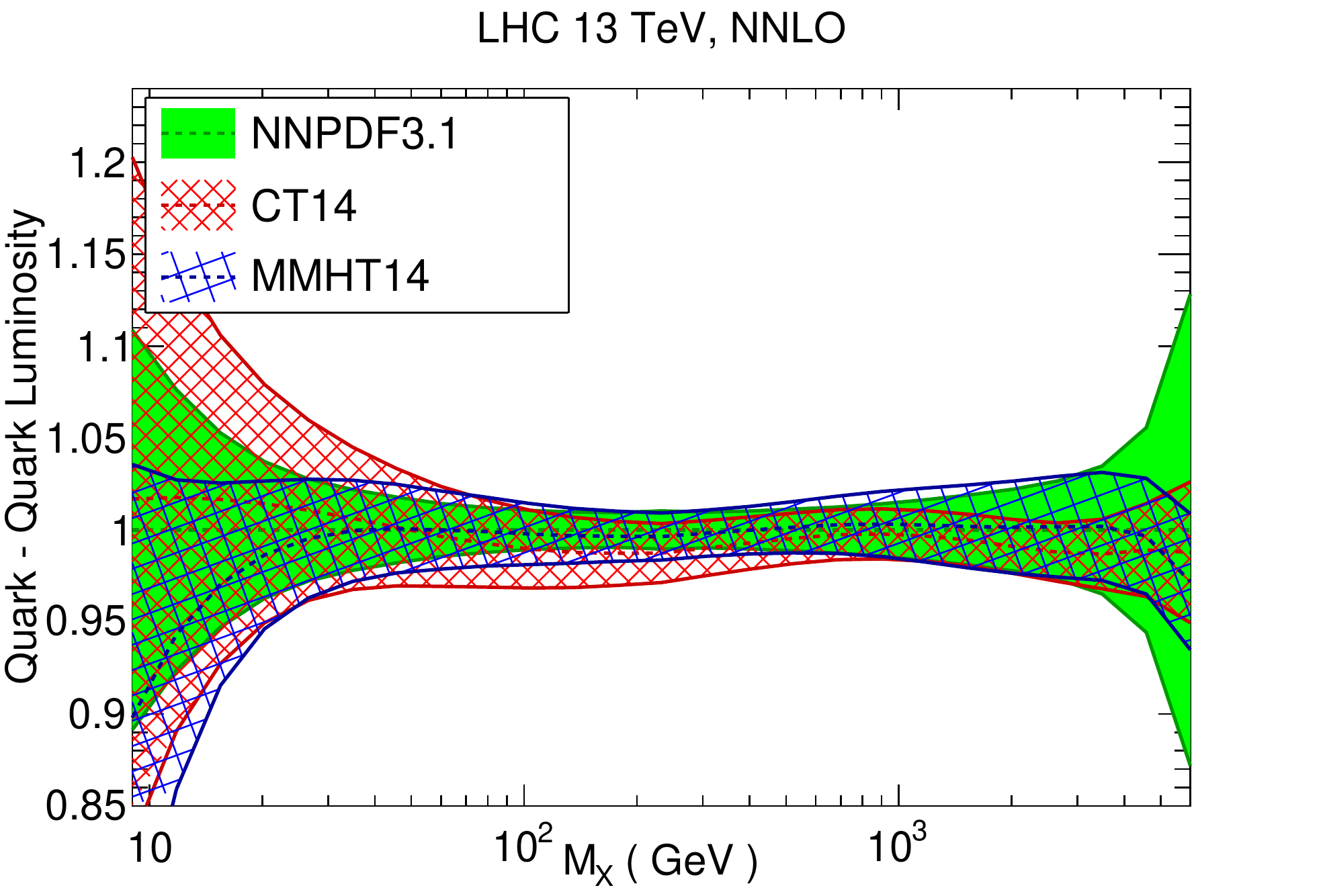}
  \includegraphics[scale=0.38]{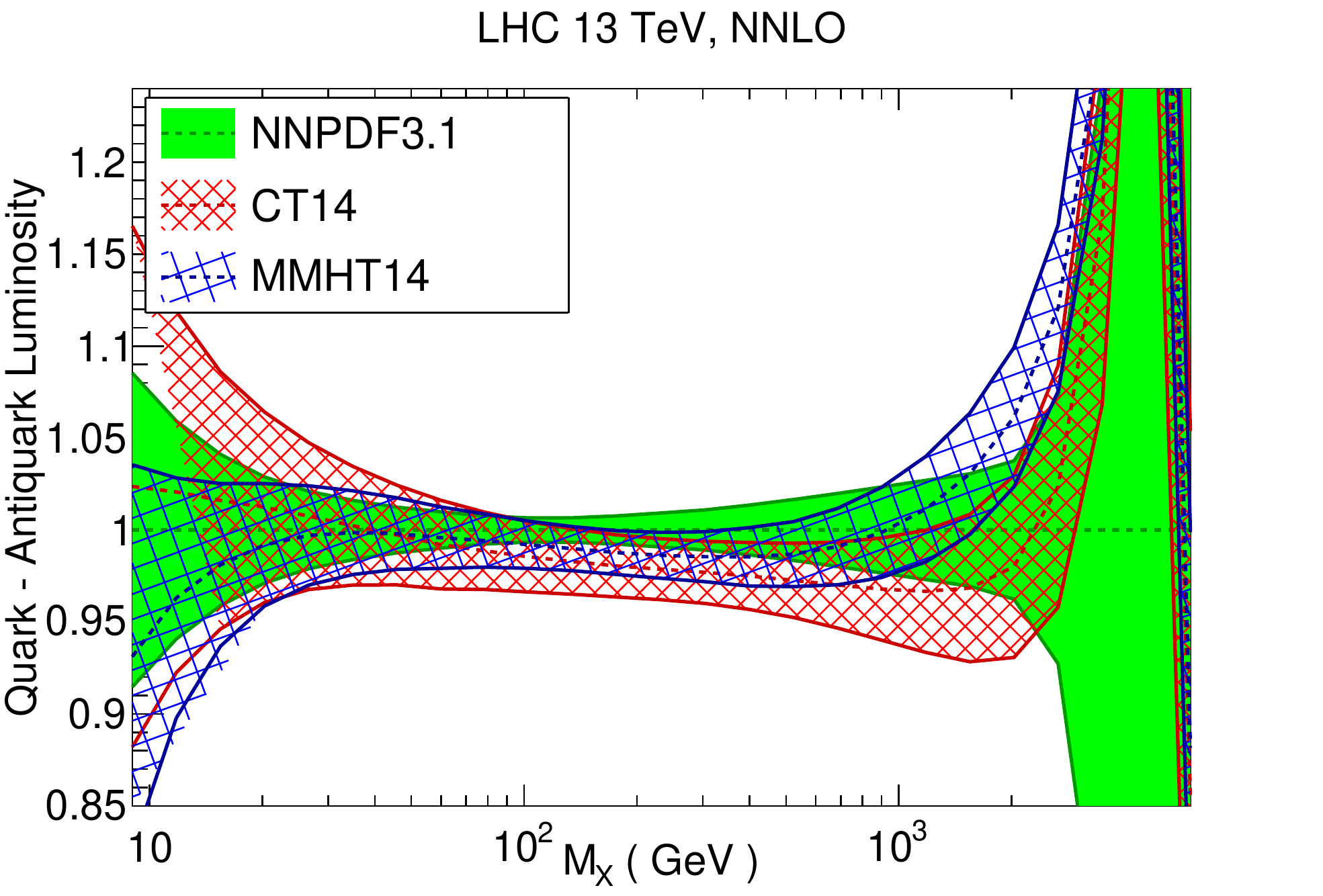}
  \includegraphics[scale=0.38]{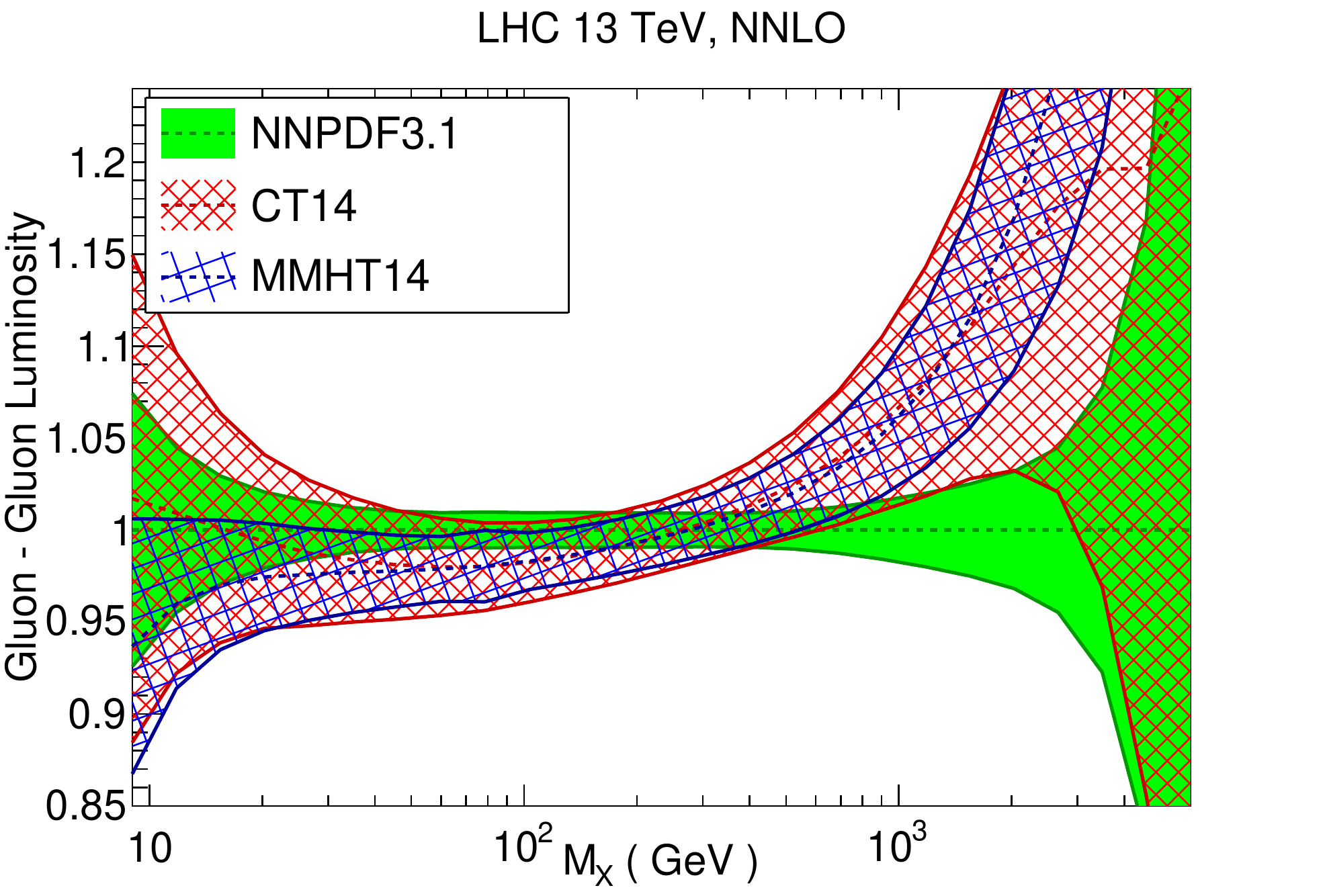}
   \includegraphics[scale=0.38]{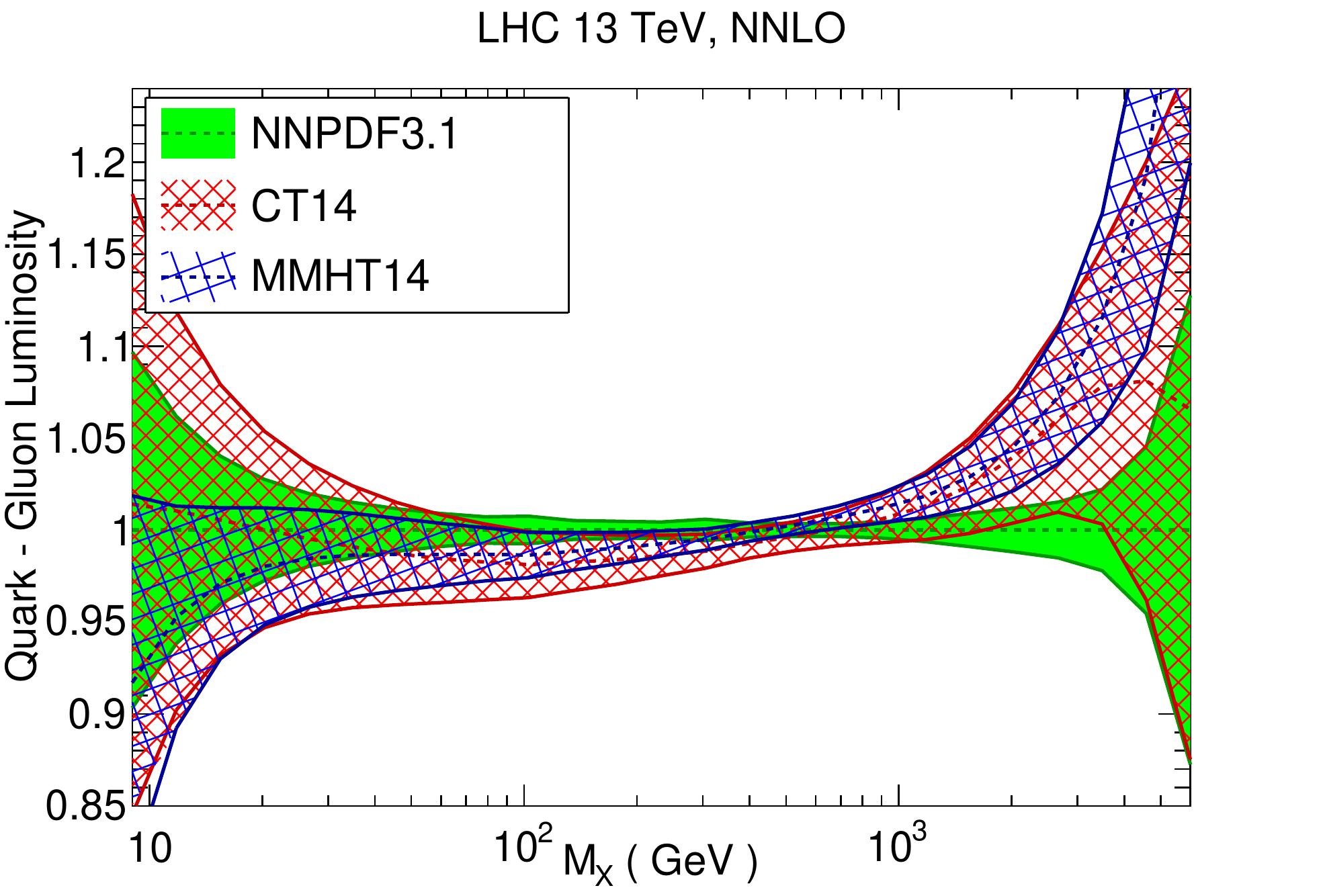}
   \caption{\small Same as Fig.~\ref{fig:lumi-31-vs-30}, now comparing
     NNPDF3.1 NNLO  to CT14 and MMHT14.
\label{fig:lumi-31-vs-global}}
\end{center}
\end{figure}

We next compare  NNPDF3.1 with CT14 and MMHT14: results are 
shown in Fig.~\ref{fig:lumi-31-vs-global}.
For the quark-quark luminosities, we find good agreement, while
for quark-antiquark  there is a somewhat bigger spread in central
values though still  agreement at the one-sigma level.
Agreement becomes marginal at large masses, $M_X\gsim 2$ TeV, reflecting
the limited knowledge of the large-$x$  PDFs. For the 
gluon-gluon and gluon-quark  channels
we find reasonable agreement for masses up to $M_X\simeq 600$ GeV, relevant for 
precision physics at the LHC, but rather worse agreement for larger
masses, relevant for BSM searches,
in particular between NNPDF3.1 and MMHT14.
Of course it should be kept in mind that NNPDF3.1 has a wider
dataset and a larger number of independently parametrized PDFs than
MMHT14 and CT14, hence the situation may change in the future once
all global PDF sets are updated.

\begin{figure}[t]
\begin{center}
  \includegraphics[scale=0.35]{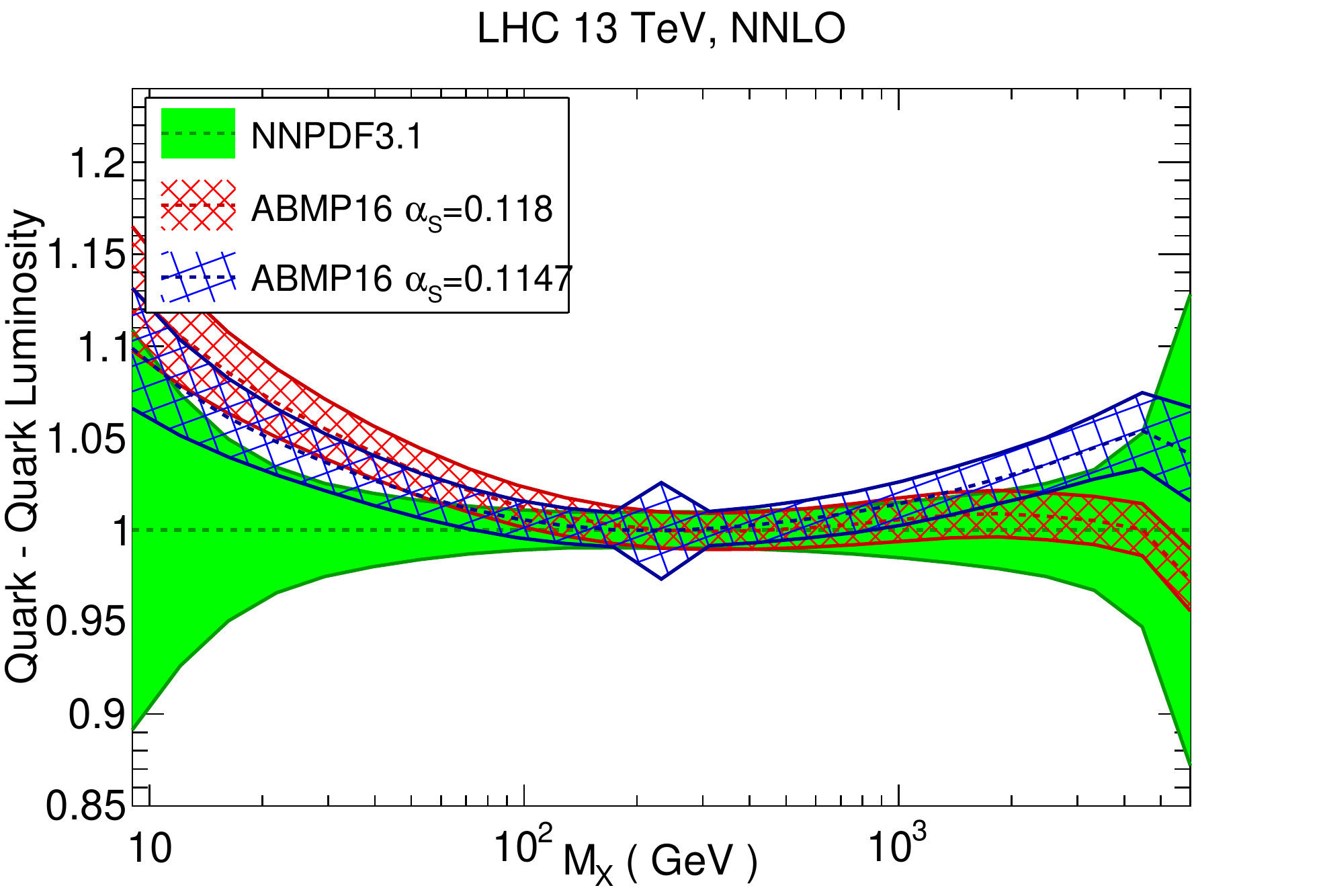}
  \includegraphics[scale=0.35]{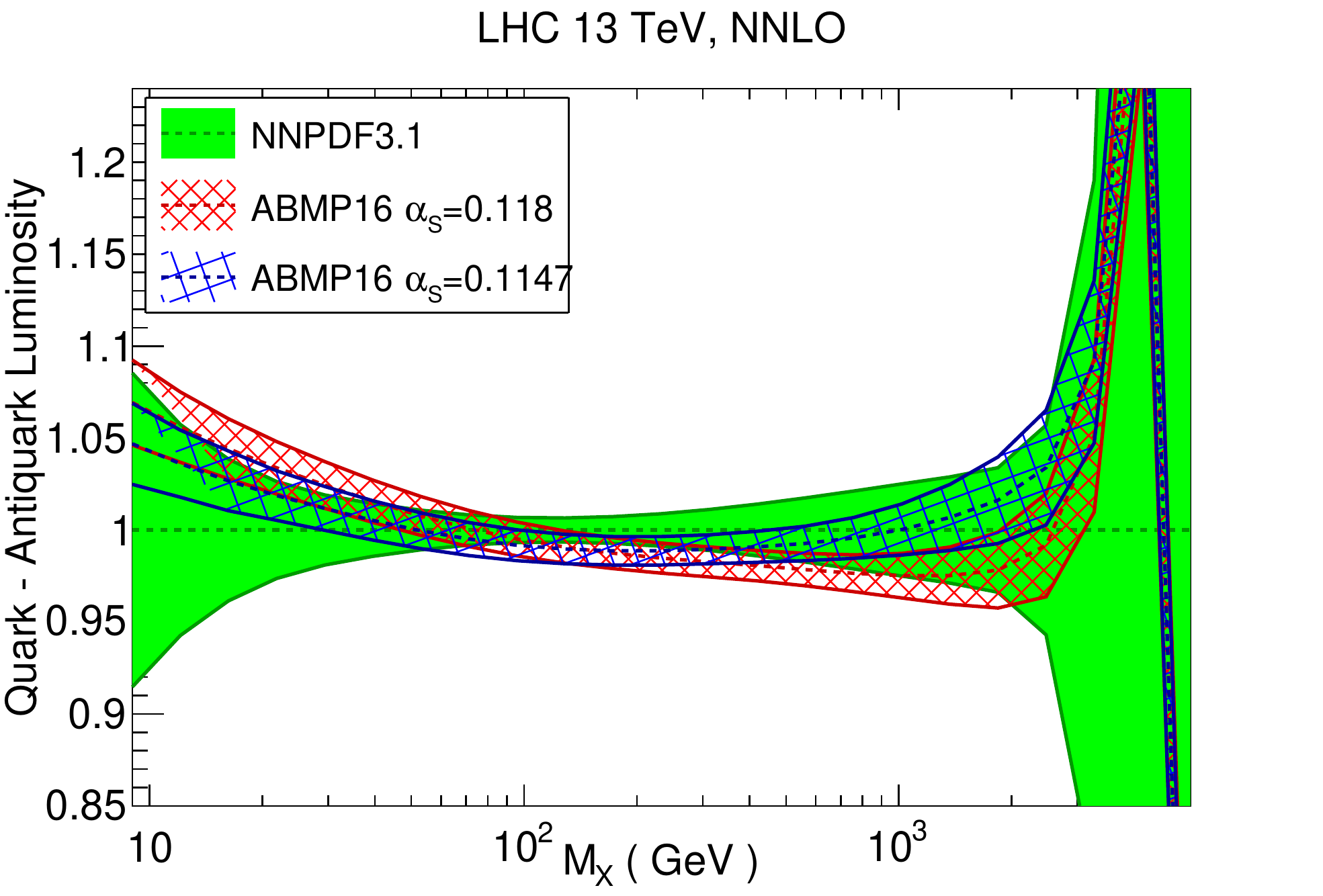}
  \includegraphics[scale=0.35]{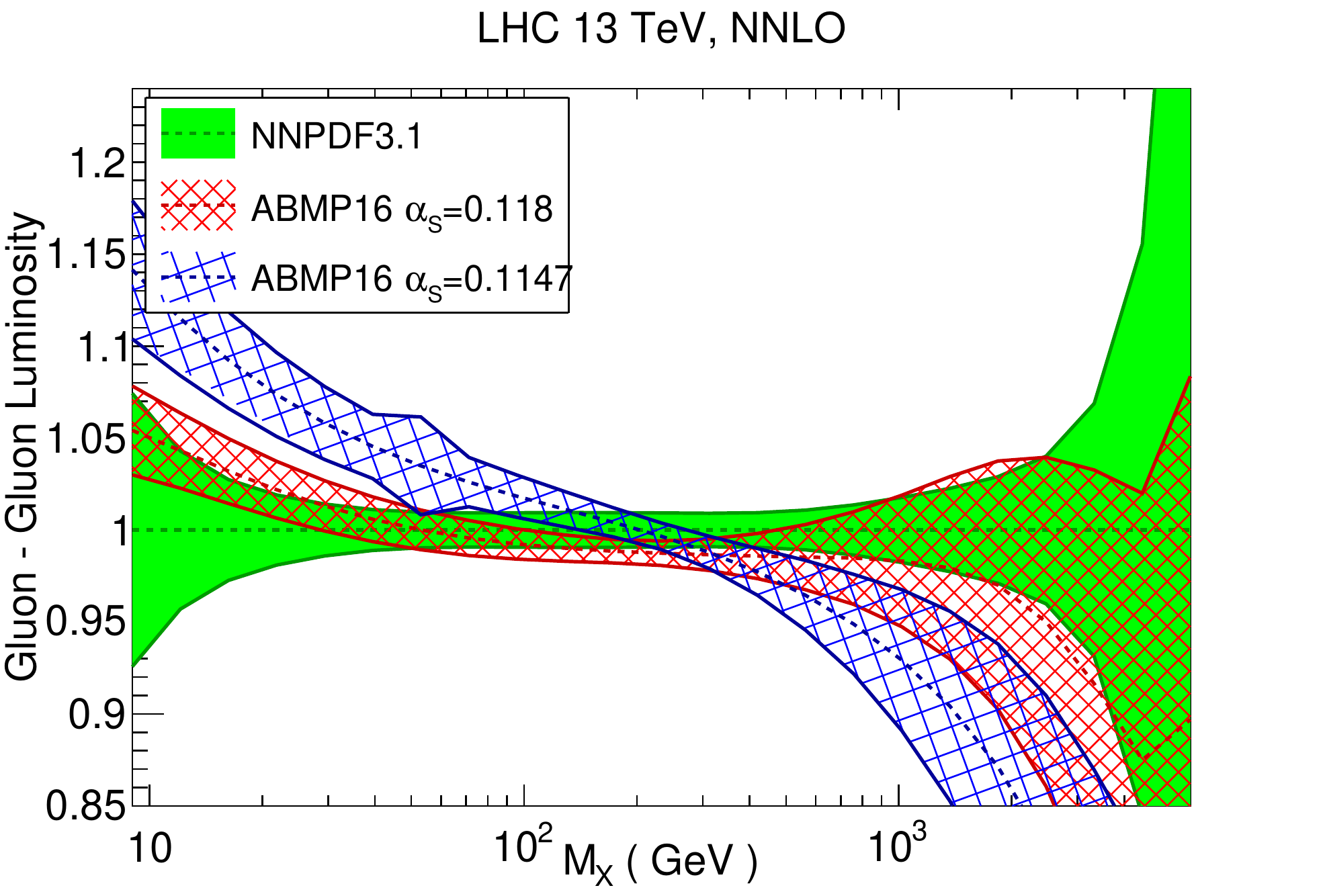}
   \includegraphics[scale=0.35]{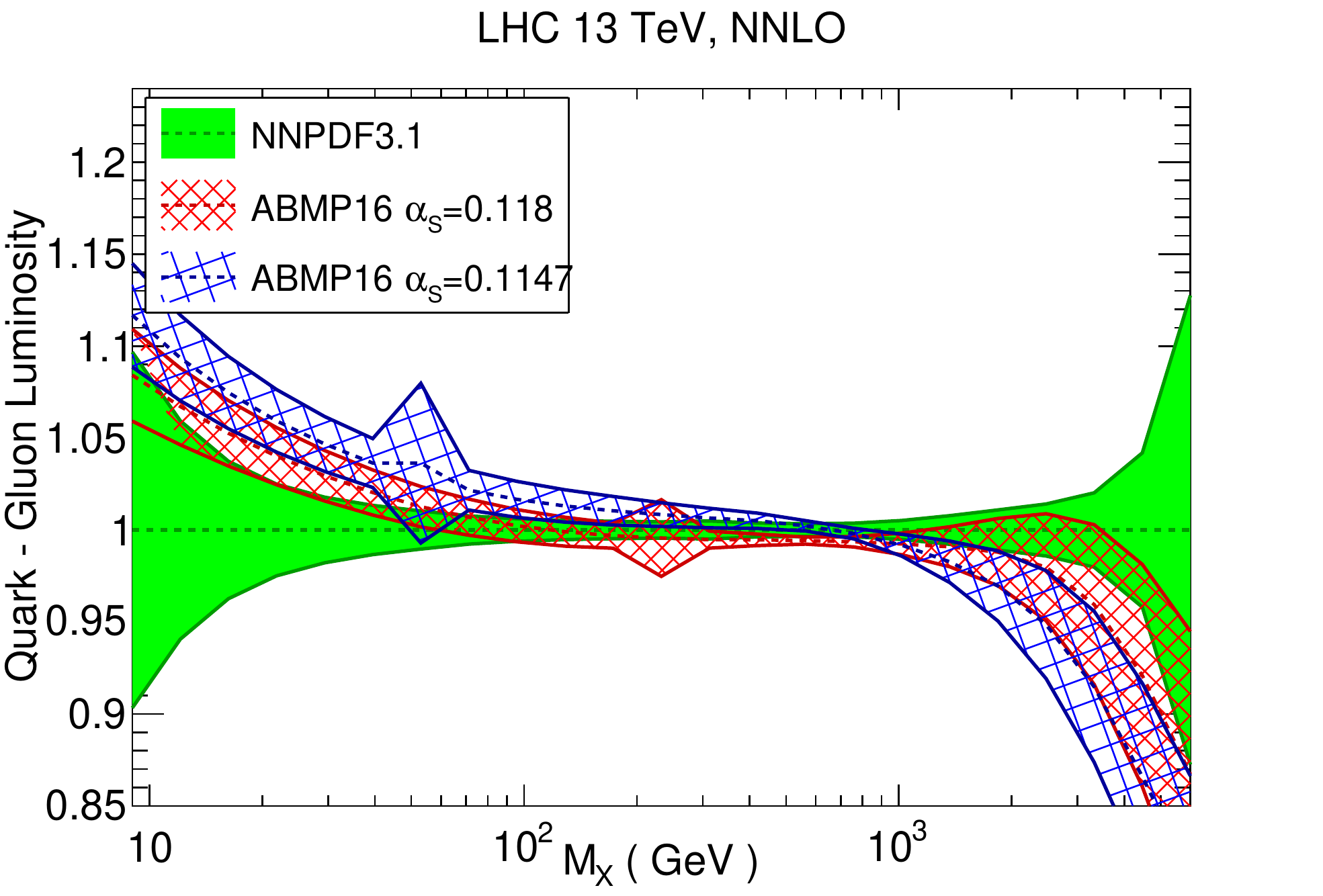}
   \caption{\small Same as Fig.~\ref{fig:lumi-31-vs-global}, now
     comparing to  ABMP16 PDFs, both with their default
     $\alpha_s(m_Z)=0.1149$  and with the common value $\alpha_s(m_Z)=0.118$.
\label{fig:lumi-31-vs-abmp}}
\end{center}
\end{figure}

Next, in Fig.~\ref{fig:lumi-31-vs-abmp} we compare to
ABMP16 PDFs. In this case, we show results corresponding both to the
default ABMP16 set, which has  $\alpha_s(m_Z)=0.1147$, and to the set with
the common  $\alpha_s(m_Z)=0.118$ adopted so far in all comparison.
     While there
     are sizable differences between NNPDF3.1 and ABMP16 when
     the default ABMP16 value  $\alpha_s(m_Z)=0.1147$ is used, especially for 
     the gluon-gluon luminosity, the agreement improves when
     $\alpha_s(m_Z)=0.118$ is adopted also for ABMP16. 
   However, ABMP16 luminosities have very small uncertainties
  at low and high $M_X$, presumably a consequence of an
  over-constrained parametrization, and of using a Hessian approach
  but with no tolerance, as discussed in Section~\ref{sec:PDFcomparisons}.

\clearpage

\begin{figure}[t]
\begin{center}
  \includegraphics[scale=0.35]{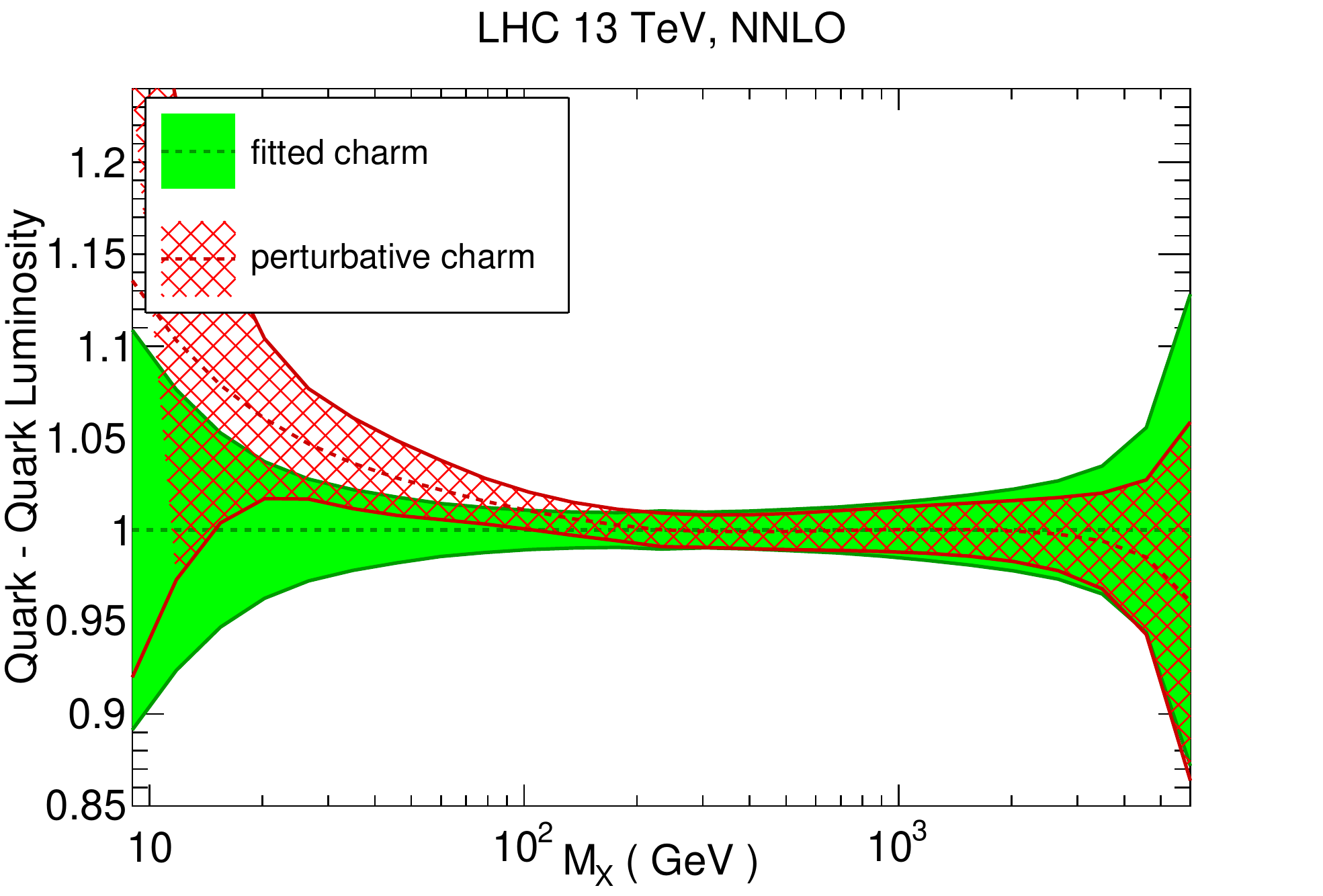}
  \includegraphics[scale=0.35]{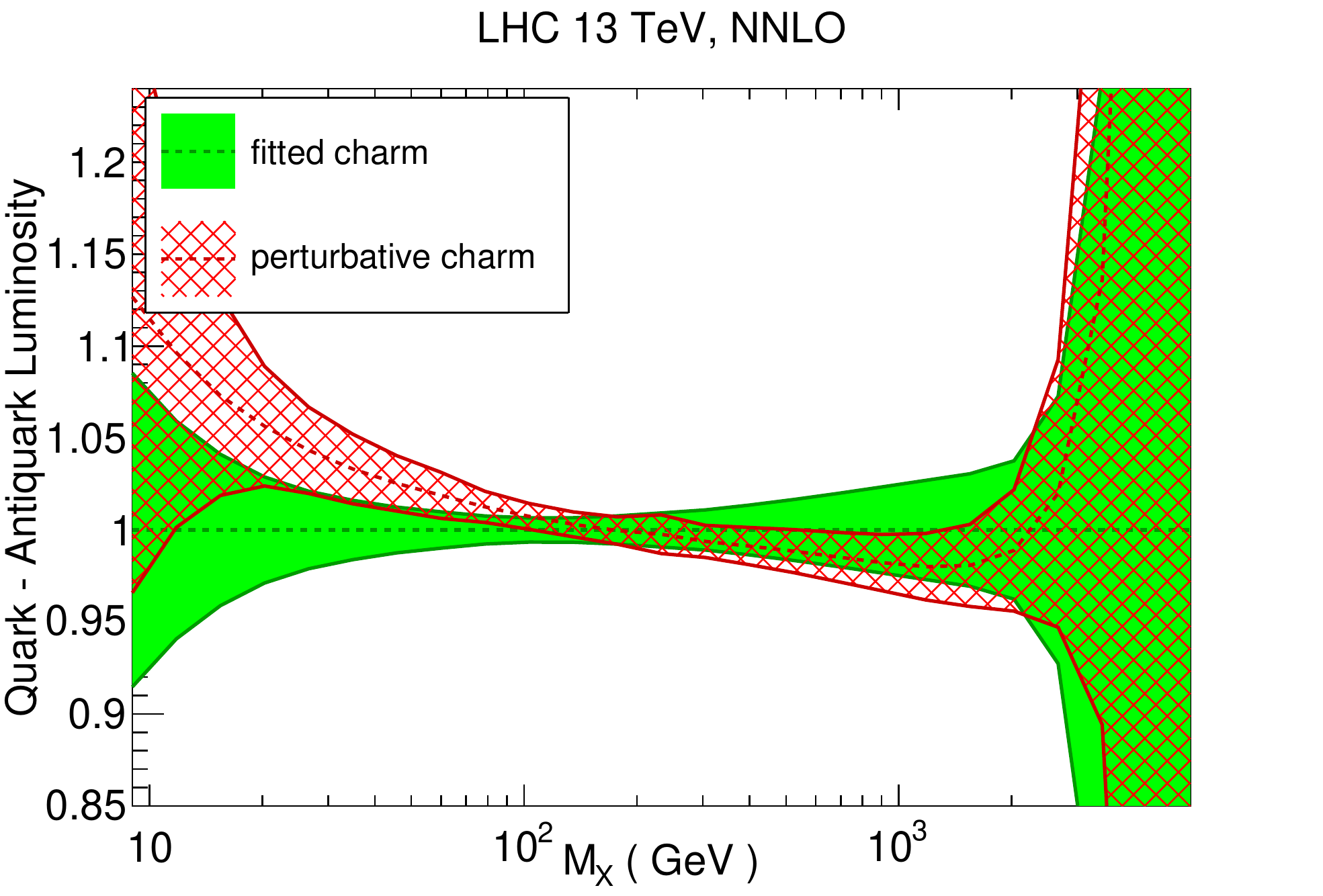}
  \includegraphics[scale=0.35]{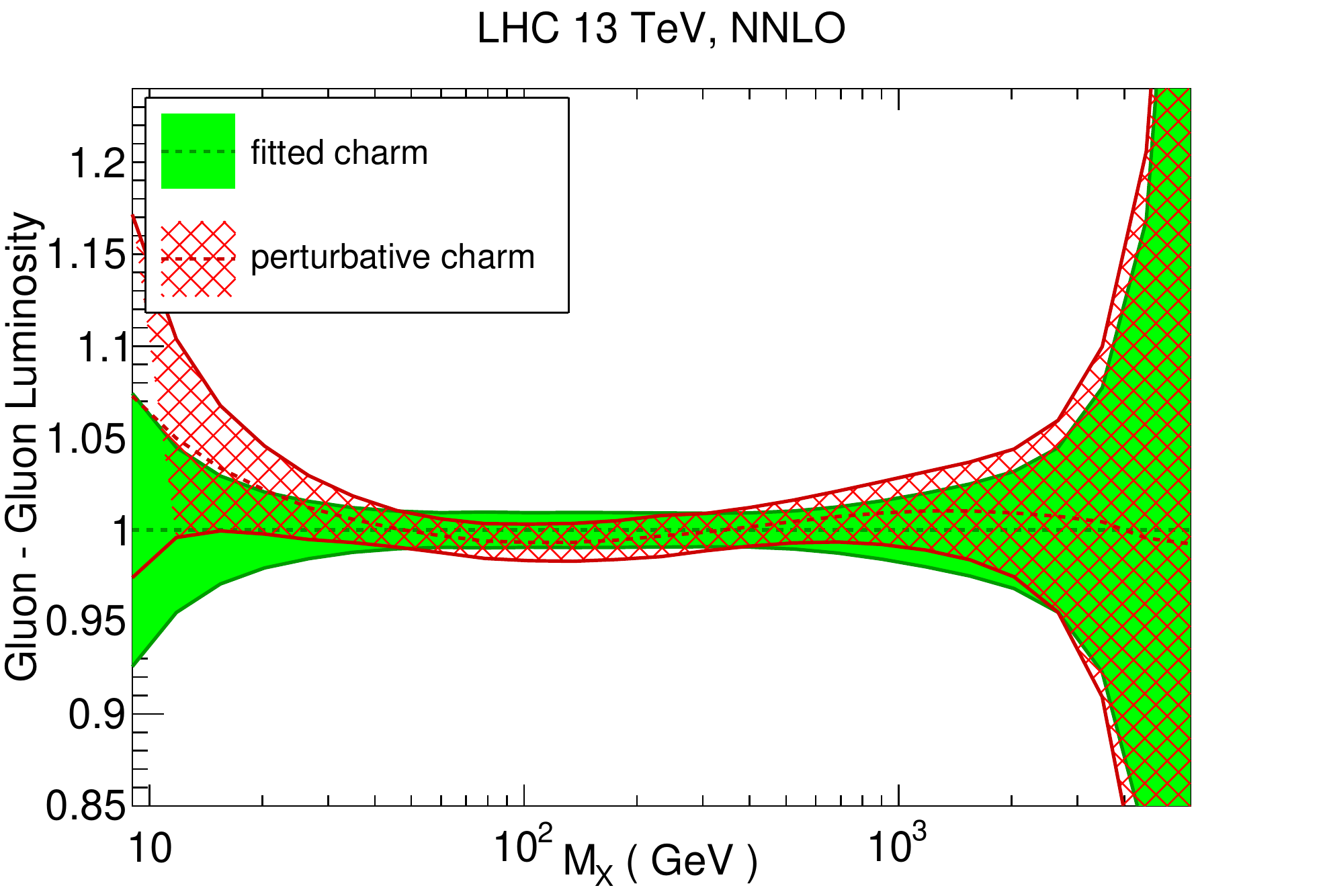}
   \includegraphics[scale=0.35]{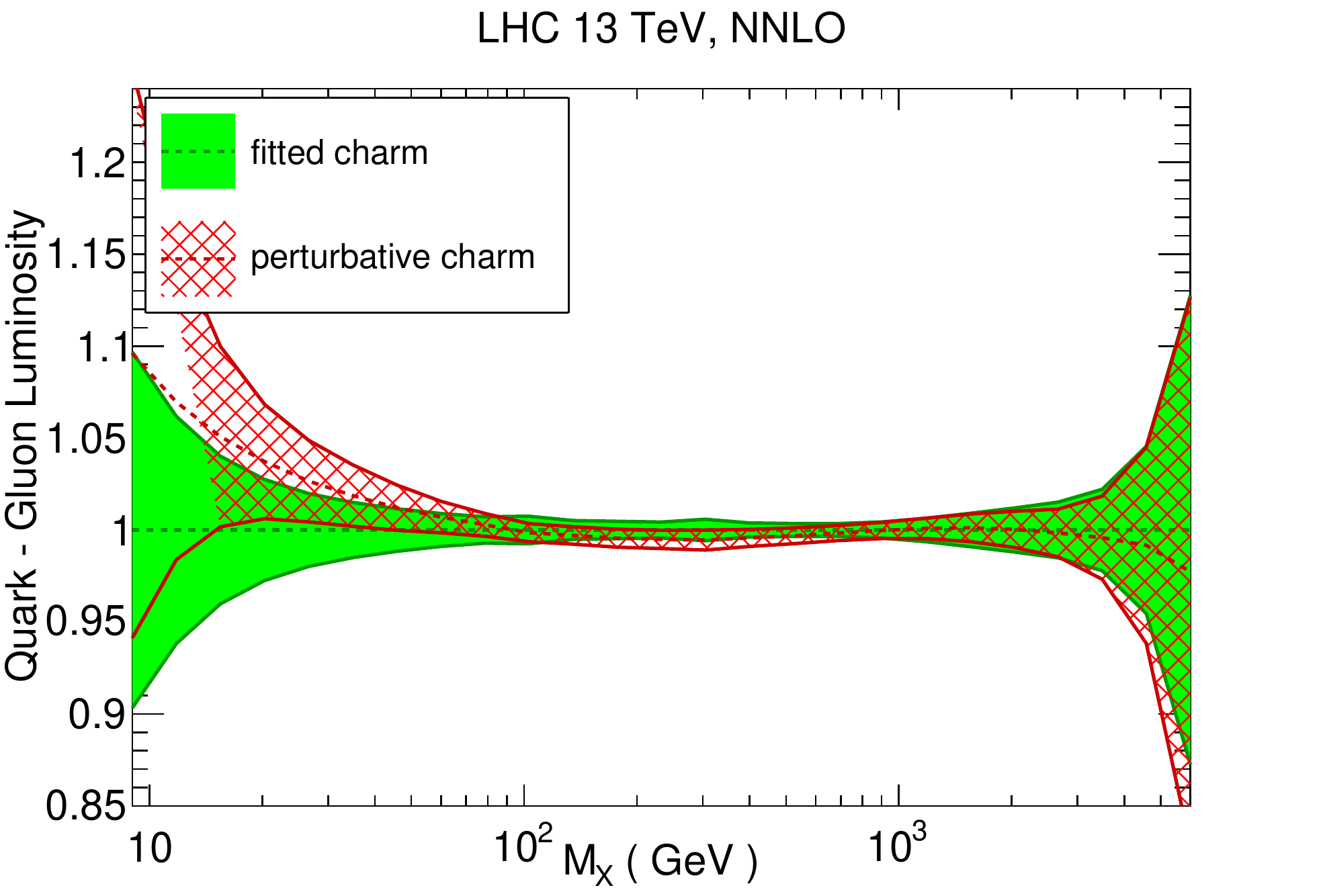}
   \caption{\small Same as Fig.~\ref{fig:lumi-31-vs-30}, now
     comparing  NNPDF3.1 to its modified version
   with perturbative charm.
\label{fig:lumi-31-fc-vs-pc}}
\end{center}
\end{figure}

Finally, in order to further emphasize the phenomenological impact of
the new NNPDF3.1 methodology, we compare NNPDF3.1 PDFs to the modified
version in which the charm PDF is perturbatively generated, already
discussed in Sects.~\ref{sec:results-mc},~\ref{sec:phenocharm}.
Results are shown in Fig.~\ref{fig:lumi-31-fc-vs-pc}.
On the one hand, we confirm that despite having one more parametrized
PDF, uncertainties are not increased. On the other hand, the effect on
central values is moderate but non-negligible. For
 gluon-gluon and quark-gluon luminosities, differences are always below
the one-sigma level, and typically rather
less.
For the quark-quark channel, results do not depend on the charm
treatment
for $M_X\gsim 200$ GeV, but for smaller invariant masses
perturbatively generated charm 
leads to a larger PDF luminosity than the best-fit parametrized charm.
For the quark-antiquark luminosity, we find a similar pattern at small
$M_X$, but  also 
some differences at medium and large $M_X$.

\clearpage

\subsection{$W$ and $Z$ production at the LHC 13 TeV}
\label{sec:pheno13TeV}

We compare 
theoretical predictions based on the NNPDF3.1 set to  $W$ and $Z$
production data at $\sqrt{s}=13$ TeV from  
ATLAS~\cite{Aad:2016naf}. Similar measurements by CMS~\cite{CMS-PAS-SMP-15-004} are not 
included in this comparison as they are still preliminary.
We compute  fiducial cross-sections using {\tt
  FEWZ}~\cite{Gavin:2012sy} at NNLO QCD 
     accuracy, using NNPDF3.1, NNPDF3.0, CT14, MMHT14
     and ABMP16 PDFs, together with the corresponding PDF uncertainty band. All calculations 
(including ABMP16) are performed with $\alpha_s=0.118$.
     Electroweak NLO corrections are  computed with
     {\tt FEWZ} for $Z$  production, and with {\tt HORACE3.2}~\cite{CarloniCalame:2007cd} for
     $W$  production.
    The fiducial phase space for the $W^{\pm}$ cross-section
    measurement in ATLAS is 
    by $p_T^l\ge 25$ GeV and $|\eta_l|\le 2.5$ for the charged lepton
    transverse momentum and pseudo-rapidity, a missing energy of
    $p_T^\nu\ge 25$ GeV 
    and a $W$ transverse mass of $m_T\ge 50$ GeV.
    For $Z$  production, 
    $p_T^l\ge 25$ GeV and $|\eta_l|\le 2.5$ for the charged leptons transverse
    momentum and rapidity and $66\le m_{ll} \le 116$ GeV for the dilepton
    invariant mass.
         
\begin{figure}[t]
\begin{center}
  \includegraphics[scale=0.88]{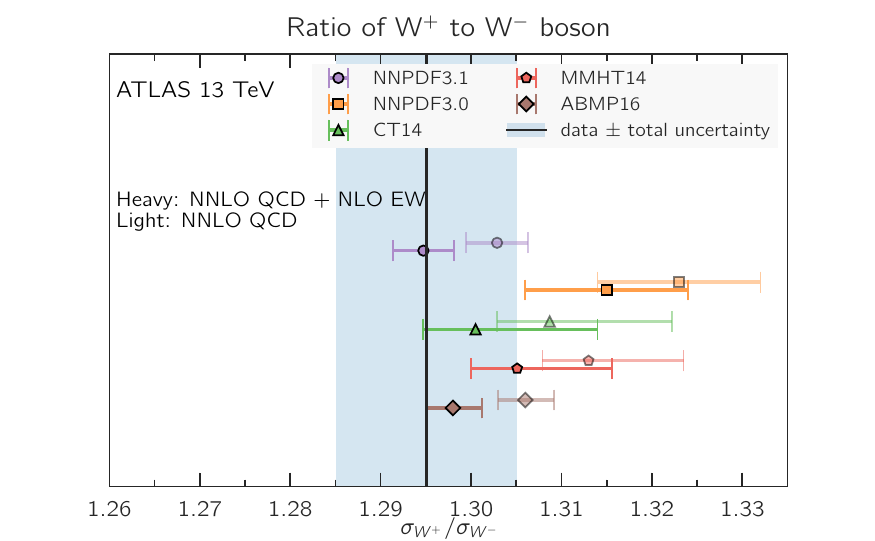}
  \includegraphics[scale=0.88]{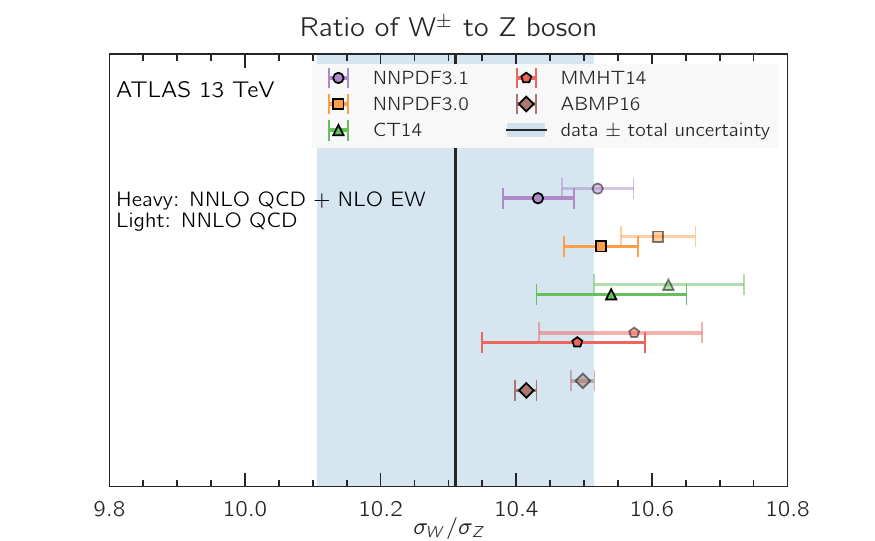}
  \caption{\small Comparison of the ATLAS 
    measurements of the $W^+/W^-$ ratio (left) 
    and the $W/Z$ ratio (right) at $\sqrt{s}=13$ TeV with theoretical
    predictions computed with different NNLO PDF sets.
    Predictions are shown with (heavy) and without (light) NLO EW corrections 
    computed with {\tt FEWZ} and {\tt HORACE}, as described in the text.
\label{fig:LHC-13tev-incxsec-1}}
\end{center}
\end{figure}

In Fig.~\ref{fig:LHC-13tev-incxsec-1} we compare the 
ATLAS~\cite{Aad:2016naf} 13~TeV
    measurements of the the $W^+/W^-$ and $W/Z$ ratios
    in the fiducial region at $\sqrt{s}=13$ TeV to theoretical
    predictions, both  with and without electroweak corrections.
We see that for both the $W^+/W^-$ ratio and the $W/Z$ ratio, all the PDF sets are in 
reasonably good agreement with the data. The uncertainty in the
theoretical prediction shown in the plot is the PDF  
uncertainty only: parametric uncertainties (in the values of $\alpha_s$ and $m_c$) and missing higher 
order QCD uncertainties are not included.
Interestingly, electroweak corrections shift the
theory predictions by around 0.5\%, and for all PDF sets they 
improve the agreement with the ATLAS 
measurements.
NNPDF3.1 results have  smaller PDF
uncertainties than NNPDF3.0, and  are in better agreement with the
ATLAS data.

\begin{figure}[t]
\begin{center}
  \includegraphics[scale=1.0]{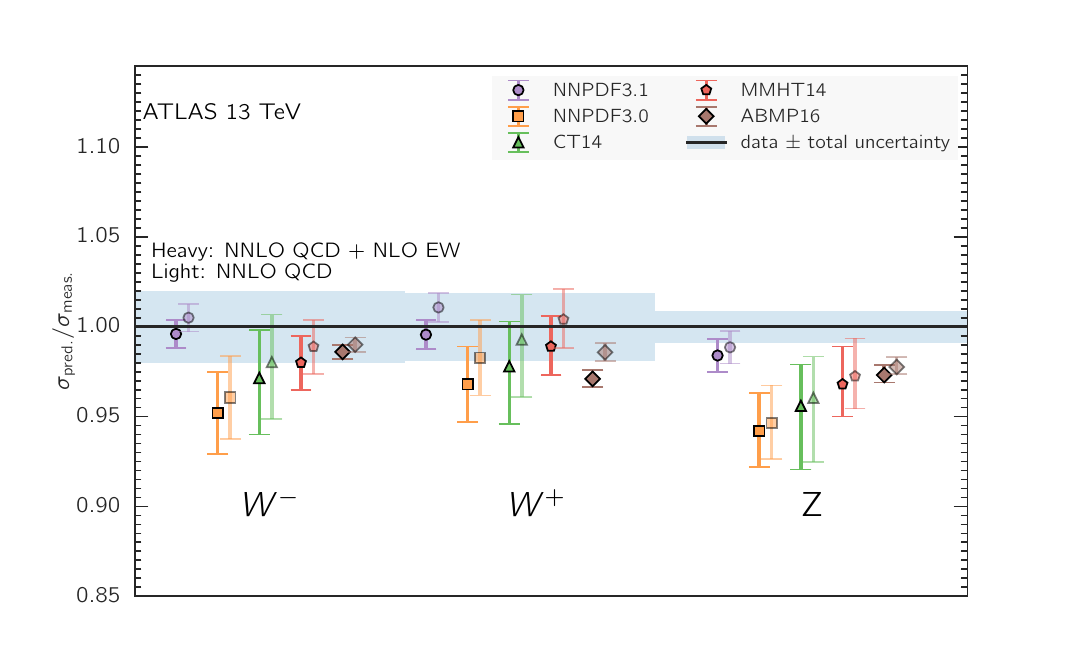}
  \caption{\small Same as Fig.~\ref{fig:LHC-13tev-incxsec-1}, now 
    for the absolute $W^+$, $W^-$ and $Z$ cross-sections.
    All predictions are normalized to the experimental central value.
\label{fig:LHC-13tev-incxsec-2}}
\end{center}
\end{figure}

The corresponding absolute $W^+$, $W^-$ and $Z$ cross-sections are
shown in Fig.~\ref{fig:LHC-13tev-incxsec-2},
    normalized in each case to the experimental central value. 
Again, predictions are generally 
    in agreement with the data, with the possible exception of 
    ABMP16 for  $Z$ production. In comparison to cross-section
    ratios, the effect of electroweak corrections on absolute
    cross-sections, around 1\% for $W^+$ and $W^-$ and around 0.5\%
for  $Z$ production, is rather less significant on the scale of
the uncertainties involved, and it does not necessarily lead to
improved agreement.
Comparing  NNPDF3.1 to NNPDF3.0 we see again considerably reduced
uncertainties and improved agreement of the prediction with data. 
This improved agreement  is particularly marked for $Z$ production,
where 3.0 was about 5\% below the data,  
while now 3.1 agrees within uncertainties.
%

\begin{figure}[t]
\begin{center}
  \includegraphics[scale=1.0]{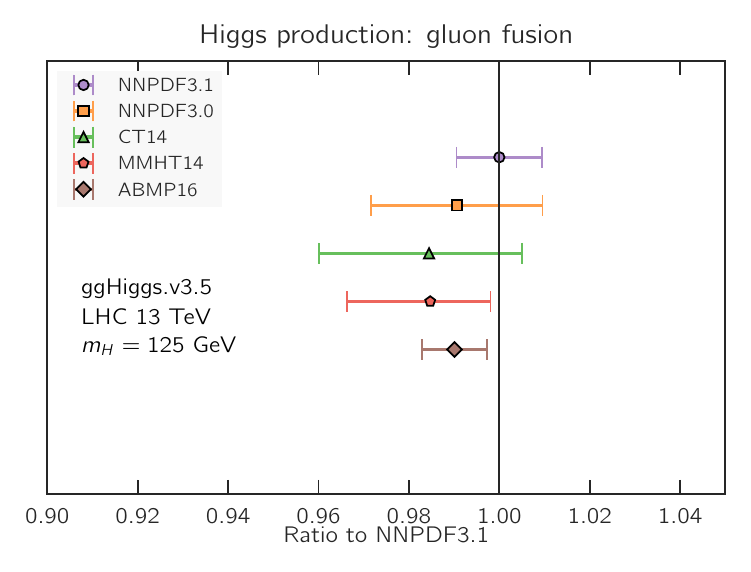}
  \includegraphics[scale=1.0]{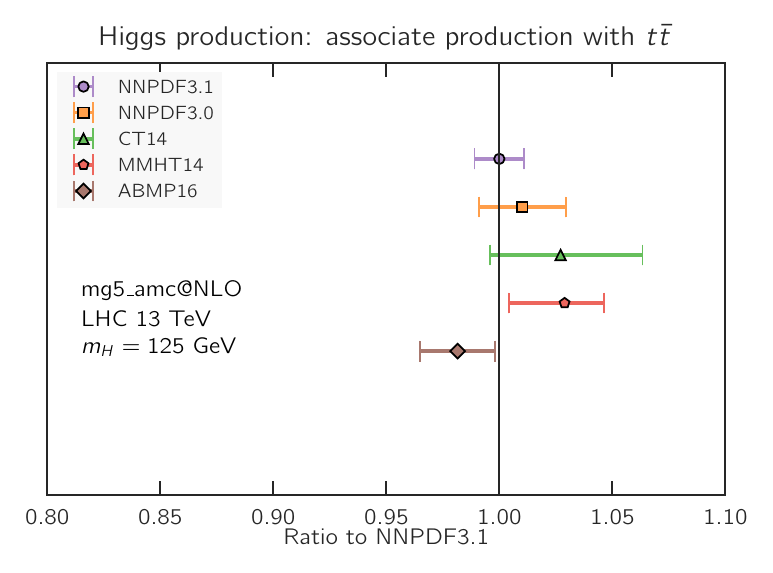}
   \includegraphics[scale=1.0]{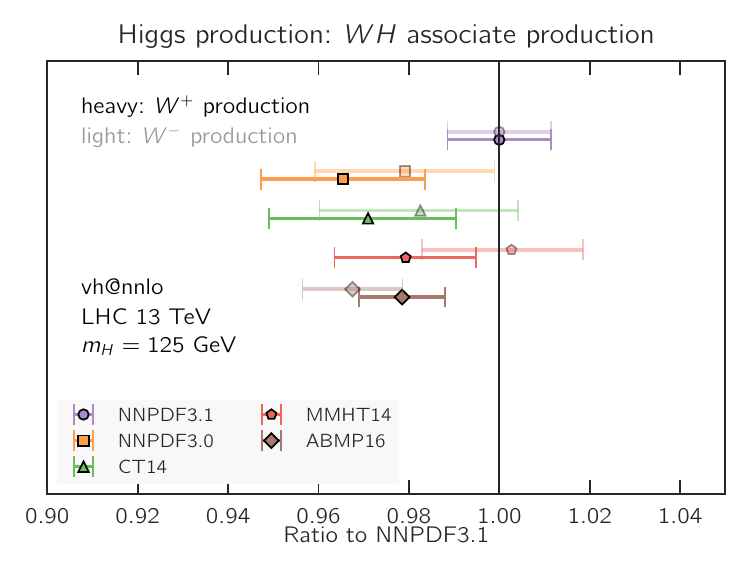}
  \includegraphics[scale=1.0]{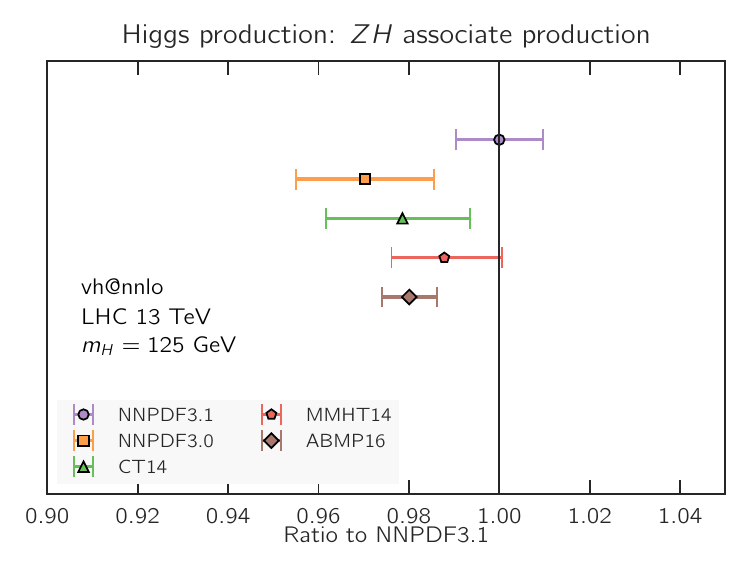}
  \caption{\small PDF dependence of the 
Higgs  production cross-sections at the LHC
    $13$ TeV for gluon fusion, $t\bar{t}$
    associated production, and $VH$ associated production.
    All results are shown as ratios to the central NNPDF3.1 result.
    Only PDF uncertainties are shown.
  \label{fig:ggH}
  }
\end{center}
\end{figure}

\begin{figure}[t]
\begin{center}
  \includegraphics[scale=1.0]{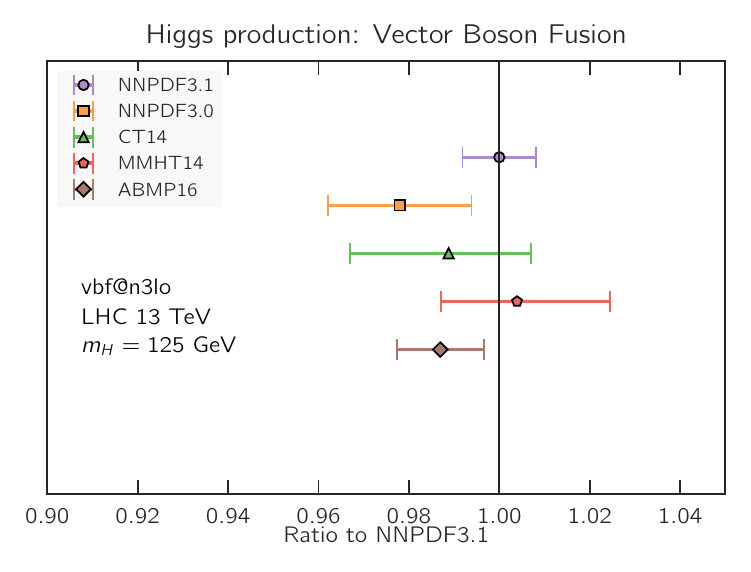}
  \includegraphics[scale=1.0]{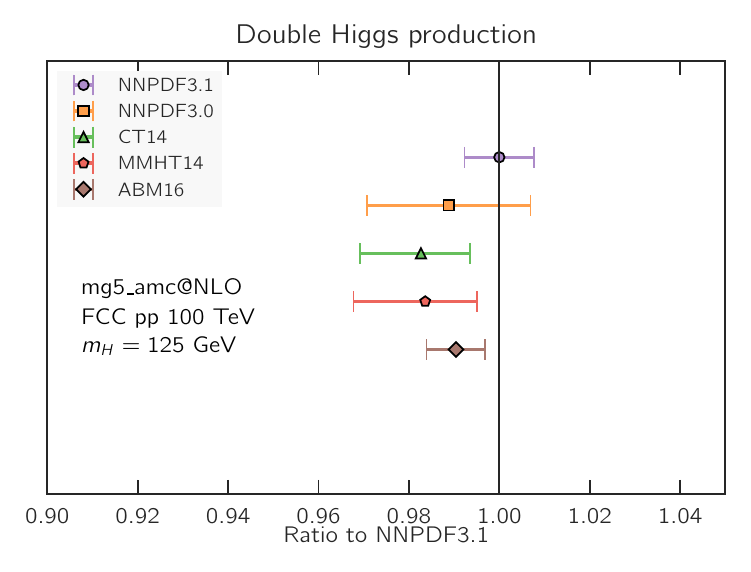}
  \caption{\small Same as Fig.~\ref{fig:ggH} for single Higgs production
    in vector boson fusion (left) and  double Higgs production
    in gluon fusion (right).
  \label{fig:ggH2}
  }
\end{center}
\end{figure}

\subsection{Higgs production}

We finally study the PDF dependence of predictions for 
inclusive Higgs production at LHC $13$ TeV, and for
Higgs pair production, which
could also be within reach of the LHC in the
near future~\cite{Bishara:2016kjn,Behr:2015oqq}.
We study single Higgs in  gluon fusion, associated production with gauge bosons and top
pairs and vector boson fusion, and double Higgs production in gluon
fusion. In each case, we show predictions normalized to the NNPDF3.1
result, and only show PDF uncertainties. All calculations 
(including ABMP16) are performed with $\alpha_s=0.118$.

The settings are the following. 
For  gluon fusion we perform the calculation at N$^3$LO using
{\tt  ggHiggs}~\cite{Ball:2013bra,Bonvini:2014jma,Bonvini:2016frm}.
Renormalization and factorization scales are set to $\mu_F=\mu_R=m_h/2$
and the computation is performed using rescaled effective theory.
For associate production with a $t\bar{t}$ pair we use
{\tt MadGraph5\_aMC@NLO}~\cite{Alwall:2014hca}, with default
factorization and renormalization scales $\mu_R=\mu_F=H_T/2$, where $H_T$ is 
the sum of the transverse masses.
For associate production with an electroweak gauge boson we use
{\tt vh@nnlo} code~\cite{Brein:2012ne} at NNLO with default scale settings.
For  vector boson fusion we perform the calculation at N$^3$LO using
{\tt proVBFH}~\cite{Dreyer:2016oyx,Cacciari:2015jma}
with the default scale settings. Finally, for double Higgs production
at the FCC~100~TeV the calculation is performed using
{\tt MadGraph\_aMC@NLO}. 

Results are shown in Figs.~\ref{fig:ggH}-\ref{fig:ggH2}.
For gluon fusion and $t\bar{t}h$, which are both driven by the gluon PDF, the
former for $x\sim10^{-2}$, and the latter for large $x$, 
results from the various  PDF sets agree within uncertainties;
NNPDF3.0 and NNPDF3.1 are also in good agreement, with the new prediction
exhibiting reduced uncertainties.
The spread of results is somewhat larger for associate production with
gauge bosons.
The NNPDF3.1 prediction is about 3\% higher than the NNPDF3.0 one,
with uncertainties reduced by a factor 2, so the two cross-sections
barely agree within uncertainties. Also, of the three PDF sets
entering the PDF4LHC15 combination, 
NNPDF3.0 gave the smallest cross-section, but NNPDF3.1 now gives the highest
one: $VH$ production is driven by the quark-antiquark luminosity,
and this enhancement for $M_X\simeq 200$ GeV between 3.0 and 3.1
could indeed be observed already in Fig~\ref{fig:lumi-31-vs-30}.
For VBF we also find that the NNPDF3.1 result is larger, by
about 2\%,  than
the NNPDF3.0 one, with smaller
uncertainties, and it is in better  agreement with other PDF sets.
Finally, for double Higgs production in gluon fusion
the central value  with  NNPDF3.1 increases slightly but is otherwise consistent
with the NNPDF3.0 prediction, and
here there  is also good agreement for all the PDF sets.

\section{Summary and outlook}
\label{sec:conclusion}
NNPDF3.1  is the new main PDF release from the
NNPDF family.
It represents a significant improvement over
NNPDF3.0, by including constraints from many new observables,
some of which are included for the first time in a global
PDF determination,  thanks to the recent availability of the corresponding
NNLO QCD corrections. Notable examples are $t\bar{t}$ differential distributions
and the $Z$ boson $p_T$ spectrum.
From the theory point of view,
the main improvement is to place the charm
PDF on an equal footing as the light quark
PDFs.
Independently parametrizing  the charm PDF  resolves a tension which would otherwise be present
between ATLAS gauge boson production and HERA inclusive structure
function data, leads to improved agreement with the 
LHC data, and turns the strong dependence of perturbatively
generated charm on the value of the pole charm mass into a PDF
uncertainty, as most of the mass dependence is reabsorbed into the initial
PDF shape. 

The NNPDF3.1 set is also the first set for which PDFs are delivered in a
variety of formats: first of all, they are released both in Hessian
and Monte Carlo form, and furthermore, the default sets are optimized
and compressed so that a smaller number of  
Monte Carlo replicas or Hessian error sets reproduces the statistical
features of much larger underlying replica sets.
We now   discuss how both Hessian and Monte
Carlo reduced sets have
been produced out of a large set of Monte Carlo
replicas;  we then summarize all PDF sets that have been made public
through the LHAPDF interface; and  finally  we
present a brief outlook on future developments.

\subsection{Validation of the NNPDF3.1 reduced sets}
\label{sec:compression}

  Default NNPDF3.1 NLO and NNLO PDFs for the central
  $\alpha_s(m_Z)=0.118$ value, as well as the modified version with
  perturbative charm discussed in Sect.~\ref{sec:results-mc}, have been produced as  $N_{\rm
    rep}=1000$ replica sets. These large replica samples have been
subsequently processed using two reduction
  strategies: the Compressed Monte Carlo (CMC) algorithm~\cite{Carrazza:2015hva},
  to obtain
  a Monte Carlo representation based on a smaller number of replicas,
  and the  MC2H algorithm~\cite{Carrazza:2015aoa},
  to achieve an optimal Hessian representation
  of the underlying PDF probability distribution with a fixed number
  of error sets. Specifically,  we have thus constructed
  CMC-PDF sets with $N_{\rm rep}=100$ replicas and
  MC2H sets with $N_{\rm eig}=100$ (symmetric) eigenvectors.

\begin{figure}[t]
  \begin{center}
     \includegraphics[scale=0.32]{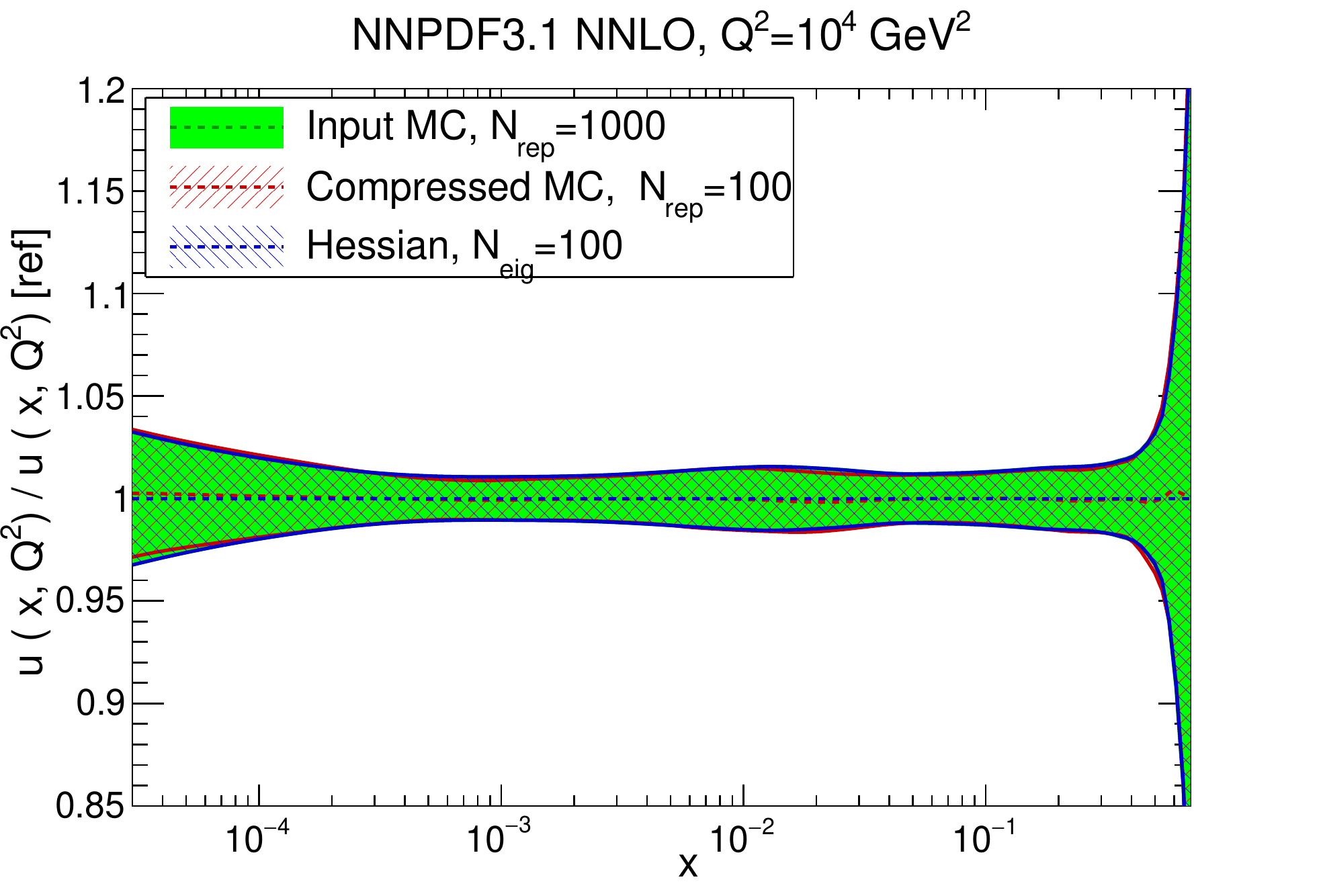}
     \includegraphics[scale=0.32]{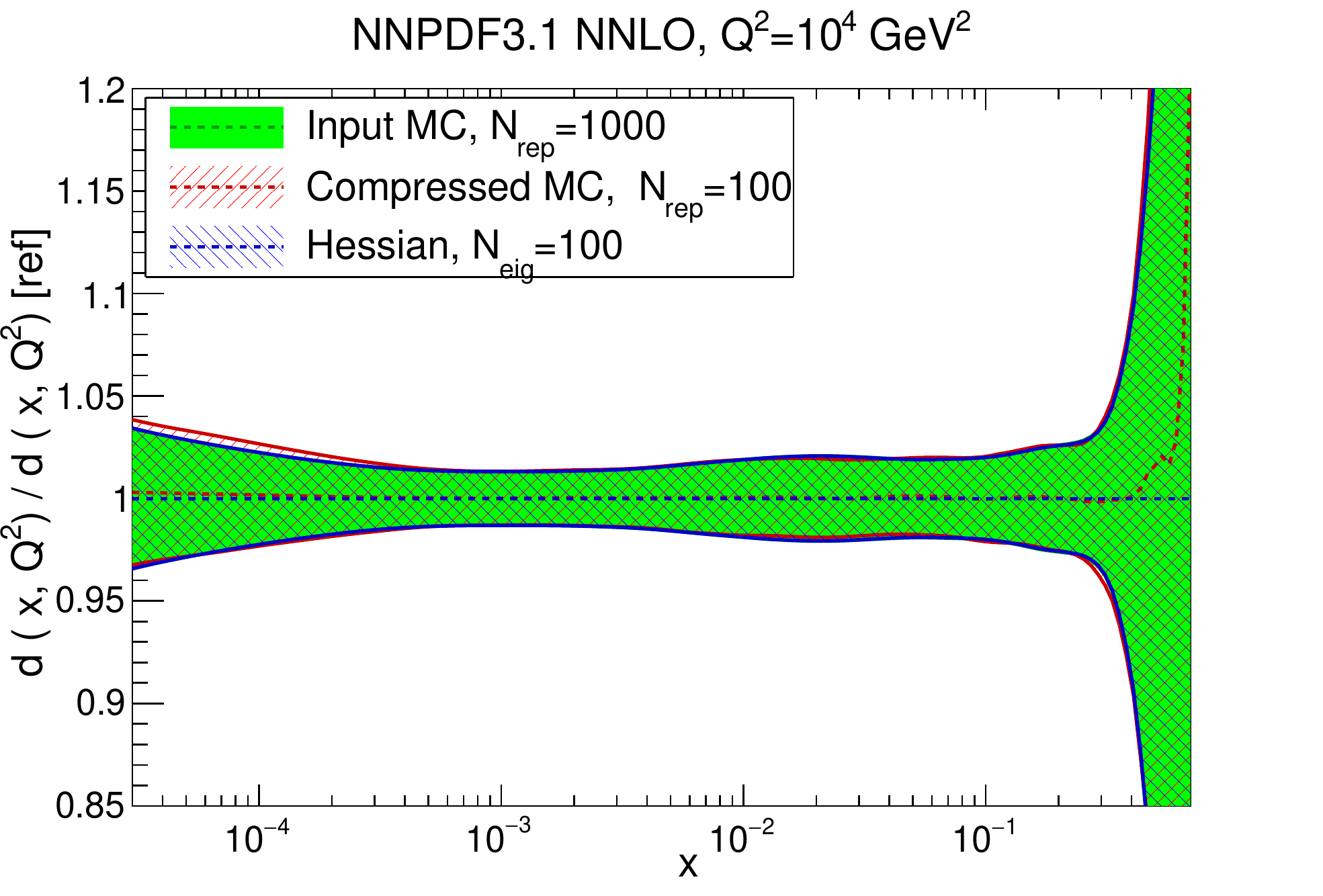}
     \includegraphics[scale=0.32]{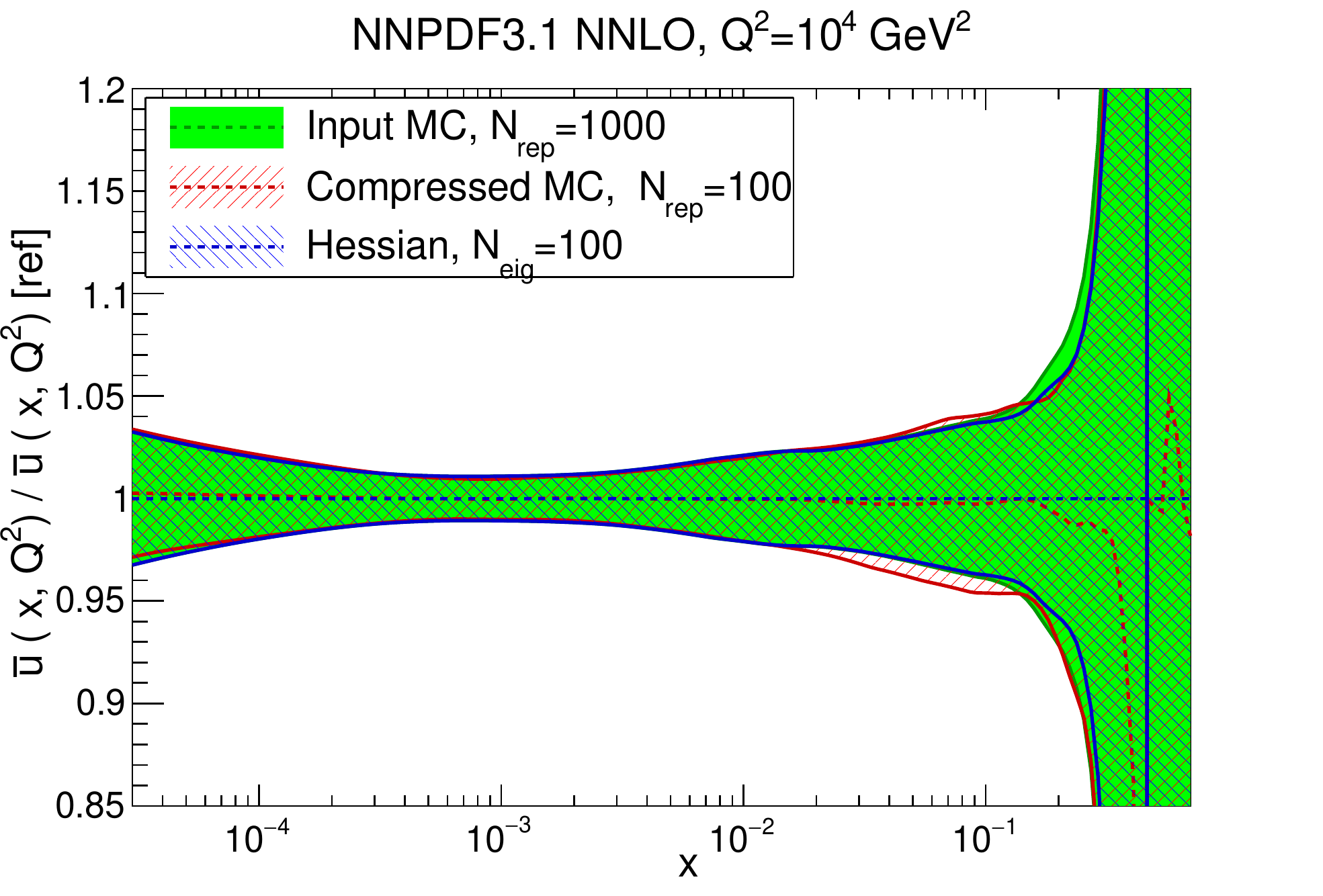}
     \includegraphics[scale=0.32]{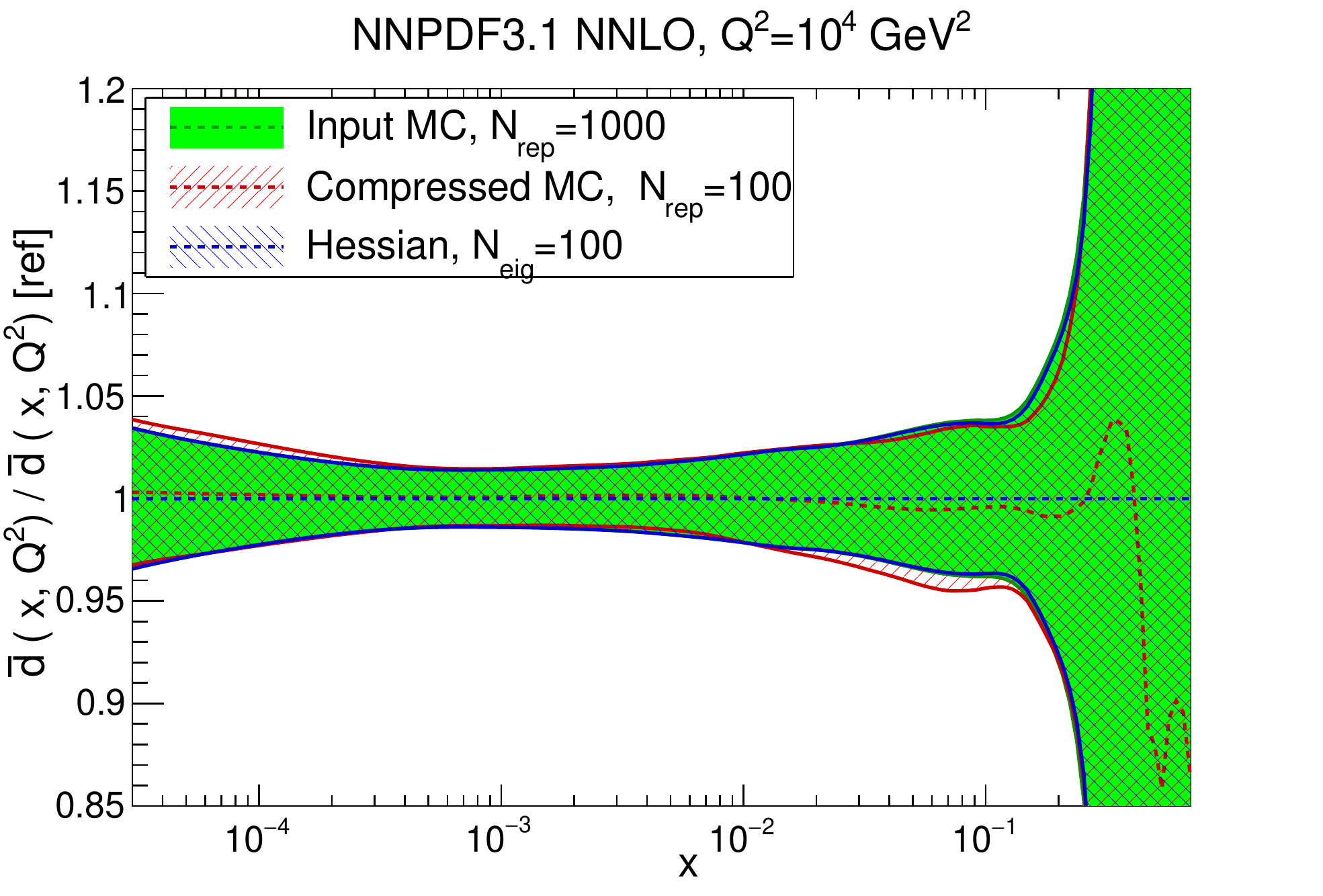}
     \includegraphics[scale=0.32]{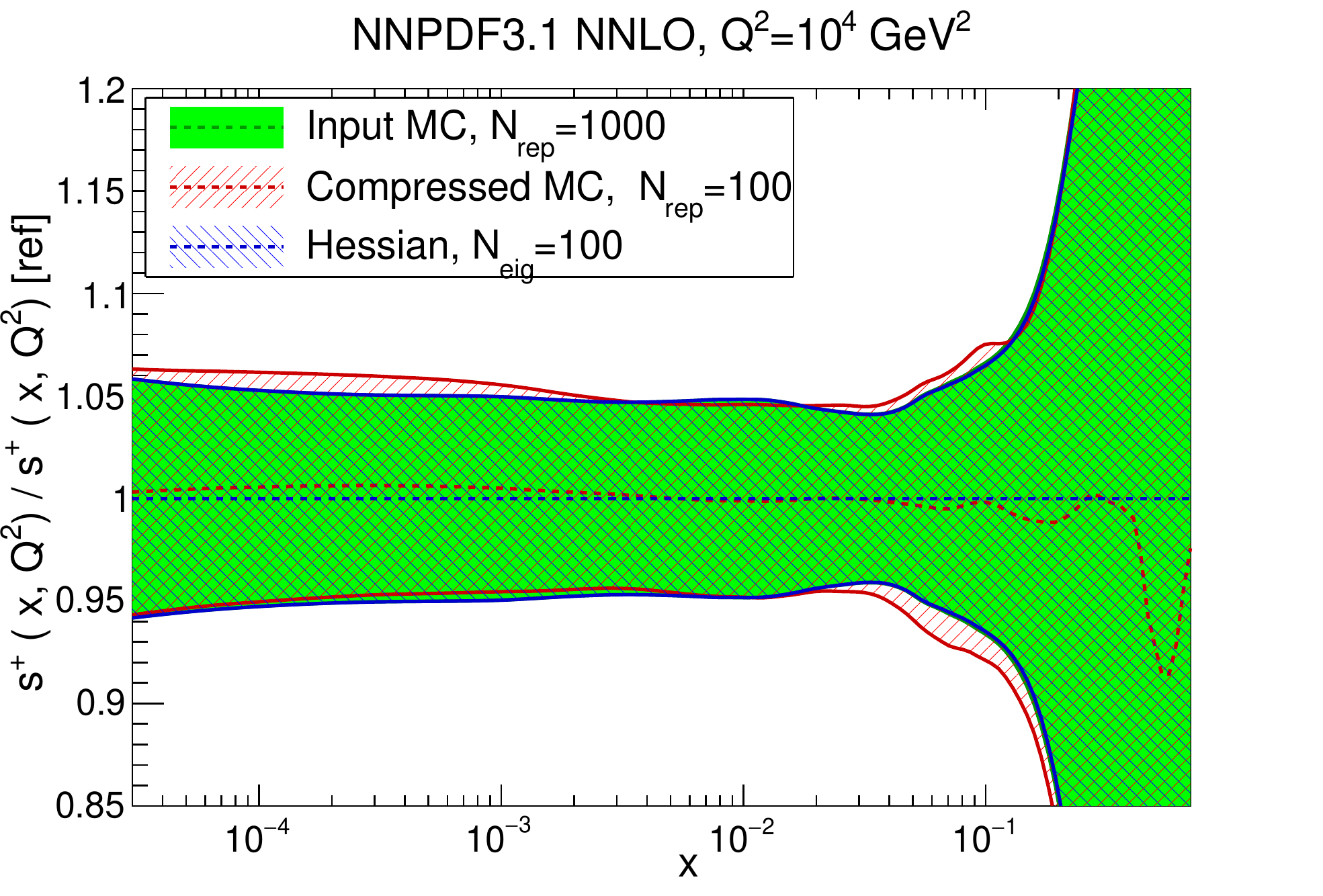}
     \includegraphics[scale=0.32]{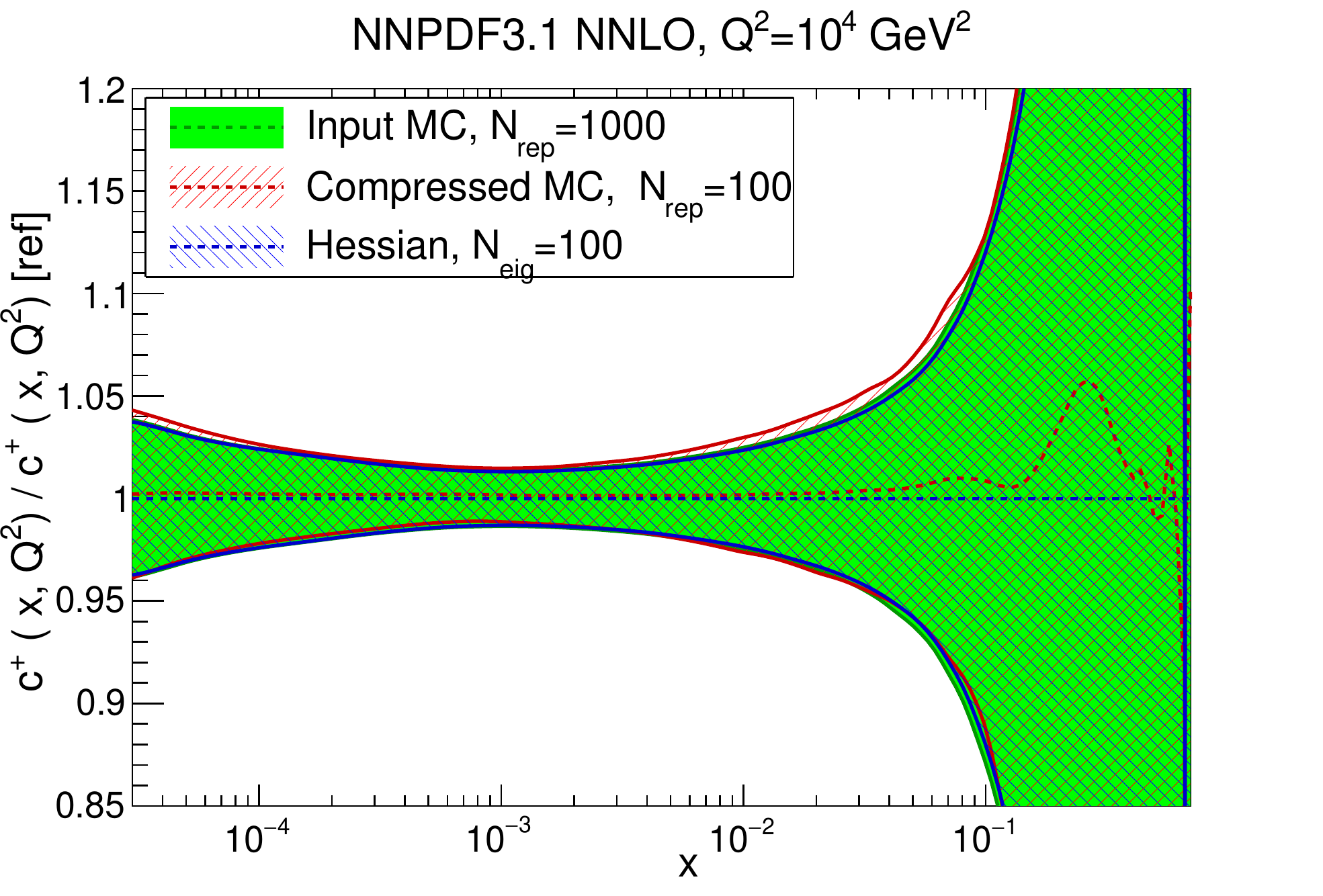}
     \includegraphics[scale=0.32]{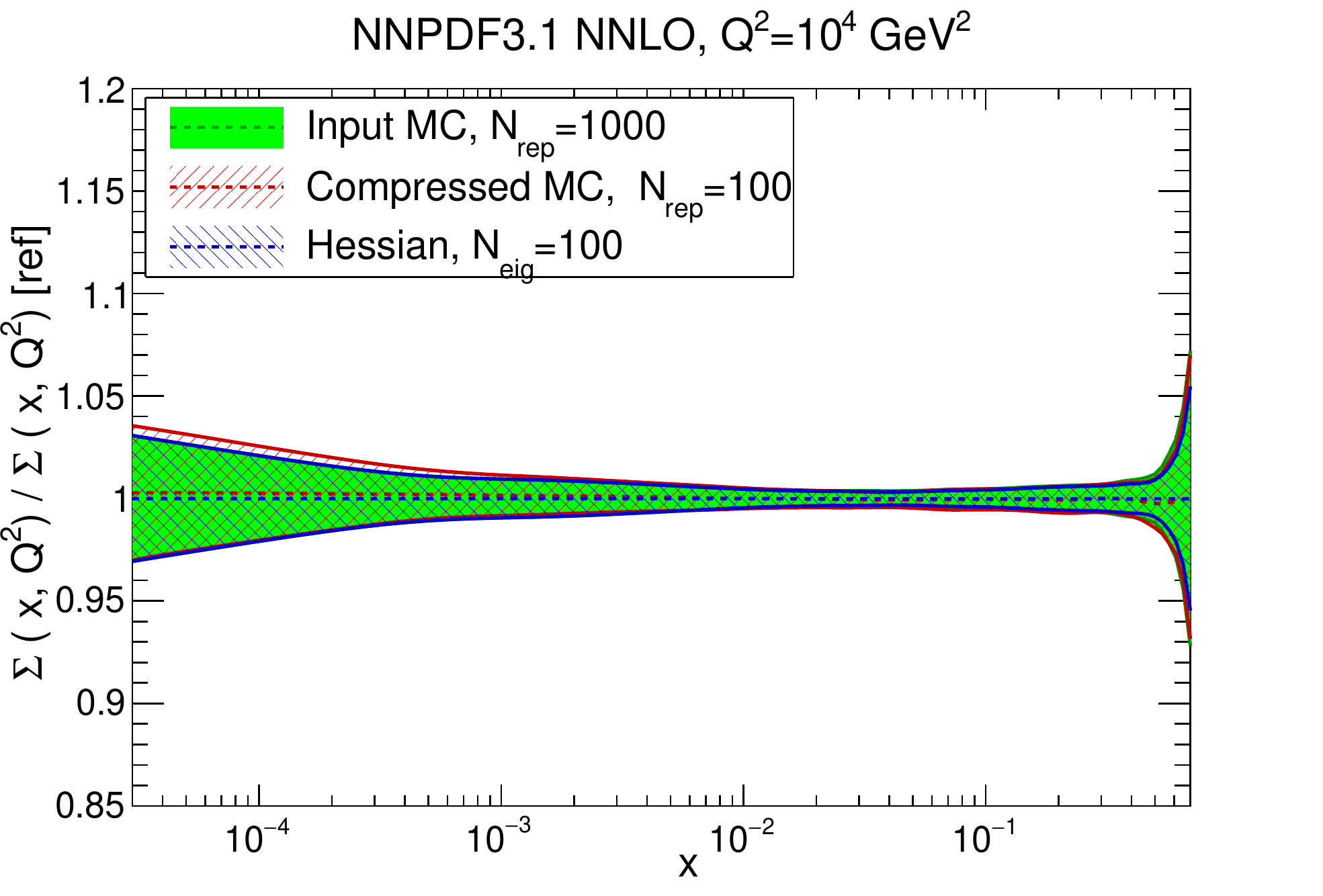}
     \includegraphics[scale=0.32]{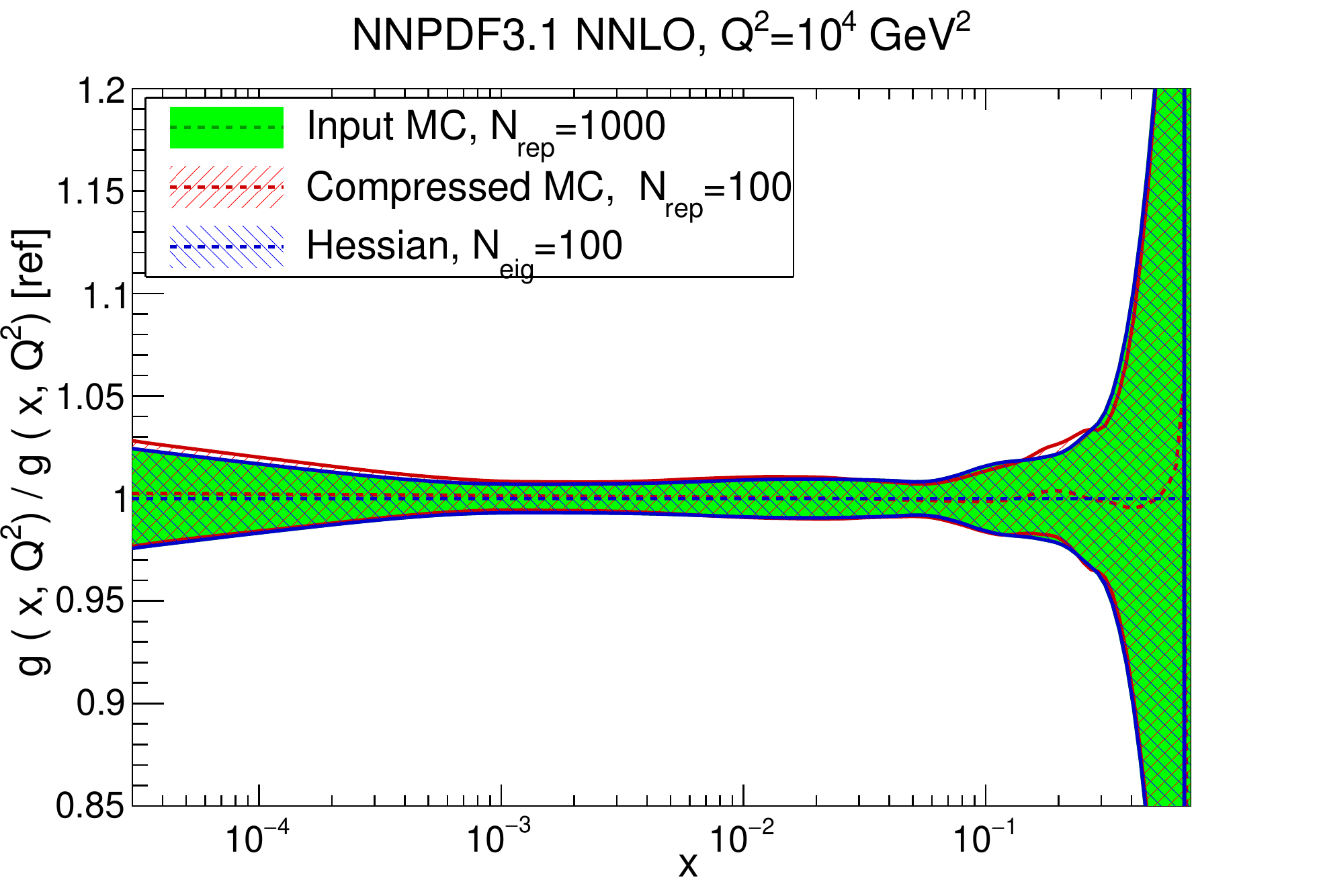}
     \caption{\small Comparison between the PDFs from the input set
       of $N_{\rm rep}=1000$ replicas of NNPDF3.1 NNLO, the 
       reduced Monte Carlo CMC-PDFs with $N_{\rm rep}=100$ replicas,
       and the MC2H hessian PDFs with $N_{\rm eig}=100$ symmetric
       eigenvalues.
    \label{fig:cmc-mc2h-validation}}
\end{center}
\end{figure}

  In Fig.~\ref{fig:cmc-mc2h-validation} we show the
  comparison between the PDFs from the input set
       of $N_{\rm rep}=1000$ replicas of NNPDF3.1 NNLO with the corresponding
       reduced sets of the CMC-PDFs with $N_{\rm rep}=100$ replicas
       and the MC2H hessian PDFs with $N_{\rm eig}=100$
       eigenvalues.
       The agreement between the input $N_{\rm rep}=1000$ replica MC
       PDFs and the two reduced sets is very good in all cases.
       By construction, the agreement in central values
       and PDF variances is slightly better for the MC2H sets,
       since the CMC-PDF sets aim to reproduce also higher
       moments in the probability distribution and thus possibly
       non-gaussian features, while Hessian sets are Gaussian by
       construction. 
       Following the analysis of~\cite{Carrazza:2015aoa,Carrazza:2015hva}
       we have verified that also the correlations between PDFs
       are reproduced to a high degree of accuracy.
       
\begin{figure}[t]
  \begin{center}
    \includegraphics[scale=0.35]{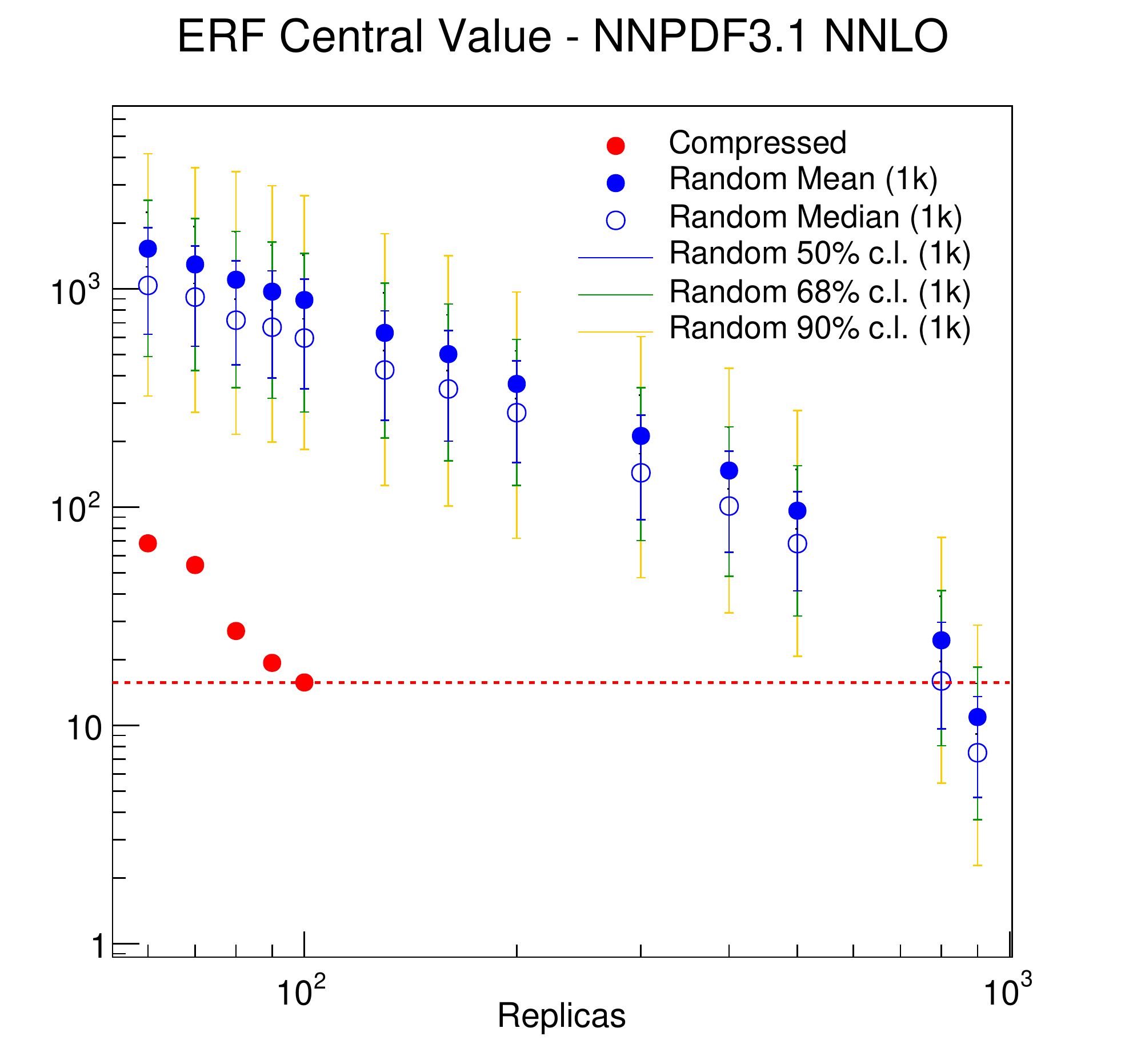}\includegraphics[scale=0.35]{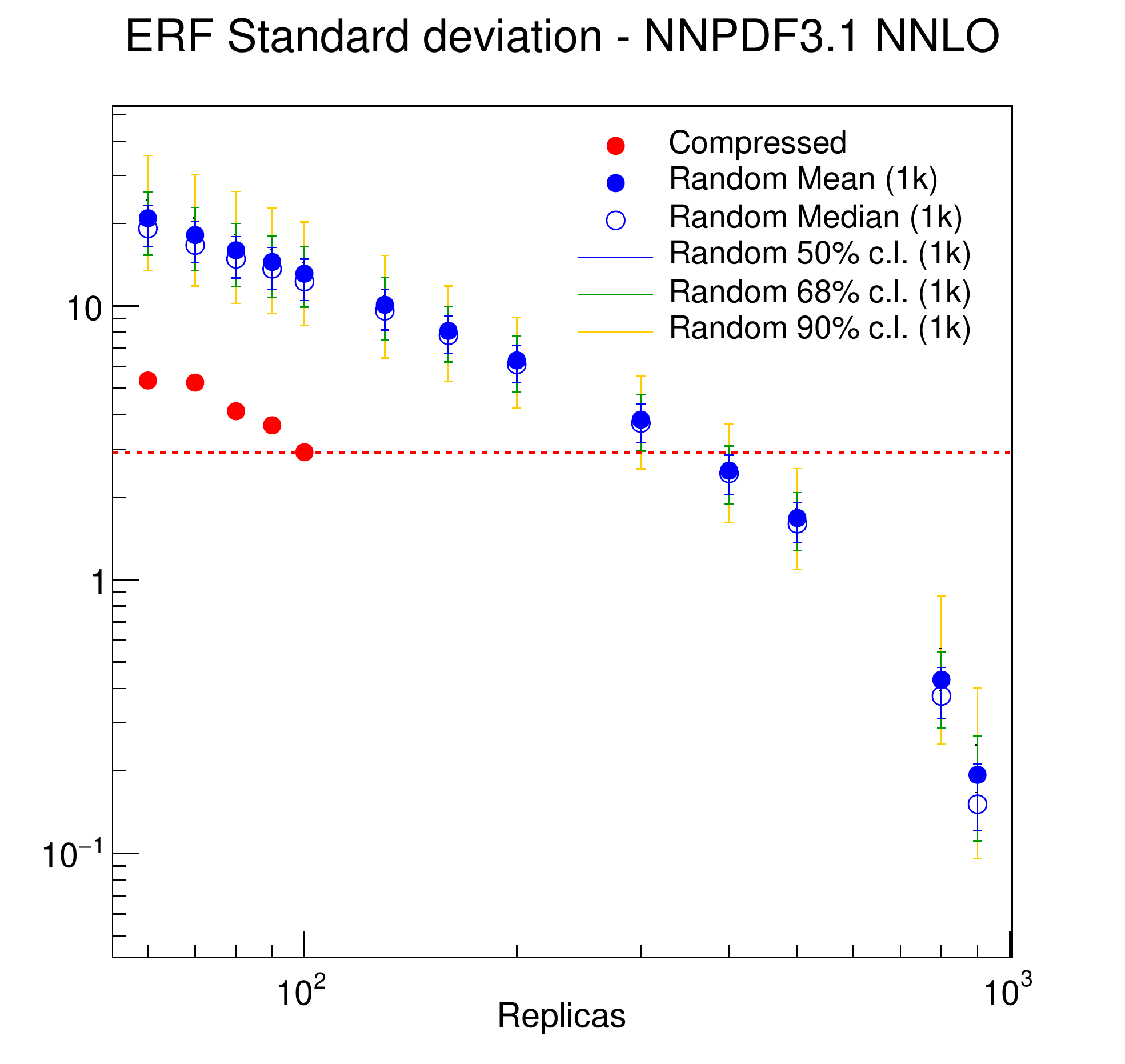}
    \includegraphics[scale=0.35]{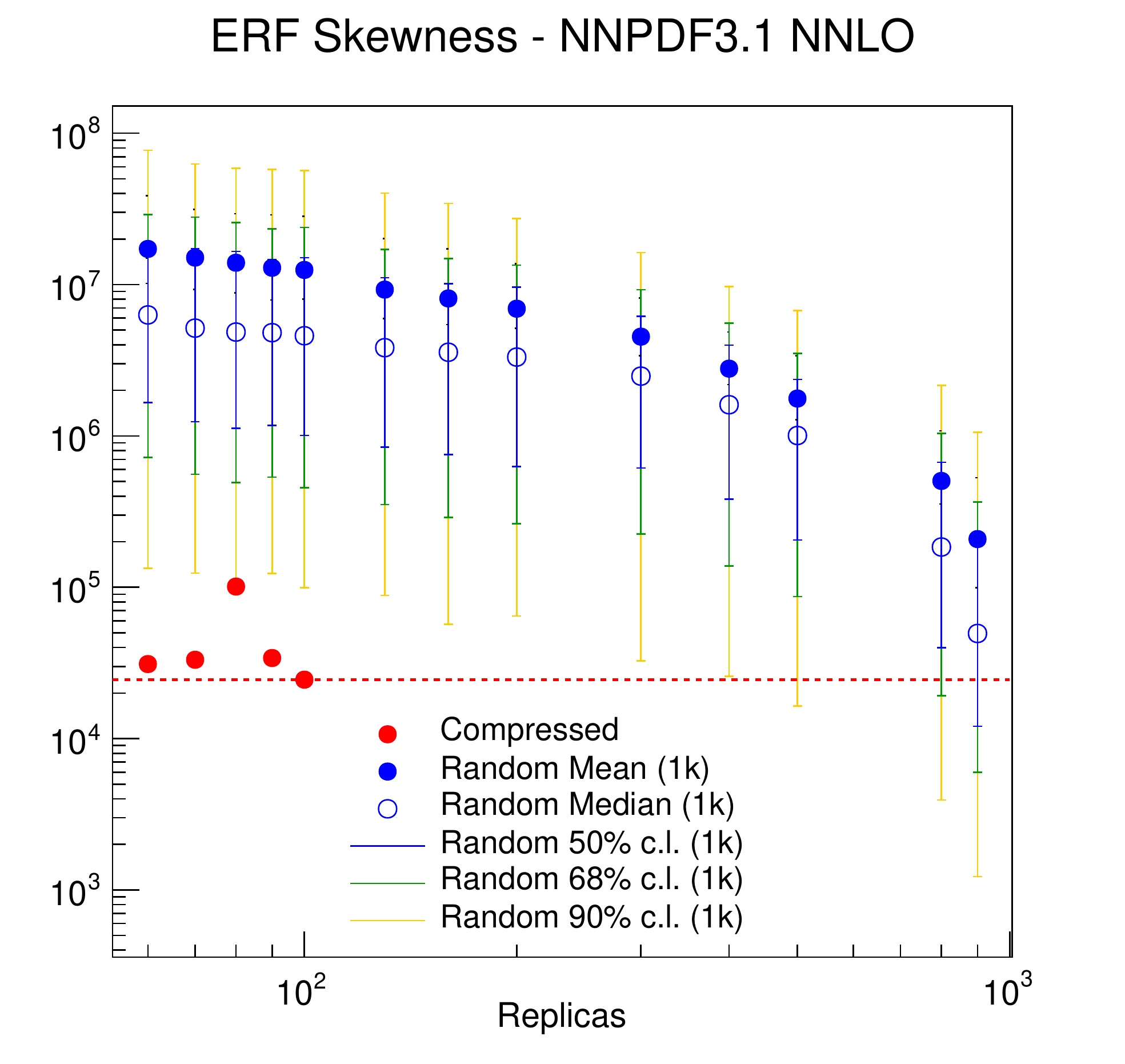}\includegraphics[scale=0.35]{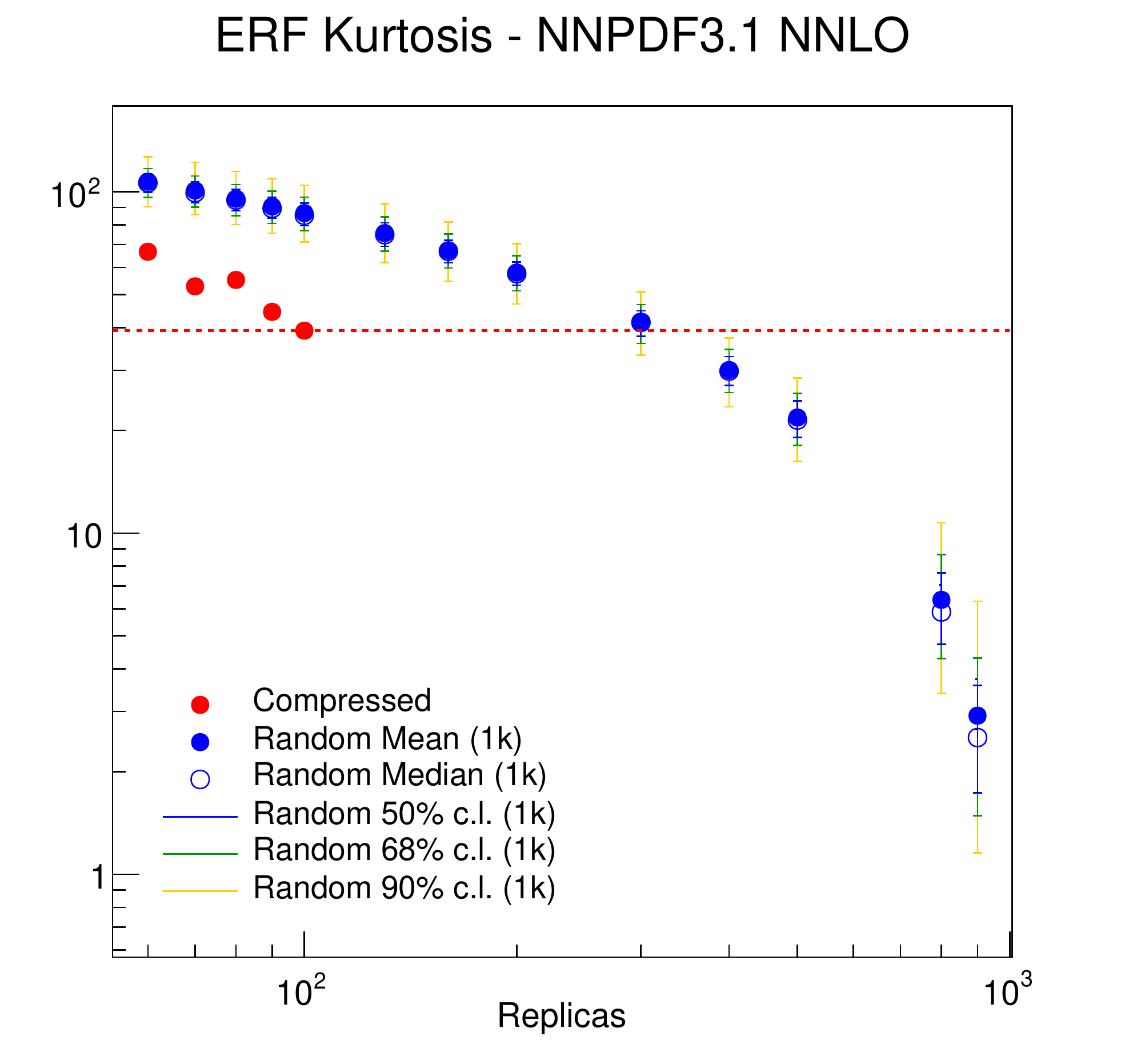}
    \includegraphics[scale=0.35]{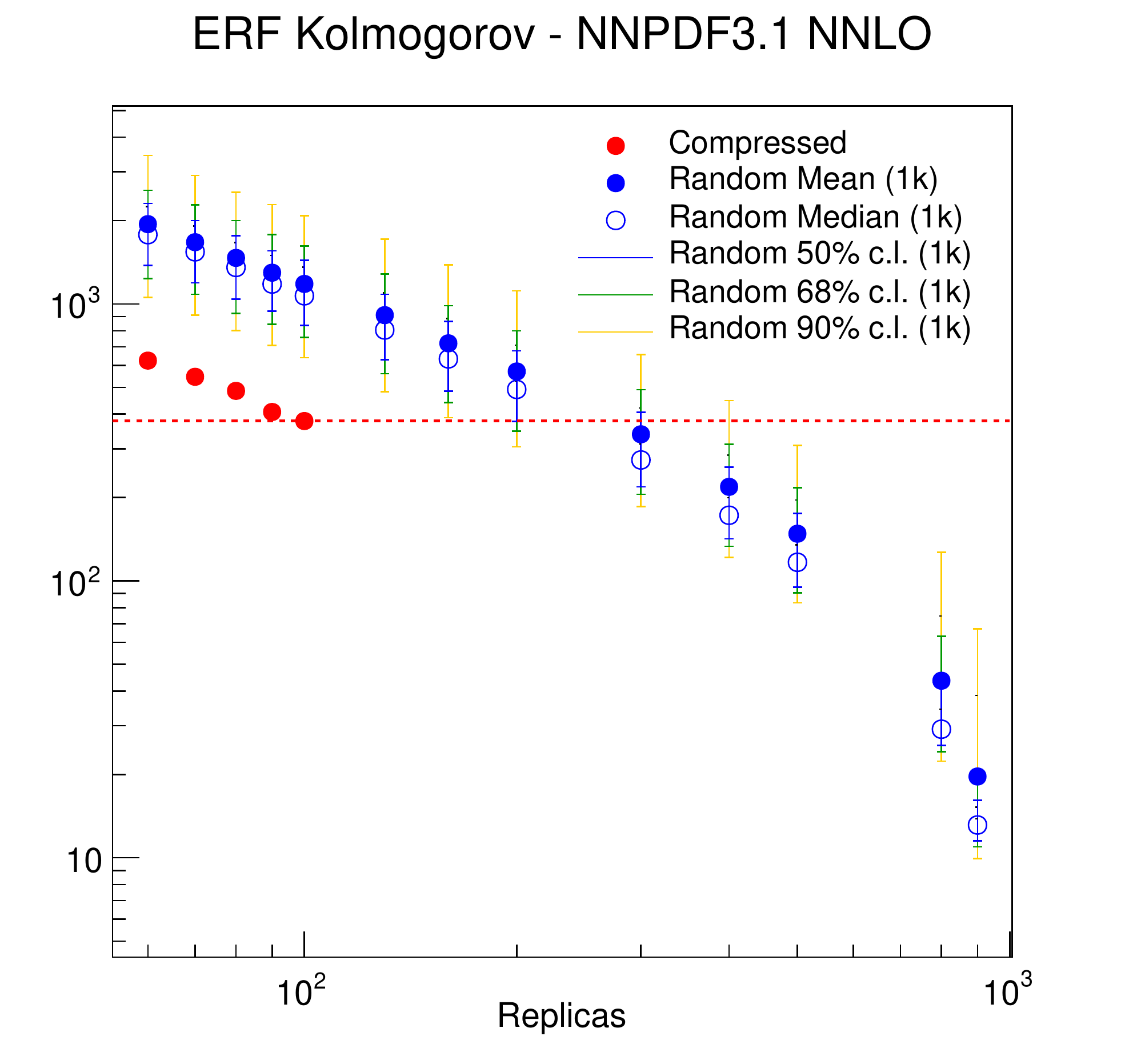}\includegraphics[scale=0.35]{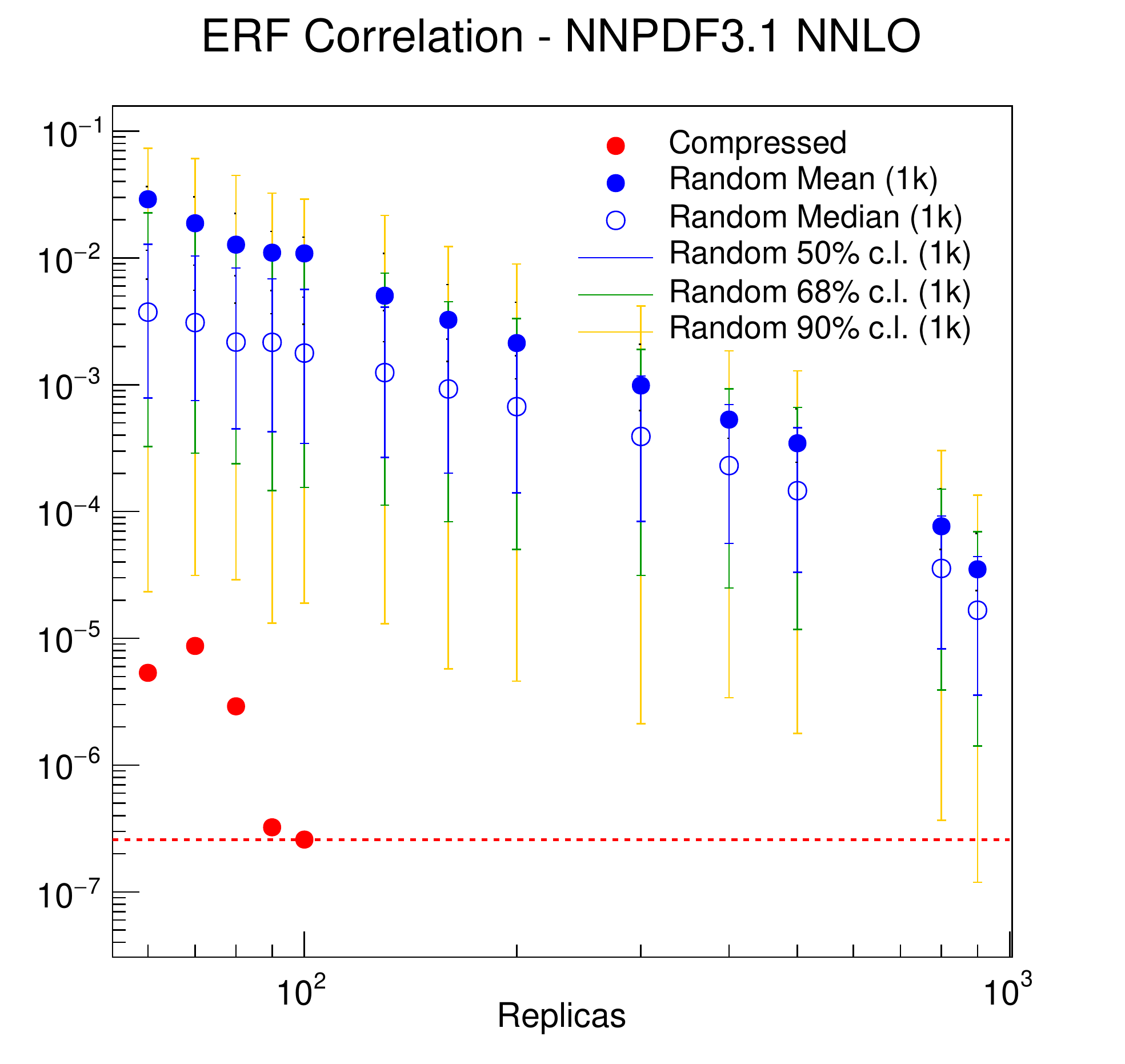}

     \caption{\small Comparison of estimators
       of the probability distributions computed using the NNPDF3.1
       NNLO input $N_{\rm
         rep}=1000$ replica set, and compressed sets of
       $\widetilde{N}_{\rm rep}$ replicas, plotted as a function of  
 $\widetilde{N}_{\rm rep}$.
       The error function (ERF) corresponding to central values, standard
       deviations, kurtosis, skewness, correlations and 
       Kolmogorov distance are shown.
    \label{fig:cmc-mc2h-validation-nrep}
  }
\end{center}
\end{figure}

In order to validate the efficiency of the CMC-PDF algorithm reduction
from the starting 
$N_{\rm rep}=1000$ replicas down to the compressed $N_{\rm rep}=100$ replicas,
in Fig.~\ref{fig:cmc-mc2h-validation-nrep} we show, following the procedure
described in~\cite{Carrazza:2015hva}, the
summary of statistical estimators
       that compare specific properties of the probability distributions defined
       by the input $N_{\rm rep}=1000$ replicas of NNPDF3.1 NNLO and the corresponding
       compressed sets as a function of $\widetilde{N}_{\rm rep}$, the number
       of replicas in the reduced set starting from $\widetilde{N}_{\rm rep}=100$.
       We compare the results of the compression algorithm with those
       of random selection of  $\widetilde{N}_{\rm rep}$ replicas out of the original
       1000 ones: the error function ERF corresponding to central
       values, standard 
       deviations, kurtosis and skewness, correlations and the
       Kolmogorov distance are all shown. 
       These results indicate that a CMC-PDF  100 replica set
       reproduces roughly the information contained in a random 
       $\widetilde{N}_{\rm rep}=400$  PDF set.

\clearpage

\subsection{Delivery}
\label{sec:delivery}

We now  provide a full list of the NNPDF3.1 PDF sets that are being 
made publicly available via the {\sc\small LHAPDF6}
interface~\cite{Buckley:2014ana}, 
\begin{center}
{\bf \url{http://lhapdf.hepforge.org/}~} .
\end{center}
As repeatedly mentioned in the paper, a very wide  set of results
concerning these PDF sets is available from the repository
\begin{center}
{\bf \url{http://nnpdf.hepforge.org/html/nnpdf31/catalog}~} .
\end{center}
All sets are made available as  $N_{\rm rep}=100$ Monte Carlo sets. 
For the baseline
sets, these are constructed out of larger  $N_{\rm rep}=1000$ replica
sets, which are also being made available. The baseline sets are also
provided as Hessian sets with 100 error sets.

The full list is the following:

\begin{itemize}

\item {\it Baseline NLO and NNLO NNPDF3.1 sets}

Baseline NLO and NNLO NNPDF3.1 sets are based on the global
dataset, with $\alpha_s(m_Z)=0.118$ and a variable-flavor number with
up to five active flavors. These sets contain $N_{\rm rep}=1000$ PDF replicas. 
\begin{flushleft}
\tt NNPDF31\_nlo\_as\_0118\_1000 \\
\tt NNPDF31\_nnlo\_as\_0118\_1000.\\
\end{flushleft}
A modified version in which charm is perturbatively generated (ad in
previous NNPDF sets) is also being made available, also with $N_{\rm rep}=1000$ PDF replicas:
\begin{flushleft}
\tt NNPDF31\_nlo\_pch\_as\_0118\_1000 \\
\tt NNPDF31\_nnlo\_pch\_as\_0118\_1000.\\
\end{flushleft}
Out of these , optimized  Monte Carlo $N_{\rm rep}=100$ 
\begin{flushleft}
\tt NNPDF31\_nlo\_as\_0118 \\
\tt NNPDF31\_nnlo\_as\_0118 \\
\tt NNPDF31\_nlo\_pch\_as\_0118 \\
\tt NNPDF31\_nnlo\_pch\_as\_0118,
\end{flushleft}
and  Hessian sets with $N_{\rm eig}=100$ eigenvectors 
\begin{flushleft}
\tt NNPDF31\_nlo\_as\_0118\_hessian \\
\tt NNPDF31\_nnlo\_as\_0118\_hessian \\
\tt NNPDF31\_nlo\_pch\_as\_0118\_hessian \\
\tt NNPDF31\_nnlo\_pch\_as\_0118\_hessian
\end{flushleft}
have been constructed as discussed in Sect.~\ref{sec:compression}
above. 

Out of these, smaller sets of eigenvectors optimized for the
computation of specific observables may be constructed using the {\tt
  SM-PDF} tool~\cite{Carrazza:2016htc}.
Specifically, sets optimized for a wide list of predefined observables
can be generated and downloaded using the web
interface~\cite{Carrazza:2016wte}  at
\begin{center}
{\bf \url{https://smpdf.mi.infn.it}~} .
\end{center}

\item {\it Flavor number variation}

We have produced sets, both at NLO and NNLO 
in which the maximum number of flavors differs
from the default 5, and it is either extended up to six, or frozen at
four:
\begin{flushleft}
  \tt NNPDF31\_nlo\_as\_0118\_nf\_4 \\
  \tt NNPDF31\_nlo\_as\_0118\_nf\_6 \\
  \tt NNPDF31\_nnlo\_as\_0118\_nf\_4 \\
  \tt NNPDF31\_nnlo\_as\_0118\_nf\_6. \\
\end{flushleft}
The variant with perturbatively generated charm is also made
available, in this case also in a $n_f=3$ fixed-flavor number scheme:
\begin{flushleft}
  \tt NNPDF31\_nlo\_pch\_as\_0118\_nf\_3 \\
  \tt NNPDF31\_nlo\_pch\_as\_0118\_nf\_4 \\
  \tt NNPDF31\_nlo\_pch\_as\_0118\_nf\_6 \\
  \tt NNPDF31\_nnlo\_pch\_as\_0118\_nf\_3 \\
  \tt NNPDF31\_nnlo\_pch\_as\_0118\_nf\_4 \\
  \tt NNPDF31\_nnlo\_pch\_as\_0118\_nf\_6. \\
\end{flushleft}

\item {\it $\alpha_s$ variation.}

We have produced NNLO sets with the following values of $\alpha_s(m_Z)$
0.108, 1.110, 0.112, 0.114, 0.116, 0.117, 0.118, 0.119, 0.120, 0.122,
0.124:
\begin{flushleft}
  \tt NNPDF31\_nnlo\_as\_0108 \\
  \tt NNPDF31\_nnlo\_as\_0110 \\
  \tt NNPDF31\_nnlo\_as\_0112 \\
  \tt NNPDF31\_nnlo\_as\_0114 \\
  \tt NNPDF31\_nnlo\_as\_0116 \\
  \tt NNPDF31\_nnlo\_as\_0117 \\
  \tt NNPDF31\_nnlo\_as\_0118 \\
  \tt NNPDF31\_nnlo\_as\_0119 \\
  \tt NNPDF31\_nnlo\_as\_0120 \\
  \tt NNPDF31\_nnlo\_as\_0122 \\
  \tt NNPDF31\_nnlo\_as\_0124. \\
\end{flushleft}

For the values $\alpha_s(m_Z)=0.116$ and $\alpha_s(m_Z)=0.120$ we have
also produced  NLO sets, 
\begin{flushleft}
  \tt NNPDF31\_nlo\_as\_0116 \\
  \tt NNPDF31\_nlo\_as\_0120, \\
\end{flushleft}
and the  variant with perturbative charm both at NLO and NNLO
\begin{flushleft}
  \tt NNPDF31\_nlo\_pch\_as\_0116 \\
  \tt NNPDF31\_nlo\_pch\_as\_0120 \\
\tt NNPDF31\_nnlo\_pch\_as\_0116\\
\tt NNPDF31\_nnlo\_pch\_as\_0120.\\
\end{flushleft}

In order to facilitate the computation of the combined PDF+$\alpha_s$
uncertainties we have also provided bundled  PDF+$\alpha_s$ variation
sets for $\alpha_s(m_Z)=0.118\pm0.002$. These are
provided both as Monte Carlo sets, and as Hessian sets. 
\begin{flushleft}
  \tt
NNPDF31\_nlo\_pdfas	\\
NNPDF31\_nnlo\_pdfas	\\
NNPDF31\_nlo\_pch\_pdfas \\	
NNPDF31\_nnlo\_pch\_pdfas  \\	
NNPDF31\_nlo\_hessian\_pdfas \\
NNPDF31\_nnlo\_hessian\_pdfas \\	
NNPDF31\_nlo\_pch\_hessian\_pdfas \\	
NNPDF31\_nnlo\_pch\_hessian\_pdfas. \\
\end{flushleft}
They are constructed as follows:
\begin{enumerate}
\item The central value (PDF member 0) is the central
  value of the corresponding $\alpha_s(m_Z)=0.118$ set.
\item The PDF members 1 to 100 correspond to the
  $N_{\rep}=100$ ($N_{\rm eig}=100$) Monte Carlo replicas
  (Hessian eigenvectors) from the $\alpha_s(m_Z)=0.118$ set.
\item The PDF members 101 and 102 are the central values
  of the sets with $\alpha_s(m_Z)=0.116$ and $\alpha_s(m_Z)=0.120$
  respectively.
\end{enumerate}
Note that, therefore, in the Hessian case member 0 is the central set,
and all remaining  bundled members $1-102$ are error sets, while in the Monte
Carlo case,  members $1-100$ are Monte Carlo replicas, while members
0,~$101$ and~$102$ are central sets (replica averages). The way they
should be used to compute combined PDF+$\alpha_s$ uncertainties is
discussed e.g. in Ref.~\cite{Butterworth:2015oua}.

\item {\it Charm mass variation}

We  provide sets with different values of the
  charm mass $m_c^{\rm pole}$.
  They are available only at NNLO
with $m_c^{\rm pole}=1.38$ GeV and $m_c^{\rm pole}=1.64$ GeV. 
  \begin{flushleft}
    {\tt NNPDF31\_nnlo\_as\_0118\_mc\_138}\\
    {\tt NNPDF31\_nnlo\_as\_0118\_mc\_164}.\\
\end{flushleft}
For
comparison, the corresponding modified version with perturbative
charm are also made available:
  \begin{flushleft}
     {\tt NNPDF31\_nnlo\_pch\_as\_0118\_mc\_138}\\
    {\tt NNPDF31\_nnlo\_pch\_as\_0118\_mc\_164}.\\
\end{flushleft}

\item {\it Forced positivity sets}

We provide sets in which PDFs are non-negative:
  \begin{flushleft}
    {\tt NNPDF31\_nlo\_as\_0118\_mc}\\
     {\tt NNPDF31\_nlo\_pch\_as\_0118\_mc}\\
    {\tt NNPDF31\_nnlo\_as\_0118\_mc}\\
     {\tt NNPDF31\_nnlo\_pch\_as\_0118\_mc}\\
\end{flushleft}
These have been constructed simply setting to zero PDFs whenever they
become negative. They are thus an approximation, provided for
convenience for use in conjunction with codes which fail when PDFs are
negative.

\item {\it LO sets} 

Leading-order PDF sets are made available 
 $\alpha_s = 0.118$ and  $\alpha_s = 0.130$: 
  \begin{flushleft}
    {\tt NNPDF31\_lo\_as\_0118}\\
    {\tt NNPDF31\_lo\_as\_0130}\\
\end{flushleft}
The corresponding variant with perturbative charm is also provided:
  \begin{flushleft}
    {\tt NNPDF31\_lo\_pch\_as\_0118}\\
    {\tt NNPDF31\_lo\_pch\_as\_0130}.
\end{flushleft}

\item {\it Reduced datasets}

PDFs determined from subsets of the full NNPDF3.1, discussed in
  Sect.~\ref{sec:impactzpt}-\ref{sec:collideronly} are also made
  available, specifically
  \begin{flushleft}
  \begin{tabular}{ll}
  {\tt NNPDF31\_nnlo\_as\_0118\_noZpt} & no $Z$ $p_T$ data, see Sect.~\ref{sec:impactzpt}; \\
  {\tt NNPDF31\_nnlo\_as\_0118\_notop} & no $t\bar{t}$ production data, see Sect.~\ref{sec:impacttop}; \\
  {\tt NNPDF31\_nnlo\_as\_0118\_nojets} & no jet data,  see Sect.~\ref{sec:jetdata}; \\
  {\tt NNPDF31\_nnlo\_as\_0118\_wEMC} & EMC charm data added, see Sect.~\ref{sec:emc};\\
  {\tt NNPDF31\_nnlo\_as\_0118\_noLHC} & no LHC data,  see Sect.~\ref{sec:nolhc};  \\
  {\tt NNPDF31\_nnlo\_as\_0118\_proton}  &  proton-target data only, see Sect.~\ref{sec:nonucl};\\
  {\tt NNPDF31\_nnlo\_as\_0118\_collider} & collider data only, see Sect.~\ref{sec:collideronly}.\\
  \end{tabular}
  \end{flushleft}
\end{itemize}

 In addition to these, any  other PDF sets  discussed in this paper  is also available upon request.

\subsection{Outlook}
\label{sec:outlook}

The NNPDF3.1 PDF determination presented here is an update of the
previous NNPDF3.0, yet it contains substantial innovations both in
terms of methodology and dataset and it leads to  substantially more
precise and accurate PDF sets. Thanks to this, several spin-offs can
be pursued, either with the goal of updating existing results
based on previous NNPDF releases, or in some cases because the greater
accuracy enables projects which  previously were either impossible or
uninteresting.

These spin-off projects include the following:
\begin{itemize}
\item A precision
determination of the strong coupling constant $\alpha_s(m_Z)$.
We expect a significant increase in both precision and accuracy compared to previous
NNPDF determinations~\cite{Lionetti:2011pw,Ball:2011us}. Specifically,
thanks to the inclusion of many collider observables
at NNLO with direct dependence on both the gluon and on
$\alpha_s(m_Z)$
 it might turn out to be advantageous to drop 
altogether data taken on nuclear targets, and at low scales, {\it i.e.} base
the $\alpha_s$ determination on the  collider-only dataset of
Sect.~\ref{sec:collideronly}, or possibly an even more conservative dataset.
\item A determination of the charm mass  $m_c$. This would be for the
  first time based on a PDF determination in which the charm PDF is
  independently parametrized, thereby avoiding bias related to the identification of $m_c$
  as a parameter which determines the size of the charm PDF.
\item A NNPDF3.1QED PDF set including the photon PDF $\gamma(x,Q^2)$, thereby updating
  the previous NNPDF2.3QED~\cite{Ball:2013hta} and
  NNPDF3.0QED~\cite{Bertone:2016ume} sets. This update should
  include recent theoretical progress: specifically the direct ``LuxQED''
  constraints which determine  the photon PDF from nucleon
  structure functions~\cite{Manohar:2016nzj};
and  NLO QED corrections to PDF evolution and DIS coefficient
functions, now included in {\tt APFEL}~\cite{Bertone:2013vaa}. 
\item PDF sets including small-$x$ resummation, based on the formalism
developed in~\cite{Altarelli:2005ni,Ball:2007ra,Altarelli:2008aj}, which
has been implemented in the {\tt HELL} code~\cite{Bonvini:2016wki}
and interfaced to {\tt APFEL}.
These  would provide an  answer the long-standing
issue of whether or not small-$x$ inclusive HERA data are adequately described
by fixed-order perturbative
evolution~\cite{Caola:2010cy,Caola:2009iy}, and may ultimately give
better control  
of theoretical uncertainties at small $x$.
\end{itemize}

On a longer timescale, further substantial improvements in dataset and
methodology are expected.
On the one hand, so far we have essentially  
restricted ourselves to LHC 8 TeV data.
A future release will include a significant
number of 13 TeV measurements, of which several,
from ATLAS, CMS and LHCb are already available. Specifically, the
inclusion of more processes is envisaged,
which are not currently part of the
dataset, which have a large potential impact on PDFs, and for which higher order corrections have become
available . These include prompt photon
production~\cite{Aaboud:2017cbm}, for which 
NNLO corrections are now available~\cite{Campbell:2016lzl}  and whose
impact on PDF is 
well-known~\cite{d'Enterria:2012yj};  single-top production
(also known at NNLO~\cite{Brucherseifer:2014ama});  and possibly forward  
$D$ meson production or more in general processes with final-state $D$
mesons, such as $W+D$ production, recently measured by
ATLAS~\cite{Aad:2014xca}, a process whose impact on PDFs has been
repeatedly emphasized~\cite{Zenaiev:2015rfa,Gauld:2015yia,Gauld:2016kpd}.

Such a further increase in dataset is likely to require substantial
methodological improvements in PDF determination. Also, it is likely
to result in a further reduction of PDF uncertainties, thereby
requiring better control of theoretical uncertainties. We
specifically expect significant progress in two different
directions. On the one hand, electroweak corrections, which are now
not included, and whose impact is kept under control through kinematic
cuts, will have to be included in a more systematic way. On the other hand,
theoretical uncertainties due to missing higher-order corrections 
will also have to be estimated. Indeed, the preliminary estimates presented in
Sect.~\ref{sec:thunc} suggest that missing higher-order uncertainties
on PDFs, currently not included in the PDF uncertainty, 
are likely to soon become non-negligible, and possibly dominant in
some kinematic regions. All of these improvements will be part of a future
major PDF release. 

\bigskip
\bigskip
\begin{center}
\rule{5cm}{.1pt}
\end{center}
\bigskip
\bigskip

\subsection*{Acknowledgments}

We thank M.~Bonvini, A.~de~Roeck, 
L.~Harland-Lang, J.~Huston, J.~Gao, P.~Nadolsky,  and
R.~Thorne for many illuminating discussions.
We thank U.~Blumenschein, M.~Boonekamp,
T.~Carli, S.~Camarda  A.~Cooper-Sarkar, J.~Kretzschmar, K.~Lohwasser,
K.~Rabbertz, V.~Radescu and M.~Schott
for help with the ATLAS measurements.
We are grateful to 
A.~Giammanco, R.~Gonzalez-Suarez, I.~Josa,  K.~Lipka, M.~Owen, L.~Perrozzi, R.~Placakyte and P.~Silva
for assistance with the CMS data.
We thank S.~Farry, P.~Ilten and R.~McNulty for help with the LHCb data.
We thank F.~Petriello and R.~Boughezal for providing us the
NNLO calculations of the $Z$ $p_T$ distributions.
We thank M.~Czakon and A.~Mitov for assistance and discussions for
the NNLO top quark pair differential distributions.
We thank J.~Currie, N.~Glover and J.~Pires for providing
us with the NNLO $K$-factors for inclusive jet production
at 7 TeV and for many discussions about jet data in PDF determination.
We thank F.~Dreyer and A.~Karlberg for sending us a private version
of the {\tt proVBFH} code.
We thank R.~Harlander for assistance with {\tt vh@nnlo}.
We thank L.~Harland-Lang for benchmarking the
NNLO calculation of the CMS 8 TeV Drell-Yan measurements.

V.~B., N.~H., J.~R., L.~R. and E.~S. are
supported by an European Research Council Starting Grant ``PDF4BSM''.
R.~D.~B. and L.~D.~D. are supported by the
UK STFC grants ST/L000458/1 and ST/P000630/1.
L.~D.~D. is supported by
the Royal Society, Wolfson Research Merit Award, grant WM140078.
S.~F. is supported by the European Research Council under the Grant Agreement
740006—NNNPDF—ERC-2016-ADG/ERC-2016-ADG.
E.~R.~N. is supported by the UK STFC grant ST/M003787/1.
S.~C. is supported by the HICCUP ERC Consolidator grant (614577).
M.~U. is
supported by a Royal Society Dorothy Hodgkin Research
Fellowship and partially supported
by the STFC grant ST/L000385/1.
S.~F and Z.~K. are supported by the Executive Research Agency (REA) of
the European Commission under the Grant Agreement PITN-GA-2012-316704
(HiggsTools).
A.~G. is supported by the European Union's Horizon 2020 research and innovation 
programme under the Marie Sk\l odowska-Curie grant agreement No 659128 - NEXTGENPDF.


\appendix

\section{Code development and benchmarking}
\label{sec:benchmarking}

In all previous NNPDF releases, including NNPDF3.0, PDF evolution and
deep-inelastic scattering were computed using the internal {\tt
  FKgenerator} code.  As discussed in Sect.~\ref{sec:expdata}, in
NNPDF3.1 PDF evolution is now performed using the public {\tt APFEL}
code. Deep-inelastic scattering and fixed-target Drell Yan are also
computed using {\tt APFEL}.
As far as perturbative evolution is concerned, {\tt APFEL} has been
extensively tested against other publicly available codes, such as
{\tt HOPPET}~\cite{Salam:2008qg} and {\tt QCDNUM}~\cite{Botje:2010ay}
for the PDF evolution and {\tt OpenQCDrad}~\cite{Alekhin:2013nda} for
the calculation of heavy-quark structure functions in the massive
scheme.

We have performed an extensive benchmarking of {\tt APFEL} and {\tt
  FKgenerator}, also involving deep-inelastic structure function (as
already mentioned in Ref.~\cite{Ball:2016neh}) and Drell-Yan cross-sections.
In the process of this benchmarking, two bugs were found in
the {\tt FKgenerator} implementation of the DIS structure functions
(one related to target-mass corrections, the other in the expressions
of the $\mathcal{O}(\alpha_s)$ charge-current massive coefficient
functions): we checked explicitly that none of them produced an effect
on NNPDF3.0 PDFs that could be distinguished from a statistical
fluctuation.

\begin{figure}[t]
\centering
\epsfig{width=0.46\textwidth,figure=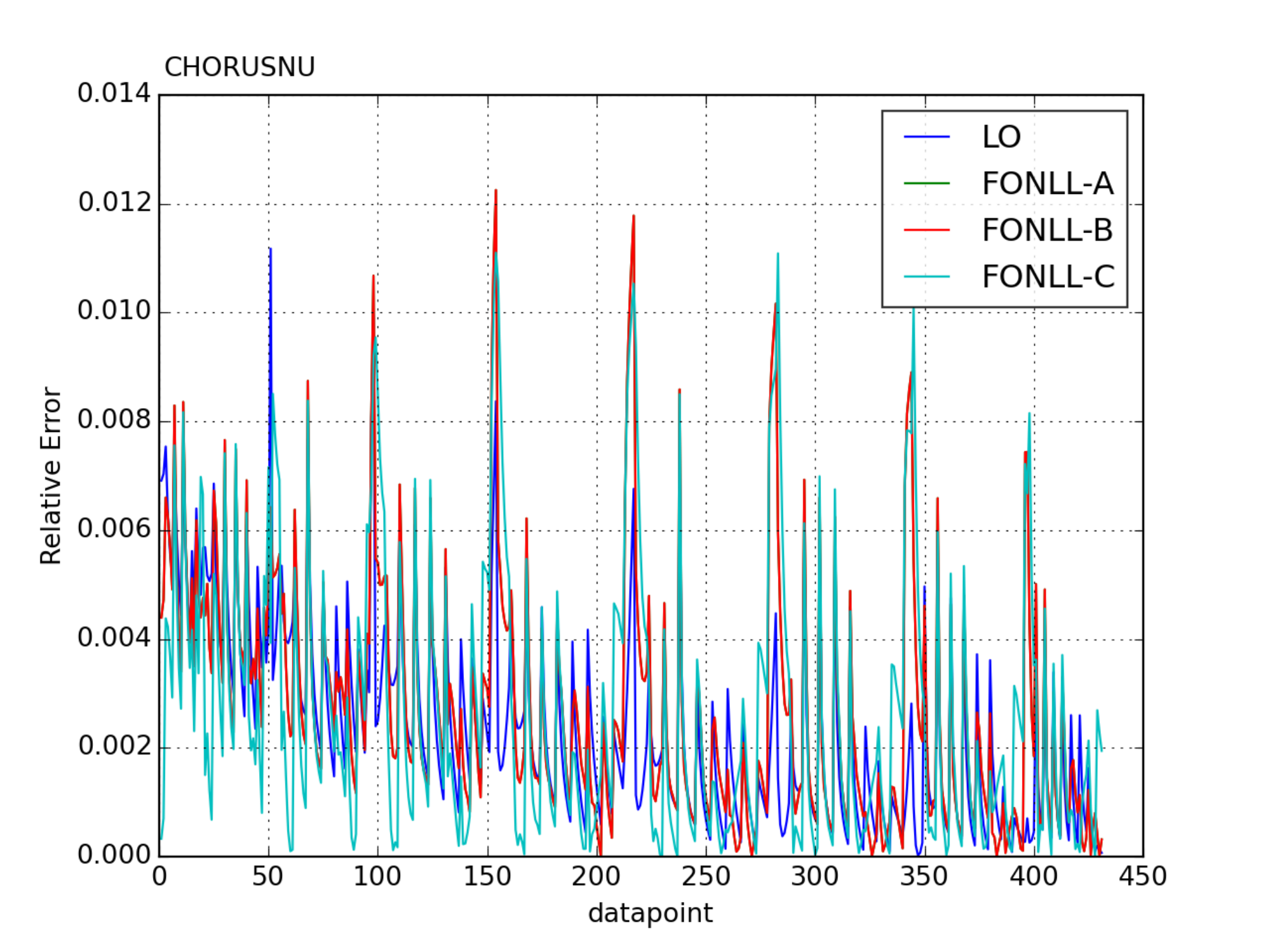}
\epsfig{width=0.46\textwidth,figure=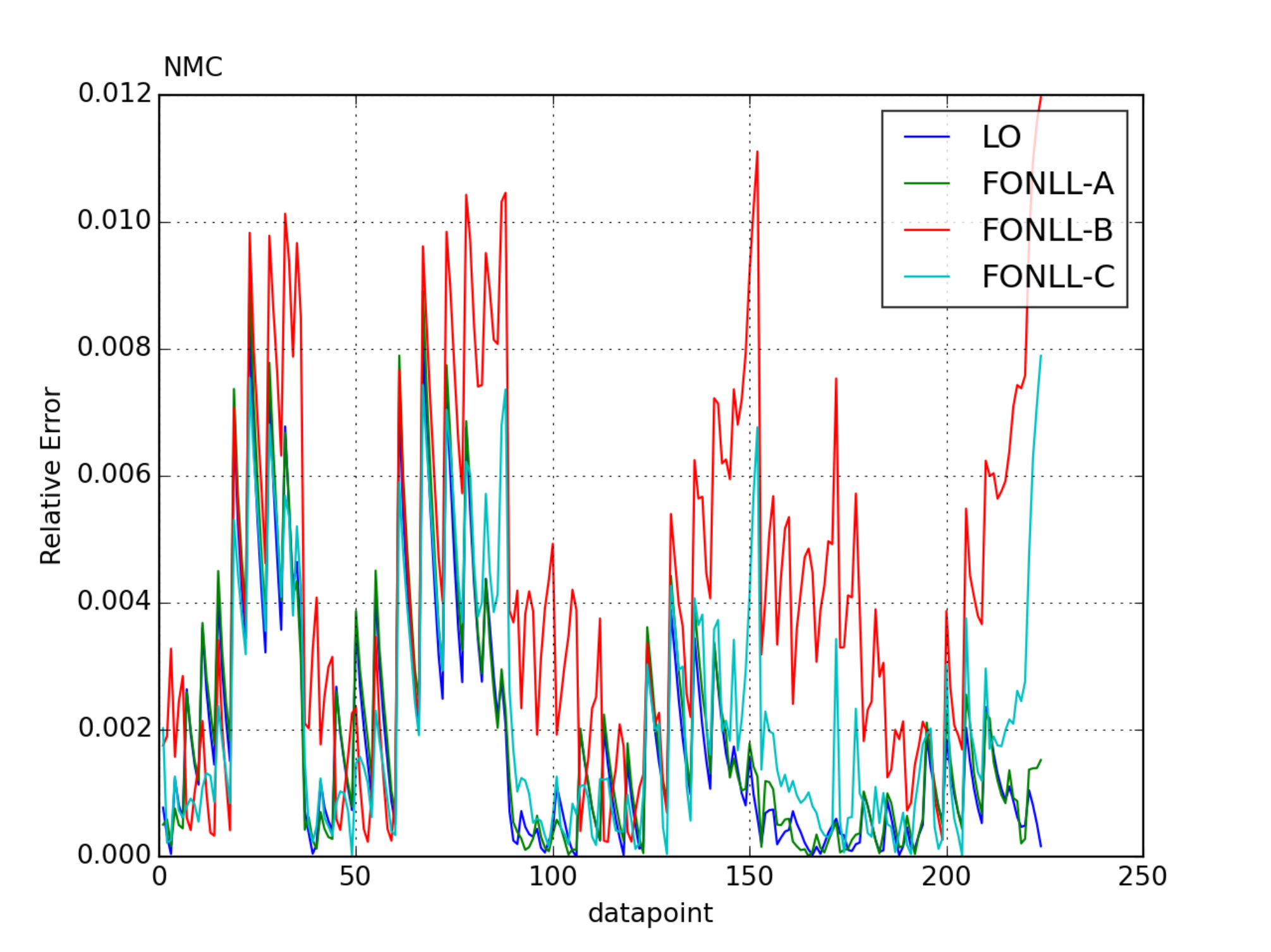}
\epsfig{width=0.46\textwidth,figure=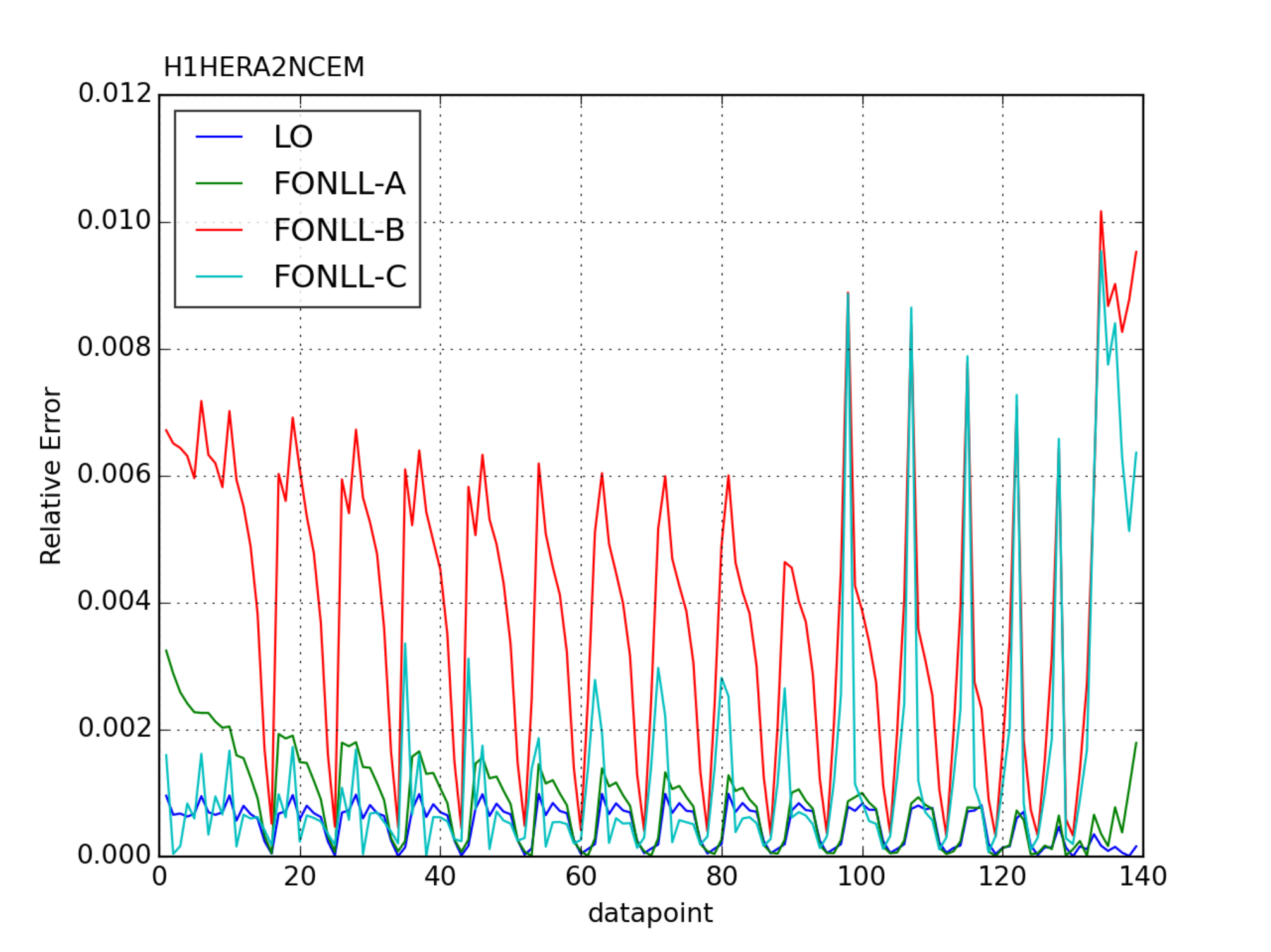}
\epsfig{width=0.46\textwidth,figure=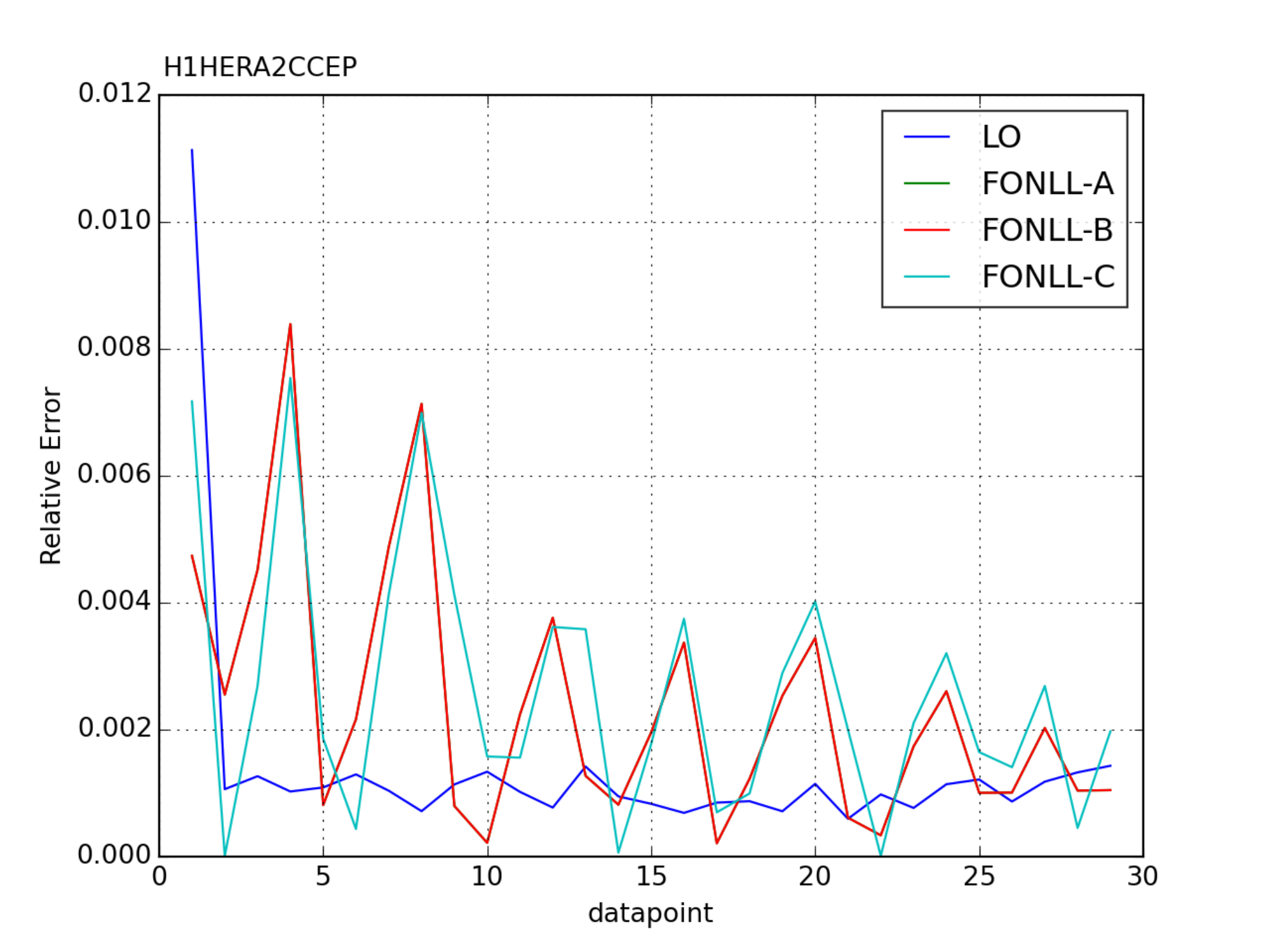}
\caption{\small Relative difference in the DIS structure functions
  computed with {\tt FKgenerator} and {\tt APFEL}, using NNPDF3.0 as
  input PDF set, for the CHORUS charged current neutrino-nucleus
  reduced cross-sections, the NMC proton reduced cross-sections and
  neutral and charged current cross-sections from the H1 experiments
  from the HERA-II dataset. Datasets are as in Tab.~1 of
  Ref.~\cite{Ball:2014uwa}.  For each dataset, we compare the
  theoretical calculations at LO and in the FONLL-A, B and
  C~\cite{Forte:2010ta} heavy-quark mass schemes.
  \label{fig:benchmarking}}
\end{figure}

Representative results of the benchmarking of deep-inelastic structure
functions are shown in Fig.~\ref{fig:benchmarking}, where we show the
relative difference between {\tt FKgenerator} and {\tt APFEL}
implementation, using NNPDF3.0 as input PDF set, for the CHORUS
charged current neutrino-nucleus reduced cross-sections, the NMC
proton reduced cross-sections and neutral and charged current cross-sections
 from the H1 experiments from the HERA-II dataset.  In each
case, we show theoretical predictions calculations at LO and in the FONLL-A,
-B and -C general-mass schemes. The two codes are in good agreement, with
differences at most being at the 1\% level, typically much smaller.
This statement holds for all perturbative orders.

\begin{figure}[t]
\centering
\epsfig{width=0.49\textwidth,figure=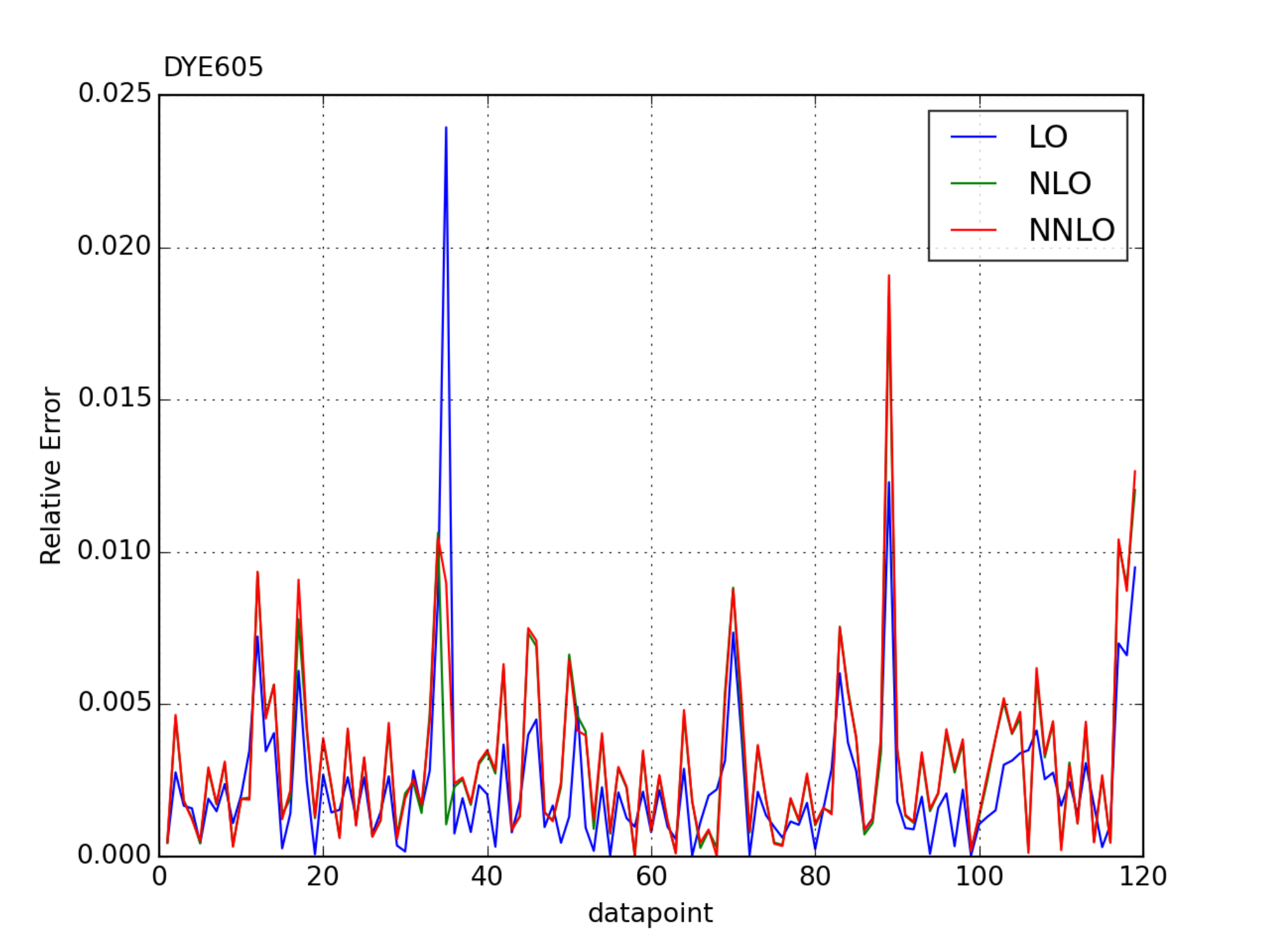}
\epsfig{width=0.49\textwidth,figure=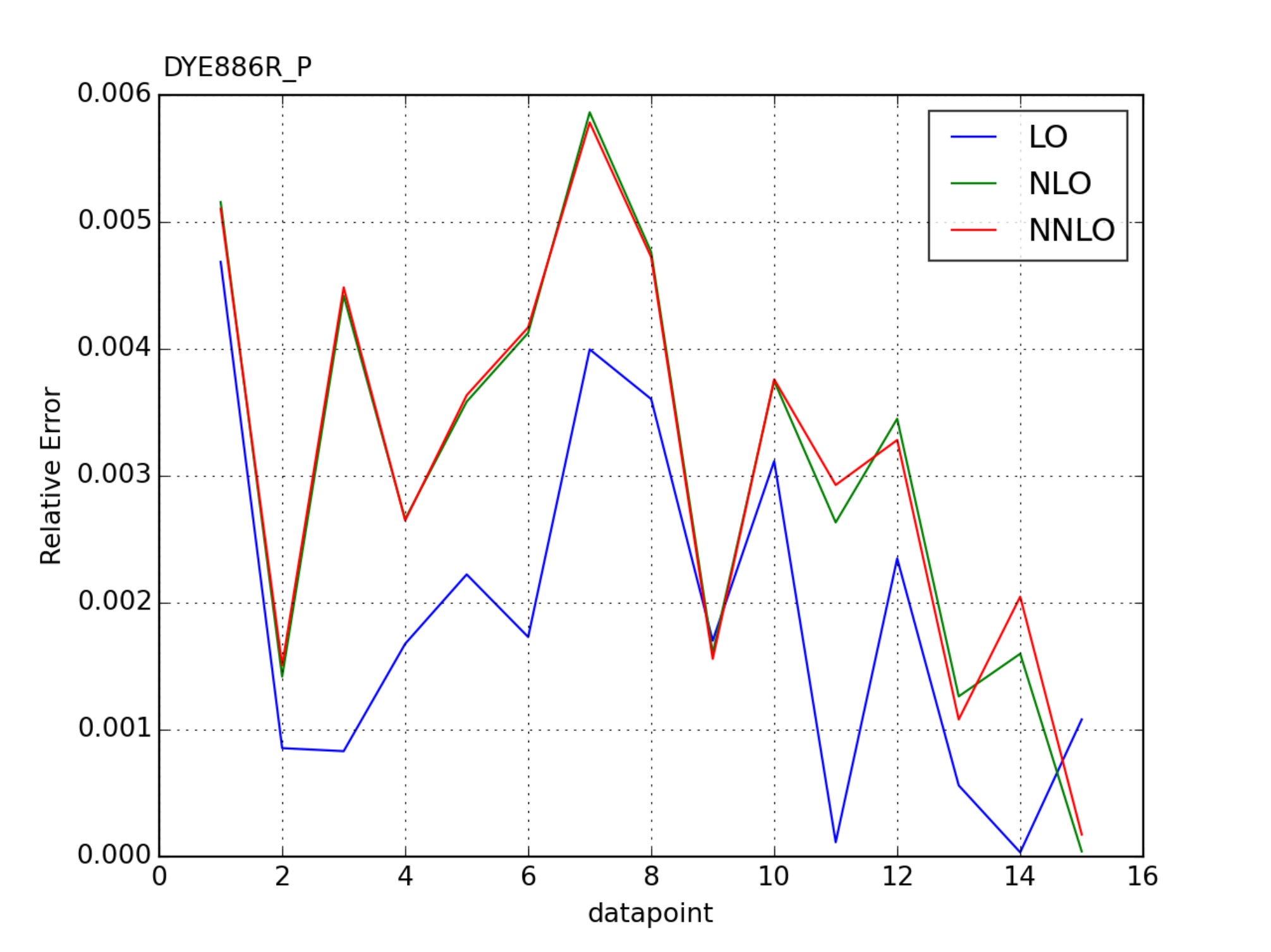}
\caption{\small Same as Fig.~\ref{fig:benchmarking} for fixed-target
  Drell-Yan cross-sections: results are shown at LO, NLO and NNLO for
  the E605 $pA$ and E866 $pp$ cross-sections datasets, as given in
  Tab.~2 of Ref.~\cite{Ball:2014uwa}.
 \label{fig:benchmarking2}} 
\end{figure}

The benchmarking of Drell-Yan cross-sections is illustrated in
Fig.~\ref{fig:benchmarking2}, where we show the relative difference
between the {\tt FKgenerator} and {\tt APFEL} calculations at LO, NLO
and NNLO for the E605 $pd$ and E866 $pp$ cross-sections, again using
NNPDF3.0 as input.  Again, differences are at most at the 2\% level
and typically much smaller. We have traced these residual differences
were to the fact that the coverage of the large-$x$ region was
sub-optimal in the {\tt FKgenerator} calculation, and is now improved
in {\tt APFEL} thanks to of a better choice of input $x$ grid.

\clearpage

\bibliography{nnpdf31}

\end{document}